\documentclass[3p, twocolumn]{elsarticle}
\usepackage[dvips]{graphicx}
\usepackage{dcolumn}
\usepackage{bm}
\usepackage{supertabular}
\usepackage{longtable}
\usepackage{color}
\usepackage{url}
\usepackage{amsmath}
\usepackage{threeparttable,caption,subcaption}
\usepackage{morefloats}[200]
\usepackage{multicol}
\usepackage[labelfont=bf]{caption}
\usepackage[resetlabels]{multibib}
\newcites{b}{appendix}
\biboptions{sort&compress}

\begin{document}
% ====================================================
% \input{title.tex}
% ===============================================
% File title.tex
% Last modified: 4 June 2017
% ===============================================
% ================================================
%              Title
% ===============================================

\title{Partial Density of States Ligand Field Theory (PDOS-LFT): Recovering
       a LFT-Like Picture and Application to Photoproperties of Ruthenium(II)
       Polypyridine Complexes}

% \author[rvt]{Denis Magero\corref{cor1}}
% \author[rvt]{Mark E. Casida}
\author{Denis Magero\corref{cor1} and Mark E.\ Casida}
\ead{magerode@gmail.com, denis.magero@univ-grenoble-alpes.fr, magerod@yahoo.com}

% \address[rvt]{
\address{
        D\'epartement de Chimie Mol\'eculaire (DCM, UMR CNRS/UGA 5250),
        Institut de Chimie Mol\'eculaire de Grenoble (ICMG, FR2607),
        Universit\'e Grenoble-Alpes,
        301 rue de la Chimie, BP 53,
        F-38041 Grenoble Cedex 9, France}   
         
%\author[focal]{Nicholas Makau}
%\author[focal]{George Amolo}
\author{George Amolo}
\address{
        Department of Physics and Space Science,
        The Technical University of Kenya,
        PO Box 52428-00200,
        Nairobi, Kenya}

\author{Nicholas Makau}

% \address[focal]{
\address{
        Computational Materials Science Group, 
        Department of Physics, 
        University of Eldoret, 
        PO Box 1125-30100
        Eldoret, Kenya}

% \author[els]{Lusweti Kituyi}
\author{Lusweti Kituyi}

% \address[els]{
\address{
        Department of Chemistry and Biochemistry, 
        University of Eldoret, 
        PO Box 1125-30100
        Eldoret, Kenya}         
   
\cortext[cor1]{Corresponding author}
        
%\address[rvt]{
%        D\'epartement de Chimie Mol\'eculaire (DCM, UMR CNRS/UJF 5250),
%        Institut de Chimie Mol\'eculaire de Grenoble (ICMG, FR2607),
%        Universit\'e Joseph Fourier (Grenoble I),
%        301 rue de la Chimie, BP 53,
%        F-38041 Grenoble Cedex 9, France}

% ================================================
%              Abstract
% ================================================
  
\begin{abstract}

% vvvvvvvvvv graphical abstract vvvvvvvvvvvvvvvv
\begin{center}
 \includegraphics[width=0.40\textwidth]{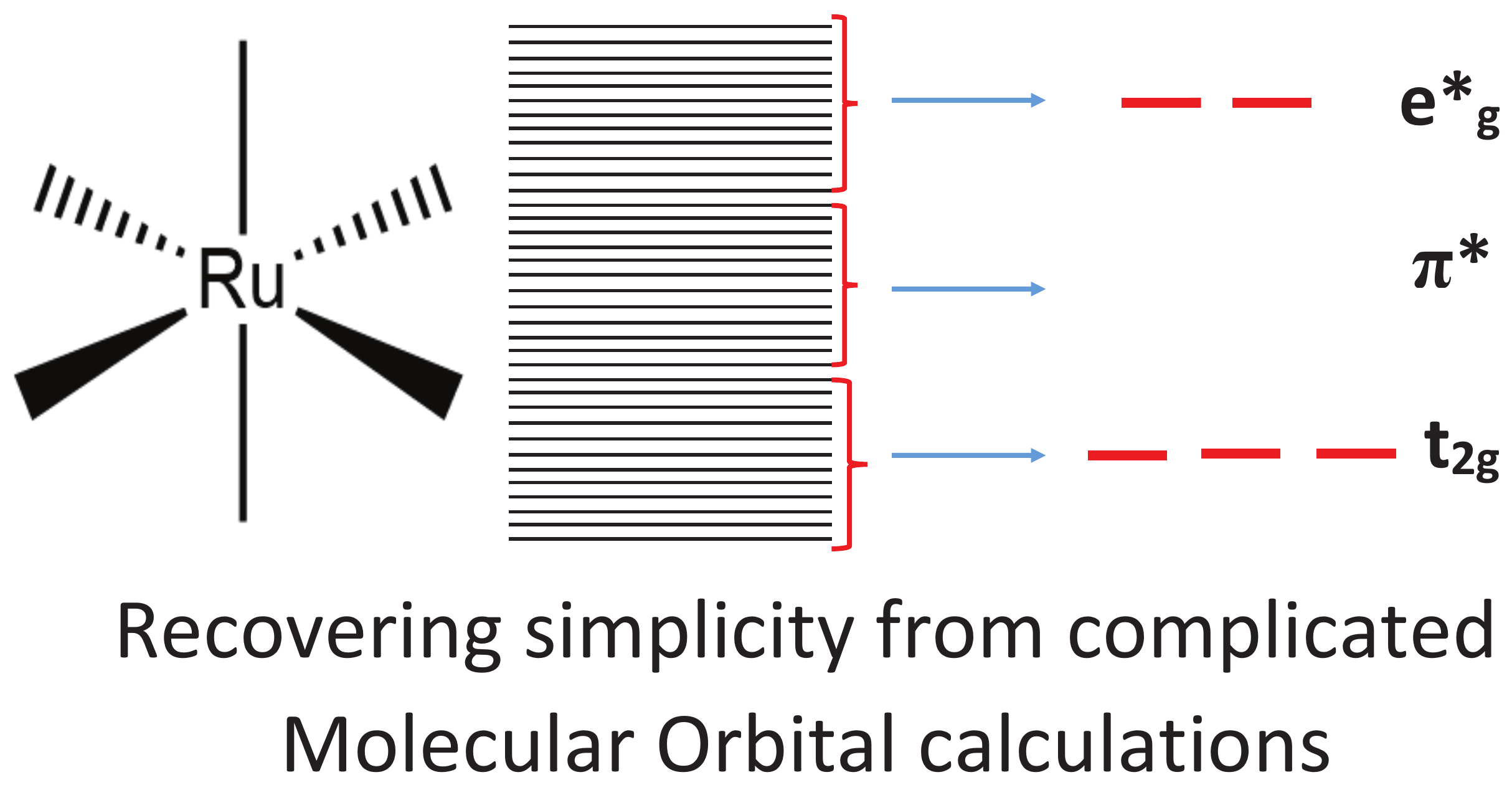} % gabstract.pdf 
\end{center}
% ^^^^^^^^^^ graphical abstract ^^^^^^^^^^^^^^^^

Gas phase density-functional theory (DFT) and time-dependent DFT (TD-DFT)
calculations are reported for a data base of 98 ruthenium(II) 
polypyridine complexes.  Comparison with X-ray crystal geometries and
with experimental absorption spectra measured in solution show an excellent
linear correlation with the results of the gas phase calculations. Comparing
this with the usual chemical understanding based upon ligand field theory (LFT)
is complicated by the large number of molecular orbitals present and especially
by the heavy mixing of the antibonding metal $e^*_g$ orbitals with ligand orbitals.
Nevertheless, we show that a deeper understanding can be obtained by a partial 
density-of-states (PDOS) analysis which allows us to extract approximate 
metal $t_{2g}$ and $e^*_g$ and ligand $\pi^*$ orbital energies in a well-defined way, 
thus providing a PDOS analogue of LFT (PDOS-LFT).  Not only do PDOS-LFT energies generate a
spectrochemical series for the ligands, but orbital energy differences provide
good estimates of TD-DFT absorption energies.  Encouraged by this success, we 
use frontier-molecular-orbital-theory-like reasoning to construct a model which
allows us in most, but not all, of the cases studied to use PDOS-LFT energies 
to provide a semiquantitative relationship between luminescence lifetimes at 
room temperature and liquid nitrogen temperature.
\end{abstract}

\begin{keyword}
polypyridine ruthenium complexes,
luminescence,
density-functional theory,
time-dependent density-functional theory,
partial density of states
\end{keyword}

\maketitle

% -----------------
% THE END 
% -----------------

% ===================================================
\section{Introduction}
\label{sec:intro}
% \input{intro.tex}
% \begin{verbatim}
% ================================
% File: intro.tex
% Last update: 11 June 2017
% ================================
% \end{verbatim}

Karl Ernst Claus had the highly unhealthy habit of tasting his chemicals, but  
(though it made him seriously sick on more than one occasion) it did help him
to discover ruthenium in 1884, in part, by following the taste from one solution
to another as he successively purified his samples \cite{L16}.
At first this newcomer to the group of platinum metals seemed to have few applications.
The situation soon changed, first with the discovery of important applications in catalysis,
and now because of the rich photochemistry of ruthenium compounds 
\cite{SCC+94,WJL+14,BJ01,N82,LKW99,HKZ03,DTL+03,MH05,MRX+07,NSF+08,SNA+08,JWW+10}. 
In particular, ruthenium complexes may be used as pigments to capture light for
drug delivery, photocatalysis, solar cells, or display applications \cite{SCC+94,WJL+14}.
Many of these applications rely on optical excitation leading to an excited state
with a long enough lifetime (typically about 1 $\mu$s) to lead to charge transfer.  
This paper concerns a relatively simple model and its use to help us to understand 
and predict the photophenomena of ruthenium complexes.  

The ideal model would be both % reasonably
quantitative and 
% reasonably 
simple.  In previous work \cite{WJL+14}, it was shown
for five complexes that gas-phase density-functional theory (DFT) and 
time-dependent DFT (TD-DFT) provide quantitative tools for predicting 
ruthenium complex crystal geometries and solution absorption spectra.  
Equally importantly, the results were reduced to a 
ligand-field theory (LFT) \cite{FH00} like
framework that can be easily related back to the usual interpretive tool used
by transition-metal-complex chemists.  This was done via the use of the concepts of the 
density-of-states (DOS) and partial DOS (PDOS) of DFT molecular orbitals (MOs)
to identify the energy range of the antibonding ruthenium $e^*_g$ orbitals whose
mixing with ligand orbitals otherwise makes them notoriously difficult to locate,
unlike the much easier case of the nonbonding ruthenium $t_{2g}$ orbitals.  
Luminescence indices were also suggested based upon this PDOS-LFT to try to 
say something about relative luminescence lifetimes of different ruthenium
complexes.  However a theory based upon only five compounds can hardly be taken
as proven. (Another approach to extracting LFT from DFT is ligand field DFT \cite{RUG+14}.
PDOS-LFT offers a complementary but simpler approach.)

Here we extend the earlier study to the large number of complexes
whose photoproperties are tabulated in the excellent, if dated, review article of
Balzini, Barigelletti, Capagna, Belser, and Von Zelewsky \cite{JBB+88}.  
Our understanding is deepened by confronting calculations for on 
the order of 100 pseudo-octahedral ruthenium complexes with experimental data.  
In particular, we are able to obtain a roughly linear correlation (albeit 
with some exceptions) between a function of PDOS-LFT energies and an average 
activation energy describing nonradiative relaxation of the luminescent excited 
state.  

% --------------------------------------------------------------------
\begin{figure}
\includegraphics[width=0.5\textwidth]{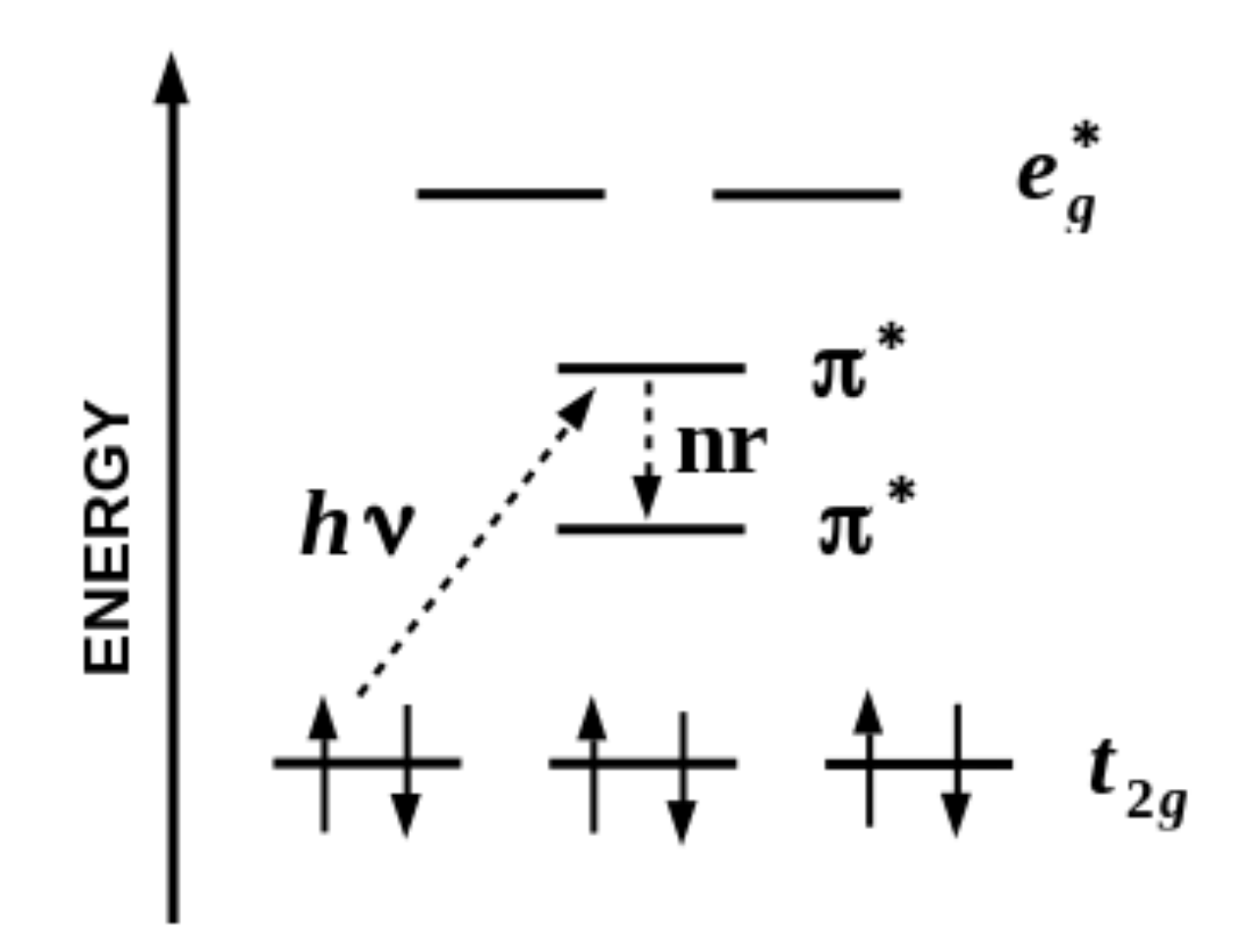}  % LFT2.pdf
\caption{\label{fig:LFT} Generic ligand field theory diagram for octahedral 
ruthenium(II) polypyridyl complexes. Note that ligand $\pi^*$ orbital energy levels 
intercalate between ruthenium $t_{2g}$ and $e^*_g$ LFT states.  The number of
$\pi^*$ levels varies depending upon the ligands (only two are shown here).
Photon absorption leads to a $t_{2g} \rightarrow \pi^*$ transition from
the ground state (GS) to a metal-to-ligand charge transfer (MLCT) state.
As $d \rightarrow d$ transitions are symmetry forbidden by the $\Delta l = \pm 1$ selection
rule in the atom, the creation of a MLCT excited state is favored over the 
formation of a metal-centered (MC) state.  Kasha's rule \cite{K50} tells us 
that nonradiative (``nr'' in the figure) transitions will take place until 
the dominant luminescence is from the lowest $\pi^*$ orbital back 
to the $t_{2g}$ orbital to reform the GS.
}
\end{figure}
% --------------------------------------------------------------------
The problem of ruthenium complex luminescence lifetimes has been well studied
in the literature 
\cite{C75,N82,JBB+88,SCC+94,LKW99,HKZ03,DTL+03,MH05,BC07a,BC07b,MRX+07,NSF+08,SNA+08,JWW+10}.
Even so, no universal detailed theory of luminescence lifetimes has emerged because
of a diversity of ligand-dependent de-excitation mechanisms.  Nevertheless there
is a commonly accepted ``generic mechanism'' \cite{JBB+88} based upon the pseudo-octahedral symmetry
LFT diagram shown in Fig.~\ref{fig:LFT}.  An initial singlet metal-ligand charge
transfer state $^1$MLCT$(t_{2g}^5 {\pi^{*1}} {e^{*0}_g})$ is formed either directly by exciting 
the ground state [$^1$GS$(t_{2g}^6 \pi^{*0} e^{*0}_g)$] or by exciting another state and 
subsequent radiationless relaxation (see the caption of Fig.~\ref{fig:LFT} for the definition of 
GS, MLCT, and MC):
% --------------------------------------------------------------------
\begin{figure}%[!htbp]
\includegraphics[width=0.4\textwidth]{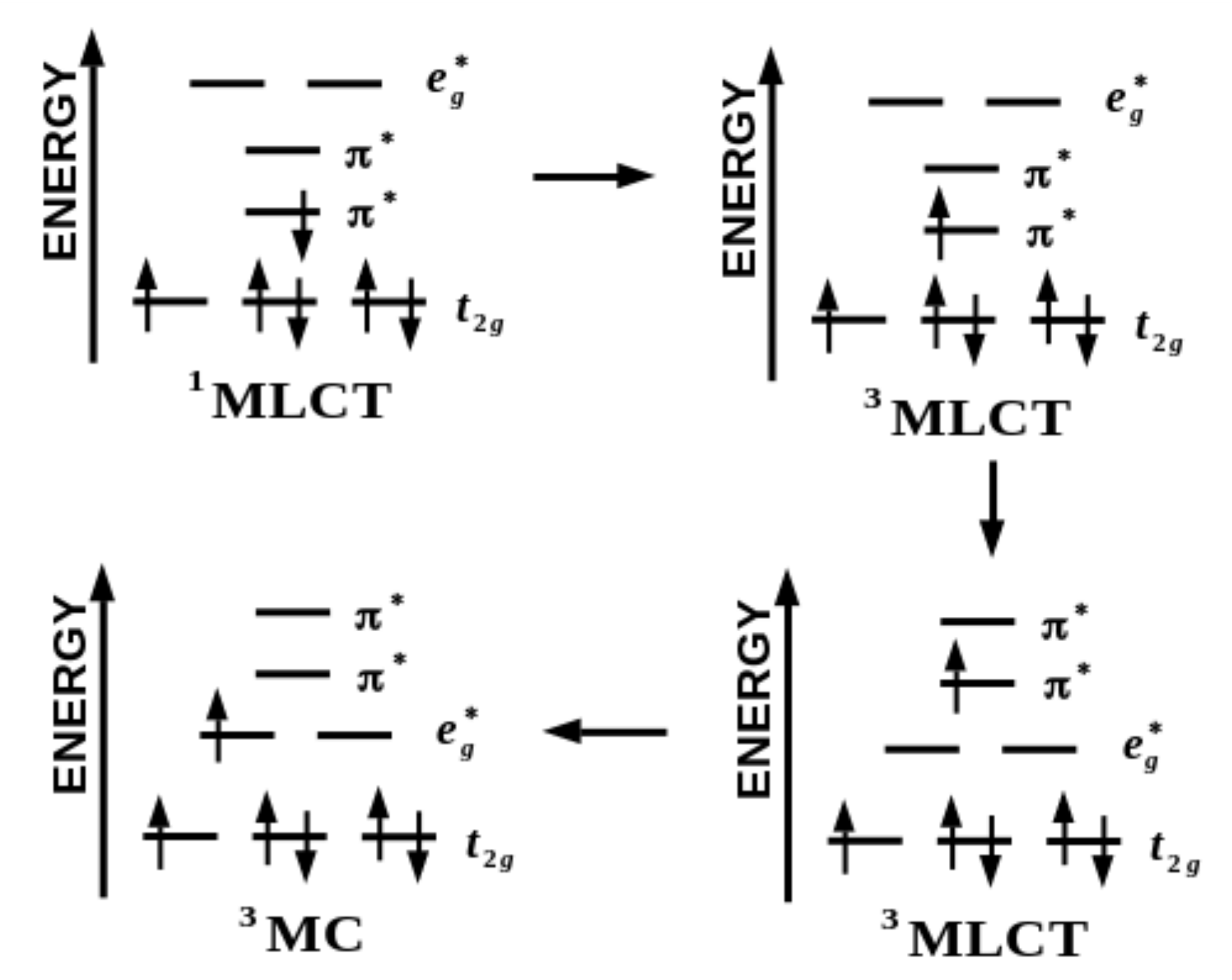} % MO2.pdf
\caption{\label{fig:MO} Orbital diagrams for the electronic GS and the most relevant 
excited states for pseudo-octahedral ruthenium(II) polypyridyl complexes. 
}
\end{figure}
% --------------------------------------------------------------------
\begin{equation}
   ^1\mbox{GS}(t_{2g}^6 \pi^{*0} e^{*0}_g) \rightarrow  
   {^1\mbox{MLCT}}(t_{2g}^5 \pi^{*1} e^{*0}_g)
   \, .
  \label{eq:intro.1}
\end{equation}
Ruthenium complex spin-orbit coupling then leads to rapid intersystem
crossing to form the corresponding triplet $^3$MLCT$(t_{2g}^5 \pi^{*1} e_g^{*0})$,
\begin{equation}
   ^1\mbox{MLCT}(t_{2g}^5 \pi^{*1} e^{*0}_g) \rightarrow {^3\mbox{MLCT}}(t_{2g}^5 \pi^{*1} e_g^{*0})
   \, .
  \label{eq:intro.2}
\end{equation}
This $^3$MLCT can phosphoresce back to the $^1$GS or it can go over an 
excited-state transition state barrier to a triplet metal center state
$^3$MC$(t_{2g}^5 e_g^{*1} \pi^{*0})$,
\begin{equation}
   ^3\mbox{MLCT}(t_{2g}^5 \pi^{*1} e_g^{*0}) 
   \begin{array}{c} k_a \\ \rightarrow \\ \leftarrow \\ k_b \end{array}
   {^3\mbox{MC}}(t_{2g}^5 e_g^{*1} \pi^{*0}) \, .
   \label{eq:intro.3}
\end{equation}
Notice how the MC $e^{*}_{g}$ MO has now presumably become lower than the 
ligand-centered (LC) $\pi^{*}$ MO (Fig.~\ref{fig:MO}).  This is possible
because of geometric relaxation as illustrated in the state diagram shown
in Fig.~\ref{fig:PES}.  The resultant state can then go through a photochemical
funnel with intersystem crossing to return to the groundstate,
\begin{equation}
  ^3\mbox{MC}(t_{2g}^5 e_g^{*1} \pi^{*0}) 
  \begin{array}{c} k_c \\ \rightarrow  \end{array}
  {^1\mbox{GS}}(t_{2g}^6 e_g^{*0} \pi^{*0}) \, .
  \label{eq:intro.4}
\end{equation}
This is presumed to involve ligands coming partially or completely off and/or
being replaced by solvent molecules.  The rate constants $k_a$, $k_b$,
and $k_c$ are the same as those defined in Ref.~\cite{BJB+87}.
Figure~\ref{fig:MO} provides a summary in the form of orbitals and Figure~\ref{fig:PES} 
in the form of potential energy curves for different states.
% --------------------------------------------------------------------
\begin{figure}
\includegraphics[width=0.5\textwidth]{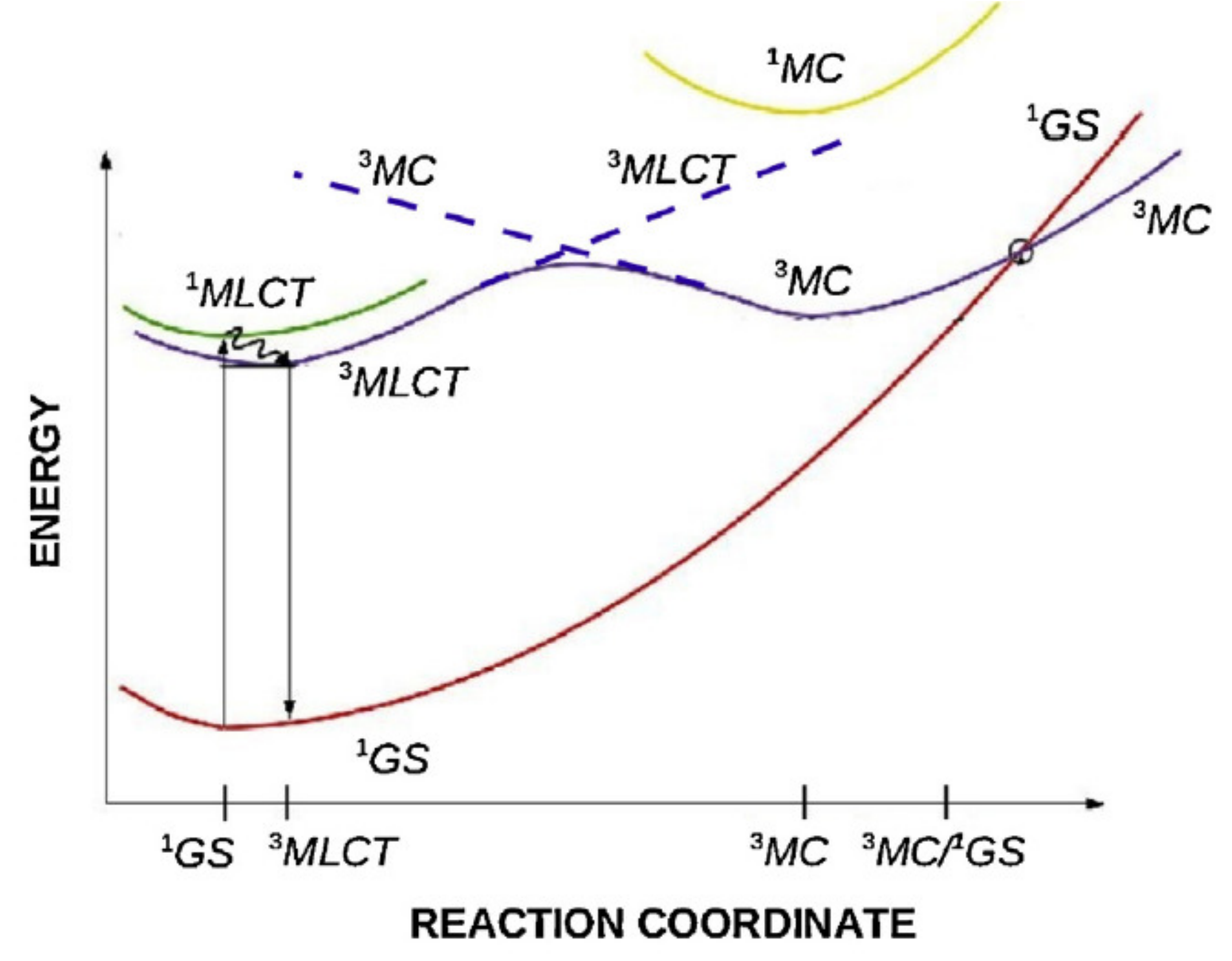}  % PES2.pdf
\caption{\label{fig:PES} The diagram shows the principle potential energy 
curves in our model. The abscissa corresponds to a reaction pathway involving
partial removal of a ligand while the ordinate represents
the state energy.  The dashed lines indicate 
diabatic states whose avoided crossing leads to the energetic barrier on the 
adiabatic surface between the $^3$MLCT and $^3$MC minima.  Figure from Ref.~\cite{WJL+14}.
}
\end{figure}
% --------------------------------------------------------------------
A key assumption is that the main luminescence quenching at room temperature is due to 
the barrier crossing [Eq.~(\ref{eq:intro.3})] followed by a rapid return to the ground state 
[Eq.~(\ref{eq:intro.4})].  For this reason, we will focus on this barrier in seeking a PDOS-LFT 
explanation for relative luminescence lifetimes, but let us admit in advance that our answer,
though general and useful, is unlikely
to be universal.  For one thing, a mixture of different types of ligands or of different
types of metal-ligand bonds, means that there is likely to be more than one path for
luminescence quenching.  Still other mechanisms might come in involving, say, unforeseen
intermediate dark states.  And, as we shall see in Sec.~\ref{sec:results}, the barrier
crossing [Eq.~(\ref{eq:intro.3})] is unlikely to be the only influence on the luminescence
lifetime at room temperature.  Nevertheless we shall be happy with a semi-quantitative 
PDOS-LFT-based theory of luminescence which works most of the time.

The rest of the paper is organized as follows: The next section discusses our choice
of molecules, theoretical methods, and computational details.  This is followed by
a results section in which evidence is first given for the ability of DFT to give 
reasonably good geometries and of TD-DFT to give reasonably good absorption spectra.
Secondly PDOS-LFT energy levels are discussed and shown to be useful for predicting
photoproperties. And thirdly a model is presented which allows us to say something
about luminescence lifetimes from PDOS-LFT energy levels. Section~\ref{sec:conclude}
concludes.  PDOS and TD-DFT spectra are presented in a separate document as supporting
information.

% ===================================================
\section{Data Base, Theoretical Method, and Computational Details}
\label{sec:details}
% \input{details.tex}
% \begin{verbatim}
% ================================
% File: details.tex
% Last update: 4 June 2017
% ================================
% \end{verbatim}

% ---------------------------------------
\subsection{Data Base}
% ---------------------------------------

% \input{./tables/names.tex}
% =======================================
% file: names.tex
% last updated: 21 March 2017
% =======================================

\begin{table}
\caption{Numbering of the compounds investigated in this paper.  With a few
         exceptions (listed but unnumbered compounds), these are the 
         mononuclear complexes with 77 K experimental luminescence lifetimes
         taken in their order of occurance from Table 1 of Ref.~\cite{JBB+88}.
         An asterisk has been added if the original table also contained
         some information about room temperature lifetimes.
         \label{tab:JBB88names1}}
\begin{tabular}{cc}
\hline \hline
number       & name                                                 \\
\hline
({\bf 1})*   & [Ru(bpy)(CN)$_4$]$^{2-}$                              \\  % p 146 
({\bf 2})    & [Ru(bpy)$_2$Cl$_2$]                                   \\  % p 146
({\bf 3})*   & [Ru(bpy)$_2$(CN)$_2$]                                 \\  % p 146-147
({\bf 4})*   & [Ru(bpy)$_2$(en)]$^{2+}$                              \\  % p 147 
({\bf 5})    & [Ru(bpy)$_2$(ox)]                                     \\ % p 147
({\bf 6})*   & [Ru(bpy)$_3$]$^{2+}$                                  \\ % p 147-150
({\bf 7})*   & [Ru(bpy)$_2$(4-n-bpy)]$^{2+}$                         \\ % p 150
({\bf 8})*   & [Ru(bpy)$_2$(3,3'-dm-bpy)]$^{2+}$                     \\ % p 151
({\bf 9})*   & [Ru(bpy)$_2$(4,4'-dm-bpy)]$^{2+}$                     \\ % p 151
({\bf 10})*  & [Ru(bpy)$_2$(4,4'-dCl-bpy)]$^{2+}$                    \\ % p 151
({\bf 11})   & [Ru(bpy)$_2$(4,4'-dn-bpy)]$^{2+}$                     \\ % p 152
({\bf 12})*  & [Ru(bpy)$_2$(4,4'-dph-bpy)]$^{2+}$                    \\ % p 152
({\bf 13})*  & [Ru(bpy)$_2$(4,4'-DTB-bpy)]$^{2+}$                    \\ % p 153
({\bf 14})*  & {\em cis}-[Ru(bpy)$_2$(m-4,4'-bpy)$_2$]$^{4+}$  \\ % p 155
({\bf 15})*  & [Ru(bpy)$_2$(bpz)]$^{2+}$                             \\ % p 155
             & [Ru(bpy)$_2$(h-phen)]$^{2+}$                          \\ % p 156
({\bf 16})*  & [Ru(bpy)$_2$(phen)]$^{2+}$                            \\ % p 156
             & [Ru(bpy)$_2$(bpym)]$^{2+}$                            \\ % p 155
({\bf 17})   & [Ru(bpy)$_2$(4,7-dm-phen)$^{2+}$                      \\ % p 156
({\bf 18})*  & [Ru(bpy)$_2$(4,7-Ph$_2$-phen)]$^{2+}$                 \\ % p 156
({\bf 19})*  & [Ru(bpy)$_2$(4,7-dhy-phen)]$^{2+}$                    \\ % p 156-157
({\bf 20})   & [Ru(bpy)$_2$(5,6-dm-phen)]$^{2+}$                     \\ % p 157 
({\bf 21})   & [Ru(bpy)$_2$(DIAF)]$^{2+}$                            \\ % p 157
({\bf 22})*  & [Ru(bpy)$_2$(DIAFO)]$^{2+}$                           \\ % p 157
({\bf 23})*  & [Ru(bpy)$_2$(taphen)]$^{2+}$                          \\ % p 157
({\bf 24})   & {\em cis}-[Ru(bpy)$_2$(py)$_2$]$^{2+}$                \\ % p 158
({\bf 25})   & {\em trans}-[Ru(bpy)$_2$(py)$_2$]$^{2+}$              \\ % p 158
({\bf 26})   & [Ru(bpy)$_2$(pic)$_2$]$^{2+}$                         \\ % p 159
({\bf 27})   & [Ru(bpy)$_2$(DPM)]$^{2+}$                             \\ % p 161
({\bf 28})   & [Ru(bpy)$_2$(DPE)]$^{2+}$                             \\ % p 161
({\bf 29})*  & [Ru(bpy)$_2$(PimH)]$^{2+}$                            \\ % p 161
({\bf 30})*  & [Ru(bpy)$_2$(PBzimH)]$^{2+}$                          \\ % p 161
({\bf 31})*  & [Ru(bpy)$_2$(biimH$_2$)]$^{2+}$                       \\ % p 161
({\bf 32})*  & [Ru(bpy)$_2$(BiBzimH$_2$)]$^{2+}$                     \\ % p 161
({\bf 33})   & [Ru(bpy)$_2$(NPP)]$^+$                                \\ % p 161
({\bf 34})   & [Ru(bpy)$_2$(piq)]$^{2+}$                             \\ % p 161
({\bf 35})   & [Ru(bpy)$_2$(hpiq)]$^{2+}$                            \\ % p 161
\hline \hline                                                 
\end{tabular}
\end{table}

% ----------------------------------------------------------------

\begin{table}
\caption{Numbering of the compounds investigated in this paper.  With a few
         exceptions (listed but unnumbered compounds), these are the 
         mononuclear complexes with 77 K experimental luminescence lifetimes
         taken in their order of occurance from Table 1 of Ref.~\cite{JBB+88}.
         An asterisk has been added if the original table also contained
         some information about room temperature lifetimes.
         \label{tab:JBB88names2}}
\begin{tabular}{cc}
\hline \hline
number    & name   \\
\hline
({\bf 36})   & [Ru(bpy)$_2$(pq)]$^{2+}$                              \\ % p 161
({\bf 37})*  & [Ru(bpy)$_2$(DMCH)]$^{2+}$                            \\ % p 163     
({\bf 38})   & [Ru(bpy)$_2$(OMCH)]$^{2+}$                            \\ % p 163     
({\bf 39})*  & [Ru(bpy)$_2$(biq)]$^{2+}$                             \\ % p 163     
({\bf 40})*  & [Ru(bpy)$_2$(i-biq)]$^{2+}$                           \\ % p 163     
({\bf 41})*  & [Ru(bpy)$_2$(BL4)]$^{2+}$                             \\ % p 163     
({\bf 42})*  & [Ru(bpy)$_2$(BL5)]$^{2+}$                             \\ % p 163     
({\bf 43})*  & [Ru(bpy)$_2$(BL6)]$^{2+}$                             \\ % p 163     
({\bf 44})*  & [Ru(bpy)$_2$(BL7)]$^{2+}$                             \\ % p 163     
({\bf 45})*  & [Ru(bpy)(3,3'-dm-bpy)$_2$]$^{2+}$                     \\ % p 163     
({\bf 46})*  & [Ru(bpy)(4,4'-DTB-bpy)$_2$]$^{2+}$                    \\ % p 164        
({\bf 47})   & [Ru(bpy)(h-phen)$_2$]$^{2+}$                          \\ % p 165    
({\bf 48})*  & [Ru(bpy)(phen)$_2$]$^{2+}$                            \\ % p 165     
({\bf 49})   & {\em cis}-[Ru(bpy)(phen)(py)$_2$]$^{2+}$              \\ % p 165         
({\bf 50})   & {\em trans}-[Ru(bpy)(phen)(py)$_2$]$^{2+}$            \\ % p 165    
({\bf 51})   & [Ru(bpy)(DIAFO)$_2$]$^{2+}$                           \\ % p 165     
({\bf 52})*  & [Ru(bpy)(taphen)$_2$]$^{2+}$                          \\ % p 165    
({\bf 53})   & [Ru(bpy)(py)$_2$(en)]$^{2+}$                          \\ % p 165     
({\bf 54})   & [Ru(bpy)(py)$_3$Cl]$^+$                               \\ % p 166     
({\bf 55})   & [Ru(bpy)(py)$_4$]$^{2+}$                              \\ % p 166        
({\bf 56})   & [Ru(bpy)(py)$_2$(PMA)]$^{2+}$                         \\ % p 166     
({\bf 57})   & [Ru(bpy)(py)$_2$(2-AEP)]$^{2+}$                       \\ % p 166         
({\bf 58})   & [Ru(bpy)(PMA)$_2$]$^{2+}$                             \\ % p 166     
({\bf 59})   & [Ru(bpy)(pq)$_2$]$^{2+}$                              \\ % p 167  
({\bf 60})*  & [Ru(bpy)(DMCH)$_2$]$^{2+}$                            \\ % p 167     
({\bf 61})*  & [Ru(bpy)(biq)$_2$]$^{2+}$                             \\ % p 167        
({\bf 62})*  & [Ru(bpy)(i-biq)$_2$]$^{2+}$                           \\ % p 167        
({\bf 63})*  & [Ru(bpy)(trpy)Cl]$^+$                                 \\ % p 167     
({\bf 64})*  & [Ru(bpy)(trpy)(CN)]$^+$                               \\ % p 167     
({\bf 65})*  & [Ru(4-n-bpy)$_3$]$^{2+}$                              \\ % p 169        
({\bf 66})   & [Ru(6-m-bpy)$_3$]$^{2+}$                              \\ % p 169       
({\bf 67})*  & [Ru(3,3'-dm-bpy)$_3$]$^{2+}$                          \\ % p 169     
({\bf 68})   & [Ru(3,3'-dm-bpy)$_2$(phen)]$^{2+}$                    \\ % p 169     
({\bf 69})   & [Ru(3,3'-dm-bpy)(phen)$_2$]$^{2+}$                    \\ % p 169     
({\bf 70})*  & [Ru(4,4'-dm-bpy)$_3$]$^{2+}$                          \\ % p 170     
\hline \hline                                                 
\end{tabular}
\end{table}

% ----------------------------------------------------------------

\begin{table}
\caption{Numbering of the compounds investigated in this paper.  With a few
         exceptions (listed but unnumbered compounds), these are the 
         mononuclear complexes with 77 K experimental luminescence lifetimes
         taken in their order of occurance from Table 1 of Ref.~\cite{JBB+88}.
         An asterisk has been added if the original table also contained
         some information about room temperature lifetimes.
         \label{tab:JBB88names3}}
\begin{tabular}{cc}
\hline \hline
number    & name   \\
\hline
({\bf 71})*  & [Ru(4,4'-dm-bpy)$_2$(4,7-dhy-phen)]$^{2+}$            \\ % p 171     
({\bf 72})*  & [Ru(4,4'-dCl-bpy)$_3$]$^{2+}$                   \\  % p 171 
({\bf 73})*  & [Ru(4,4'-dph-bpy)$_3$]$^{2+}$                   \\  % p 172 
({\bf 74})*  & [Ru(4,4'-DTB-bpy)$_3$]$^{2+}$                   \\  % p 173
({\bf 75})   & [Ru(6,6'-dm-bpy)$_3$]$^{2+}$                    \\  % p 174
({\bf 76})   & [Ru(h-phen)$_3$]$^{2+}$                         \\  % p 176
%             & [Ru(h-phen)$_2$(CN)$_2$]$^{2+}$                 \\  % p 177
({\bf 77})*  & [Ru(phen)$_3$]$^{2+}$                           \\  % p 177-179
({\bf 78})*  & [Ru(phen)$_2$(4,7-dhy-phen)]$^{2+}$             \\  % p 180
({\bf 79})   & [Ru(phen)$_2$(pq)]$^{2+}$                       \\  % p 180
({\bf 80})   & [Ru(phen)$_2$(DMCH)]$^{2+}$                     \\  % p 180
({\bf 81})   & [Ru(phen)$_2$(biq)]$^{2+}$                      \\  % p 180
({\bf 82})   & [Ru(phen)(pq)$_2$]$^{2+}$                       \\  % p 181
({\bf 83})   & [Ru(phen)(biq)$_2$]$^{2+}$                      \\  % p 181 
({\bf 84})   & [Ru(2-m-phen)$_3$]$^{2+}$                       \\  % p 181
({\bf 85})   & [Ru(2,9-dm-phen)$_3$]$^{2+}$                    \\  % p 183
({\bf 86})*  & [Ru(4,7-Ph$_2$-phen)$_3$]$^{2+}$                \\  % p 184
({\bf 87})*  & [Ru(4,7-dhy-phen)(tm1-phen)$_2$]$^{2+}$         \\  % p 185
%             & [Ru(taphen)$_3$]$^{2+}$                         \\  % p 186
({\bf 88})   & [Ru(DPA)$_3$]$^-$                               \\  % p 187
%             & $\{$[Ru(DPA)$_3$]$\cdot$H$_2$O$\}^-$            \\  % p 187
({\bf 89})   & [Ru(DPA)(DPAH)$_2$]$^{+}$                       \\  % p 187
({\bf 90})   & [Ru(DPAH)$_3$]$^{2+}$                           \\  % p 187
({\bf 91})   & [Ru(Azpy)$_3$]$^{2+}$                           \\  % p 187
({\bf 92})   & [Ru(NA)$_3$]$^{2+}$                             \\  % p 187
%              & [Ru(bt)$_3$]$^{2+}$                             \\  % p 187
({\bf 93})   & [Ru(hpiq)$_3$]$^{2+}$                           \\  % p 188
({\bf 94})   & [Ru(pq)$_3$]$^{2+}$                             \\  % p 188
({\bf 95})   & [Ru(pq)$_2$(biq)]$^{2+}$                        \\  % p 188
({\bf 96})   & [Ru(pq)(biq)$_2$]$^{2+}$                        \\  % p 188
({\bf 97})   & [Ru(pynapy)$_3$]$^{2+}$                         \\  % p 188
({\bf 98})*  & [Ru(DMCH)$_2$Cl$_2$]                            \\  % p 188
({\bf 99})*  & [Ru(DMCH)$_2$(CN)$_2$]                          \\  % p 188
({\bf 100})  & [Ru(DMCH)$_3$]$^{2+}$                           \\  % p 189
({\bf 101})  & [Ru(dinapy)$_3$]$^{2+}$                         \\  % p 189
({\bf 102})  & [Ru(biq)$_2$Cl$_2$]                             \\  % p 189
({\bf 103})* & [Ru(biq)$_2$(CN)$_2$]                           \\  % p 189
({\bf 104})  & [Ru(biq)$_3$]$^{2+}$                            \\  % p 189
({\bf 105})  & [Ru(i-biq)$_2$Cl$_2$]                           \\  % p 189
\hline \hline                                                 
\end{tabular}
\end{table}

% ----------------------------------------------------------------

\begin{table}
\caption{Numbering of the compounds investigated in this paper.  With a few
         exceptions (listed but unnumbered compounds), these are the 
         mononuclear complexes with 77 K experimental luminescence lifetimes
         taken in their order of occurance from Table 1 of Ref.~\cite{JBB+88}.
         An asterisk has been added if the original table also contained
         some information about room temperature lifetimes.
         \label{tab:JBB88names4}}
\begin{tabular}{cc}
\hline \hline
number    & name   \\
\hline
({\bf 106})* & [Ru(i-biq)$_2$(CN)$_2$]                         \\  % p 189 
({\bf 107})* & [Ru(i-biq)$_3$]$^{2+}$                          \\  % p 189
({\bf 108})* & [Ru(trpy)$_2$]$^{2+}$                           \\  % p 190
({\bf 109})  & [Ru(tro)$_2$]$^{2+}$                            \\  % p 190
({\bf 110})  & [Ru(tsite)$_2$]$^{2+}$                          \\  % p 190
({\bf 111})  & [Ru(dqp)$_2$]$^{2+}$                            \\  % p 190
\hline \hline                                                 
\end{tabular}
\end{table}

%%%%%
% EOF
%%%%%
Our theoretical calculations are based upon an old but unusually extensive list
of the photoproperties of ruthenium complexes.  In particular, our calculations 
are based upon the photoproperties of about 300 mononuclear 
ruthenium complexes reported in Table 1 of the 1988 review article 
of Juris, Balzini, Barigelletti, Capagna, Belser, and Von~Zelewsky \cite{JBB+88}.  Of 
these, the 111 complexes shown in Tables~\ref{tab:JBB88names1}, \ref{tab:JBB88names2},
\ref{tab:JBB88names3}, and \ref{tab:JBB88names4} have luminescence data either at
room temperature (RT) or at the boiling point of liquid nitrogen (77 K).  
For convenience we have numbered them in the same order as they appear in Table 1 
of review article~\cite{JBB+88}.  Note that this luminescence data was not necessarily measured 
in the same solvent for different compounds, or even for any given compound, and 
that the reported precision of the measurements vary.  The ligand abbreviations are given in 
Appendix~\ref{sec:ligands}.  With a few exceptions (CN$^-$, Cl$^-$, ox, NPP, NA, bt, en),
the ligands are pyridine and polypyridine N-type ligands, many of are found in
common lists of the well-known spectrochemical series governing the ligand field 
splitting $\Delta$,
\begin{equation}
  \Delta : \mbox{ Cl$^-$ } < \mbox{ py } < \mbox{ en } < \mbox{ bpy } < \mbox{ phen } <  \mbox{ CN$^-$ }
  \, .
  \label{eq:details.1}
\end{equation}
Calculations have been carried out on 98 of these 111 complexes, with 14 left
untreated either because of lack of a good initial guess for the complex structure,
convergence difficulties, or simple lack of time.  Not every calculation is necessarily
useful as some needed to be discarded for theoretical reasons (an unbound $e^*_g$ orbital)
and not every property could be calculated for every compound.
% , which still leaves
% a complete set of calculations for around 80 compounds.  
Furthermore we were not 
able to find comparison data for every property of every complex but we think that the 
extensiveness of our calculations and of the comparison with experiment for a broad 
range of complexes and properties should be highly useful.

% ---------------------------------------------
\subsection{Computational Methods and Details}
% ---------------------------------------------

The calculations reported in this paper are very similar to those reported in Ref.~\cite{WJL+14}.
Version B.05 of the {\sc Gaussian 03} \cite{g03} quantum chemistry package was used in Ref.~\cite{WJL+14}.  
Here we use version D.01 of {\sc Gaussian 09} \cite{g09}.  Density-functional theory (DFT) and 
time-dependent (TD-)DFT calculations were carried out using the same B3LYP 
functional.  This is a three-parameter hybrid functional using 
Hartree-Fock (HF) exchange,
the usual analytical form of the local density approximation (LDAx) for
exchange \cite{KS65}, Becke's 1988 generalized gradient approximation (GGA) exchange 
B88x \cite{B88}, the Vosko-Wilk-Nusair parameterization of the LDA
correlation (LDAc) \cite{VWN80}, and Lee, Yang, and Parr's GGA for correlation (LYP88c) 
\cite{LYP88},
\begin{eqnarray}
  E_{xc}^{\text{B3LYP}} &= &(1-a_0) E_x^{\text{LDA}} \nonumber \\
      & + & a_0 E_x^{\text{HF}} + a_x E_x^{\text{B88x}} \nonumber \\
   & + &  a_c E_c^{\text{LYP88c}} + (1-a_c) E_c^{\mbox{VWN80c}} \, , \nonumber \\
  \label{eq:details.1.5}
\end{eqnarray}
where $a_0 = 0.20$, $a_x = 0.72$, and $a_c = 0.81$ are taken from Becke's
B3P functional \cite{B93}.

These calculations require us to choose a Gaussian-type basis set.  As in Ref.~\cite{WJL+14},
we used the double-zeta quality LANL2DZ basis set for ruthenium along with the corresponding
effective core potential (ECP) \cite{HW85, HW85a}.  All-electron 6-31G and 6-31G(d) basis 
sets \cite{DHP71,HDP72,HP73,HP74,FPH+82,RRP+01,PA91} were used for all the elements in the 
first three periods of the periodic table.  Note that Ref.~\cite{WJL+14} only used the
smaller 6-31G basis set, while the present work is able to verify basis set convergence by
also reporting results with the larger 6-31G(d) compounds.  However, due to the very large
number of calculations carried out and the size of the molecules, calculations with still
larger basis sets were judged to fall outside of the scope of the present study.
Unless otherwise mentioned, extensive use of program defaults was used for many 
of the computational parameters.  Neither explicit nor dielectric cavity models were 
used in our calculations, so that all calculations reported in this article
are technically for gas-phase molecules.

% ----------------------------------------------------------------
\begin{table}
\caption{List of 39 compounds with crystal structures. An asterisk indicates
         that the CCDC structure was modified.
         \label{tab:geom3}}
\begin{tabular}{ccc}
\hline \hline
Number      & CCDC\cite{CCDC} & Citation \\
            & reference code  &          \\
\hline
%MEC ({\bf 1})   &               & Ref.~\cite{MSK+03} \\
({\bf 2})   & AHEHIF        & Ref.~\cite{PYD+07,PYD+08} \\
({\bf 3})   & LESLEB        & Ref.~\cite{OEH06}  \\
({\bf 4})   & SAXCIE        & Ref.~\cite{F05}     \\
({\bf 5})   & YAQJOP        & Ref.~\cite{SBL+99} \\
({\bf 6})   & BPYRUF        & Ref.~\cite{RJ79}    \\
({\bf 7})   & DIXVEL        & Ref.~\cite{RSC+85}  \\
% ({\bf 8})   &             & Q1.8           &           \\
({\bf 9})   & JUQHEI        & Ref.~\cite{HIE+99a,HIE+99b}  \\
({\bf 10})  & BAQYEY        & Ref.~\cite{HGA+03a,HGA+03b}  \\
% ({\bf 11})  & Q1.11       & Q1.11          &           \\
% ({\bf 12})  & Q1.12       & Q1.12          &           \\
% ({\bf 13})  & Q1.13       & Q1.13          &           \\
({\bf 14})* & OBITIC01      & Ref.~\cite{DGLB01,DGLB02} \\
% ({\bf 15})  & Q1.14       & Q1.14          &           \\
({\bf 16})  & TIXFOV        & Ref.~\cite{YCZJ95} \\
% ({\bf 16})  & INIYOT        & Ref.~\cite{SF04a,SF04b} \\
% ({\bf 16})  & UGOZEX        & Ref.~\cite{YYS+14,YYS+15}  \\
({\bf 17})  & XOFQEO        & Ref.~\cite{YLZ02a,YLZ02b}  \\
% ({\bf 18})  & Q1.17       & Q1.17          &           \\
% ({\bf 19})  &             &                & Ref.~\cite{KBD+04}   \\
({\bf 20})  & IBAGAU        & Ref.~\cite{KBD+04a,KBD+04b} \\
({\bf 21})  & COMVIJ        & Ref.~\cite{HFC84} \\
({\bf 22})  & YAGJAR10      & Ref.~\cite{WJWR95} \\
% ({\bf 23})  & Q2.3        & Q2.3           &           \\
({\bf 24})  & GEBHEA        & Ref.~\cite{HST+88}   \\
({\bf 25})  & QUBRIO        & Ref.~\cite{KVZA01a,KVZA01b}          \\
({\bf 26})  & MESWUC        & Ref.~\cite{VMU+00,VMU+02}   \\
% ({\bf 27})  &             &                &           \\
% ({\bf 28})  &             &                &           \\
% ({\bf 29})  & Q1.23       & Q1.23          &           \\
% ({\bf 30})  & Q1.24       & Q1.24          &           \\
({\bf 31})  & KEWQOT        & Ref.~\cite{DAW06,DAW07}     \\
({\bf 32})  & NUYKIC        & Ref.~\cite{RPG+10a,RPG+10b}    \\
({\bf 33})  & XOCXIW        & Ref.~\cite{WOK+01} \\
% ({\bf 34})  &             &                & Q1.34??   \\
% ({\bf 35})  &             &                &           \\
({\bf 36})  & HUWGEL        & Ref.~\cite{FSP00,FSP03} \\
% ({\bf 37})  &             &                &           \\
% ({\bf 38})  &             &                & Q1.29 Ref.~\cite{BMKH12} \\
%MEC ({\bf 39})  &               & Ref.~\cite{JBB+82} \\
% ({\bf 40})  & Q1.31       &                &           \\
% ({\bf 41})  &             &                &           \\
% ({\bf 42})  &             &                &           \\
%( {\bf 43})  &             &                &           \\
%( {\bf 44})  &             &                &           \\
%( {\bf 45})  & Q1.32       & Q1.32          &           \\
({\bf 46})  & QOMYEX        & Ref.~\cite{RSG+04a,RSG+04b}  \\
%( {\bf 47})  &             &                &           \\
({\bf 48})  & JEMWAA        &  Ref.~\cite{MRP+06a,MRP+06b}   \\
%( {\bf 49})  &             &                &           \\
%( {\bf 50})  &             &                &           \\
({\bf 51})*  & YAGJAR10     & Ref.~\cite{WJWR95}   \\
% {\bf 52})  & Q2.4        & Q2.4           &           \\
% ({\bf 53})  &             &                &           \\
% ({\bf 54})  & Q2.5        & Q2.5           &           \\
% ({\bf 55})  & Q2.2        & Q2.2           &           \\
% ({\bf 56})  &             &                &           \\
% ({\bf 57})  &             &                &           \\
% ({\bf 58})  &             &                &           \\
% ({\bf 59})  &             &                &           \\
% ({\bf 60})  &             &                &           \\
%MEC ({\bf 61})  &               & Ref.~\cite{JBB+82} \\
({\bf 62})* & PATLAX        & Ref.~\cite{KSKY92} \\
({\bf 63})  & WAKRUX        & Ref.~\cite{TKK10a,TKK10b} \\
({\bf 64})  & NAMFOY        & Ref.~\cite{TAL+11,TAL+12} \\
% ({\bf 65})  & Q2.1        & Q2.1           &           \\
({\bf 66})  & FINREA        & Ref.~\cite{OSCG03,OSCG05} \\
% ({\bf 67})  &             &                & Q1.41??   \\
% ({\bf 68})  & Q1.42       & Q1.42          &           \\
% ({\bf 69})  & Q1.43       & Q1.43          &           \\
% ({\bf 70})  & Q1.44       & Q1.44          &           \\
% ({\bf 71})  & Q1.45       & Q1.45          &           \\
% ({\bf 72})  &             & Q1.46          &           \\
% ({\bf 73})  & Q1.47       & Q1.47          &           \\
({\bf 74})  & NOFPII        & Ref.~\cite{SSG+06,SSG+08}  \\
% ({\bf 75})  & QICHEQ        & Ref.~\cite{IYE+07a,IYE+07b} \\
({\bf 75})  & FINRIE        & Ref.~\cite{OSCG03c,OSCG05}  \\
% ({\bf 76})  &             &                &           \\
({\bf 77})  & ZIFCAU        & Ref.~\cite{GCT+12,GCT+13}  \\
% ({\bf 78})  & Q1.50       & Q1.50          &           \\
% ({\bf 79})  &             & Q1.51          &           \\
% ({\bf 80})  &             & Q1.52          &           \\
({\bf 81})  & IFAXUI        & Ref.~\cite{BCF+02a,BCF+02b} \\
% ({\bf 82})  &             & Q1.54          &           \\
({\bf 83})  & GEYZOB        & Ref.~\cite{WHH+12a,WHH+12b} \\
({\bf 84})  & FINRAW        & Ref.~\cite{OSCG03d,OSCG05} \\
% ({\bf 85})  &             &                & Q1.57??   \\
% ({\bf 86})  &             & Q1.58          &           \\
% ({\bf 87})  &             & Q1.59          &           \\
% ({\bf 88})  &             &                &           \\
% ({\bf 89})  &             &                &           \\
% ({\bf 90})  &             &                &           \\
({\bf 91})  & MARVAD        & Ref.~\cite{HGK+04,HGK+05}  \\
% ({\bf 92})  &             &                &           \\
% ({\bf 93})  & Q2.6        & Q2.6           &           \\
({\bf 94})  & VAJLUO        & Ref.~\cite{FP02,FP03} \\
% ({\bf 95})  &             &                &           \\
% ({\bf 96})  &             &                &           \\
% ({\bf 97})  &             &                &           \\
% ({\bf 98})  &             & Q1.60          &           \\
% ({\bf 99})  &             &                &           \\
% ({\bf 100}) &             &                &           \\
% ({\bf 101}) &             &                &           \\
% ({\bf 102}) &             &                &           \\
% ({\bf 103}) &             &                &           \\
%MEC ({\bf 104}) &               & Ref.~\cite{JBB+82} \\
% ({\bf 105}) &             &                &           \\
% ({\bf 106}) &             &                &           \\
({\bf 107}) & PATLAX        & Ref.~\cite{KSKY92} \\
({\bf 108}) & BENHUZ        & Ref.~\cite{KRB+12a,KRB+12b} \\
({\bf 109}) & BOFGEJ        & Ref.~\cite{KHM+08a,KHM+08b} \\
% ({\bf 110}) &             &                &           \\
% ({\bf 111}) &             & Q1.63          &           \\
%%%%%%%%
% STOP %
%%%%%%%%
\hline \hline                                                 
\end{tabular}
\end{table}
% -----------------------------------------------------------------------

The geometries of all the complexes were optimized and (local) minima were confirmed
by the absence of imaginary vibrational frequencies.  Whenever possible, the
geometry optimizations began from X-ray crystal structural data obtained from the Cambridge
Crystallographic Data Centre (CCDC) \cite{A02,CCDC}.  This was the case for the compounds
listed in Table~\ref{tab:geom3}.  Start geometries indicated with an asterisk in 
Table~\ref{tab:geom3} were constructed from the CCDC data of a related compound.
Otherwise crystal coordinates were generated from the {\sc Gaussview} 
program \cite{gv5}, taking into account  specific symmetries and crystallographic volumes.
The threshold for optimization was set to ultrafine with self consistent field 
(SCF) convergence being being set to very tight.

Time-dependent DFT \cite{C09,CH12} gas-phase absorption spectra were calculated at the optimized
ground-state geometries using the same functional and basis sets as for the ground-state calculation.
In all cases, at least 100 singlet states were included in calculations of spectra.
As in Ref.~\cite{WJL+14}, a theoretical molar extinction spectrum is calculated via,
\begin{equation}
  \epsilon(\omega)=\frac{\pi{N_A} e^2}{2 \epsilon_0{m_e} c \, ln(10)} S (\omega)
  \, ,
  \label{eq:details.2}
\end{equation}
from the corresponding spectral function,
\begin{equation}
  S(\omega)=\sum_I f_I \delta(\omega-\omega_I) \, ,
  \label{eq:details.3}
\end{equation}
using an in-house python program {\sc Spectrum.py}.  The result is a theoretical spectrum with the
same units and the same order of magnitude as the experimentally-measured absorption spectrum,
thereby allowing easy comparison of theory and experiment, albeit at the expense of introducing
a single empirical parameter which accounts for spectral broadening due to vibrational structure, 
solvent broadening (but not solvent shifts), and finite experimental resolution.  
This is the full width at half maximum (FWHM) which has been set to 40 nm throughout.

Density-of-states (DOS) and partial DOS (PDOS) were obtained using another in-house
python program called {\sc Pdos.py} previously described in the supplementary information
associated with Ref.~\cite{WJL+14}.  This allows us to identify the positions of ligand-field
theory (LFT) like ruthenium $d$ states as well as ligand $\pi$ states after suitable broadening.
At a practical level, using {\sc Pdos.py} involves carrying out another single point calculation 
with the option {\tt (pop=full gfinput iop(6/7=3,3/33=1,3/36=-1)}, thereby causing 
{\sc Gaussian} to output the number of basis functions {\tt Nbasis}, the overlap matrix, 
the eigenvalues, and the MO coefficients. {\sc Pdos.py} then takes this information from the 
{\sc Gaussian} output files and calculates the (P)DOS.  We used a FWHM of 0.25 eV  with 40000 points 
for graphing.

% ===================================================
\section{Results}
\label{sec:results}
% \input{results.tex}
% \begin{verbatim}
% ================================
% File: results.tex
% Last update: 4 June 2017
% ================================
% \end{verbatim}

The results of our calculations are divided into three subsections.
In the first subsection (Sec.~\ref{sec:valid}), we validate the 
ability of DFT to be able to determine ground-state structures and
the ability of TD-DFT to be able to simulate experimental absorption
spectra.  The second subsection (Sec.~\ref{sec:PDOS-LFT}) 
extracts $t_{2g}$, $e^*_g$, and $\pi^*$ energies from PDOS-LFT and 
shows that these correlate with peaks in measured absorption spectra.
The final subsection (Sec.~\ref{sec:luminesce}) discusses the extent
to which PDOS-LFT can be used to predict which compounds may have long
luminescence lifetimes.

% ------------------------------------
\subsection{Structure and Properties}
\label{sec:valid}
% ------------------------------------

% ----------------------------
\subsubsection{Geometries}
% ----------------------------

We first test whether our DFT calculations are consistent with observed
X-ray crystallography geometries by seeing how much typical bond lengths
and bond angles change when the geometry is re-optimized in gas phase using
the X-ray geometries as start geometries. Naturally we expect some expansion
of the molecule as there are fewer constraints in the gas phase than in the
solid phase but, nevertheless, we expect gas-phase and solid-state geometries
to be correlated. 

The need to judge correlation requires us to make a short review of linear
regression as some of the concepts that we use are expected to be unfamiliar
to even expert readers.  Linear regression is just a least squares fit
of $N$ $(x_i,y_i)$ data points to the familiar equation,
\begin{equation}
  y = mx + b \, . 
  \label{eq:results.1}
\end{equation}
Minimizing the error
\begin{equation}
   {\cal E} = \sum_{i=1,N} \left( y_i - m x_i - b \right)^2 \, ,
   \label{eq:results.2}
\end{equation}
gives the usual formulae for the slope and intercept,
\begin{eqnarray}
  m & = & \frac{\langle xy \rangle - \langle x \rangle \langle y \rangle}
          {\langle x^2 \rangle - \langle x \rangle^2 } \nonumber \\
  b & = & \frac{\langle y \rangle \langle xy \rangle - \langle x \rangle
         \langle xy \rangle}{\langle x^2 \rangle - \langle x \rangle^2 }
  \, ,
  \label{eq:results.3}
\end{eqnarray}
where we have introduced the notation,
\begin{equation}
  \langle f(x,y) \rangle = \frac{1}{N} \sum_{i=1,N} f(x_i,y_i) \, .
  \label{eq:results.4}
\end{equation}
for the average of the $N$ $f(x_i,y_i)$ values.
The goodness of fit is usually judged by the correlation coefficient 
defined as,
\begin{equation}
  R^2 = \frac{\left( \langle x y \rangle - \langle x \rangle \langle y \rangle \right)^2} 
  {\left( \langle x^2 \rangle - \langle x \rangle^2 \right)
   \left( \langle y^2 \rangle - \langle y \rangle^2 \right) }
  \, ,
  \label{eq:results.5}
\end{equation}
which is close to unity for a good fit.  Up to this point, everything 
corresponds to the standard formulae implemented in typical spreadsheet 
programs.

However, we need to go a little further because the correlation coefficient
is {\em not} a good measure of the error in the sense that the correlation 
coefficient calculated over a small range of $x_i$ values may be very 
different from the correlation coefficient obtained when all the data is 
taken into consideration.  That is why it is often better to calculate 
the standard error which is defined as the standard deviation of 
the $y_i$ values from those obtained from the fit.  It may be calculated as,
\begin{equation}
  \Delta y = \sqrt{ \left( \frac{N}{N-2} \right) \left( \langle y^2 \rangle
    - \langle y \rangle^2 \right) \left( 1 - R^2 \right) }
  \, ,
  \label{eq:results.6}
\end{equation}
which also shows the relation of the standard error to the correlation
coefficient.  Furthermore, following Ref.~\cite{DCT+15}, it is often more 
interesting to invert the fit so that,
\begin{equation}
  x = \frac{y}{m} - \frac{b}{m} \, .
  \label{eq:results.7}
\end{equation}
The {\em predictability},
\begin{equation}
  \Delta x = \frac{\Delta y}{\vert m \vert} \, ,
  \label{eq:results.8}
\end{equation}
then represents the expected error in predicting the experimental results
using our theoretical model.  In reporting the results of our fits, we will
give the slope $m$, the intercept $b$, the correlation coefficient $R$,
and the predictability $\Delta x$.

In order to see how they are correlated, theoretical and experimental 
bond distances and angles were compared for 35 of the 39 complexes in 
Table~\ref{tab:geom3}.  Complexes {\bf (14)} and {\bf (62)} are excluded because
their start geometries were a modified version of the original X-ray
crystal structures.  Complexes {\bf (33)} and {\bf (51)} are excluded
because we were unable to converge the gas-phase geometry optimizations.

% ------------------------------------------------------
\begin{figure}
\begin{tabular}{l}
(a) \\
\includegraphics[width=0.45\textwidth]{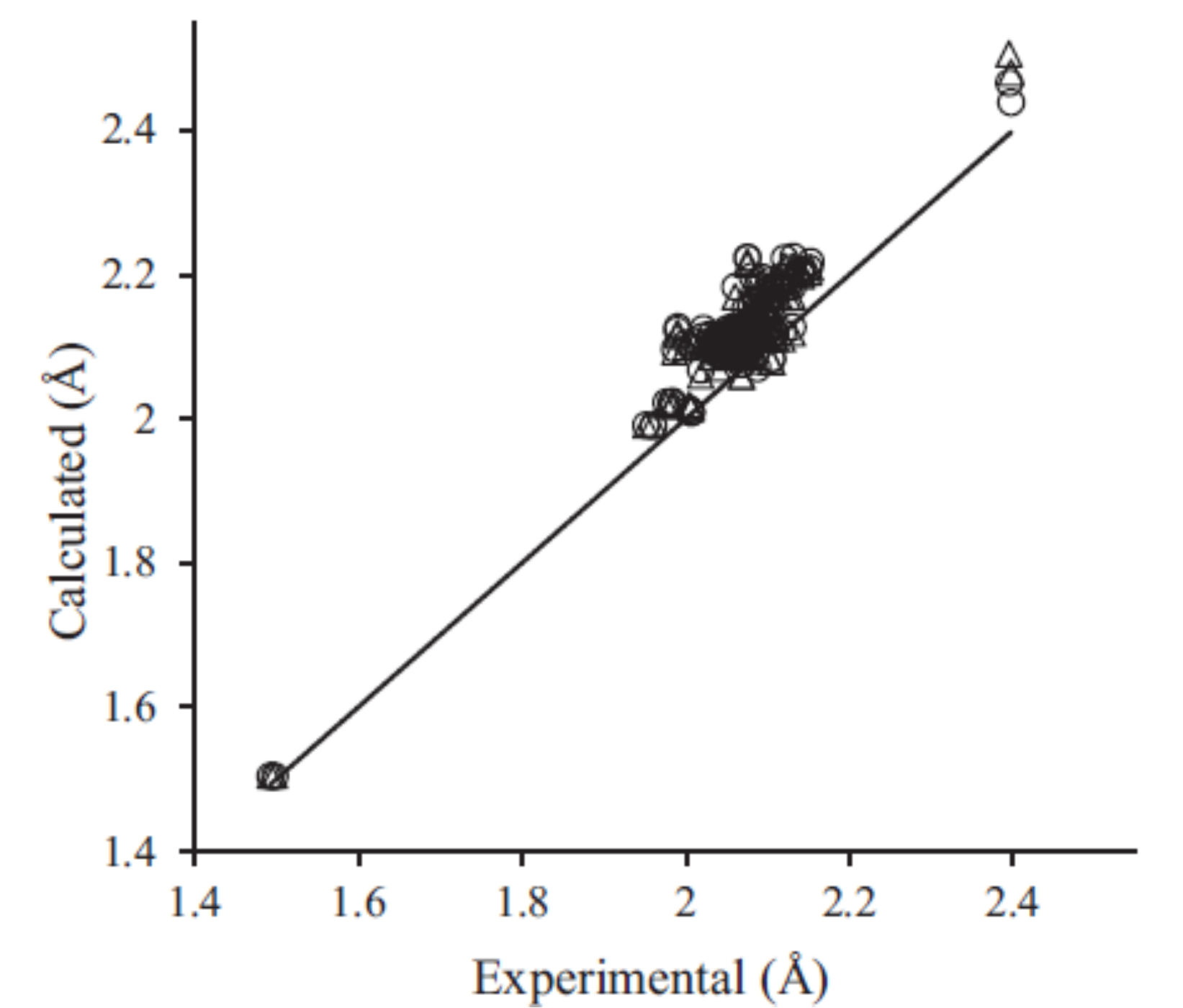} \\  % geometries1d.pdf
(b) \\
\includegraphics[width=0.45\textwidth]{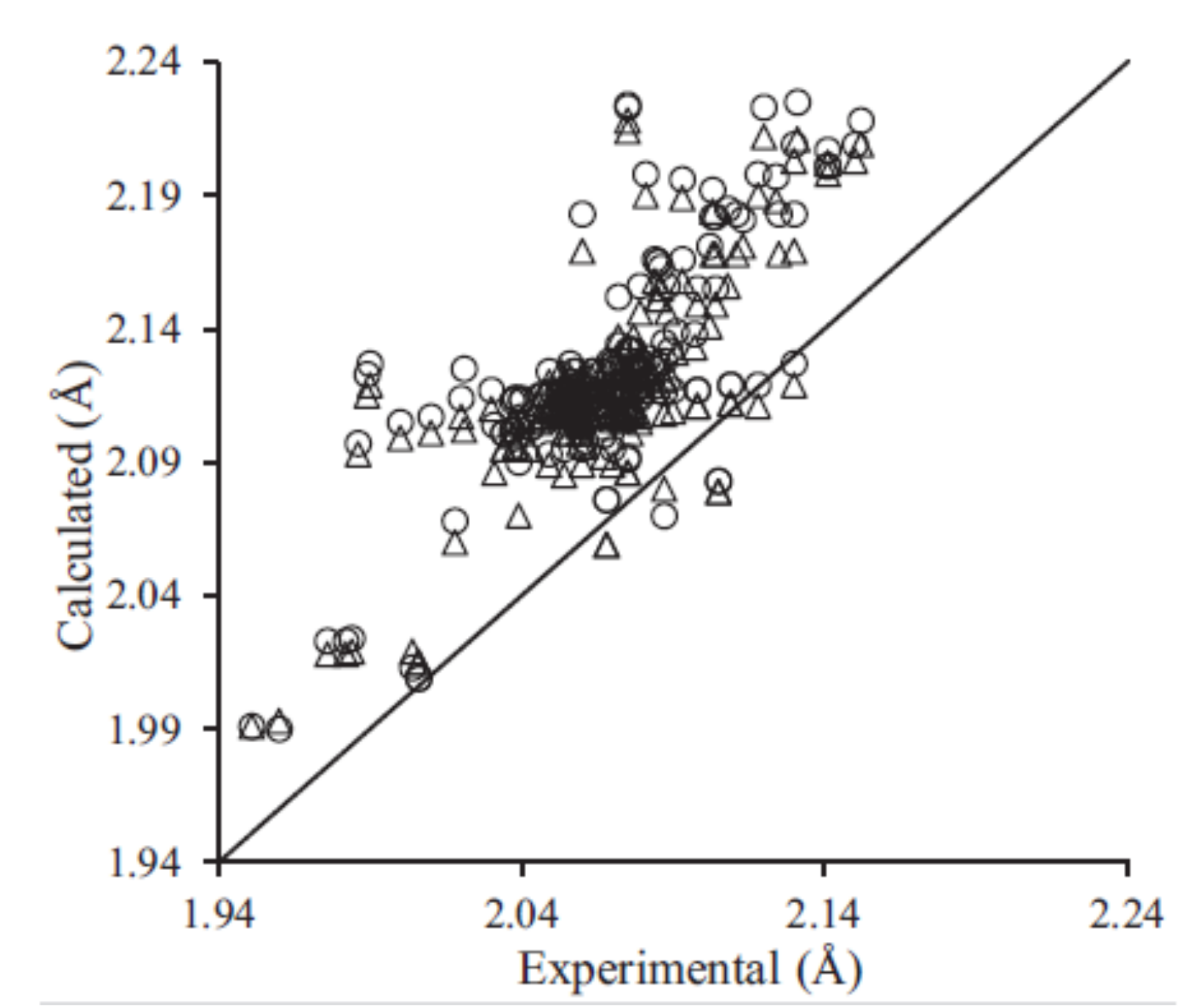} \\  % geometries2d.pdf
\end{tabular}
\caption{
(a) Correlation graph between calculated DFT bond lengths for the 6-31G ($\triangle$)
and 6-31G(d) ($\circ$) and 184  measured X-ray crystallographic bond lengths. 
(b) Enlargement. The 45$^\circ$ line indicates 
perfect agreement with experiment.
\label{fig:geometries}
}
\end{figure}
% ------------------------------------------------------
Figure~\ref{fig:geometries} shows how calculated gas-phase bond 
lengths compare with X-ray crystal structure geometries.  Only 
ligand-metal bond lengths have been considered.  As expected the 
calculated gas-phase bond lengths are typically longer than those 
in the X-ray crystal structures.  However Table~\ref{tab:fit} shows 
that the correlation is actually excellent with a predictability 
of 0.0251 {\AA} for the 6-31G basis set and 0.0262 {\AA} for the 
6-31G(d) basis set.  This may be compared with the typical error 
of 0.005 {\AA} obtained for 20 organic molecules with the same 
functional and the 6-31G(d) basis set (p.~124 of Ref.~\cite{KH00}).  
Note, however, that the comparison made there is against
gas phase data and that predicting the geometries of transition metal
complexes is in general more challenging than predicting the geometries
of purely organic molecules.  
% It is also interesting to notice that,
% while the 6-31G(d) basis set
% results seems to overestimate the experimental results a bit more than
% do the 6-31G basis set results and while the 6-31G(d) basis set results
% seem to correlate slightly less well than do the 6-31G basis set results
% with the measured X-ray crystal bond lengths, the differences between 
% the results with the two different basis sets are not really significant.
It is interesting to note that geometries predicted using the 6-31G basis set
are better correlated with experimental X-ray geometries obtained using the 
seemingly better 6-31G(d) basis set.  This could be an indication that
the 6-31G(d) basis set is less well balanced than is the 6-31G basis set.
However the differences in the results obtained with the two basis sets
are not really significant.
% \input{./tables/fit.tex}
% ==============================================
% File: fit.tex
% Last modified: 24 March 2017
% ==============================================
\begin{table}
\caption{Least squares fit parameters.
\label{tab:fit}}
\begin{tabular}{cccccc}
\hline \hline
Basis Set & $m$  &  $b$ & $R^2$ & $\Delta x$ \\
\hline
\multicolumn{5}{c}{bond length/{\AA}} \\
6-31G     & 1.04387 & -0.04195 & 0.90450 & 0.02505  \\
6-31G(d)  & 1.02988 & -0.00602 & 0.89674 & 0.02617  \\
\multicolumn{5}{c}{bond angles/degrees} \\
6-31G     & 1.07230 & -6.39743 & 0.93055 & 1.10263 \\
6-31G(d)  & 1.07424 & -6.79970 & 0.92682 & 1.13407 \\
\multicolumn{5}{c}{$\lambda$/nm} \\
6-31G     & 0.78134 & 81.20545 & 0.47982 & 33.6319 \\ % about 0.17 eV at 500 nm
6-31G(d)  & 0.77617 & 67.65579 & 0.40762 & 53.5672 \\ % about 0.27 eV at 500 nm
\hline \hline
\end{tabular}
\end{table}

% ------------------------------------------------------
\begin{figure}
\begin{tabular}{l}
(a) \\
\includegraphics[width=0.45\textwidth]{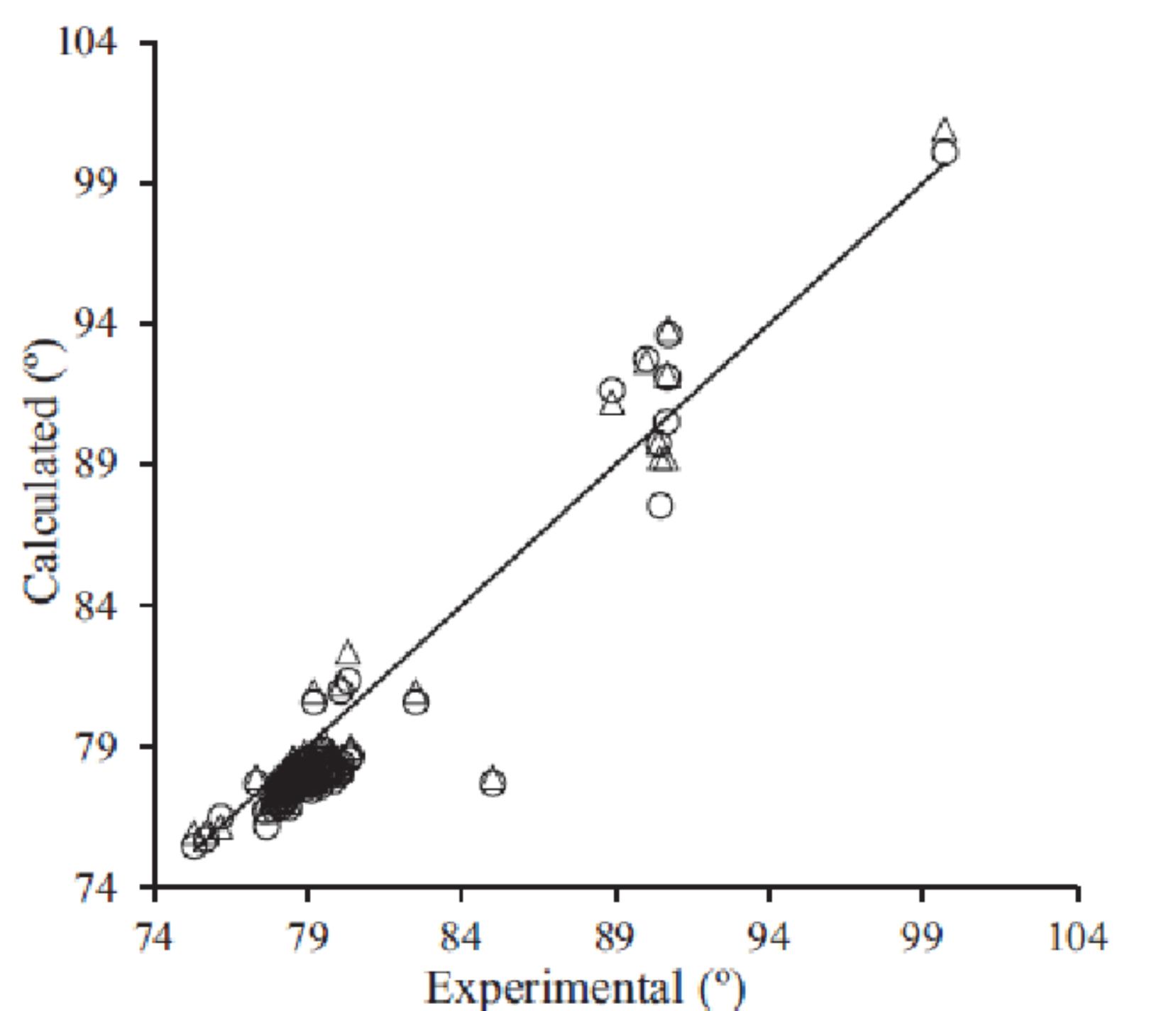} \\  % bondangles1d.pdf
(b) \\
\includegraphics[width=0.45\textwidth]{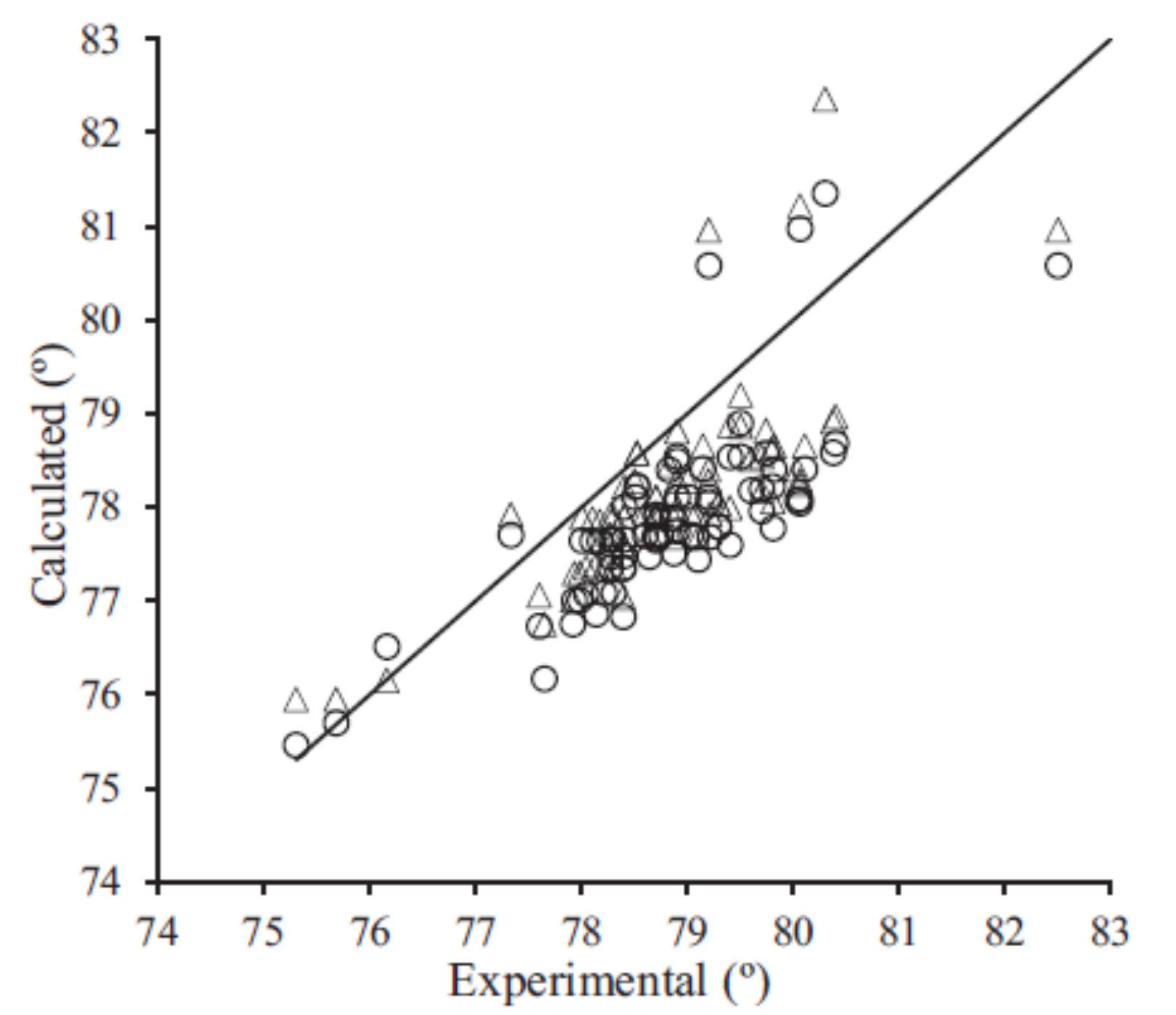} \\  % bondangles2d.pdf
\end{tabular}
\caption{
(a) Correlation graph between calculated DFT bond angles for the 6-31G ($\triangle$)
and 6-31G(d) ($\circ$) and 85 measured X-ray crystallographic bond
angles.  (b) Enlargement.  The 45$^\circ$ line indicates perfect 
agreement with experiment.
\label{fig:bondangles}
}
\end{figure}
% ------------------------------------------------------
Figure~\ref{fig:bondangles} shows how calculated
gas-phase bond angles compare with X-ray crystal structure geometries.
Only ligand-metal-ligand angles near 90$^\circ$ have been considered.
The calculated bond angles tend to be smaller than the X-ray
crystal structure bond angles.  
Table~\ref{tab:fit} shows that the correlation is actually excellent
with a predictability of 1.103$^\circ$ for the 6-31G basis set and
1.134$^\circ$ for the 6-31G(d) basis set.  This may be compared with
the typical error quoted as being on the order of a few tenths of
a degree obtained for 20 organic molecules 
with the same functional and the 6-31G(d) basis set (p.~124
of Ref.~\cite{KH00}).  Of course, once again, this is not unexpected
because the comparison is against
gas phase data and that predicting the geometries of transition metal
complexes is in general more challenging than predicting the geometries
of purely organic molecules.  It is also interesting to notice that,
% while 
the 6-31G(d) basis set results correlated slightly less well
with experiment than do the 6-31G basis set results, but the
difference is not really significant.

% ----------------------------
\subsubsection{Absorption Spectra}
% ----------------------------
% ------------------------------------------------------
\begin{figure}
\begin{tabular}{l}
(a) \\
\includegraphics[width=0.45\textwidth]{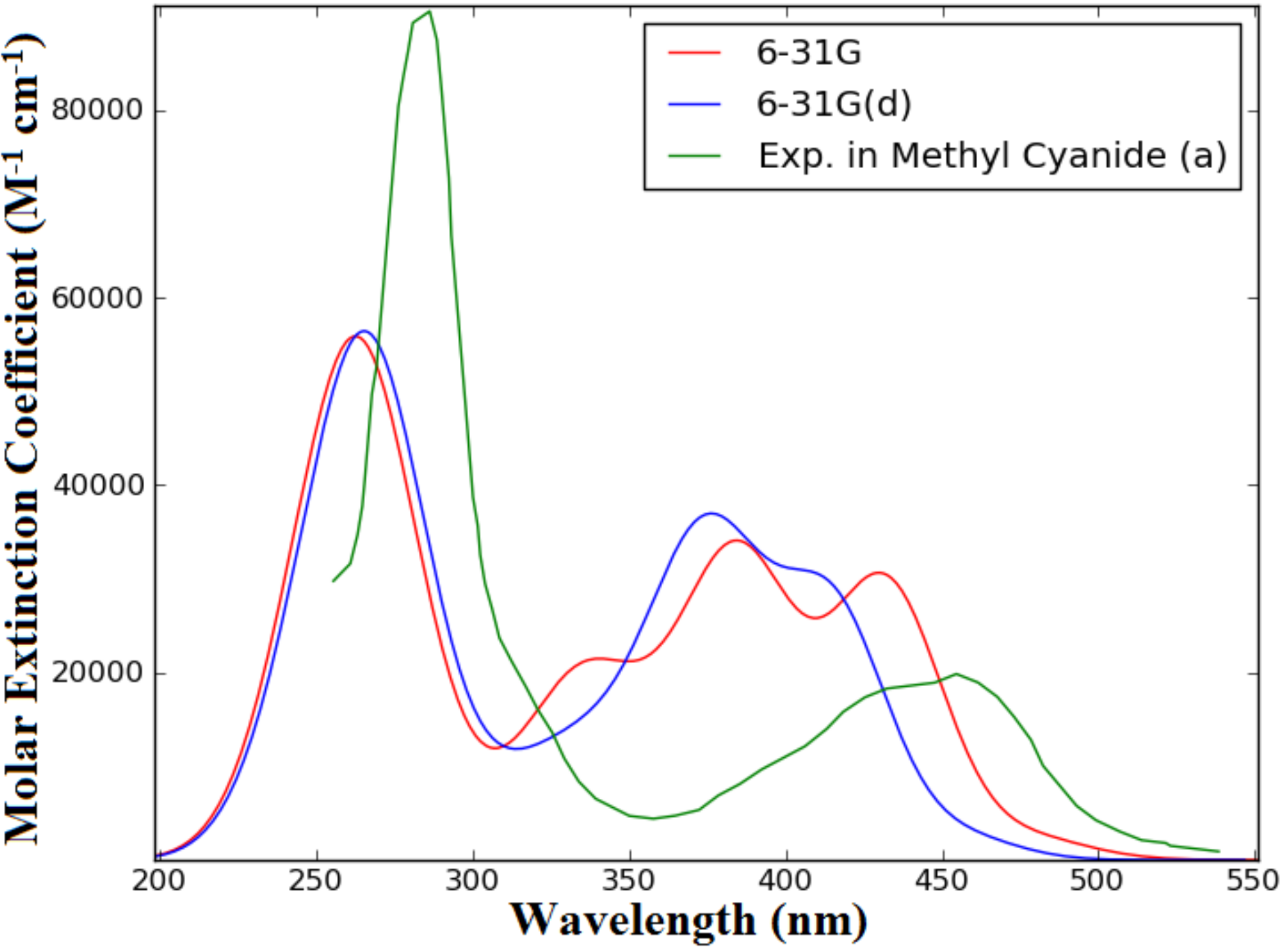} \\  % cmplx018_spectra.pdf
(b) \\
\includegraphics[width=0.45\textwidth]{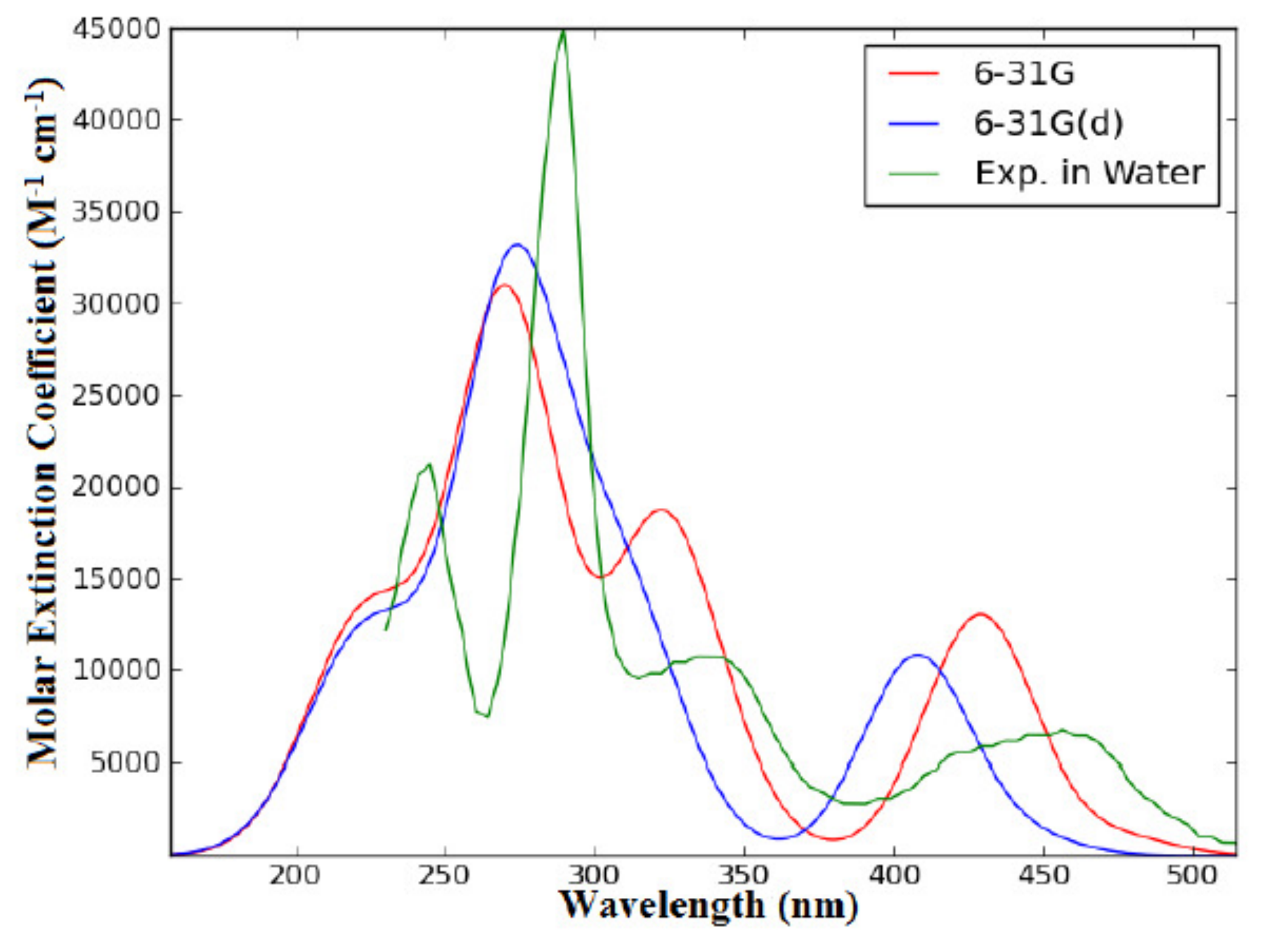} \\ % cmplx024_spectra.pdf
\end{tabular}
\caption{
Comparison of calculated gas-phase absorption spectra (6-31G, red; 6-31G(d) blue)
with an experimental spectrum: (a) complex {\bf (18)}, experimental spectrum
in acetonitrile from Ref.~\cite{YYS+15};
(b) complex {\bf (24)}, experimental spectrum in water from Ref.~\cite{BGS+13}.
\label{fig:spectra}
}
\end{figure}
% ------------------------------------------------------
We now wish to see if (TD-)DFT is able to give absorption spectra in 
reasonable agreement with experiment.  Some example comparisons of
spectra are given in Fig.~\ref{fig:spectra}.  
Many other TD-DFT spectra are given in the Supplementary Information.
Note that no adjustable
parameters have been used other than the FWHM (Sec.~\ref{sec:details}).
Such spectra are expected to be accurate to about 0.2 eV, which is
not extremely accurate but which is often adequate for qualitative 
assignments of spectral features.  Typical complexes show two to four 
peaks, where some of the peaks are only visible as shoulders.
Other spectra are given in the Supplementary Information.
We have noticed that the lowest energy 6-31G(d) peak is often 
blue-shifted with respect to the corresponding 6-31G peak, with much 
less differences between the basis sets for higher-energy features
in the TD-B3LYP spectra.  The shift of the lower energy peak may
indicate that the addition of $d$ functions is improving the description
of the ground-state (i.e., lowering its energy) more than it is 
improving the description of the excited state.  If so, this effect is
less marked for higher excited states.

% ------------------------------------------------------
\begin{figure}
\includegraphics[width=0.45\textwidth]{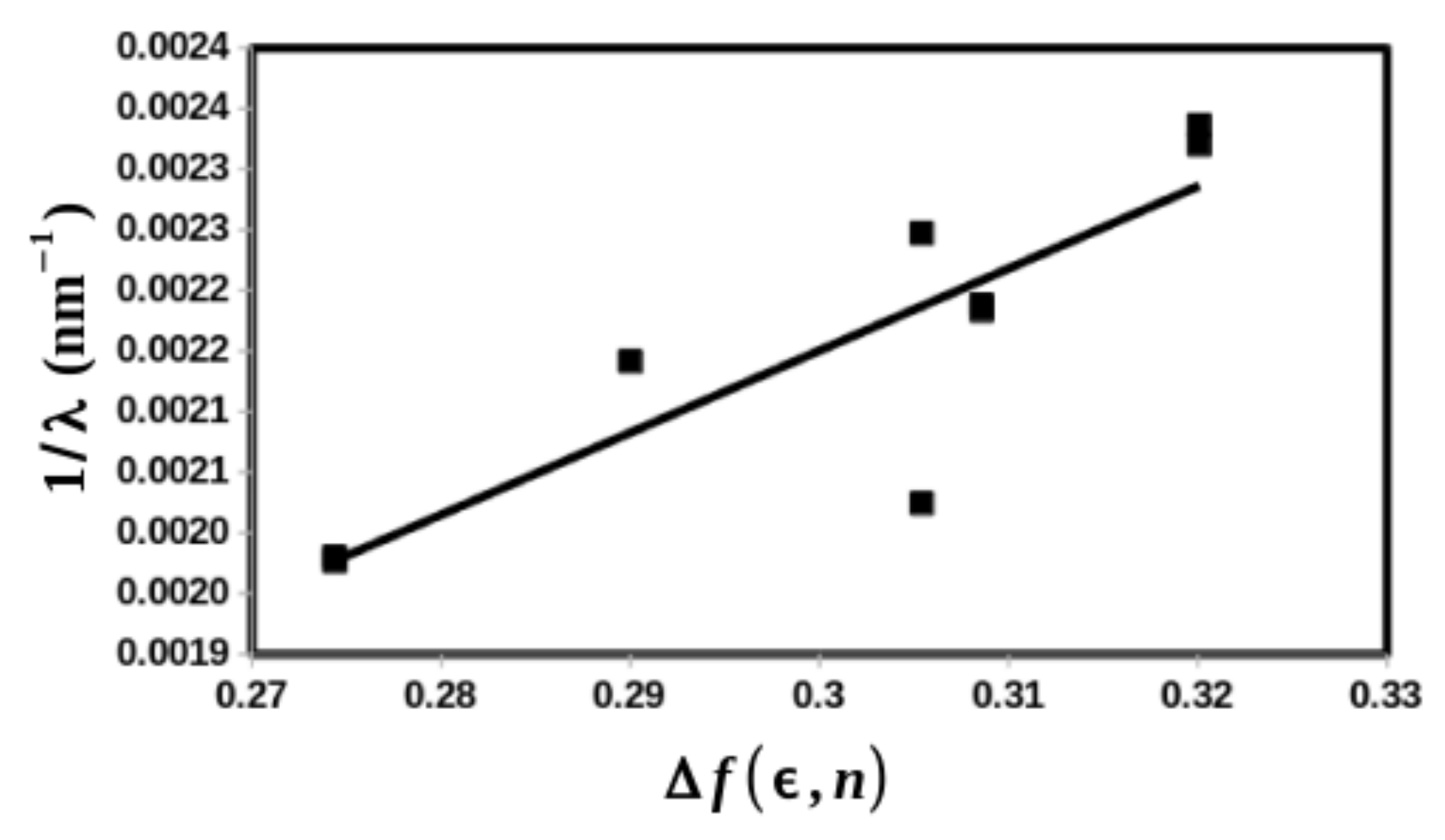}  % solventshift.pdf
\caption{
Inverse of the wavelength of the lowest energy absorption plotted 
against the orientational polarization for complex {\bf (3)} and 
various solvents listed in Table 1 of Ref.~\cite{JBB+88}, except
for chloroform which did not fit the trend estabilished by the other
solvents.  Note that $\Delta f(\epsilon,n)=0.3054$ for acetonitrile.
\label{fig:solventshift}
}
\end{figure}
% ------------------------------------------------------
In order to compare theory and experiment for several molecules, it
is useful to focus on the lowest energy transition.  Data for this 
has been collected from several references and is conveniently
provided in Table 1 of Ref.~\cite{JBB+88} for several solvents.
Since the lowest energy transition is expected to be of 
$t_{2g} \rightarrow \pi^*$ charge-transfer type, we can anticipate
some solvent dependence, though it is often relatively small.  We have
tried to minimize solvent effects by estimating a best value in 
acetonitrile, a common solvent for polypyridinal ruthenium complexes.  
As discussed in Ref.~\cite{CPCK09}, there are several ways to estimate 
solvent shifts in spectra and all involve approximations.  The one
we chose consists of seeking the best linear relationship between the
inverse wavelength and the orientation polarizability,
\begin{equation}
  \Delta f(\epsilon,n) = \frac{\epsilon-1}{2\epsilon+1} - \frac{n^2-1}{2n^2+1}
  \, ,
  \label{eq:results.9}
\end{equation}
which comes out of Onsager's reaction field theory.  Here $\epsilon$ is the 
solvent dielectric constant and $n$ is the solvent refractive index.  
Note that this is only valid for a given transition weakly interacting with
a dielectric medium.  An example plot is shown in Fig.~\ref{fig:solventshift} 
for complex {\bf (3)} where the solvent shift is particularly marked and there 
are two experimental values for the absorption in acetonitrile.  The value from 
the linear plot for this compound and best estimates in acetonitrile where they 
could be extracted are shown in Table~\ref{tab:AN}.
% \input{./tables/AN.tex}
% =======================================
% file: AN.tex
% last updated: 26 May 2017
% =======================================

\begin{table}
\caption{Best estimates of the lowest energy absorption maximum in acetonitrile
         based upon data from Table 1 of Ref.~\cite{JBB+88}.
         \label{tab:AN}}
\begin{tabular}{cccc}
\hline \hline
number       & wavelength & number   & wavelength  \\
             & (nm)            &          &  (nm)            \\
\hline
({\bf 1})    & 431.2           & ({\bf 53})   & 476.0           \\  
({\bf 2})    & 524.0           & ({\bf 56})   & 465.0           \\
({\bf 3})    & 457.2           & ({\bf 57})   & 455.0           \\
({\bf 6})    & 451.5           & ({\bf 58})   & 478.0           \\ 
({\bf 7})    & 493.0           & ({\bf 59})   & 479.1           \\
({\bf 8})    & 448.0           & ({\bf 60})   & 559.0           \\
({\bf 9})    & 445.0           & ({\bf 61})   & 547.0           \\
({\bf 10})   & 448.0           & ({\bf 62})   & 450.0           \\
({\bf 11})   & 514.0           & ({\bf 63})   & 506.0           \\
({\bf 12})   & 458.0           & ({\bf 64})   & 487.3           \\
({\bf 13})   & 450.0           & ({\bf 65})   & 480.0           \\
({\bf 14})   & 450.0           & ({\bf 66})   & 448.0           \\
({\bf 15})   & 472.7           & ({\bf 67})   & 456.0           \\
({\bf 22})   & 440.0           & ({\bf 70})   & 456.0           \\
({\bf 23})   & 440.0           & ({\bf 72})   & 462.0           \\
({\bf 24})   & 450.0           & ({\bf 73})   & 474.0           \\
({\bf 25})   & 474.0           & ({\bf 74})   & 456.0           \\
({\bf 29})   & 460.0           & ({\bf 76})   & 453.0           \\
({\bf 30})   & 458.0           & ({\bf 77})   & 467.4           \\
({\bf 31})   & 473.0           & ({\bf 79})   & 483.0           \\ 
({\bf 32})   & 463.0           & ({\bf 80})   & 522.0           \\
({\bf 34})   & 483.0           & ({\bf 81})   & 522.0           \\
({\bf 35})   & 480.0           & ({\bf 90})   & 375.0           \\
({\bf 36})   & 478.0           & ({\bf 92})   & 505.0           \\
({\bf 37})   & 528.0           & ({\bf 93})   & 494.0           \\
({\bf 38})   & 562.0           & ({\bf 94})   & 483.5           \\
({\bf 39})   & 526.0           & ({\bf 97})   & 526.0           \\
({\bf 40})   & 540.0           & ({\bf 99})   & 605.1           \\
({\bf 45})   & 453.0           & ({\bf 100})  & 540.0           \\
({\bf 46})   & 454.0           & ({\bf 101})  & 585.0           \\
({\bf 47})   & 450.0           & ({\bf 103})  & 605.1           \\
({\bf 48})   & 442.0           & ({\bf 104})  & 524.0           \\
({\bf 49})   & 448.0           & ({\bf 105})  & 436.0           \\
({\bf 50})   & 476.0           & ({\bf 106})  & 414.8           \\
({\bf 51})   & 446.0           & ({\bf 107})  & 392.0           \\
({\bf 52})   & 438.0           & ({\bf 108})  & 473.1           \\
\hline \hline                                                 
\end{tabular}
\end{table}

%%%%%
% EOF
%%%%%

% ------------------------------------------------------
\begin{figure}
\includegraphics[width=0.45\textwidth]{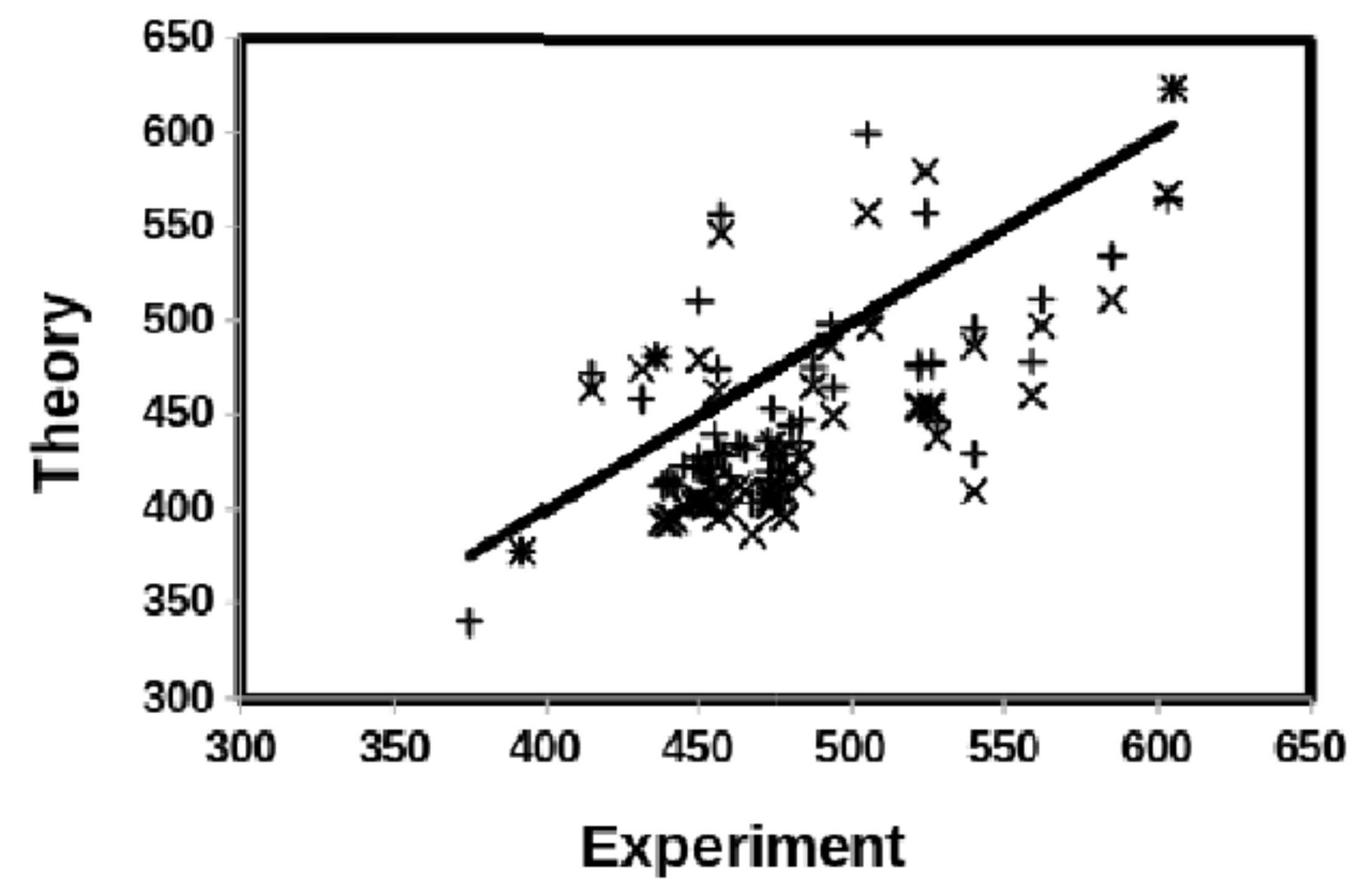}  % spectracorrel.pdf
\caption{
Correlation graph between calculated lowest energy absorption wavelengths 
for the 6-31G ($+$) and 6-31G(d) ($\times$) and 59 best estimates for the 
best estimate of the experimental lowest energy absorption wavelengths 
in acetonitrile, all in nm.  The 45$^\circ$ line indicates perfect 
agreement with experiment.
\label{fig:spectracorrel}
}
\end{figure}
% ------------------------------------------------------
Figure~\ref{fig:spectracorrel} shows how our TD-DFT spectra compare with experimental
spectra for the placement of the lowest energy absorption maximum.  
The calculated predictability shown in Table~\ref{tab:fit} corresponds to
0.17 eV for the 6-31G basis set and to 0.27 eV for the 6-31G(d) basis set
at 500 nm.  This is the sort of accuracy we normally expect from TD-DFT
in the absence of any particular problems such as, say, strong density
relaxation upon excitation.  We conclude that our DFT model is a reasonably
good descriptor of the experimental situation.

% -----------------------------------
\subsection{PDOS-LFT}
\label{sec:PDOS-LFT}
% -----------------------------------

We now come to the heart of this paper, namely the partial density-of-states (PDOS)
technique for extracting ligand field theory (LFT) like information from DFT 
calculations.  This is needed by chemists as it is their traditional tool for
thinking about and discussing spectra (e.g., Fig.~\ref{fig:spectra}) and other
photoprocesses in transition metal complexes.  It is also nontrivial because the
usual pseudo-octahedral orbitals $t_{2g}$ and $e^*_g$ do not emerge automatically
from DFT calculations.  This statement is less true of the nonbonding $t_{2g}$ which
can often be identified by direct visualization of DFT molecular orbitals, but
it is very true of the antibonding $e^*_g$ orbitals which (because they are 
antibonding) mix heavily with ligand orbitals, making it impossible to identify
individual $e^*_g$ orbitals among the DFT molecular orbitals in the absence of
special tools.  The tool we have used here is the very simple one used in
Ref.~\cite{WJL+14}, namely a PDOS analysis based upon Mulliken charges.
Other (P)DOS graphs may be found in the Supplementary Information.

% ------------------------------------------------------
\begin{figure}
\begin{tabular}{l}
(a) \\
\includegraphics[width=0.45\textwidth]{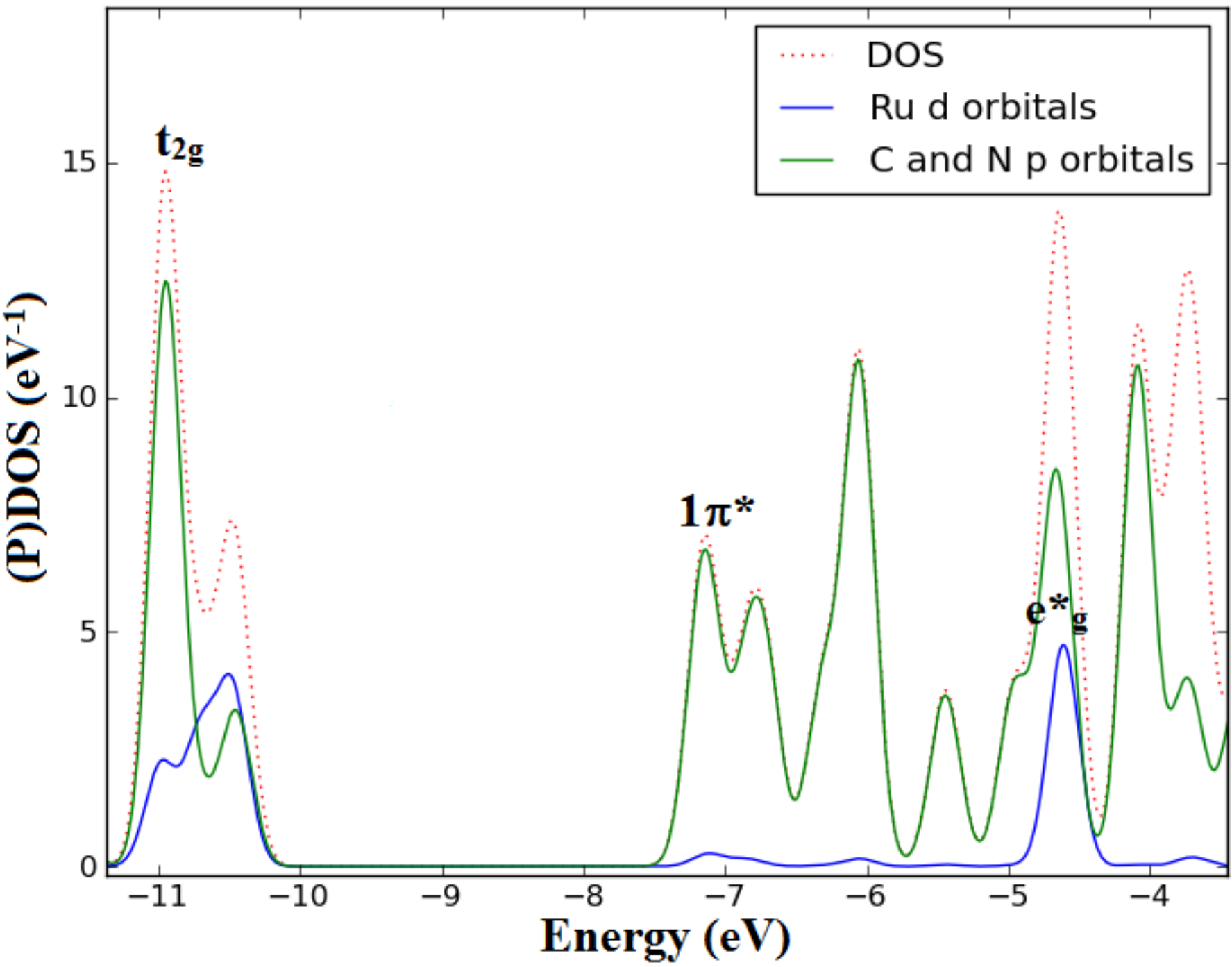} \\ % cmplx018_pdos_631g.pdf
(b) \\
\includegraphics[width=0.45\textwidth]{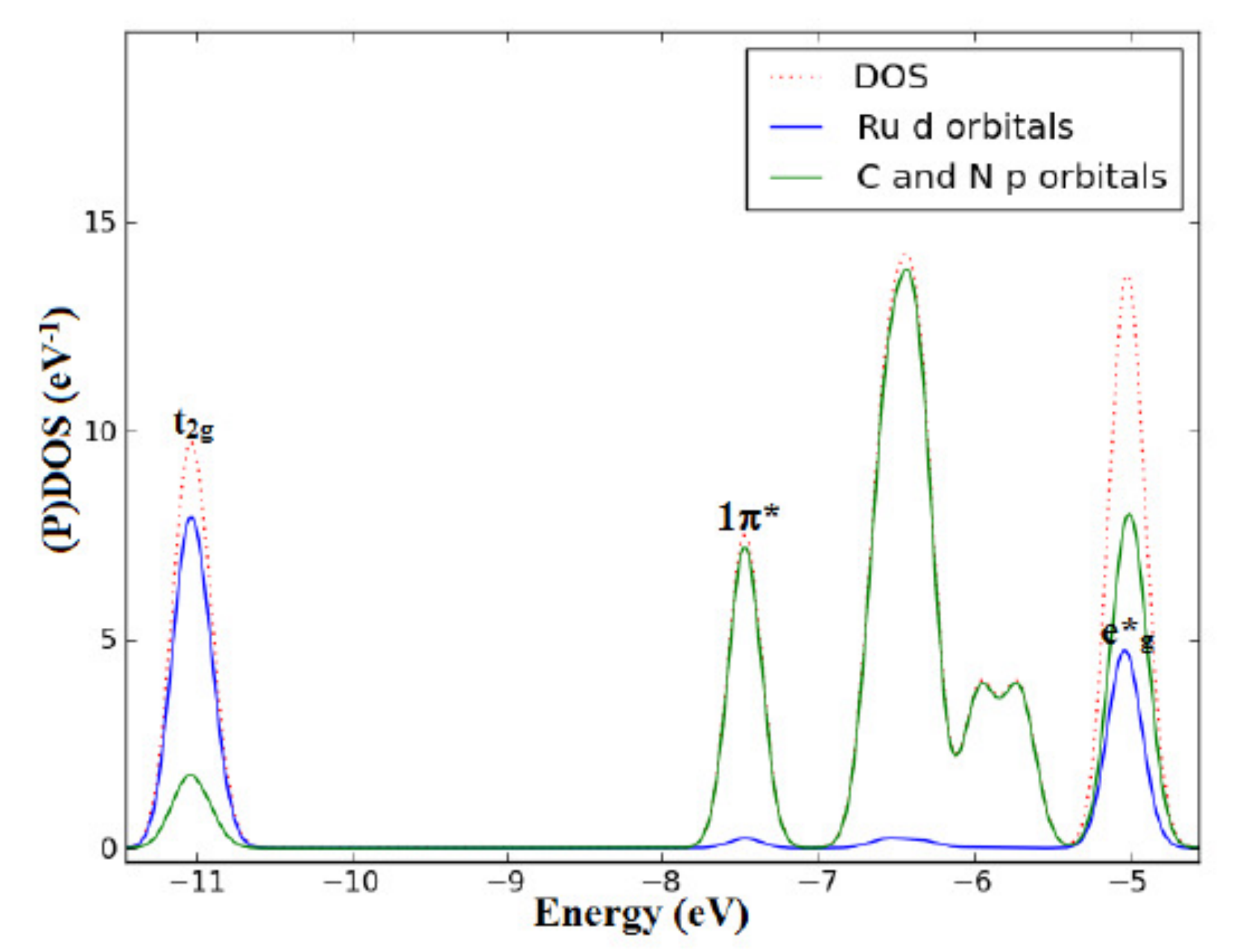} \\ % cmplx024_pdos_631g.pdf
\end{tabular}
\caption{
B3LYP/6-31G (P)DOS calculated for (a) complex {\bf (18)} and (b) complex {\bf (24)}.
Note that the corresponding highest-occupied molecular orbital energies are
-10.43 eV for complex {\bf (18)} and  -10.95 for complex {\bf (24)}.
\label{fig:pdos}
}
\end{figure}
% ------------------------------------------------------
The concept of the density-of-states (DOS) is borrowed from solid-state physics.
The idea is to replace the orbital energy levels, which have become too dense
for convenient interpretation, with a gaussian-broadened stick spectrum,
\begin{equation}
  \mbox{DOS}(\epsilon) = \sum_i g(\epsilon - \epsilon_i) \, ,
  \label{eq:results.10}
\end{equation}
where the gaussian,
\begin{equation}
  g(\epsilon) = \sqrt{\frac{\alpha}{\pi}} e^{-\alpha x^2} \, ,
  \label{eq:results.11}
\end{equation}
is normalized to unity.  The parameter $\alpha$ is fixed by the FWHM according to the
relation,
\begin{equation}
  \mbox{FWHM} = 2 \sqrt{ \frac{\ln 2}{\alpha}} \, .
  \label{eq:results.12}
\end{equation}
% We lose the concept of individual orbital energy levels when using the DOS but
% peak heights tell us in which energy ranges contain the highest density of orbital
% energy levels.  
We loose the concept of individual orbital energy levels when using the DOS.
Nevertheless an isolated DOS peak of unit area corresponds to a single
underlying orbital energy level, a DOS peak integrating to an area of two
corresponds to two closely spaced underlying orbital energy levels, etc.
Figure~\ref{fig:pdos} provides an example of the DOS of two complexes.
Note that each peak represents one to several underlying molecular orbital levels.

The partial density-of-states (PDOS) goes a step further by introducing an atomic
orbital decomposition of the DOS.  Thus the PDOS for the $\mu$th atomic orbital
is,
\begin{equation}
  \mbox{PDOS}_\mu(\epsilon) = \sum_i q_{\mu,i} g(\epsilon-\epsilon_i) \, .
  \label{eq:results.13}
\end{equation} 
Here the quantity $q_{\mu,i}$ is the Mulliken atomic charge of atomic orbital $\mu$
in molecular orbital $i$.  We obtain the ruthenium $d$ PDOS by summing PDOS$_\mu$
over all $d$-type atomic orbitals on ruthenium.  Similarly we obtain the $\pi$ PDOS
by summing PDOS$_\mu$ over all the $p$-type atomic orbitals on the heavy atoms (e.g., 
on C, N, and O) on the ligands.  As seen in Fig.~\ref{fig:pdos}, the approximate energies 
of the $t_{2g}$ and $e^*_g$ orbitals on the ruthenium clearly emerge for complex
{\bf (24)} with the expected peak height ratio of 3:2.  We also see a loss of 
$t_{2g}$ degeneracy for complex {\bf (18)} due to breaking of perfect octahedral
symmetry in this complex as well as some small seemingly random $d$-orbital density
contributing to molecular orbitals at other energies.

% ------------------------------------------------------
\begin{figure}
\begin{tabular}{l}
(a) \\
\includegraphics[width=0.45\textwidth]{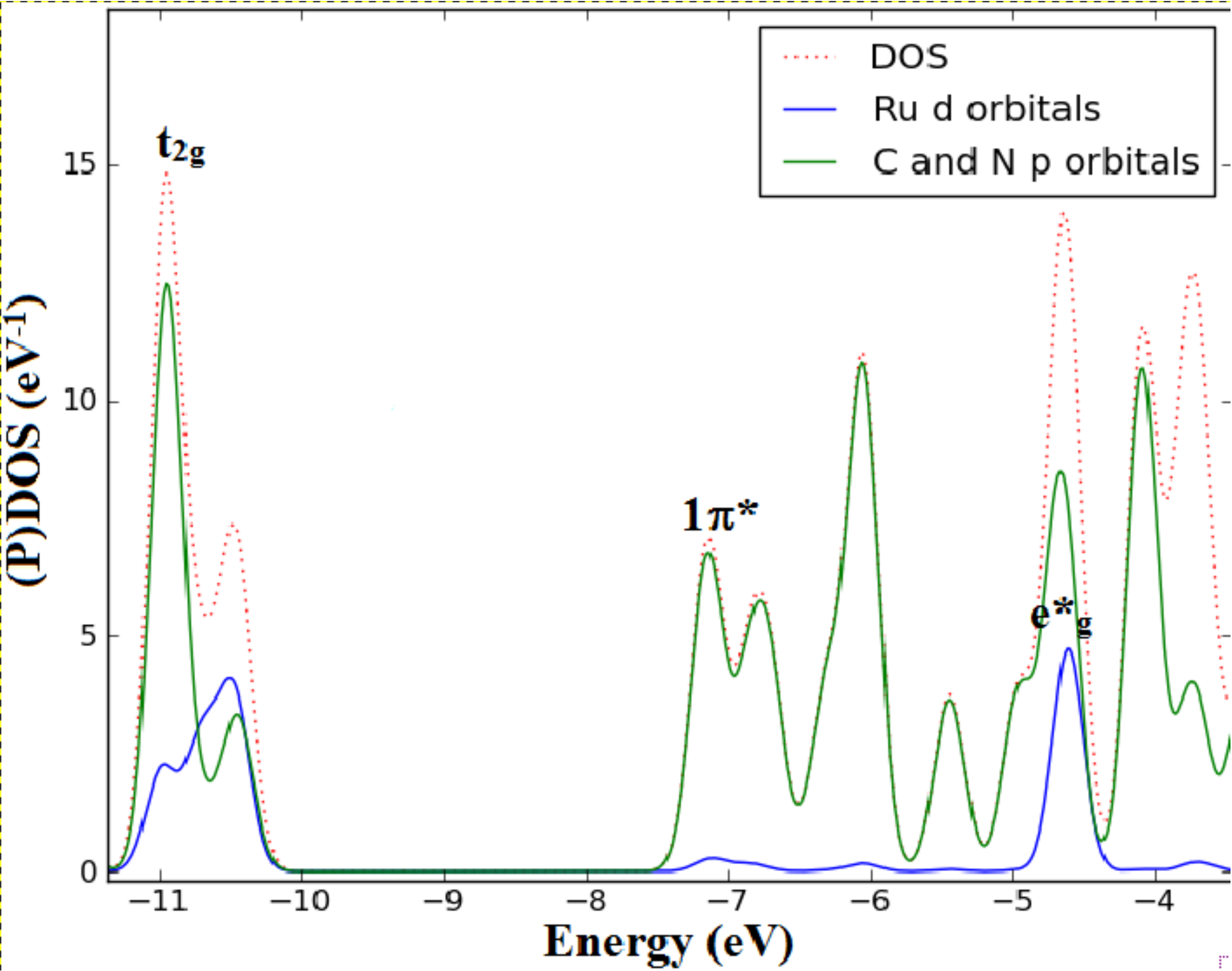} \\  % cmplx018_pdos_631g.pdf
(b) \\
\includegraphics[width=0.45\textwidth]{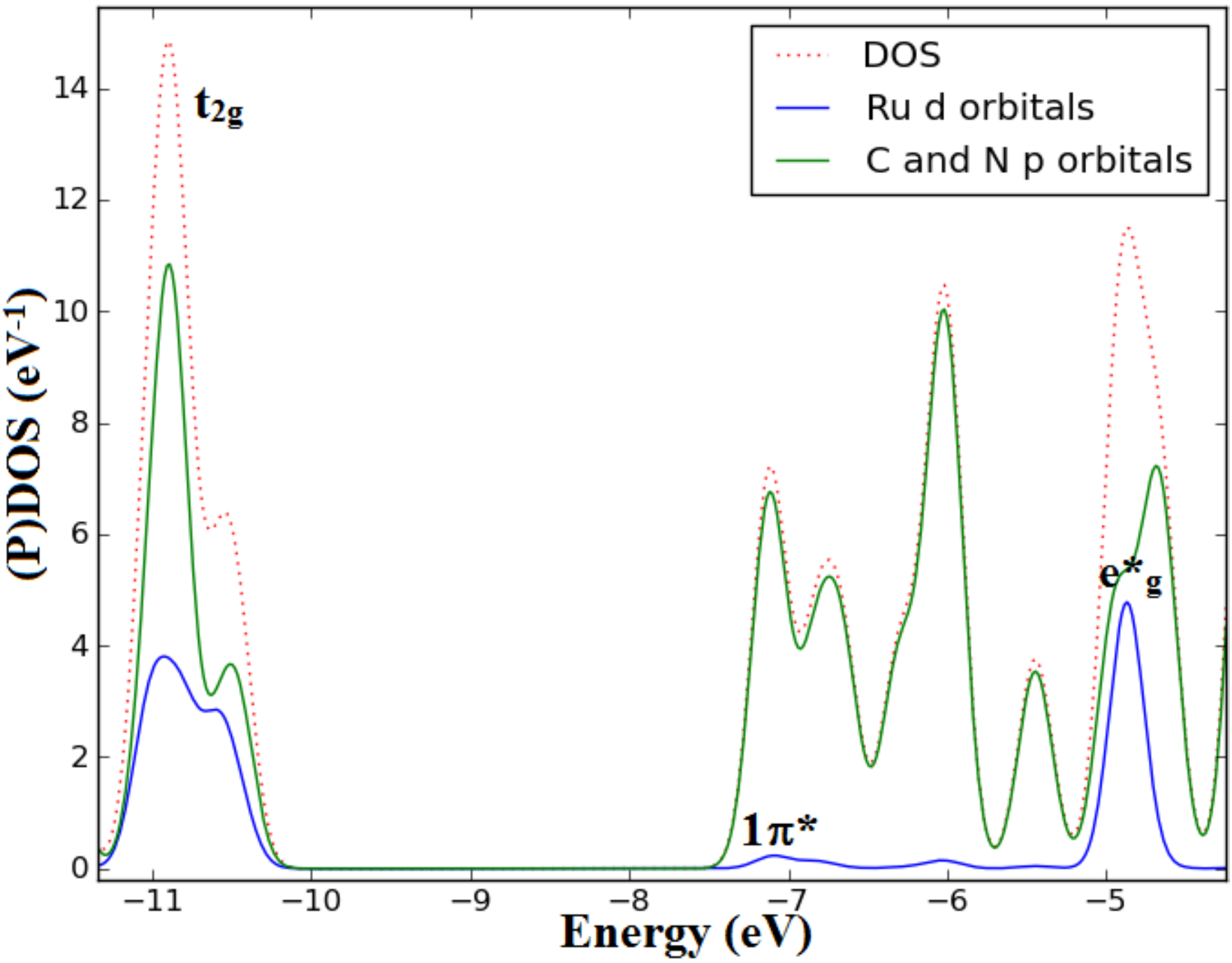} \\ % cmplx018_pdos_631gd.pdf
\end{tabular}
\caption{
B3LYP (P)DOS calculated for complex {\bf (18)}: (a) 6-31G basis and (b) 6-31G(d) basis.
\label{fig:pdos2}
}
\end{figure}
% ------------------------------------------------------
It should be noted that the PDOS analysis, while highly useful, also contains
a degree of arbitrariness.  In the first place, the precise picture will vary as
the FWHM is varied.  This is why it is best to use a fixed value of the FWHM as
we do in this paper.  Also, the PDOS shares the basis-dependence of the Mulliken
analysis.  This is illustrated in Fig.~\ref{fig:pdos2} where the $e^*_g$ peak 
shifts slightly relative to the $\pi^*$ peaks when going from the 6-31G to 
the 6-31G(d) basis set.  Many other examples allowing the comparison of the 
PDOS calculated with the two different basis sets for a wide variety of 
complexes may be found in the Supplementary Information and provide further 
evidence for slight basis-set dependent shifts in the PDOS.  However an 
important exception is in the case of unbound (i.e., positive energy) 
orbitals where a finite basis set is trying to describe a continuum.  
These cases are marked with an asterisk (*) in the Supplementary Information 
and can show very great differences between the position and character of 
the $t_{2g}$ and $e^*_g$ PDOS peaks in going from the 6-31G to the 6-31G(d) 
basis sets, such as is the case for complex {\bf (7)}* where there is a 
simple $t_{2g}$ peak in the PDOS calculated with the 6-31G basis set
and a triple $t_{2g}$ peak in the PDOS calculated with the 6-31G(d) basis set.

% ------------------------------------------------------
\begin{figure}
\includegraphics[width=0.45\textwidth]{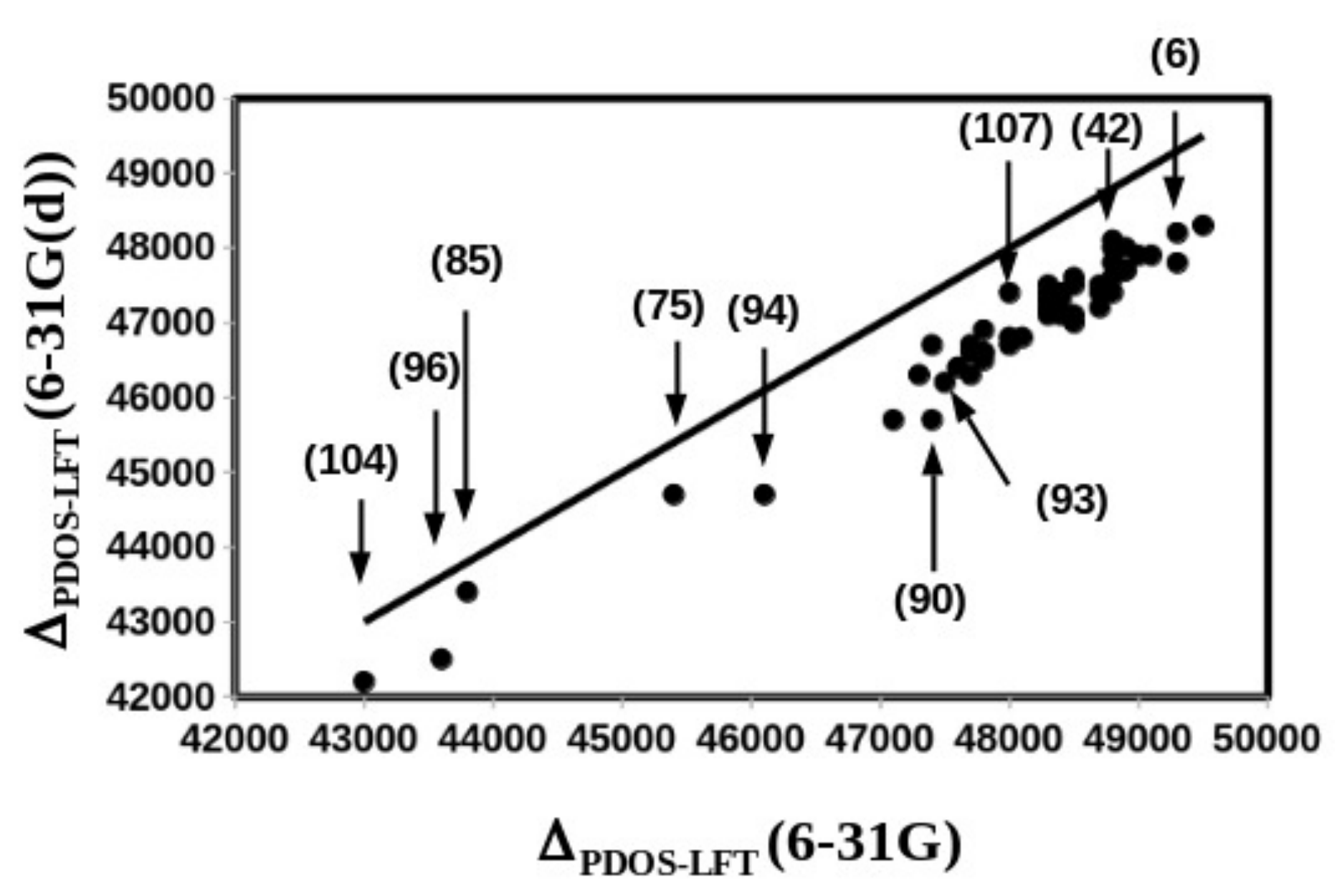}  % DeltaGap.pdf
\caption{
Correlation plot between $\Delta_{\mbox{PDOS-LFT}}$ calculated in cm$^{-1}$
with the 6-31G and 6-31G(d) basis sets for 55 complexes.  
The diagonal line indicates where 
points should lie in the event of hypothetical perfect agreement between 
the two sets of results.  A least squares fit to the calculated points gives
the equation $\Delta_{\mbox{PDOS-LFT}}$(6-31G(d)) = 
0.940 $\Delta_{\mbox{PDOS-LFT}}$(6-31G) + 1740. cm$^{-1}$.  
Complexes whose data points are marked: {\bf (6)} [Ru(bpy)$_3$]$^{2+}$, 
{\bf (42)} [Ru(bpy)$_2$(BL5)]$^{2+}$, {\bf (107)} [Ru(i-biq)$_2$(BL5)]$^{2+}$,
{\bf (93)} [Ru(hpiq)$_3$]$^{2+}$, {\bf (75)} [Ru(6,6'-dm-bpy)$_3$]$^{2+}$,
{\bf (85)} [Ru(2,9-dm-phen)$_3$]$^{2+}$, {\bf (96)} [Ru(pq)(biq)$_2$]$^{2+}$,
and {\bf (104)} [Ru(biq)$_3$]$^{2+}$. 
\label{fig:DeltaGap}
}
\end{figure}
% ------------------------------------------------------
% \input{tables/LFT.tex}
% =======================================
% file: LFT.tex
% last updated: 21 April 2017
% =======================================

\begin{table}
\caption{$\Delta_{\text{PDOS-LFT}}$ (cm$^{-1}$) for complexes with 
         simple $t_{2g}$ and $e^*_g$ peaks.
         \label{tab:LFT_1}}
\begin{tabular}{ccc}
\hline \hline
number      & 6-31G       & 6-31G(d) \\  % 1 eV = 8 065.73 cm-1
\hline
({\bf 6})   & 49 300.     & 48 200.  \\
({\bf 8})   & 48 800.     & 48 100.  \\
({\bf 9})   & 48 900.     & 48 000.  \\
({\bf 11})  & 49 300.     & 47 800.  \\
({\bf 13})  & 49 100.     & 47 900.  \\
({\bf 14})  & 47 800.     & 46 900.  \\
({\bf 15})  & 49 500.     & 48 300.  \\
({\bf 16})  & 48 800.     & 47 800.  \\
({\bf 17})  & 48 300.     & 47 400.  \\
({\bf 19})  & 47 800.     & 46 500.  \\
({\bf 20})  & 48 800.     & 47 400.  \\
({\bf 23})  & 48 400.     & 47 400.  \\
({\bf 24})  & 48 400.     & 47 300.  \\
({\bf 25})  & 48 400.     & 47 100.  \\
({\bf 26})  & 48 300.     & 47 200.  \\
({\bf 27})  & 48 000.     & 46 800.  \\
({\bf 28})  & 47 700.     & 46 700.  \\
({\bf 29})  & 48 700.     & 47 500.  \\
({\bf 30})  & 48 500.     & 47 500.  \\
({\bf 31})  & 48 500.     & 47 000.  \\
({\bf 32})  & 48 000.     & 46 700.  \\
({\bf 34})  & 49 000.     & 47 900.  \\
({\bf 40})  & 48 700.     & 47 400.  \\
({\bf 41})  & 48 900.     & 47 700.  \\
({\bf 42})  & 48 800.     & 48 000.  \\
({\bf 46})  & 48 900.     & 47 700.  \\
({\bf 47})  & 48 800.     & 47 600.  \\
({\bf 48})  & 48 700.     & 47 200.  \\
({\bf 50})  & 48 300.     & 47 100.  \\
\hline \hline                                                 
\end{tabular}
\end{table}

% ----------------------------------------------------------------

\begin{table}
\caption{$\Delta_{\text{PDOS-LFT}}$ (cm$^{-1}$) for complexes with 
         simple $t_{2g}$ and $e^*_g$ peaks.
         \label{tab:LFT_2}}
\begin{tabular}{ccc}
\hline \hline
number       & 6-31G & 6-31G(d) \\
\hline
\hline \hline                                                 
({\bf 52})  & 48 100.     & 46 800.  \\
({\bf 53})  & 47 400.     & 46 700.  \\
({\bf 55})  & 47 100.     & 45 700.  \\
({\bf 56})  & 47 700.     & 46 600.  \\
({\bf 57})  & 47 300.     & 46 300.  \\
({\bf 58})  & 47 600.     & 46 400.  \\
({\bf 70})  & 48 300.     & 47 500.  \\
({\bf 71})  & 47 800.     & 46 600.  \\
({\bf 75})  & 45 400.     & 44 700.  \\
({\bf 76})  & 48 400.     & 47 600.  \\
({\bf 77})  & 48 300.     & 47 300.  \\
({\bf 78})  & 47 800.     & 46 600.  \\
({\bf 85})  & 43 800.     & 43 400.  \\
({\bf 87})  & 47 700.     & 46 300.  \\
({\bf 90})  & 47 400.     & 45 700.  \\
({\bf 93})  & 47 500.     & 46 600.  \\
({\bf 94})  & 46 100.     & 44 700.  \\
({\bf 96})  & 43 600.     & 42 500.  \\
({\bf 97})  & 48 500.     & 47 100.  \\
({\bf 104}) & 43 000.     & 42 200.  \\
({\bf 107}) & 48 000.     & 47 400.  \\
\hline \hline
\end{tabular}
\end{table}

%%%%%
% EOF
%%%%%
We thus have a ligand-field theory (LFT) like PDOS-LFT picture.  However it 
is {\em not} LFT as the PDOS-LFT splitting $\Delta_{\text{PDOS-LFT}}$ 
calculated as the energy difference between the $e^*_g$ and $t_{2g}$ PDOS 
peaks is not the same as the $\Delta_{\text{LFT}}$ expected from LFT.   
We can see this by comparing numbers for the much studied complex {\bf (6)}.
According to our calculations, 
$\Delta_{\text{PDOS-LFT}} = $49 300 cm$^{-1}$ calculated with the 6-31G basis
set and 48 200 cm$^{-1}$ calculated with the 6-31G(d) basis set. This can
be compared with the value $\Delta_{\text{PDOS-LFT}} = $48 000 cm$^{-1}$ and
with $\Delta_{\text{LFT}} = $28 600 cm$^{-1}$ reported previously \cite{WJL+14}.
Clearly $\Delta_{\text{PDOS-LFT}}$ is much larger than $\Delta_{\text{LFT}}$
so that PDOS-LFT is different from the usual LFT.

Tables~\ref{tab:LFT_1} and \ref{tab:LFT_2} show the values
of $\Delta_{\text{PDOS-LFT}}$ for complexes sufficiently close to octahedral
symmetry to show simple $t_{2g}$ and $e^*_g$ PDOS peaks, excluding complexes
where the $e^*_g$ peak is unbound.  Figure~\ref{fig:DeltaGap} provides a
graphical comparison of how $\Delta_{\text{PDOS-LFT}}$ changes in going from the
6-31G to the 6-31G(d) basis set.  The correlation is linear up to some residual
scatter which can be explained by the precision of the graphical measurement
of the distance between peaks.  In general, the $\Delta_{\text{PDOS-LFT}}$ splitting % gap 
closes a bit when the larger basis set is used compared to the smaller basis 
set.  Although $\Delta_{\text{PDOS-LFT}}\neq\Delta_{\text{LFT}}$, we do 
expect them to have the same trends and so to be able to establish 
spectrochemical series.  Thus from Fig.~\ref{fig:DeltaGap} we may,
for example, deduce the following relationship for ligand field strength:
\begin{eqnarray}
 \Delta &:& \mbox{bpy $>$ i-biq $>$ hpiq $>$ DPAH $>$ pq} \nonumber \\
            &>& \mbox{6,6'-dm-bpy $>$ 2,9-dm-phen $>$ biq}
 \, . \nonumber \\
 \label{eq:results.14}
\end{eqnarray}
Ligand abbreviations are defined in Appendix~\ref{sec:ligands}.

% ------------------------------------------------------
\begin{figure}
\includegraphics[width=0.45\textwidth]{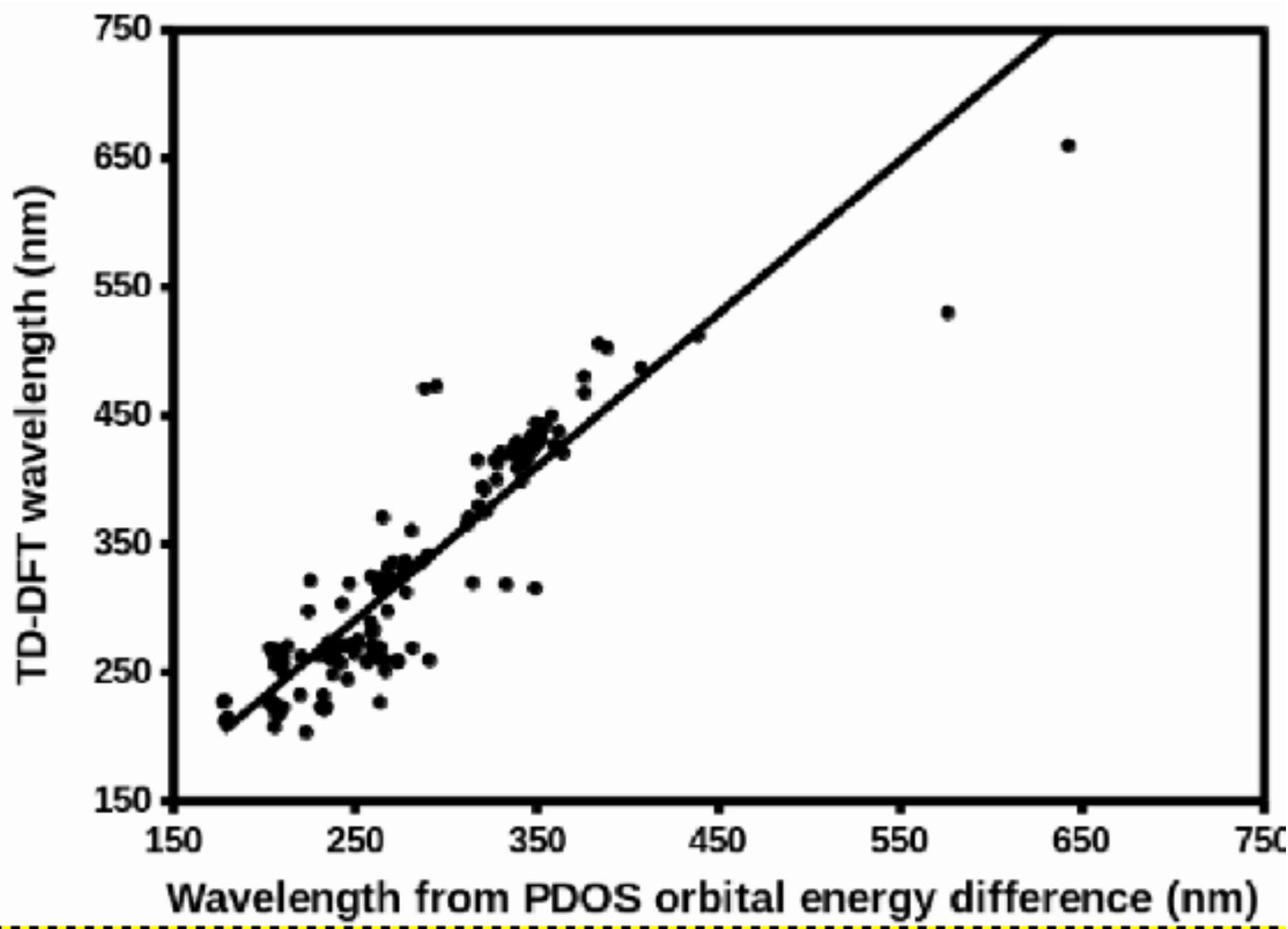} %  orbitalstatecorr.pdf
\caption{
Correlation plot between $t_{2g} \rightarrow \pi^*$ PDOS orbital energy
differences calculated using the 6-31G basis set and TD-B3LYP(6-31G)
absorption spectra peaks.  A least squares fit to the 161 data points
gives the line indicated on the graph whose equation is 
$\lambda(\omega_S) = 1.19 \lambda(\epsilon_{\pi^*} - \epsilon_{t_{2g}}) - 
+ \mbox{7.63 nm}$.
\label{fig:orbitalstatecorr}
}
\end{figure}
% ------------------------------------------------------
Since the usual LFT splitting $\Delta_{\text{LFT}}$ is extracted
from absorption spectra \cite{FH00}, it is interesting to see to what extent
energy differences between PDOS-LFT peaks correlate with the 
position of TD-DFT absorption spectra peaks.  Often times, TD-DFT
results may be analyzed within the two-orbital two-electron model
(TOTEM) (See, e.g., the review Ref.~\cite{CH12}).  Let us consider a simpler hybrid functional,
\begin{equation}
  E_{xc}^{\text{Hybrid}} = (1-a_0) E_x^{\text{GGA}} + a_0 E_x^{\text{HF}}
    + E_c^{\text{GGA}} \, ,
 \label{eq:results.15}
\end{equation}
than the B3LYP functional [Eq.~(\ref{eq:details.1.5})] as it already
captures all the essential features which are of interest to us here.
In the Tamm-Dancoff approximation (See, e.g., the review Ref.~\cite{CH12}), 
the TOTEM model gives the following formulae for the singlet $\omega_S$ and 
triplet $\omega_T$ excitation energies:
\begin{eqnarray}
 \omega_S^{\text{Hybrid}} & = & \epsilon_a - \epsilon_i + 
  2 ( ia \vert f_H \vert ai ) \nonumber \\
  & - & a_0 (ii \vert f_H \vert aa) 
   +  (1-a_0) (ia \vert f_x^{\alpha,\alpha} \vert ai) \nonumber \\
  & + & 
  a_c (ia \vert f_c^{\alpha, \alpha} + f_c^{\alpha,\beta} \vert ai) \nonumber \\
 \omega_T^{\text{Hybrid}} & = & \epsilon_a - \epsilon_i \nonumber \\
  & + & (1-a_0) (ia \vert f_x^{\alpha,\alpha} \vert ai) \nonumber \\
  & + & 
  a_c (ia \vert f_c^{\alpha, \alpha} - f_c^{\alpha,\beta} \vert ai) \, ,
 \label{eq:results.16}
\end{eqnarray}
where we follow the notation of Ref.~\cite{CH12}.  If the two-electron
integrals are (or their sum is) sufficiently small, then we may expect that excitation energies
may be approximated, albeit rather roughly, as orbital energy differences:
\begin{equation}
  \omega \approx \epsilon_a - \epsilon_i \, .
  \label{eq:results.17}
\end{equation}
This was checked by taking the same complexes treated in 
Fig.~\ref{fig:DeltaGap} and comparing the wavelength corresponding to the
$\epsilon_{t_{2g}} \rightarrow \epsilon_{\pi^*}$ transitions with the 
wavelength of the corresponding peaks in the TD-B3LYP absorption spectra  
in Fig.~\ref{fig:orbitalstatecorr}.
A least squares fit indicates quite a good correlation in the sense that 
the slope is only slightly greater than unity and the intercept is small.
However there is a large scatter of the data points around the fit line
which may be due to neglect of two-electron integrals but may equally well
be due to difficulty assigning the precise positions of PDOS peaks and of
peaks (and particularly of shoulders) in the TD-B3LYP spectra.  Nevertheless
we find the figure to be quite encouraging  in that the figure suggests that a PDOS-LFT
orbital model may provide useful insight into the behavior of excited states.

% -----------------------------------
\subsection{Luminescence Lifetimes}
\label{sec:luminesce}
% -----------------------------------

% ------------------------------------------------------
\begin{figure}
\includegraphics[width=0.45\textwidth]{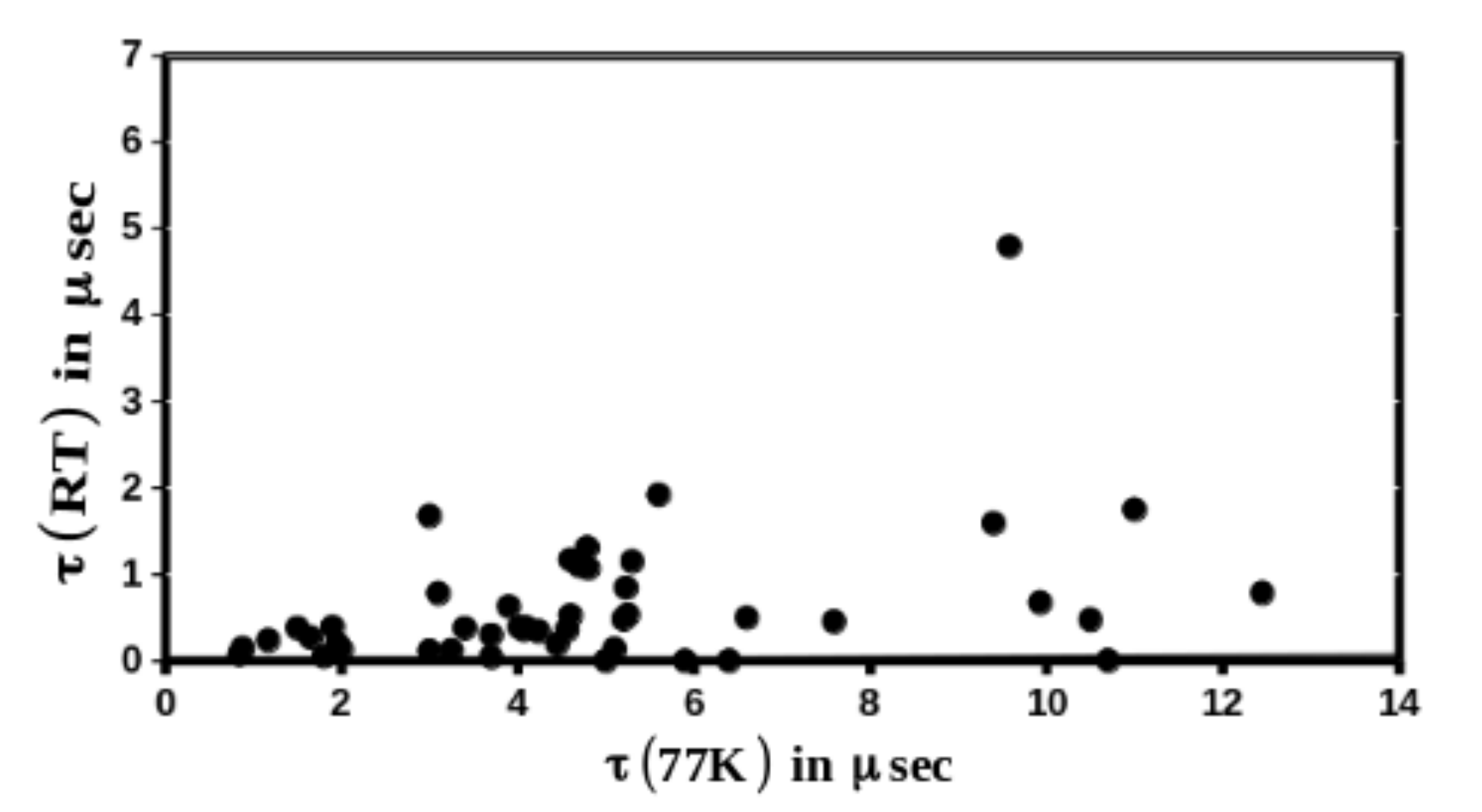}  % lifetimes1.pdf
\caption{
Correlation between luminescence lifetimes at room temperature (RT)
and at liquid nitrogen temperature (77K).
\label{fig:lifetimes}
}
\end{figure}
% ------------------------------------------------------
% \input{./tables/lumdata.tex}
% =======================================
% file: lumdata.tex
% last updated: 22 May 2017   
% =======================================

\begin{table}
\caption{Compounds with both room temperature (RT) and liquid nitrogen 
         temperature (77K) data from Table 1 of Ref.~\cite{JBB+88}.
         An asterisk has been added if the PDOS $e_g^*$ orbital is 
         unbound. Luminescence times are averages over different measurements
         in different solvents.  See text for the definition of 
         $\Delta E_{\text{ave}}$.
         \label{tab:lumdata1}}
\begin{tabular}{cccc}
\hline \hline
number         & $\tau$(77K) & $\tau$(RT) & $\Delta E_{\text{ave}}$  \\
               & $\mu$s      & $\mu$s     & cm$^{-1}$            \\
\hline
({\bf 1})*     & 3.7         & 0.043      & 321.             \\
% ({\bf 2)}    & 0.43        &            &                         \\
({\bf 3)}*     & 3.7         & 0.30       & 181.              \\
({\bf 4)}      & 0.84        & 0.070      & 179.              \\
% ({\bf 5})*   & 0.60        &            &                         \\
({\bf 6)}      & 5.23        & 0.845      & 132.             \\
({\bf 7)}*     & 3.1         & 0.78       & 100.              \\
({\bf 8})      & 5.25        & 0.533      & 165.              \\
({\bf 9})      & 5.2         & 0.48       & 172.              \\
% ({\bf 11})   & 2.5         &            &                         \\
({\bf 12})     & 5.6         & 1.92       & 77.               \\
({\bf 13})     & 4.6         & 1.17       & 99.              \\
({\bf 14})     & 7.59        & 0.454      & 203.             \\
({\bf 15})     & 3.4         & 0.378      & 158.              \\
({\bf 16})     & 6.6         & 0.497      & 186.             \\
% ({\bf 17})   & 5.6         &            &                         \\
({\bf 18})     & 9.4         & 1.591      & 128.             \\
({\bf 19})     & 3.9         & 0.628      & 132.              \\
% ({\bf 20})   & 6.1         &            &                         \\
({\bf 21})     & 5.9         & 0.00007    & 818.              \\
({\bf 22})     & 5.1         & 0.137      & 261.             \\
({\bf 23})     & 1.8         & 0.05       & 259.              \\
% ({\bf 24})   & 11.4        &            &                         \\
% ({\bf 25})   & 10.6        &            &                         \\
% ({\bf 26})   & 5.6         &            &                         \\
% ({\bf 27})   & 7.2         &            &                         \\
% ({\bf 28})   & 7.5         &            &                         \\
({\bf 29})     & 4.450       & 0.192      & 227.             \\
({\bf 30})     & 4.240       & 0.340      & 182.             \\
({\bf 31})     & 3.25        & 0.121      & 237.             \\
({\bf 32})     & 3.000       & 0.115      & 235.              \\
% ({\bf 34})   & 1.7         &            &                         \\
% ({\bf 35})   & 2.3         &            &                         \\
% ({\bf 36})   & 3.05        &            &                         \\
({\bf 37})     & 1.5         & 0.38       & 99.              \\
% ({\bf 38})   & 2.6         &            &                         \\
({\bf 39})     & 1.65        & 0.27       & 131.              \\
({\bf 40})     & 4.7         & 1.1        & 105.            \\
\hline \hline
\end{tabular}
\end{table}

% ---------------------------------------------

\begin{table}
\caption{Compounds with both room temperature (RT) and liquid nitrogen 
         temperature (77K) data from Table 1 of Ref.~\cite{JBB+88}.
         An asterisk has been added if the PDOS $e_g^*$ orbital is 
         unbound. Luminescence times are averages over different measurements
         in different solvents. See text for the definition of 
         $\Delta E_{\text{ave}}$.
         \label{tab:lumdata2}}
\begin{tabular}{cccc}
\hline \hline
number         & $\tau$(77K) & $\tau$(RT) & $\Delta E_{\text{ave}}$  \\
               & $\mu$s      & $\mu$s     & cm$^{-1}$            \\
\hline
({\bf 41})     & 4.560       & 0.356      & 184.                   \\
({\bf 42})     & 4.090       & 0.390      & 170.                    \\
({\bf 43})     & 4.020       & 0.389      & 169.                   \\
({\bf 44})     & 4.070       & 0.360      & 175.              \\
({\bf 46})     & 4.8         & 1.07       & 108.              \\
% ({\bf 47})   & 4.9         &            &                         \\
({\bf 48})     & 12.45       & 0.784      & 200.                  \\
% ({\bf 50})   &             &            &                         \\
({\bf 52})     & 2.0         & 0.13       & 197.                 \\
% ({\bf 53})   & 10.9        &            &                         \\
% ({\bf 55})   & 10.9        &            &                         \\
% ({\bf 56})   & 10.6        &            &                         \\
% ({\bf 57})   & 11.3        &            &                         \\
% ({\bf 58})   & 9.0         &            &                         \\
({\bf 60})     & 1.9         & 0.39       & 114.             \\
({\bf 61})     & 1.95        & 0.20       & 164.              \\
% ({\bf 63})   & 2.4         &            &                         \\
({\bf 64})*    & 5.0         & 0.001      & 615.              \\
% ({\bf 66})*  & 4.1         &            &                         \\
({\bf 67})*    & 6.4         & 0.21       & 632.              \\
% ({\bf 69})   & 9.4         &            &                         \\
({\bf 70})     & 4.6         & 0.525      & 157.              \\
({\bf 71})     & 10.50       & 0.475      & 223.              \\
({\bf 73})     & 4.79        & 1.31       & 94.               \\
({\bf 74})     & 5.3         & 1.15       & 110.              \\
% ({\bf 75})   & 2.5         &            &                         \\
% ({\bf 76})   & 5.5         &            &                         \\
({\bf 77})     & 9.93        & 0.673      & 194.              \\
({\bf 78})     & 3.0         & 1.675      & 42.                \\
% ({\bf 79})   & 3.45        &            &                         \\
% ({\bf 80})   & 2.0         &            &                         \\
% ({\bf 81})   & 1.95        &            &                         \\
% ({\bf 82})   & 3.8         &            &                         \\
% ({\bf 83})   & 2.3         &            &                         \\
% ({\bf 84})   & 5.9         &            &                         \\
% ({\bf 85})   & 2.3         &            &                         \\
({\bf 86})     & 9.58        & 4.796      & 50.               \\
({\bf 87})     & 11.0        & 1.750      & 132.                   \\
% ({\bf 88})*  & 5.9         &            &                         \\
% ({\bf 89)}   & 6.          &            &                         \\
% ({\bf 90})   & 2.0         &            &                         \\
% ({\bf 91})   & 0.01        &            &                         \\
% ({\bf 92})   & 0.01        &            &                         \\
% ({\bf 93})   & 1.1         &            &                         \\
% ({\bf 94})   & 3.4         &            &                         \\
% ({\bf 95})   & 2.4         &            &                         \\
% ({\bf 96})   & 2.0         &            &                         \\
% ({\bf 97})   & 0.50        &            &                         \\
% ({\bf 98})   & 0.72        &            &                         \\
({\bf 99})*    & 1.17        & 0.167      & 115.               \\
% ({\bf 100})  & 1.8         &            &                         \\
% ({\bf 101})  & 0.18        &            &                         \\
% ({\bf 102})  & 0.46        &            &                         \\
({\bf 103})*   & 0.88        & 0.147      & 129.             \\
% ({\bf 104})  & 2.6         &            &                         \\
% ({\bf 105})  & 2.2         &            &                         \\
({\bf 106})*   & 177.        & 0.237      & 477.              \\
({\bf 107})    & 96.0        & 0.1475     & 467.              \\
({\bf 108})    & 10.7        & 0.0037     & 575.              \\
% ({\bf 109})  & 12.3        &            &                         \\
% ({\bf 110})  & 7.8         &            &                         \\
% ({\bf 111})* & 5.185       &            &                         \\
\hline \hline
\end{tabular}
\end{table}

%%%%%
% EOF
%%%%%
Since gas-phase B3LYP geometries are a good indicator of ruthenium complex
crystal geometries, gas-phase TD-B3LYP geometries are a reasonable indicator
of ruthenium complex absorption spectra in solution, and PDOS-LFT energies
provide a first approximation to absorption spectra energies, then we may
also hope to be able to say something about ruthenium complex luminescence
lifetimes on the basis of PDOS-LFT information.  Indeed this was the reasoning
given in the seminal paper \cite{WJL+14} where PDOS-LFT luminescence indices
were proposed upon the basis of the idea that the room temperature (RT) 
luminescence lifetime should increase with the height of the 
$^3$MLCT $\rightarrow$ $^3$MC barrier shown in Fig.~\ref{fig:PES}.
This barrier-height dependence would also imply a strong temperature
dependence which is indeed seen in the 48 liquid nitrogen (77 K) and
room temperature (RT) values in Tables~\ref{tab:lumdata1} and \ref{tab:lumdata2} 
and the 46 points in Fig.~\ref{fig:lifetimes}.

In order to see where PDOS-LFT-derived luminescence indices may be able to
say something about luminescence lifetimes, we need first to understand the various
contributions to luminescence lifetimes.
Luminescence lifetime experiments measure the decay rate of the intensity
of light luminescing at a particular wavelength as a function of time.
This gives a temperature ($T$) dependent decay constant $k(T)$ which is
related to the decay lifetime $\tau(T)$ by,
\begin{equation}
   k(T) = \frac{1}{\tau(T)} \, .
   \label{eq:results.18}
\end{equation}
The luminescence lifetime determined from the decay rate of measured
intensity is a measure of the rate of disappearance of the luminescent
species --- in this case, the phosphorescent $^3$MLCT state.  In addition
to phosphorescence, other physical phenomena are also included in the
decay lifetime $\tau(T)$ which generally depend  upon the temperature $T$.
The decay rate constant may be separated,
\begin{equation}
   k(T) = k_0 + k_b^{nr}(T) \, ,
   \label{eq:results.19}
\end{equation}
into a temperature-independent part,
\begin{equation}
   k_0 = k^{r} + k_a^{nr} \, ,
   \label{eq:results.20}
\end{equation}
where the superscript ``r'' refers to ``radiative'' and the superscript
``nr'' refers to ``nonradiative'' \cite{MAB+95}.  The temperature-independent
part describes processes which continue to be operational even at very low
temperatures.  ($k_0$ is assumed to be equal to $k(T)$ at $T$ = 84 K 
in Ref.~\cite{BBZ+85}.)  The temperature-dependent part may be
further separated as \cite{JBB+88},
\begin{equation}
  k_b^{nr}(T) = k_{\text{melt}}(T) + k_{\text{equlib}}(T)
       + k_{\text{barrier}}(T) \, ,
  \label{eq:results.21}
\end{equation}
where,
\begin{equation}
  k_{\text{melt}}(T) = \frac{B}{1+\exp \left[ C \left( \frac{1}{T} 
    - \frac{1}{T_B} \right) \right]}
  \label{eq:results.22}
\end{equation}
describes the melting of the solid matrix of the solution at low
temperature, where $k_{\text{melt}}(T)$ = constant $B$ for T$\rightarrow \infty$ and 
$k_{\text{melt}}(T)=0$ for T$\rightarrow$0. $T_{B}$ is the temperature at which 
$k_{\text{melt}}(T)$=B/2 and C is a temperature related to the viscosity effect;
\begin{equation}
  k_{\text{equlib}}(T) = A_1 e^{-\Delta E_1/RT} \, ,
  \label{eq:results.23}
\end{equation}
describes thermal equilibrium with higher energy states of the same
electronic nature (e.g., states with the same symmetry in an octahedral
complex according to LFT but which are split with ligands giving only
pseudo-octahedral symmetry), and
\begin{equation}
  k_{\text{barrier}}(T) = A_2 e^{-\Delta E_2/RT} \, ,
  \label{eq:results.24}
\end{equation}
is an Arrhenius term describing crossing of the $^3$MLCT $\rightarrow$ $^3$MC
barrier prior to subsequent de-activation to $^1$GS.  As pointed out
in Ref.~\cite{BJB+87}, $\Delta E_2$ is only the $^3$MLCT $\rightarrow$
$^3$MC activation energy barrier when $k_c >> k_b$ [see Eqs.~(\ref{eq:intro.3}) and
(\ref{eq:intro.4})], but the situation becomes more complicated if (for example) 
$k_b >> k_c$.  Putting it altogether results in,
\begin{eqnarray}
  k & = & k_0 + k_{\text{melt}}(T) + k_{\text{equlib}}(T) + k_{\text{barrier}}(T)
           \nonumber \\
           & = & k_0 + \frac{B}{1+\exp\left[C\left(\frac{1}{T}-\frac{1}{T_B}\right)\right]} 
           \nonumber \\
           & + & A_1 e^{-\Delta E_1/RT} + A_2 e^{-\Delta E_2/RT}
  \, .
  \label{eq:results.25}
\end{eqnarray}
The barrier term is commonly believed to dominate over the other
terms at high-enough temperatures.  If so, then we may hope to be able
to relate $\Delta E_2$ to the features of the PDOS-LFT theory.  

% \input{./tables/Rubpy3param.tex}
% ==============================================
% File: Rubpy3param.tex
% Last modified: 22 May 2016
% ==============================================

\begin{table}
\centering
\caption{
Parameters describing the temperature dependence of the luminescence
decay rate of [Ru(bpy)$_3$]$^{2+}$ in propionitrile/buylronitile 
(4:5 v/v) from p.~108 of Ref.~\cite{JBB+88}, except: $C$ and $T_B$
were determined by variation within the recommended range until we 
obtained results similar to those in Fig.~6 of Ref.~\cite{BBZ+85}.
$A_{\text{ave}}$ and $\Delta E_{\text{ave}}$ are calculated as
explained in the text.
\label{tab:Rubpy3param}
}
\begin{tabular}{cc}
\hline 

\hline
parameter & value \\
\hline

\hline

$k_0$ & 2 $\times$ 10$^5$ s$^{-1}$ \\
$B$   & 2.1 $\times$ 10$^5$ s$^{-1}$ \\
$C$   & 1900 \\
$T_B$ & 125 K \\
$A_1$ & 5.6 $\times$ 10$^5$ s$^{-1}$ \\
$\Delta E_1$ & 90 cm$^{-1}$ \\
$A_2$ & 1.3 $\times$ 10$^{14}$ s$^{-1}$ \\
$\Delta E_2$ & 3960 cm$^{-1}$ \\
$A_{ave}$ & 2.707 $\times$ 10$^6$ s$^{-1}$ \\
$\Delta E_{ave}$ & 159.98 cm$^{-1}$ \\ 
\hline 

\hline
\end{tabular}
\end{table}

%%%%%
% EOF
%%%%%
% ------------------------------------------------------
\begin{figure}[!h]
\includegraphics[width=0.45\textwidth]{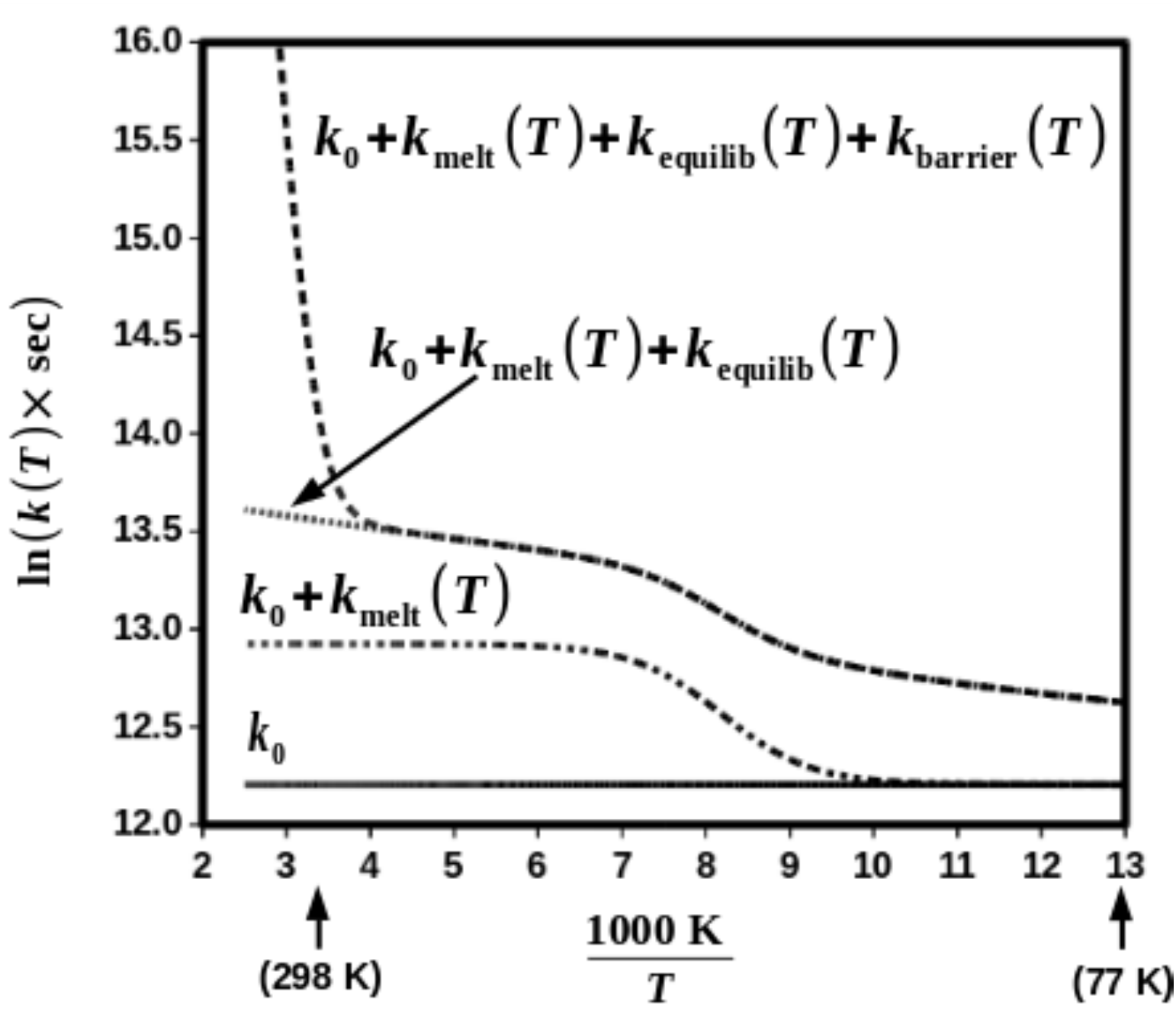} % loginverseT2.pdf
\caption{
\label{fig:loginverseT}
Plot of $\ln (k)$ versus $1/T$ for luminescence decay rates for 
[Ru(bpy)$_3$]$^{2+}$ in propionitile/butyronitrile (4:5 v/v).
}
\end{figure}
% ------------------------------------------------------
But how high a temperature is high-enough to make this hope reasonable?  
% It will be seen that it is not generally possible to obtain the barrier activation 
% energy $\Delta E_2$ based upon what is typically experimentally available and so 
% there will be need to have to accept a very crude compromise based upon what is typically 
% available namely decay constants (or times) available only at room temperature 
% (RT) or liquid nitrogen temperature (77 K).
% The first step to seeing this entails 
We can get some idea of the answer to this question by
examination of the relative importance of 
the different terms in Eq.~(\ref{eq:results.21}) for [Ru(bpy)$_3$]$^{2+}$ in 
propionitrile/butyronitrile (4:5 v/v) using the parameters given
in Table~\ref{tab:Rubpy3param}.  
Data is often plotted as $\ln (k)$ versus $1/T$ as shown in Fig.~\ref{fig:loginverseT}.  
A look at the different contributions on the excited state lifetimes is also shown
on the same plot.  It looks very different on different scales as different physical 
effects come into play in different temperature regimes.  Only $k_0$ is important
below about 30 K. From about 30 K to 100 K, $k_{\text{equilib}}$ becomes important.  The 
melting term $k_{\text{melt}}$ switches on from about 100 K to about 250 K.  After 250 K, 
$k_{\text{barrier}}$ rapidly begins to dominate.  Unfortunately $k_{\text{barrier}}$ is 
not the single overwhelmingly dominant term at RT (298 K).
% % ------------------------------------------------------
% \begin{figure} 
% \includegraphics[width=0.50\textwidth]{graphics/decaycurve300new.pdf}
% \caption{
% \label{fig:Rubpy3decay}
% Contributions of different terms to the temperature dependance of
% luminescence decay rates for [Ru(bpy)$_3$]$^{2+}$ in propionitrile/butyronitrile
% (4:5 v/v).
% }
% \end{figure}
% % ------------------------------------------------------

This means that it is very difficult to extract an accurate value of the
triplet barrier energy $\Delta E_2$ from only the luminescence decay
constants at 77 K and at RT.  We have tried various ways to do so, but
all of them suffer from some sort of numerical instability resulting from trying to
get a relatively small number from taking the difference of two large numbers.
Improved computational precision would not solve this problem because the
accuracy of the two large numbers is limited by experimental precision.
We therefore choose a different route and simply {\em assume} that RT is a high-enough
temperature to neglect all but the barrier term.  Figure~\ref{fig:loginverseT}
shows that this is only a very rough approximation at best. However we have
little alternative but to make this approximation given the nature of the
primary readily available data.  That is, the best that can be done if the 
only data available is the luminescence decay
constants at 77 K and at RT, is to fit to the very simple equation,
\begin{equation}
  k(T) = A_{\text{ave}} e^{-\Delta E_{\text{ave}} /RT} \, .
  \label{eq:results.26}
\end{equation}
This may also be regarded as an alternative (overly simplistic) model.
If we can use this model to explain how $\Delta E_{\text{ave}}$ may be estimated 
from PDOS-LFT, then we will nevertheless have access to information about
the interrelationship of luminescence lifetimes at 77 K and at RT.
% The $\Delta E_{\text{ave}}$s are also given in Tables~\ref{tab:lumdata1} 
% and \ref{tab:lumdata2}.  Figure~\ref{fig:EaveEst} shows that $\Delta E_{\text{ave}}$
% and $\Delta E_{\text{est}}$ behave very similarly.  This is encouraging in the
% sense that no numerical difficulties are encountered in calculating $\Delta E_{\text{ave}}$.
With this caveat, we will confine subsequent discussion to luminescence indices for
predicting $\Delta E_{\text{ave}}$.

% ------------------------------------------------------
\begin{figure}[!h]
\includegraphics[width=0.45\textwidth]{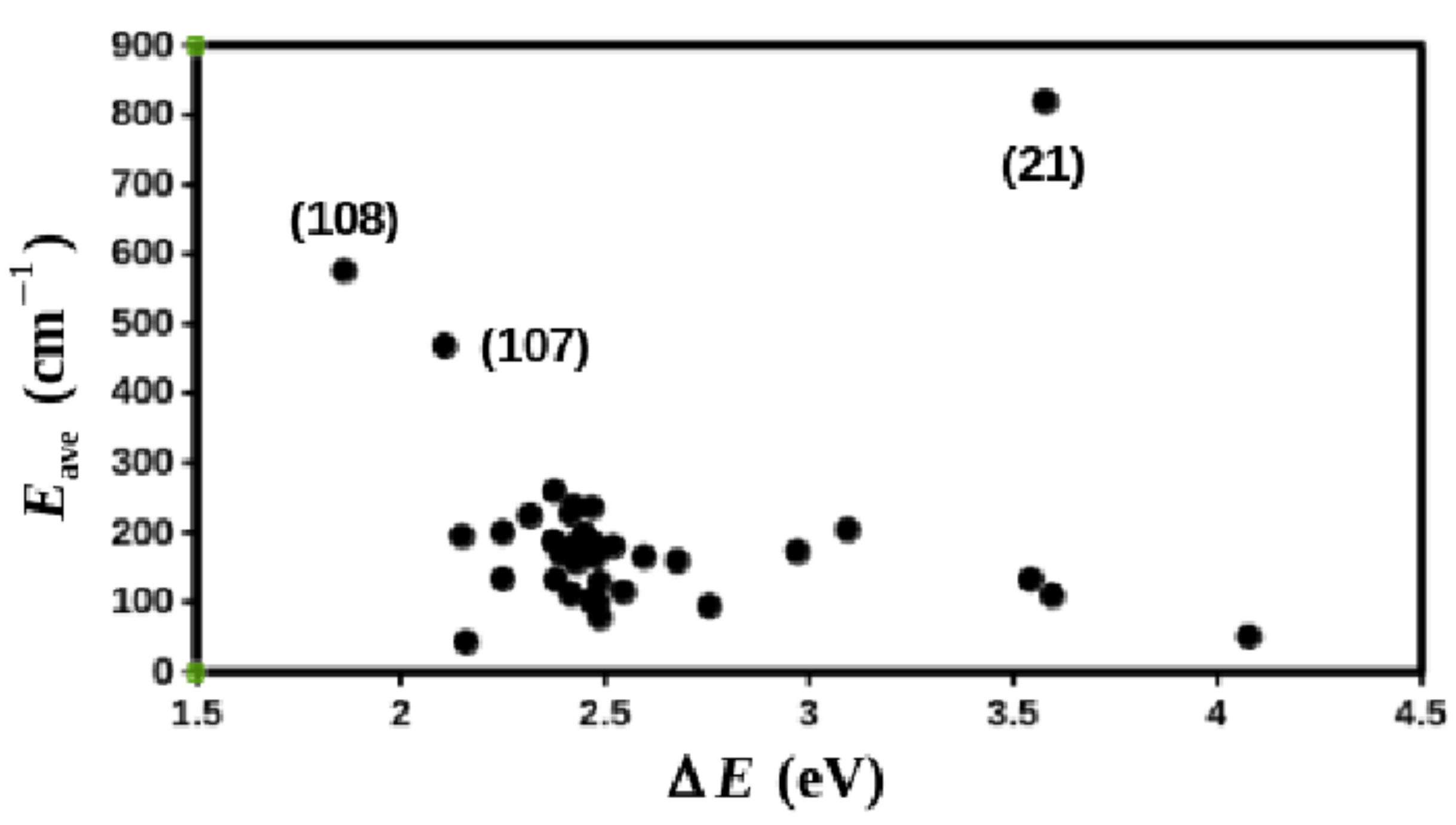}  % EaveDeltaE.pdf
\caption{
\label{fig:EaveDeltaE}
Correlation between $\Delta E$ and $E_{\text{ave}}$.
}
\end{figure}
% ------------------------------------------------------
Let us now turn to the challenge posed in Ref.~\cite{WJL+14}, namely that of
coming up with MO-based indices (or, more exactly, PDOS-LFT-based 
indices) for predicting luminescence lifetimes.  The argument was made
that the $\Delta E_{\text{ave}}$ should be largest when the $^3$MC-$^3$MLCT
state energy difference is smallest.  In LFT-PDOS terms, this corresponds
to $\Delta E_{\text{ave}}$ being smallest when the MO energy difference,
\begin{equation}
  \Delta E = \epsilon_{e_g^*} - \epsilon_{\pi^*} \, ,
  \label{eq:results.27}
\end{equation}
is smallest.  We can check this using the PDOS corresponding to the complexes
listed in Tables~\ref{tab:lumdata1} and \ref{tab:lumdata2}.  A few complexes
have to be eliminated when the known underbinding of DFT has led to unbound
$e^*_g$ orbitals.  Nevertheless, this still leaves 36 data points.  The correlation
between $\Delta E$ and $E_{\text{ave}}$ is shown in Fig.~\ref{fig:EaveDeltaE}.
The correlation is surprisingly bad.  

% ------------------------------------------------------
\begin{figure}[!h]
\includegraphics[width=0.45\textwidth]{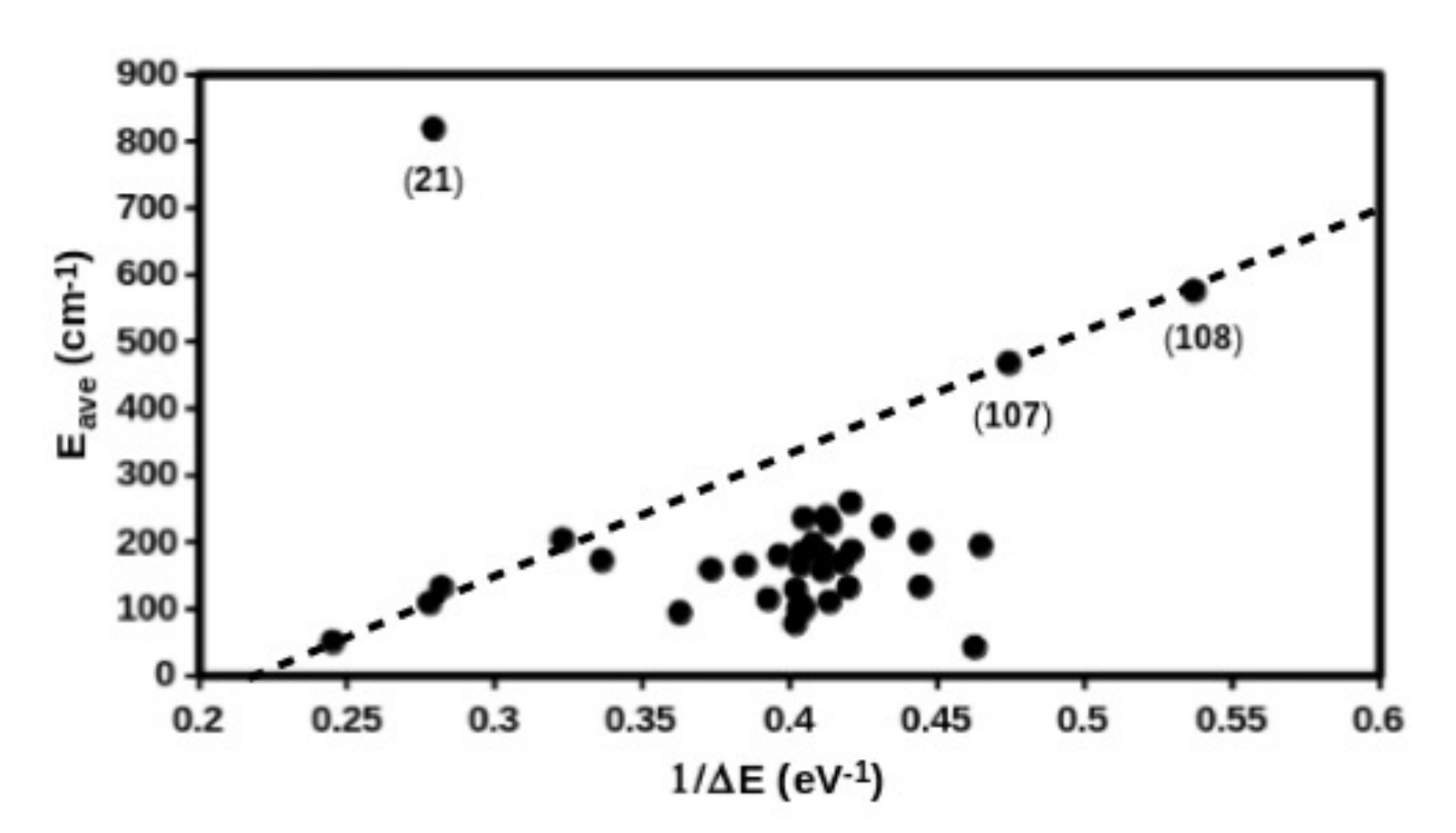}  % EaveInvDeltaE.pdf
\caption{
\label{fig:EaveInvDeltaE}
Correlation between $1/\Delta E$ and $E_{\text{ave}}$.
The dashed line is only a guide to the eye.
}
\end{figure}
% ------------------------------------------------------
We are thus led to think more deeply about the avoided crossing of two states
with diabatic energies $E_1$ and $E_2$ and coupling matrix element $W$.  The
adiabatic energies may be found by diagonalizing the two-state hamiltonian
matrix,
\begin{equation}
  {\bf H} = \left[ \begin{array}{cc} E_1 & W \\
                                     W   & E_2 \end{array} \right] 
  \, .
  \label{eq:results.28}
\end{equation}
The exact and perturbative solutions are,
\begin{eqnarray}
  E_+ & = & \bar{E} + \frac{1}{2} \sqrt{\left(\Delta E\right)^2 + 4 W^2} \nonumber \\
      & \approx & E_2 - \frac{W^2}{\Delta E} \nonumber \\
  E_- & = & \bar{E} -  \frac{1}{2} \sqrt{\left(\Delta E\right)^2 + 4 W^2} \nonumber \\
      & \approx & E_1 + \frac{W^2}{\Delta E} \, ,
  \label{eq:results.29}
\end{eqnarray}
where 
\begin{equation}
  \bar{E} = \frac{E_1+E_2}{2} \, .
  \label{eq:results.30}
\end{equation}
is the average of the diagonal elements.  Following ideas very similar to those
found in frontier MO theory (FMOT) \cite{F76,A07}, we will adapt the pertubative 
formulae {\em evaluated at the ground state geometry},
\begin{equation}
  E_- - E_1 \approx \frac{W^2}{\Delta E} \, .
  \label{eq:results.31}
\end{equation}
as the estimate of the triplet state energy barrier.  More exactly, the slope
of the potential energy curve for the excited state at the ground-state
equilibrium geometry provides a rough indication of trends in the height of the excited-state
energy barrier.  Although we are not actually doing FMOT, but rather presenting
something which we suppose to be novel, it should be born in mind that our
theory resembles FMOT and so is subject to criticism similar to that which
Dewar so reasonably leveled at FMOT \cite{D89}.  Nevertheless FMOT continues
to be used and indeed was honored by the 1981 Nobel Prize in Chemistry because, 
occasional failures set aside, FMOT frequently provides a simple explanation 
of chemical reactivity.  Likewise we seek a simple explanation of luminesence
lifetimes but expect there to be occasional exceptions.
Equation~\ref{eq:results.31} suggests that $E_{\text{ave}}$ should 
correlate better with $1/\Delta E$ than with $\Delta E$.  This hypothesis is tested 
in Fig.~\ref{fig:EaveInvDeltaE}.  Figure~\ref{fig:EaveInvDeltaE} does
indeed seem more linear than does Fig.~\ref{fig:EaveDeltaE}, but 
the line in Fig.~\ref{fig:EaveInvDeltaE} seems to take the form of
an upperbound to a scatter of $E_{\text{ave}}$ values.

% ------------------------------------------------------
\begin{figure}[!h]
\includegraphics[width=0.45\textwidth]{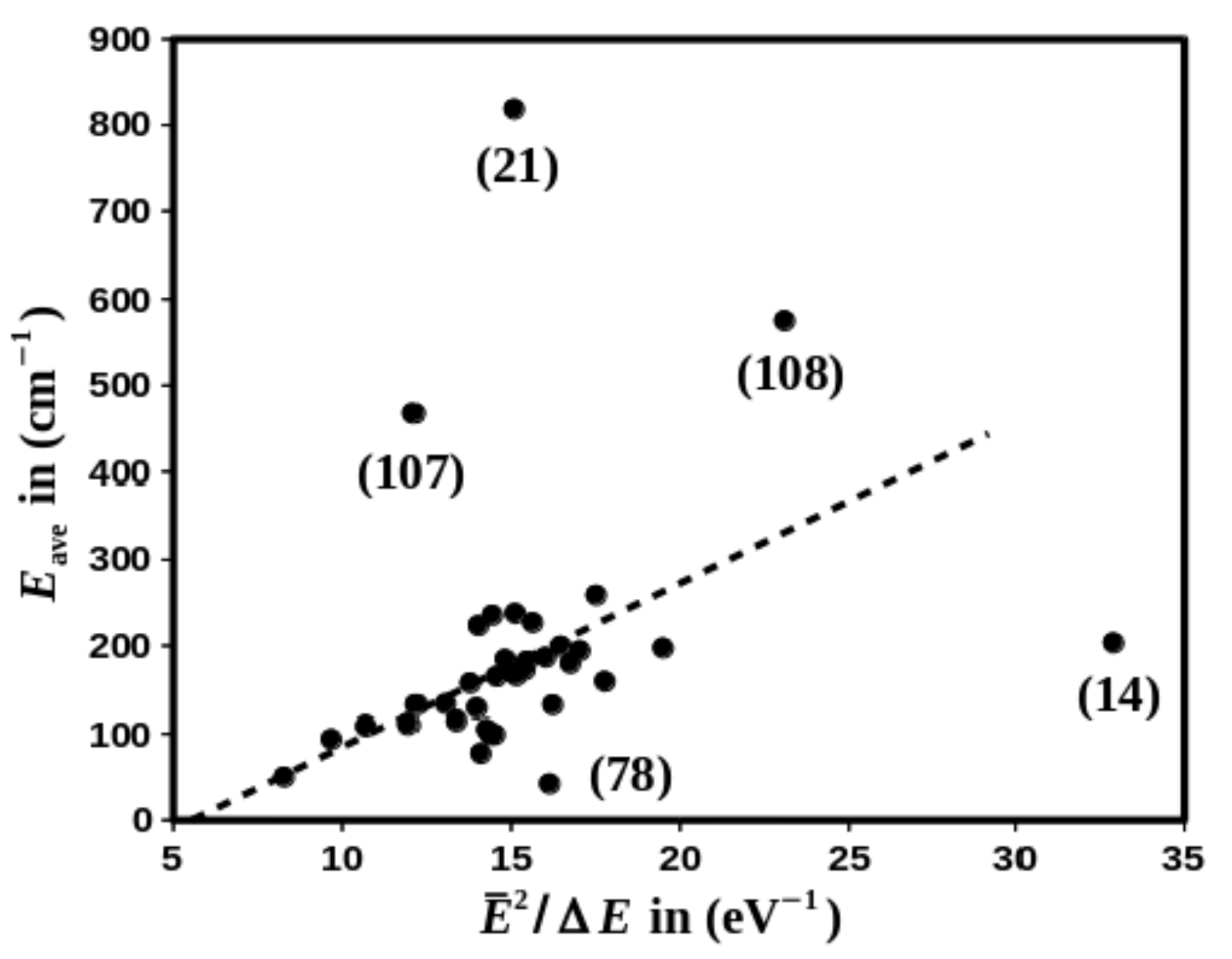} % EaveEbar2InvDeltaE.pdf
\caption{
\label{fig:EaveEbarInvDeltaE}
Correlation between $\bar{E}^2/\Delta E$ and $E_{\text{ave}}$.
The dashed line is only a guide to the eye.
}
\end{figure}
% ------------------------------------------------------
A clue as to how to further improve our theory is to notice that
while $E_{\text{ave}}$ has units of energy, $1/\Delta E$ has units
of inverse energy.  This should be corrected by the quantity $W$
which also has units of energy, but which is not obviously related
to the PDOS-LFT picture from which we seek to extract clues about
luminescence lifetimes.  Again, we take our lead from Roald Hoffmann
(one of the fathers of FMOT), and estimate $W$ by
the Wolfberg-Helmholtz-like formula \cite{WH52},
\begin{equation}
  W = S \bar{E}  \, ,
  \label{eq:results.32}
\end{equation}
where $\bar{E}$ was defined in Eq.~(\ref{eq:results.30})
% \begin{equation}
%   \bar{E} = \frac{E_1+E_2}{2} \, 
%   \label{eq:results.33}
% \end{equation}
and $S$ is some sort of overlap matrix element.  Let us assume that
$S \approx$ constant, and so compare $E_{\text{ave}}$ against $\bar{E}^2/\Delta E$
which both have the energy units.  The result is shown in 
Fig.~\ref{fig:EaveEbarInvDeltaE}.  Except for a few complexes [({\bf 14}), 
({\bf 21}), ({\bf 107}), ({\bf 108}), and possibly ({\bf (78)}], the result
is finally a reasonably good linear correlation.  Indeed a least squares fit
[$E_{\text{ave}} = (\mbox{9.348 cm$^{-1}$/eV})(\bar{E}^2/\Delta E) + 
\mbox{50.764 cm$^{-1}$}$] indicates that the line passes pretty nearly through
the origin as might be expected from our simple FMOT-like theory.

In principle we might be able to do better by being able to provide some
suitable estimate of the overlap $S$.  One suggestion was given in 
Ref.~\cite{WJL+14} which involved the percentage of $d$ contribution to the
$\pi^*$ peak times the percentage of $\pi$ contribution to the $t_{2g}$
peak.  We have tried this and several other similar ideas as a way to 
construct an estimate of $S$ and have found no way to improve upon
$\bar{E}^2/\Delta E$ as the best estimator of $E_{\text{ave}}$.  We therefore
conclude that this is the best we are going to obtain.  The outliers in
Fig.~\ref{fig:EaveEbarInvDeltaE} (i.e., those far from the line correlating
$E_{\text{ave}}$ with $\bar{E}^2/\Delta E$ might easily be accounted for by such
things as the roughness of the 
estimates of luminescence lifetimes which, on the one hand, are not always
reported very accurately and which, on the other hand, have been averaged
over different values in different solvents.  It is also possible that not
all ruthenium complexes have the same type of decay mechanism --- and varying
the ligands is an excellent way to increase the number of ways a ligand can
come off and go on again, leading us back to the ground state.
Indeed, as explained above, we do not even expect our FMOT-like approach to 
work 100\% of the time and so are happy that it works as well as it seems to work.

% ===================================================
\section{Conclusion}
\label{sec:conclude}
% \input{conclude.tex}
% \begin{verbatim}
% ================================
% File: conclude.tex
% Last update: 4 June 2017
% ================================
% \end{verbatim}

We have shown that gas-phase DFT and TD-DFT calculations give results that
correlate well with crystal geometries and with solution absorption spectra
of ruthenium complex spectra.  This is not really a surprise.  It has been 
noticed before and has even been treated in review articles focusing on the 
spectra of transition metal complexes \cite{C03,RRGB04,VZ07}.  However quantifying
this relationship for a very large group of ruthenium(II) polypyridine 
complexes is already useful.

Also important for present purposes, we have shown that PDOS-LFT provides an
interpretational tool, different from, but similar to traditional LFT.  It allows
a semiquantitative prediction of trends in absorption specta and it allows us
to generate spectrochemical series based upon calculated $t_{2g}$-$e_g^*$
energy differences.  This is far from easy to do by other means because TD-DFT
calculations provide more information than is otherwise easily mapped onto 
LFT concepts.  In particular, while the nonbonding $t_{2g}$ orbitals may
often be identified by visualization of specific individual molecular orbitals
of the metal complex, the antibonding $e_g^*$ orbitals mix too heavily with
ligand orbitals to extract their energies by direct visualization of metal 
complex orbitals.  On the other hand, approximate $e_g^*$ orbital energies 
may be obtained in a well-defined manner using the PDOS technique.

This led us to believe that we might be able to develop a simple PDOS-LFT model
which could be useful for understanding and hence for helping to design ligands
to tailor specific photochemical properties of the ligands of ruthenium(II)
polypyridine complexes.  Indeed we were able to use ideas reminicent of frontier
molecular orbital theory to build a simple model which provides a linear correlation
in many, but not all cases, between an average triplet state transition barrier
energy and the square of the average of the $e_g^*$ and lowest $\pi^*$ PDOS-LFT
energies divided by their difference.  
% This 
Exceptions might be due to insufficiently precise experimental data, approximations
inherent in a FMOT-like approach, or real differences in the luminescence decay mechanisms
of different complexes.

Our simple PDOS-LFT model will not replace more elaborate modeling,
but it provides a relatively quick and easy way to relate luminescence lifetimes
at room temperature and liquid nitrogen temperature.  In so doing, it becomes
possible to explore many more complexes than would be possible with a more detailed
model.  

% Indeed the agreement is good enough that it suggests that it suggests that
% molecules having a behavior significantly different from our model may well have 
% unique photochemical features not present in the other molecules in our data set.

In the future, we plan to calculate triplet state energy barriers from
explicit searches of TD-DFT excited-state potential energy surfaces for at least
a few 
% molecules 
ruthenium(II) polypyridine complexes
and compare them with our PDOS-LFT model.
% ==================================================
\section*{Acknowledgement}
% \input{thanks.tex}
%\begin{verbatim}
% ================================
% File: thanks.hxv
% Last modified: 28 May 2017
% ================================
%\end{verbatim}

This work is part of the Franco-Kenyan ELEPHOX (ELEctrochemical and 
PHOtochemical Properties of Some Remarkable Ruthenium and Iron CompleXes) 
project.  DM thanks the French Embassy in Kenya for his doctoral scholarship.
DM and MEC would also like to acknowledge useful training and exchanges
made possible through the African School on Electronic Structure Methods and
Applications (ASESMA, {\tt https://asesma.ictp.it/}).
We would like to thank Pierre Girard, S\'ebastien Morin, and Denis Charapoff
for technical support in the context of the Grenoble 
{\em Centre d'Exp\'erimentation du Calcul Intensif en Chimie} ({\em CECIC})
computers used for the calculations reported here. DM acknowledges the Computational Material Science Group (CMSG)  lab  facilities, {\tt http://www.uoeld.ac.ke/cmsg/}.
DM and MEC acknowledge useful conversations with  Cleophas Muhavini Wawire,
Lat\'evi Max Lawson Daku, Chantal Daniel, Qingchao Sun, Andreas Hauser, and Xiuwen Zhou. 
MEC acknowledges useful discussions with Fr\'ed\'erique Loiseau and with 
Damien Jouvenot.

% =================================================
\section*{Author Contributions}
% \input{contrib.tex}
% \begin{verbatim}
% ================================
% File: contrib.hxv
% Last modified: 2 June 2017
% ================================
% \end{verbatim}

Calculations were carried out by Denis Magero under the direction of
Mark E.\ Casida (50\%), George Amolo (16.67\%), Nicholas Makau (16.67\%), 
and Lusweti Kituyi (16.67\%).  
The writing of the manuscript is the result of a joint effort based upon
detailed progress reports by Denis Magero, commented by all the authors,
and amalgamated into the present form by Mark E.\ Casida.  
All authors have read and approved the final manuscript.

% ------------
% EOF
% ------------
% =================================================
\section*{Conflicts of Interest}
% \input{conflict.tex}
%\begin{verbatim}
% ================================
% File: conflict.hxv
% Last modified: 12 May 2016
% ================================
%\end{verbatim}

The authors declare no conflict of interest.

% =================================================
\section*{Supplementary Material}
% \input{supplmat.tex}
%\begin{verbatim}
% ================================
% File: supplmat.tex
% Last modified: 18 April 2018
% ================================
%\end{verbatim}

The Supplementary Material contains our gas-phase calculated 
(partial) density-of-states [(P)DOS] and TD-B3LYP absorption spectra.

% ===================================================
\appendix
% ---------------------------------------------------
\section{Some Common Abbreviations}
\label{sec:common}
% \input{common.tex}
% \begin{verbatim}
% ================================
% File: common.tex
% Last update: 4 June 2017
% ================================
% \end{verbatim}

This paper contains a large number of abbreviations in order to keep the text
from becoming too cumbersome.  For the reader's convenience, we summarize some
of these abbreviations in this appendix.  Ligand abbreviations are given in 
the next appendix.  Common solvent abbreviations are:
\begin{description}

\item[AN] acetylnitrile, CH$_3$CN.
\item[D$_2$O] heavy water, $^2$H$_2$O.
\item[DMF] dimethylformamide, (CH$_3$)$_2$N-CHO.
\item[eglc] ethyleneglycol,  HOCH$_2$CH$_2$OH.
\item[en] ethylenediamine, H$_2$NCH$_2$CH$_2$NH$_2$.
\item[EPA] ether/iso-pentane/ethanol (5:5:2).
\item[EtOH] ethanol, CH$_3$CH$_2$OH.
\item[H$_2$O] water, H$_2$O.
\item[MeOH] methanol, CH$_3$OH.
\item[PC] propylene carbonate,

\end{description}
Some other abbrevations used in the text are:
\begin{description}

\item[AO] Atomic orbital.
\item[B3LYP] Three-parameter hybrid Becke exchange plus Lee-Yang-Parr
             correlation density functional.
\item[DFT] Density-functional theory.
\item[DOS] Density-of-states.
\item[ECP] Effective core potential.
\item[FMOT] Frontier molecular orbital theory.
\item[GS] Ground state.
\item[LFT] Ligand field theory.
\item[MC] Metal centered.
\item[MLCT] Metal-ligand charge transfer.
\item[MO] Molecular orbital.
\item[PDOS] Partial density of states.
\item[RT] Room temperature.
\item[SCF] Self-consistent field.
\item[TD] Time dependent.
\item[v/v] Volume to volume.

\end{description}

% ========================================================
\section{List of Ligand Abbreviations}
\label{sec:ligands}
% \input{ligandlist.tex}
% \begin{verbatim}
% ================================
% File: ligandlist.tex
% Last update: 3 February 2017
% ================================
% \end{verbatim}

\begin{figure}
\begin{tabular}{c}
\includegraphics[width=0.20\textwidth]{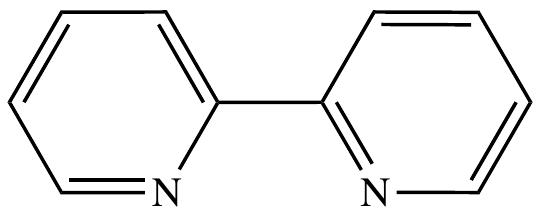}\\  % molecules/bpy.pdf
{\bf bpy}: 2,2'-bipyridine \\
\includegraphics[width=0.15\textwidth]{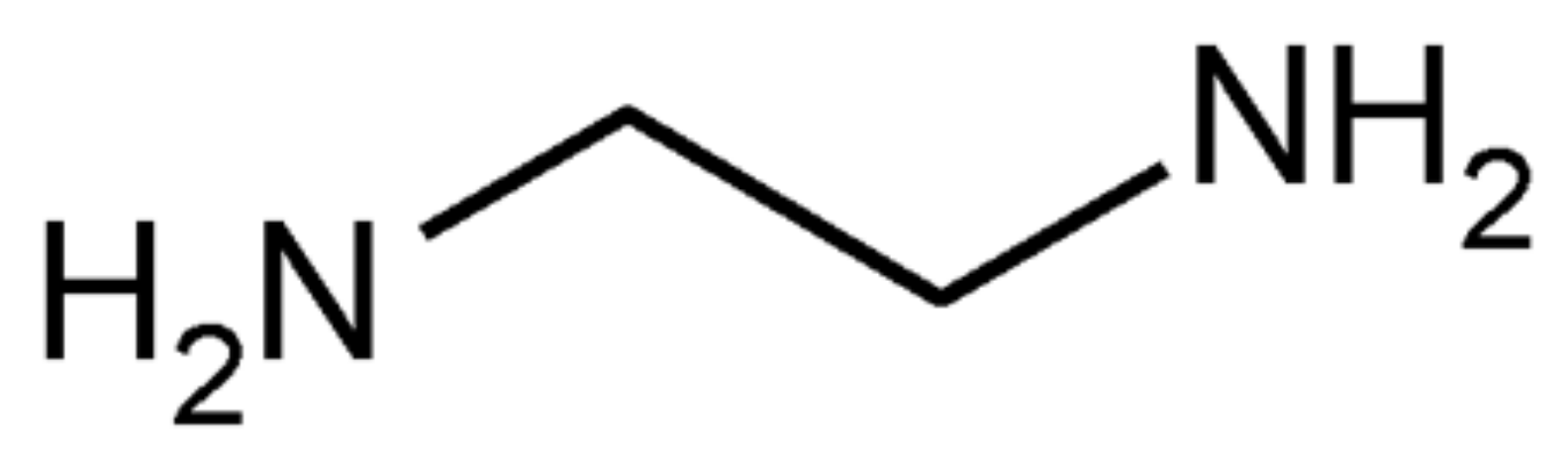}\\  % molecules/en.pdf
{\bf en}: 1,2-ethylenediamine \\
\includegraphics[width=0.10\textwidth]{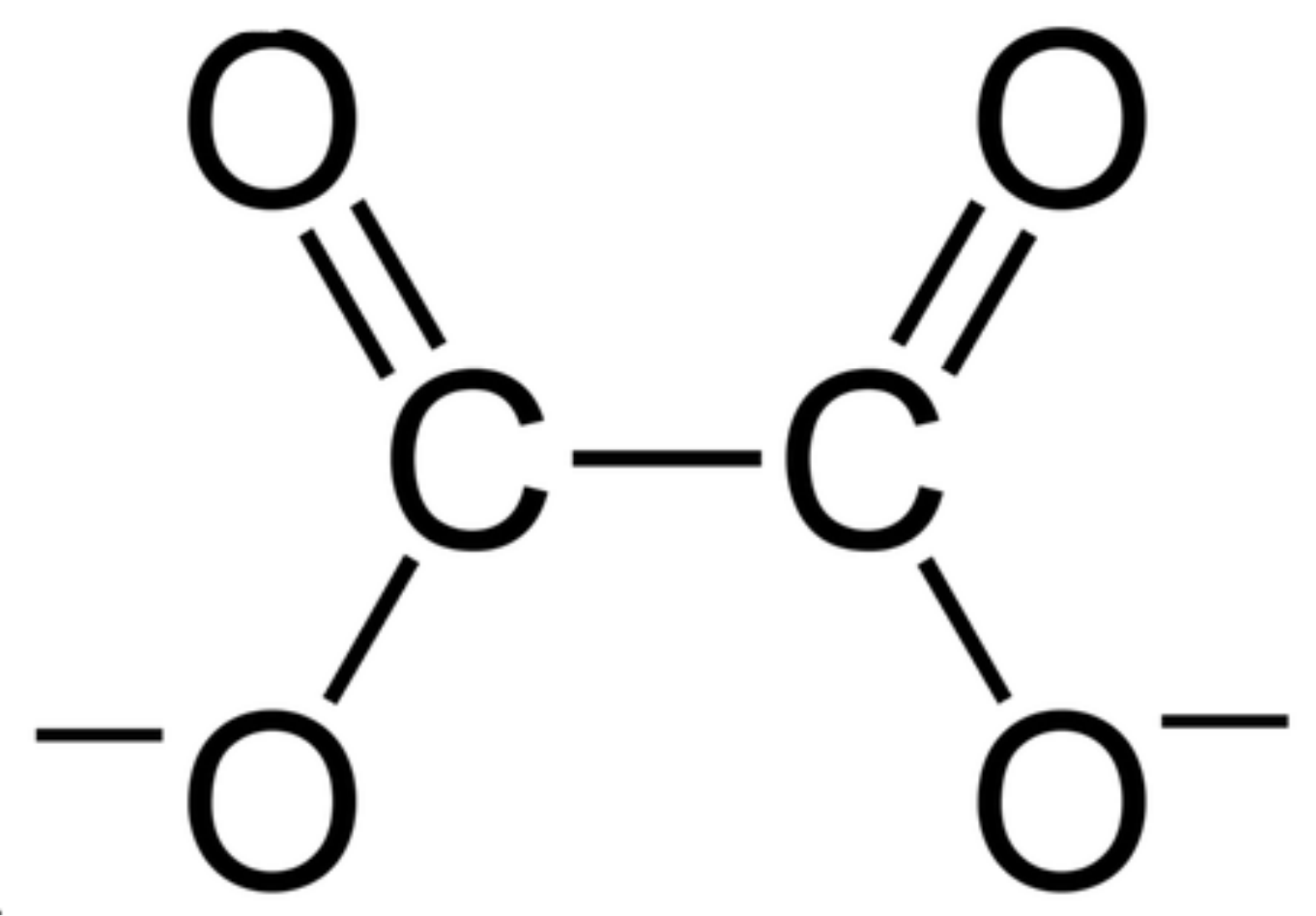}\\ % molecules/ox.pdf
{\bf ox}: oxylate ion \\
\includegraphics[width=0.24\textwidth]{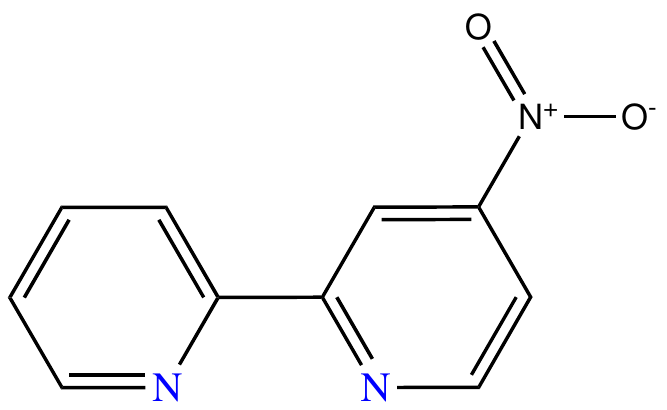}\\  % molecules/4-n-bpy.pdf
{\bf 4-n-bpy}: 4-nitro-2,2'-bipyridine \\
\includegraphics[width=0.20\textwidth]{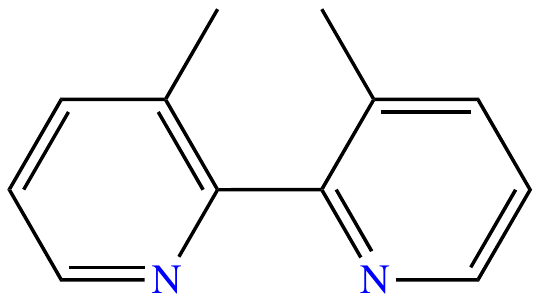} \\ % molecules/33-dm-bpy.pdf
{\bf 3,3'-dm-bpy}: 3,3'-dimethyl-2,2'-bipyridine \\
\includegraphics[width=0.24\textwidth]{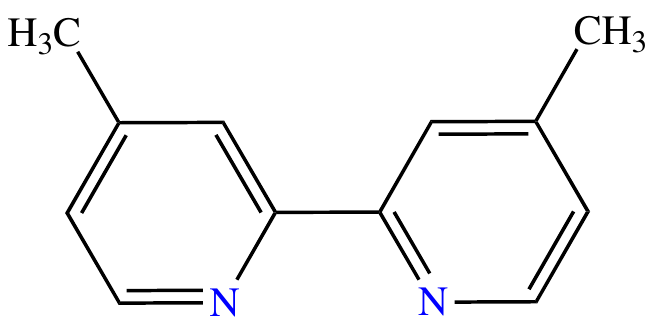} \\ % molecules/44-dm-bpy.pdf
{\bf 4,4'-dm-bpy}: 4,4'-dimethyl-2,2'-bipyridine \\
\includegraphics[width=0.24\textwidth]{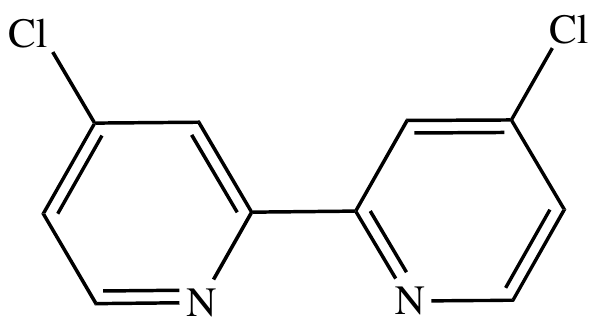} \\ % molecules/44-dCl-bpy.pdf
{\bf 4,4'-dCl-bpy}: 4,4'-dichloro-2,2'-bipyridine \\
\includegraphics[width=0.25\textwidth]{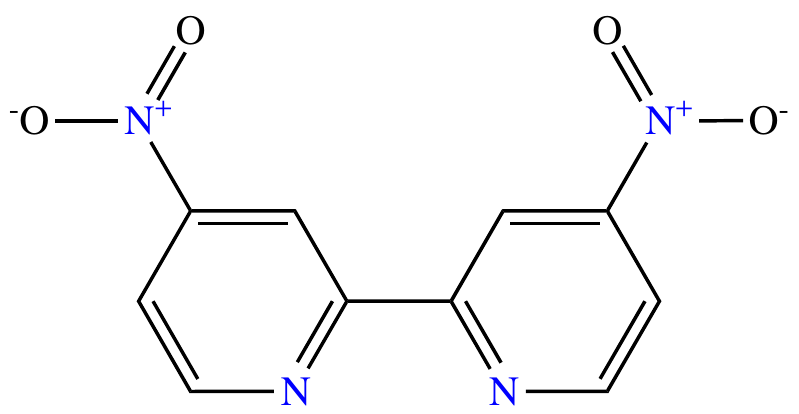} \\ % molecules/44-dn-bpy.pdf
{\bf 4,4'-dn-bpy}: 4,4'-dinitro-2,2'-bipyridine \\
\includegraphics[width=0.24\textwidth]{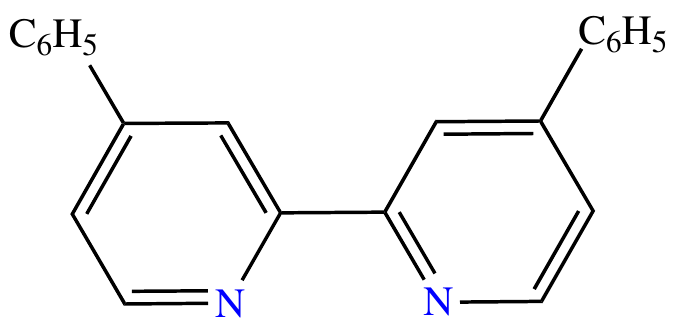} \\ % molecules/44-dph-bpy.pdf
{\bf 4,4'-dph-bpy}: 4,4'-diphenyl-2,2'-bipyridine \\
\end{tabular}
\caption{
Ligand list (part I).  \label{fig:liglist.1}
}
\end{figure}

\begin{figure}
\begin{tabular}{c}
\includegraphics[width=0.27\textwidth]{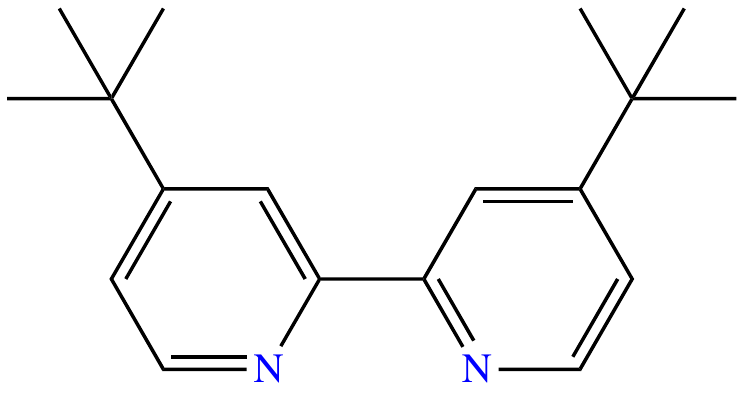} \\ % molecules/44-dtb-bpy.pdf
{\bf 4,4'-DTB-bpy}: 4,4'-di-tert-butyl-2,2'-bipyridine\\
\includegraphics[width=0.19\textwidth]{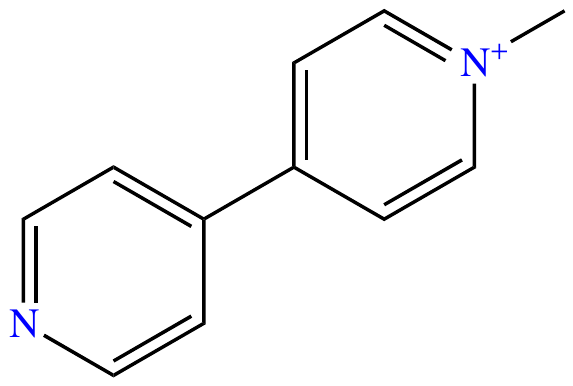} \\ % molecules/m-44-bpy.pdf
{\bf m-4,4'-bpy}: $N$-methyl-4,4'-bipyridine \\
\includegraphics[width=0.20\textwidth]{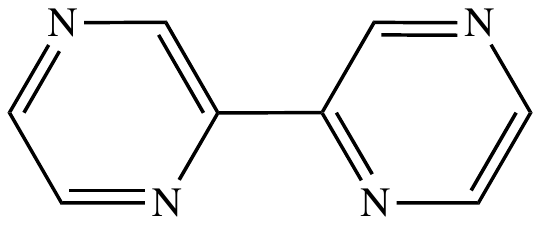} \\ % molecules/bpz.pdf
{\bf bpz}: 2,2'-bipyrazine \\
\includegraphics[width=0.15\textwidth]{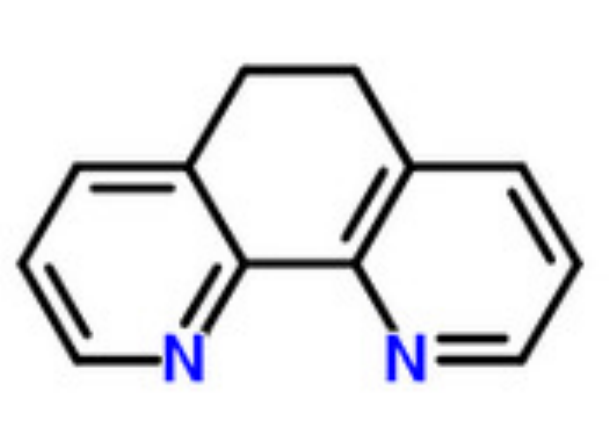} \\ % molecules/h-phen.pdf
{\bf h-phen}: 5,6-dihydro-1,10-phenanthroline \\
\includegraphics[width=0.15\textwidth]{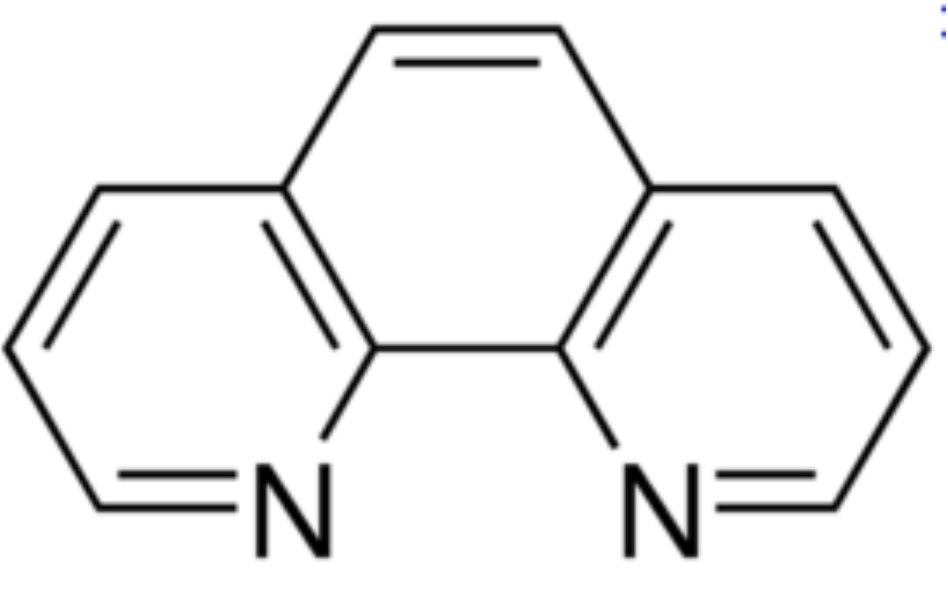} \\ % molecules/phen.pdf
{\bf phen}: 1,10-phenanthroline \\
\includegraphics[width=0.15\textwidth]{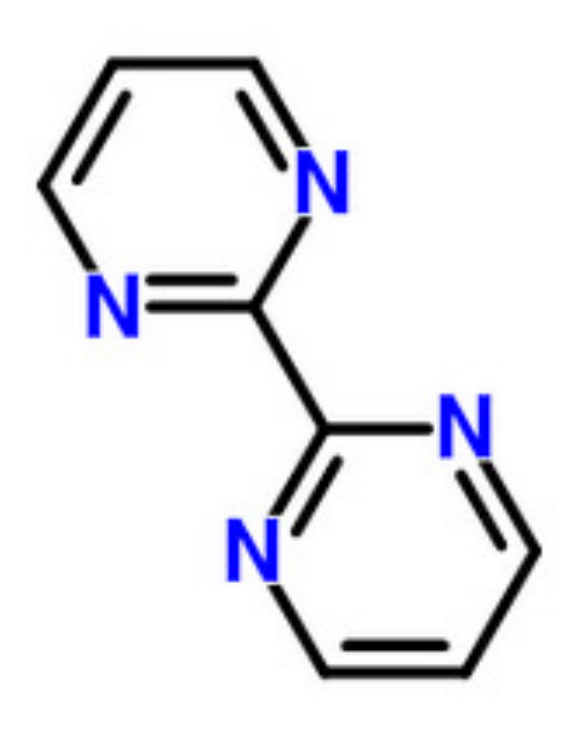} \\ % molecules/bpym.pdf
{\bf bpym}: 2,2'-bipyrimidine \\
\end{tabular}
\caption{
Ligand list (part II).  \label{fig:liglist.2}
}
\end{figure}

\begin{figure}
\begin{tabular}{c}
\includegraphics[width=0.20\textwidth]{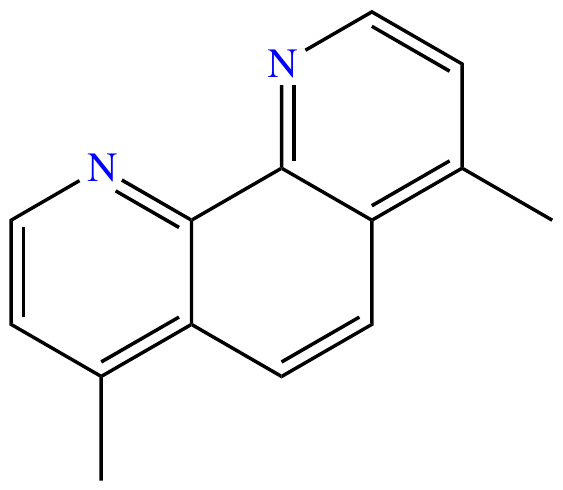} \\ % molecules/47-dm-phen.pdf
{\bf 4,7-dm-phen}: 4,7-dimethyl-1,10-phenanthroline \\
\includegraphics[width=0.25\textwidth]{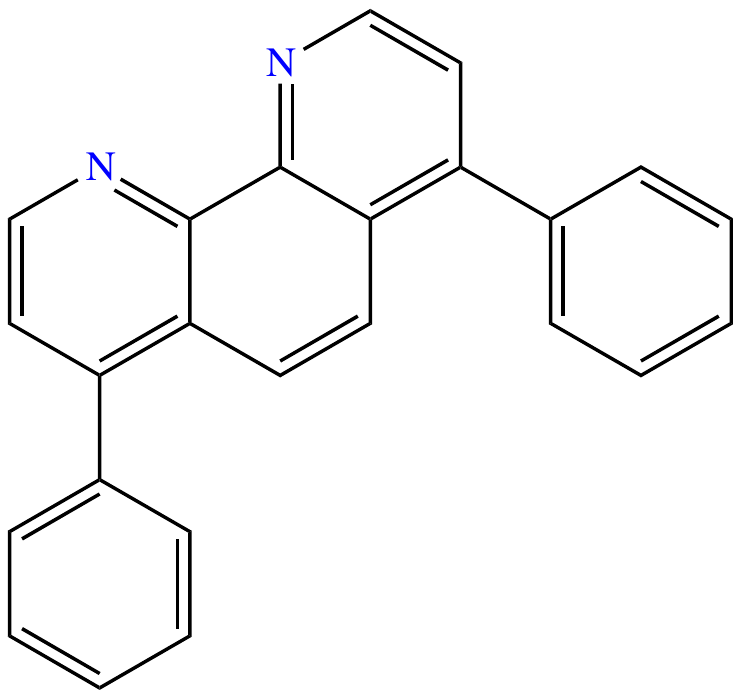} \\ % molecules/47-ph2-phen.pdf
{\bf 4,7-Ph$_2$-phen}: 4,7-diphenyl-1,10-phenanthroline \\
\includegraphics[width=0.23\textwidth]{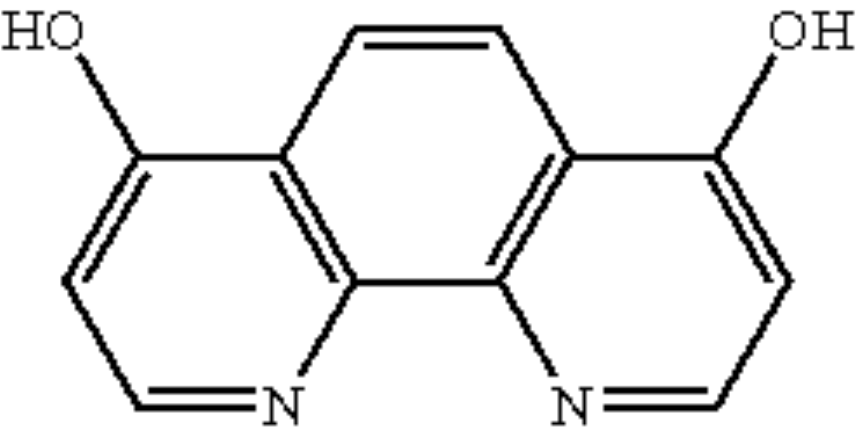} \\ % molecules/47-dhy-phen.pdf
{\bf 4,7-dhy-phen}: 4,7-dihydroxy-1,10-phenanthroline \\
\includegraphics[width=0.17\textwidth]{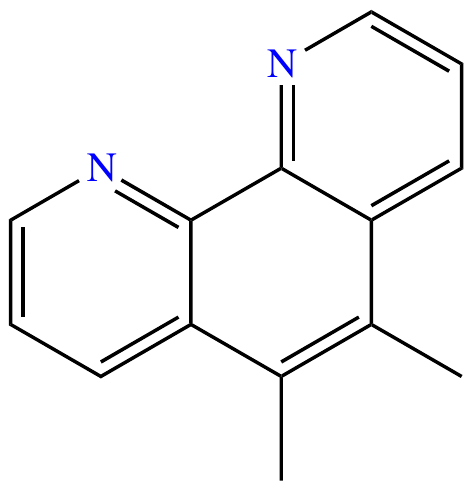} \\ % molecules/56-dm-phen.pdf
{\bf 5,6-dm-phen}: 5,6-dimethyl-1,10-phenanthroline \\
\includegraphics[width=0.15\textwidth]{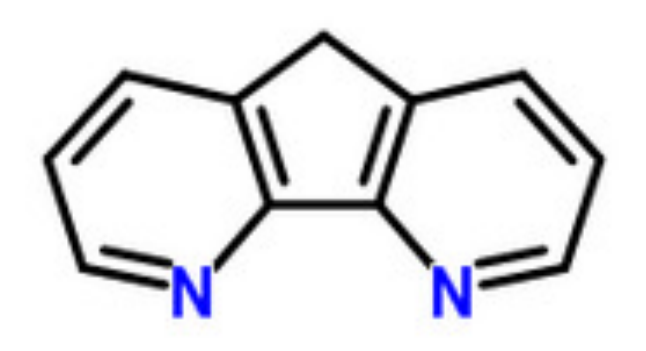} \\ % molecules/DIAF.pdf
{\bf DIAF}: 4,5-diazafluorene \\
\includegraphics[width=0.15\textwidth]{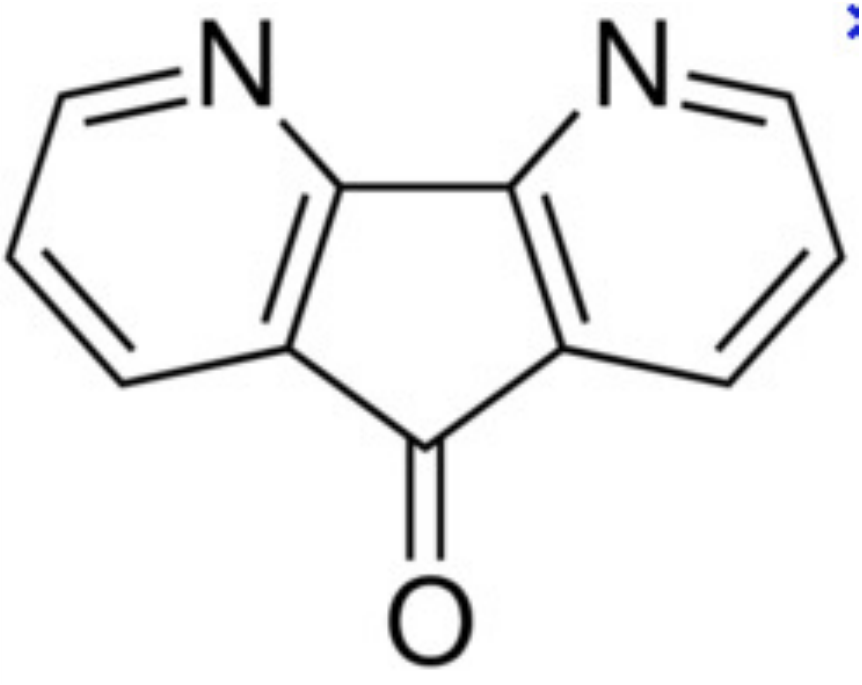} \\  % molecules/DIAFO.pdf
{\bf DIAFO}: 4,5-diazafluoren-9-one \\
\end{tabular}
\caption{
Ligand list (part III).  \label{fig:liglist.3}
}
\end{figure}

\begin{figure}
\begin{tabular}{c}
\includegraphics[width=0.15\textwidth]{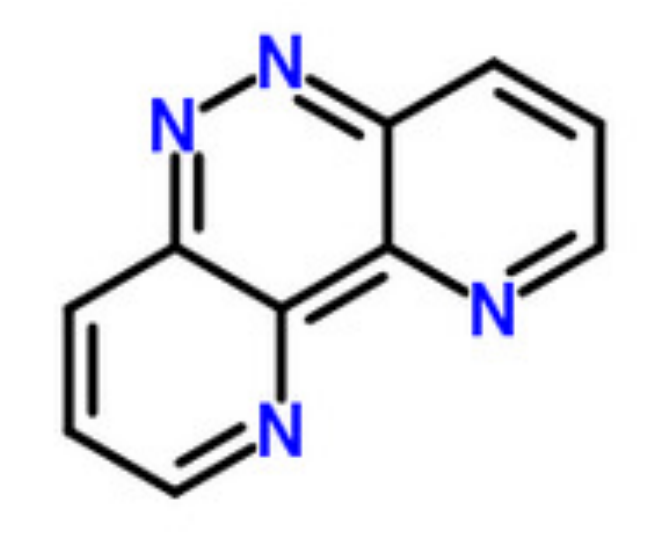} \\ % molecules/taphen.pdf
{\bf taphen}: dipyrido[3,2-{\em c}:2',3'-{\em e}]pyridazine \\
\includegraphics[width=0.05\textwidth]{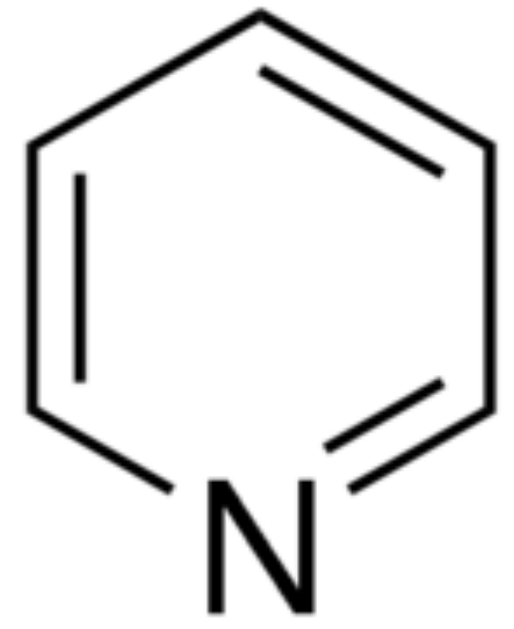} \\ % molecules/py.pdf
{\bf py}: pyridine \\
\includegraphics[width=0.05\textwidth]{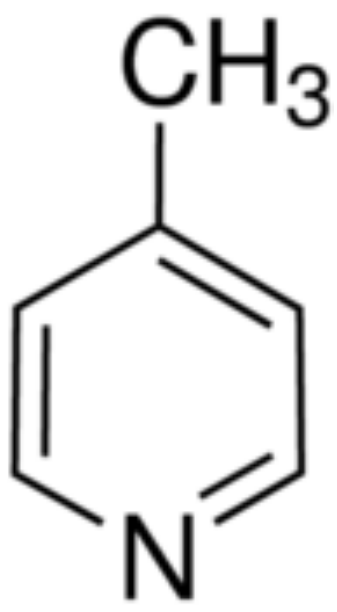} \\ % molecules/pic.pdf
{\bf pic}: 4-methyl-pyridine \\
\includegraphics[width=0.15\textwidth]{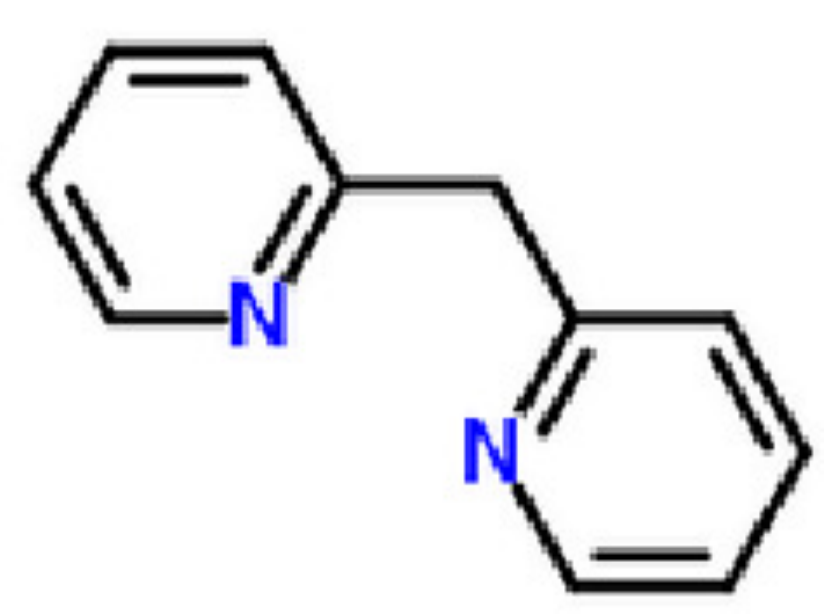} \\ % molecules/DPM.pdf
{\bf DPM}: di-(2-pyridyl)-methane \\
\includegraphics[width=0.15\textwidth]{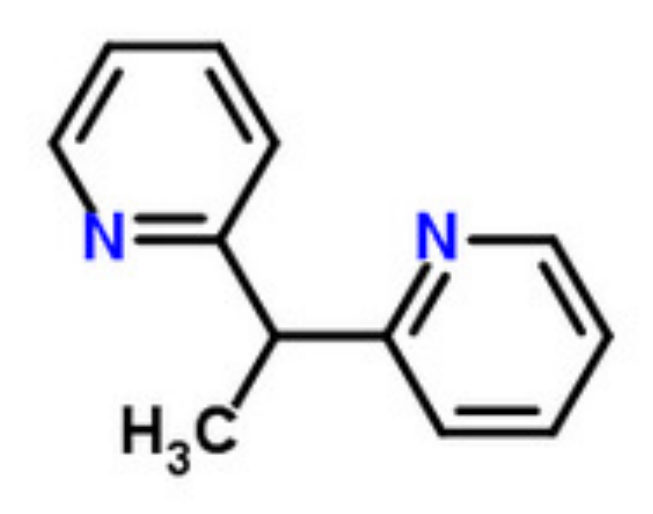} \\ % molecules/DPE.pdf
{\bf DPE}: di-(2-pyridyl)-ethane \\
\includegraphics[width=0.18\textwidth]{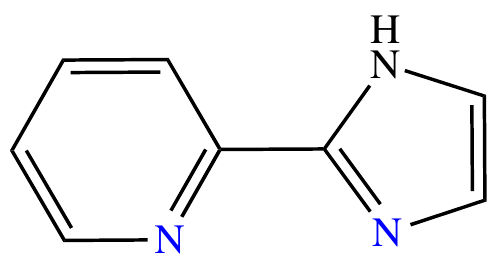}\\ % molecules/PimH.pdf
{\bf PimH}: 2-(2-pyridyl)imidazole\\
\includegraphics[width=0.25\textwidth]{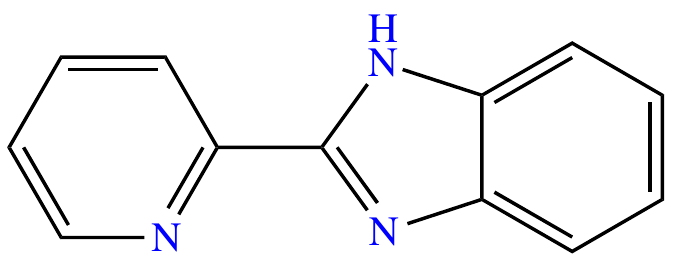} \\ % molecules/PBzimH.pdf
{\bf PBzimH}: 2-(2-pyridyl)benzimidazole \\
\includegraphics[width=0.19\textwidth]{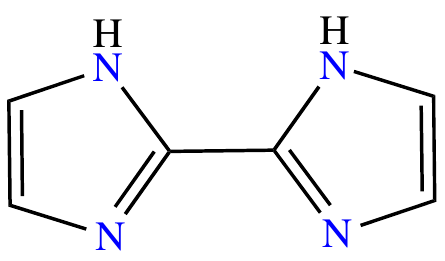} \\ % molecules/biimH2.pdf
{\bf biimH$_{2}$}: 2,2'-biimidazole \\
\includegraphics[width=0.25\textwidth]{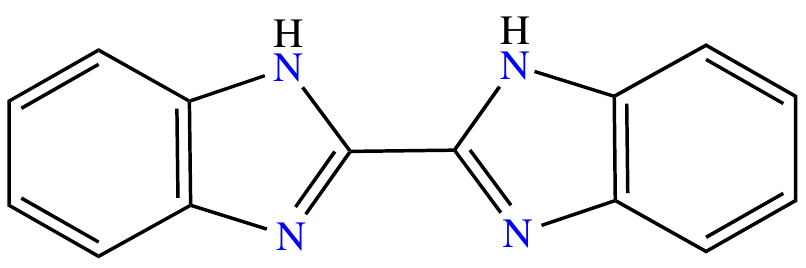} \\ % molecules/BiBzimH2.pdf
{\bf BiBzimH$_{2}$}: 1H,1'H-2,2'-bibenzo[d]imidazole \\
\end{tabular}
\caption{
Ligand list (part IV).  \label{fig:liglist.4}
}
\end{figure}

\begin{figure}
\begin{tabular}{c}
\includegraphics[width=0.10\textwidth]{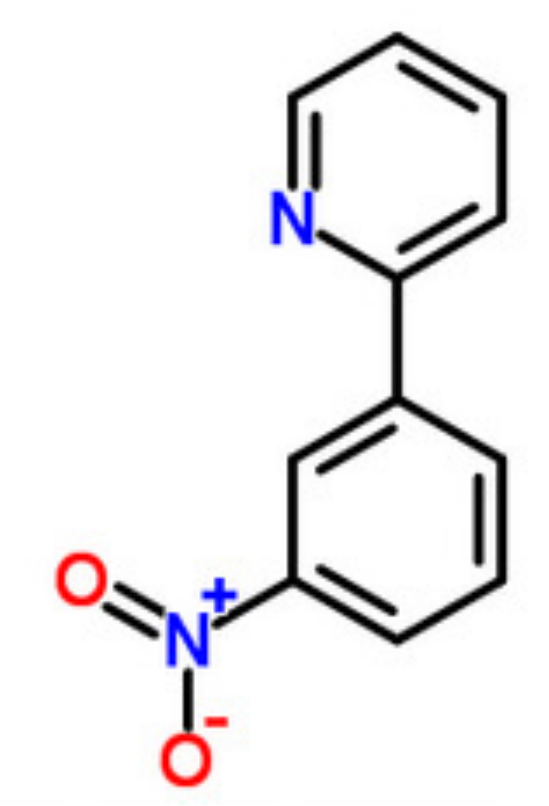} \\ % molecules/NPP.pdf
{\bf NPP}: 2-(3-nitrophenyl)-pyridine \\
\includegraphics[width=0.10\textwidth]{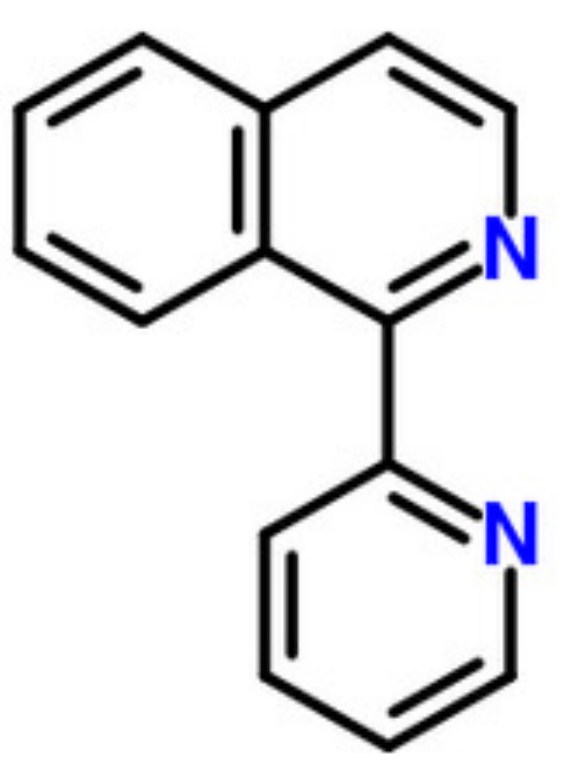} \\ % molecules/piq.pdf
{\bf piq}: 1-(2-pyridyl)-isoquinoline \\
\includegraphics[width=0.17\textwidth]{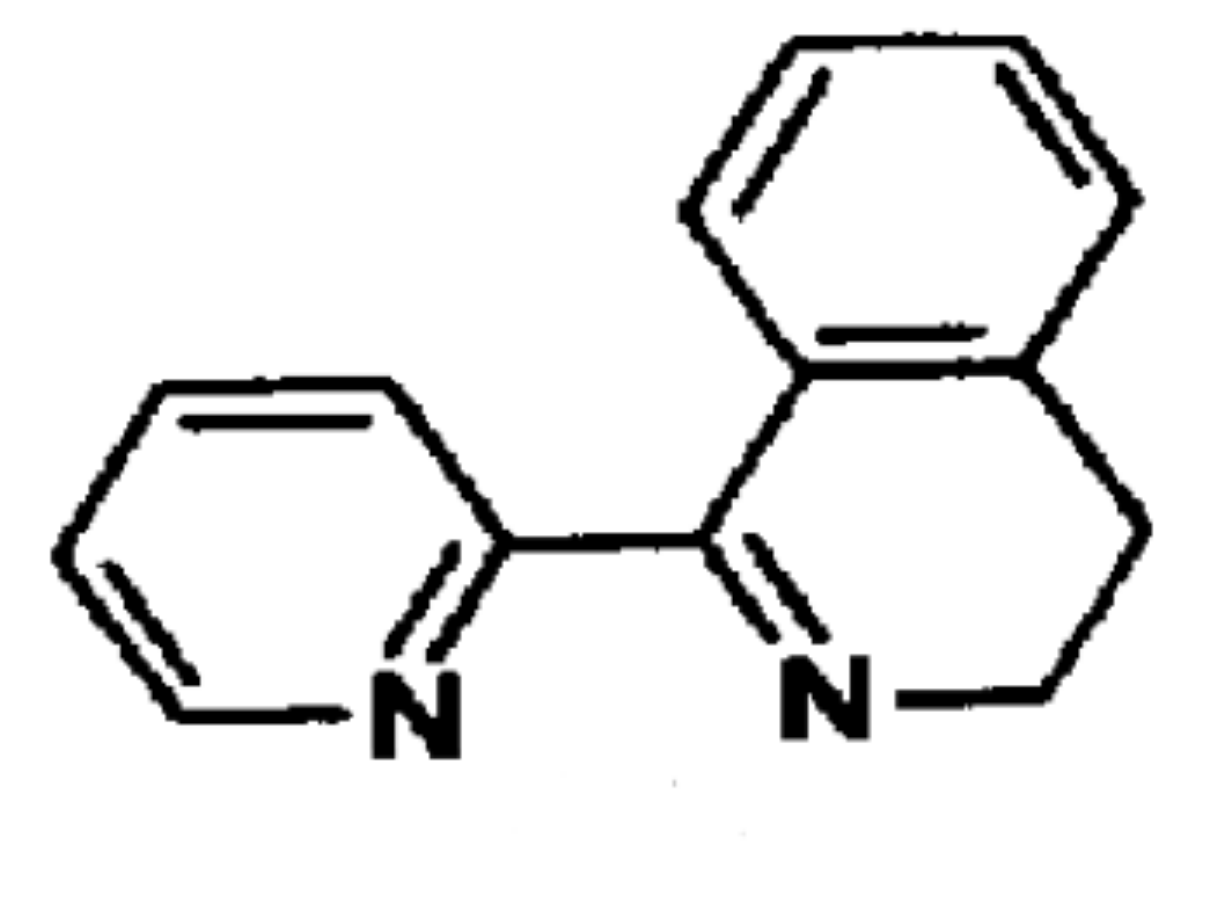} \\ % molecules/hpiq.pdf
{\bf hpiq}: 3,4-dihydro-1(2-pyridyl)-isoquinoline \\
\includegraphics[width=0.20\textwidth]{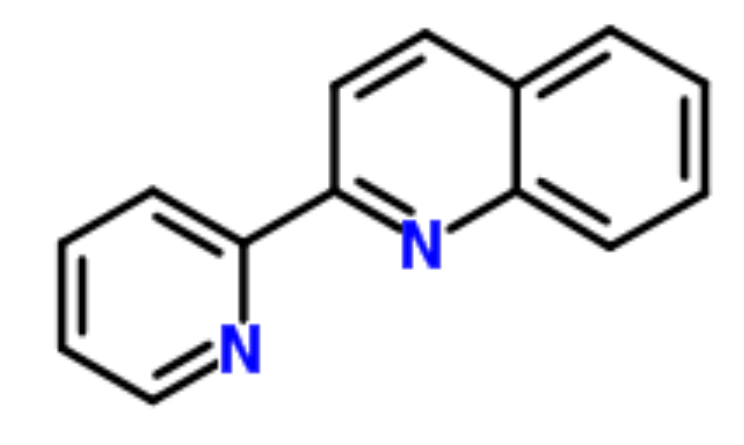} \\ % molecules/pq.pdf
{\bf pq}: 2-(2-pyridyl)quinoline \\
\includegraphics[width=0.25\textwidth]{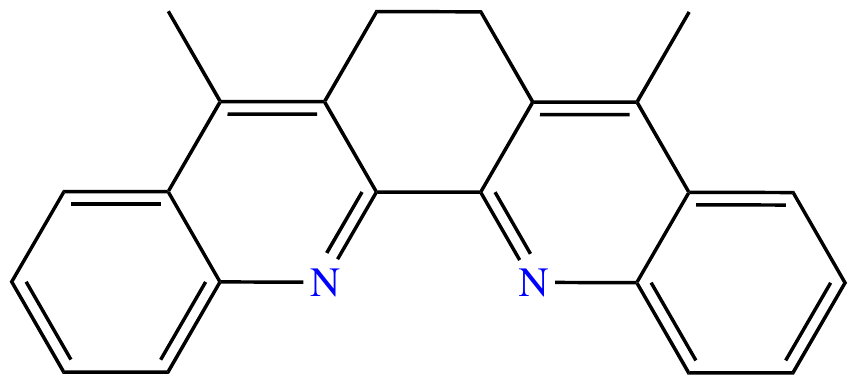}\\ % molecules/DMCH2CN2.pdf
{\bf DMCH}: 5,6-dihydro-4,7-dimethyl-dibenzo[3,2-{\em b}:2'3'-{\em j}] \\ 
$[1,10]$phenanthroline\\
\includegraphics[width=0.25\textwidth]{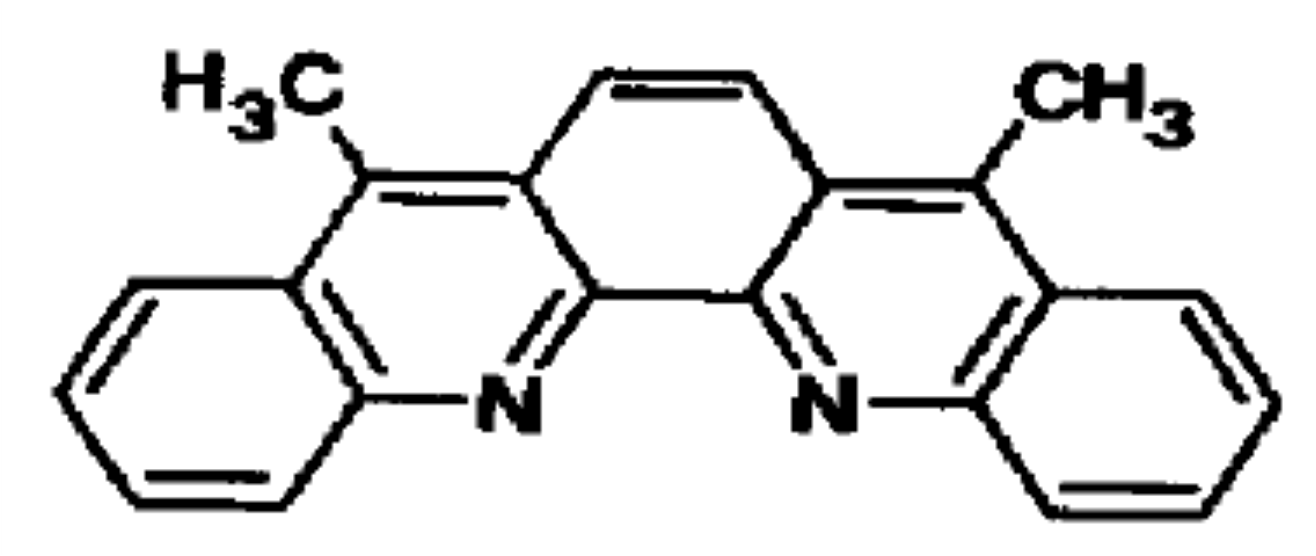}\\ % molecules/OMCH.pdf
{\bf OMCH}: 4,7-dimethyldibenzo[3,1-{\em b}:2',3'-{\em j}]\\
$[1,10]$phenanthroline \\
\includegraphics[width=0.25\textwidth]{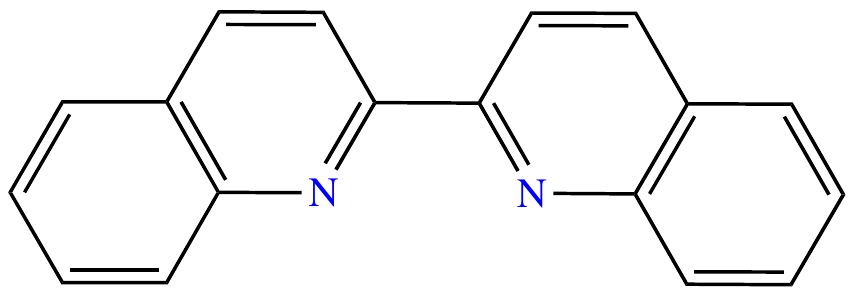} \\ % molecules/biq.pdf
{\bf biq}: 2,2'-biquinoline \\
\includegraphics[width=0.25\textwidth]{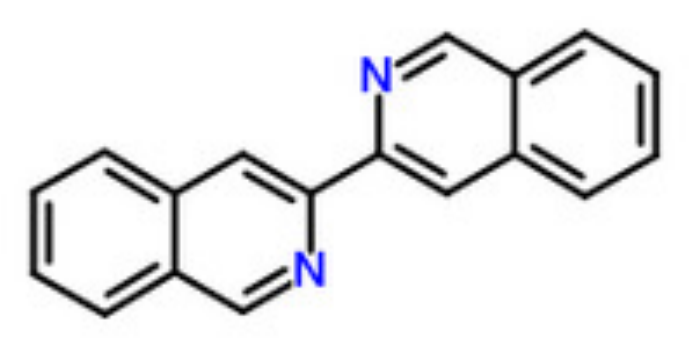} \\ % molecules/i-biq.pdf
{\bf i-biq}: 3,3'-biisoquinoline \\
\end{tabular}
\caption{
Ligand list (part V).  \label{fig:liglist.5}
}
\end{figure}

\begin{figure}
\begin{tabular}{c}
\includegraphics[width=0.25\textwidth]{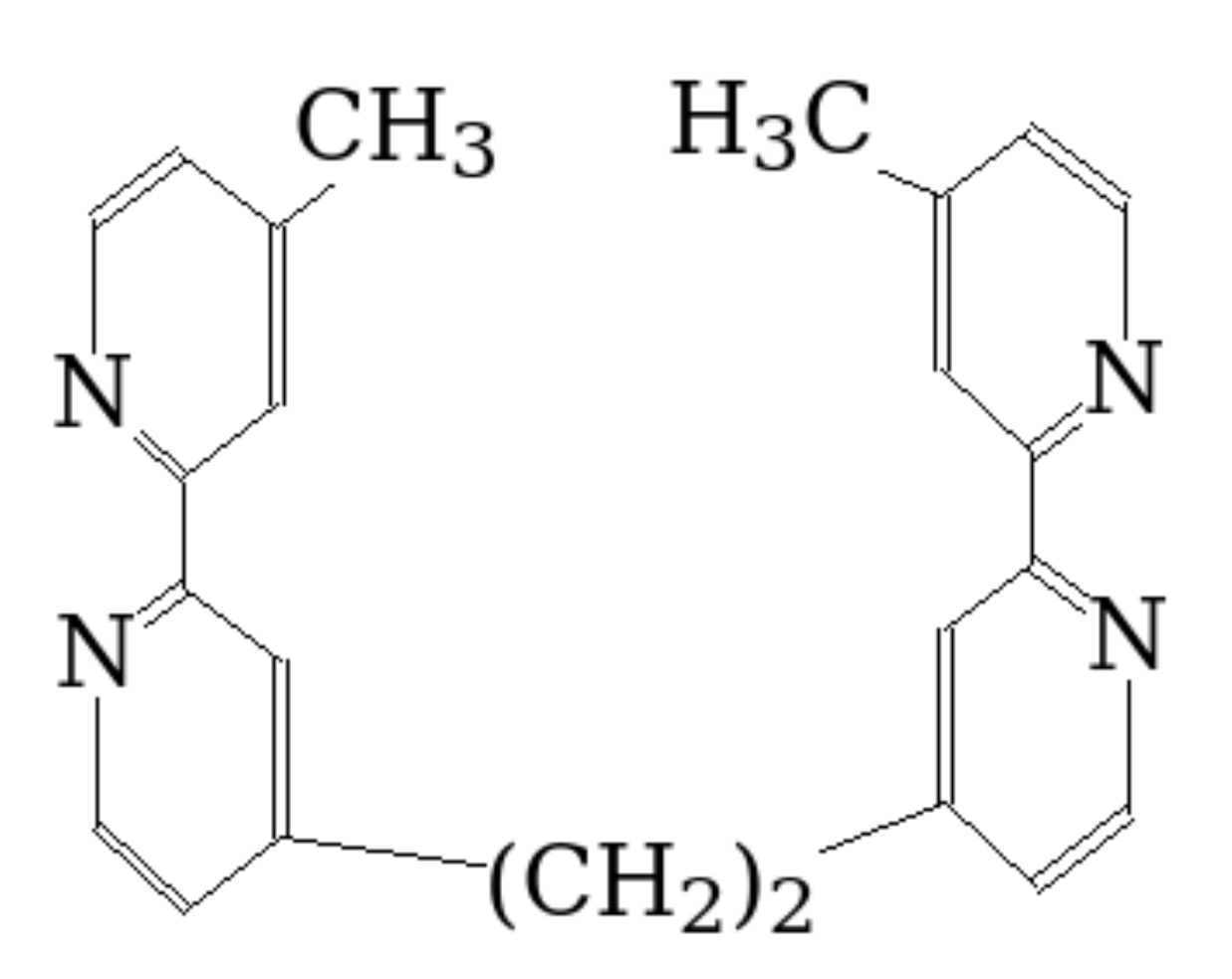} \\  % molecules/BL4.pdf
{\bf BL4}: 1,2-bis[4-(4'-methyl-2,2'-bipyridyl)]ethane  \\
\includegraphics[width=0.25\textwidth]{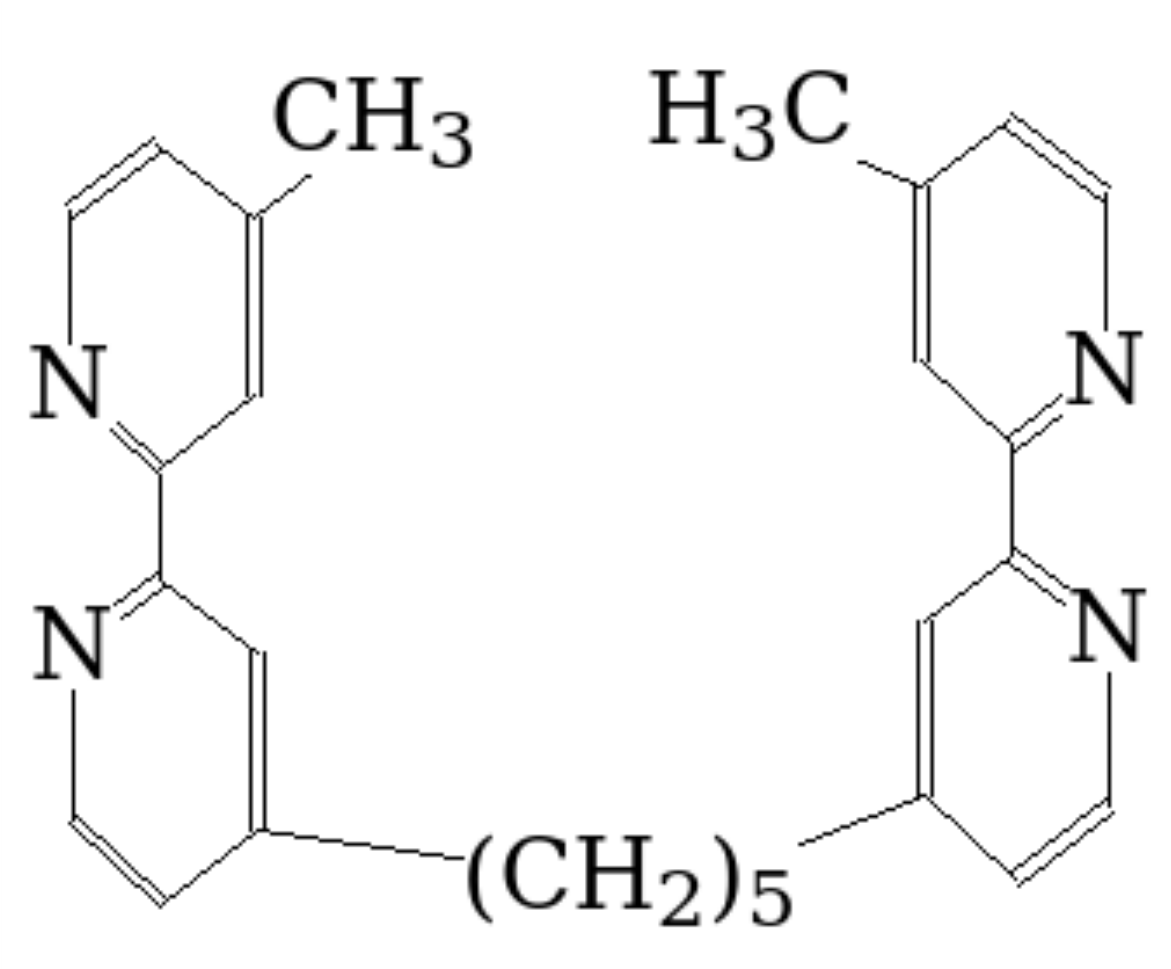} \\ % molecules/BL5.pdf
{\bf BL5}: 1,5-bis[4-(4'-methyl-2,2'-bipyridyl)]pentane  \\
\includegraphics[width=0.40\textwidth]{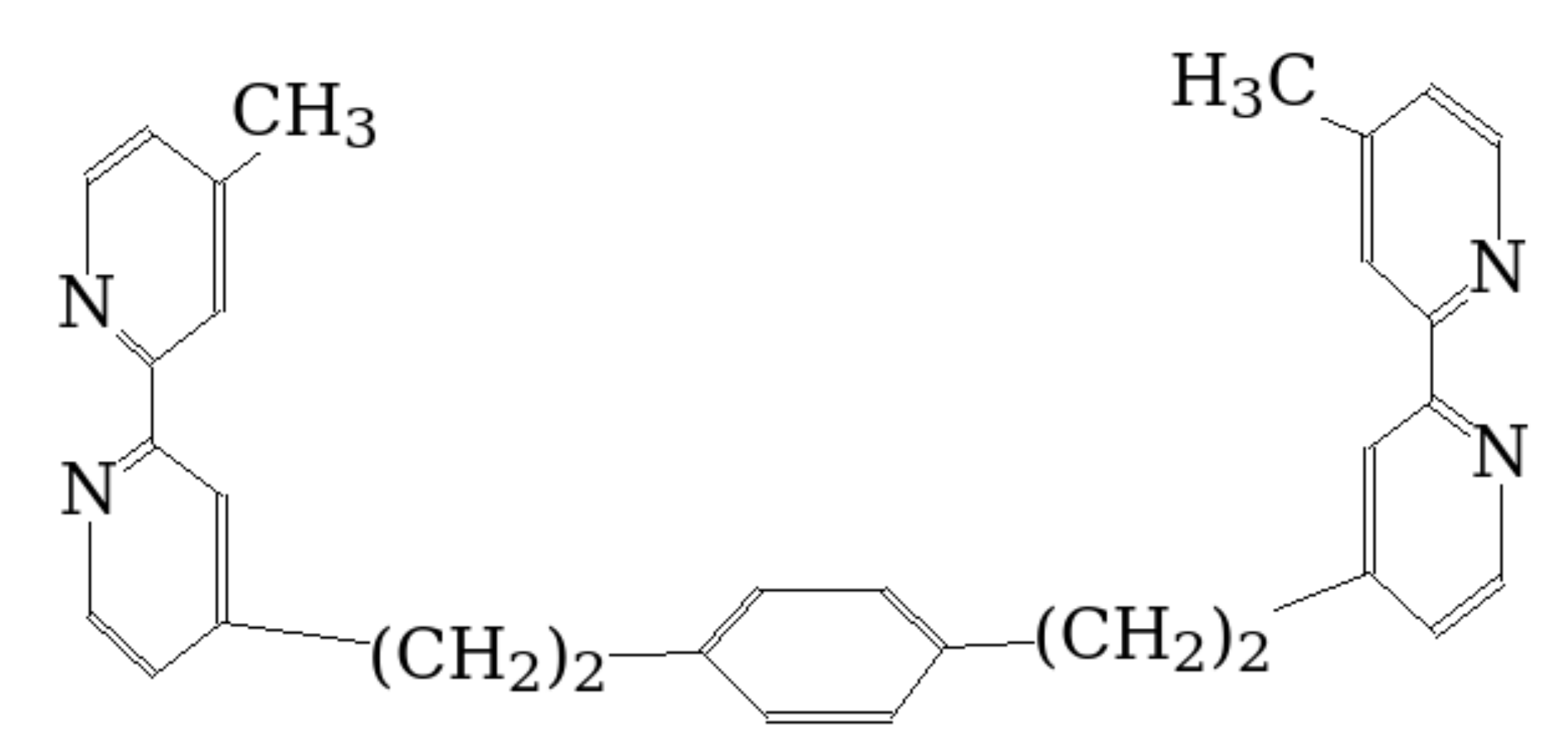} \\ % molecules/BL6.pdf
{\bf BL6}: 1,4,-bis[4-(4'-methyl-2,2'-bipyridyl)]benzene  \\
\includegraphics[width=0.25\textwidth]{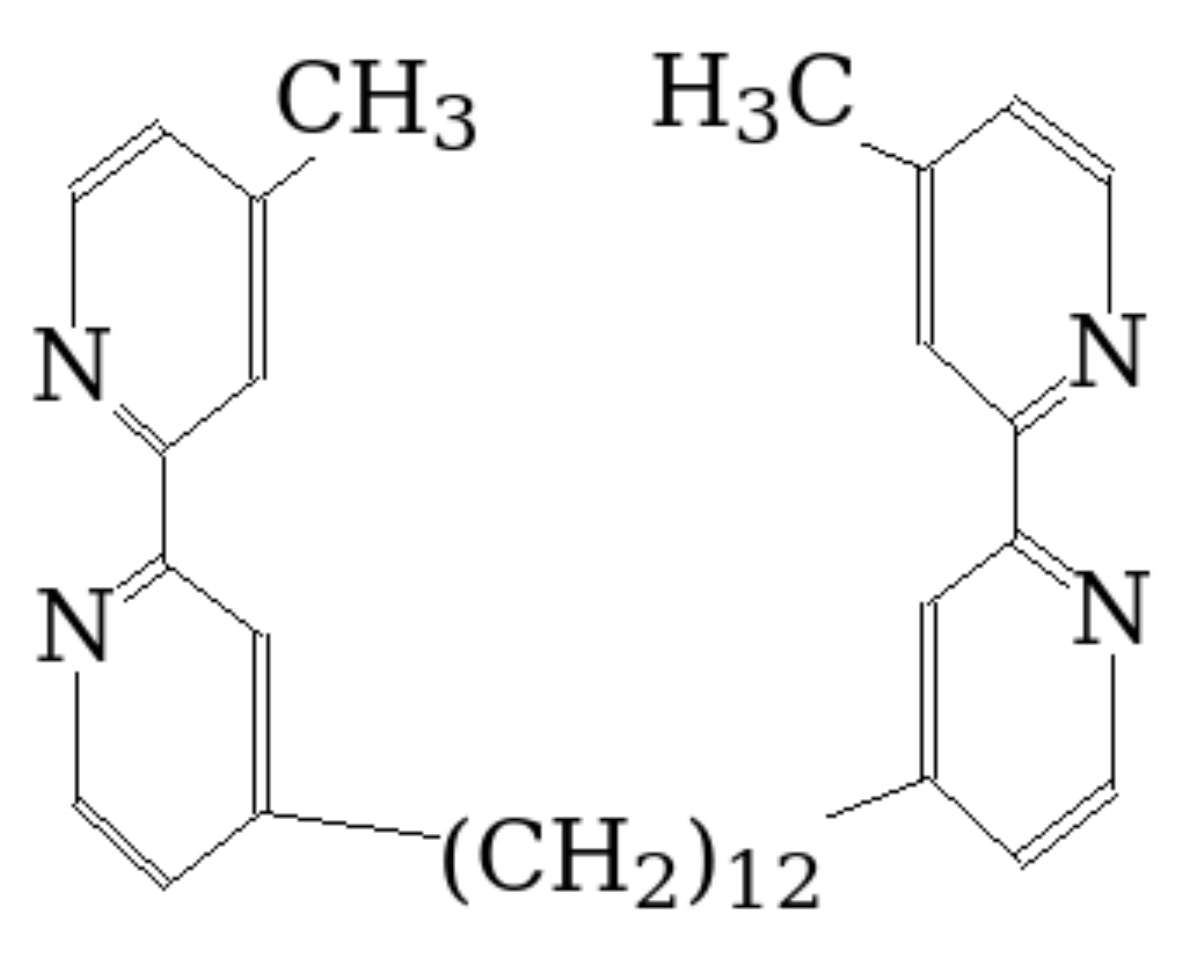} \\ % molecules/BL7.pdf
{\bf BL7}: 1,12-bis[4-r'-methyl-2,2'-bipyridyl)]dodecane \\
\includegraphics[width=0.15\textwidth]{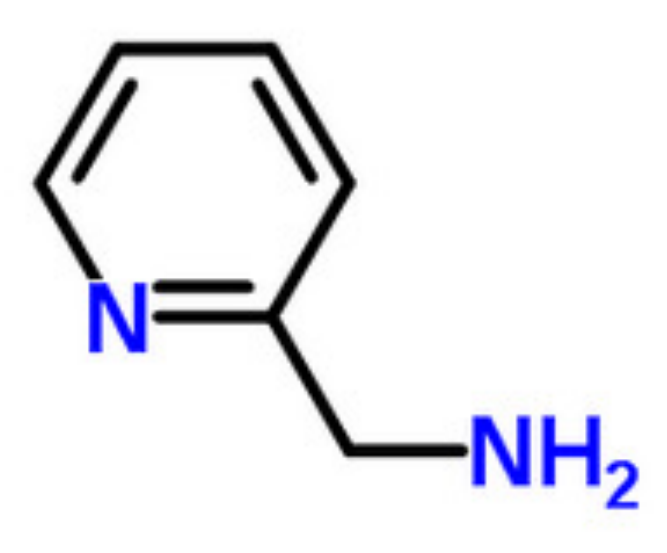} \\ % molecules/PMA.pdf
{\bf PMA}: 2-aminomethylpyridine \\
\end{tabular}
\caption{
Ligand list (part VI).  \label{fig:liglist.6}
}
\end{figure}

\begin{figure}
\begin{tabular}{c}
\includegraphics[width=0.15\textwidth]{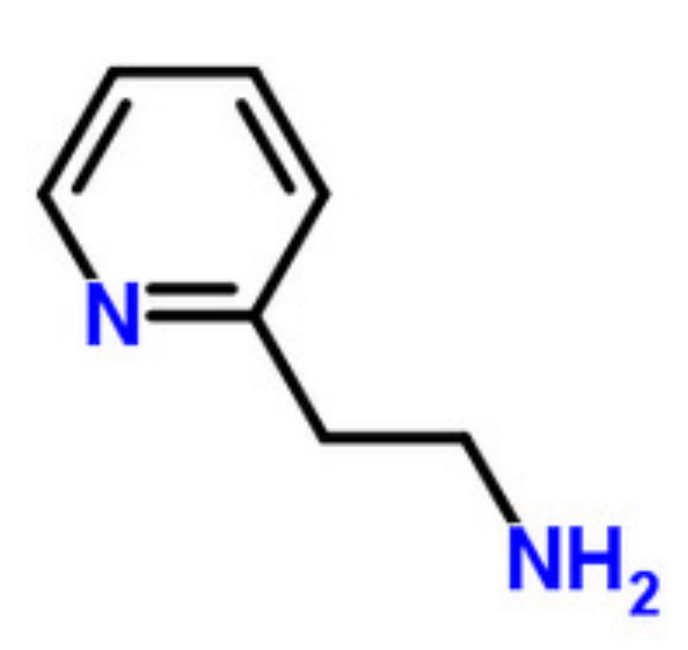} \\ % molecules/2-AEP.pdf
{\bf 2-AEP}: 2-(2-aminoethyl)pyridine \\
\includegraphics[width=0.25\textwidth]{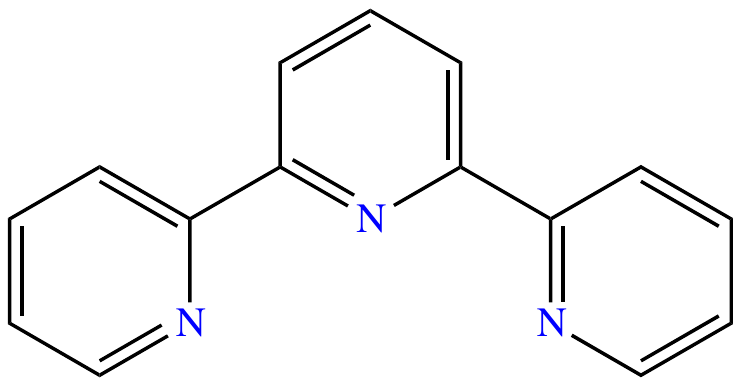} \\ % molecules/trpy.pdf
{\bf trpy}: 2,2';6',2"-terpyridine \\
\includegraphics[width=0.15\textwidth]{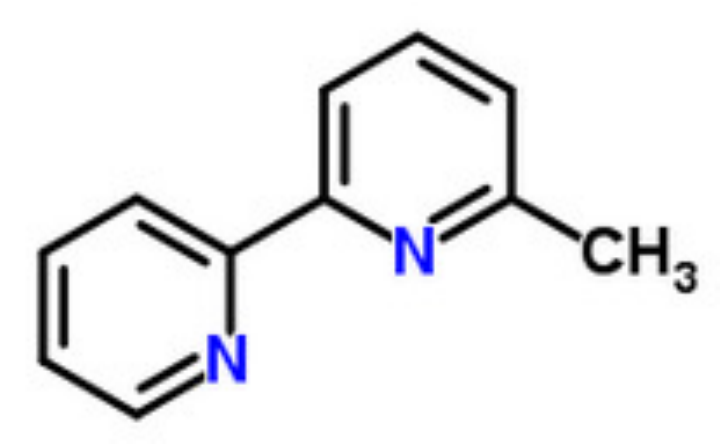} \\ % molecules/6-m-bpy.pdf
{\bf 6-m-bpy}: 6-methyl-2,2'-bipyridine \\
\includegraphics[width=0.15\textwidth]{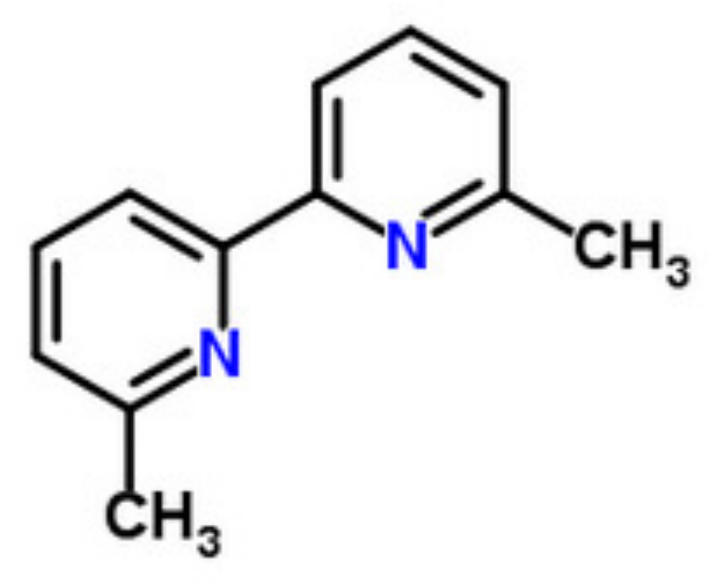} \\ % molecules/66-dm-bpy.pdf
{\bf 6,6'-dm-bpy}: 6,6'-dimethyl-2,2'-bipyridine \\
\includegraphics[width=0.15\textwidth]{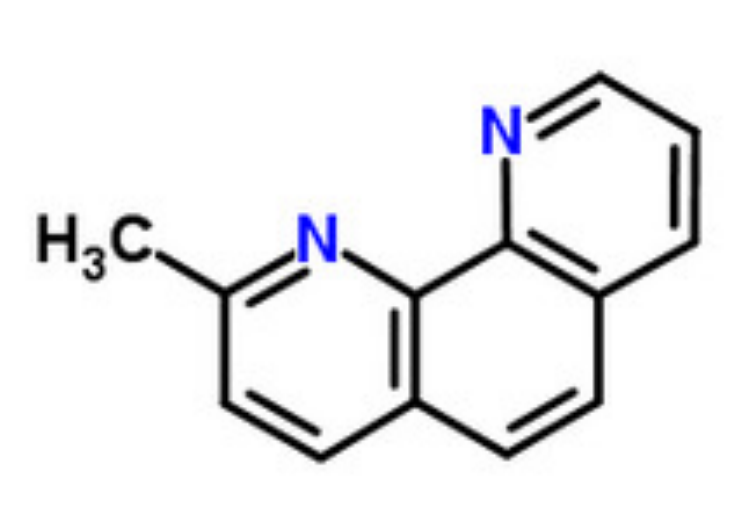} \\ % molecules/2-m-phen.pdf
{\bf 2-m-phen}: 2-methyl-1,10-phenanthroline \\
\includegraphics[width=0.25\textwidth]{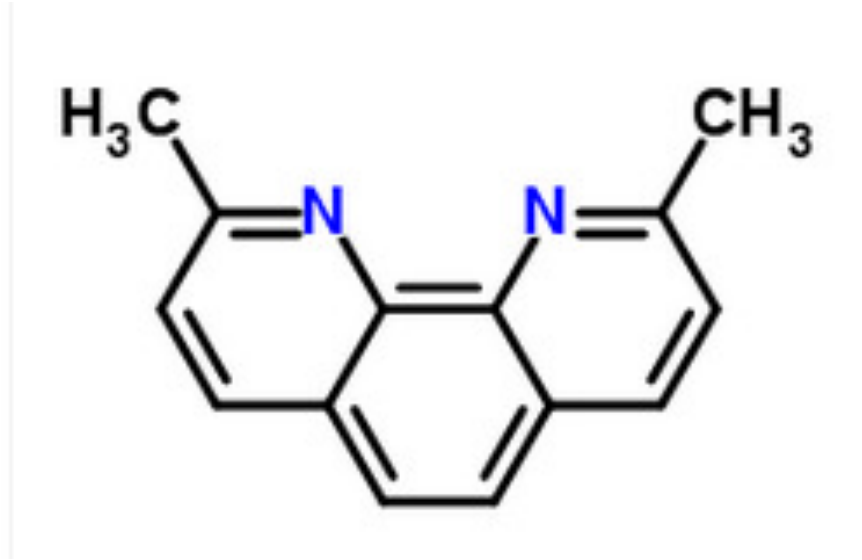} \\ % molecules/29-dm-phen.pdf
{\bf 2,9-dm-phen}: 2,9-dimethyl-1,10-phenanthroline \\
\end{tabular}
\caption{
Ligand list (part VII).  \label{fig:liglist.7}
}
\end{figure}

\begin{figure}
\begin{tabular}{c}
\includegraphics[width=0.25\textwidth]{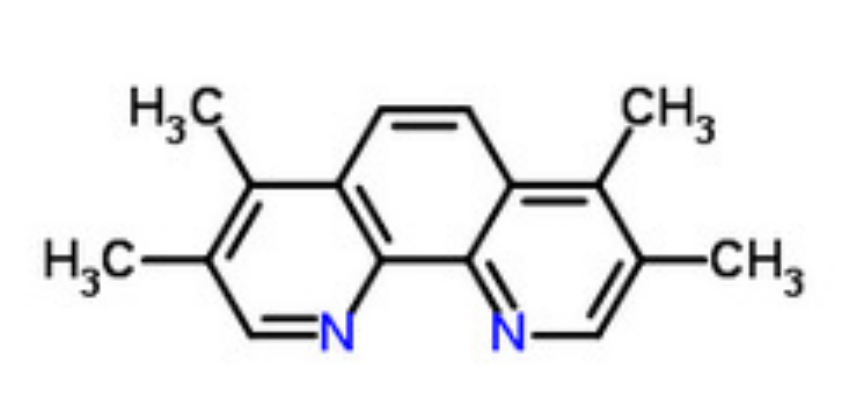} \\ % molecules/tm1-phen.pdf
{\bf tm1-phen}: 3,4,7,8-tetramethyl-1,10-phenanthroline \\
\includegraphics[width=0.25\textwidth]{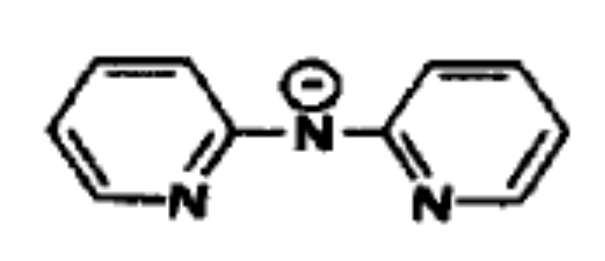} \\  % molecules/DPA.pdf
{\bf DPA}: di-2-pyridylamine anion \\
\includegraphics[width=0.15\textwidth]{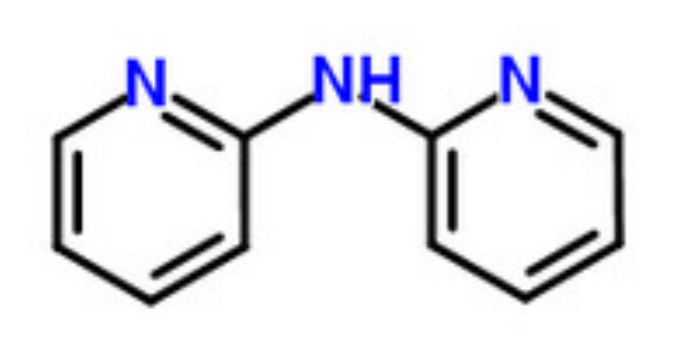} \\ % molecules/DPAH.pdf
{\bf DPAH}: di-2-pyridylamine \\
\includegraphics[width=0.10\textwidth]{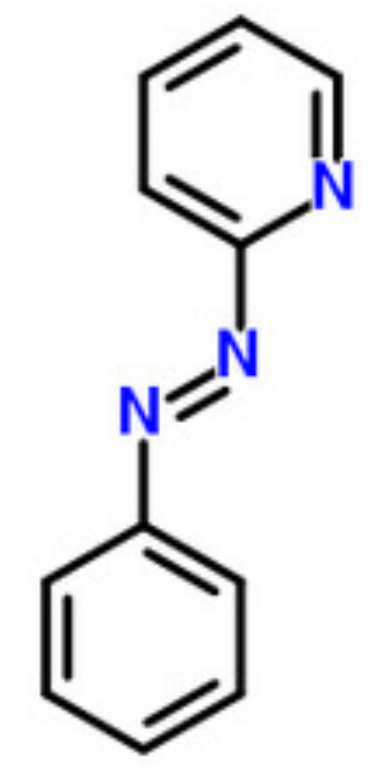} \\ % molecules/Azpy.pdf
{\bf Azpy}: 2-(phenylazo)pyridine \\
\includegraphics[width=0.30\textwidth]{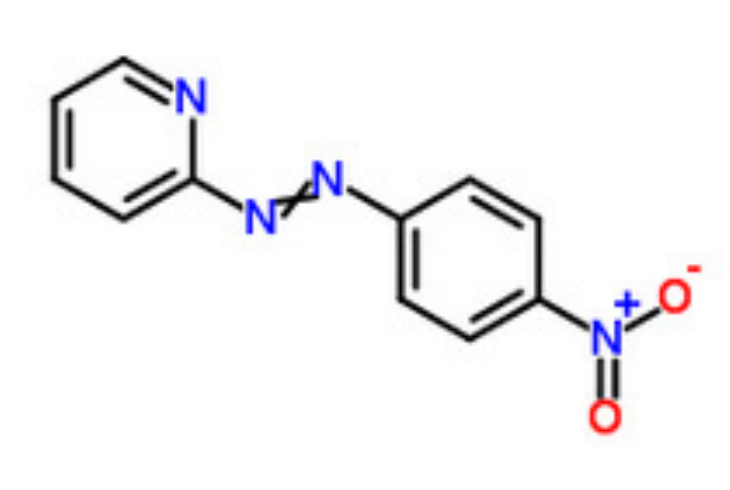} \\ % molecules/NA.pdf
{\bf NA}: 2-((4-nitrophenyl)azo)pyridine \\
\includegraphics[width=0.15\textwidth]{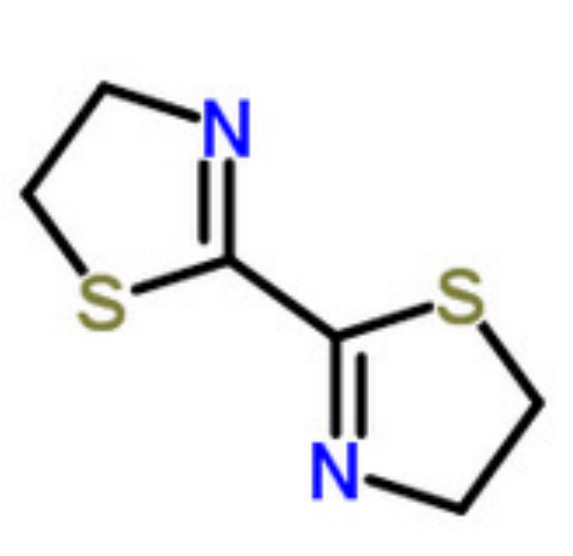} \\ % molecules/bt.pdf
{\bf bt}: 2,2'-bi-2-thiazoline \\
\end{tabular}
\caption{
Ligand list (part VIII).  \label{fig:liglist.8}
}
\end{figure}

\begin{figure}
\begin{tabular}{c}
\includegraphics[width=0.20\textwidth]{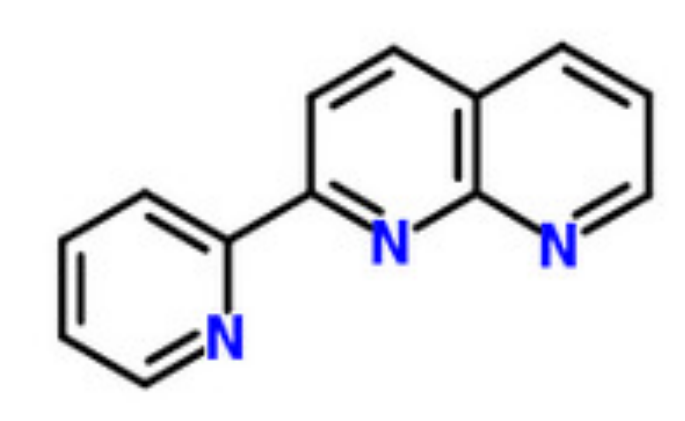} \\ % molecules/pynapy.pdf
{\bf pynapy}: 2-(2-pyridyl)-1,8-naphthyridine \\
\includegraphics[width=0.30\textwidth]{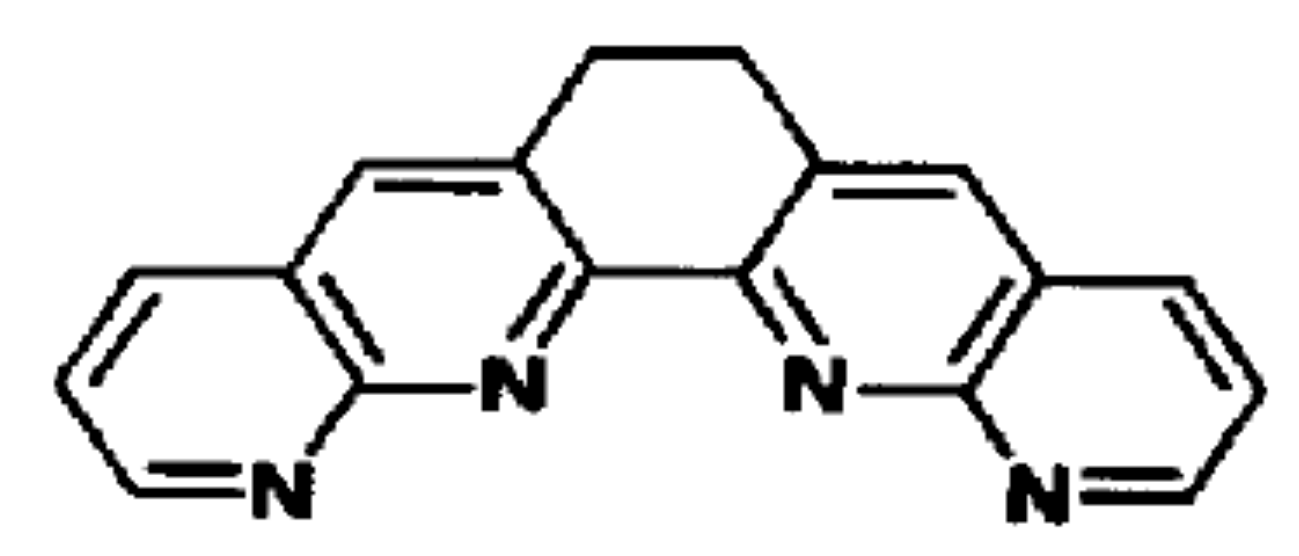} \\ % molecules/dinapy.pdf
{\bf dinapy}: 5,6-dihydro-dipyrido[3,2-{\em b}:2',3'-{\em j}]\\
$[1,10]$phenanthroline \\
\includegraphics[width=0.30\textwidth]{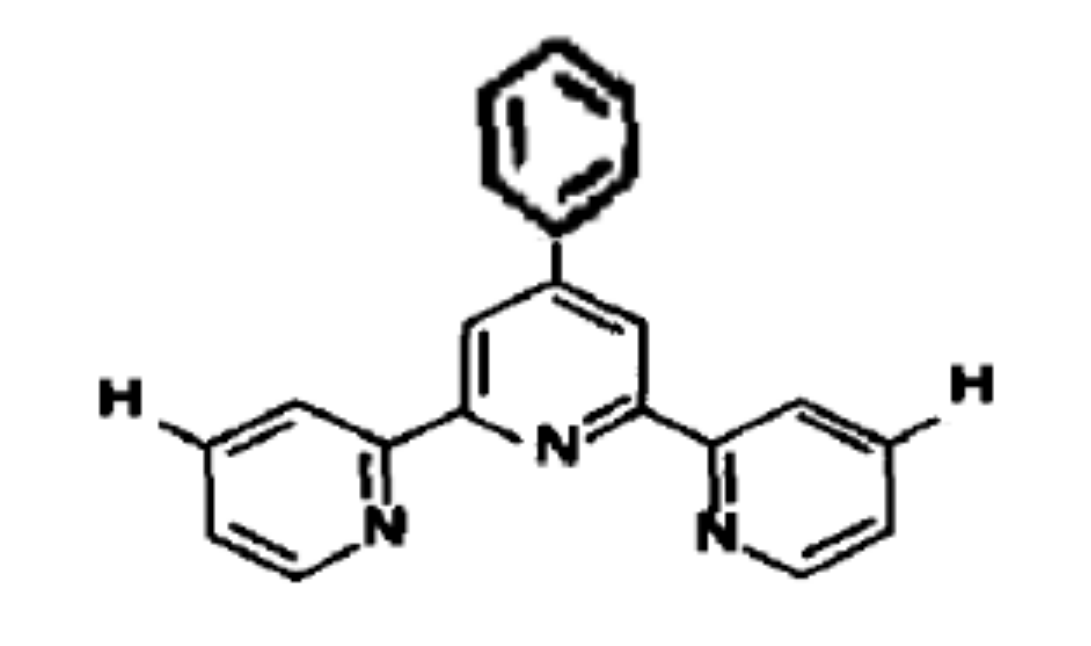} \\ % molecules/tro.pdf
{\bf tro}: 4'-phenyl-2,2',2''-tripyridine \\
\includegraphics[width=0.30\textwidth]{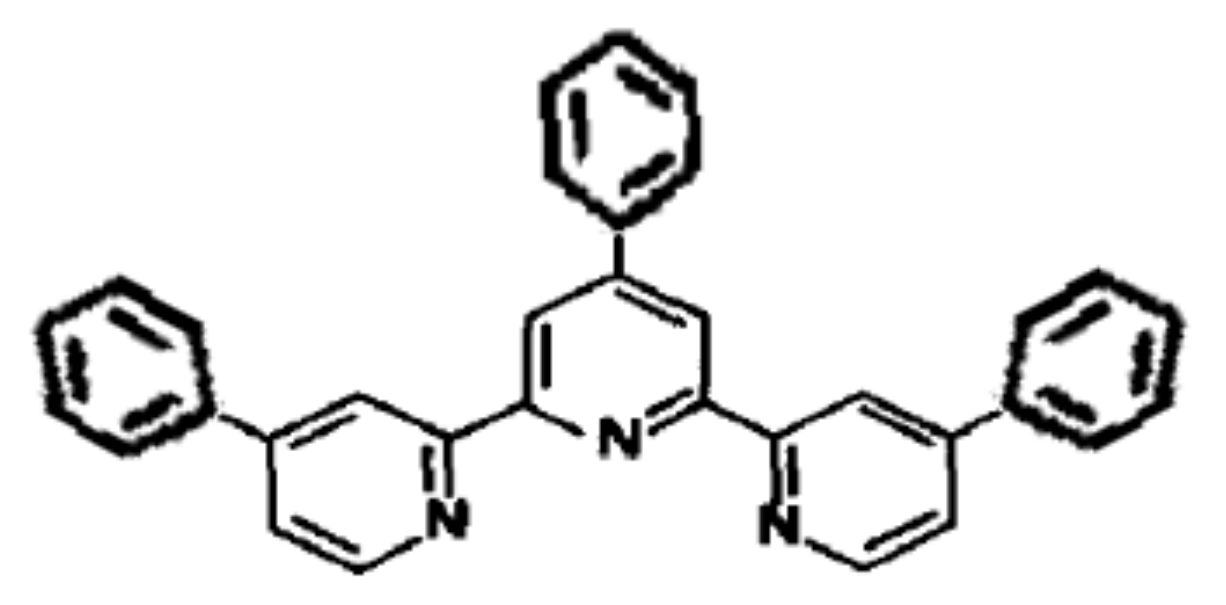} \\ % molecules/tsite.pdf
{\bf tsite}: 4,4',4''-triphenyl-2,2',2''-tripyridine \\
\includegraphics[width=0.30\textwidth]{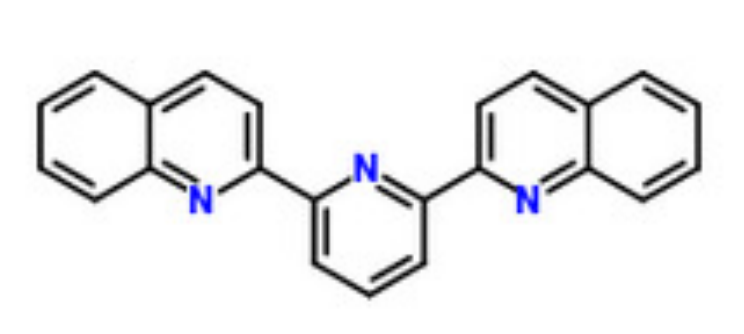} \\ % molecules/dqp.pdf
{\bf dqp}: 2,6-di-(2'-quinolyl)pyridine \\
\end{tabular}
\caption{
Ligand list (part IX).  \label{fig:liglist.9}
}
\end{figure}

The ligand abbreviations used in this paper are the same as those used
in Ref.~\cite{JBB+88}.  For the readers convenience, these ligands are 
shown in Figs.~\ref{fig:liglist.1}, \ref{fig:liglist.2}, \ref{fig:liglist.3}, 
\ref{fig:liglist.4}, \ref{fig:liglist.5}, \ref{fig:liglist.6}, 
\ref{fig:liglist.7}, \ref{fig:liglist.8}, and \ref{fig:liglist.9} % ,
% and \ref{fig:liglist.10}
and in order of their appearance in the 
Tables~\ref{tab:JBB88names1}, \ref{tab:JBB88names2}, \ref{tab:JBB88names3}, 
and \ref{tab:JBB88names4}.
% --------------------------------------------
% --------------------------------------------

% --------------------------------------------
% --------------------------------------------

% --------------------------------------------
% --------------------------------------------

% --------------------------------------------
% --------------------------------------------

% --------------------------------------------
% --------------------------------------------

% --------------------------------------------
% --------------------------------------------

% --------------------------------------------
% --------------------------------------------

% --------------------------------------------
% --------------------------------------------

% --------------------------------------------
% --------------------------------------------
% ------------------------------
\bibliographystyle{myaip}
\bibliography{refs}

% -----------------------------------------------
\end{document}

% ====================================================
% \input{titlesupp.tex}
% ===============================================
% File titlesupp.tex
% Last modified: 11 April 2017
% ===============================================
% ================================================
%              Title
% ===============================================
\onecolumn
\begin{center}
\textbf{\large Supplementary Material: Partial Density of States Ligand Field
Theory (PDOS-LFT): Recovering a LFT-Like Picture and Application to the
Photoproperties of Ruthenium Polypyridine Complexes}\\
by Denis Magero, Mark E.\ Casida, Nicholas Makau, George Amolo, and 
Lusweti Kituyi\\
Last update: \today
\end{center}

This supplementary material consists of a systematic collection of 
our calculated partial density of states (PDOS) and time-dependent B3LYP 
(TD-B3LYP) spectra for the complexes treated in the main paper.  

B3LYP highest-occupied molecular orbital (HOMO) energies, taken directly
from the {\sc Gaussian} outputs, are also given.  These provide an indication 
of the start of the HOMO-LUMO (lowest unoccupied molecular orbital) gap.
The corresponding notion in solid-state physics is the Fermi energy (roughly
equal to the average of the HOMO and LUMO energies) which is an alternative
way to indicate the position of the HOMO-LUMO gap.
% Calculated B3LYP highest-occupied molecular orbital (HOMO) energies
% are also given.  

Complexes indicated with an asterisk (*) have 
unbound (i.e., postive energy) $e_g^*$ orbitals in their PDOS. 
Some complexes could not be included
because of difficulty optimizing their geometries.  The PDOS could
not always be calculated because of current program limitations.  Complexes
with only TD-B3LYP spectra are indicated with a dagger ($\dagger$).

% \begin{verbatim}
% 
% Message from Denis of 13 April 2017:
% 
% Dear Mark, Dear All,
% 
% Indeed there is much work to be done on the supplementary information 
% regarding the issues that you had raised. At the moment, I would 
% suggest that if the complexes have the spectra and not the PDOS, 
% we can keep it in the database and justify the reason why we could 
% not find the PDOS. A good example is the complexes that contained a 
% Cl atom as one of the ligands, you realize that for all these complexes, 
% there was no PDOS convoluted and the program crushed without a definite 
% error that could be traced back to. For the opposite case, where we 
% have the PDOS and not the spectra, in the calculations such a case does 
% not exist.
% 
% For the bug that was discovered, it did not affect the calculations. 
% I remember redoing the spectrum plots and all of them were giving the 
% same spectra with peaks at the same point and with the same molar 
% extinction coefficient.
% 
% About the issue of luminescence lifetimes falling again once Delta E is 
% too large, I will have to check on this and give a reply as soon as possible.
% 
% Best regards.
% 
% 
% \end{verbatim}

% From {\tt spectrum\_v3.py}:
% \begin{verbatim}
% % Last updated: 6 July 2015                                        #
% # The program is now translated into English (from French).        #
% # There was a bug in the routine convers4 which is now corrected.  #
% # This bug led to wrong results for the conversion cm-1 or eV to   #
% # nm.                                                              #
% \end{verbatim}

\tableofcontents

% -----------------
% THE END 
% -----------------

% \maketitle
% ---------------------------------------------------

% \input{./tables/PDOS.tex}
% \input{./tables/spectra.tex}

% ================================================
\newpage
\section{Complex {\bf (1)}*: [Ru(bpy)(CN)$_4$]$^{2-}$}
% ================================================

\begin{center}
   {\bf PDOS}
\end{center}

\begin{center}
\begin{tabular}{cc}
\includegraphics[width=0.4\textwidth]{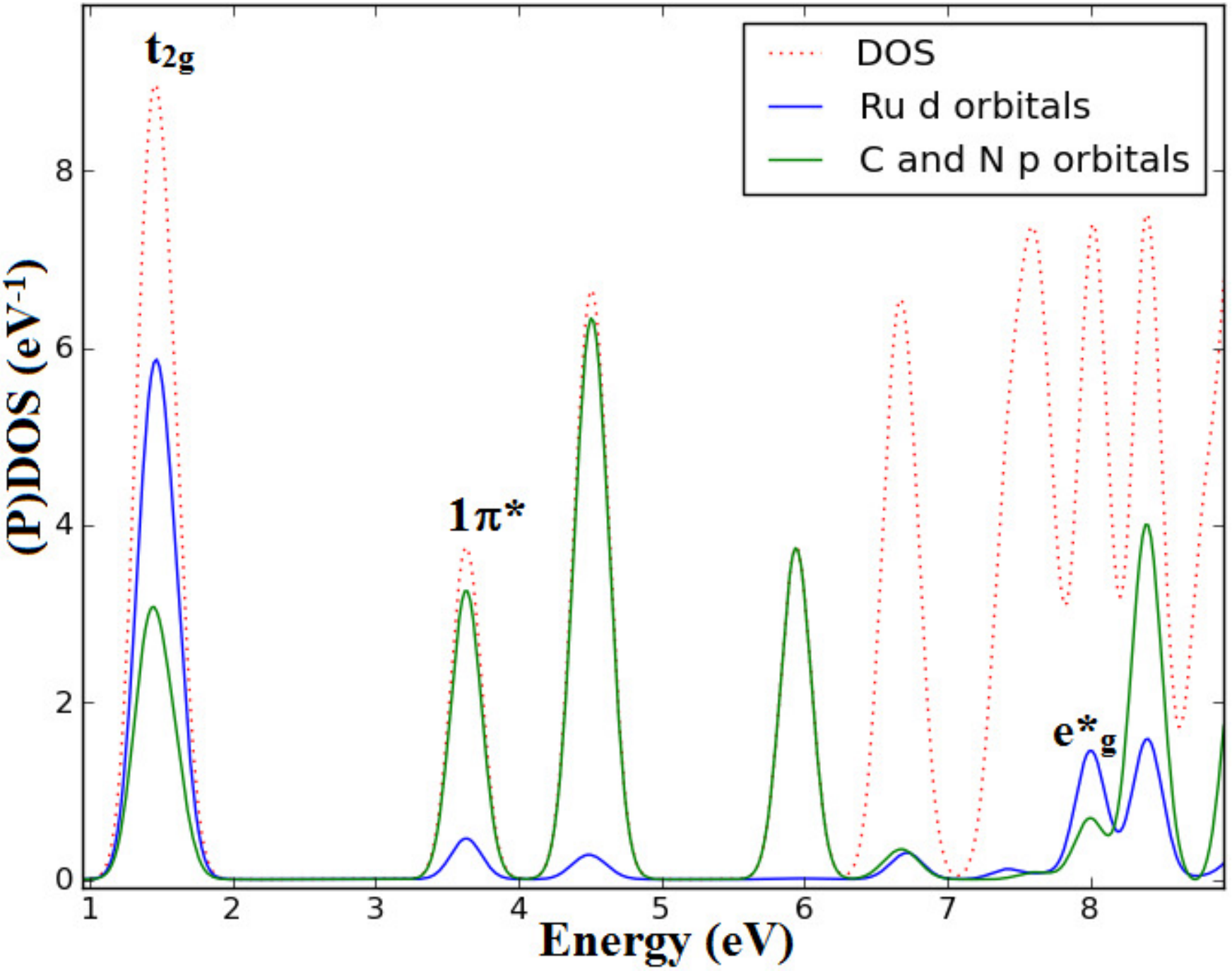} &
\includegraphics[width=0.4\textwidth]{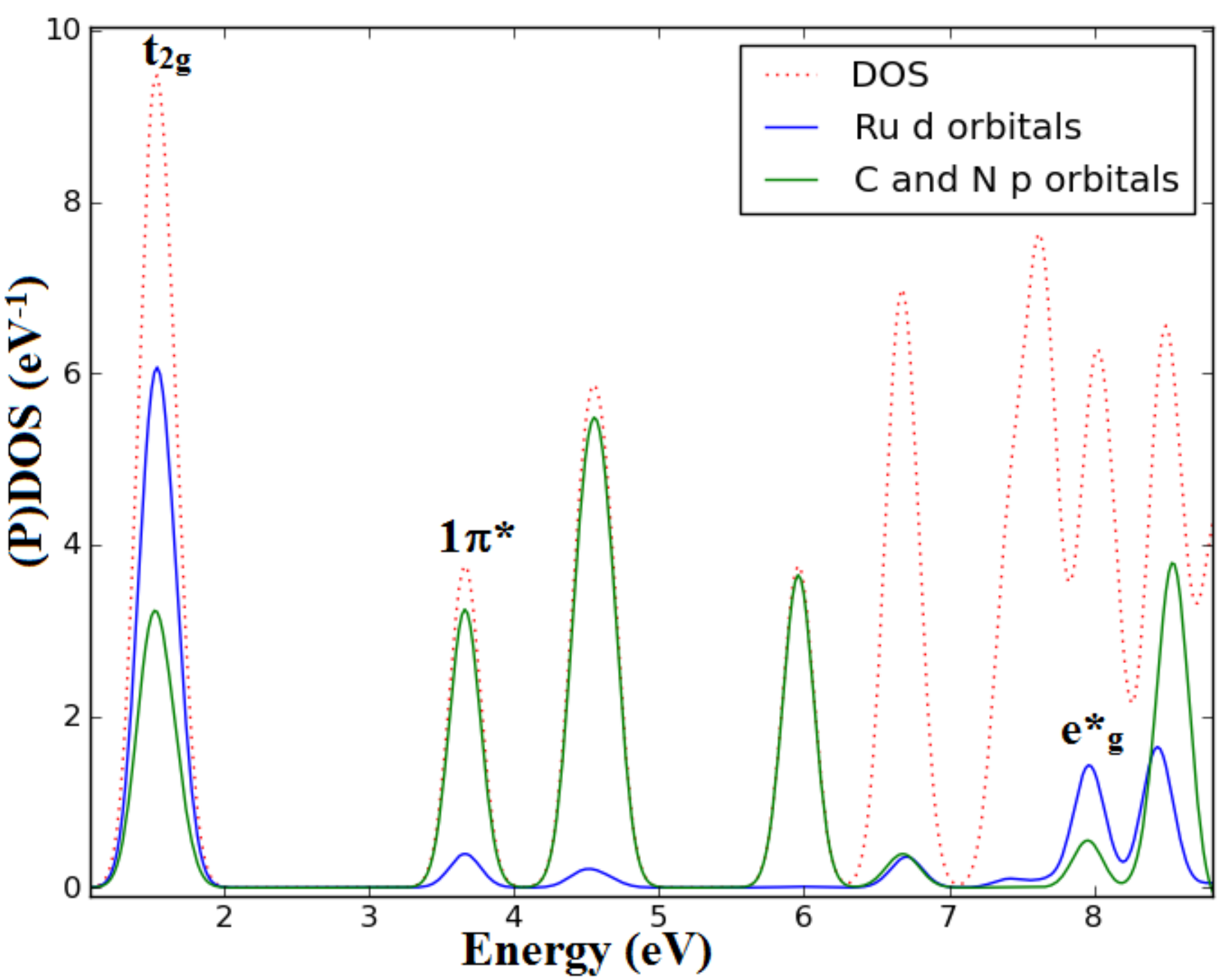} \\
B3LYP/6-31G & B3LYP/6-31G(d) \\
$\epsilon_{\text{HOMO}} = \mbox{1.57 eV}$ & 
$\epsilon_{\text{HOMO}} = \mbox{1.63 eV}$ 
\end{tabular}
\end{center}
Total and partial density of states of [Ru(bpy)(CN)$_{4}$]$^{2-}$ partitioned 
over Ru d orbitals and ligand C and N p orbitals. 
% for the 6-31G (left-hand side) and 6-31G(d) (right-hand side) basis sets.

% \begin{center}
% \begin{tabular}{cc}
% \hline \hline 
% 6-31G & 6-31G(d) \\
% \hline
% 1.57 eV & 1.63 eV \\
% \hline \hline
% \end{tabular}
% HOMO energy.
% \end{center}

\begin{center}
   {\bf Absorption Spectrum}
\end{center}

\begin{center}
\includegraphics[width=0.8\textwidth]{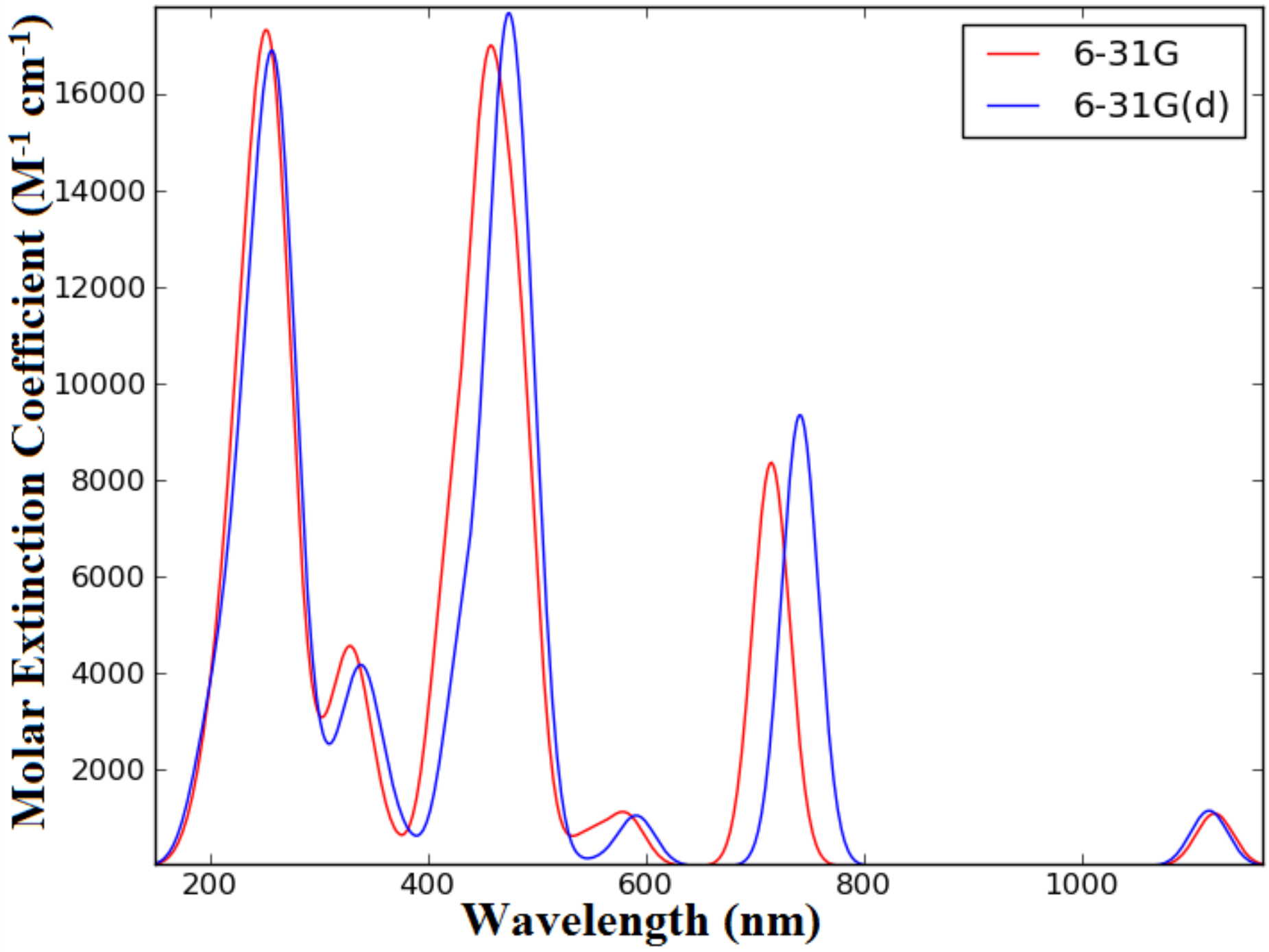}
\end{center}
[Ru(bpy)(CN)$_{4}$]$^{2-}$ TD-B3LYP/6-31G and TD-B3LYP/6-31G(d) spectra.

% ================================================
\newpage
\section{Complex {\bf (2)}$^\dagger$: [Ru(bpy)$_2$Cl$_2$]}
% ================================================

\begin{center}
\begin{tabular}{cc}
B3LYP/6-31G & B3LYP/6-31G(d) \\
$\epsilon_{\text{HOMO}} = \mbox{-4.52 eV}$ & 
$\epsilon_{\text{HOMO}} = \mbox{-4.47 eV}$ 
\end{tabular}
\end{center}

\begin{center}
   {\bf Absorption Spectrum}
\end{center}

\begin{center}
\includegraphics[width=0.8\textwidth]{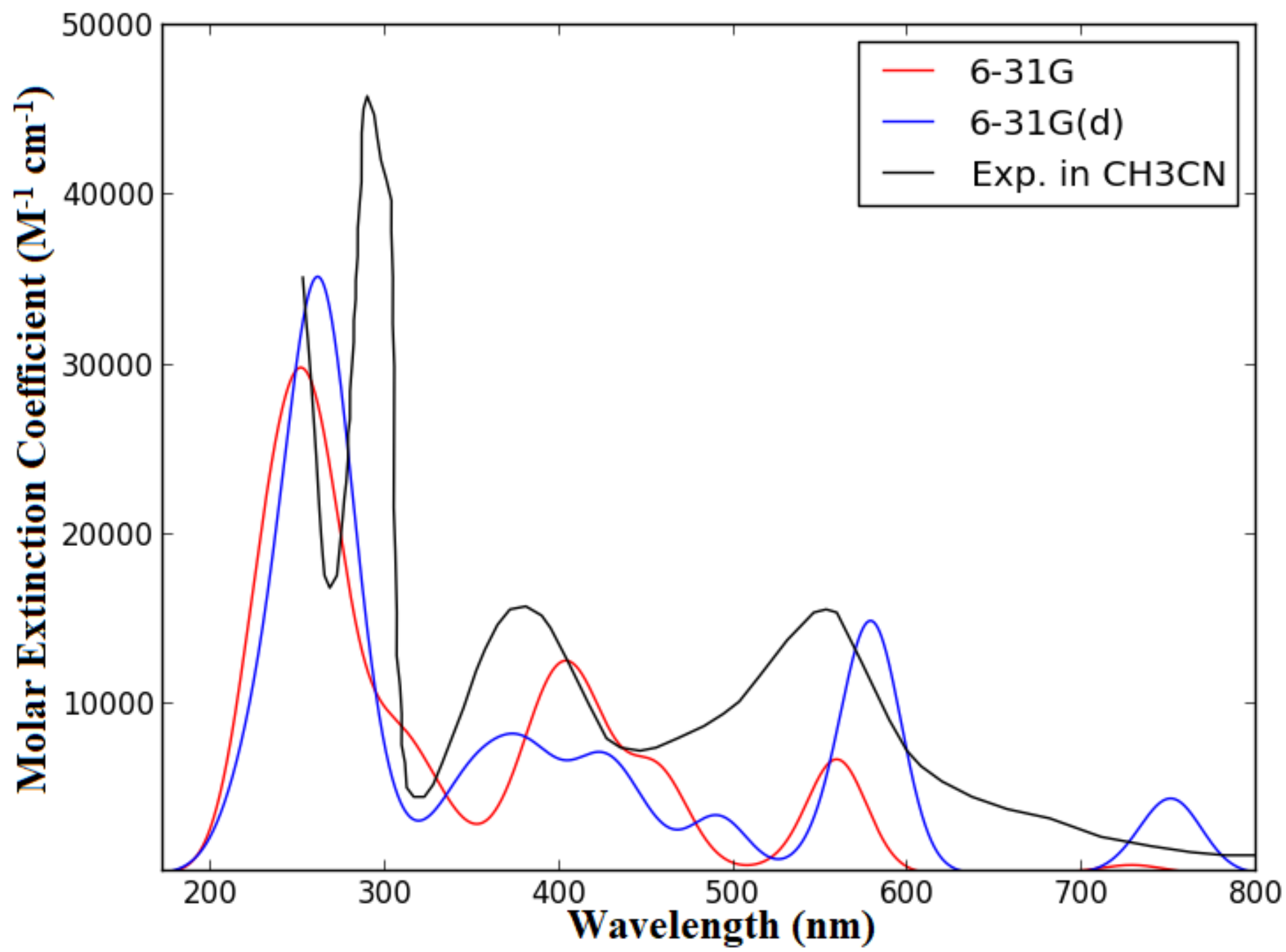}
\end{center}
[Ru(bpy)$_2$Cl$_2$]
TD-B3LYP/6-31G, TD-B3LYP/6-31G(d), and experimental spectra.
Experimental curve measured at room temperature in acetonitrile \cite{GCS10}.

% ================================================
\newpage
\section{Complex {\bf (3)}*: [Ru(bpy)$_2$(CN)$_2$]}
% ================================================

\begin{center}
   {\bf PDOS}
\end{center}

\begin{center}
\begin{tabular}{cc}
\includegraphics[width=0.4\textwidth]{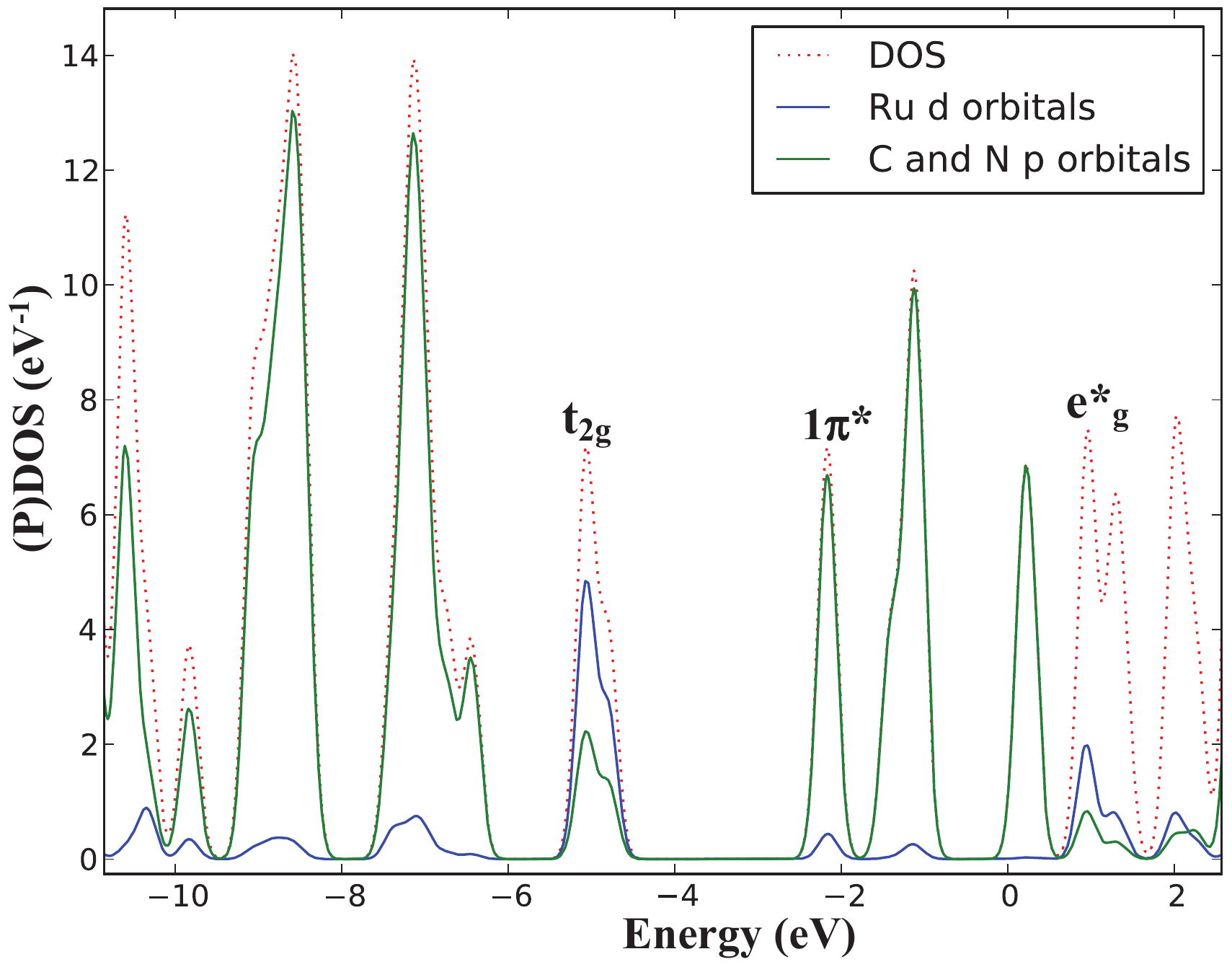} &
\includegraphics[width=0.4\textwidth]{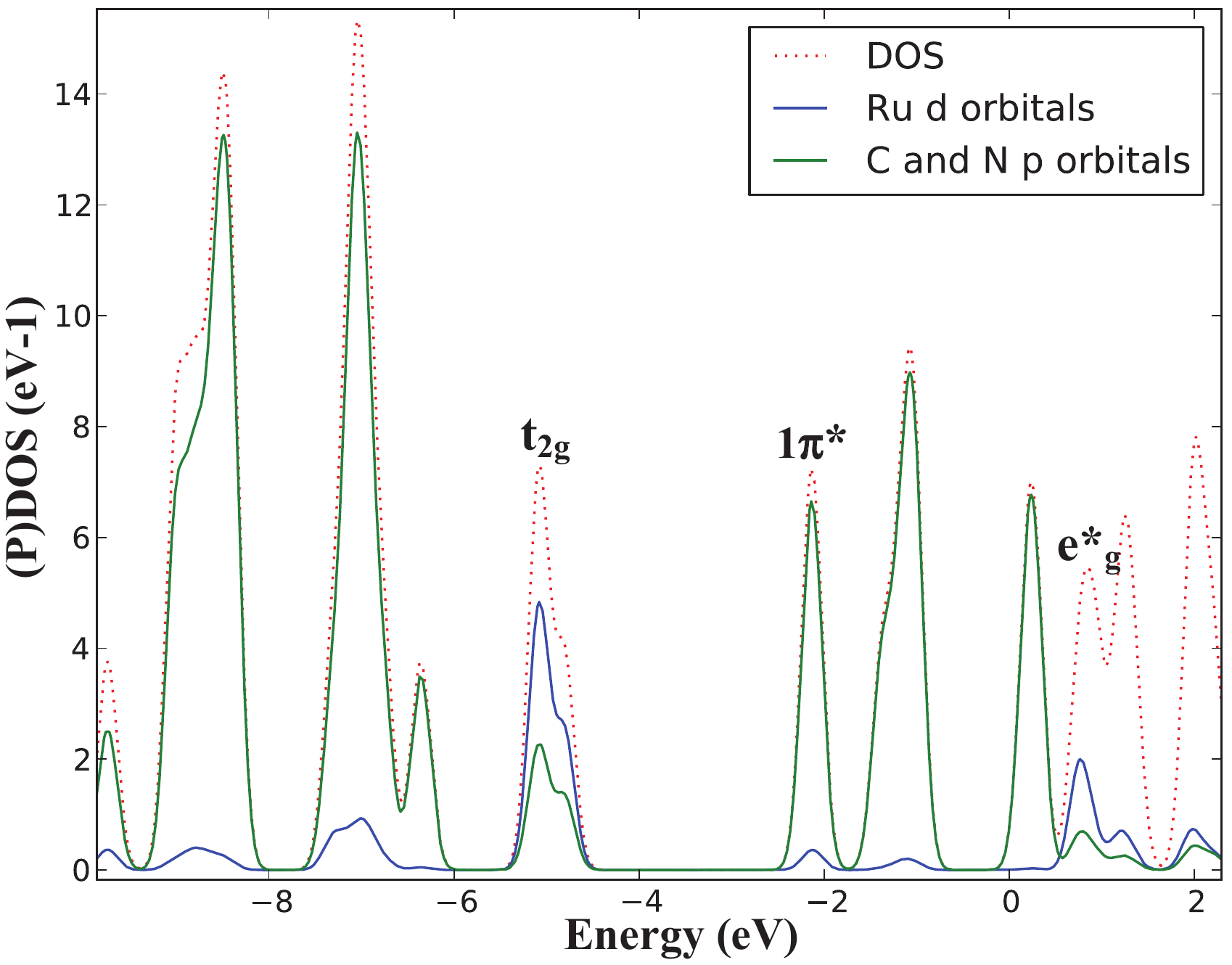} \\
B3LYP/6-31G & B3LYP/6-31G(d) \\
$\epsilon_{\text{HOMO}} = \mbox{-4.80 eV}$ & 
$\epsilon_{\text{HOMO}} = \mbox{-4.80 eV}$ 
\end{tabular}
\end{center}
Total and partial density of states of [Ru(bpy)$_2$(CN)$_2]$ partitioned over Ru 
d orbitals and ligand C and N p orbitals. 
% for the 6-31G (left-hand side) and 6-31G(d) (right-hand side) basis sets.

\begin{center}
   {\bf Absorption Spectrum}
\end{center}

\begin{center}
\includegraphics[width=0.8\textwidth]{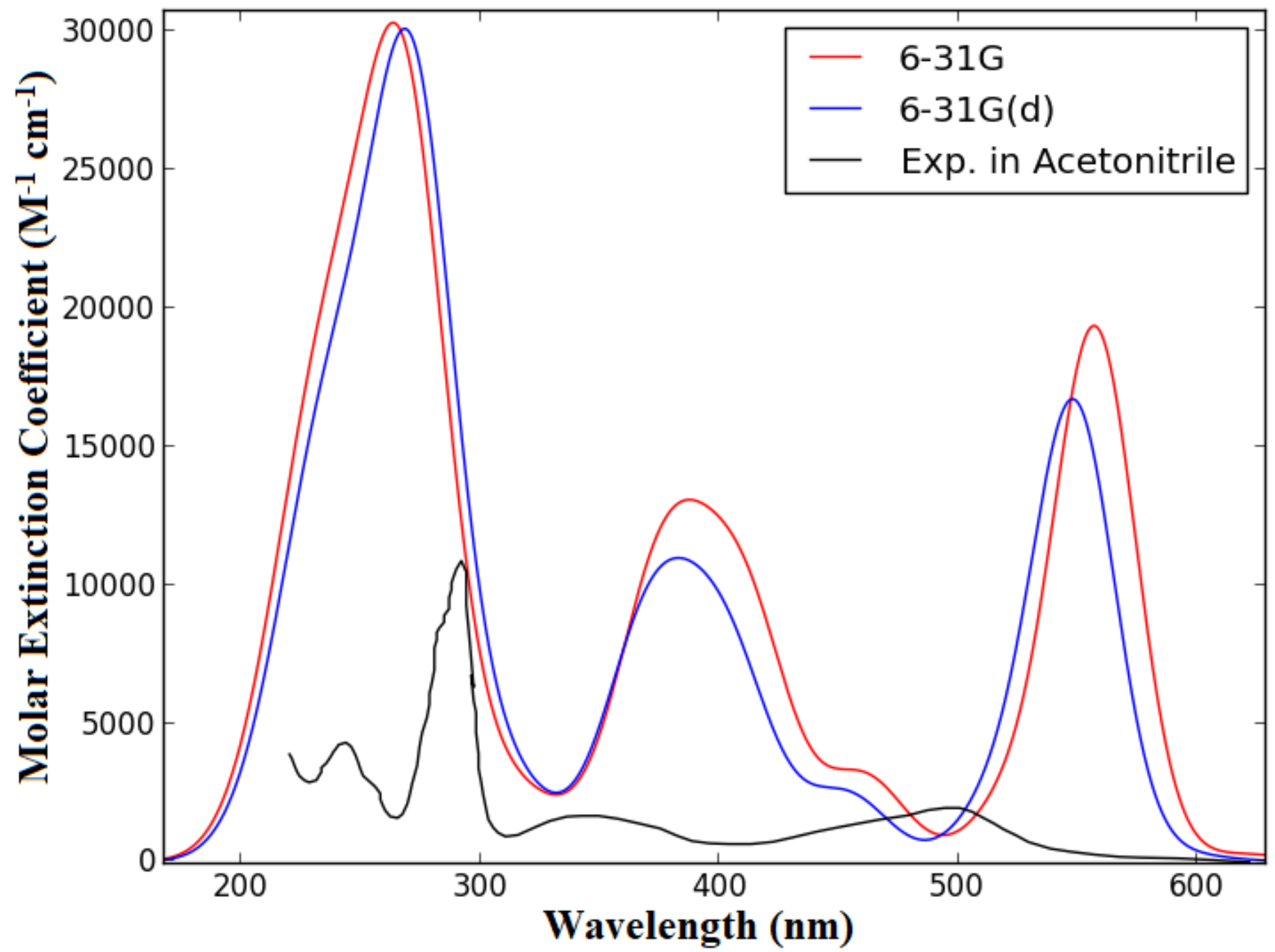}
\end{center}
[Ru(bpy)$_2$(CN)$_2$]
TD-B3LYP/6-31G, TD-B3LYP/6-31G(d), and experimental spectra.
Experimental curve measured at room temperature in acetonitrile \cite{FLH07}.

% ================================================
\newpage
\section{Complex {\bf (4)}: [Ru(bpy)$_2$(en)]}
% ================================================

\begin{center}
   {\bf PDOS}
\end{center}

\begin{center}
\begin{tabular}{cc}
\includegraphics[width=0.4\textwidth]{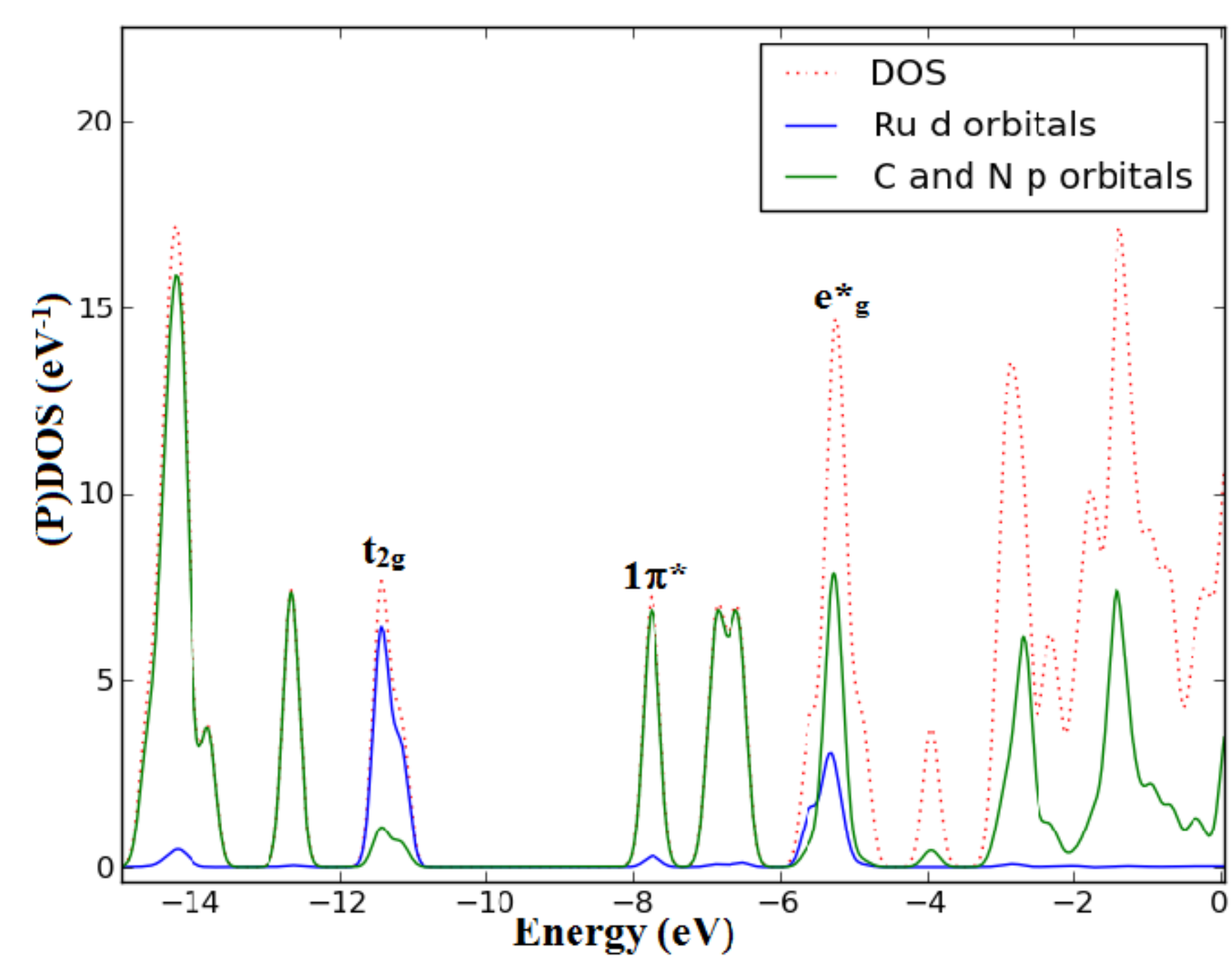} &
\includegraphics[width=0.4\textwidth]{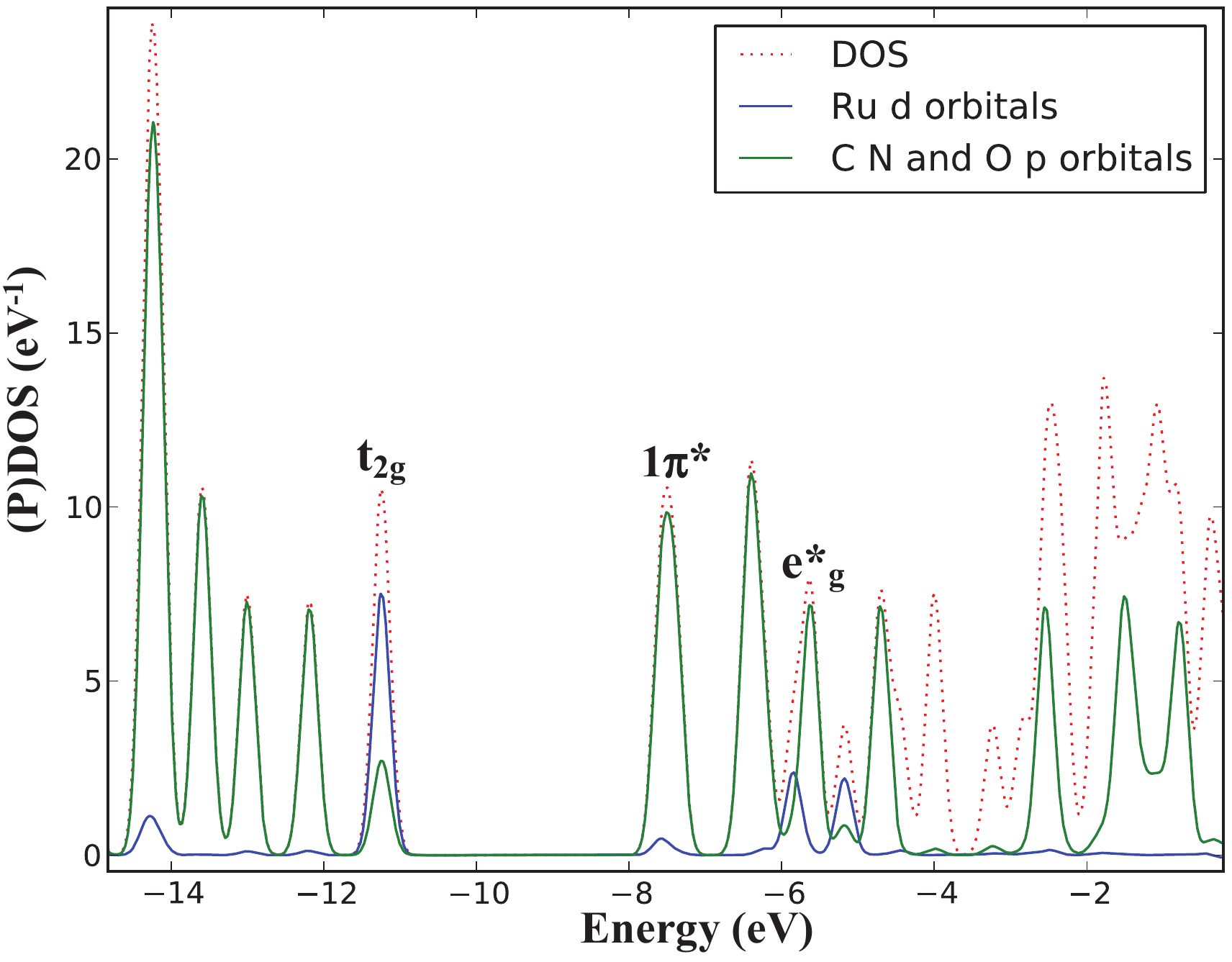} \\
B3LYP/6-31G & B3LYP/6-31G(d) \\
$\epsilon_{\text{HOMO}} = \mbox{-11.16 eV}$ & 
$\epsilon_{\text{HOMO}} = \mbox{-11.32 eV}$ 
\end{tabular}
\end{center}
Total and partial density of states of [Ru(bpy)$_2$(en)]$^{2+}$ partitioned 
over Ru d orbitals and ligand C and N p orbitals.
% for the 6-31G (left-hand side) and 6-31G(d) (right-hand side) basis sets.

\begin{center}
   {\bf Absorption Spectrum}
\end{center}

\begin{center}
\includegraphics[width=0.8\textwidth]{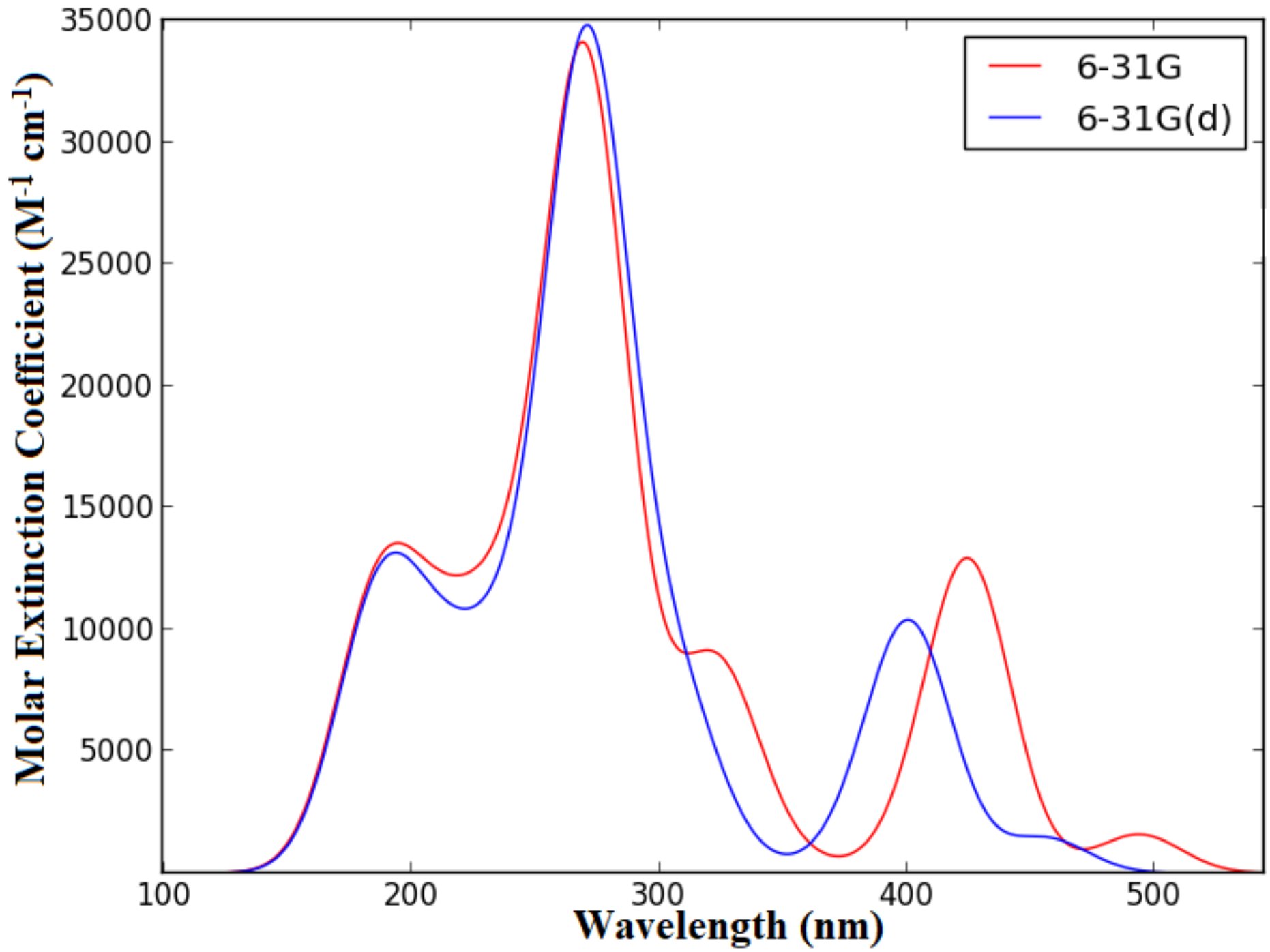}
\end{center}
[Ru(bpy)$_2$)(en)]$^{2+}$ 
TD-B3LYP/6-31G and TD-B3LYP/6-31G(d) spectra.

% ================================================
\newpage
\section{Complex {\bf (5)}*: [Ru(bpy)$_2$(ox)]}
% ================================================

\begin{center}
   {\bf PDOS}
\end{center}

\begin{center}
\begin{tabular}{cc}
\includegraphics[width=0.4\textwidth]{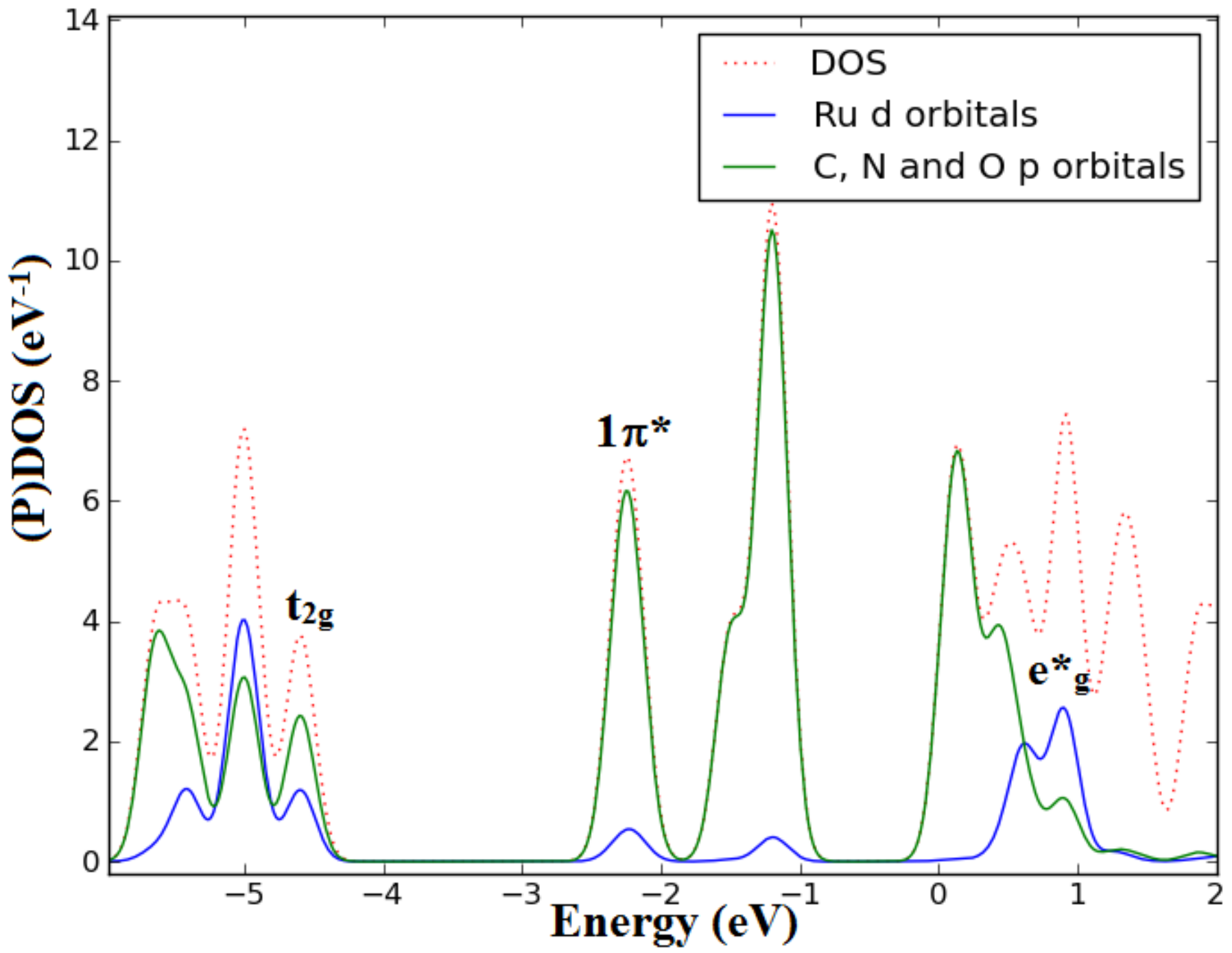} &
\includegraphics[width=0.4\textwidth]{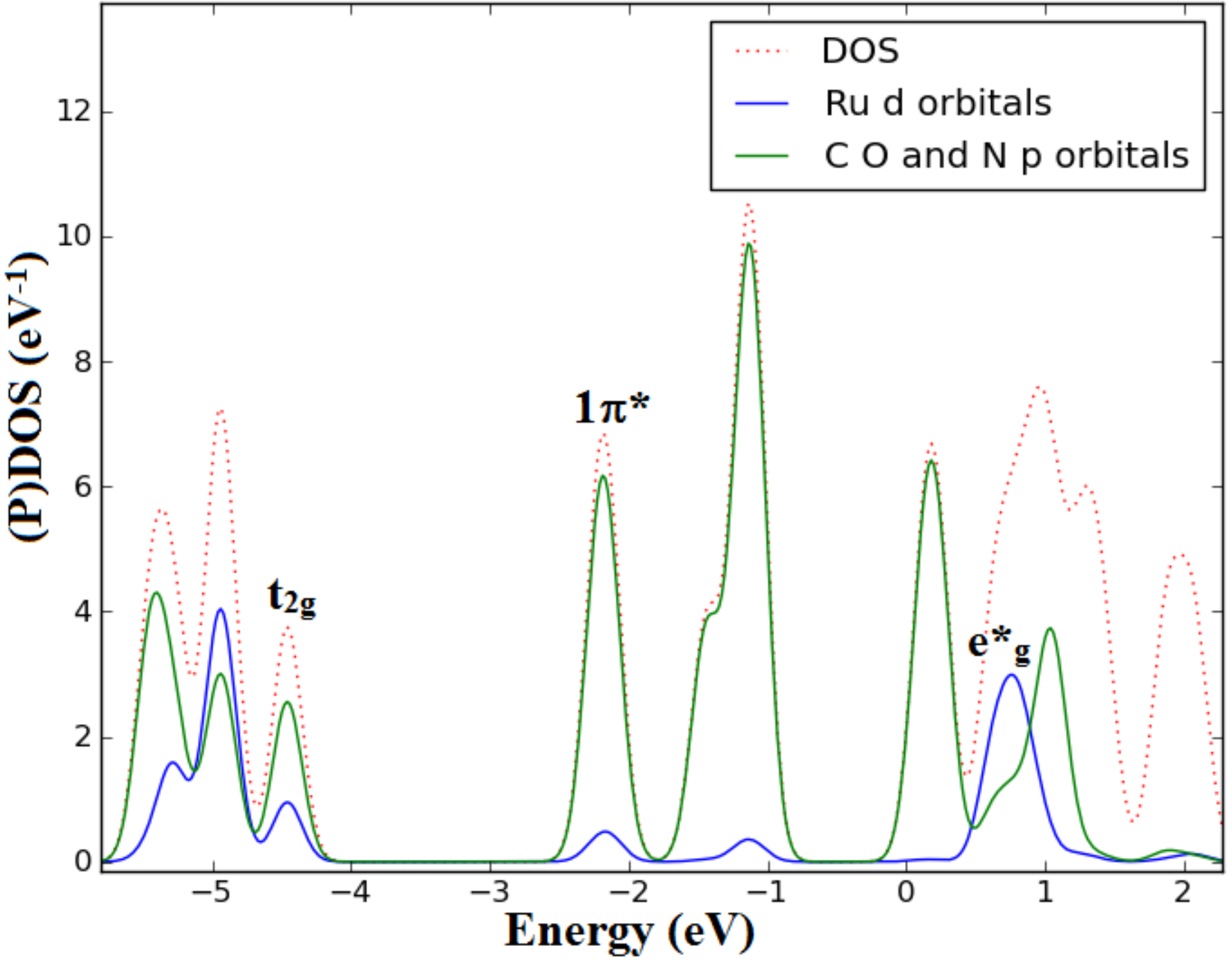} \\
B3LYP/6-31G & B3LYP/6-31G(d) \\
$\epsilon_{\text{HOMO}} = \mbox{-4.60 eV}$ & 
$\epsilon_{\text{HOMO}} = \mbox{-4.46 eV}$ 
\end{tabular}
\end{center}
Total and partial density of states of [Ru(bpy)$_2$(ox)] partitioned over 
Ru d orbitals and ligand C, O and N p orbitals. 
% for the 6-31G (left-hand side) and 6-31G(d) (right-hand side) basis sets.

\begin{center}
   {\bf Absorption Spectrum}
\end{center}

\begin{center}
\includegraphics[width=0.8\textwidth]{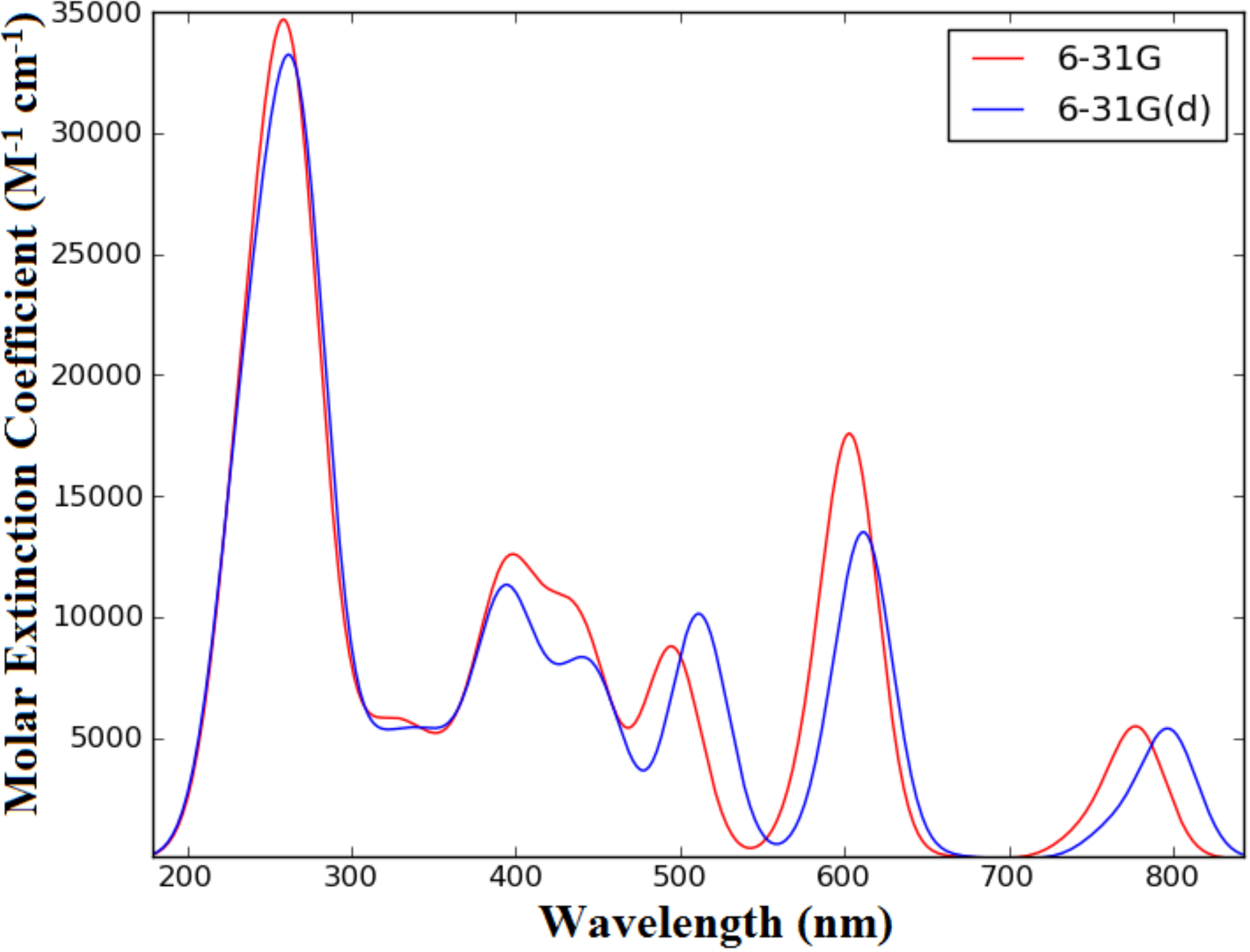}
\end{center}
[Ru(bpy)$_{2}$(ox)] 
TD-B3LYP/6-31G and TD-B3LYP/6-31G(d) spectra.

% ================================================
\newpage
\section{Complex {\bf (6)}: [Ru(bpy)$_3]^{2+}$}
% ================================================

\begin{center}
   {\bf PDOS}
\end{center}

\begin{center}
\begin{tabular}{cc}
\includegraphics[width=0.4\textwidth]{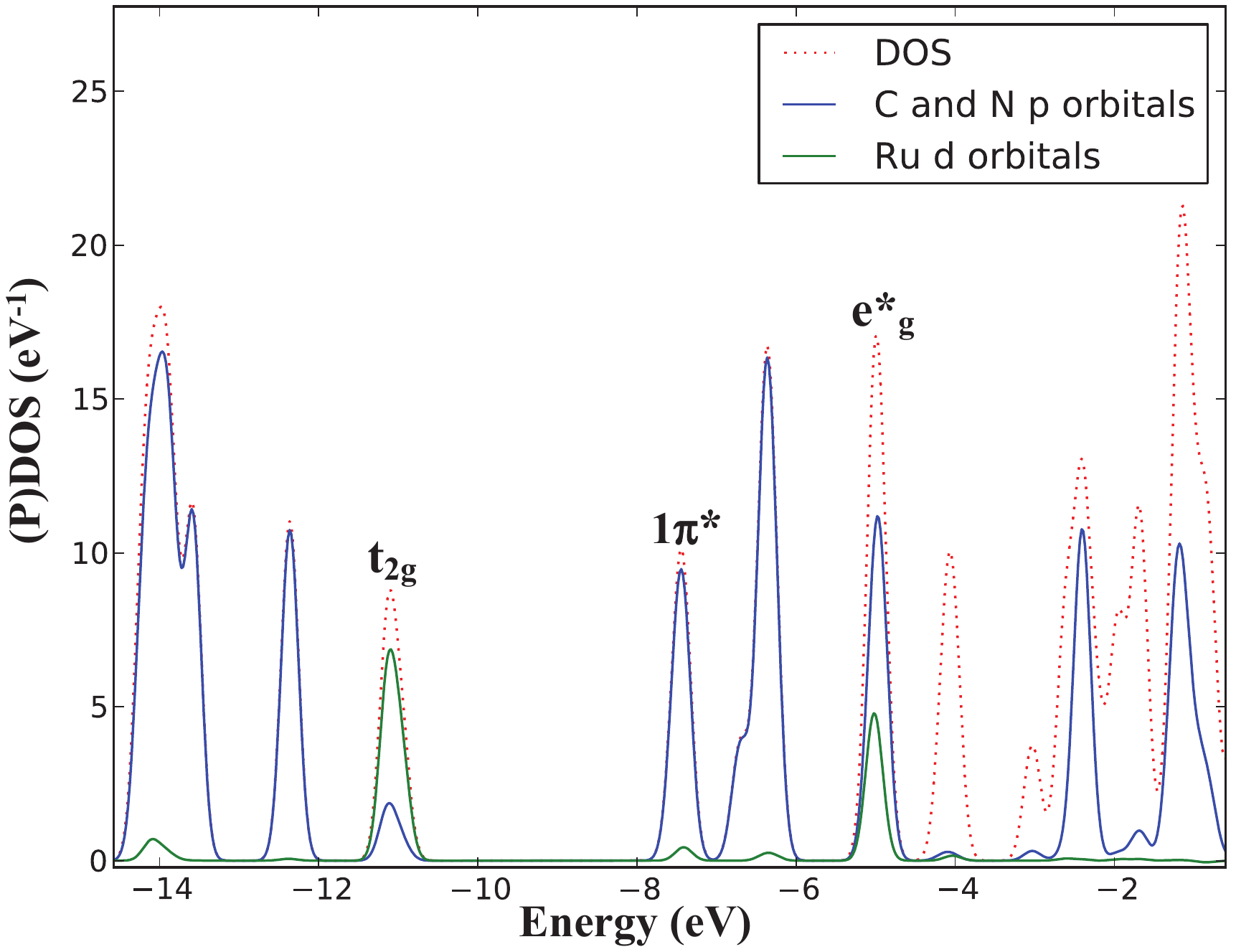} &
\includegraphics[width=0.4\textwidth]{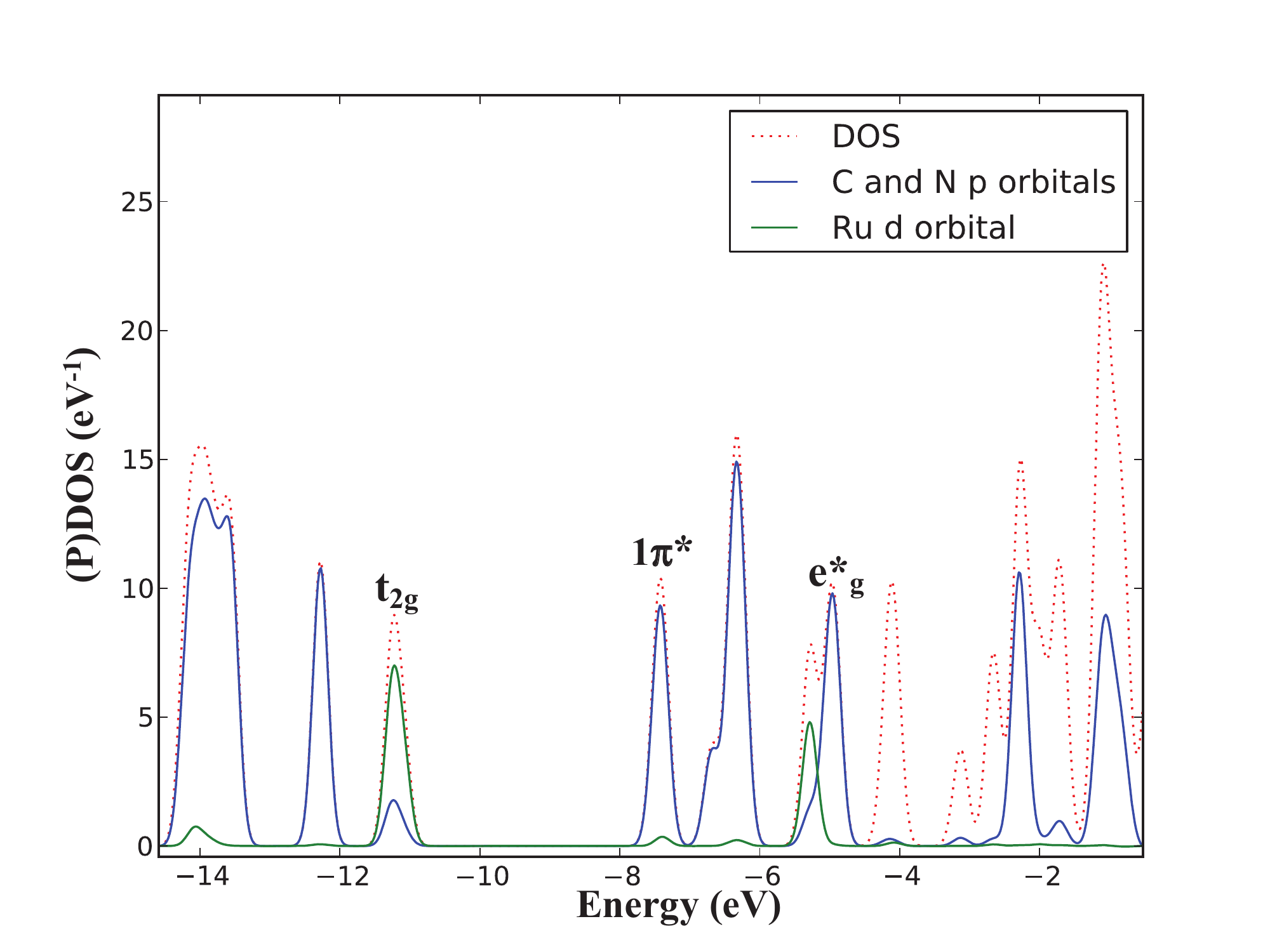} \\
B3LYP/6-31G & B3LYP/6-31G(d) \\
$\epsilon_{\text{HOMO}} = \mbox{-11.20 eV}$ & 
$\epsilon_{\text{HOMO}} = \mbox{-11.31 eV}$ 
\end{tabular}
\end{center}
Total and partial density of states of [Ru(bpy)$_3]^{2+}$ partitioned over 
Ru d orbitals and ligand C and N p orbitals. 
% for the 6-31G (left-hand side) and 6-31G(d) (right-hand side) basis sets.

\begin{center}
   {\bf Absorption Spectrum}
\end{center}

\begin{center}
\includegraphics[width=0.8\textwidth]{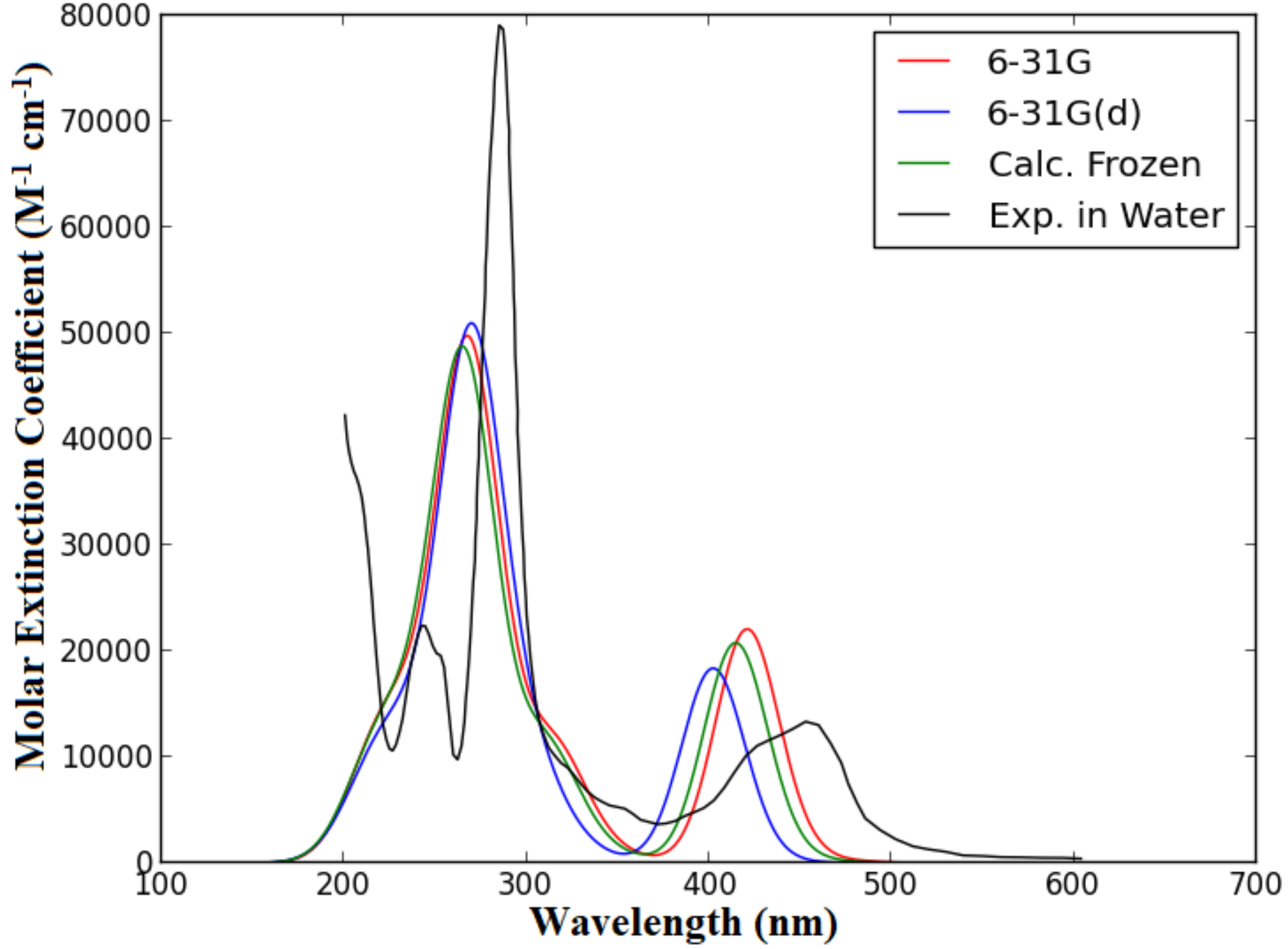}
\end{center}
[Ru(bpy)$_3$]$^{2+}$
TD-B3LYP/6-31G, TD-B3LYP/6-31G(d), and experimental spectra.
``Frozen'' means a calculation at the X-ray crystallography geometry 
without further optimization.
Experimental curve measured at room temperature in water \cite{YHS97}.

% ================================================
\newpage
\section{Complex {\bf (7)}*: [Ru(bpy)$_2$(4-n-bpy)]$^{2+}$}
% ================================================

\begin{center}
   {\bf PDOS}
\end{center}

\begin{center}
\begin{tabular}{cc}
\includegraphics[width=0.4\textwidth]{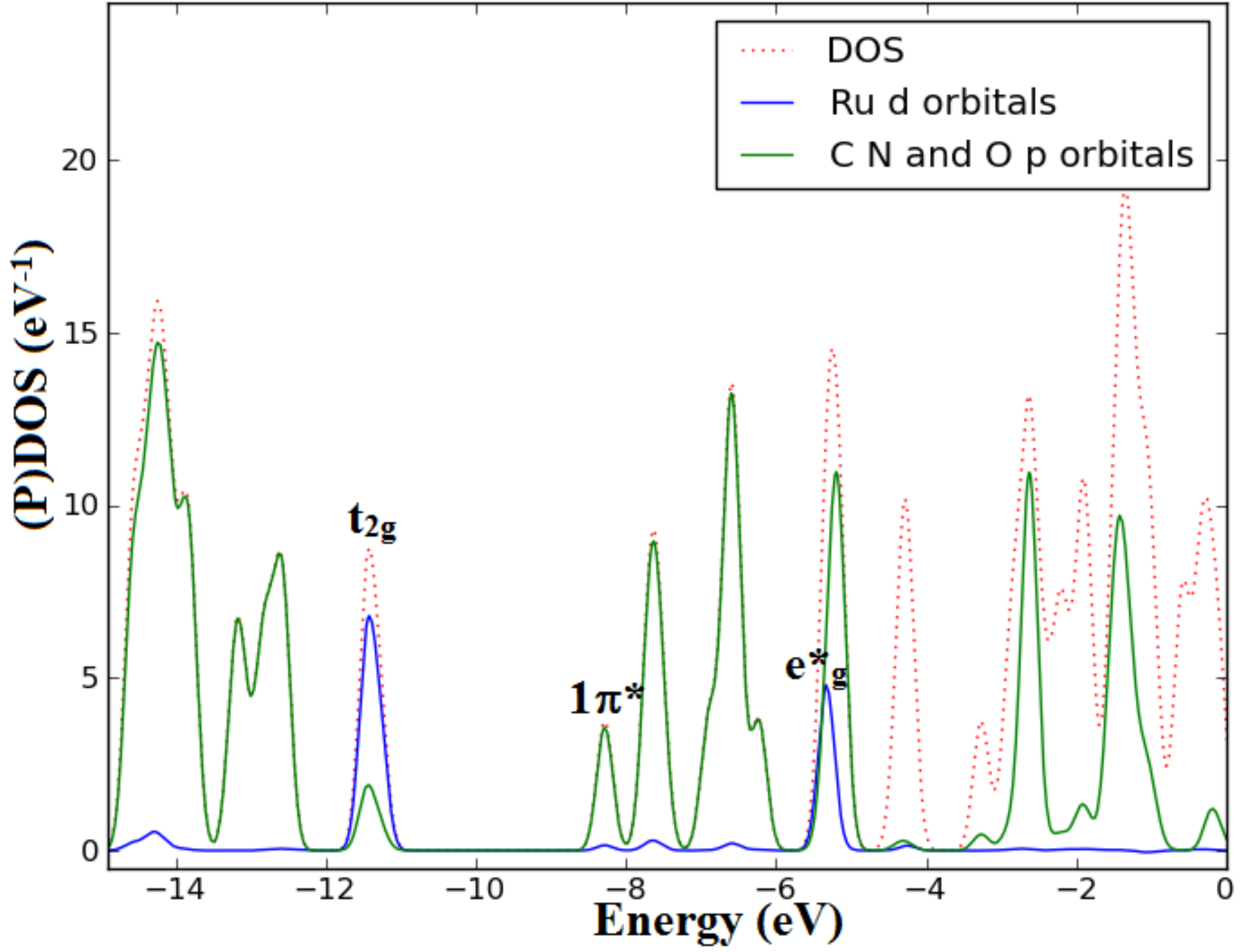} &
\includegraphics[width=0.4\textwidth]{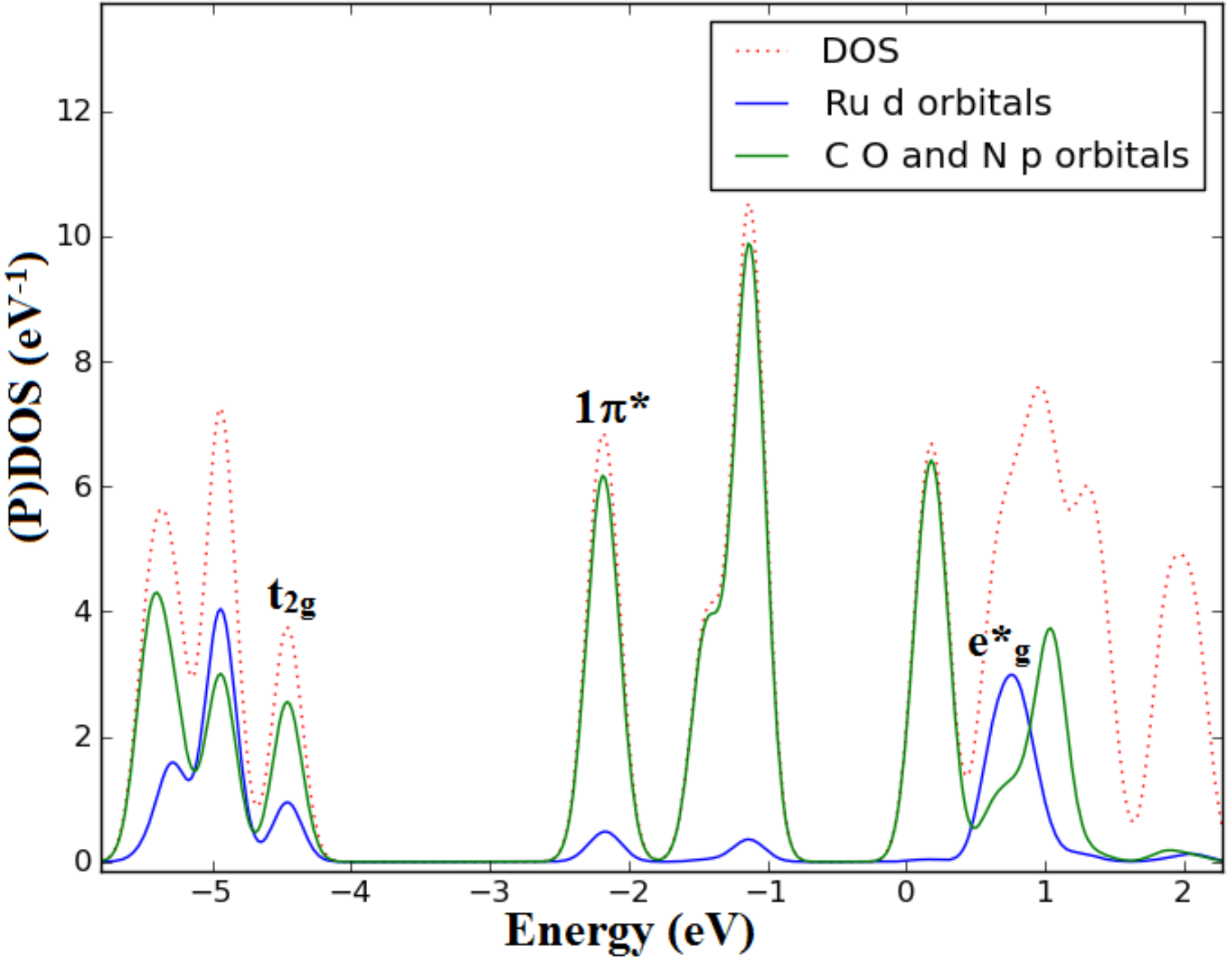} \\
B3LYP/6-31G & B3LYP/6-31G(d) \\
$\epsilon_{\text{HOMO}} = \mbox{-11.28 eV}$ & 
$\epsilon_{\text{HOMO}} = \mbox{-11.36 eV}$ 
\end{tabular}
\end{center}
Total and partial density of states of [Ru(bpy)$_2$(4-n-bpy)]$^{2+}$ 
partitioned over Ru d orbitals and ligand C, O and N p orbitals.
% for the 6-31G (left-hand side) and 6-31G(d) (right-hand side) basis sets.

\begin{center}
   {\bf Absorption Spectrum}
\end{center}

\begin{center}
\includegraphics[width=0.8\textwidth]{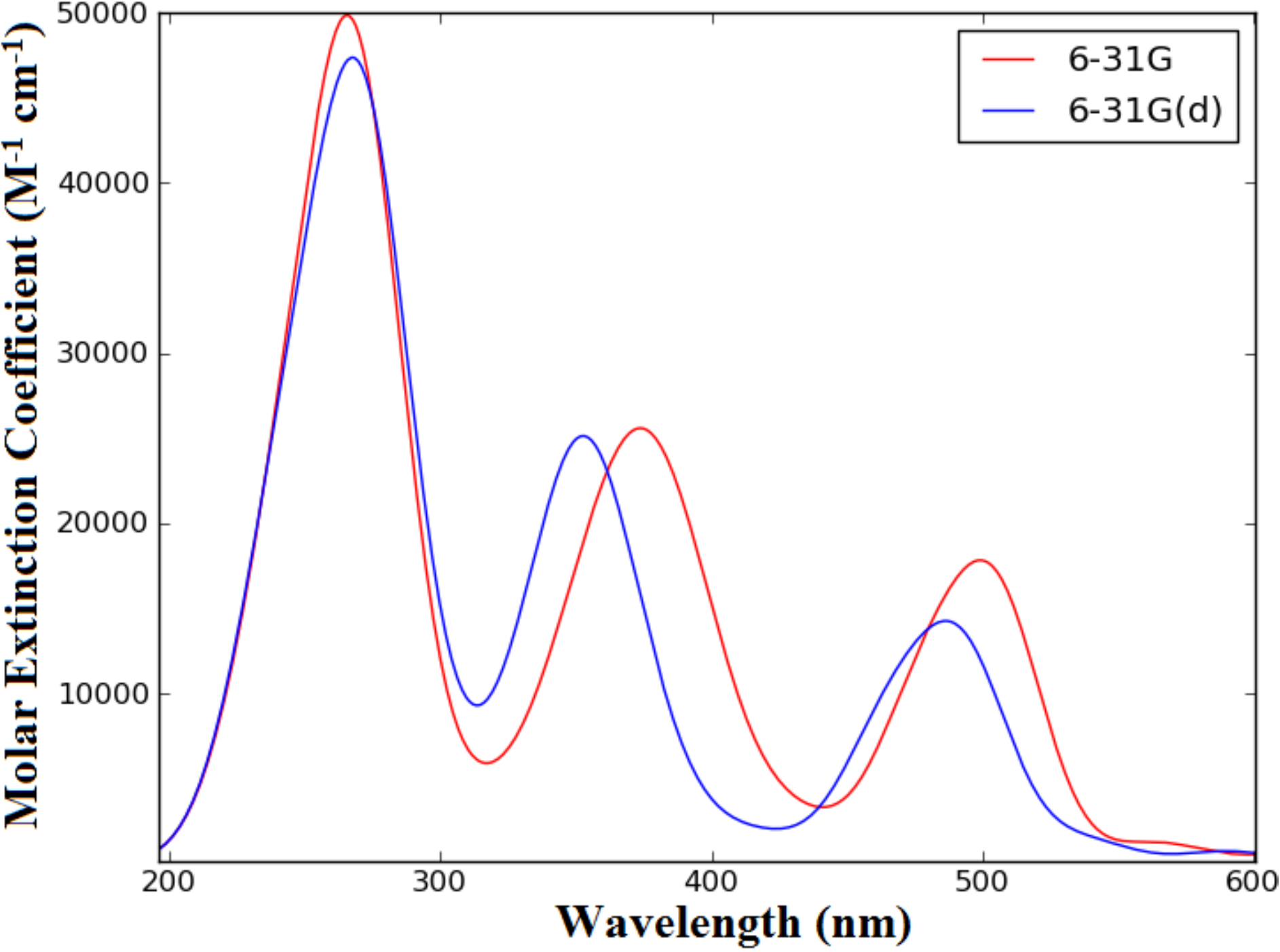}
\end{center}
[Ru(bpy)$_2$(4-n-bpy)]$^{+}$
TD-B3LYP/6-31G and TD-B3LYP/6-31G(d) spectra.

% ================================================
\newpage
\section{Complex {\bf (8)}: [Ru(bpy)$_2$(3,3'-dm-bpy)]$^{2+}$}
% ================================================

\begin{center}
   {\bf PDOS}
\end{center}

\begin{center}
\begin{tabular}{cc}
\includegraphics[width=0.4\textwidth]{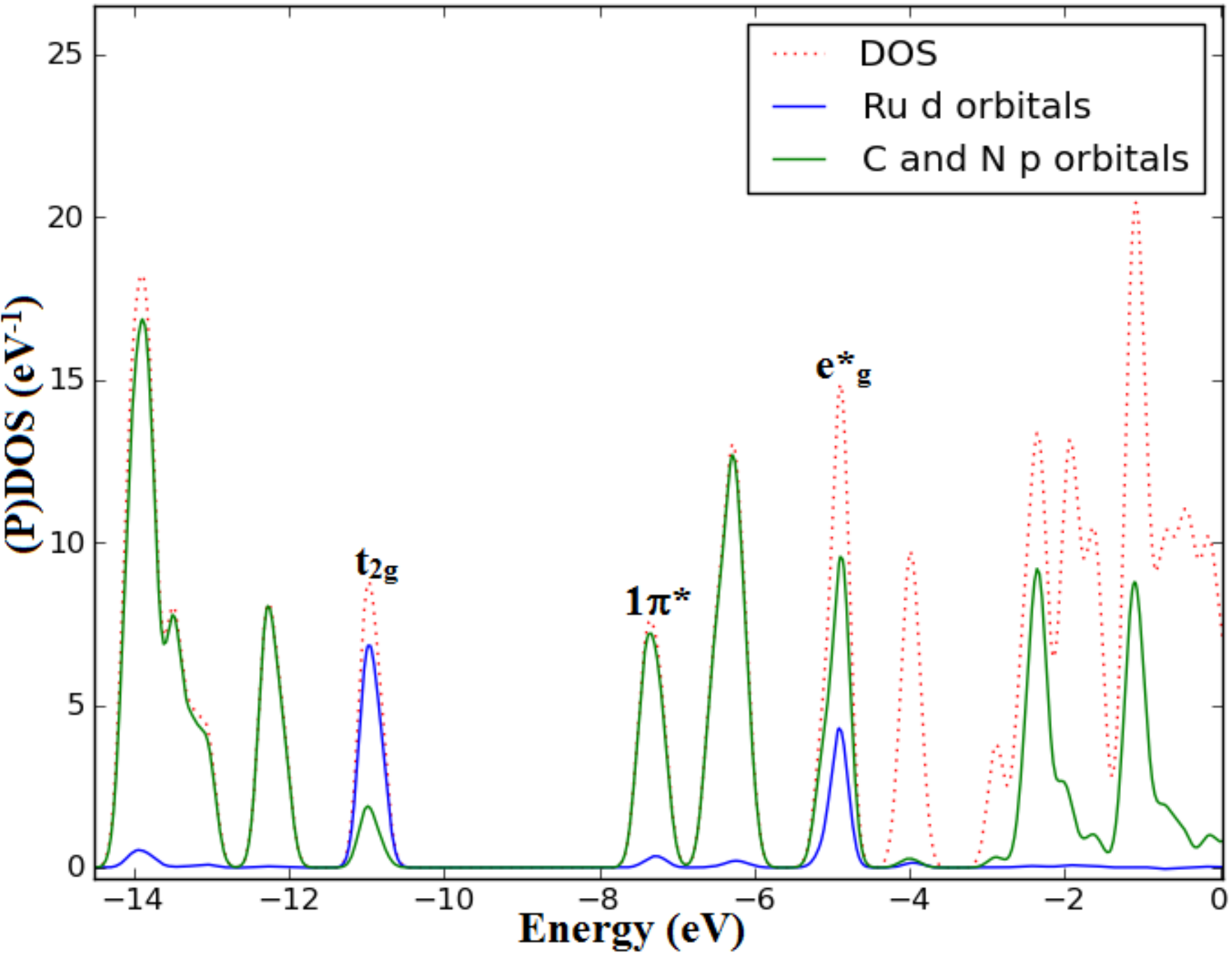} &
\includegraphics[width=0.4\textwidth]{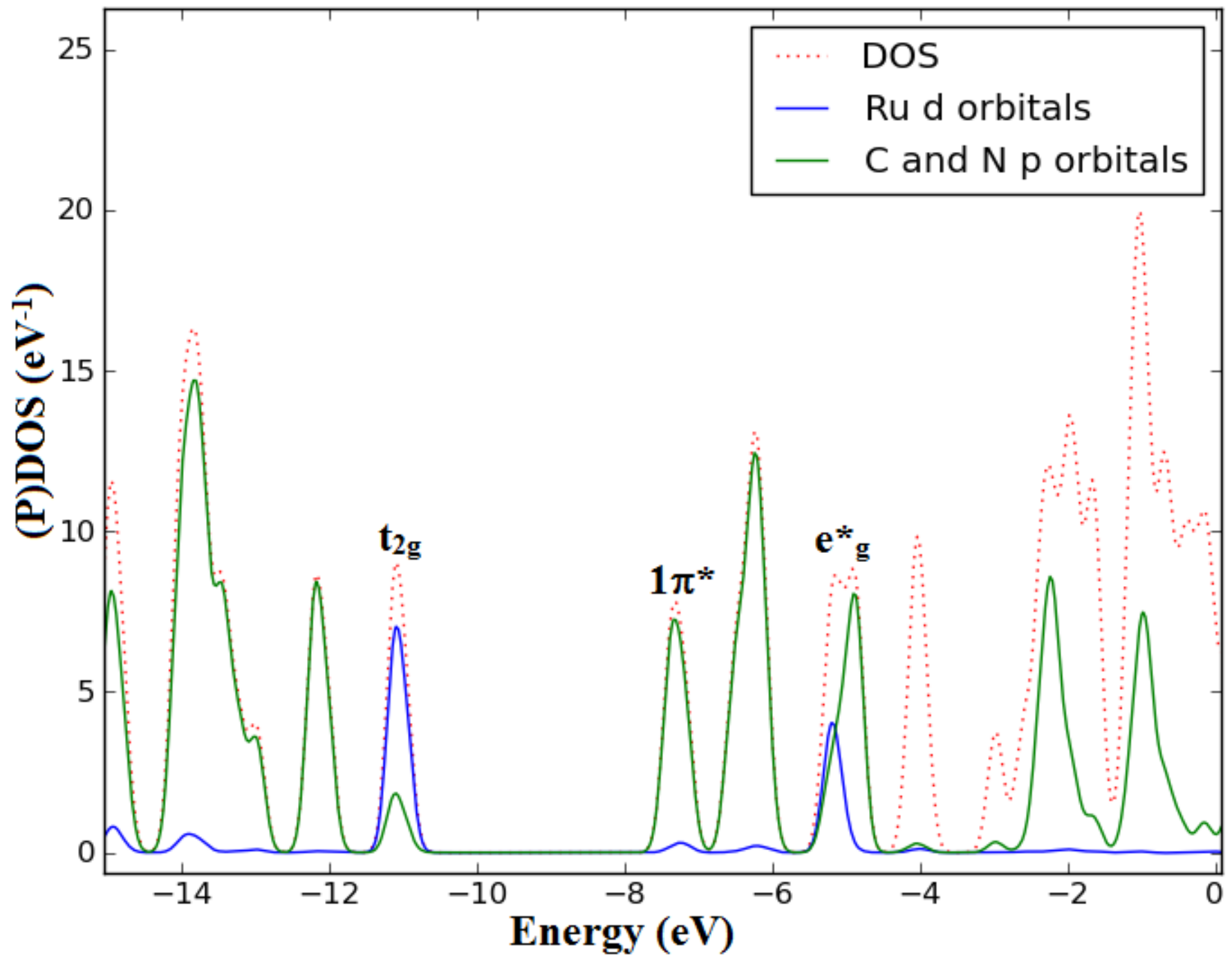}\\
B3LYP/6-31G & B3LYP/6-31G(d) \\
$\epsilon_{\text{HOMO}} = \mbox{-10.84 eV}$ & 
$\epsilon_{\text{HOMO}} = \mbox{-10.97 eV}$ 
\end{tabular}
\end{center}
Total and partial density of states of [Ru(bpy)$_2$(3,3'-dm-bpy)]$^{2+}$ 
partitioned over Ru d orbitals and ligand C and N p orbitals.
% for the 6-31G (left-hand side) and 6-31G(d) (right-hand side) basis sets.

\begin{center}
   {\bf Absorption Spectrum}
\end{center}

\begin{center}
\includegraphics[width=0.8\textwidth]{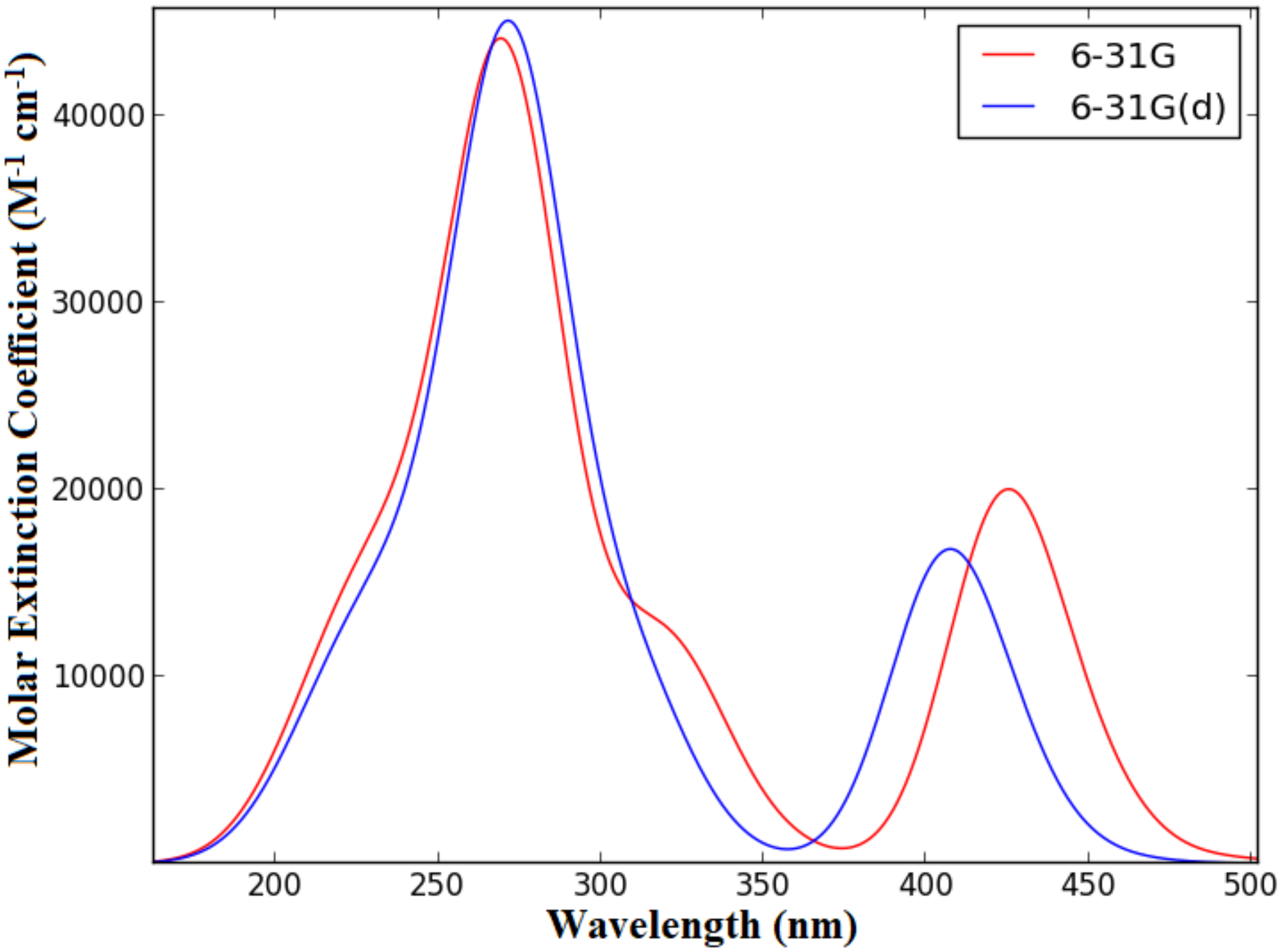}
\end{center}
[Ru(bpy)$_2$(3,3'-dm-bpy)]$^{2+}$ 
TD-B3LYP/6-31G and TD-B3LYP/6-31G(d) spectra.

% ================================================
\newpage
\section{Complex {\bf (9)}: [Ru(bpy)$_2$(4,4'-dm-bpy)]$^{2+}$}
% ================================================

\begin{center}
   {\bf PDOS}
\end{center}

\begin{center}
\begin{tabular}{cc}
\includegraphics[width=0.4\textwidth]{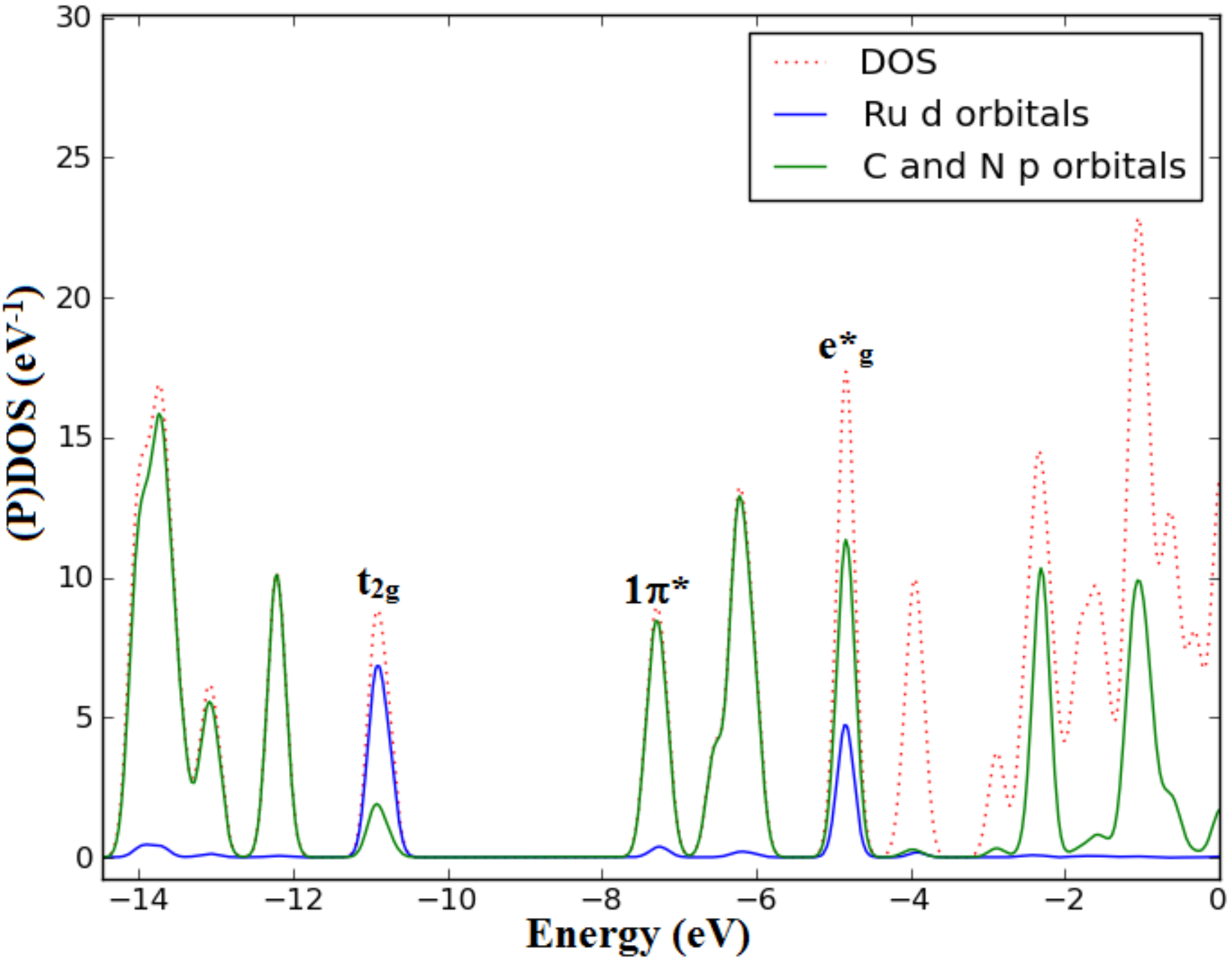} &
\includegraphics[width=0.4\textwidth]{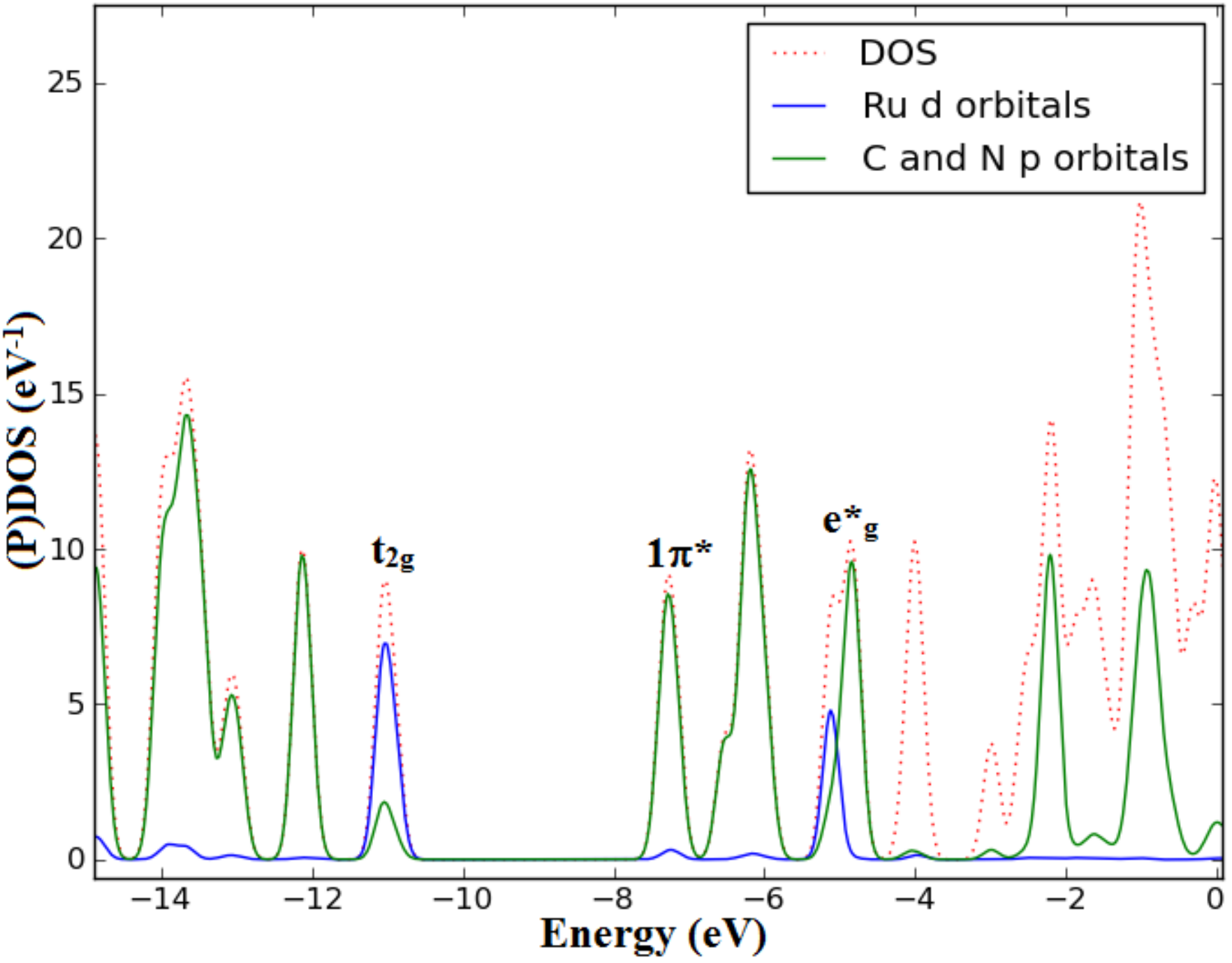} \\
B3LYP/6-31G & B3LYP/6-31G(d) \\
$\epsilon_{\text{HOMO}} = \mbox{-10.78 eV}$ & 
$\epsilon_{\text{HOMO}} = \mbox{-10.91 eV}$ 
\end{tabular}
\end{center}
Total and partial density of states of [Ru(bpy)$_2$(4,4'-dm-bpy)]$^{2+}$ 
partitioned over Ru d orbitals and ligand C and N p orbitals.
% for the 6-31G (left-hand side) and 6-31G(d) (right-hand side) basis sets.

\begin{center}
   {\bf Absorption Spectrum}
\end{center}

\begin{center}
\includegraphics[width=0.8\textwidth]{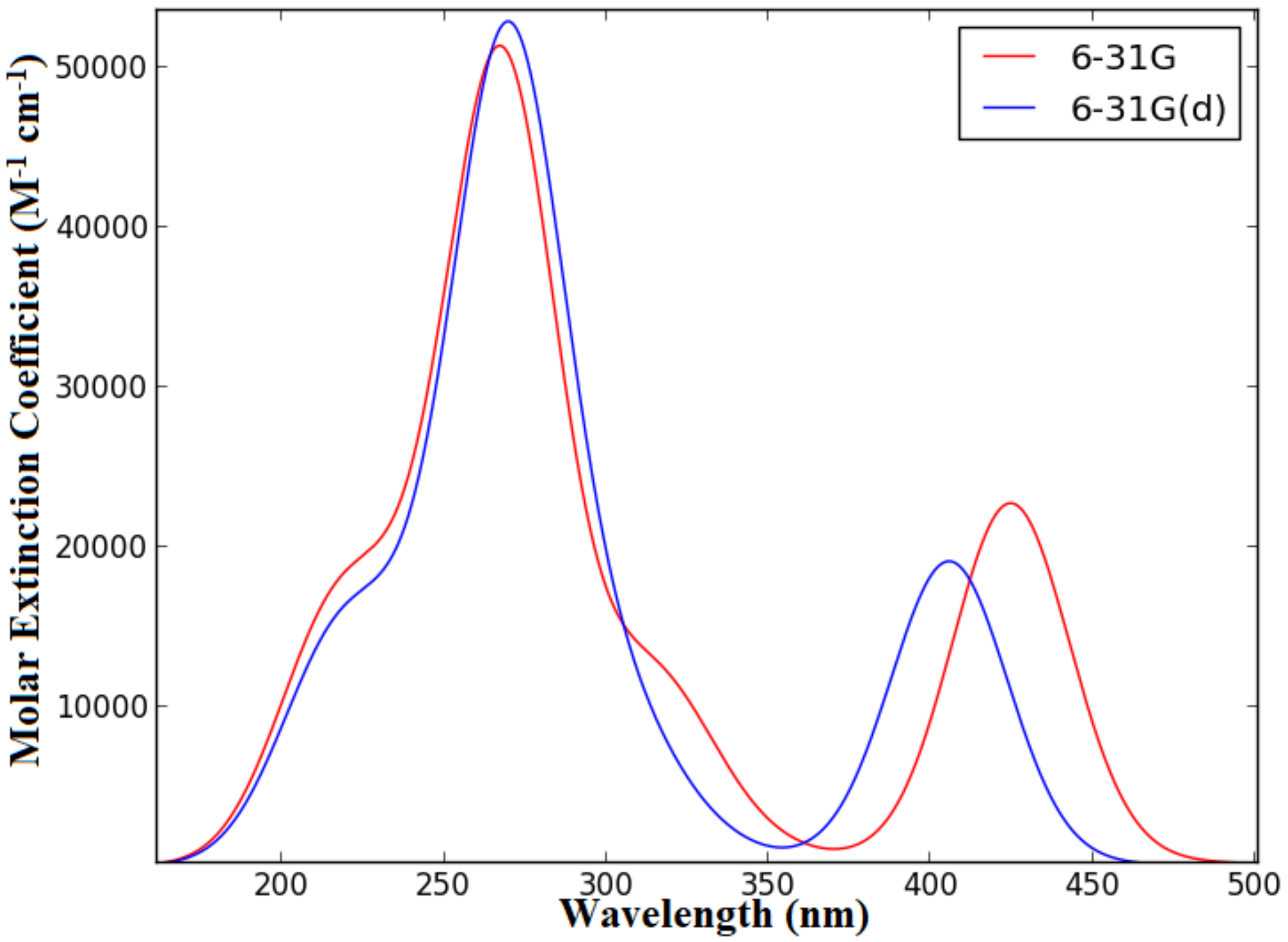}
\end{center}
[Ru(bpy)$_2$(4,4'-dm-bpy)]$^{2+}$ 
TD-B3LYP/6-31G and TD-B3LYP/6-31G(d) spectra.

% % ================================================
% \newpage
% \section{Complex {\bf (10)}: [Ru(bpy)$_2$(4,4'-dCl-bpy)]$^{2+}$}
% % ================================================
% 
% \begin{center}
%    {\bf PDOS}
% \end{center}
% 
% \begin{center}
% \includegraphics[width=0.4\textwidth]{graphics1/framedquestionmark.pdf}
% \includegraphics[width=0.4\textwidth]{graphics1/framedquestionmark.pdf}
% \end{center}
% {\color{magenta} PDOS could not be calculated for complexes containing Cl.}
% 
% \begin{center}
%    {\bf Absorption Spectrum}
% \end{center}
% 
% \begin{center}
% \includegraphics[width=0.4\textwidth]{graphics1/framedquestionmark.pdf}
% \end{center}
% {\color{red} Do we have this?}

% ================================================
\newpage
\section{Complex {\bf (11)}: [Ru(bpy)$_2$(4,4'-dn-bpy)]$^{2+}$}
% ================================================

\begin{center}
   {\bf PDOS}
\end{center}

\begin{center}
\begin{tabular}{cc}
\includegraphics[width=0.4\textwidth]{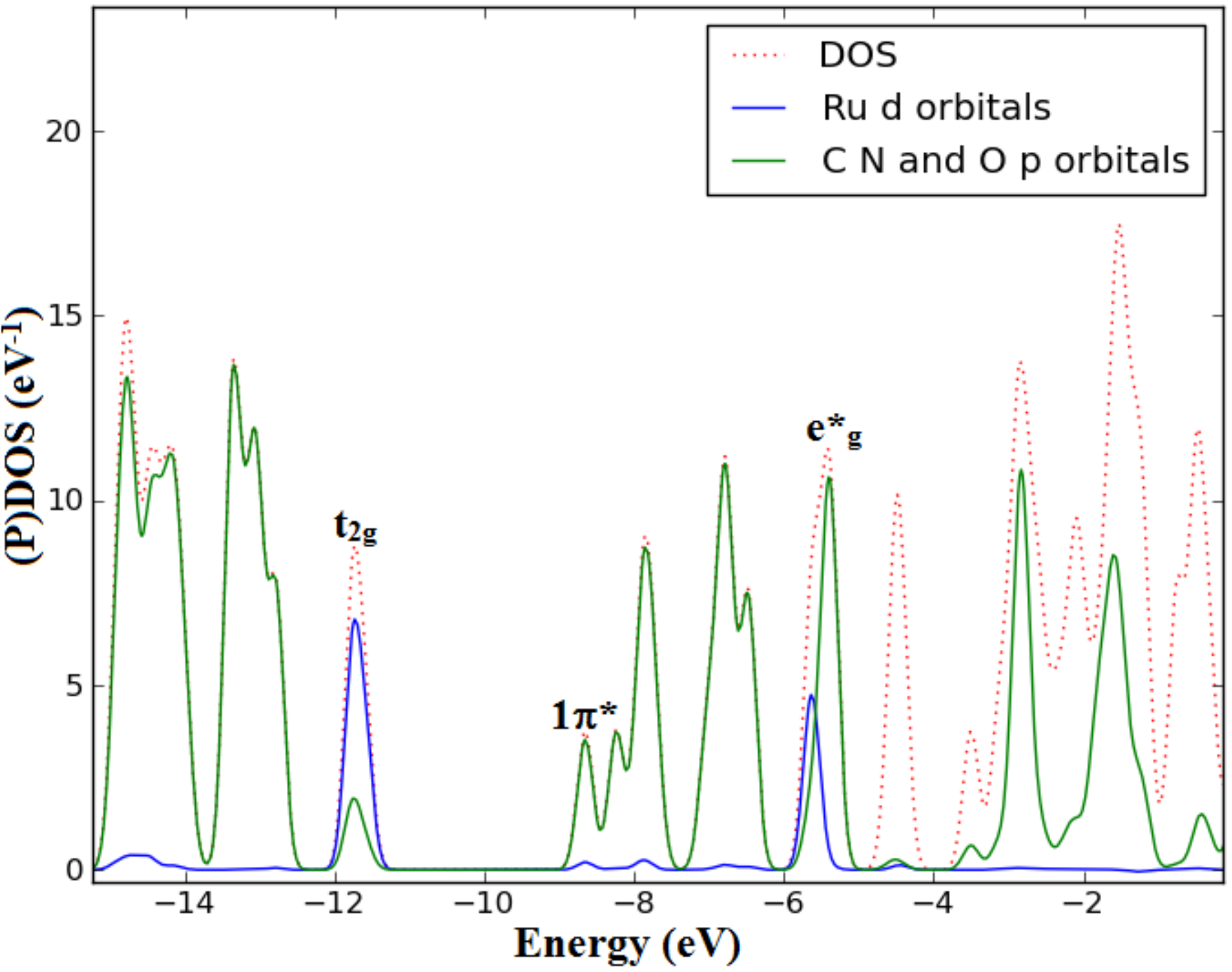} &
\includegraphics[width=0.4\textwidth]{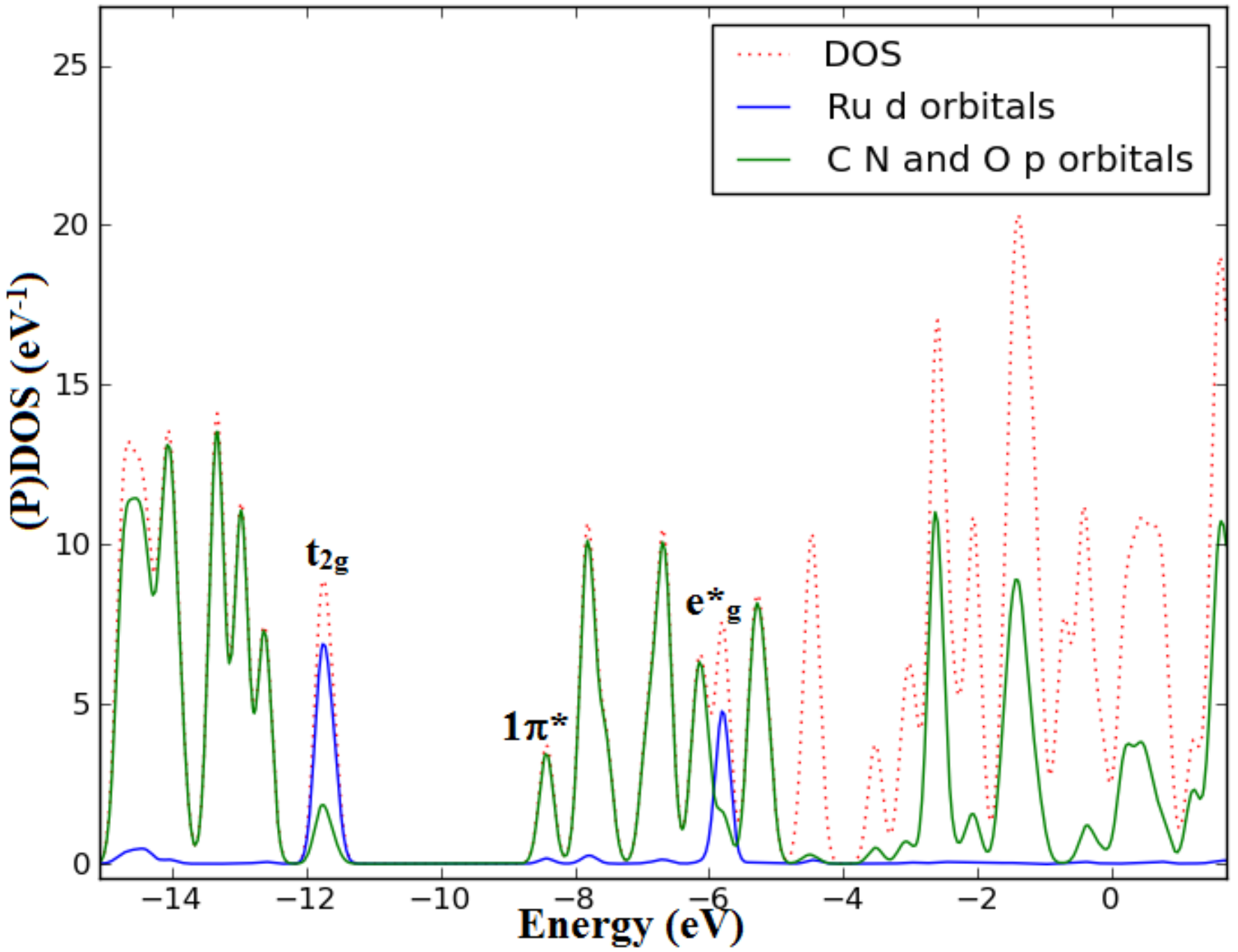} \\
B3LYP/6-31G & B3LYP/6-31G(d) \\
$\epsilon_{\text{HOMO}} = \mbox{-10.59 eV}$ & 
$\epsilon_{\text{HOMO}} = \mbox{-10.62 eV}$ 
\end{tabular}
\end{center}
Total and partial density of states of [Ru(bpy)$_2$(4,4'-dn-bpy)]$^{2+}$ 
partitioned over Ru d orbitals and ligand C, O and N p orbitals.
% for the 6-31G (left-hand side) and 6-31G(d) (right-hand side) basis sets.

\begin{center}
   {\bf Absorption Spectrum}
\end{center}

\begin{center}
\includegraphics[width=0.8\textwidth]{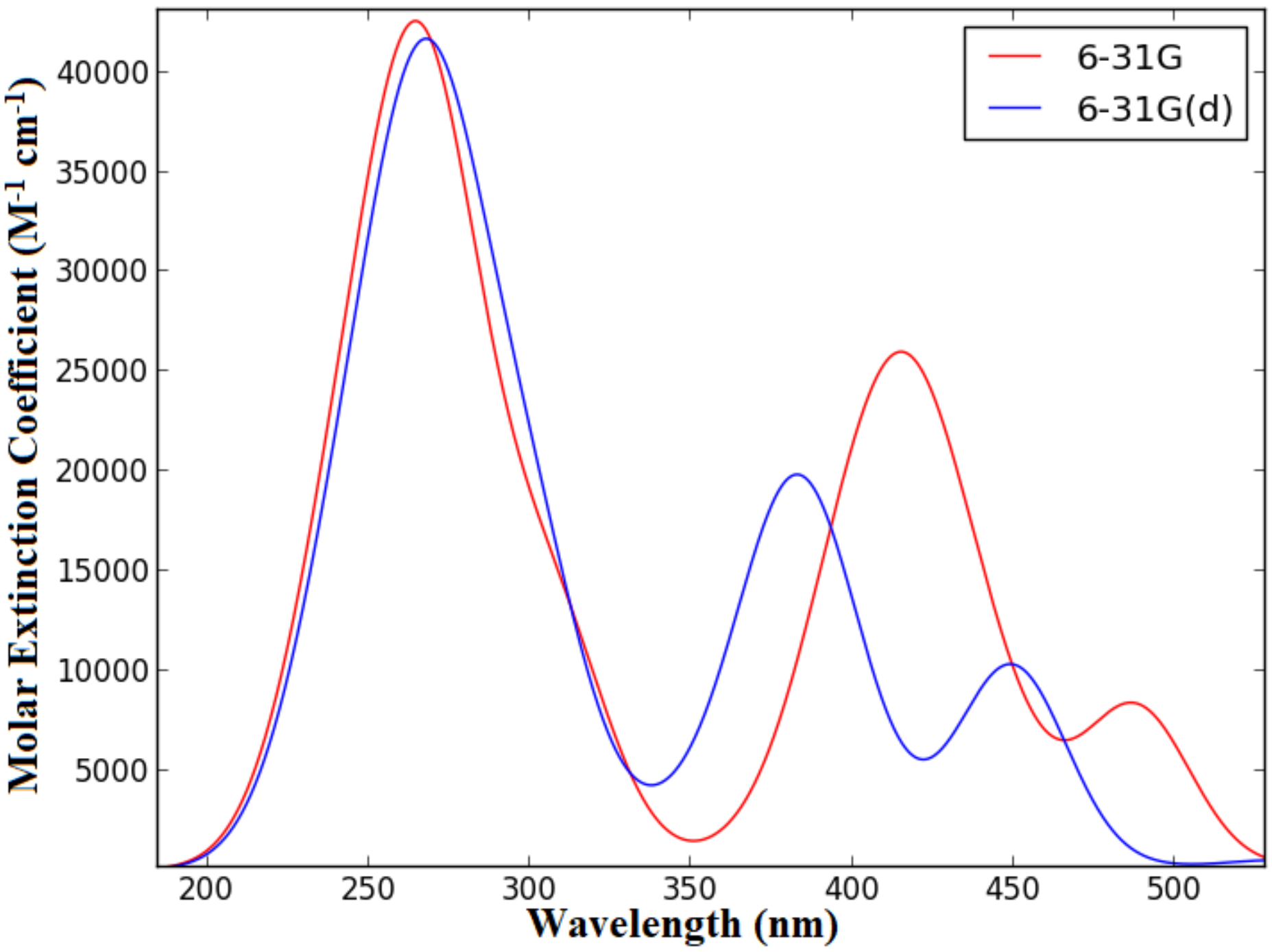}
\end{center}
[Ru(bpy)$_2$(4,4'-dn-bpy)]$^{2+}$
TD-B3LYP/6-31G and TD-B3LYP/6-31G(d) spectra.

% ================================================
\newpage
\section{Complex {\bf (12)}: [Ru(bpy)$_2$(4,4'-dph-bpy)]$^{2+}$}
% ================================================

\begin{center}
   {\bf PDOS}
\end{center}

\begin{center}
\begin{tabular}{cc}
\includegraphics[width=0.4\textwidth]{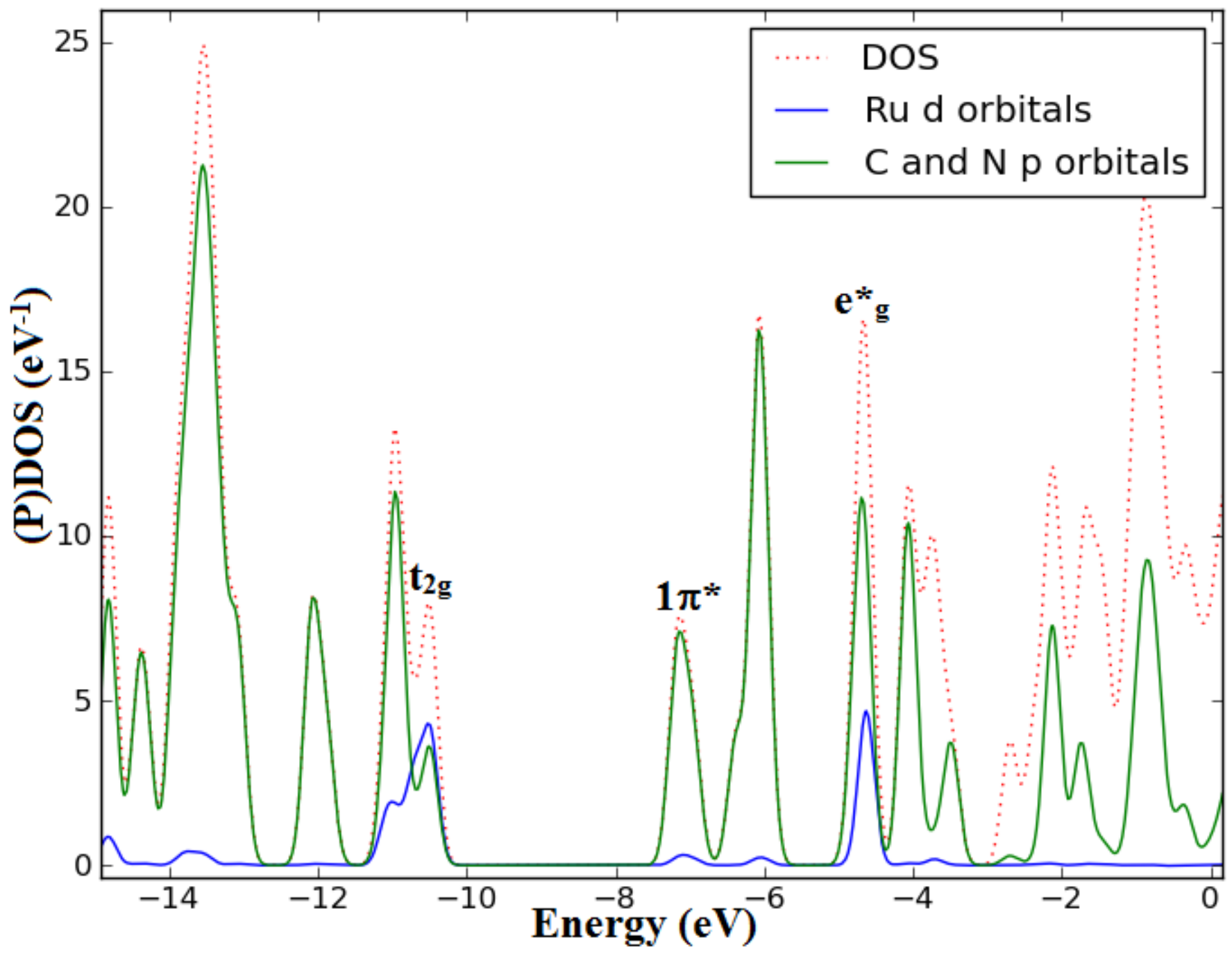} &
\includegraphics[width=0.4\textwidth]{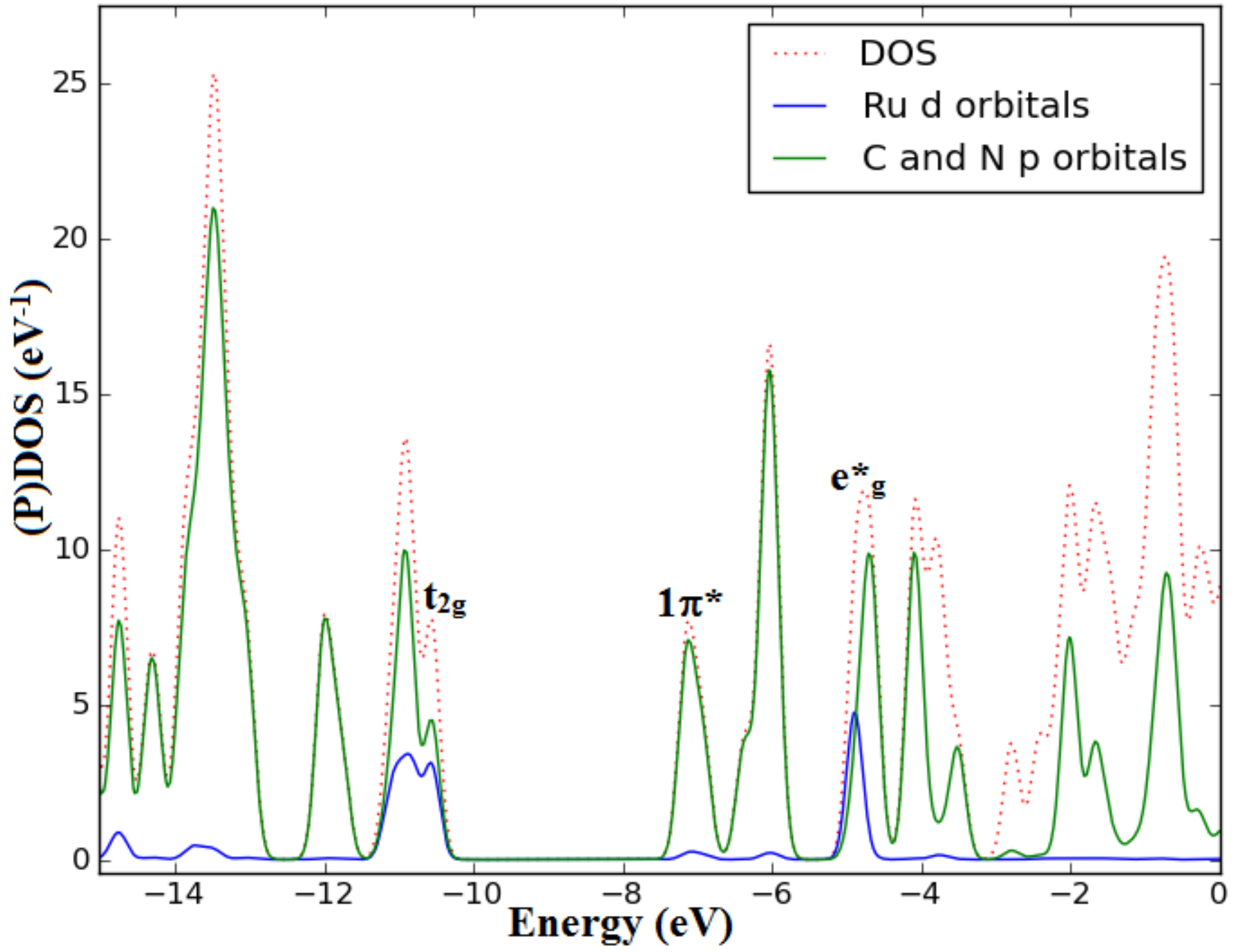} \\
B3LYP/6-31G & B3LYP/6-31G(d) \\
$\epsilon_{\text{HOMO}} = \mbox{-10.48 eV}$ & 
$\epsilon_{\text{HOMO}} = \mbox{-10.56 eV}$ 
\end{tabular}
\end{center}
Total and partial density of states of [Ru(bpy)$_2$(4,4'-dph-bpy)]$^{2+}$ 
partitioned over Ru d orbitals and ligand C and N p orbitals.
% for the 6-31G (left-hand side) and 6-31G(d) (right-hand side) basis sets.

\begin{center}
   {\bf Absorption Spectrum}
\end{center}

\begin{center}
\includegraphics[width=0.8\textwidth]{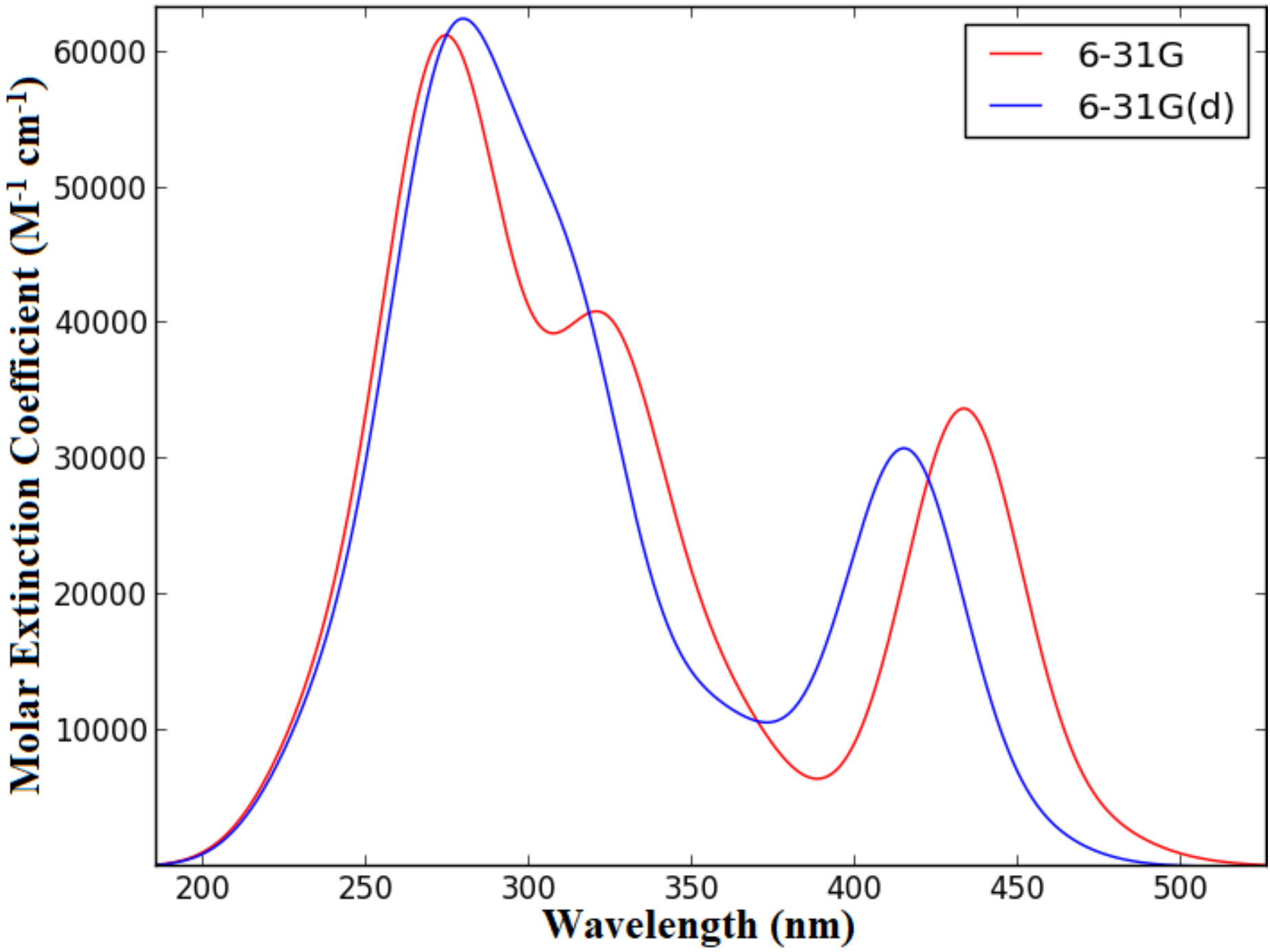}
\end{center}
[Ru(bpy)$_2$(4,4'-dph-bpy)]$^{2+}$
TD-B3LYP/6-31G and TD-B3LYP/6-31G(d) spectra.

% ================================================
\newpage
\section{Complex {\bf (13)}: [Ru(bpy)$_2$(4,4'-DTB-bpy)]$^{2+}$}
% ================================================

\begin{center}
   {\bf PDOS}
\end{center}

\begin{center}
\begin{tabular}{cc}
\includegraphics[width=0.4\textwidth]{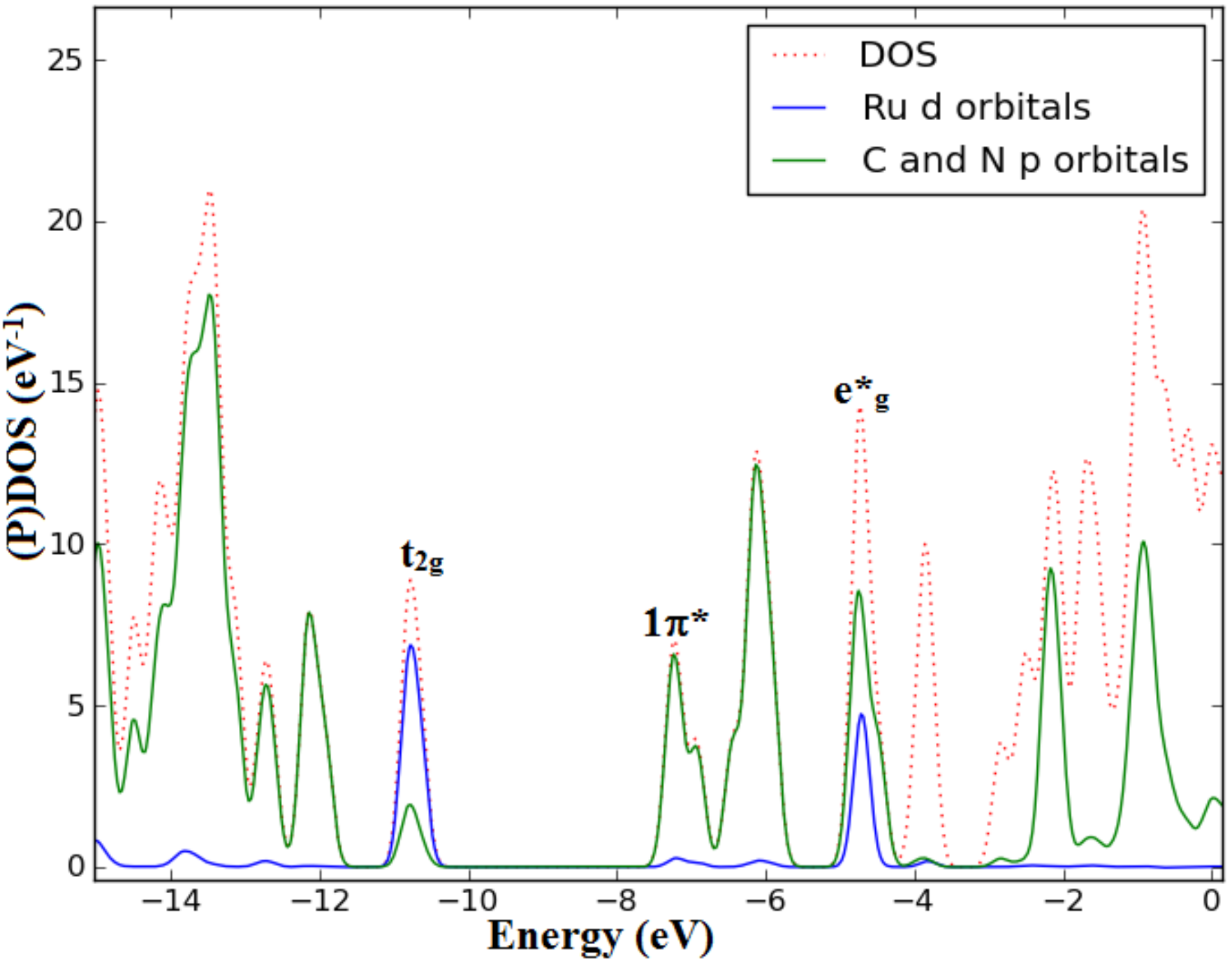} &
\includegraphics[width=0.4\textwidth]{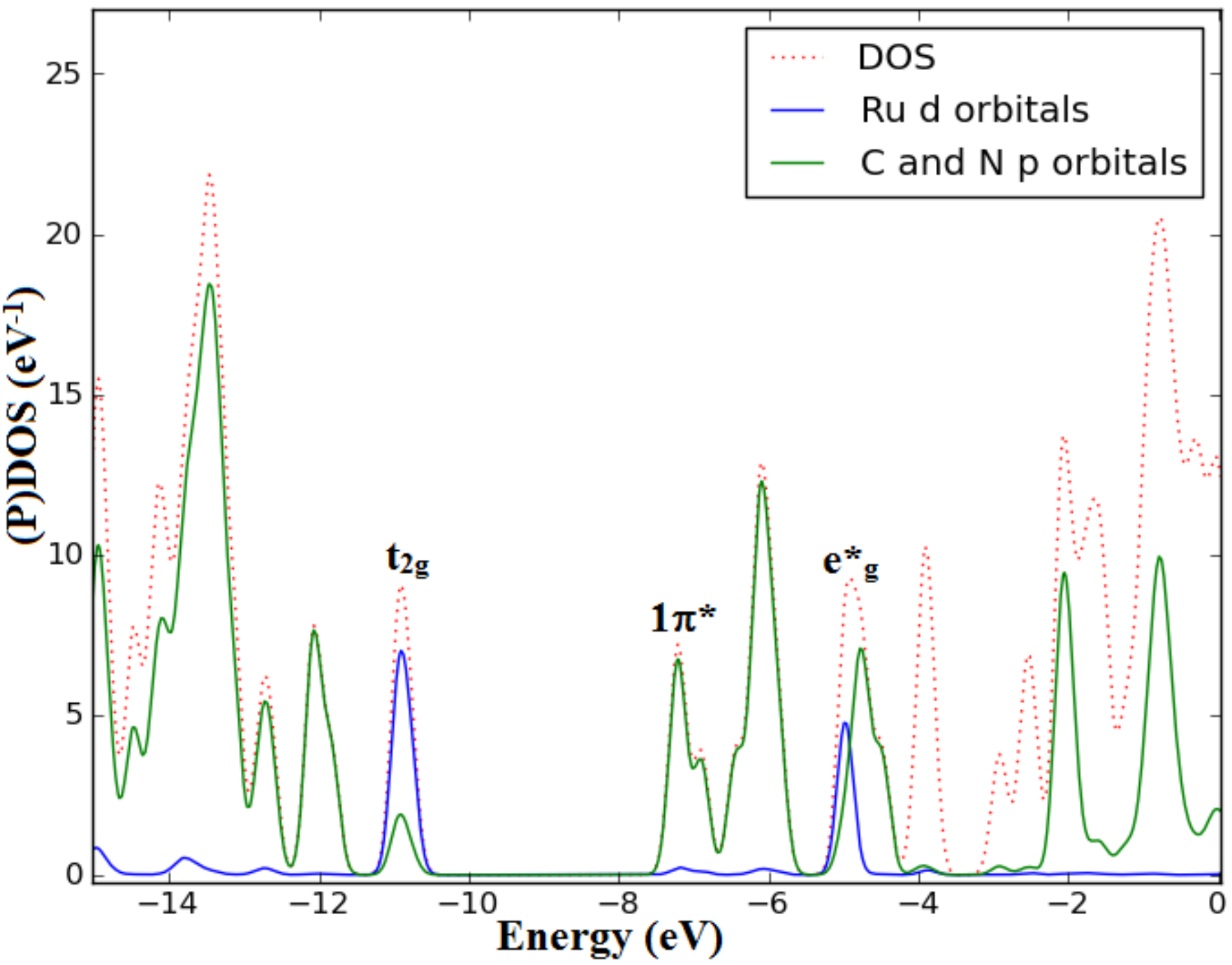} \\
B3LYP/6-31G & B3LYP/6-31G(d) \\
$\epsilon_{\text{HOMO}} = \mbox{-10.65 eV}$ & 
$\epsilon_{\text{HOMO}} = \mbox{-10.78 eV}$ 
\end{tabular}
\end{center}
Total and partial density of states of [Ru(bpy)$_2$(4,4'-DTB-bpy)]$^{2+}$ 
partitioned over Ru d orbitals and ligand C and N p orbitals. 
% for the 6-31G (left-hand side) and 6-31G(d) (right-hand side) basis sets.

\begin{center}
   {\bf Absorption Spectrum}
\end{center}

\begin{center}
\includegraphics[width=0.8\textwidth]{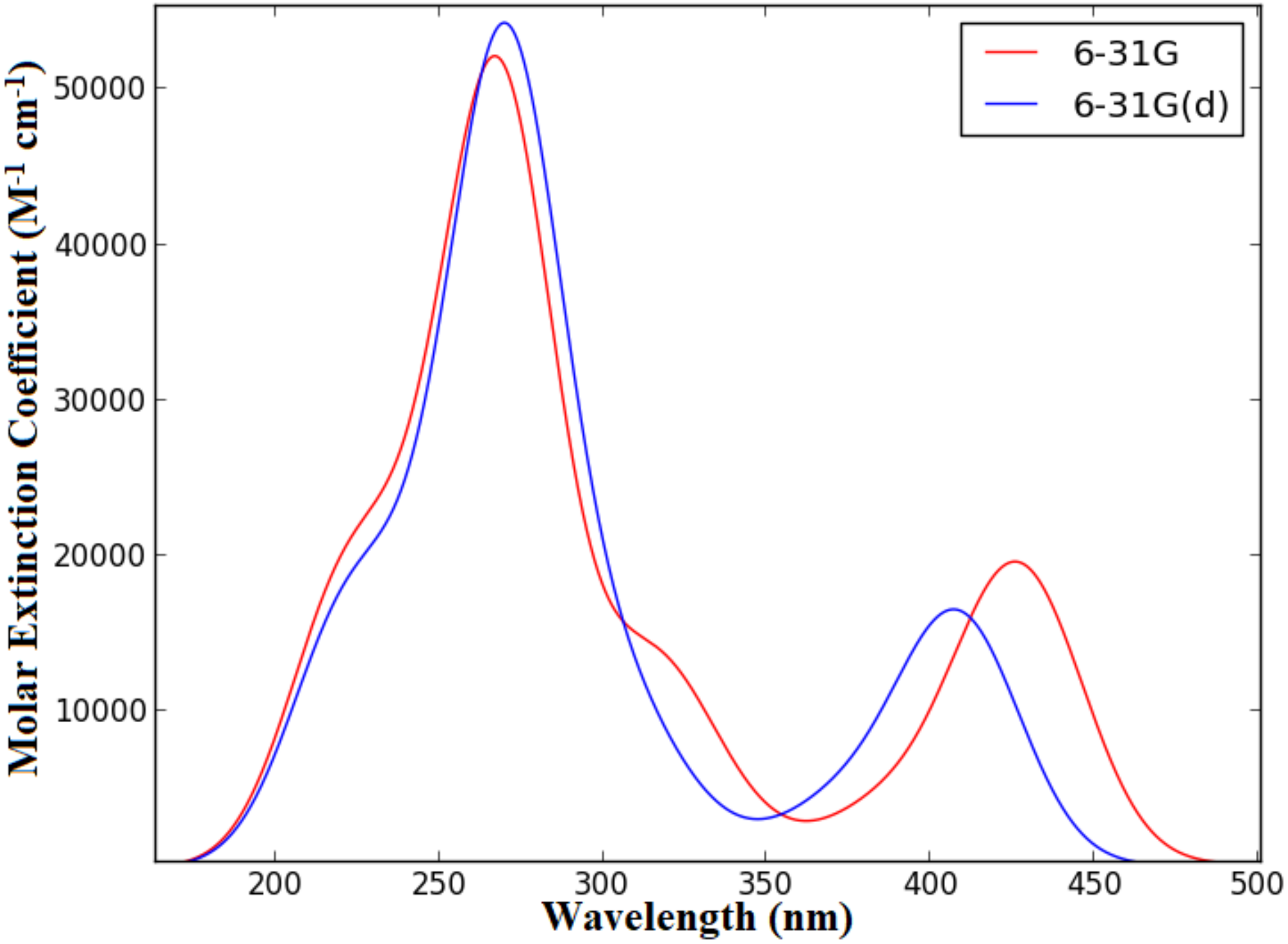}
\end{center}
[Ru(bpy)$_2$(4,4'-DTB-bpy)]$^{2+}$ 
TD-B3LYP/6-31G and TD-B3LYP/6-31G(d) spectra.

% ================================================
\newpage
\section{Complex {\bf (14)}: {\em cis}-[Ru(bpy)$_2$(m-4,4'-bpy)$_2$]$^{4+}$}
% ================================================

\begin{center}
   {\bf PDOS}
\end{center}

\begin{center}
\begin{tabular}{cc}
\includegraphics[width=0.4\textwidth]{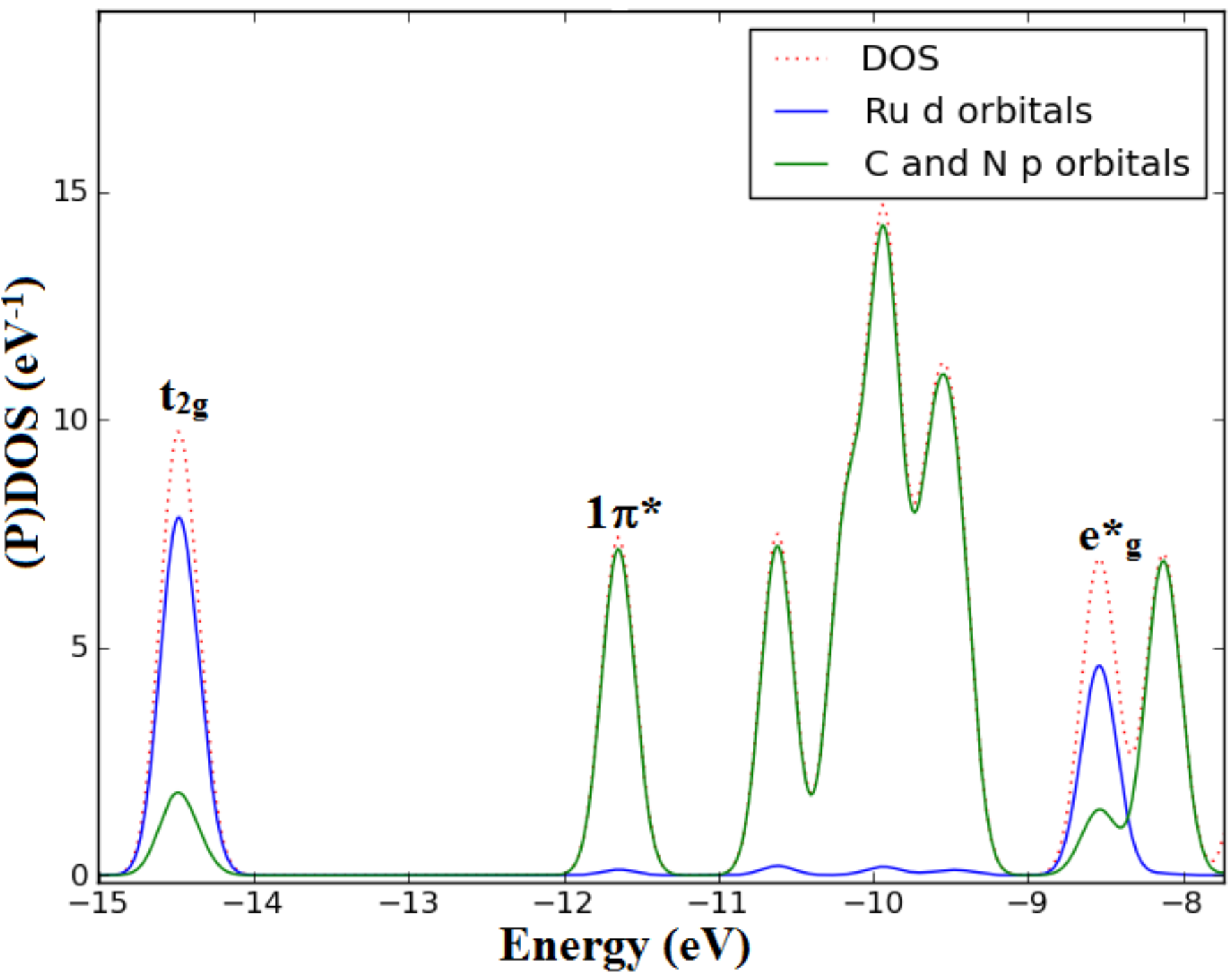} &
\includegraphics[width=0.4\textwidth]{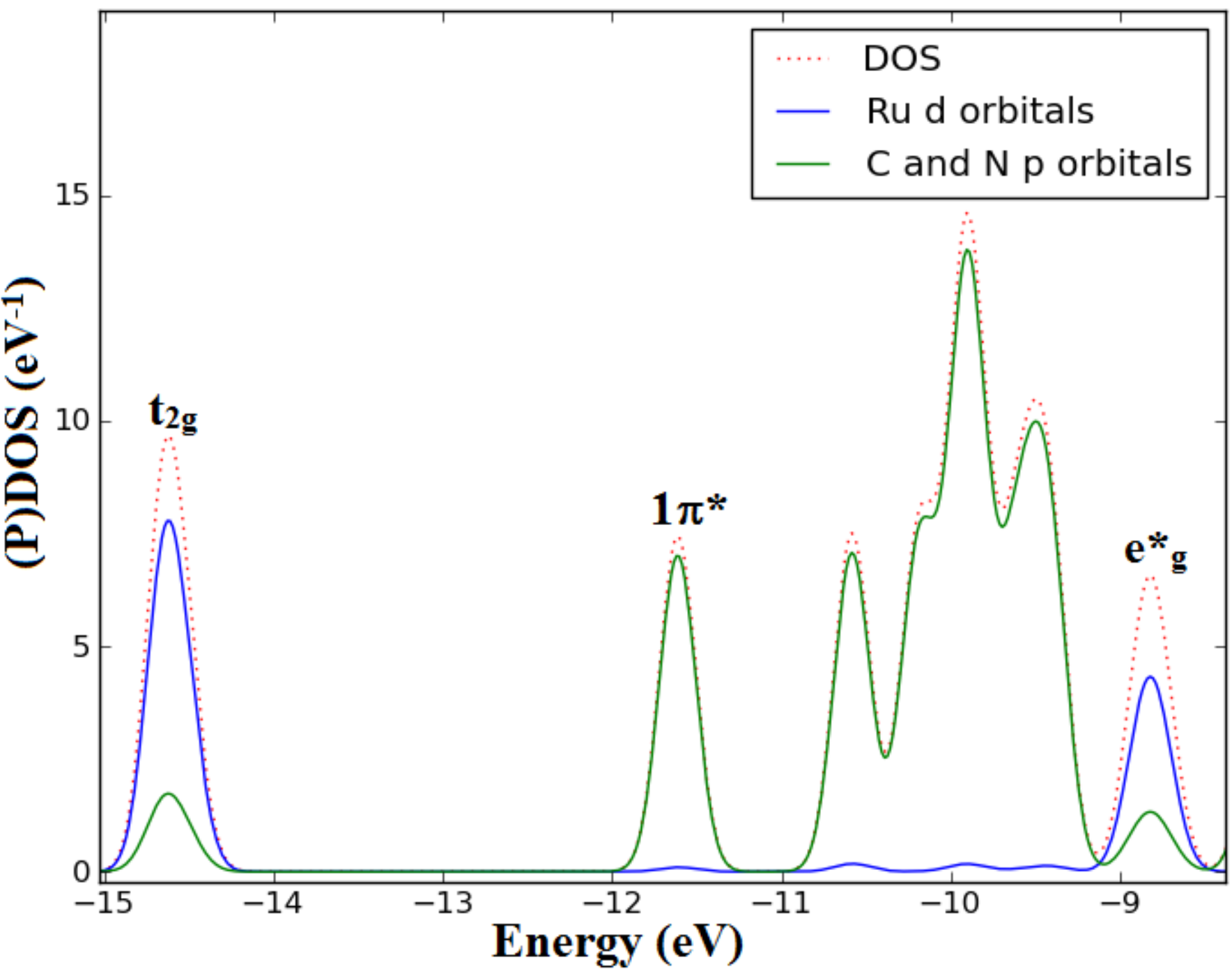} \\
B3LYP/6-31G & B3LYP/6-31G(d) \\
$\epsilon_{\text{HOMO}} = \mbox{-14.40 eV}$ & 
$\epsilon_{\text{HOMO}} = \mbox{-14.52 eV}$ 
\end{tabular}
\end{center}
Total and partial density of states of {\em cis}-[Ru(bpy)$_2$(m-4,4'-bpy)$_2$)]$^{4+}$  
partitioned over Ru d orbitals and ligand C and N p orbitals.
% for the 6-31G (left-hand side) and 6-31G(d) (right-hand side) basis sets.

\begin{center}
   {\bf Absorption Spectrum}
\end{center}

\begin{center}
\includegraphics[width=0.8\textwidth]{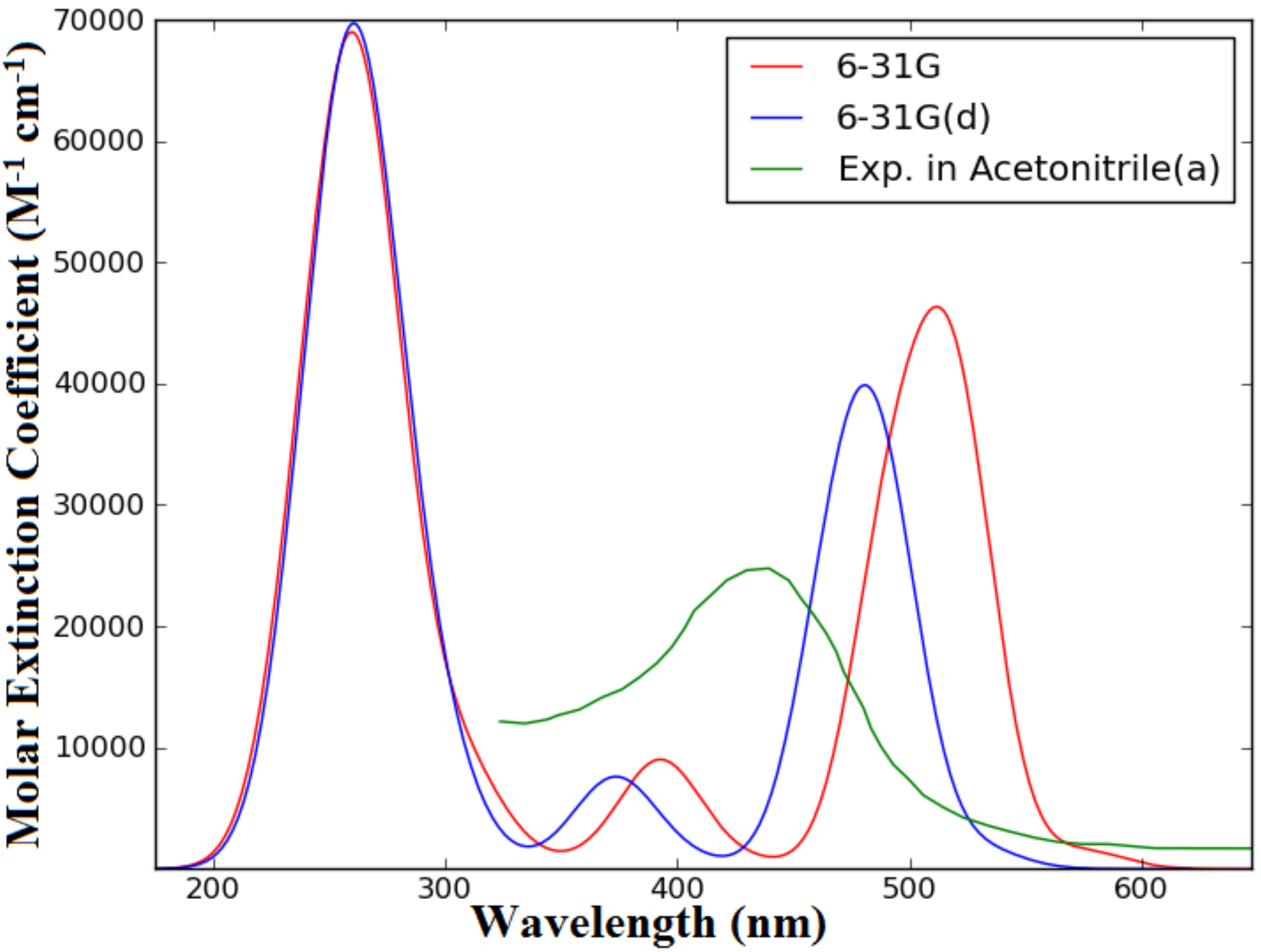}
\end{center}
{\em Cis}-[Ru(bpy)$_2$(m-4,4'-bpy)$_2$)]$^{4+}$ 
TD-B3LYP/6-31G, TD-B3LYP/6-31G(d), and experimental spectra.
Experimental curve at 25$^\circ$C in acetonitrile\cite{SAF+78}.

% ================================================
\newpage
\section{Complex {\bf (15)}: [Ru(bpy)$_2$(bpz)]$^{2+}$}
% ================================================

\begin{center}
   {\bf PDOS}
\end{center}

\begin{center}
\begin{tabular}{cc}
\includegraphics[width=0.4\textwidth]{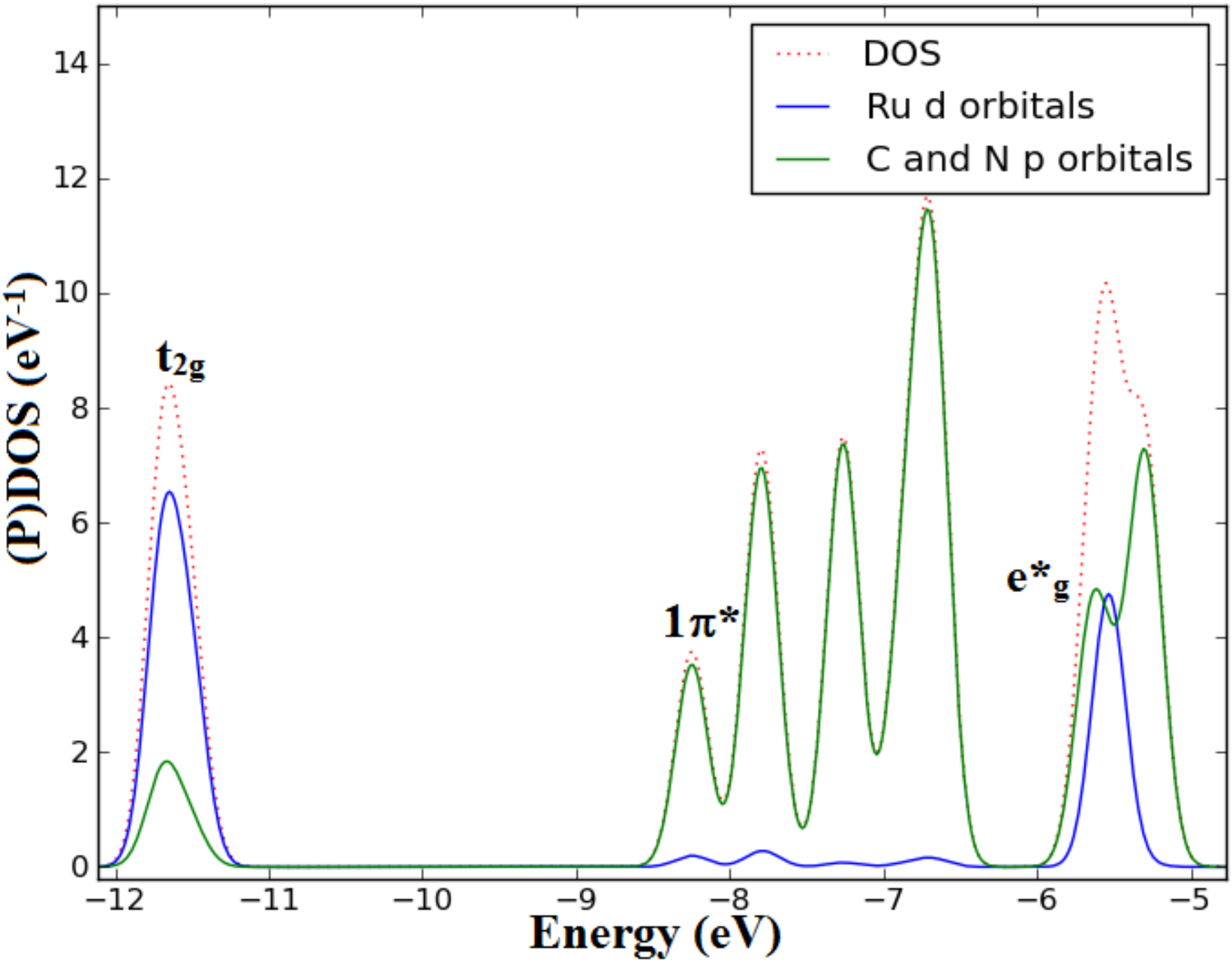} &
\includegraphics[width=0.4\textwidth]{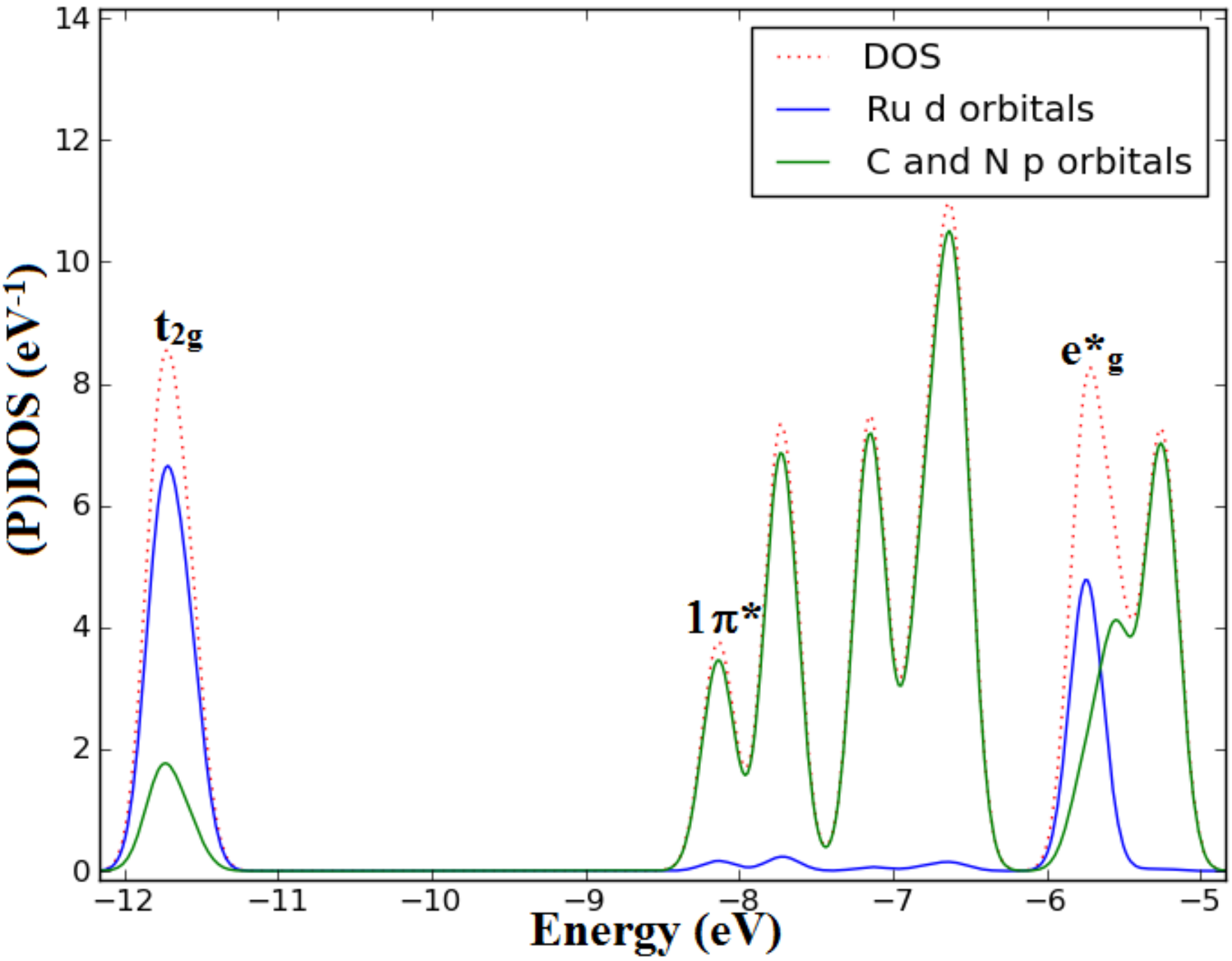} \\
B3LYP/6-31G & B3LYP/6-31G(d) \\
$\epsilon_{\text{HOMO}} = \mbox{-11.51 eV}$ & 
$\epsilon_{\text{HOMO}} = \mbox{-11.58 eV}$ 
\end{tabular}
\end{center}
Total and partial density of states of [Ru(bpy)$_2$(bpz)]$^{2+}$ 
partitioned over Ru d orbitals and ligand C and N p orbitals. 
% for the 6-31G (left-hand side) and 6-31G(d) (right-hand side) basis sets.

\begin{center}
   {\bf Absorption Spectrum}
\end{center}

\begin{center}
\includegraphics[width=0.8\textwidth]{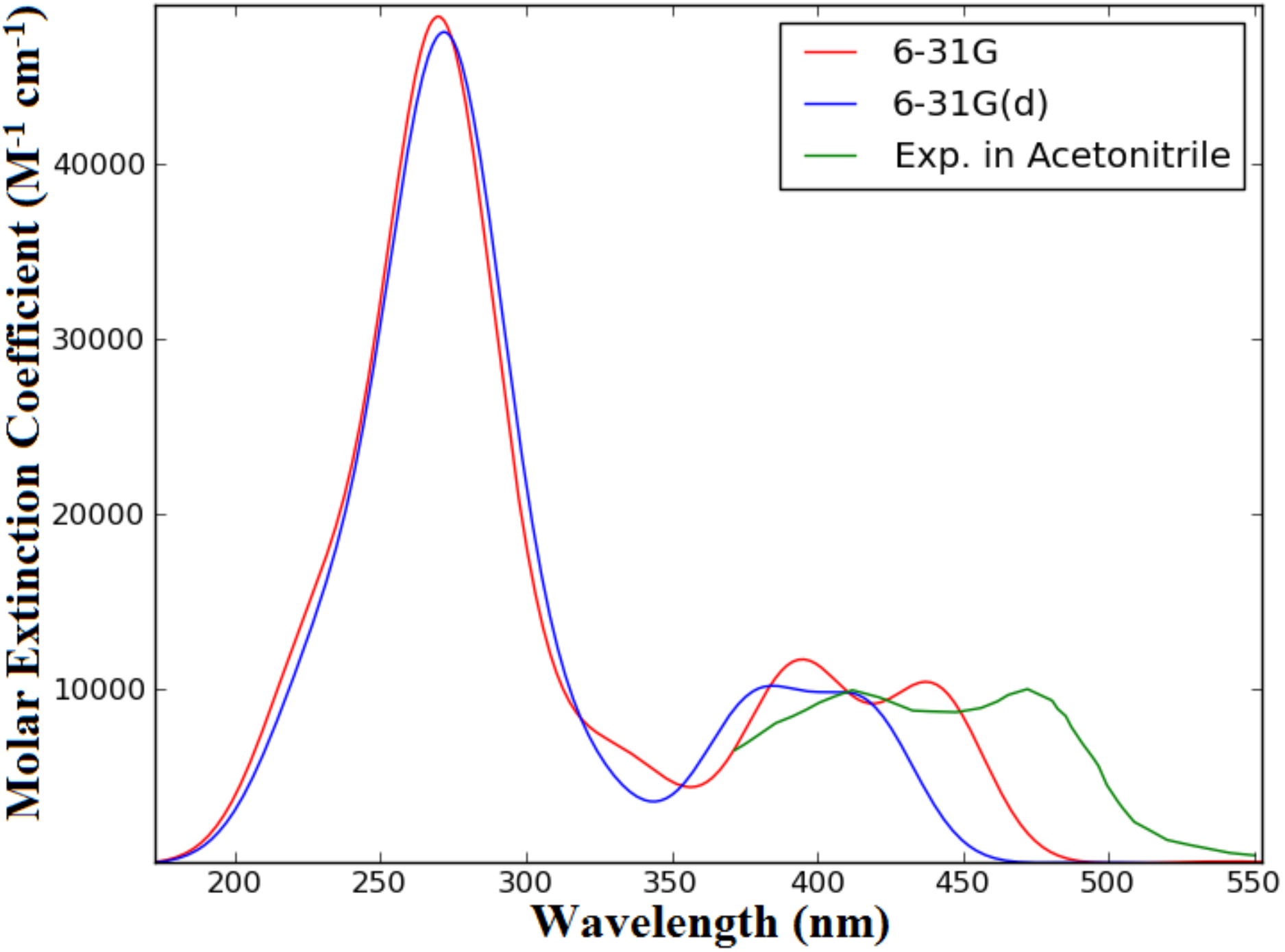}
\end{center}
[Ru(bpy)$_2$(bpz)]$^{2+}$
TD-B3LYP/6-31G, TD-B3LYP/6-31G(d), and experimental spectra.
Experimental curve from \cite{RAMC83}.

% ================================================
\newpage
\section{Complex {\bf (16)}: [Ru(bpy)$_2$(phen)]$^{2+}$}
% ================================================

\begin{center}
   {\bf PDOS}
\end{center}

\begin{center}
\begin{tabular}{cc}
\includegraphics[width=0.4\textwidth]{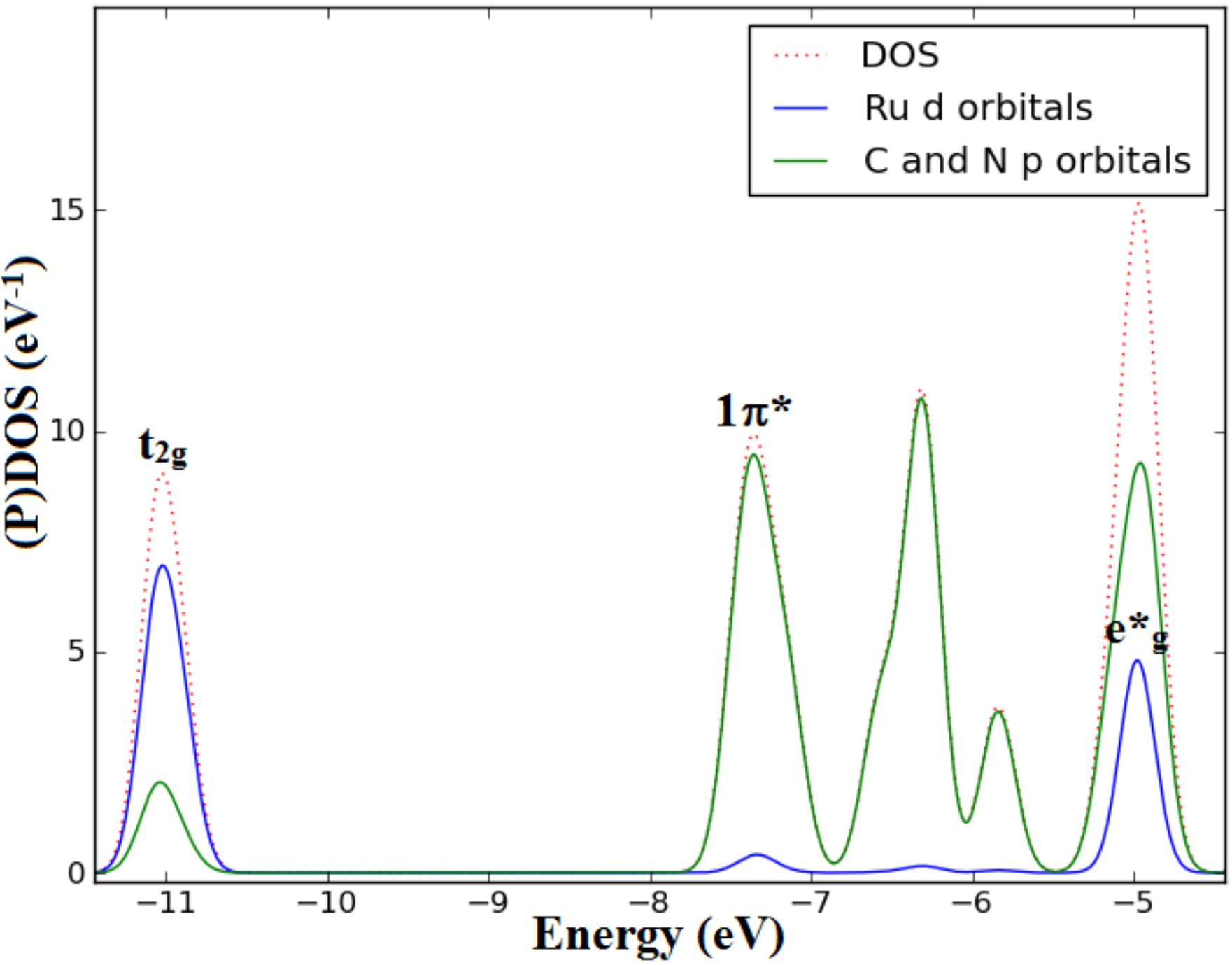} &
\includegraphics[width=0.4\textwidth]{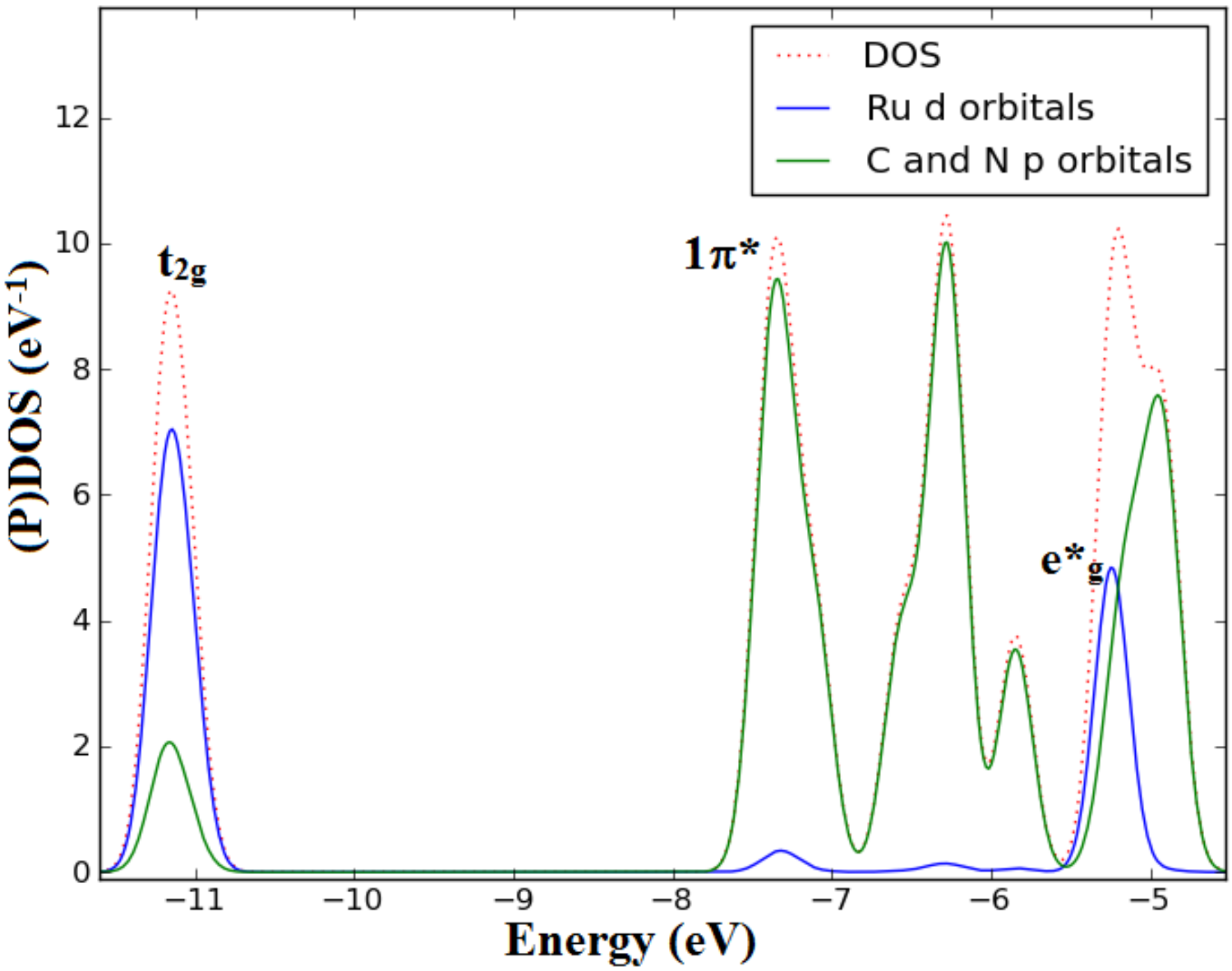} \\
B3LYP/6-31G & B3LYP/6-31G(d) \\
$\epsilon_{\text{HOMO}} = \mbox{-10.91 eV}$ & 
$\epsilon_{\text{HOMO}} = \mbox{-11.05 eV}$ 
\end{tabular}
\end{center}
Total and partial density of states of [Ru(bpy)$_2$(phen)]$^{2+}$ 
partitioned over Ru d orbitals and ligand C and N p orbitals.
% for the 6-31G (left-hand side) and 6-31G(d) (right-hand side) basis sets.

\begin{center}
   {\bf Absorption Spectrum}
\end{center}

\begin{center}
\includegraphics[width=0.8\textwidth]{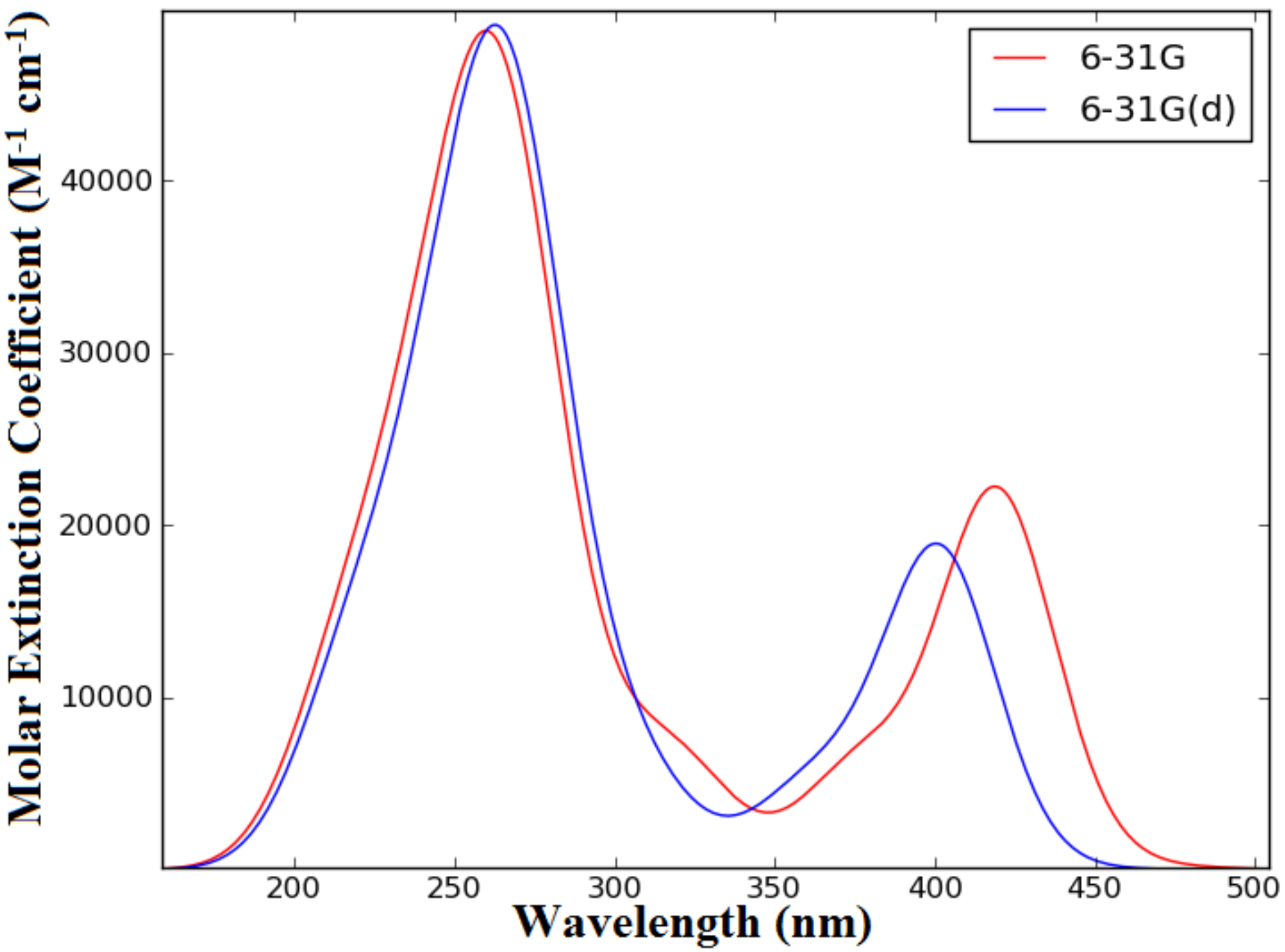}
\end{center}
[Ru(bpy)$_2$(phen)]$^{2+}$
TD-B3LYP/6-31G and TD-B3LYP/6-31G(d) spectra.

% ================================================
\newpage
\section{Complex {\bf (17)}: [Ru(bpy)$_2$(4,7-dm-phen)]$^{2+}$}
% ================================================

\begin{center}
   {\bf PDOS}
\end{center}

\begin{center}
\begin{tabular}{cc}
\includegraphics[width=0.4\textwidth]{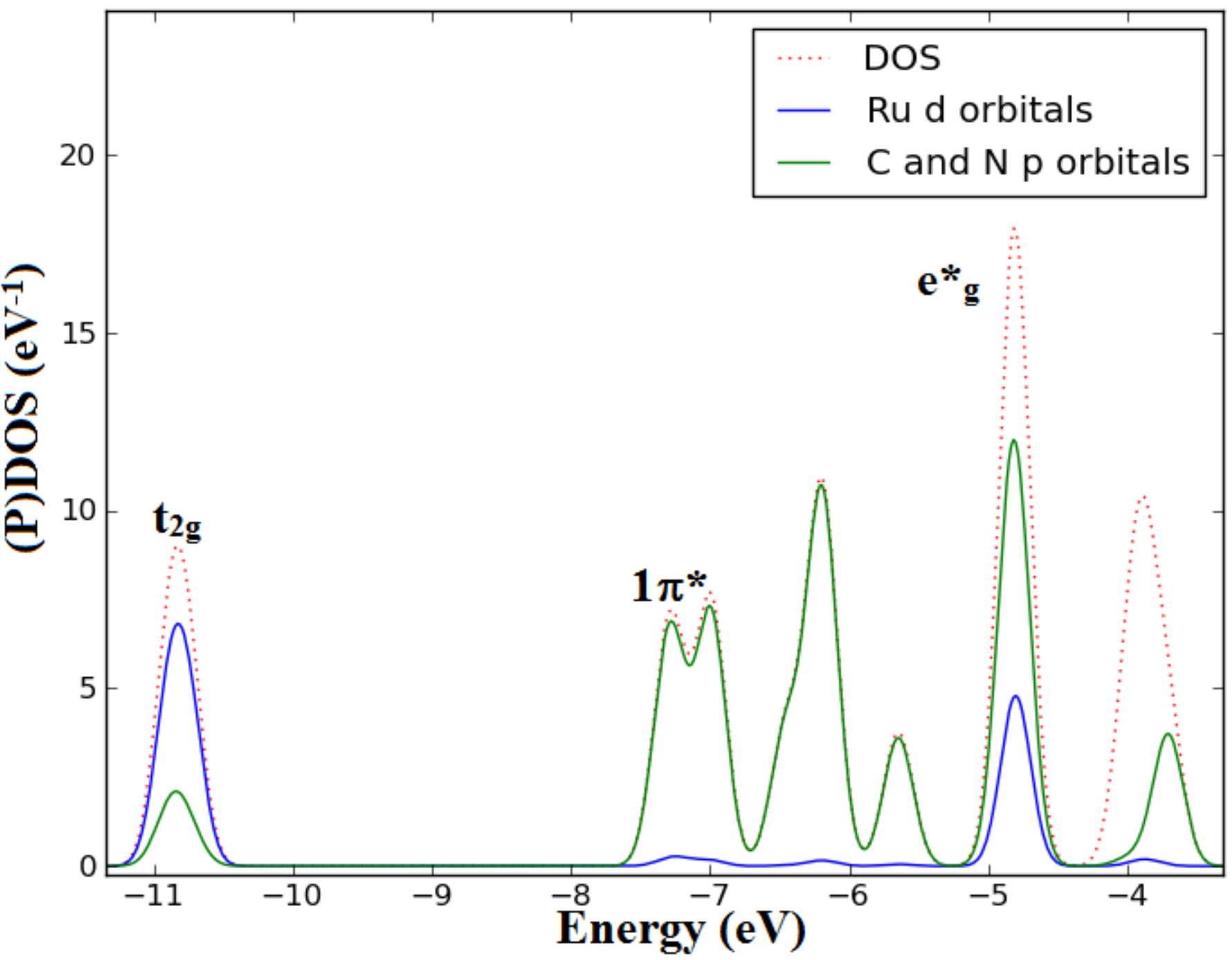} &
\includegraphics[width=0.4\textwidth]{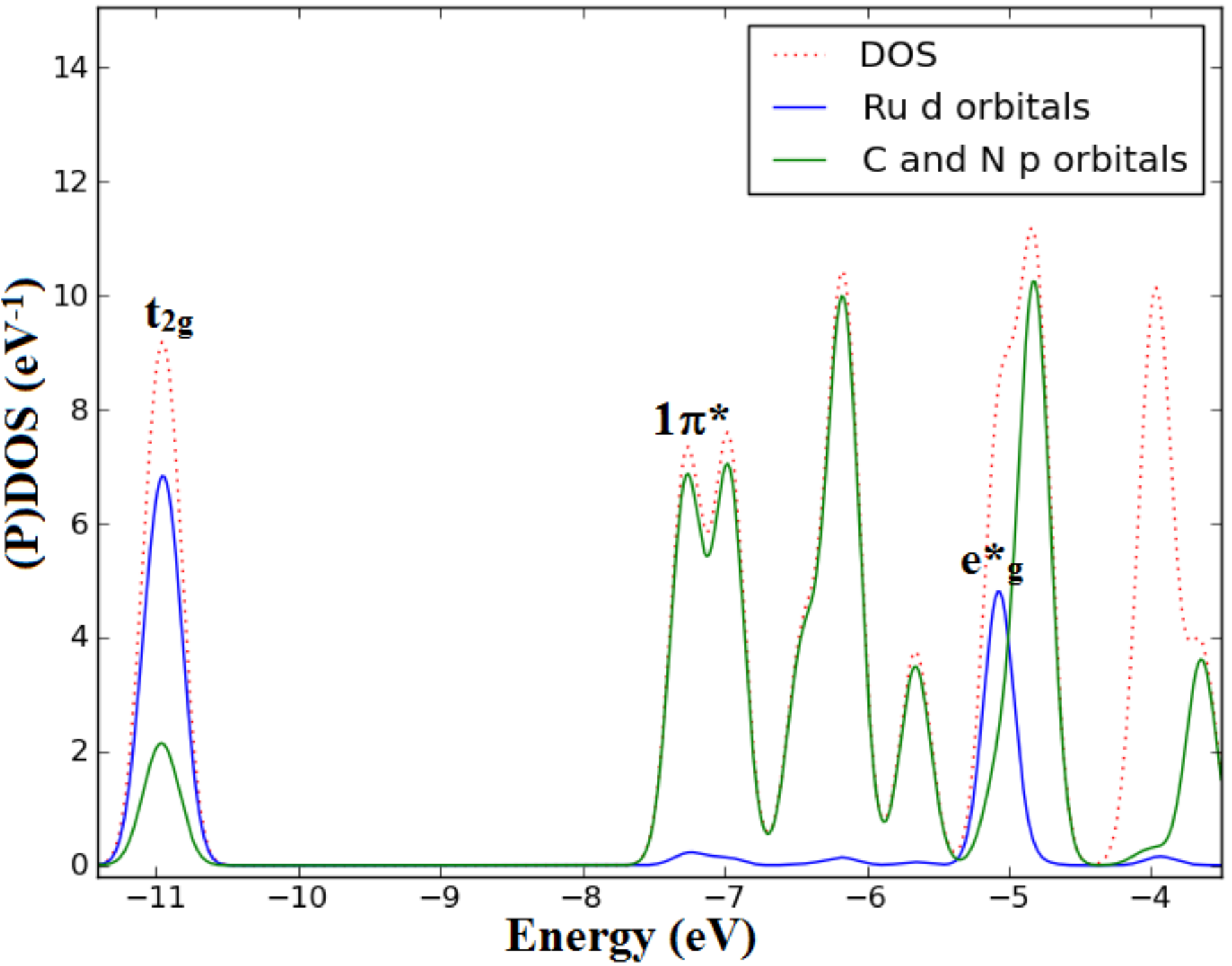} \\
B3LYP/6-31G & B3LYP/6-31G(d) \\
$\epsilon_{\text{HOMO}} = \mbox{-10.73 eV}$ & 
$\epsilon_{\text{HOMO}} = \mbox{-10.87 eV}$ 
\end{tabular}
\end{center}
Total and partial density of states of [Ru(bpy)$_2$(4,7-dm-phen)]$^{2+}$ 
partitioned over Ru d orbitals and ligand C and N p orbitals. 
% for the 6-31G (left-hand side) and 6-31G(d) (right-hand side) basis sets.

\begin{center}
   {\bf Absorption Spectrum}
\end{center}

\begin{center}
\includegraphics[width=0.8\textwidth]{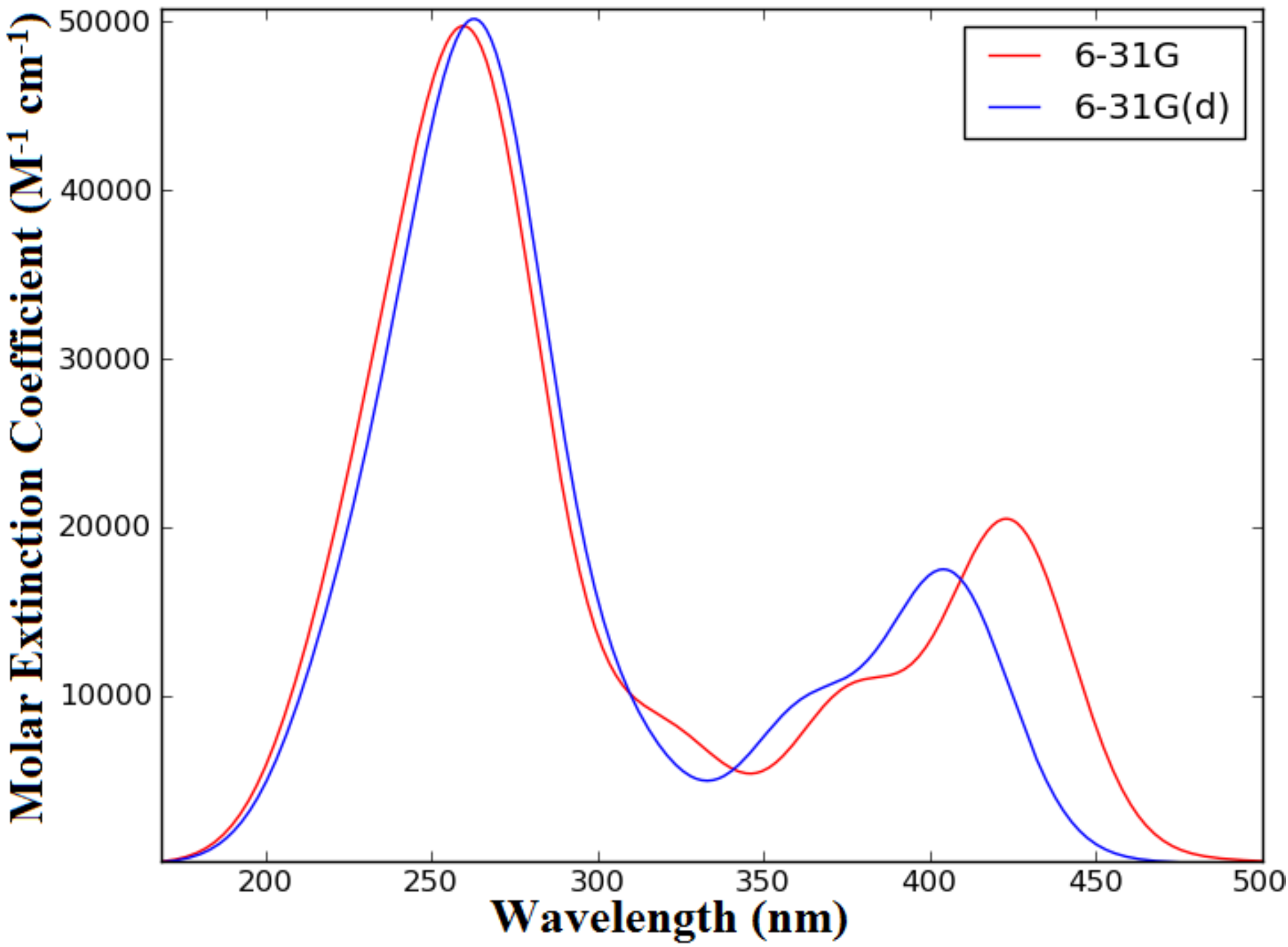}
\end{center}
[Ru(bpy)$_2$(4,7-dm-phen)]$^{2+}$
TD-B3LYP/6-31G and TD-B3LYP/6-31G(d) spectra.

% ================================================
\newpage
\section{Complex {\bf (18)}: [Ru(bpy)$_2$(4,7-Ph$_2$-phen)]$^{2+}$}
% ================================================

\begin{center}
   {\bf PDOS}
\end{center}

\begin{center}
\begin{tabular}{cc}
\includegraphics[width=0.4\textwidth]{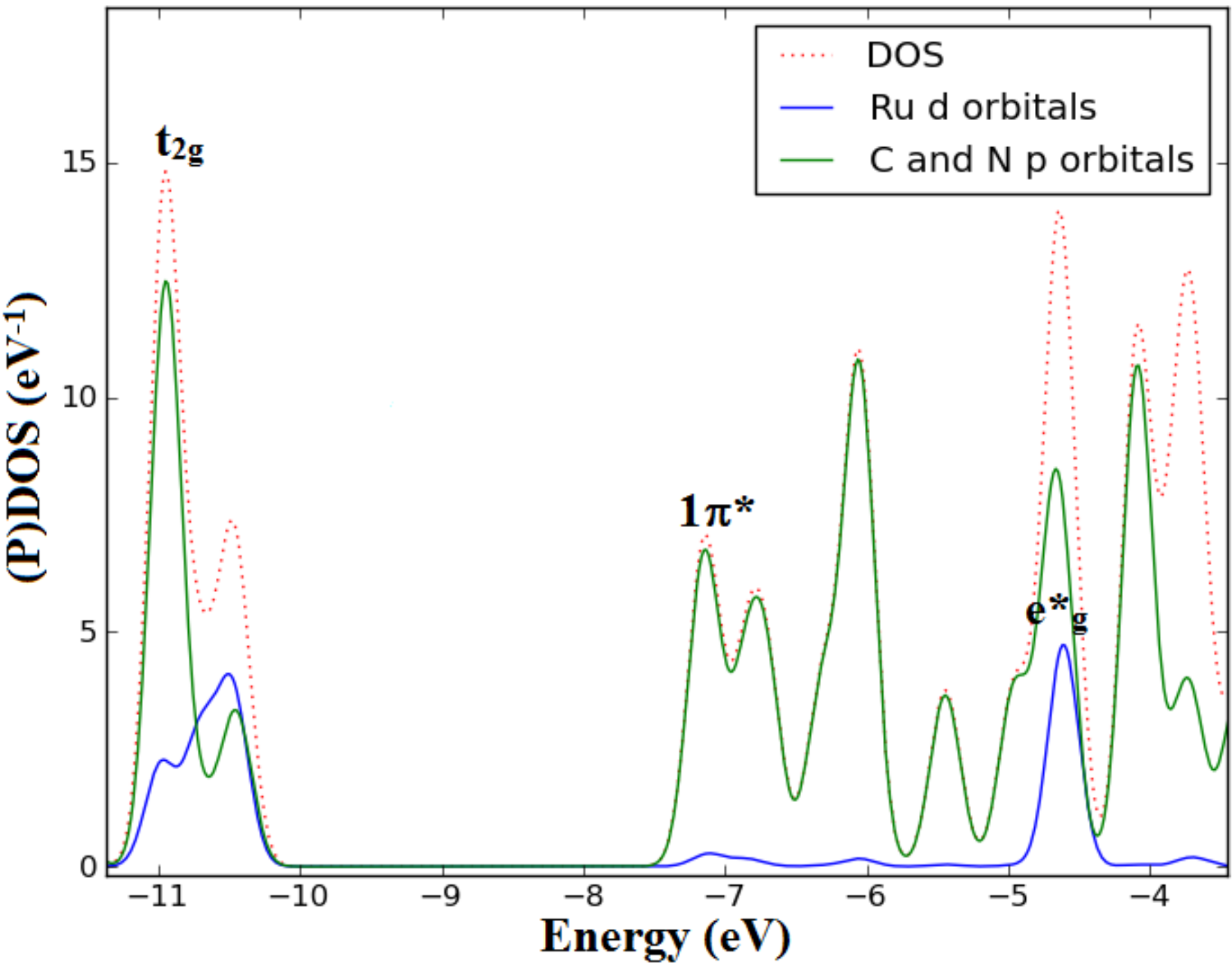} &
\includegraphics[width=0.4\textwidth]{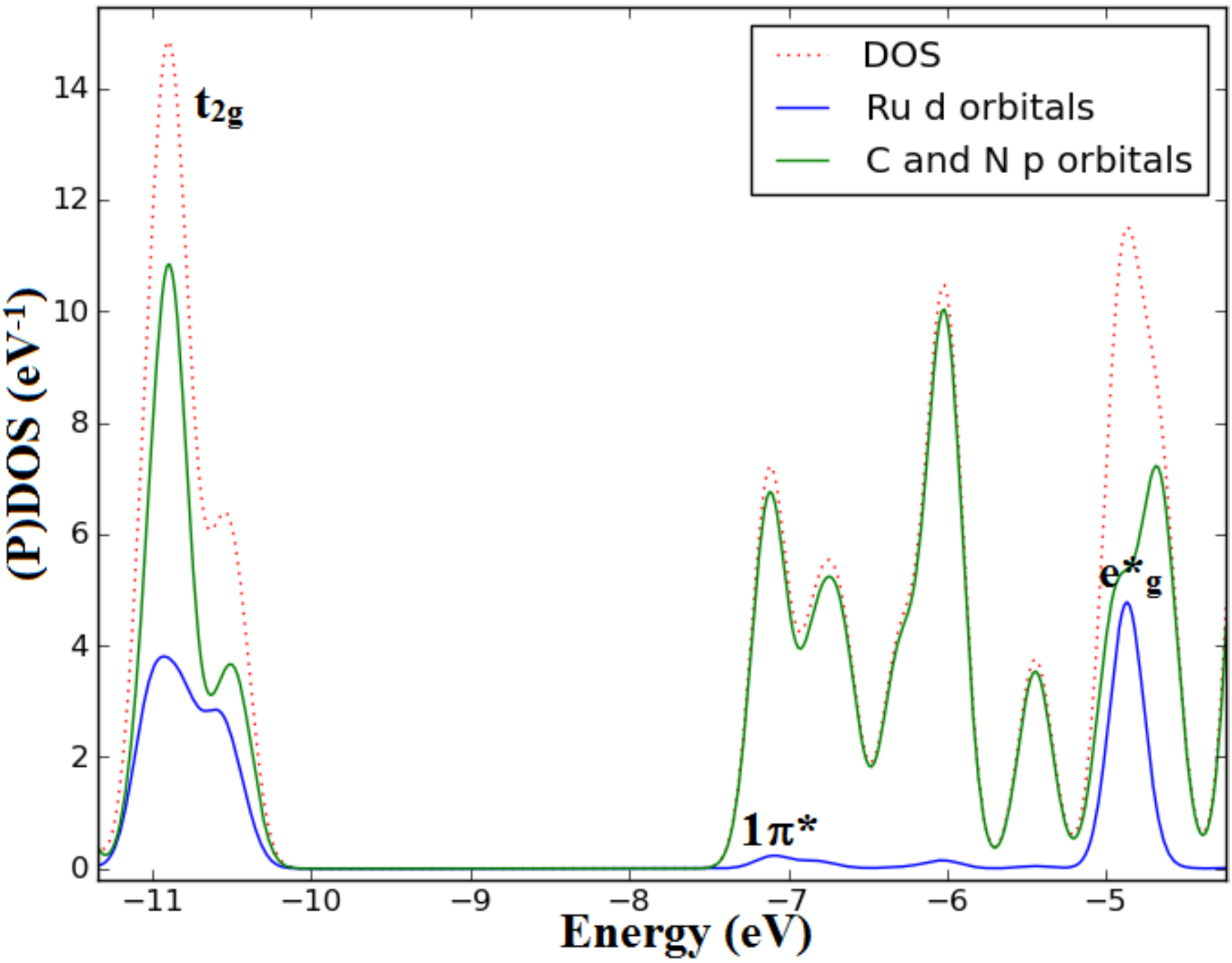} \\
B3LYP/6-31G & B3LYP/6-31G(d) \\
$\epsilon_{\text{HOMO}} = \mbox{-10.43 eV}$ & 
$\epsilon_{\text{HOMO}} = \mbox{-10.46 eV}$ 
\end{tabular}
\end{center}
Total and partial density of states of [Ru(bpy)$_2$(4,7-Ph$_{2}$-phen)]$^{2+}$ 
partitioned over Ru d orbitals and ligand C and N p orbitals.
% for the 6-31G (left-hand side) and 6-31G(d) (right-hand side) basis sets.

\begin{center}
   {\bf Absorption Spectrum}
\end{center}

\begin{center}
\includegraphics[width=0.8\textwidth]{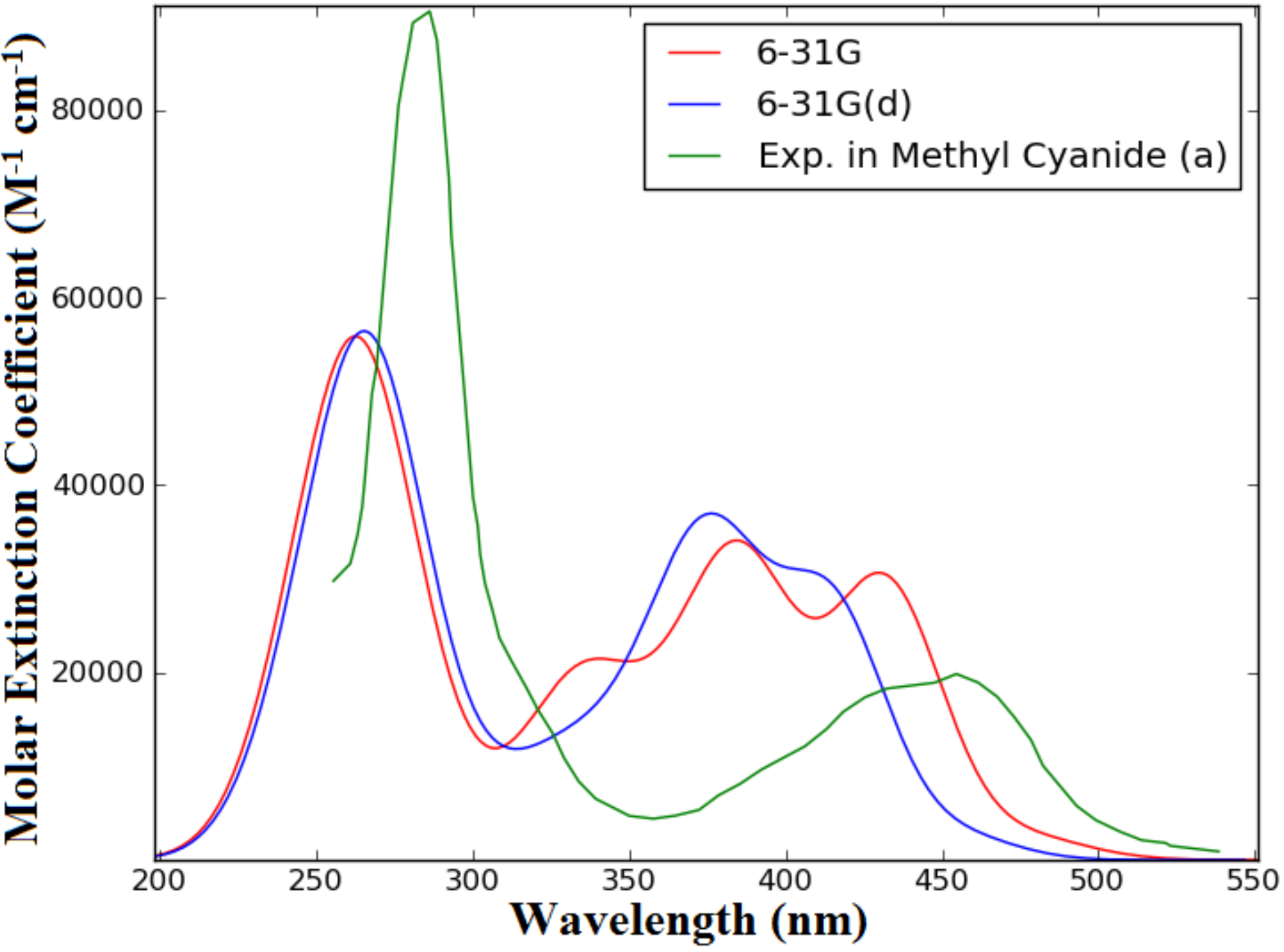}
\end{center}
[Ru(bpy)$_2$(4,7-Ph$_{2}$-phen)]$^{2+}$ 
TD-B3LYP/6-31G, TD-B3LYP/6-31G(d), and experimental spectra.
Experimental spectrum measured in acetonitrile \cite{YYS+15}.

% ================================================
\newpage
\section{Complex {\bf (19)}: [Ru(bpy)$_2$(4,7-dhy-phen)]$^{2+}$}
% ================================================

\begin{center}
   {\bf PDOS}
\end{center}

\begin{center}
\begin{tabular}{cc}
\includegraphics[width=0.4\textwidth]{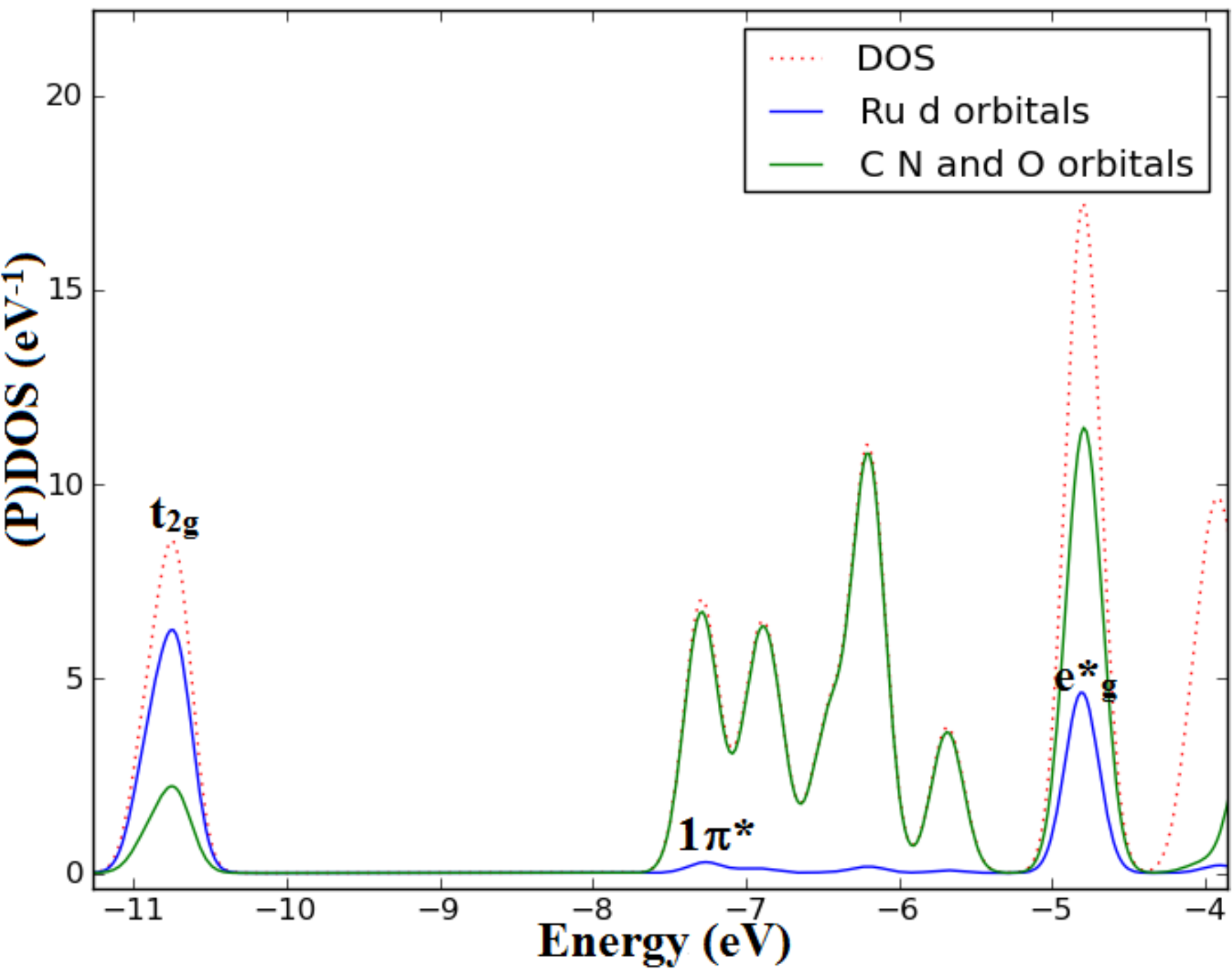} &
\includegraphics[width=0.4\textwidth]{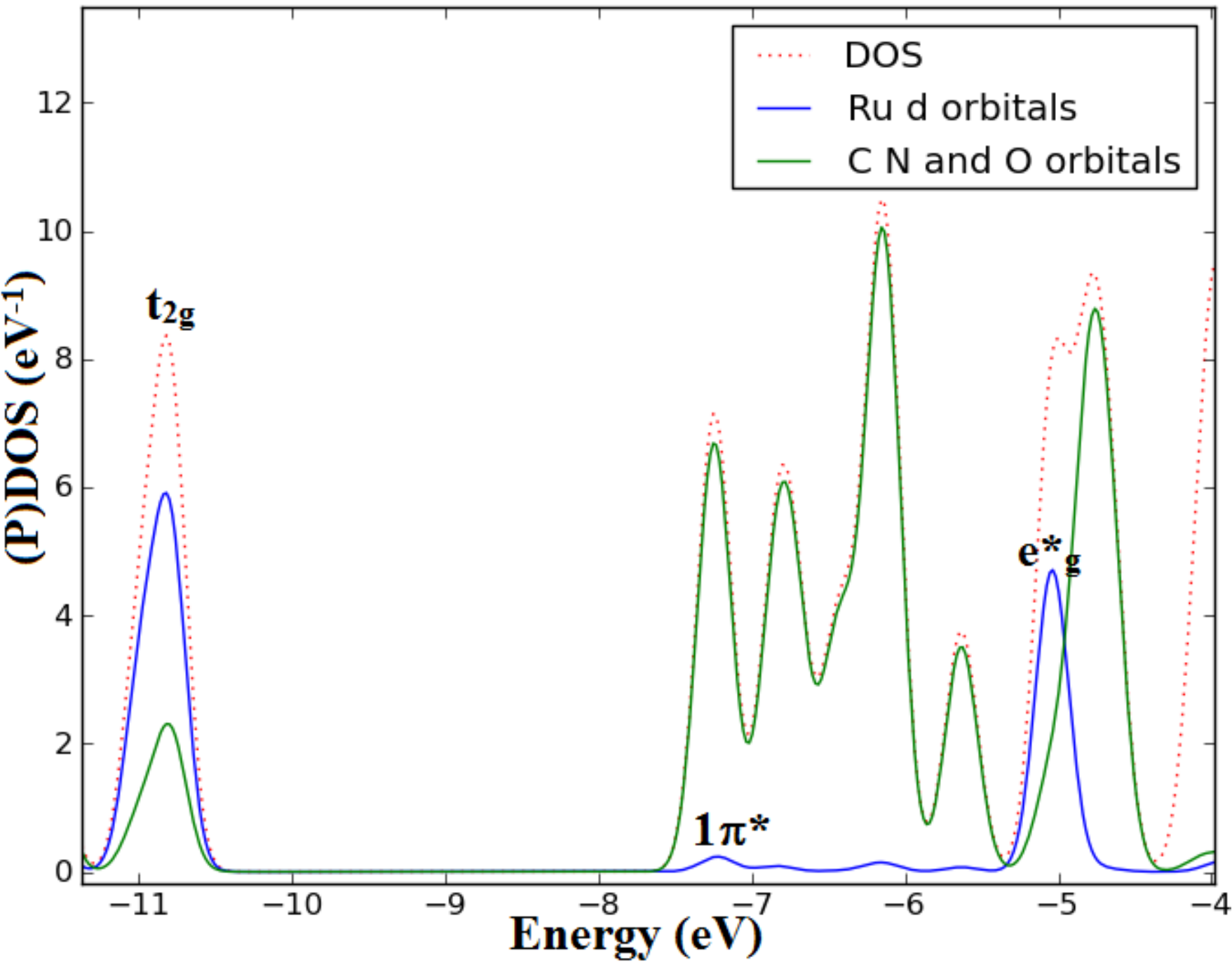} \\
B3LYP/6-31G & B3LYP/6-31G(d) \\
$\epsilon_{\text{HOMO}} = \mbox{-10.71 eV}$ & 
$\epsilon_{\text{HOMO}} = \mbox{-10.79 eV}$ 
\end{tabular}
\end{center}
Total and partial density of states of [Ru(bpy)$_2$(4,7-dhy-phen)]$^{2+}$  
partitioned over Ru d orbitals and ligand C,O, and N p orbitals. 
% for the 6-31G (left-hand side) and 6-31G(d) (right-hand side) basis sets.

\begin{center}
   {\bf Absorption Spectrum}
\end{center}

\begin{center}
\includegraphics[width=0.8\textwidth]{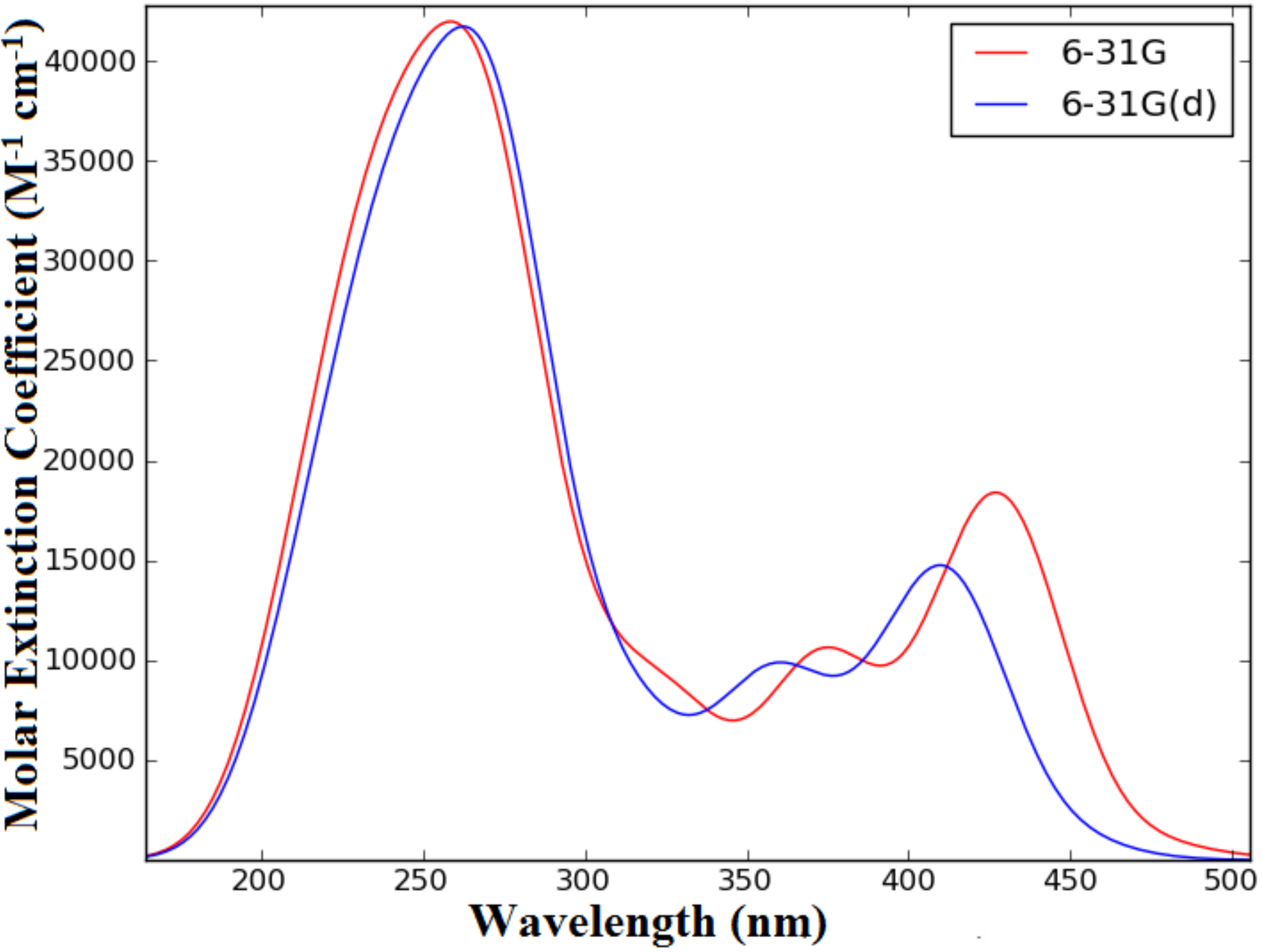}
\end{center}
[Ru(bpy)$_2$(4,7-dhy-phen)]$^{2+}$ 
TD-B3LYP/6-31G and TD-B3LYP/6-31G(d) spectra.

% ================================================
\newpage
\section{Complex {\bf (20)}: [Ru(bpy)$_2$(5,6-dm-phen)]$^{2+}$}
% ================================================

\begin{center}
   {\bf PDOS}
\end{center}

\begin{center}
\begin{tabular}{cc}
\includegraphics[width=0.4\textwidth]{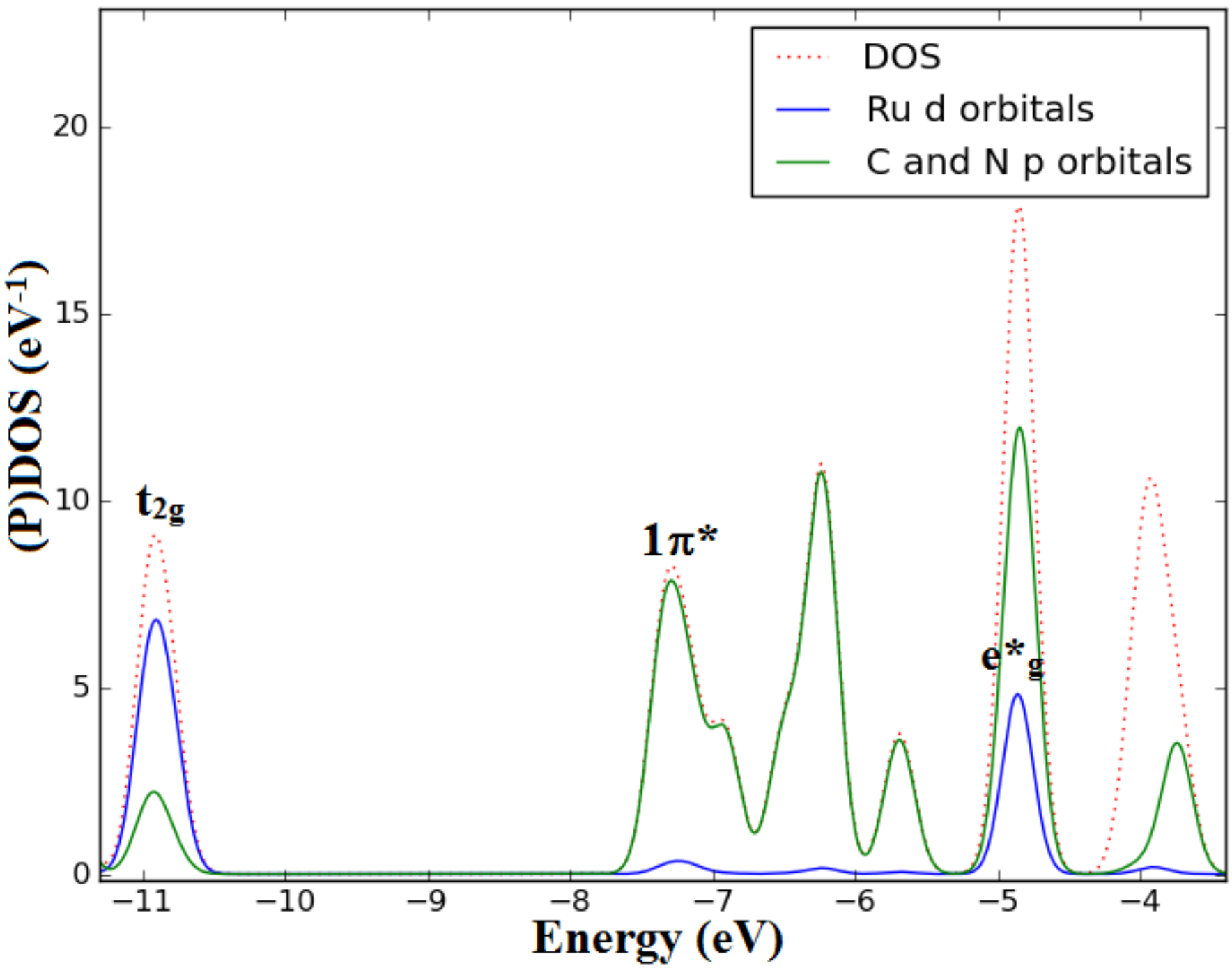} &
\includegraphics[width=0.4\textwidth]{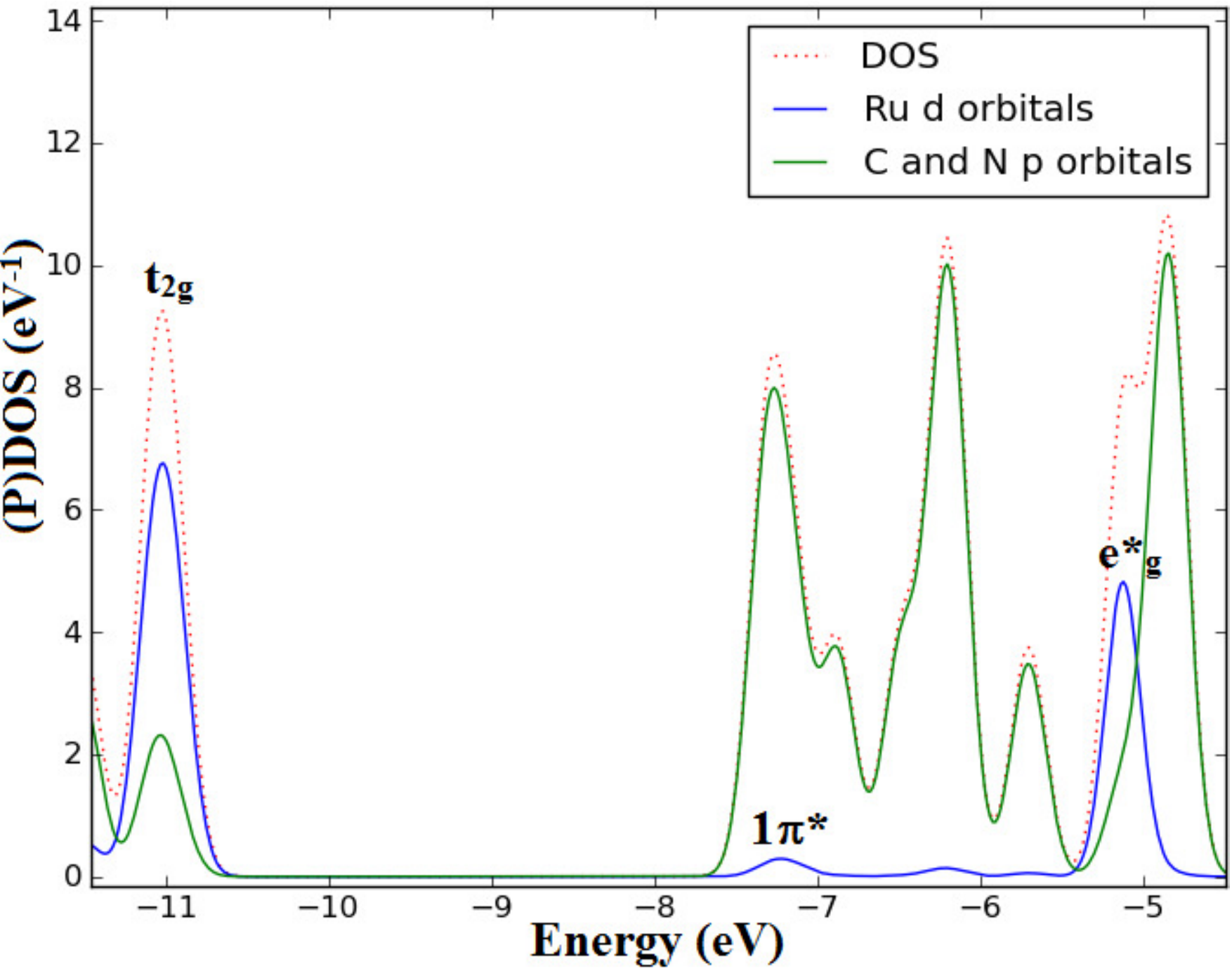} \\
B3LYP/6-31G & B3LYP/6-31G(d) \\
$\epsilon_{\text{HOMO}} = \mbox{-10.80 eV}$ & 
$\epsilon_{\text{HOMO}} = \mbox{-10.94 eV}$ 
\end{tabular}
\end{center}
Total and partial density of states of [Ru(bpy)$_{2}$(5,6-dm-phen)]$^{2+}$ 
partitioned over Ru d orbitals and ligand C and N p orbitals. 
% for the 6-31G (left-hand side) and 6-31G(d) (right-hand side) basis sets.

\begin{center}
   {\bf Absorption Spectrum}
\end{center}

\begin{center}
\includegraphics[width=0.8\textwidth]{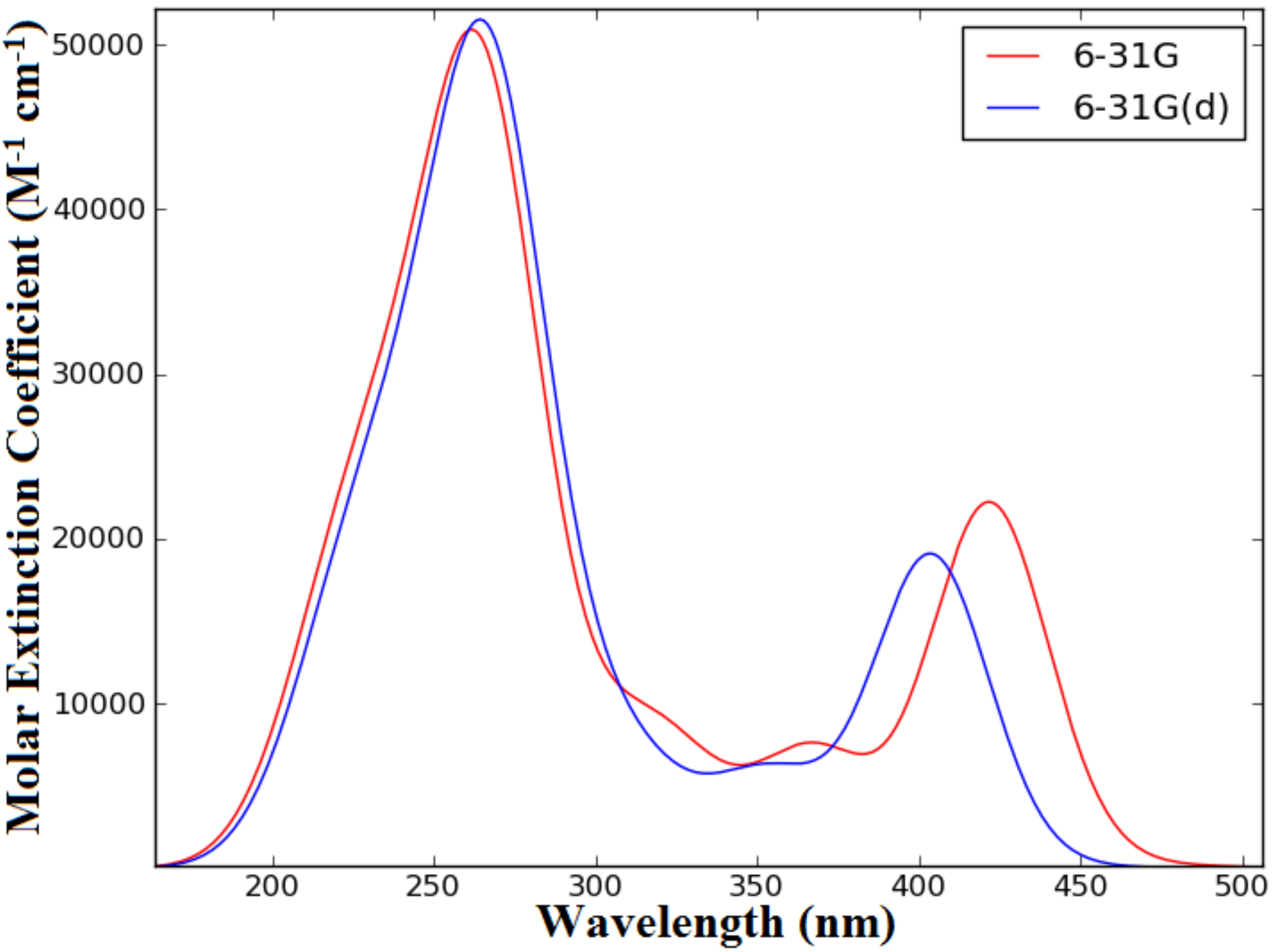}
\end{center}
[Ru(bpy)$_{2}$(5,6-dm-phen)]$^{2+}$
TD-B3LYP/6-31G and TD-B3LYP/6-31G(d) spectra.

% ================================================
\newpage
\section{Complex {\bf (21)}: [Ru(bpy)$_2$(DIAF)]$^{2+}$}
% ================================================

\begin{center}
   {\bf PDOS}
\end{center}

\begin{center}
\begin{tabular}{cc}
\includegraphics[width=0.4\textwidth]{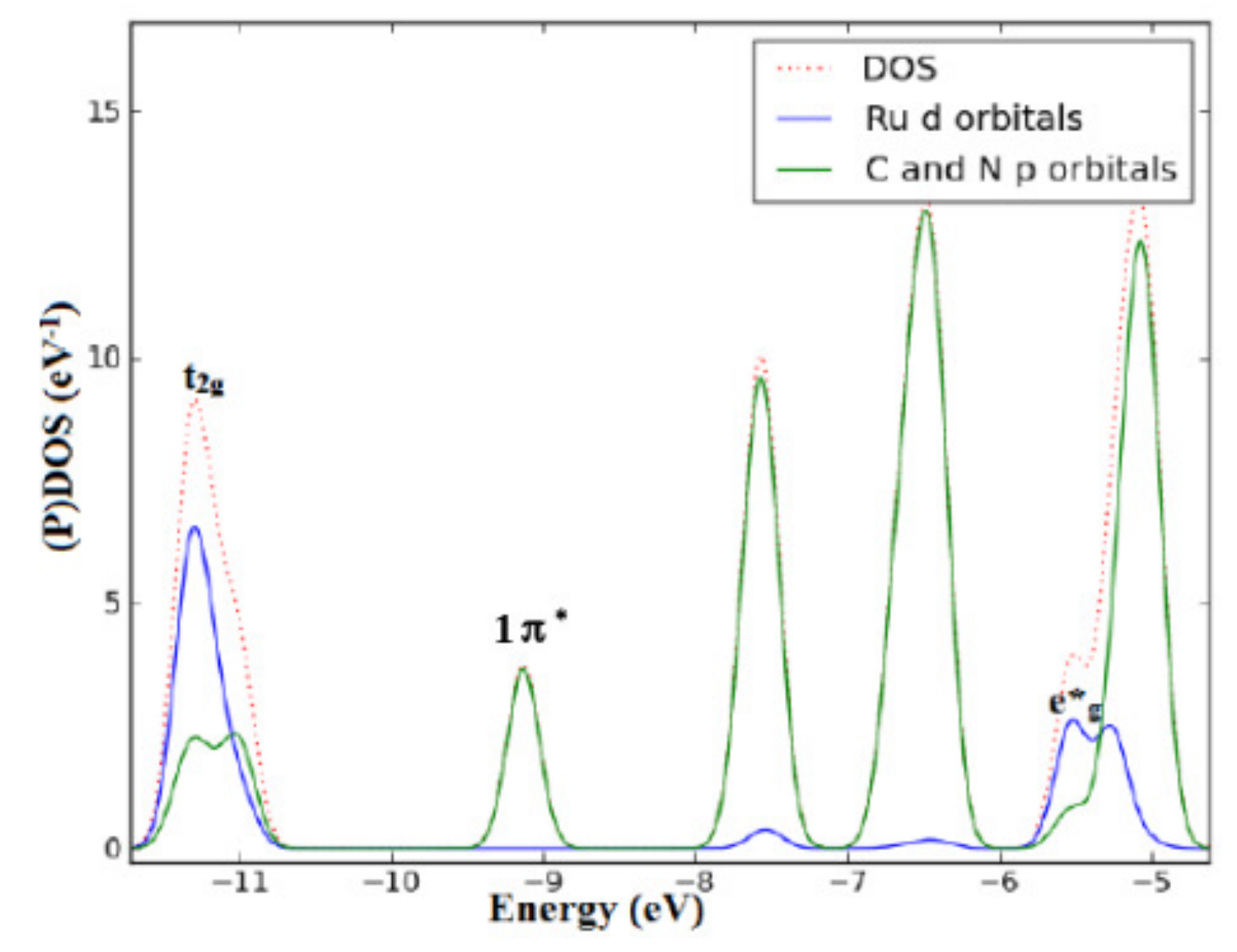} &
\includegraphics[width=0.4\textwidth]{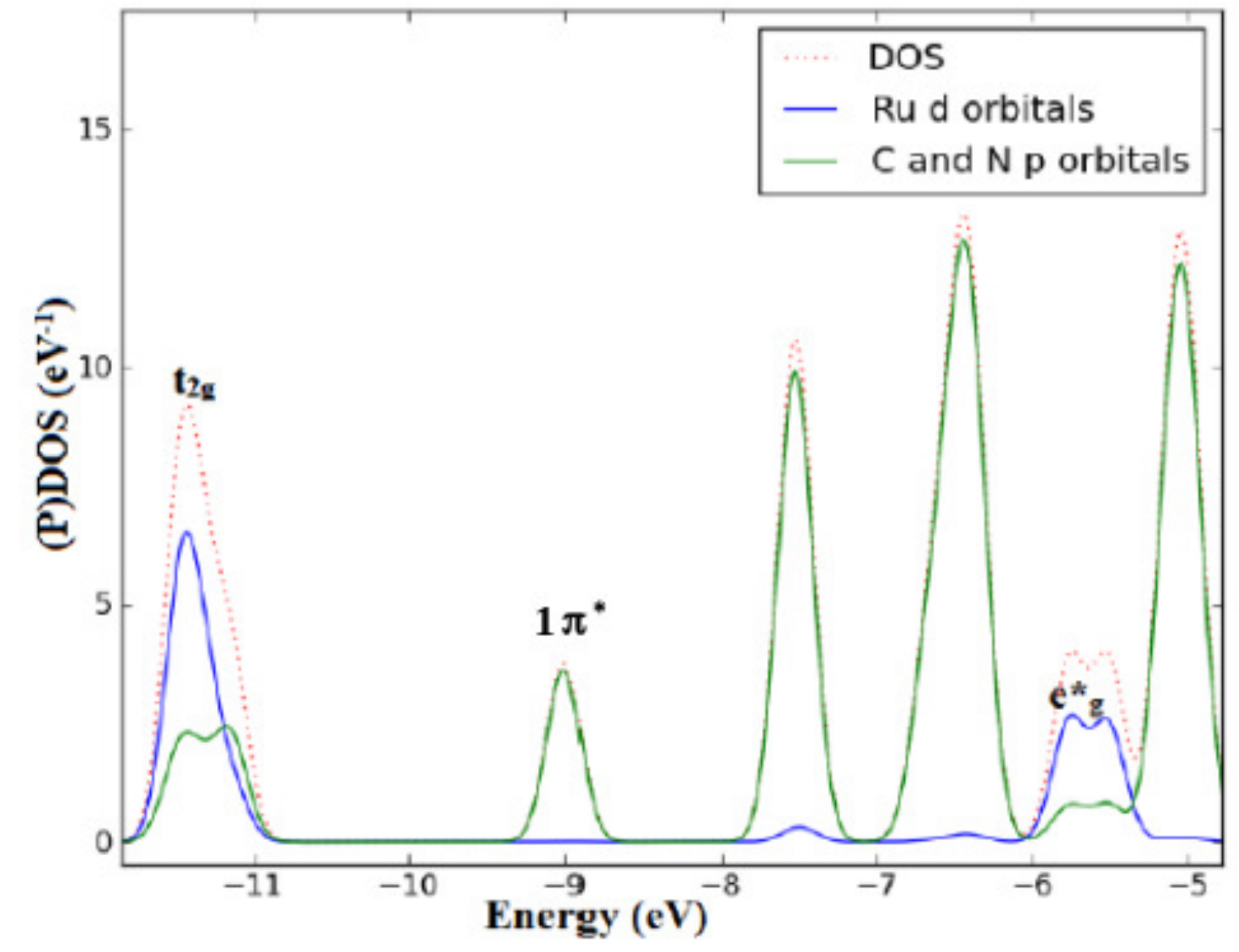} \\
B3LYP/6-31G & B3LYP/6-31G(d) \\
$\epsilon_{\text{HOMO}} = \mbox{-11.00 eV}$ & 
$\epsilon_{\text{HOMO}} = \mbox{-11.14 eV}$ 
\end{tabular}
\end{center}
Total and partial density of states of [Ru(bpy)$_2$(DIAF)]$^{2+}$
partitioned over Ru d orbitals and ligand C and N p orbitals. 
% for the 6-31G (left-hand side) and 6-31G(d) (right-hand side) basis sets.

\begin{center}
   {\bf Absorption Spectrum}
\end{center}

\begin{center}
\includegraphics[width=0.8\textwidth]{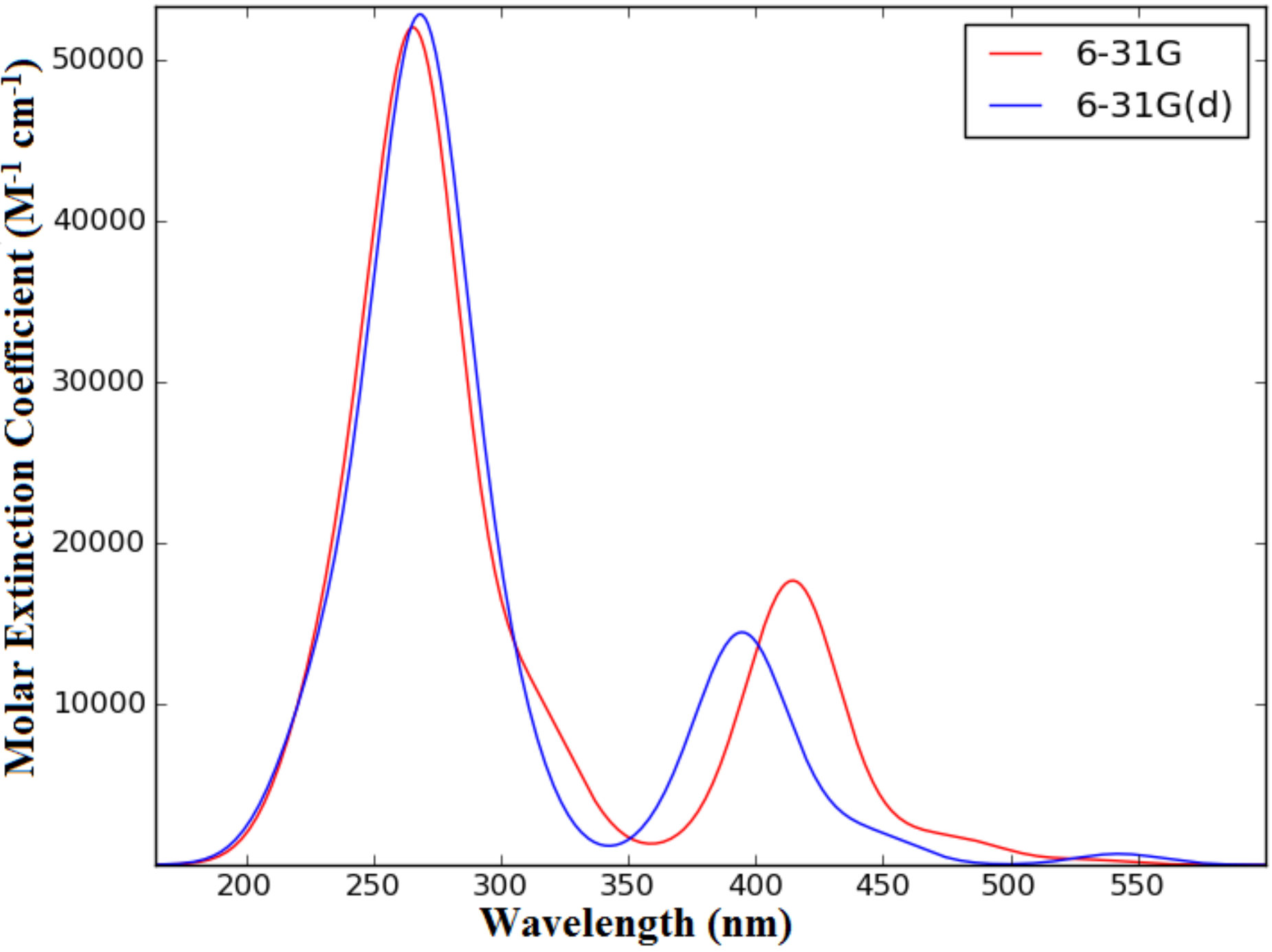}
\end{center}
[Ru(bpy)$_2$(DIAF)]$^{2+}$
TD-B3LYP/6-31G and TD-B3LYP/6-31G(d) spectra.

% ================================================
\newpage
\section{Complex {\bf (22)}$^\dagger$: [Ru(bpy)$_2$(DIAFO)]$^{2+}$}
% ================================================

\begin{center}
\begin{tabular}{cc}
B3LYP/6-31G & B3LYP/6-31G(d) \\
$\epsilon_{\text{HOMO}} = \mbox{-11.17 eV}$ & 
$\epsilon_{\text{HOMO}} = \mbox{-11.30 eV}$ 
\end{tabular}
\end{center}

% \begin{center}
%    {\bf PDOS}
% \end{center}
% 
% \begin{center}
% \includegraphics[width=0.4\textwidth]{graphics1/framedquestionmark.pdf}
% \includegraphics[width=0.4\textwidth]{graphics1/framedquestionmark.pdf}
% \end{center}
% {\color{red} Do we have this?}

\begin{center}
   {\bf Absorption Spectrum}
\end{center}

\begin{center}
\includegraphics[width=0.8\textwidth]{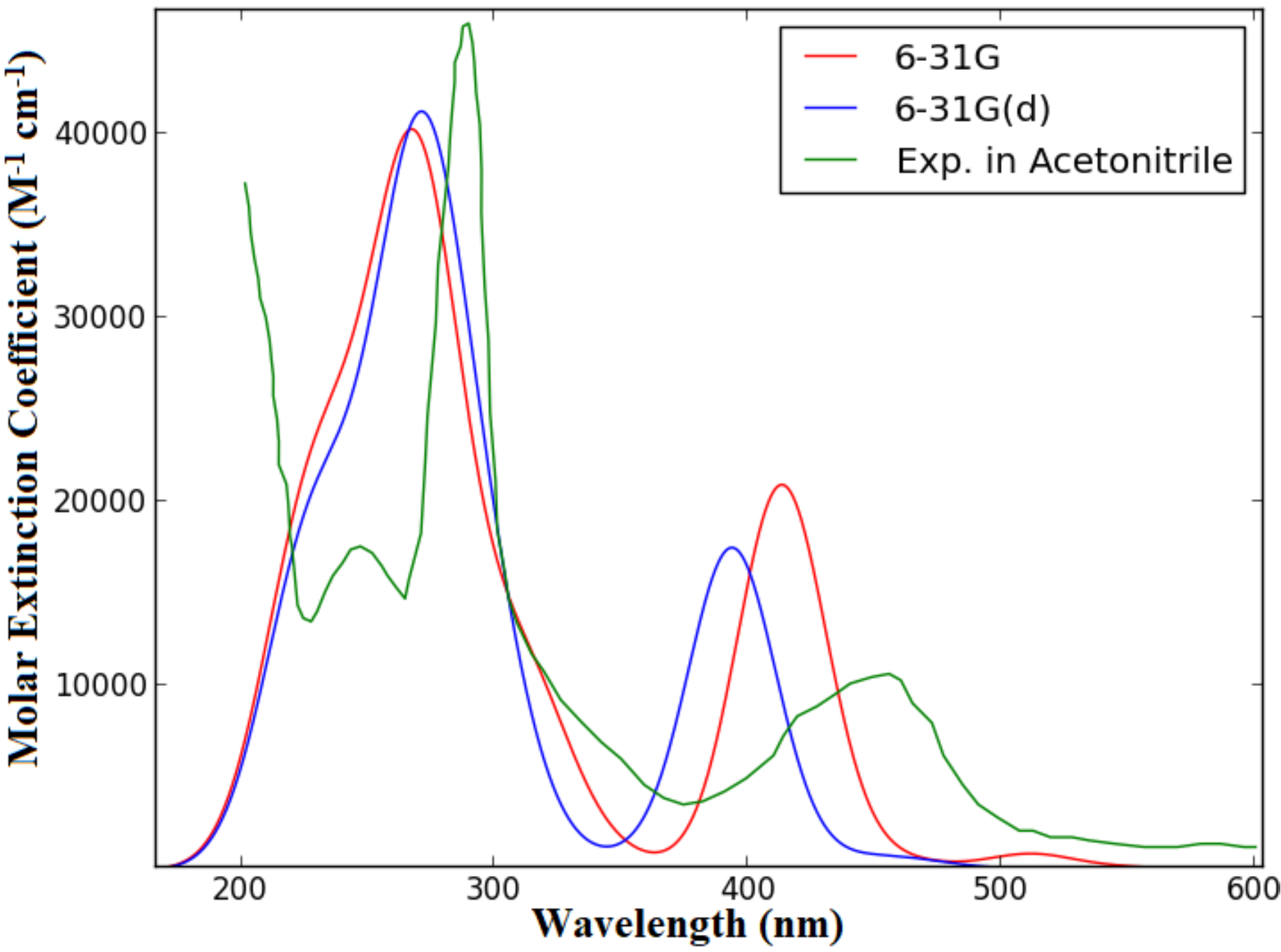}
\end{center}
[Ru(bpy)$_2$(DIAFO)]$^{2+}$
TD-B3LYP/6-31G, TD-B3LYP/6-31G(d), and experimental spectra.
Experimental spectrum measured at room temperature in 
acetonitrile \cite{HXA+10}.

% ================================================
\newpage
\section{Complex {\bf (23)}: [Ru(bpy)$_2$(taphen)]$^{2+}$}
% ================================================

\begin{center}
   {\bf PDOS}
\end{center}

\begin{center}
\begin{tabular}{cc}
\includegraphics[width=0.4\textwidth]{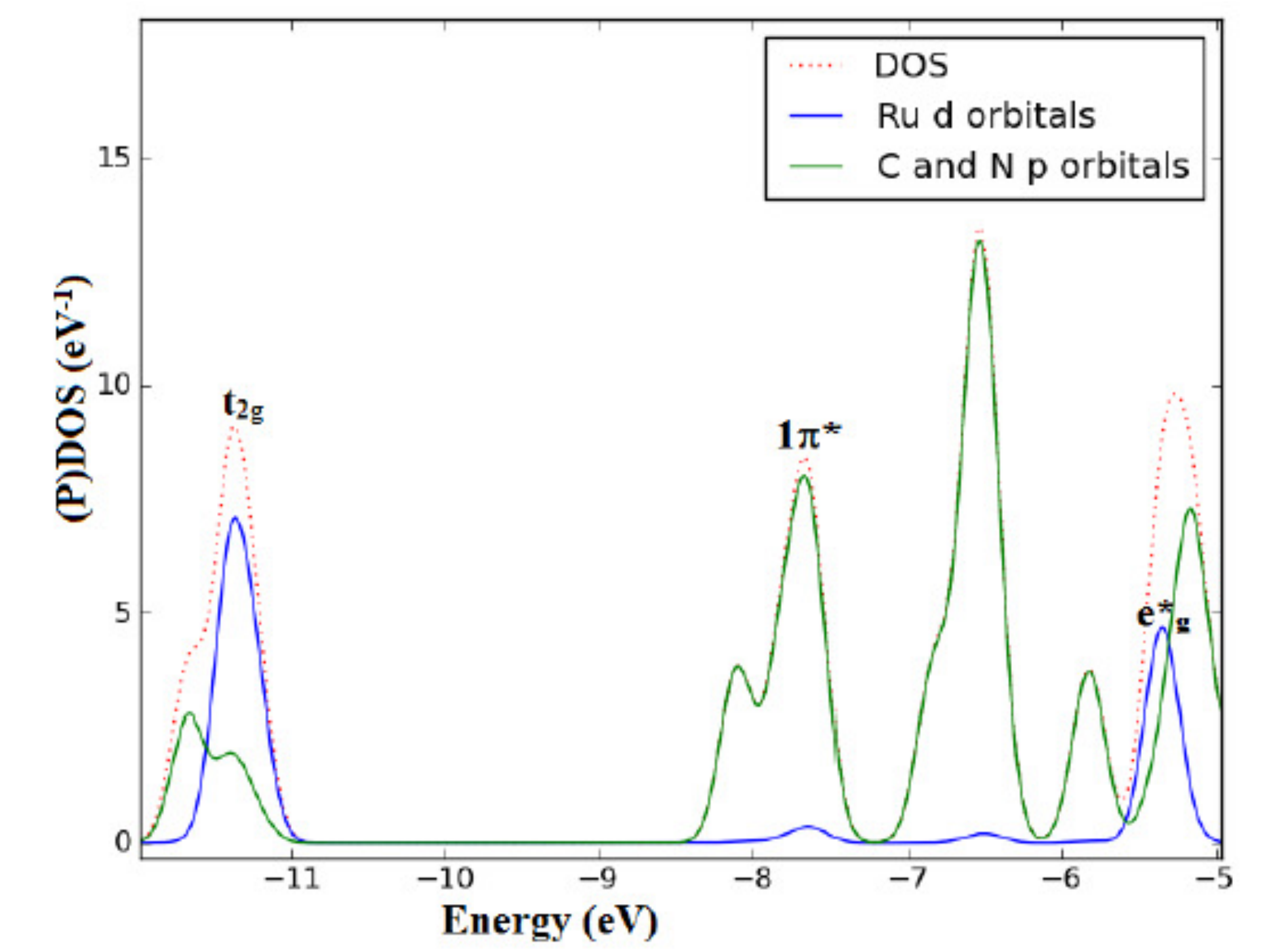} &
\includegraphics[width=0.4\textwidth]{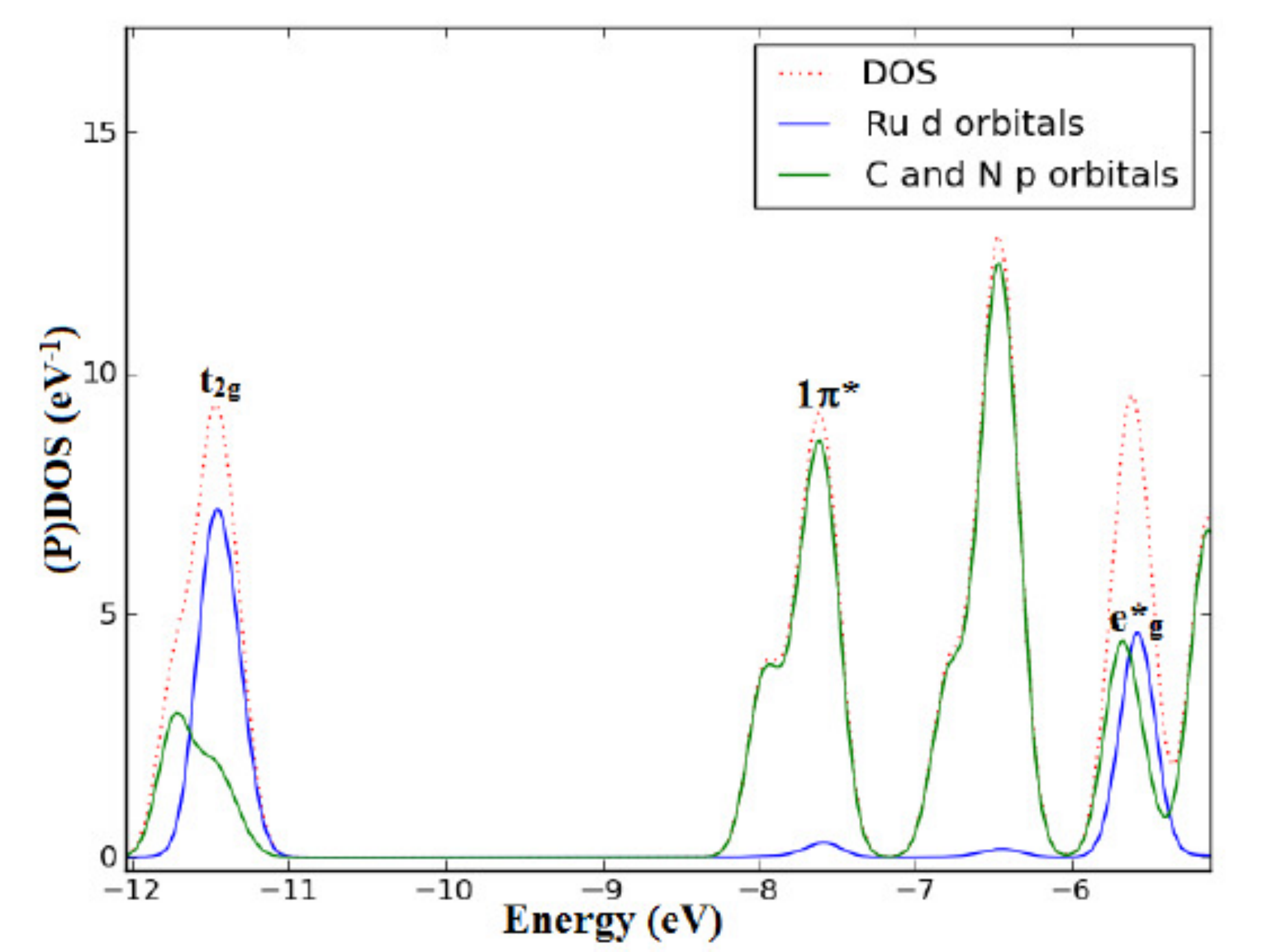} \\
B3LYP/6-31G & B3LYP/6-31G(d) \\
$\epsilon_{\text{HOMO}} = \mbox{-11.25 eV}$ & 
$\epsilon_{\text{HOMO}} = \mbox{-11.35 eV}$ 
\end{tabular}
\end{center}
Total and partial density of states of [Ru(bpy)$_2$(taphen)]$^{2+}$
partitioned over Ru d orbitals and ligand C and N p orbitals. 
% for the 6-31G (left-hand side) and 6-31G(d) (right-hand side) basis sets.

\begin{center}
   {\bf Absorption Spectrum}
\end{center}

\begin{center}
\includegraphics[width=0.8\textwidth]{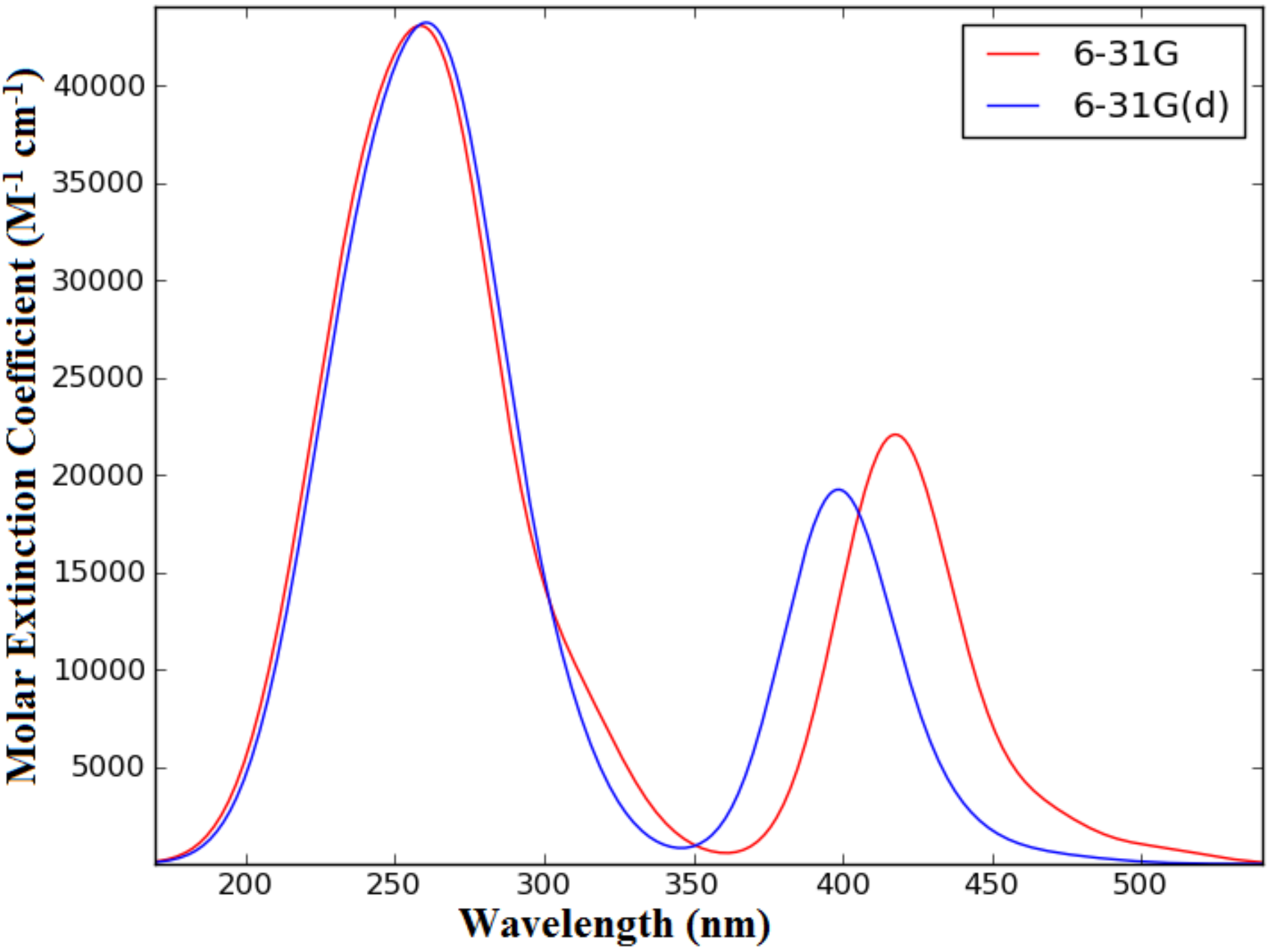}
\end{center}
[Ru(bpy)$_2$(taphen)]$^{2+}$
TD-B3LYP/6-31G and TD-B3LYP/6-31G(d) spectra.

% ================================================
\newpage
\section{Complex {\bf (24)}: {\em cis}-[Ru(bpy)$_2$(py)$_2$]$^{2+}$}
% ================================================

\begin{center}
   {\bf PDOS}
\end{center}

\begin{center}
\begin{tabular}{cc}
\includegraphics[width=0.4\textwidth]{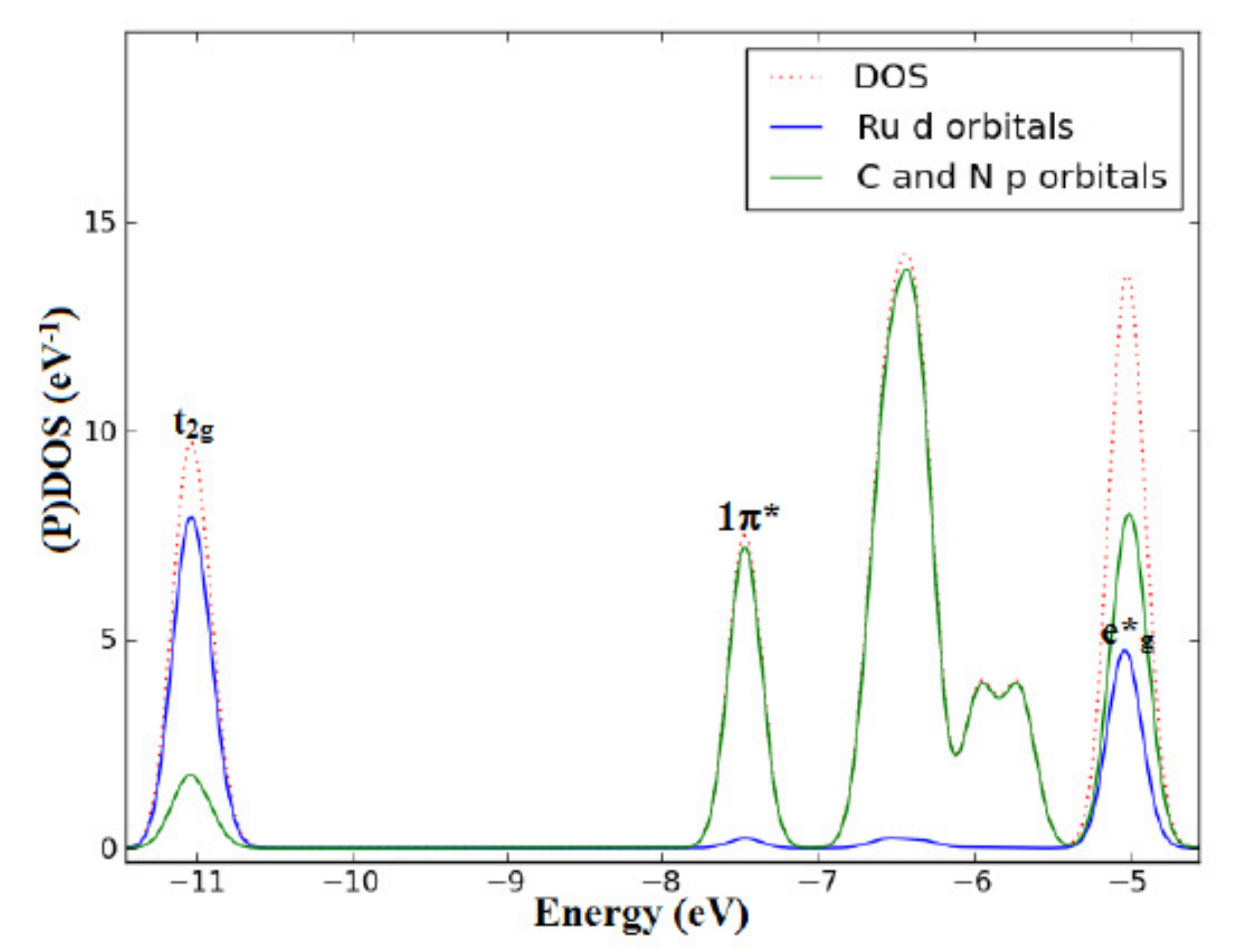} &
\includegraphics[width=0.4\textwidth]{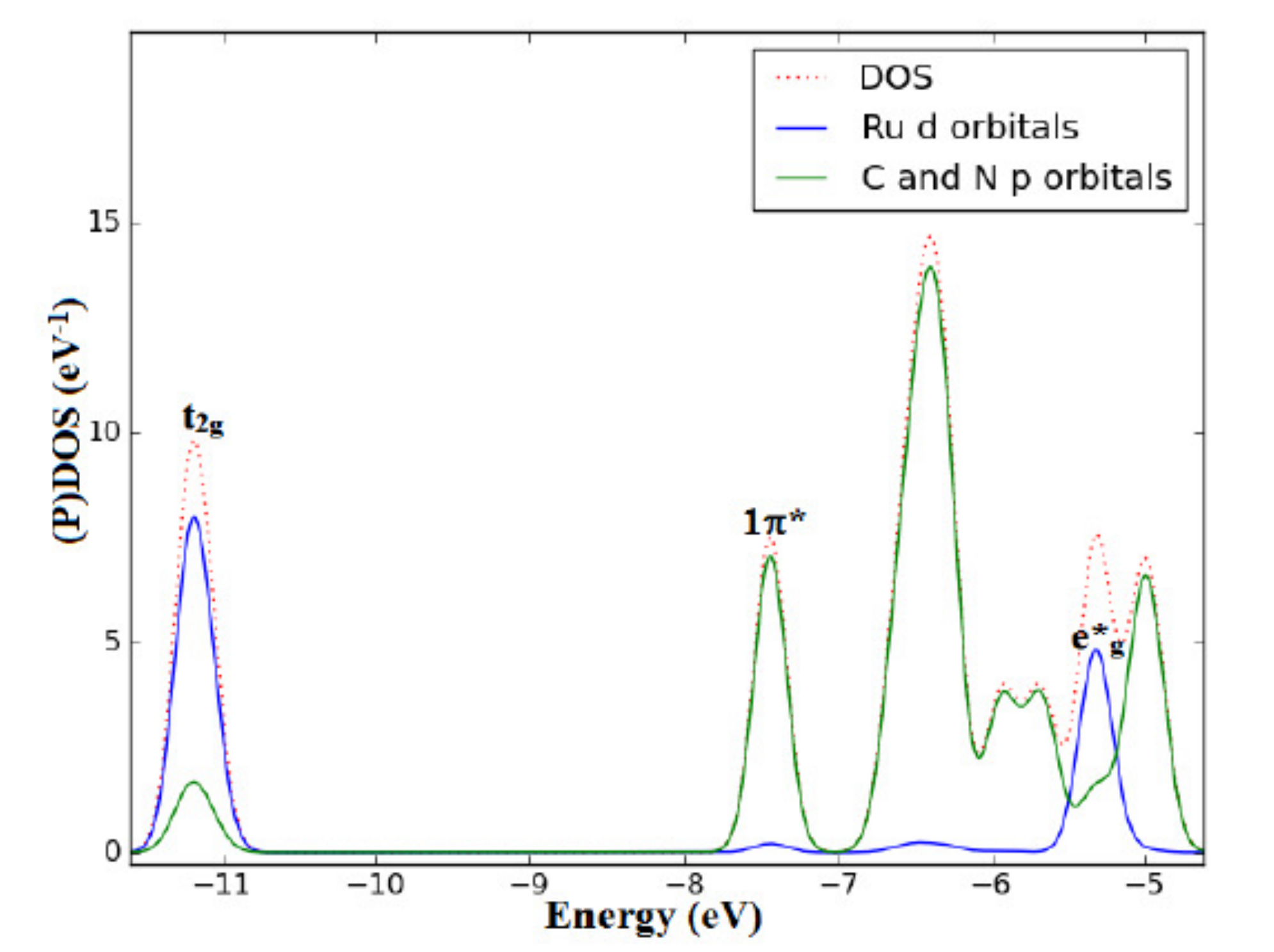} \\
B3LYP/6-31G & B3LYP/6-31G(d) \\
$\epsilon_{\text{HOMO}} = \mbox{-10.95 eV}$ & 
$\epsilon_{\text{HOMO}} = \mbox{-11.10 eV}$ 
\end{tabular}
\end{center}
Total and partial density of states of {\em cis}-[Ru(bpy)$_2$(py)$_2$]$^{2+}$
partitioned over Ru d orbitals and ligand C and N p orbitals. 
% for the 6-31G (left-hand side) and 6-31G* (right-hand side) basis sets.

\begin{center}
   {\bf Absorption Spectrum}
\end{center}

\begin{center}
\includegraphics[width=0.8\textwidth]{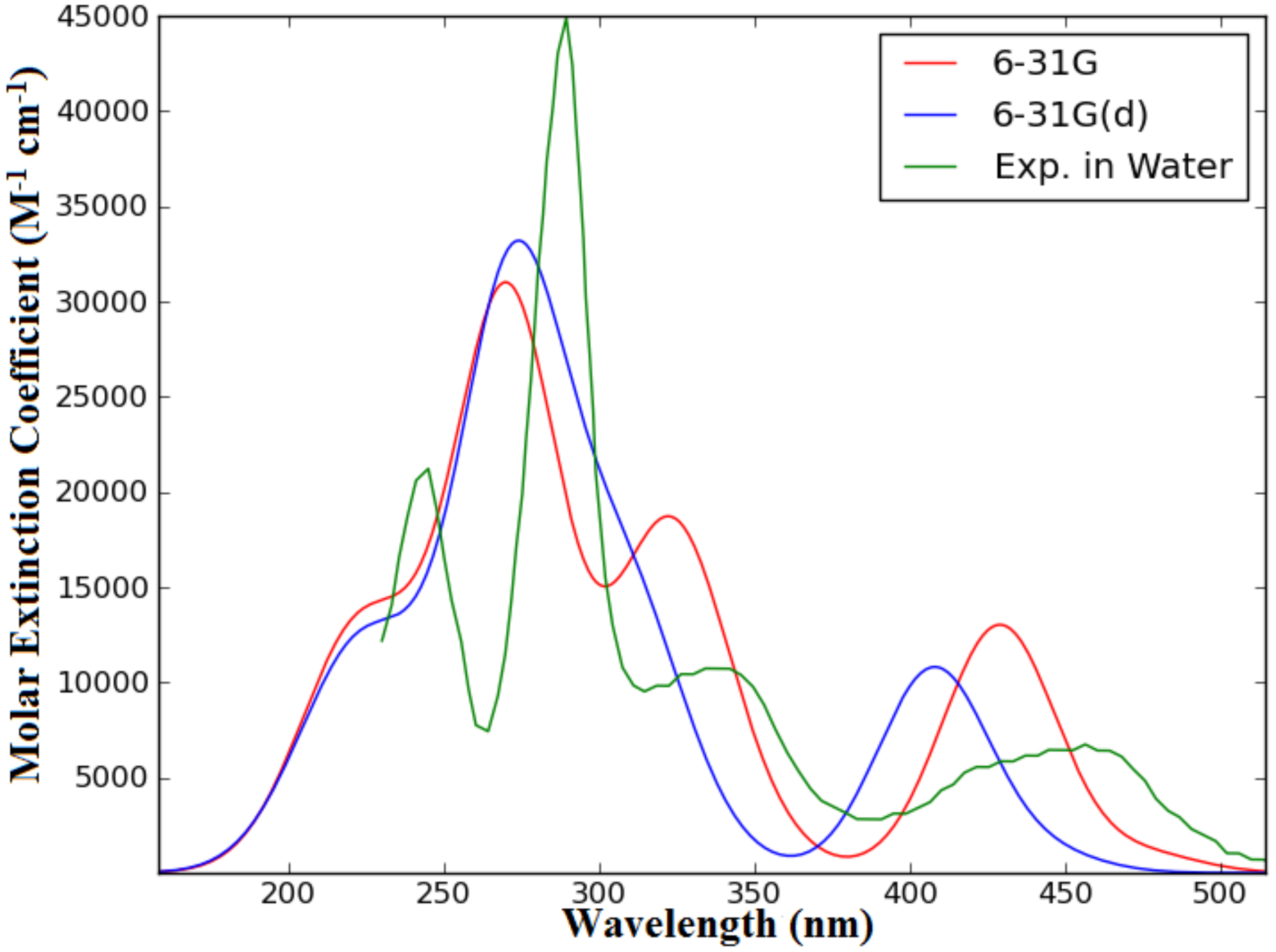}
\end{center}
{\em Cis}-[Ru(bpy)$_2$(py)$_2$]$^{2+}$ 
TD-B3LYP/6-31G, TD-B3LYP/6-31G(d), and experimental spectra.
Experimental spectrum measured in water \cite{BGS+13}.

% ================================================
\newpage
\section{Complex {\bf (25)}: {\em trans}-[Ru(bpy)$_2$(py)$_2$]$^{2+}$}
% ================================================

\begin{center}
   {\bf PDOS}
\end{center}

\begin{center}
\begin{tabular}{cc}
\includegraphics[width=0.4\textwidth]{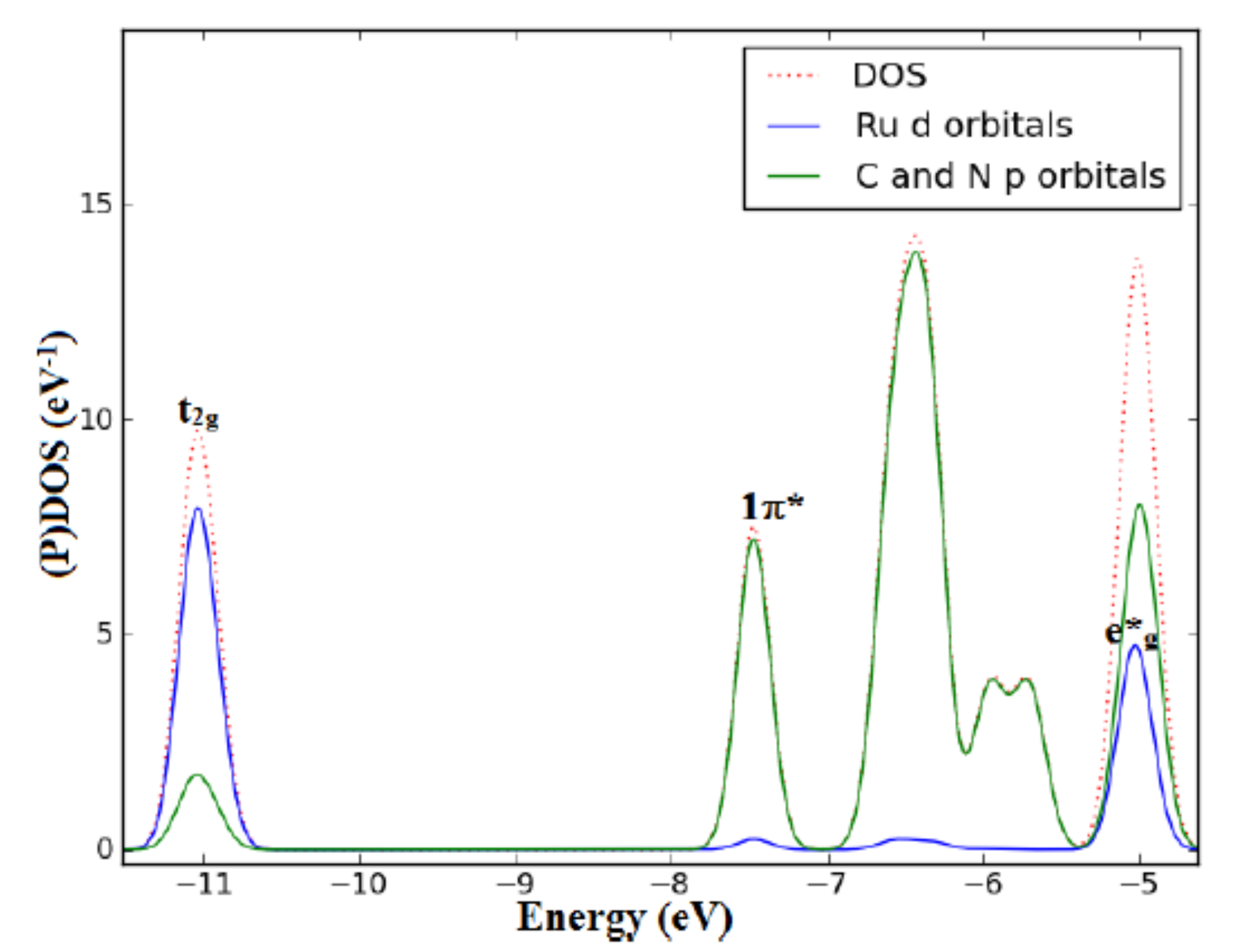} &
\includegraphics[width=0.4\textwidth]{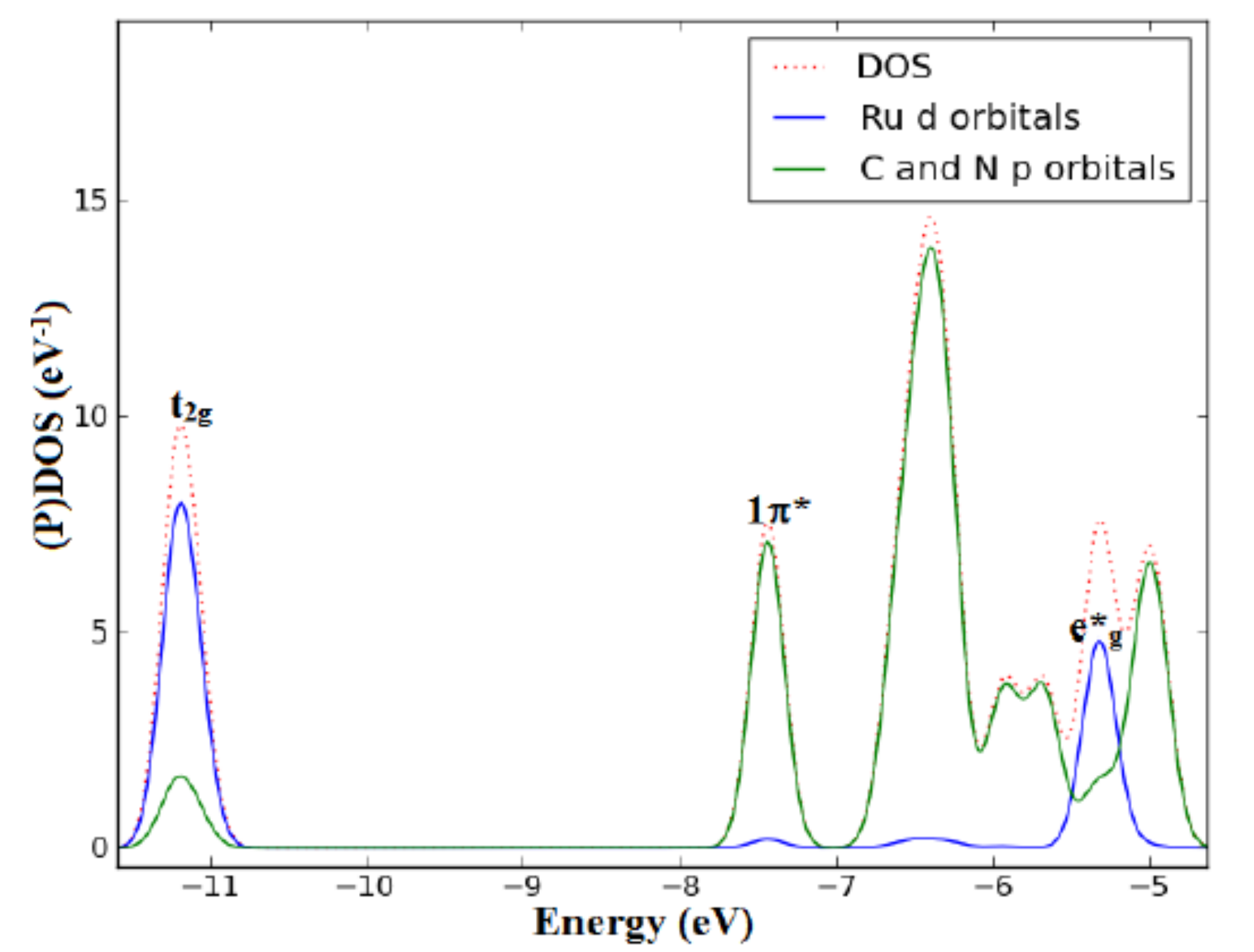} \\
B3LYP/6-31G & B3LYP/6-31G(d) \\
$\epsilon_{\text{HOMO}} = \mbox{-10.95 eV}$ & 
$\epsilon_{\text{HOMO}} = \mbox{-11.10 eV}$ 
\end{tabular}
\end{center}
Total and partial density of states of {\em trans}-[Ru(bpy)$_2$(py)$_2$]$^{2+}$
partitioned over Ru d orbitals and ligand C and N p orbitals. 
% for the 6-31G (left-hand side) and 6-31G* (right-hand side) basis sets.

\begin{center}
   {\bf Absorption Spectrum}
\end{center}

\begin{center}
\includegraphics[width=0.8\textwidth]{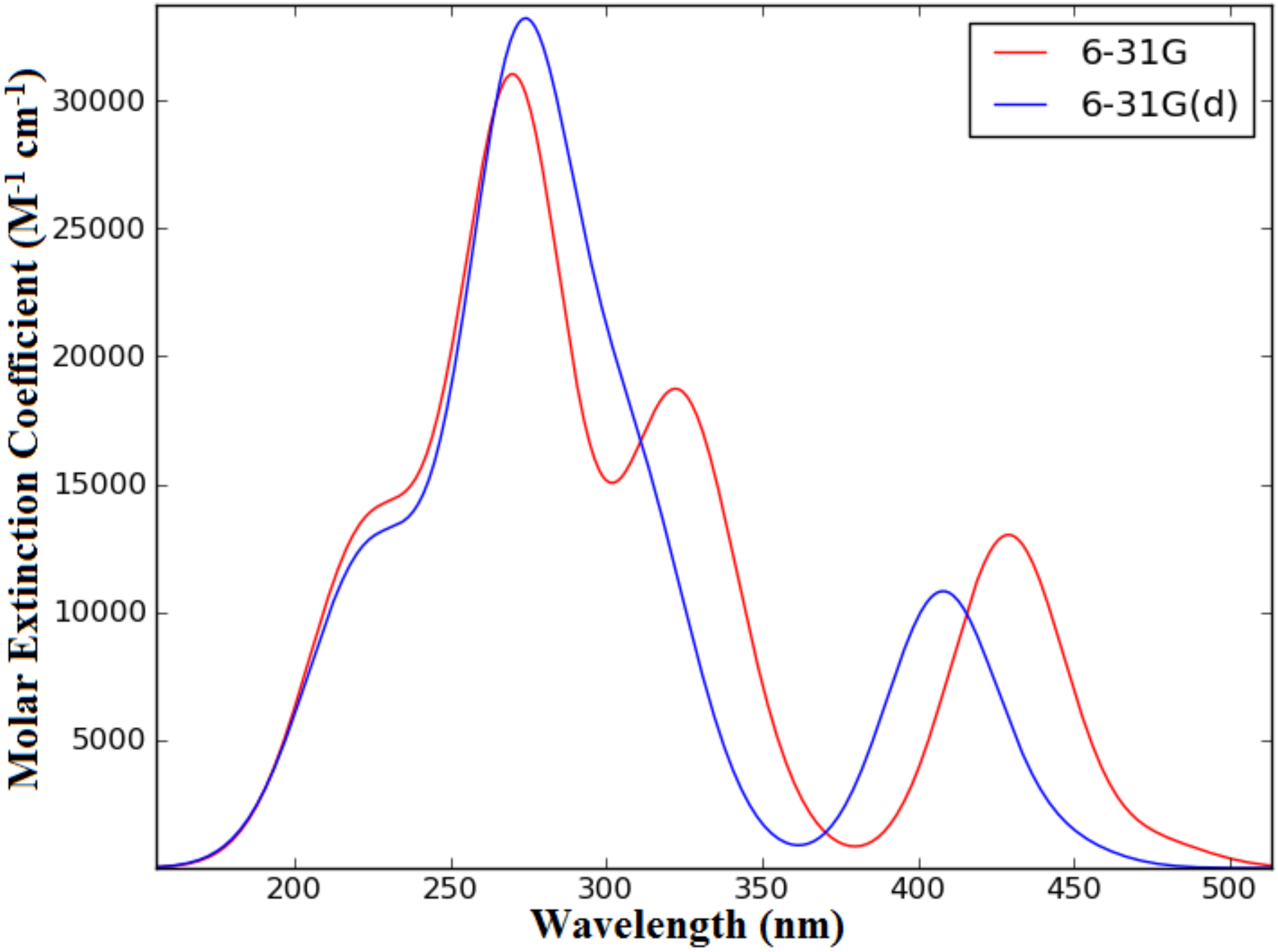}
\end{center}
{\em Trans}-[Ru(bpy)$_2$(py)$_2$]$^{2+}$
TD-B3LYP/6-31G and TD-B3LYP/6-31G(d) spectra.

% ================================================
\newpage
\section{Complex {\bf (26)}: [Ru(bpy)$_2$(pic)$_2$]$^{2+}$}
% ================================================

\begin{center}
   {\bf PDOS}
\end{center}

\begin{center}
\begin{tabular}{cc}
\includegraphics[width=0.4\textwidth]{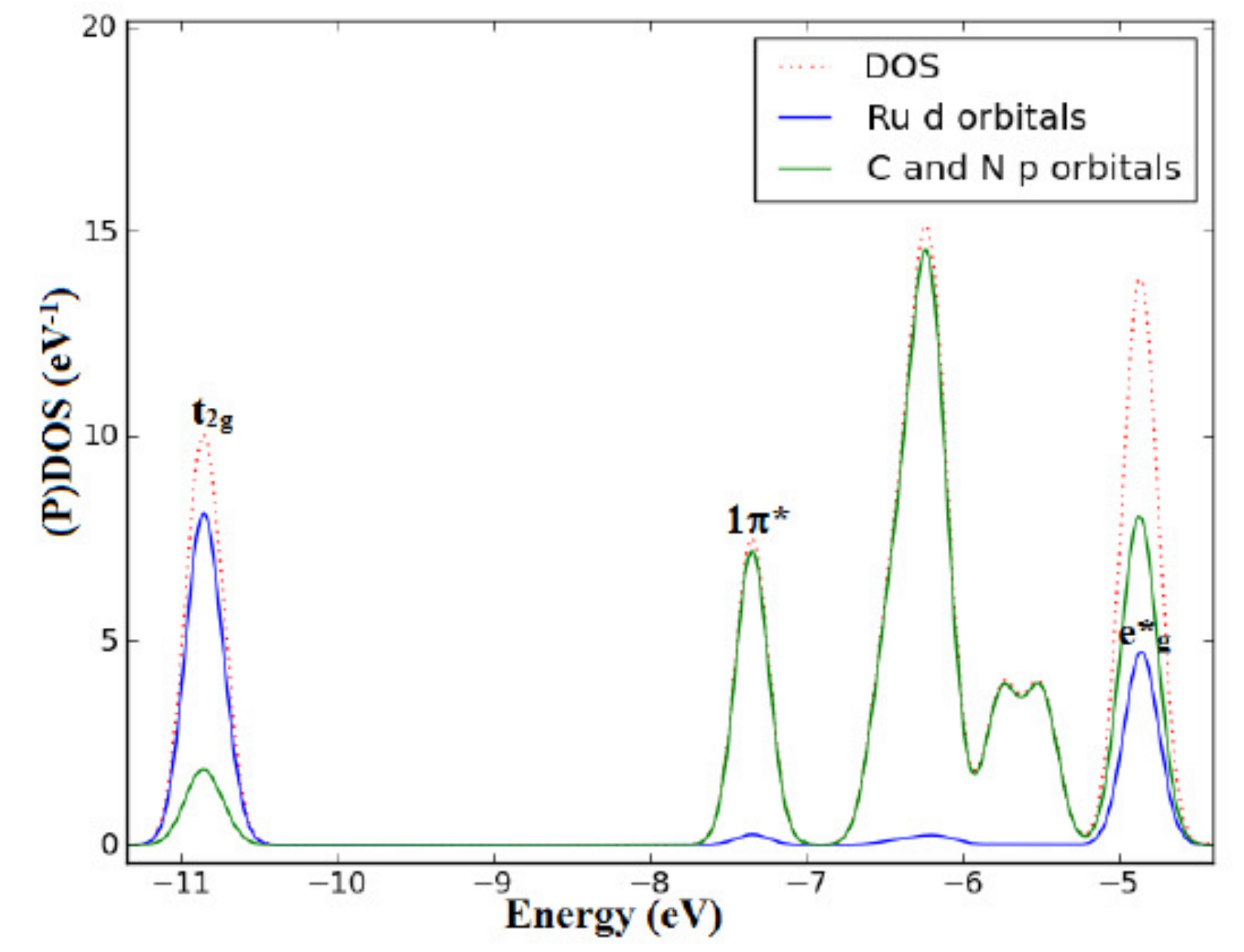} &
\includegraphics[width=0.4\textwidth]{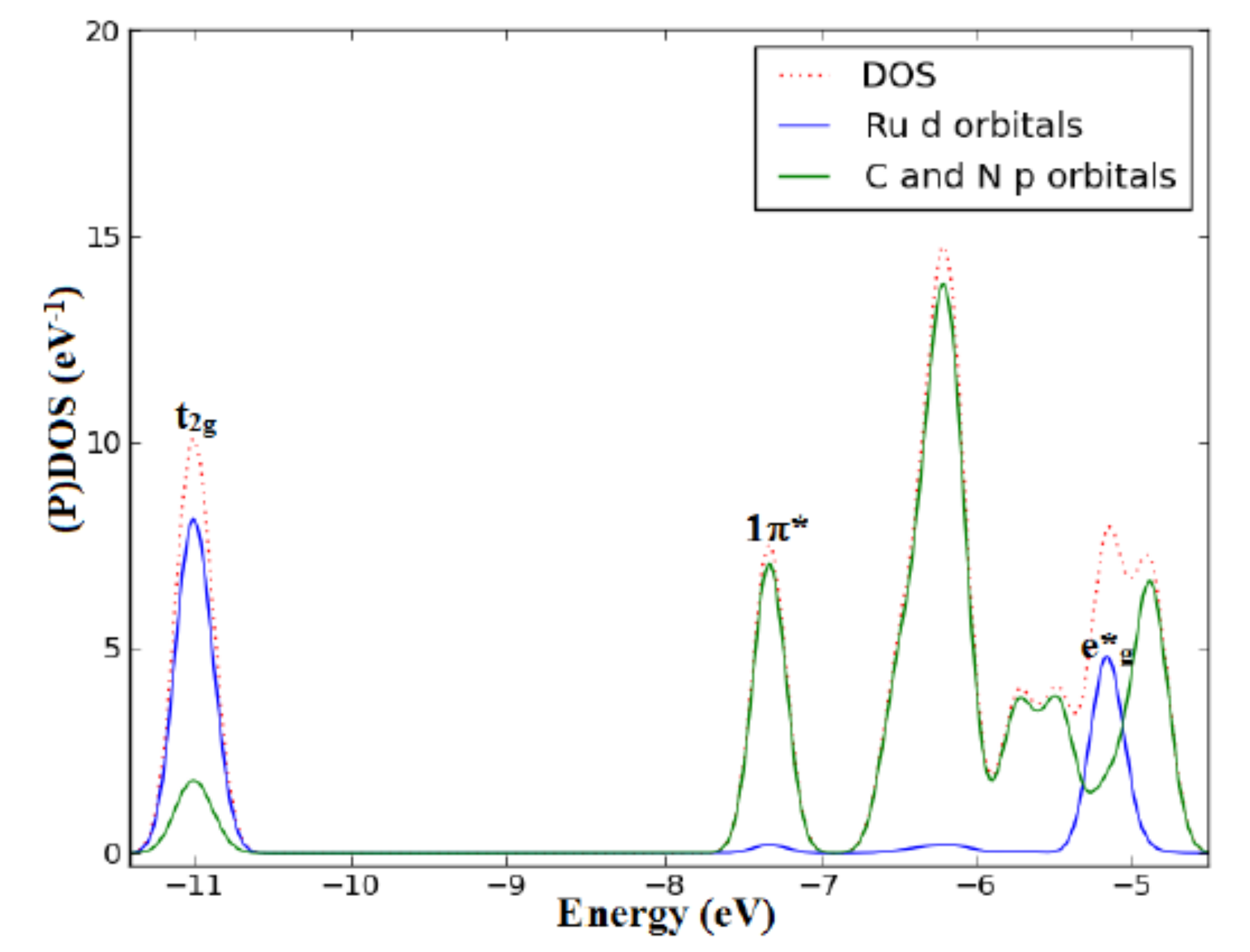} \\
B3LYP/6-31G & B3LYP/6-31G(d) \\
$\epsilon_{\text{HOMO}} = \mbox{-10.78 eV}$ & 
$\epsilon_{\text{HOMO}} = \mbox{-10.93 eV}$ 
\end{tabular}
\end{center}
Total and partial density of states of [Ru(bpy)$_2$(pic)$_2$]$^{2+}$
partitioned over Ru d orbitals and ligand C and N p orbitals. 
% for the 6-31G (left-hand side) and 6-31G* (right-hand side) basis sets.

\begin{center}
   {\bf Absorption Spectrum}
\end{center}

\begin{center}
\includegraphics[width=0.8\textwidth]{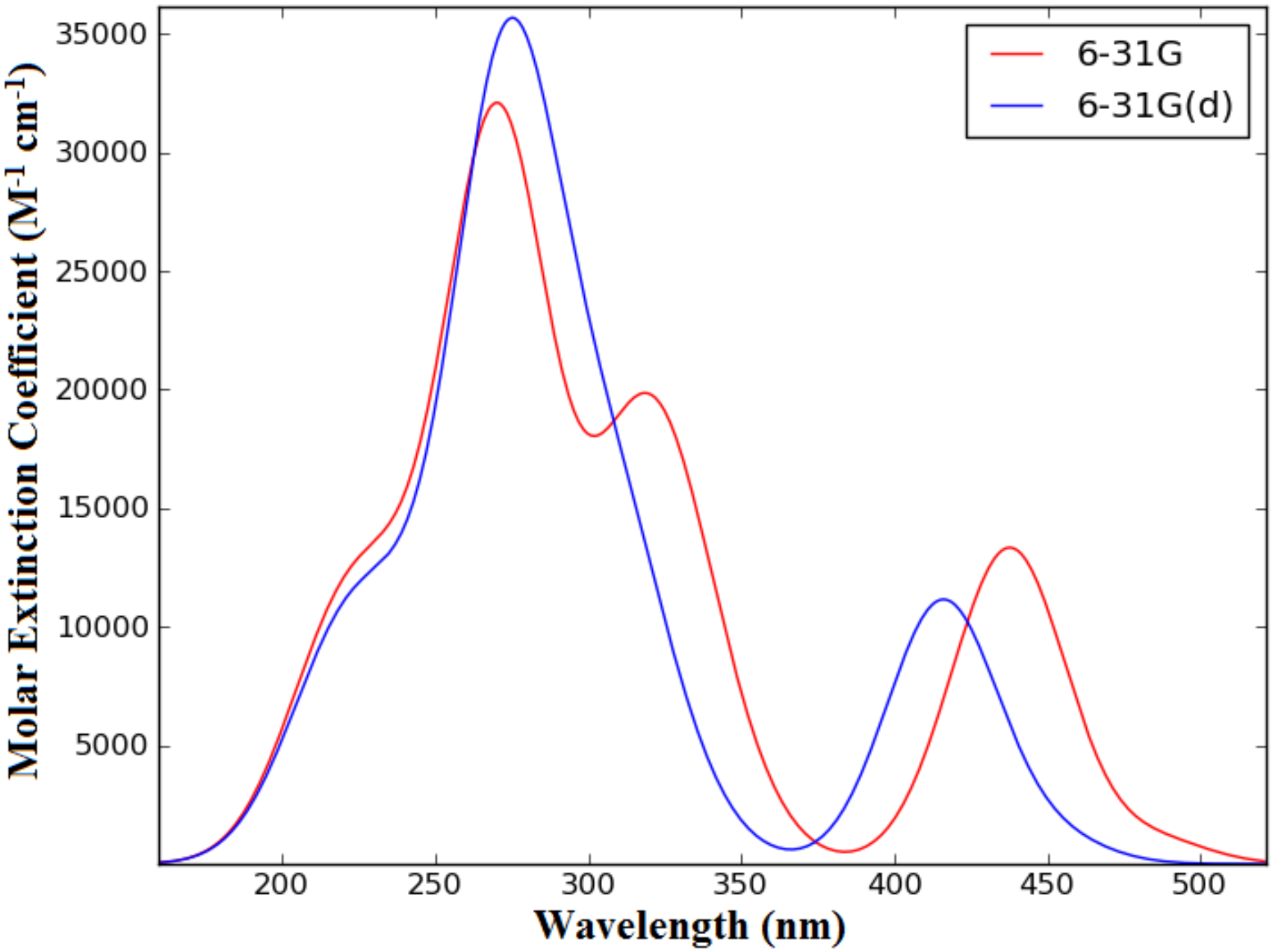}
\end{center}
[Ru(bpy)$_2$(pic)$_2$]$^{2+}$
TD-B3LYP/6-31G and TD-B3LYP/6-31G(d) spectra.

% ================================================
\newpage
\section{Complex {\bf (27)}: [Ru(bpy)$_2$(DPM)]$^{2+}$}
% ================================================

\begin{center}
   {\bf PDOS}
\end{center}

\begin{center}
\begin{tabular}{cc}
\includegraphics[width=0.4\textwidth]{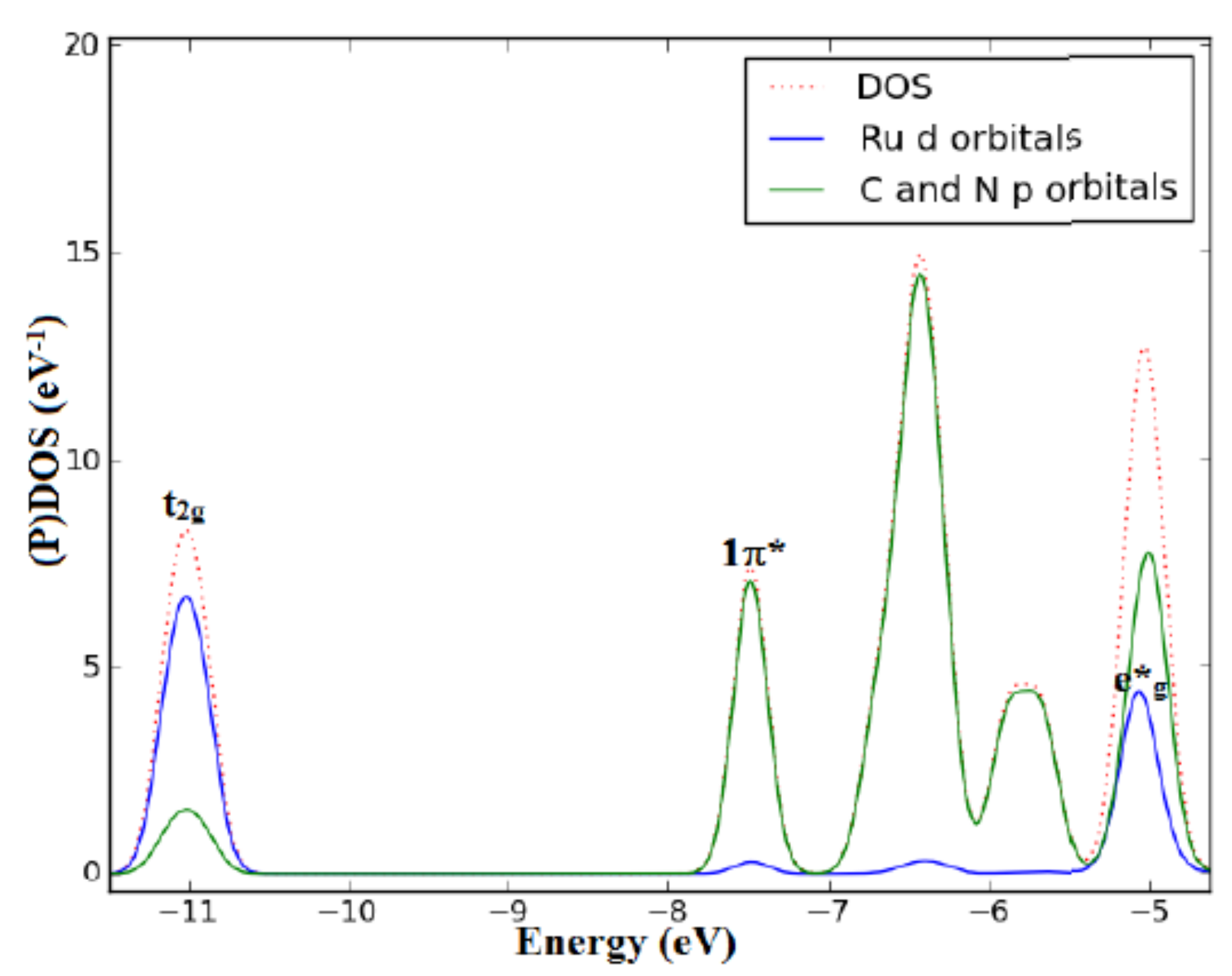} &
\includegraphics[width=0.4\textwidth]{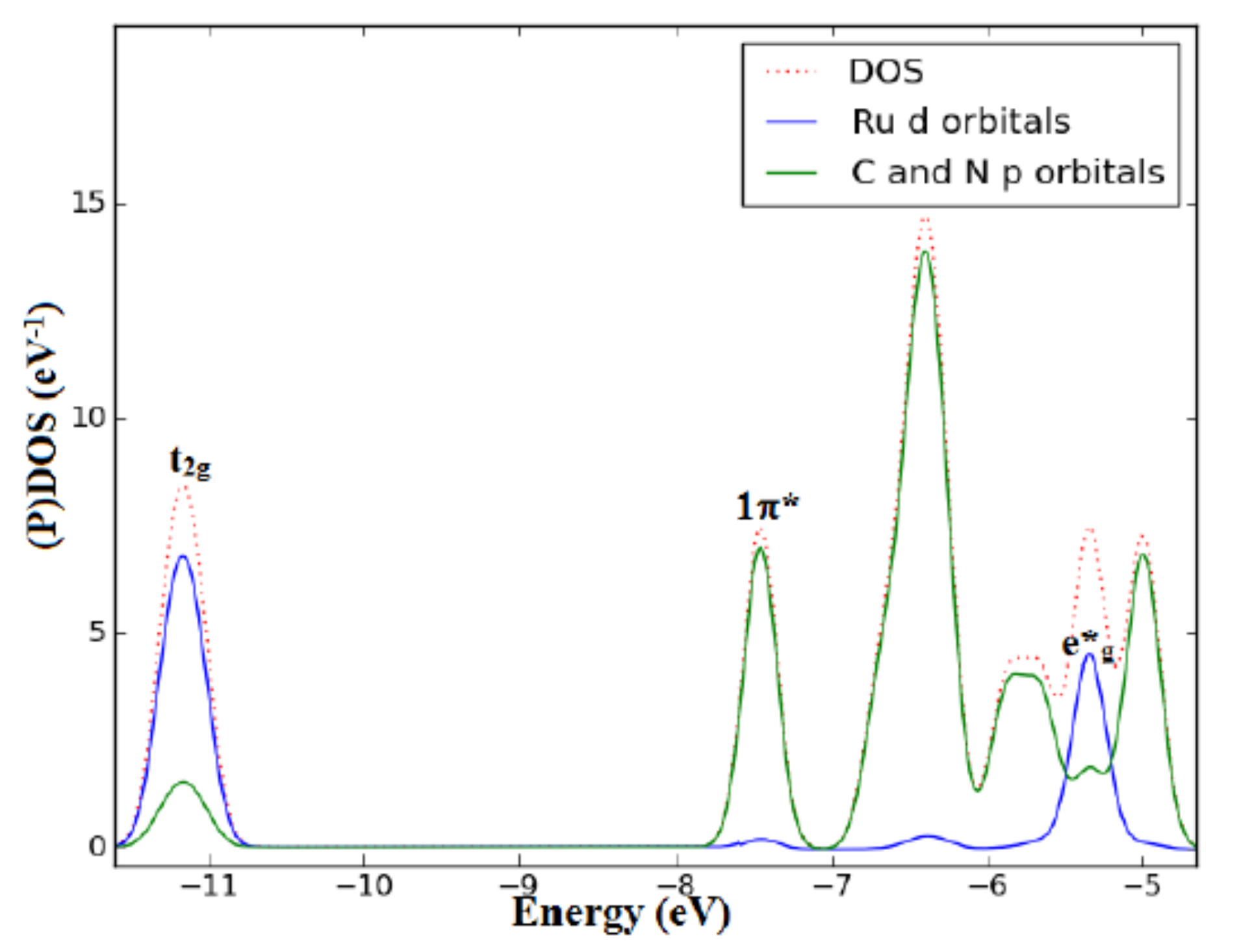} \\
B3LYP/6-31G & B3LYP/6-31G(d) \\
$\epsilon_{\text{HOMO}} = \mbox{-10.91 eV}$ & 
$\epsilon_{\text{HOMO}} = \mbox{-11.06 eV}$ 
\end{tabular}
\end{center}
Total and partial density of states of [Ru(bpy)$_2$(DPM)]$^{2+}$
partitioned over Ru d orbitals and ligand C and N p orbitals. 
% for the 6-31G (left-hand side) and 6-31G* (right-hand side) basis sets.

\begin{center}
   {\bf Absorption Spectrum}
\end{center}

\begin{center}
\includegraphics[width=0.8\textwidth]{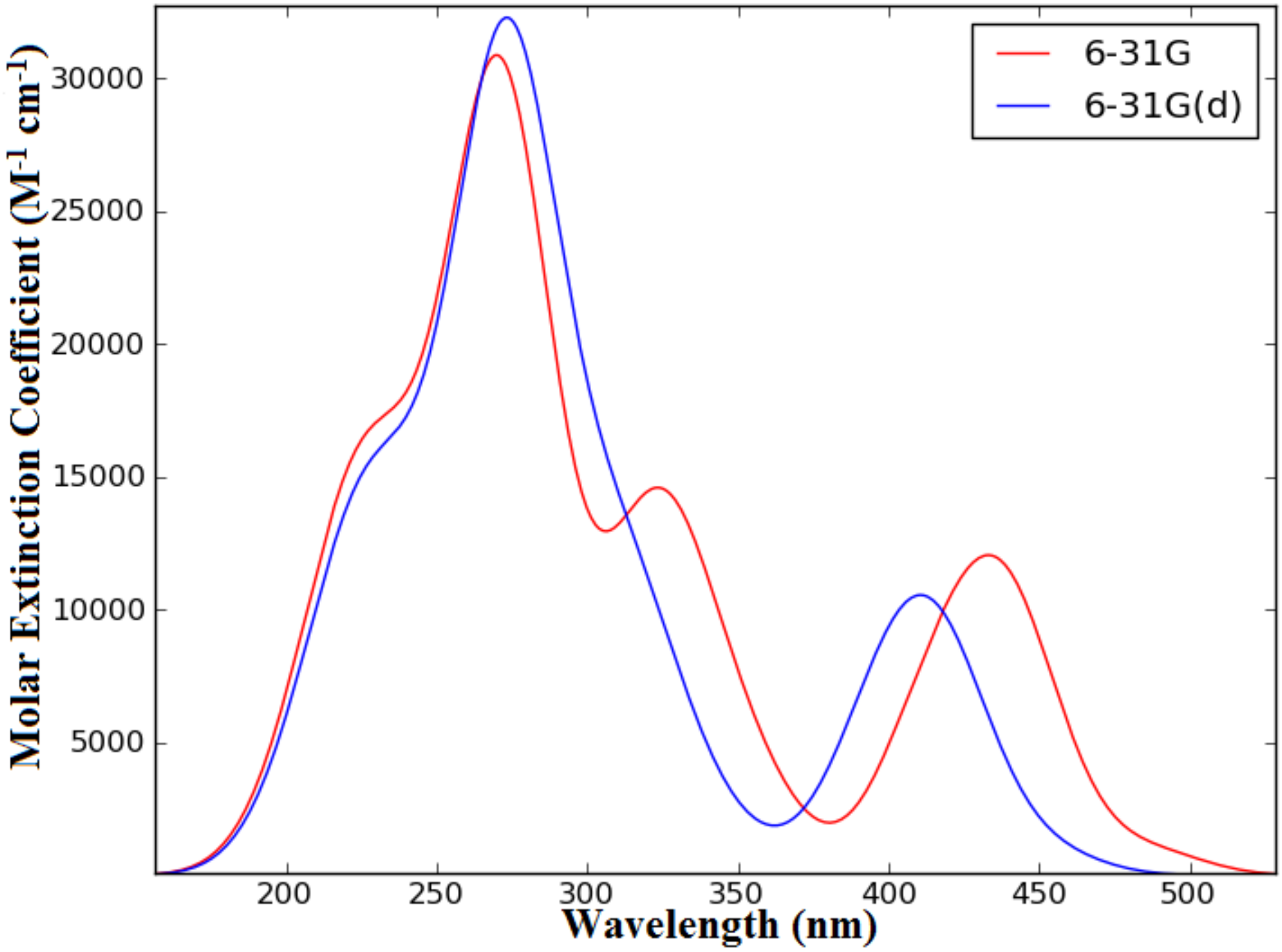}
\end{center}
[Ru(bpy)$_2$(DPM)]$^{2+}$
TD-B3LYP/6-31G and TD-B3LYP/6-31G(d) spectra.

% ================================================
\newpage
\section{Complex {\bf (28)}: [Ru(bpy)$_2$(DPE)]$^{2+}$}
% ================================================

\begin{center}
   {\bf PDOS}
\end{center}

\begin{center}
\begin{tabular}{cc}
\includegraphics[width=0.4\textwidth]{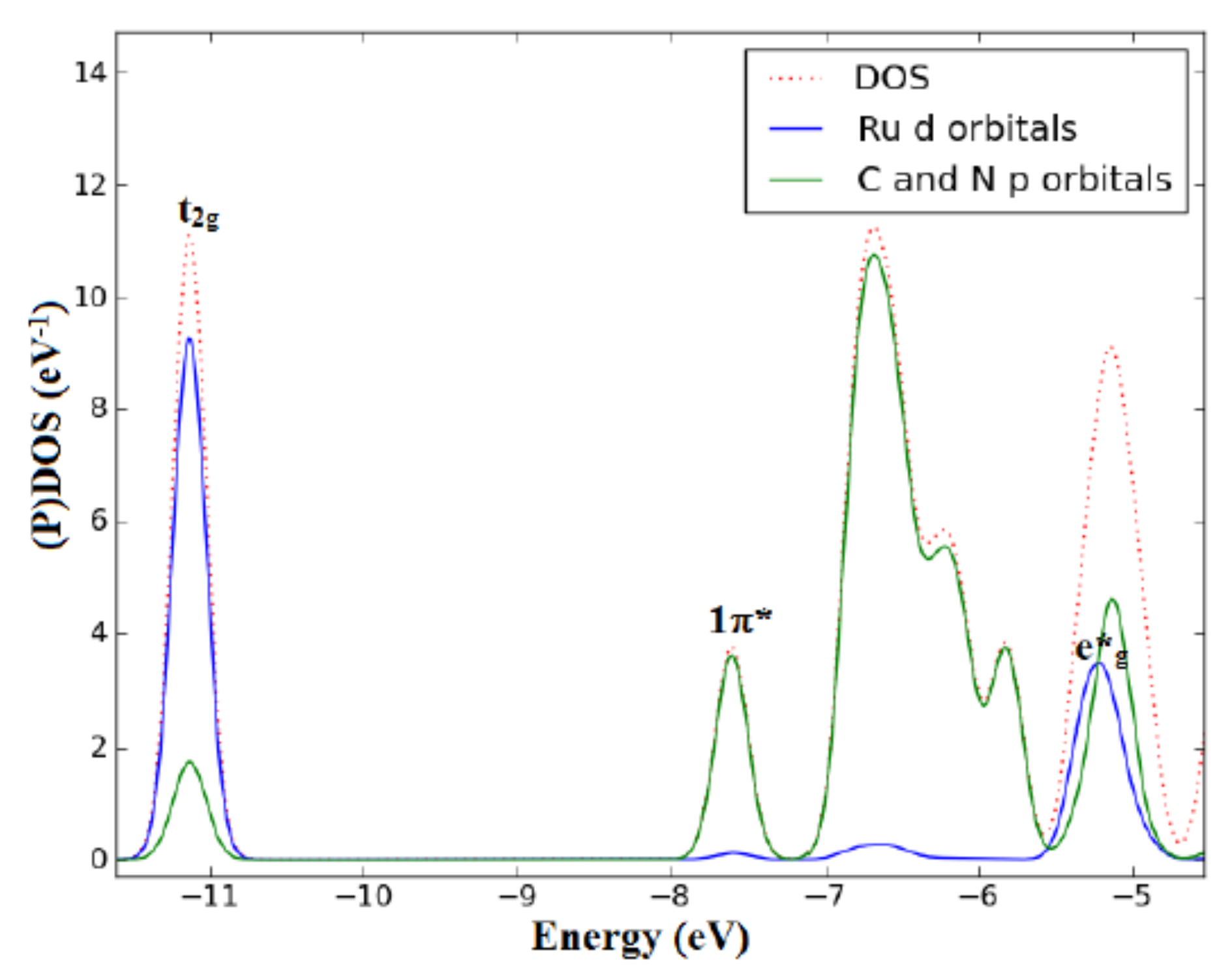} &
\includegraphics[width=0.4\textwidth]{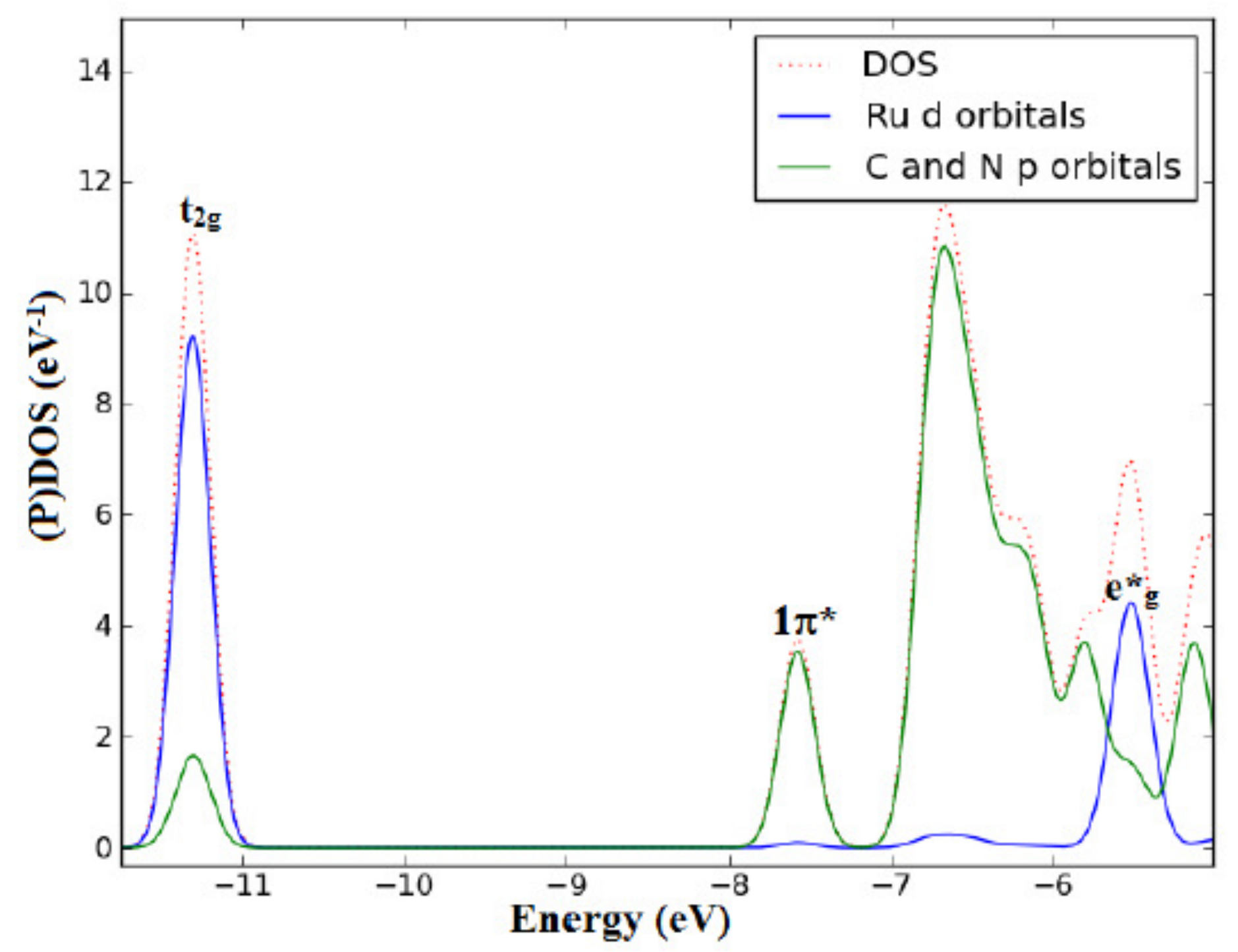} \\
B3LYP/6-31G & B3LYP/6-31G(d) \\
$\epsilon_{\text{HOMO}} = \mbox{-10.91 eV}$ & 
$\epsilon_{\text{HOMO}} = \mbox{-11.08 eV}$ 
\end{tabular}
\end{center}
Total and partial density of states of [Ru(bpy)$_2$(DPE)]$^{2+}$
partitioned over Ru d orbitals and ligand C and N p orbitals.
% for the 6-31G (left-hand side) and 6-31G* (right-hand side) basis sets.

\begin{center}
   {\bf Absorption Spectrum}
\end{center}

\begin{center}
\includegraphics[width=0.8\textwidth]{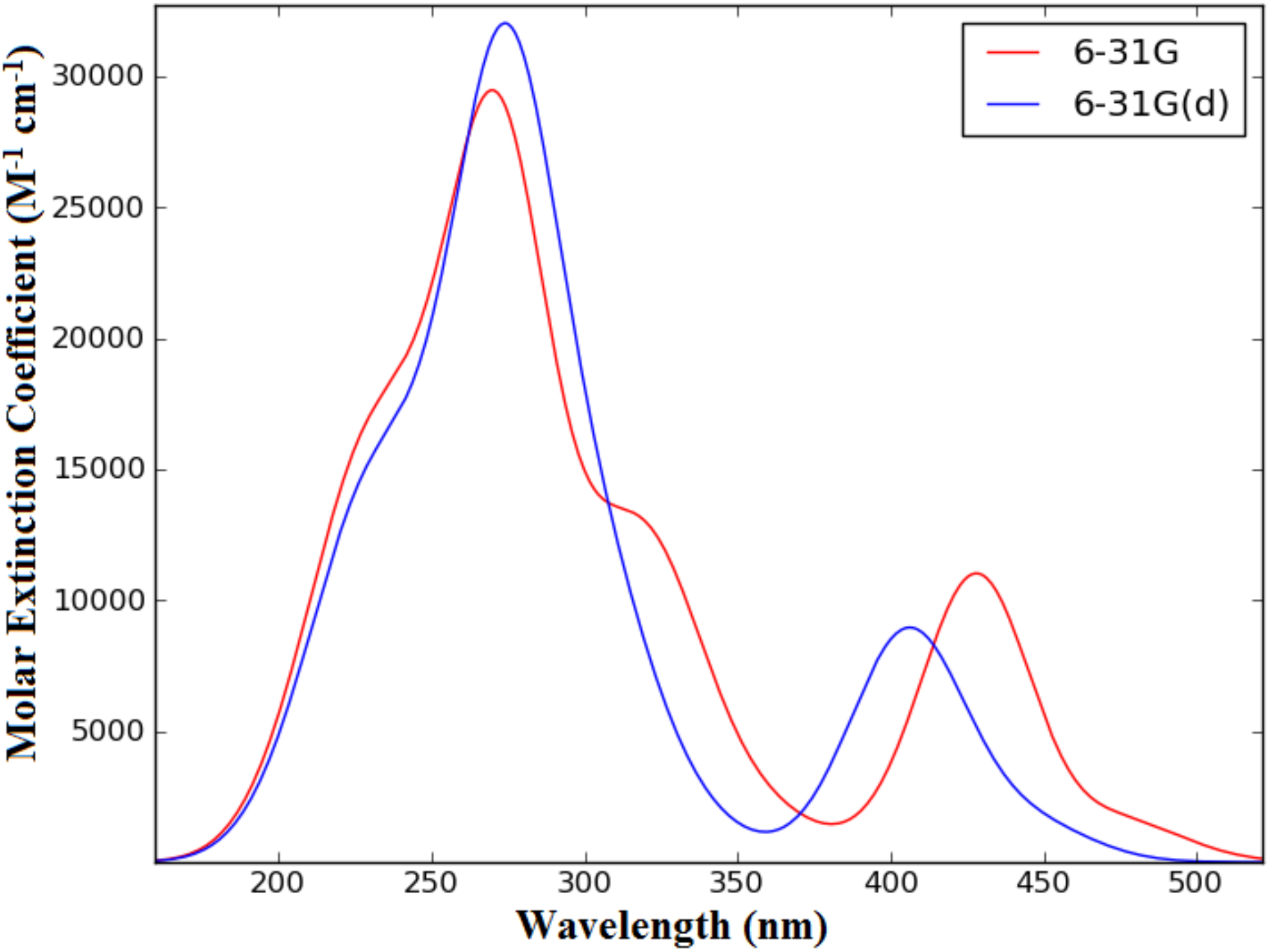}
\end{center}
[Ru(bpy)$_2$(DPE)]$^{2+}$
TD-B3LYP/6-31G and TD-B3LYP/6-31G(d) spectra.

% ================================================
\newpage
\section{Complex {\bf (29)}: [Ru(bpy)$_2$(PimH)]$^{2+}$}
% ================================================

\begin{center}
   {\bf PDOS}
\end{center}

\begin{center}
\begin{tabular}{cc}
\includegraphics[width=0.4\textwidth]{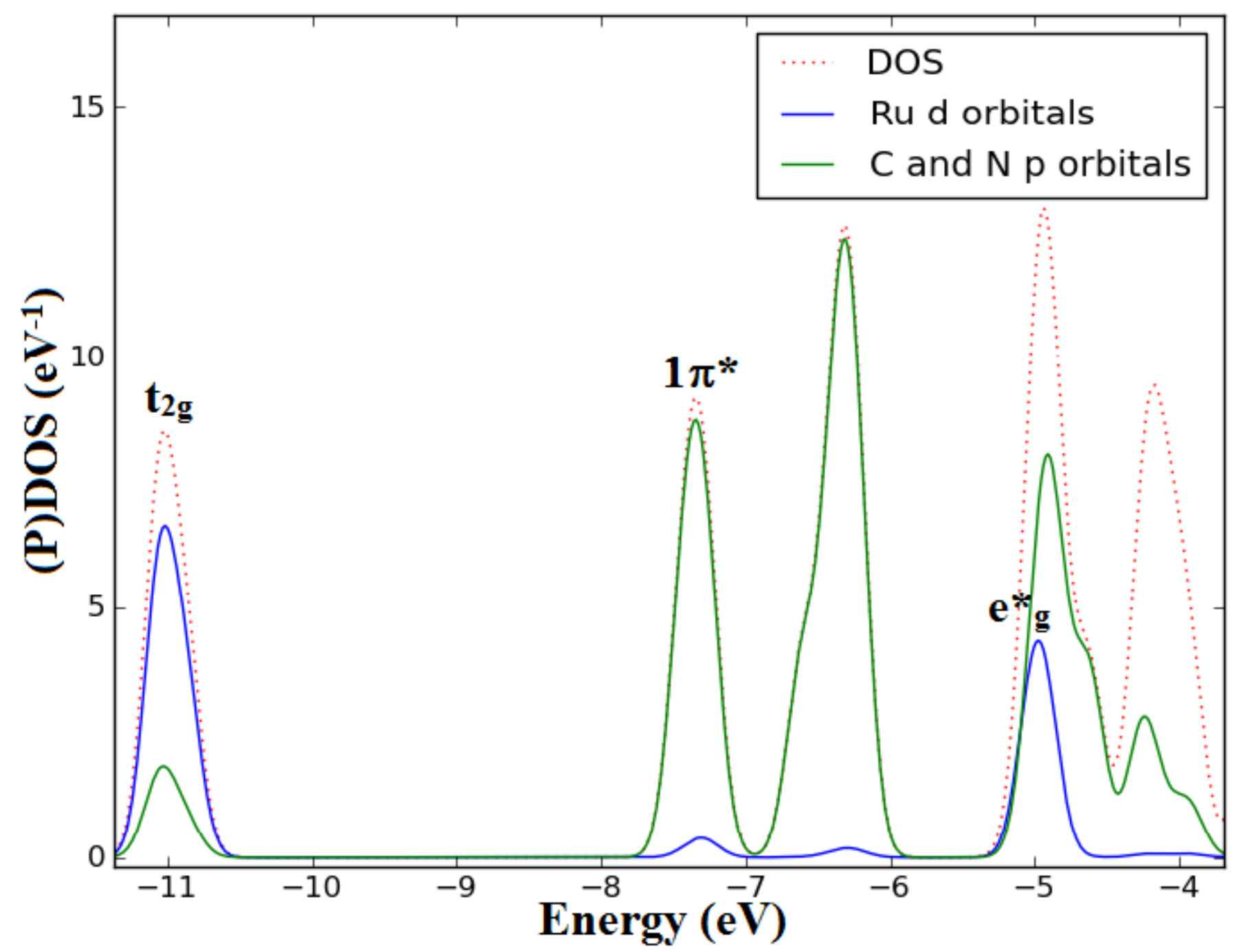} &
\includegraphics[width=0.4\textwidth]{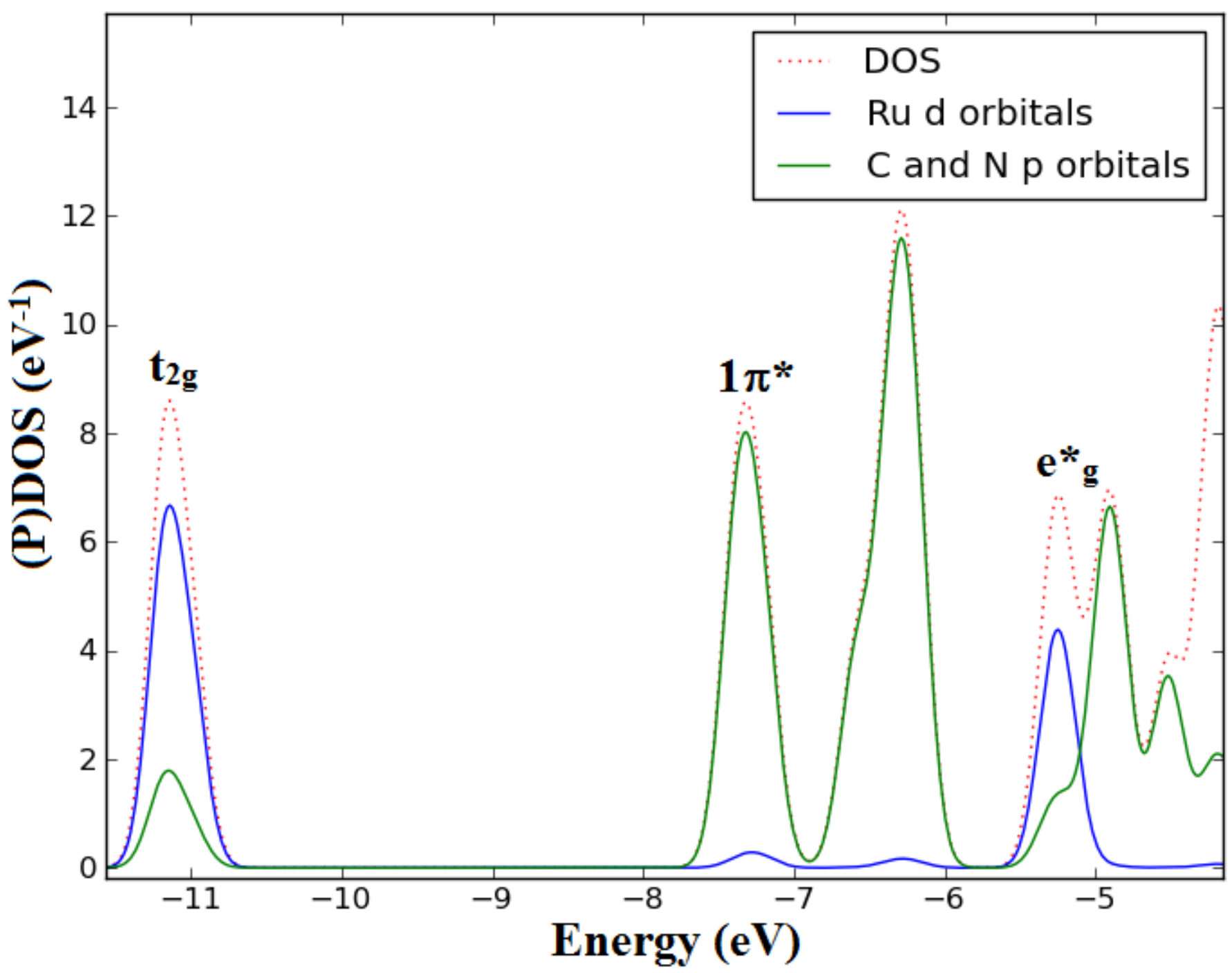} \\
B3LYP/6-31G & B3LYP/6-31G(d) \\
$\epsilon_{\text{HOMO}} = \mbox{-10.87 eV}$ & 
$\epsilon_{\text{HOMO}} = \mbox{-11.00 eV}$ 
\end{tabular}
\end{center}
Total and partial density of states of [Ru(bpy)$_2$(PimH)]$^{_2+}$ 
partitioned over Ru d orbitals and ligand C and N p orbitals. 
% for the 6-31G (left-hand side) and 6-31G(d) (right-hand side) basis sets.

\begin{center}
   {\bf Absorption Spectrum}
\end{center}

\begin{center}
\includegraphics[width=0.8\textwidth]{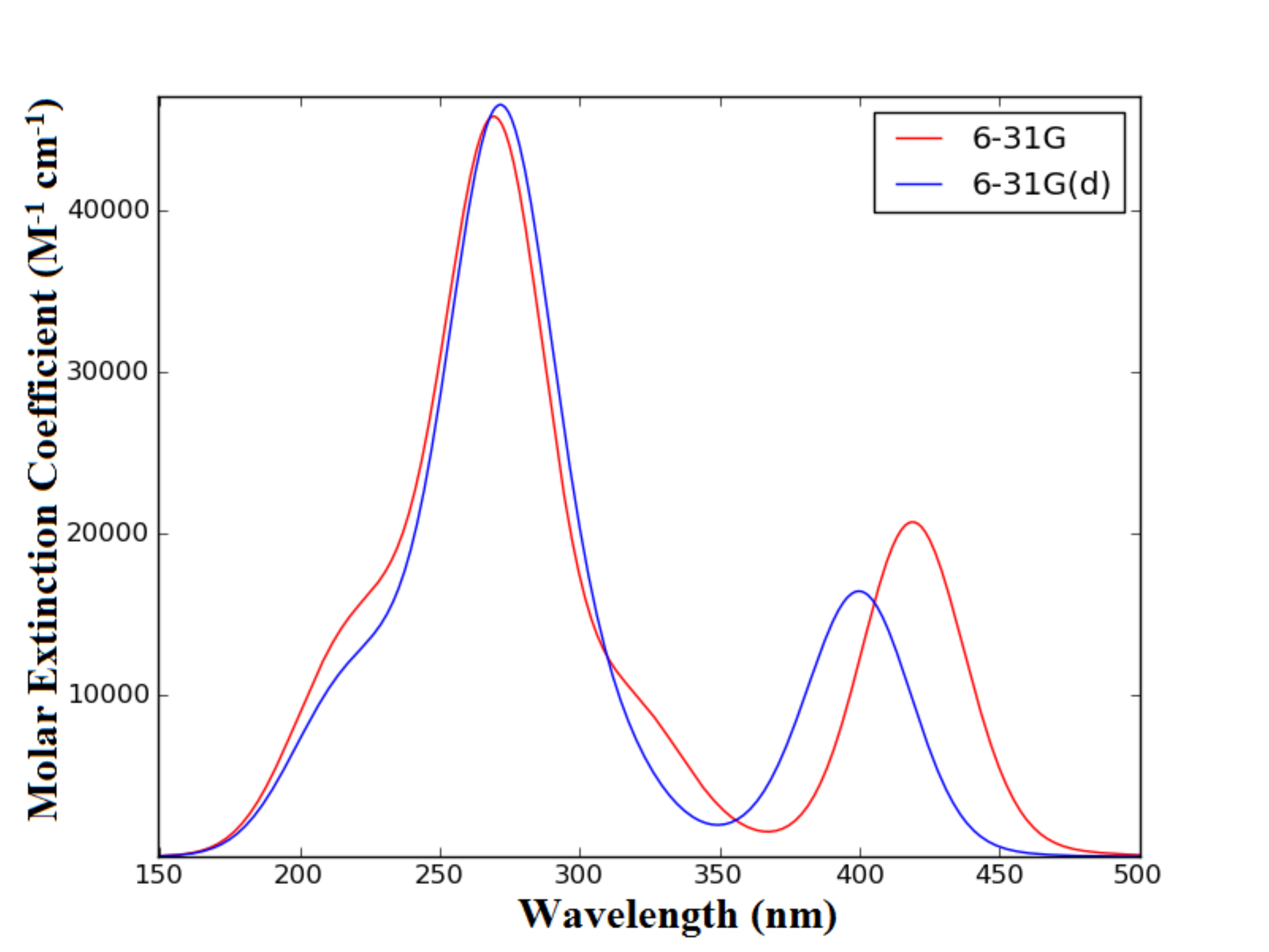}
\end{center}
[Ru(bpy)$_{2}$(PimH)]$^{2+}$
TD-B3LYP/6-31G and TD-B3LYP/6-31G(d) spectra.

% ================================================
\newpage
\section{Complex {\bf (30)}: [Ru(bpy)$_2$(PBzimH)]$^{2+}$}
% ================================================

\begin{center}
   {\bf PDOS}
\end{center}

\begin{center}
\begin{tabular}{cc}
\includegraphics[width=0.4\textwidth]{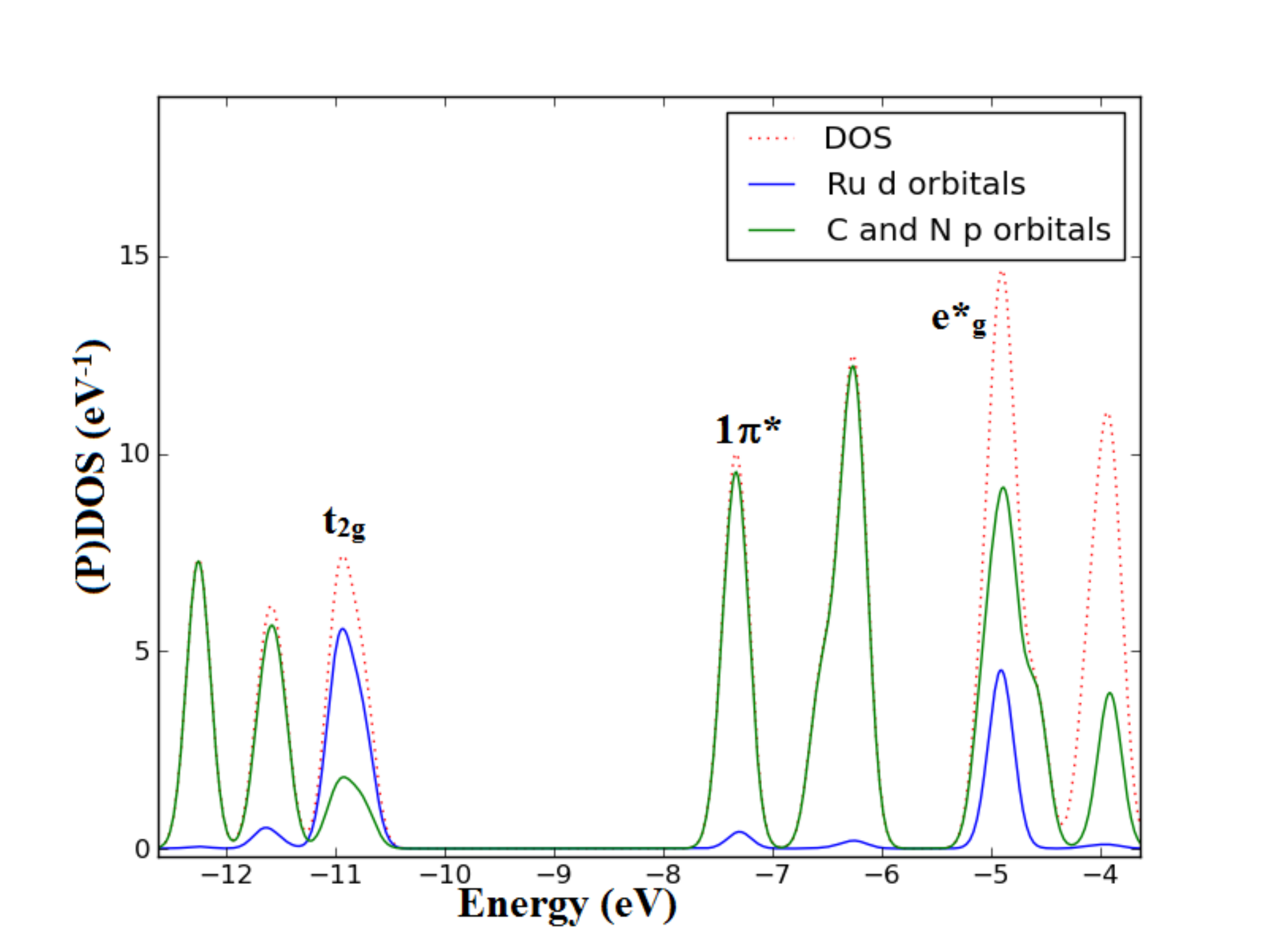} &
\includegraphics[width=0.4\textwidth]{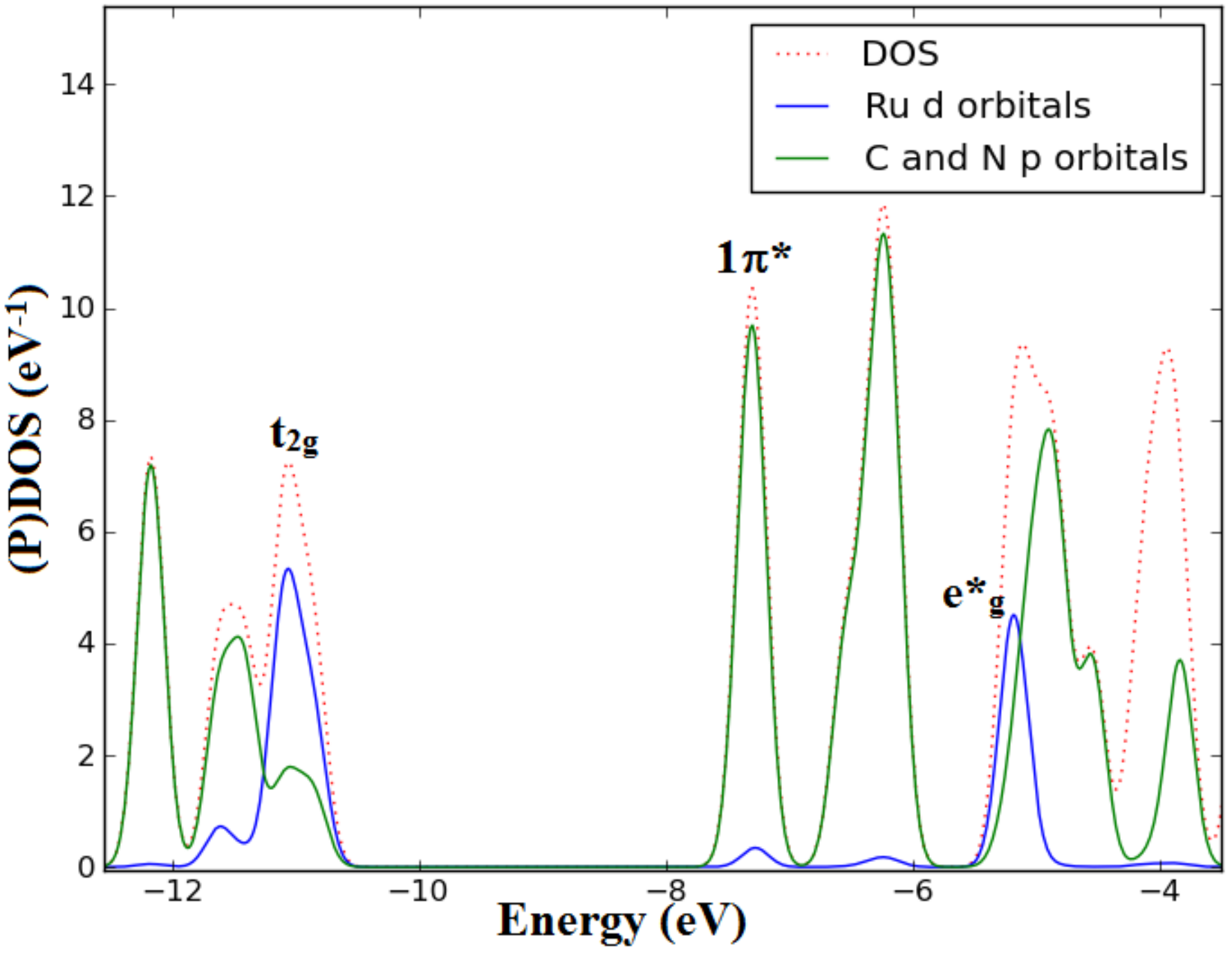} \\
B3LYP/6-31G & B3LYP/6-31G(d) \\
$\epsilon_{\text{HOMO}} = \mbox{-10.75 eV}$ & 
$\epsilon_{\text{HOMO}} = \mbox{-10.86 eV}$ 
\end{tabular}
\end{center}
Total and partial density of states of [Ru(bpy)$_{2}$(PBzimH)]$^{2+}$ 
partitioned over Ru d orbitals and ligand C and N p orbitals. 
% for the 6-31G (left-hand side) and 6-31G(d) (right-hand side) basis sets.

\begin{center}
   {\bf Absorption Spectrum}
\end{center}

\begin{center}
\includegraphics[width=0.8\textwidth]{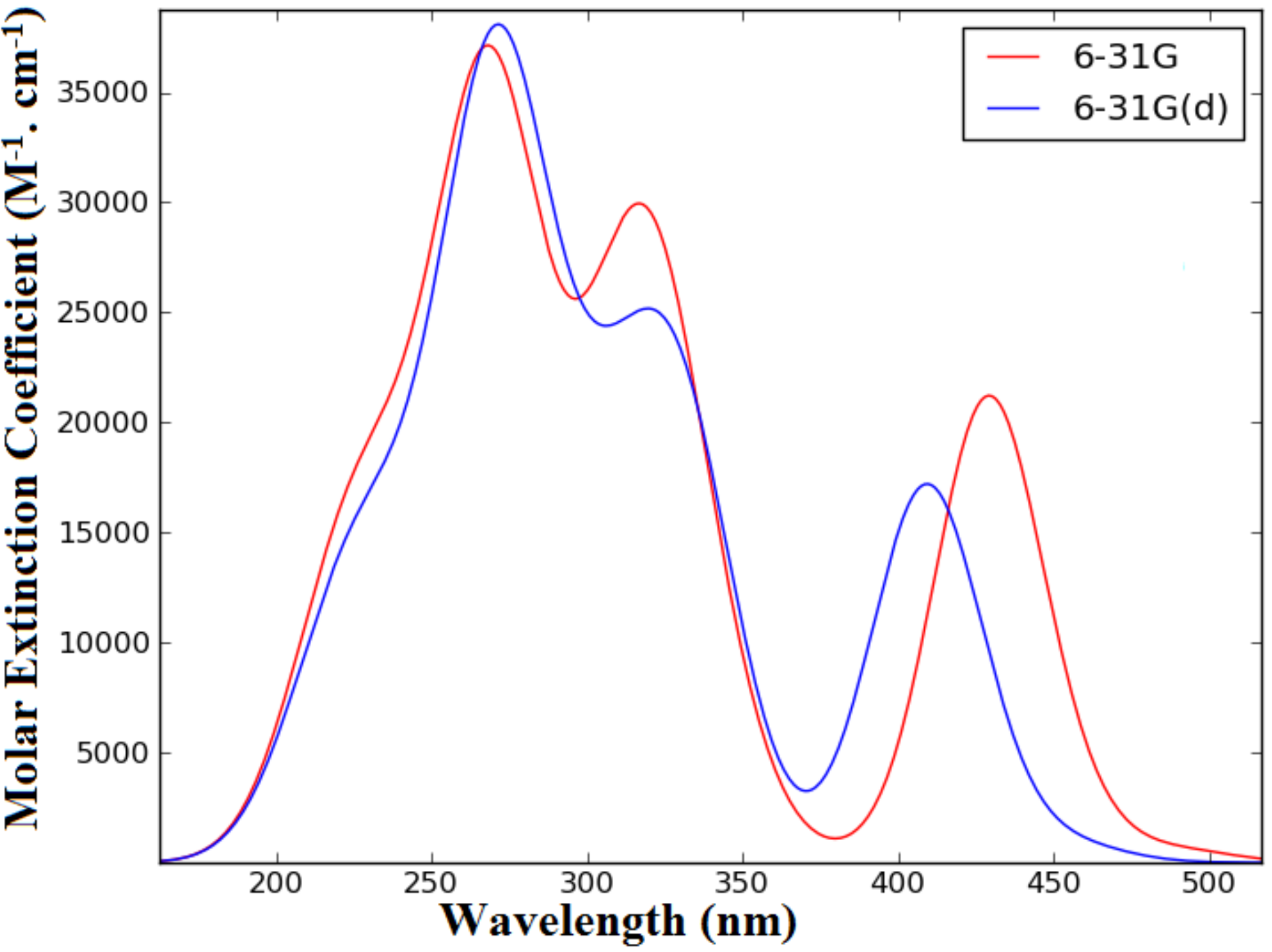}
\end{center}
[Ru(bpy)$_{2}$(PBzimH)]$^{2+}$
TD-B3LYP/6-31G and TD-B3LYP/6-31G(d) spectra.

% ================================================
\newpage
\section{Complex {\bf (31)}: [Ru(bpy)$_2$(biimH$_2$)]$^{2+}$}
% ================================================

\begin{center}
   {\bf PDOS}
\end{center}

\begin{center}
\begin{tabular}{cc}
\includegraphics[width=0.4\textwidth]{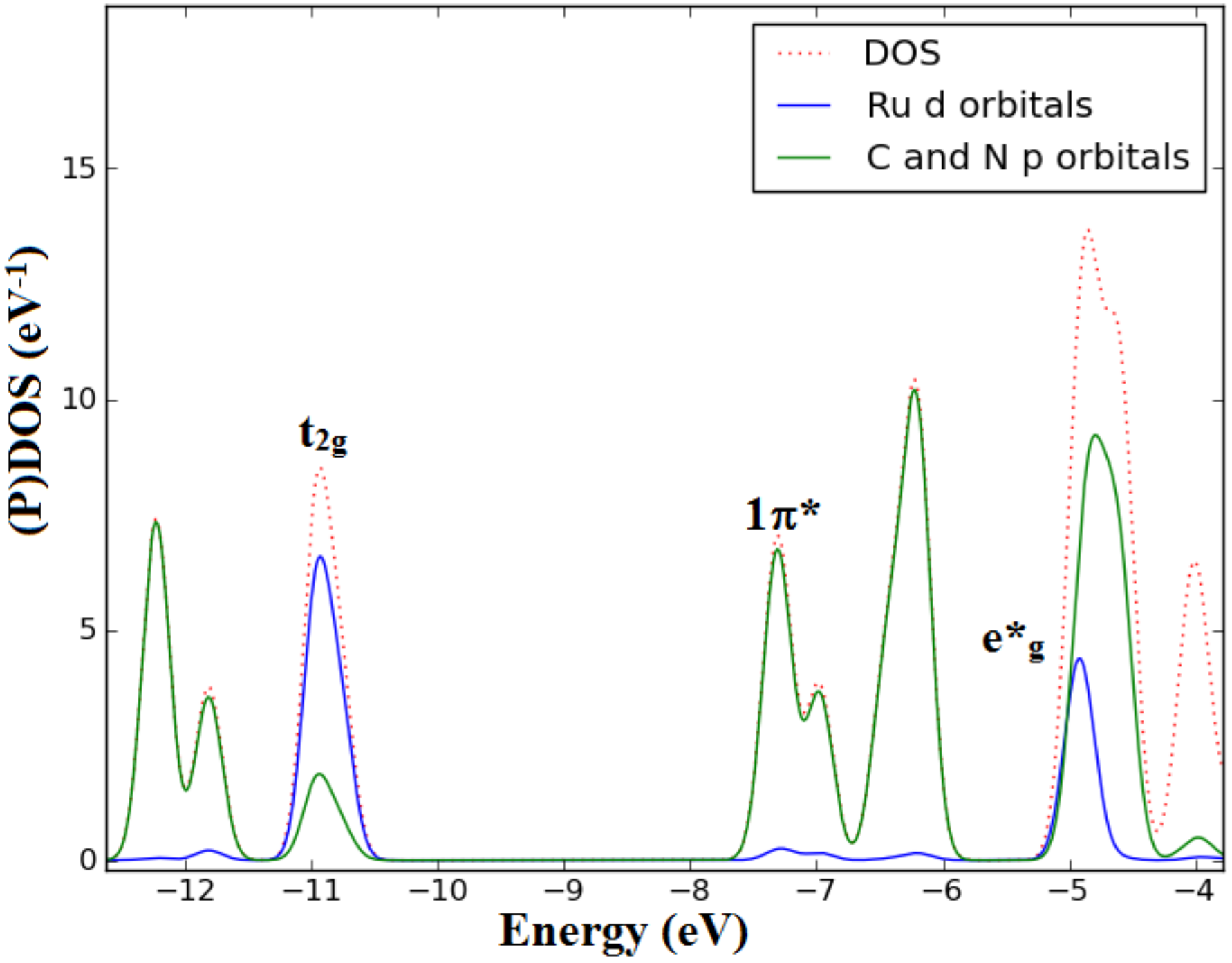} &
\includegraphics[width=0.4\textwidth]{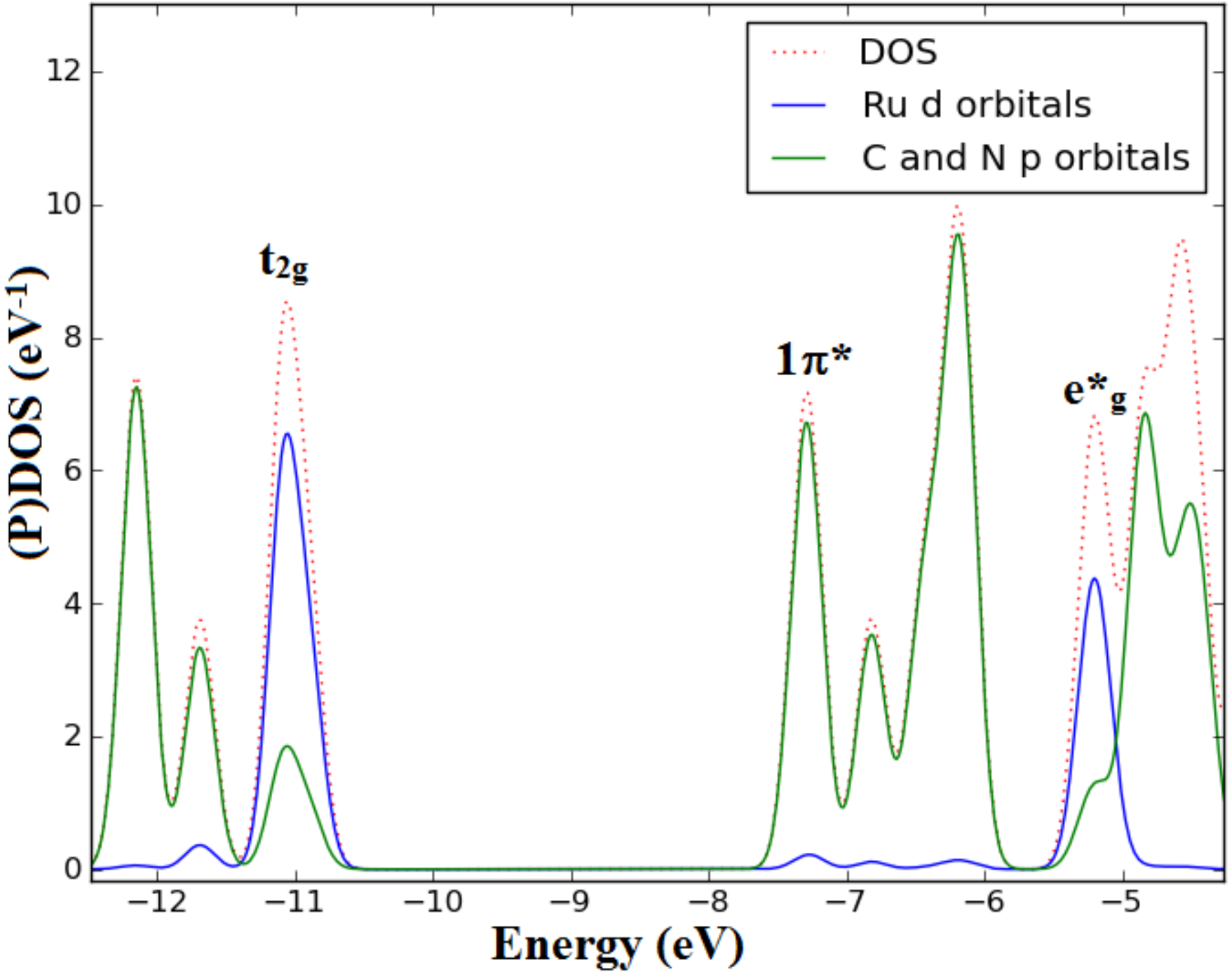} \\
B3LYP/6-31G & B3LYP/6-31G(d) \\
$\epsilon_{\text{HOMO}} = \mbox{-10.77 eV}$ & 
$\epsilon_{\text{HOMO}} = \mbox{-10.90 eV}$ 
\end{tabular}
\end{center}
Total and partial density of states of [Ru(bpy)$_{2}$(biimH$_2$)]$^{2+}$ 
partitioned over Ru d orbitals and ligand C and N p orbitals. 
% for the 6-31G (left-hand side) and 6-31G(d) (right-hand side) basis sets.

\begin{center}
   {\bf Absorption Spectrum}
\end{center}

\begin{center}
\includegraphics[width=0.8\textwidth]{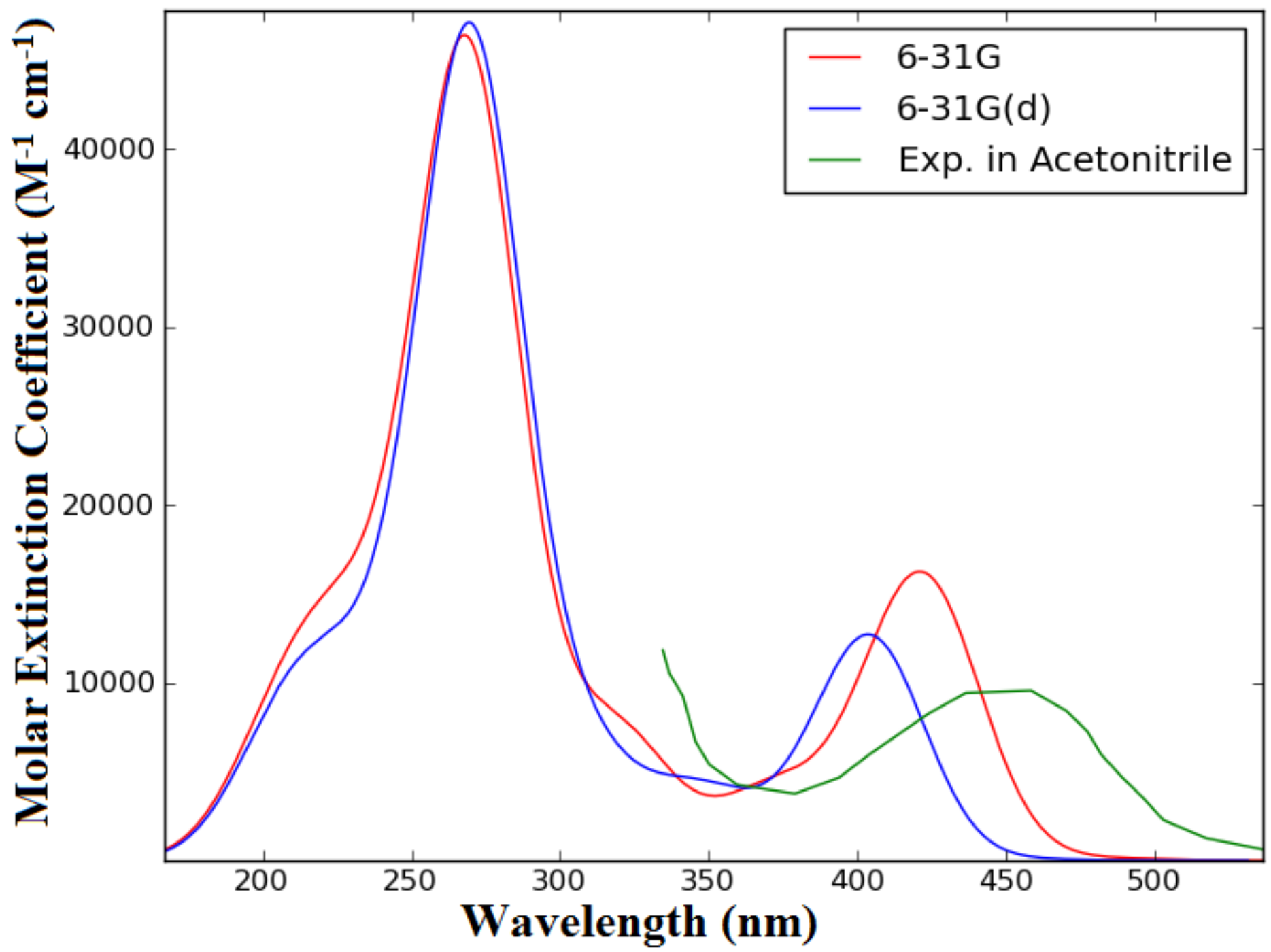}
\end{center}
[Ru(bpy)$_{2}$(biimH$_{2}$)]$^{2+}$
TD-B3LYP/6-31G and TD-B3LYP/6-31G(d) spectra.

% ================================================
\newpage
\section{Complex {\bf (32)}: [Ru(bpy)$_2$(BiBzimH$_2$)]$^{2+}$}
% ================================================

\begin{center}
   {\bf PDOS}
\end{center}

\begin{center}
\begin{tabular}{cc}
\includegraphics[width=0.4\textwidth]{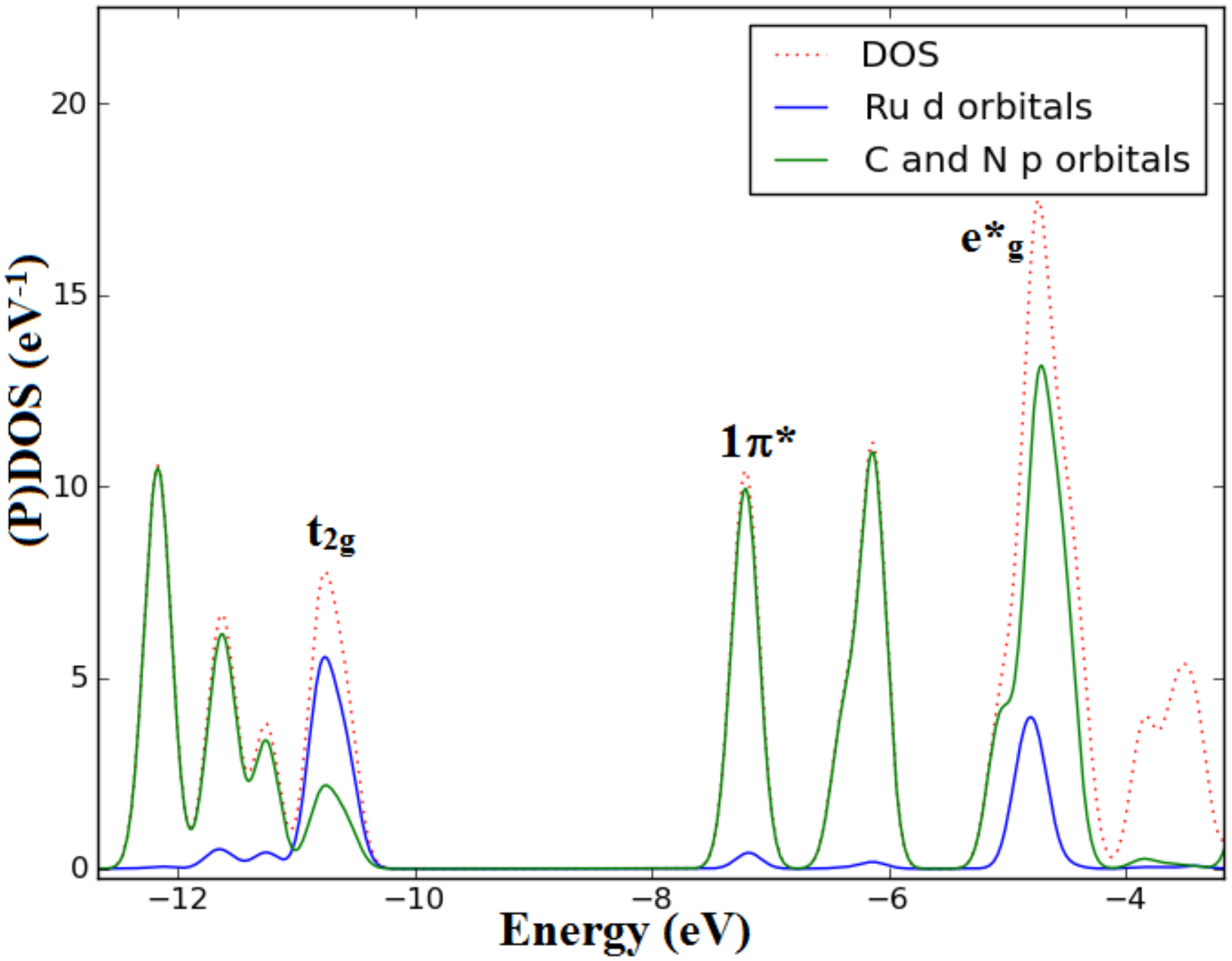} &
\includegraphics[width=0.4\textwidth]{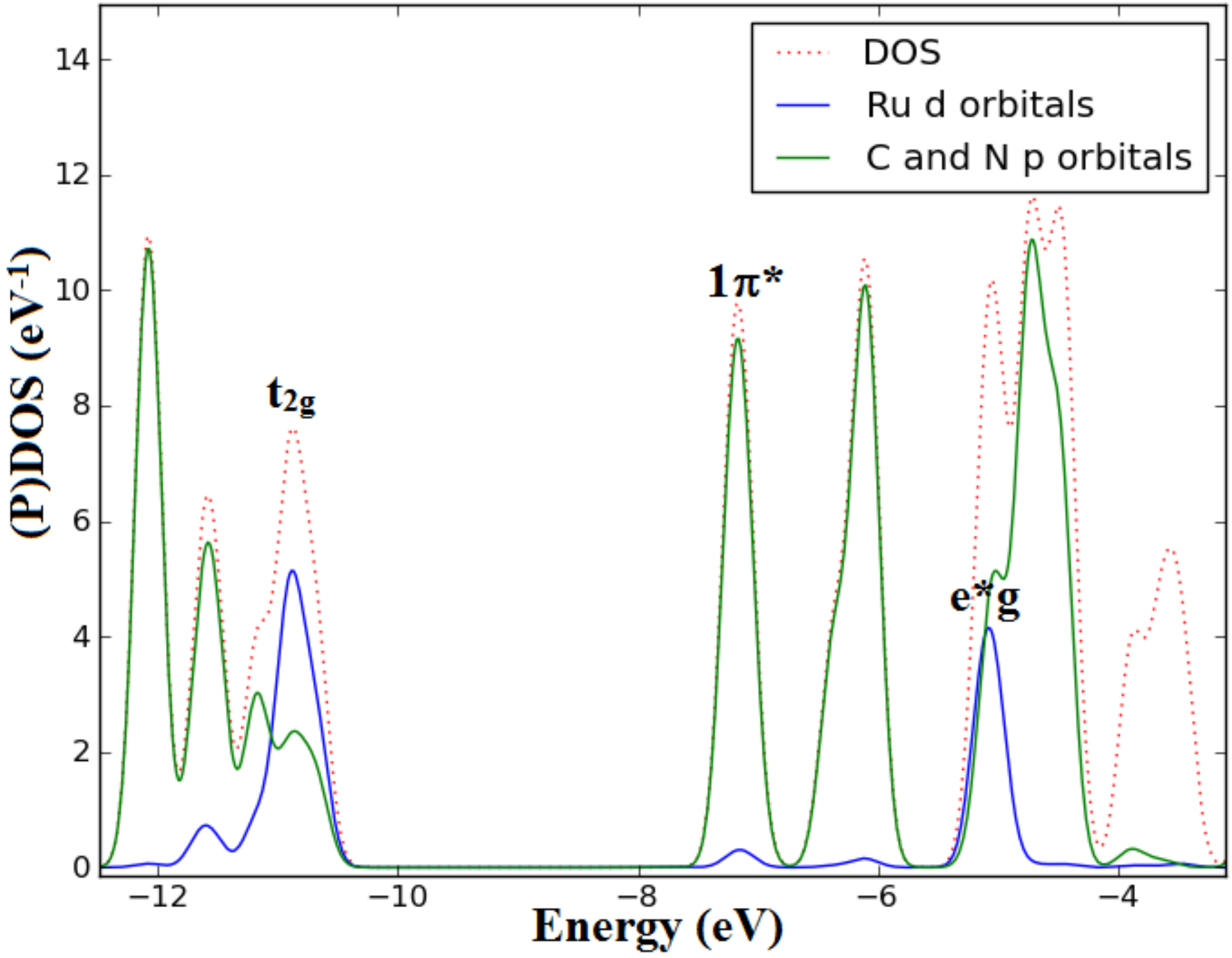} \\
B3LYP/6-31G & B3LYP/6-31G(d) \\
$\epsilon_{\text{HOMO}} = \mbox{-10.58 eV}$ & 
$\epsilon_{\text{HOMO}} = \mbox{-10.68 eV}$ 
\end{tabular}
\end{center}
Total and partial density of states of [Ru(bpy)$_{2}$(BiBzimH$_2$)]$^{2+}$ 
partitioned over Ru d orbitals and ligand C and N p orbitals. 
% for the 6-31G (left-hand side) and 6-31G(d) (right-hand side) basis sets.

\begin{center}
   {\bf Absorption Spectrum}
\end{center}

\begin{center}
\includegraphics[width=0.8\textwidth]{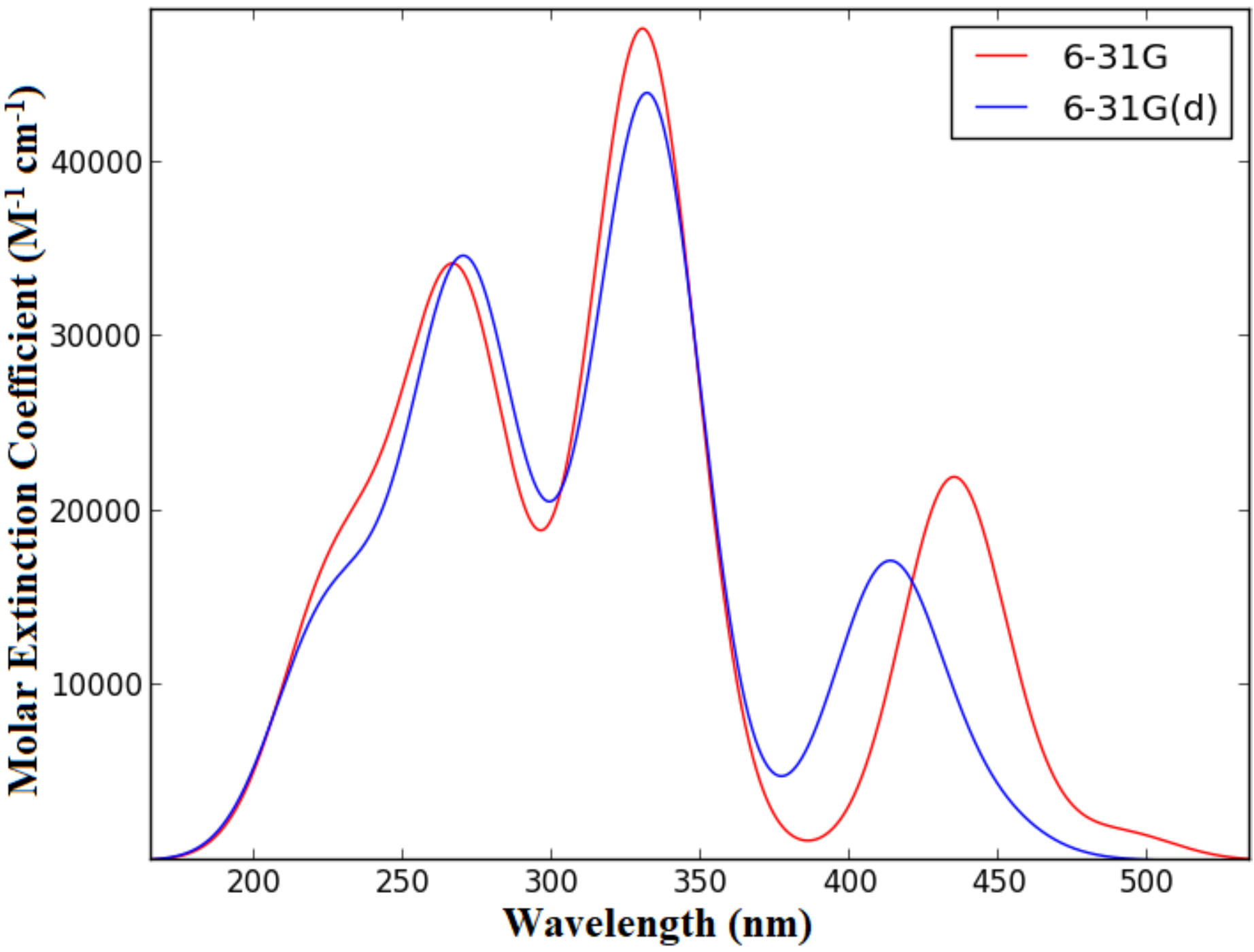}
\end{center}
[Ru(bpy)$_{2}$(BiBzimH$_{2}$)]$^{2+}$
TD-B3LYP/6-31G, TD-B3LYP/6-31G(d), and experimental spectra.
Experimental curve measured at room temperature in acetonitrile\cite{LNZ+13}.

% % ================================================
% \newpage
% \section{Complex {\bf (33)}: [Ru(bpy)$_2$(NPP)]$^{2+}$}
% % ================================================
% 
% {\color{magenta} \sf The geometry optimzation was unsuccessful for this complex.}
% 
% % \begin{center}
% %    {\bf PDOS}
% % \end{center}
% 
% % \begin{center}
% % \includegraphics[width=0.4\textwidth]{graphics1/framedquestionmark.pdf}
% % \includegraphics[width=0.4\textwidth]{graphics1/framedquestionmark.pdf}
% % \end{center}
% % {\color{red} Do we have this?}
%  
% % \begin{center}
% %    {\bf Absorption Spectrum}
% % \end{center}
% 
% % \begin{center}
% % \includegraphics[width=0.4\textwidth]{graphics1/framedquestionmark.pdf}
% % \end{center}
% % {\color{red} Do we have this?}

% ================================================
\newpage
\section{Complex {\bf (34)}: [Ru(bpy)$_2$(piq)]$^{2+}$}
% ================================================

\begin{center}
   {\bf PDOS}
\end{center}

\begin{center}
\begin{tabular}{cc}
\includegraphics[width=0.4\textwidth]{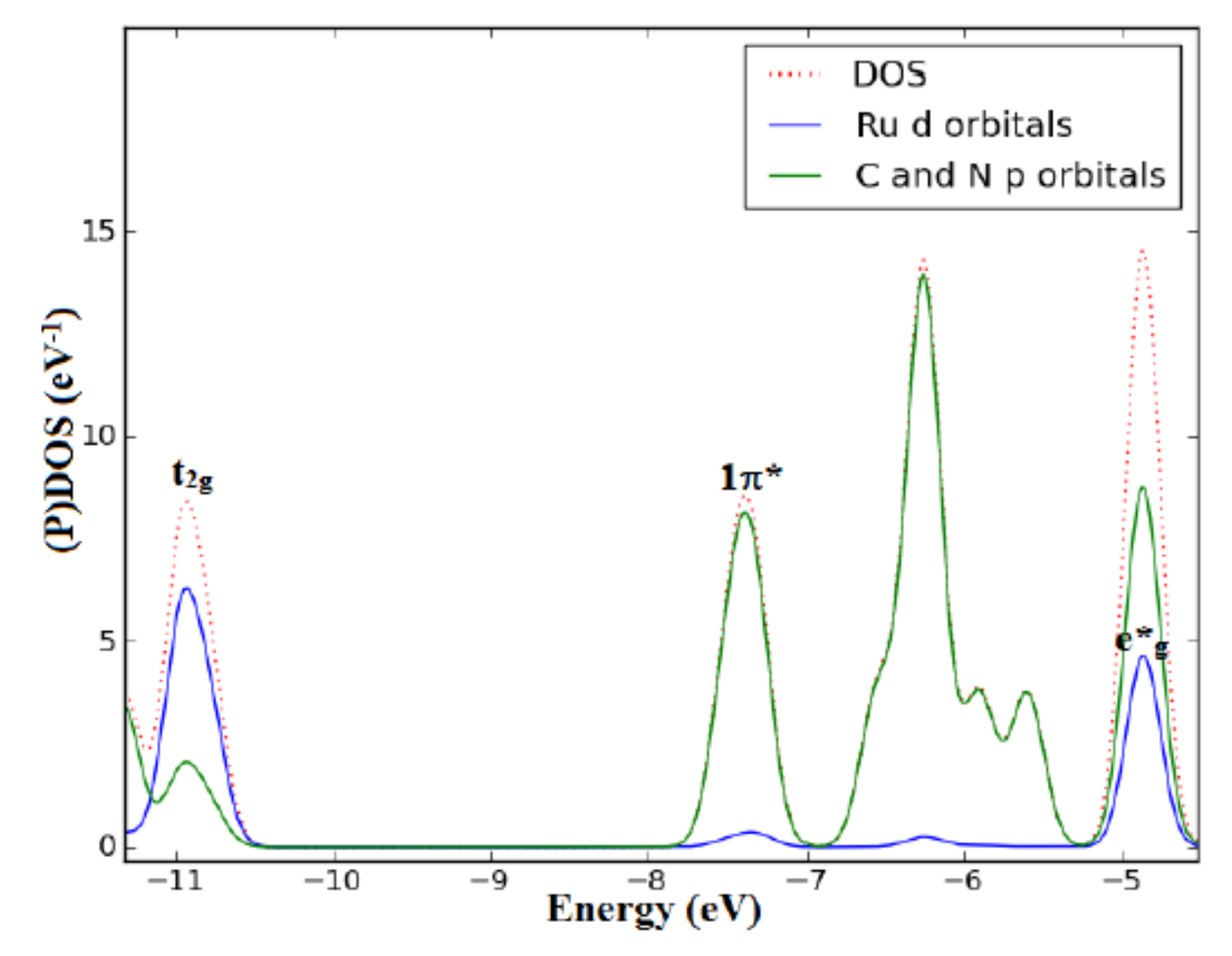} &
\includegraphics[width=0.4\textwidth]{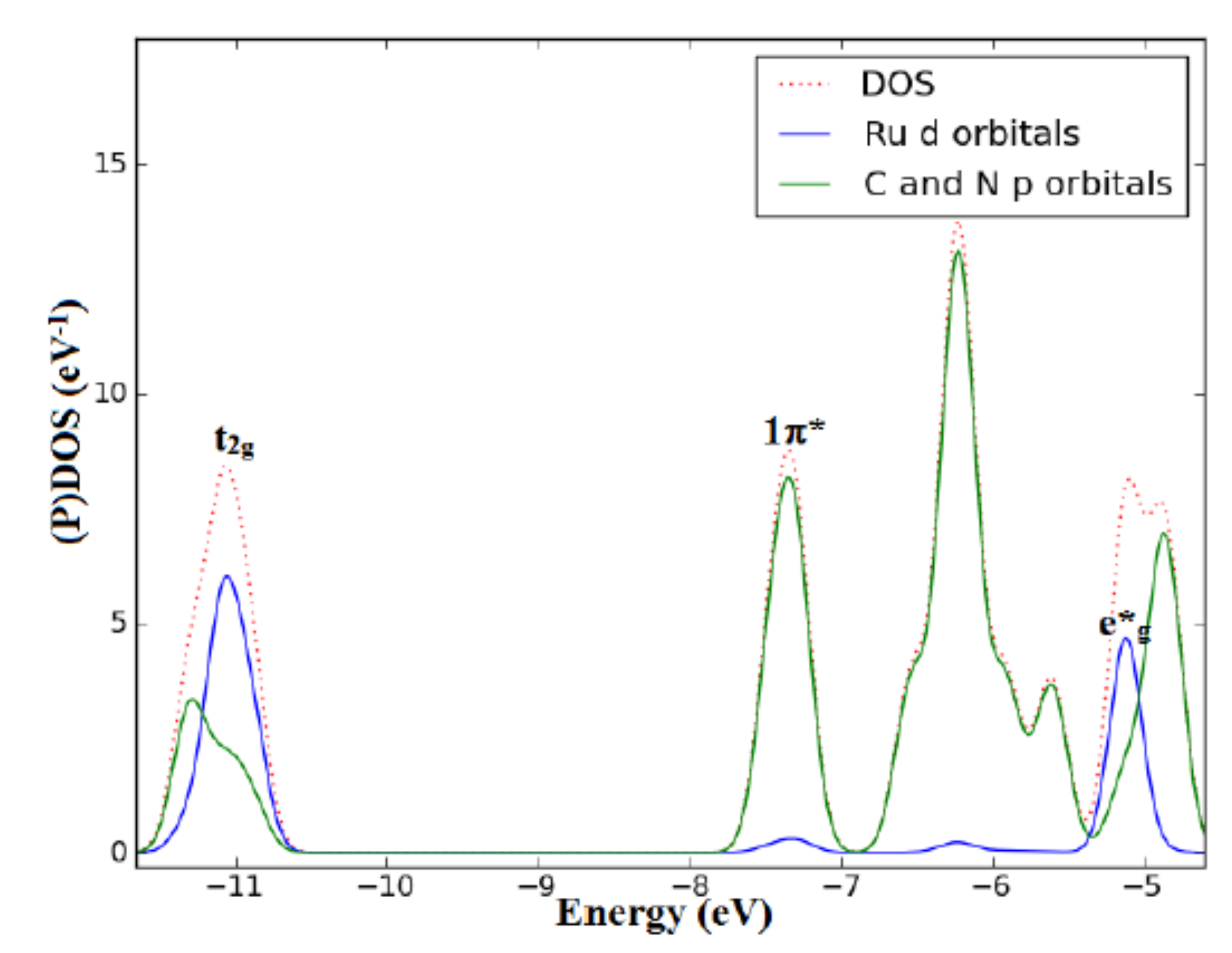} \\
B3LYP/6-31G & B3LYP/6-31G(d) \\
$\epsilon_{\text{HOMO}} = \mbox{-10.79 eV}$ & 
$\epsilon_{\text{HOMO}} = \mbox{-10.89 eV}$ 
\end{tabular}
\end{center}
Total and partial density of states of [Ru(bpy)$_2$(piq)]$^{2+}$
partitioned over Ru d orbitals and ligand C and N p orbitals.
%  for the 6-31G (left-hand side) and 6-31G* (right-hand side) basis sets.

\begin{center}
   {\bf Absorption Spectrum}
\end{center}

\begin{center}
\includegraphics[width=0.8\textwidth]{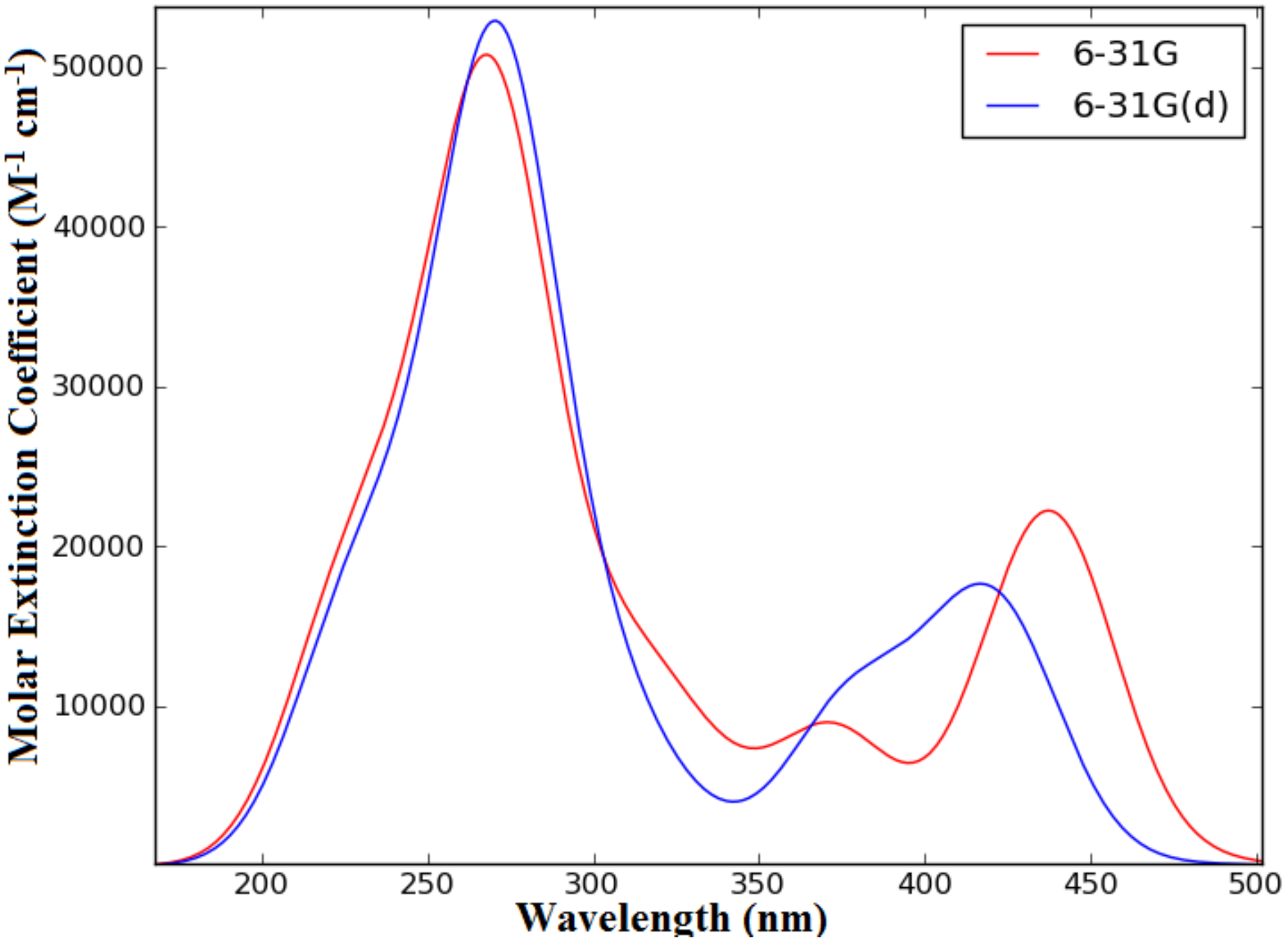}
\end{center}
[Ru(bpy)$_2$(piq)]$^{2+}$
TD-B3LYP/6-31G and TD-B3LYP/6-31G(d) spectra.

% ================================================
\newpage
\section{Complex {\bf (35)}: [Ru(bpy)$_2$(hpiq)]$^{2+}$}
% ================================================

\begin{center}
   {\bf PDOS}
\end{center}

\begin{center}
\begin{tabular}{cc}
\includegraphics[width=0.4\textwidth]{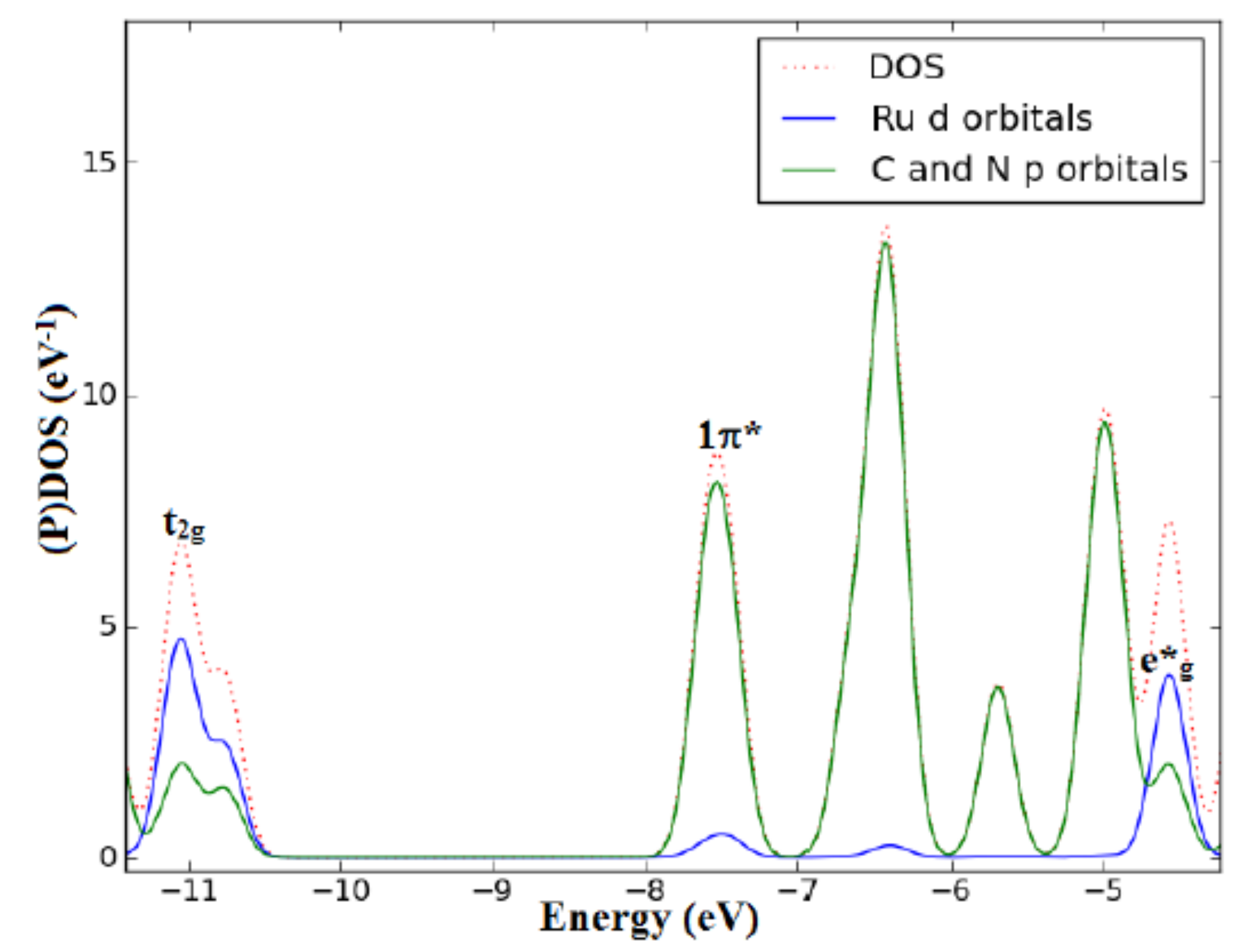} &
\includegraphics[width=0.4\textwidth]{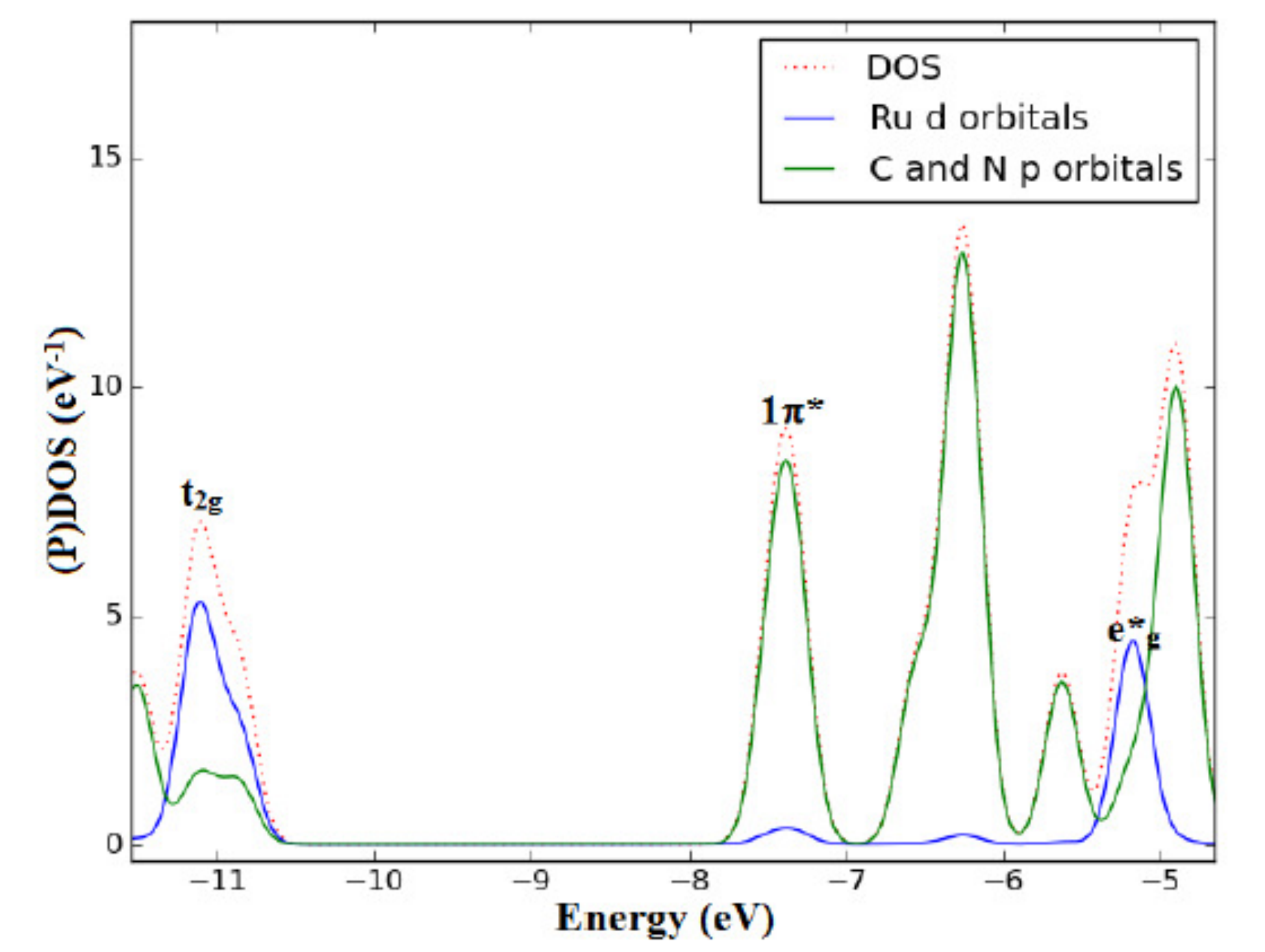} \\
B3LYP/6-31G & B3LYP/6-31G(d) \\
$\epsilon_{\text{HOMO}} = \mbox{-10.76 eV}$ & 
$\epsilon_{\text{HOMO}} = \mbox{-10.86 eV}$ 
\end{tabular}
\end{center}
Total and partial density of states of [Ru(bpy)$_2$(hpiq)]$^{2+}$
partitioned over Ru d orbitals and ligand C and N p orbitals.
% for the 6-31G (left-hand side) and 6-31G* (right-hand side) basis sets.

\begin{center}
   {\bf Absorption Spectrum}
\end{center}

\begin{center}
\includegraphics[width=0.8\textwidth]{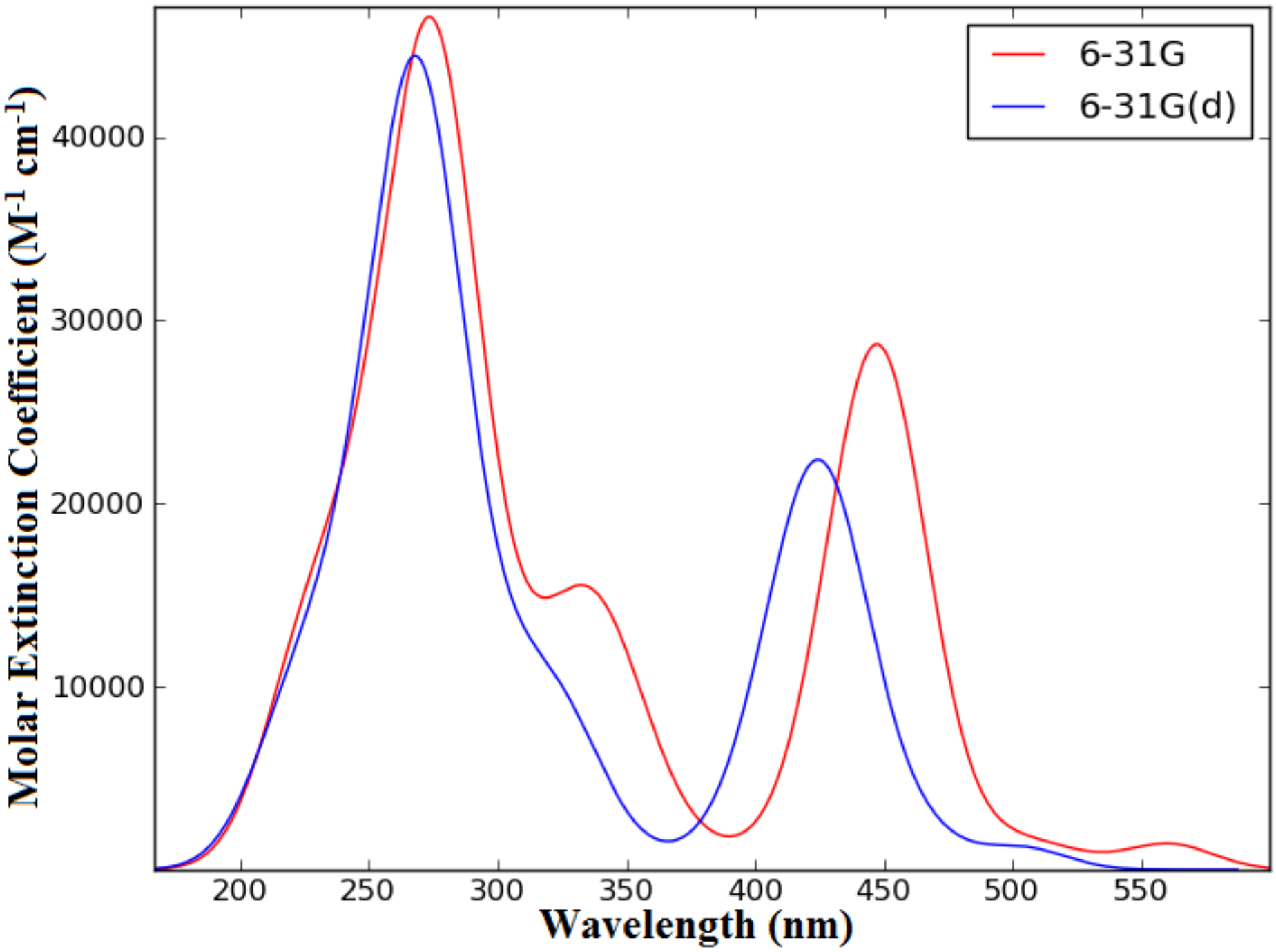}
\end{center}
[Ru(bpy)$_2$(hpiq)]$^{2+}$
TD-B3LYP/6-31G and TD-B3LYP/6-31G(d) spectra.

% ================================================
\newpage
\section{Complex {\bf (36)}: [Ru(bpy)$_2$(pq)]$^{2+}$}
% ================================================

\begin{center}
   {\bf PDOS}
\end{center}

\begin{center}
\begin{tabular}{cc}
\includegraphics[width=0.4\textwidth]{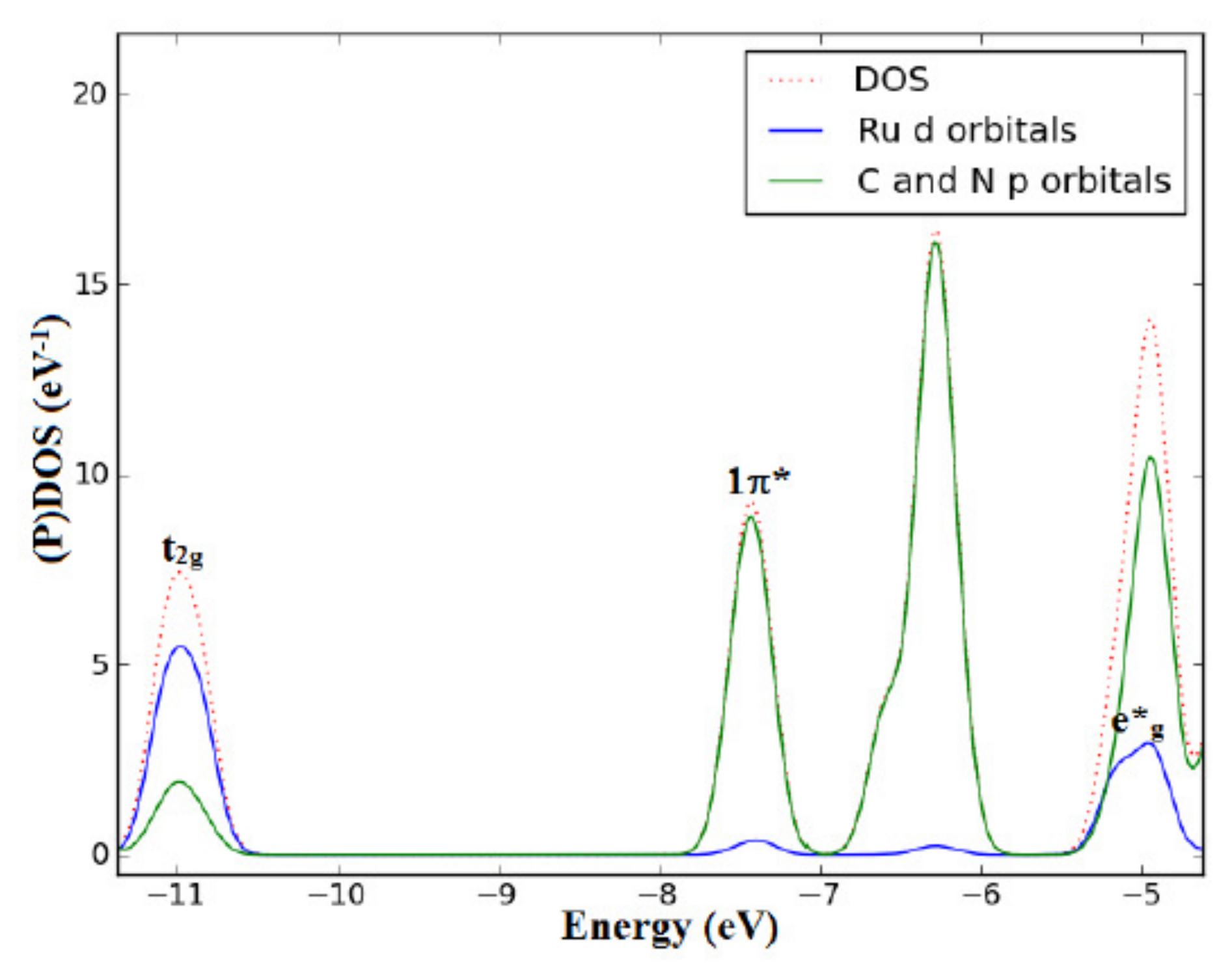} &
\includegraphics[width=0.4\textwidth]{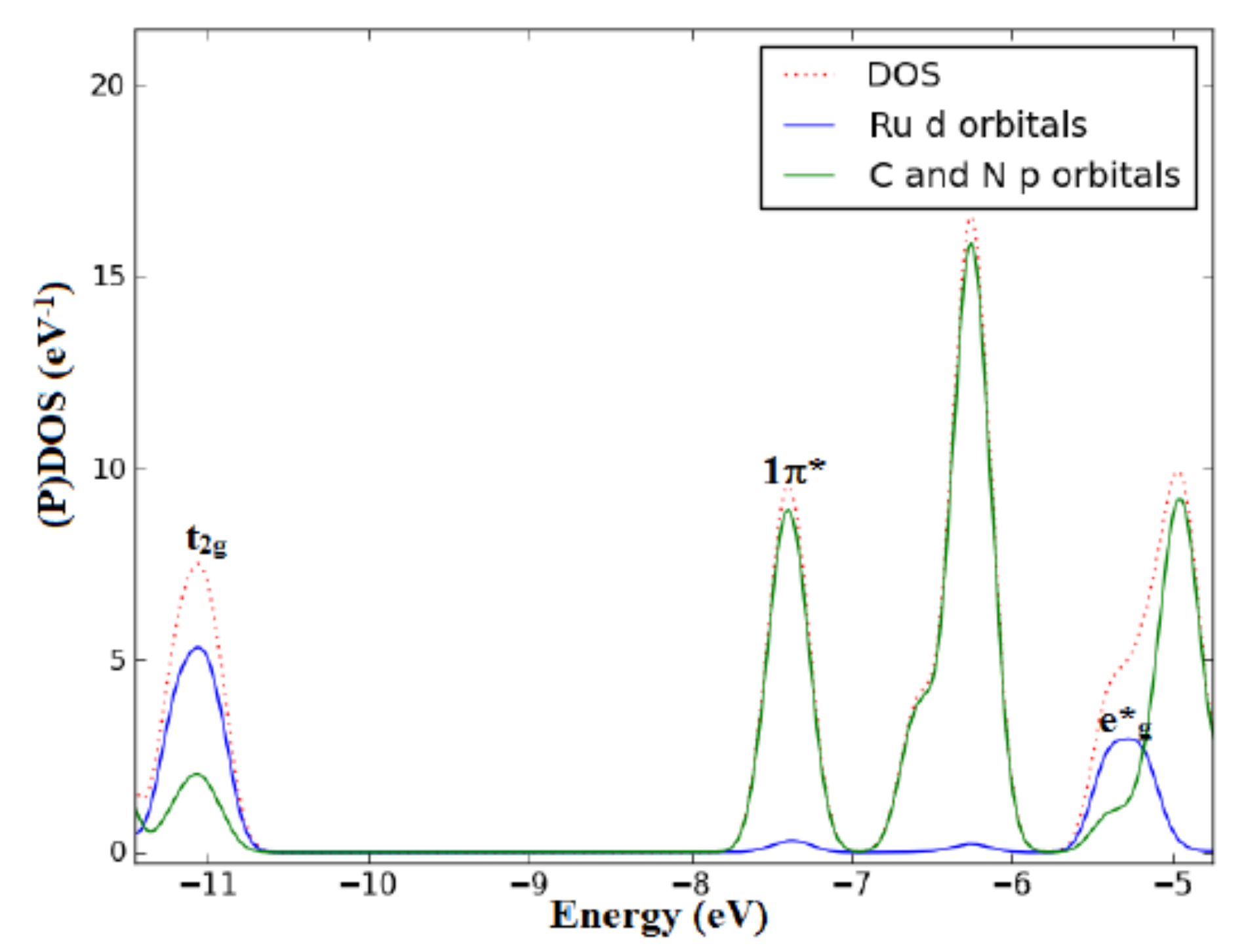} \\
B3LYP/6-31G & B3LYP/6-31G(d) \\
$\epsilon_{\text{HOMO}} = \mbox{-10.84 eV}$ & 
$\epsilon_{\text{HOMO}} = \mbox{-10.96 eV}$ 
\end{tabular}
\end{center}
Total and partial density of states of [Ru(bpy)$_2$(pq)]$^{2+}$
partitioned over Ru d orbitals and ligand C and N p orbitals. 
% for the 6-31G (left-hand side) and 6-31G* (right-hand side) basis sets.

\begin{center}
   {\bf Absorption Spectrum}
\end{center}

\begin{center}
\includegraphics[width=0.8\textwidth]{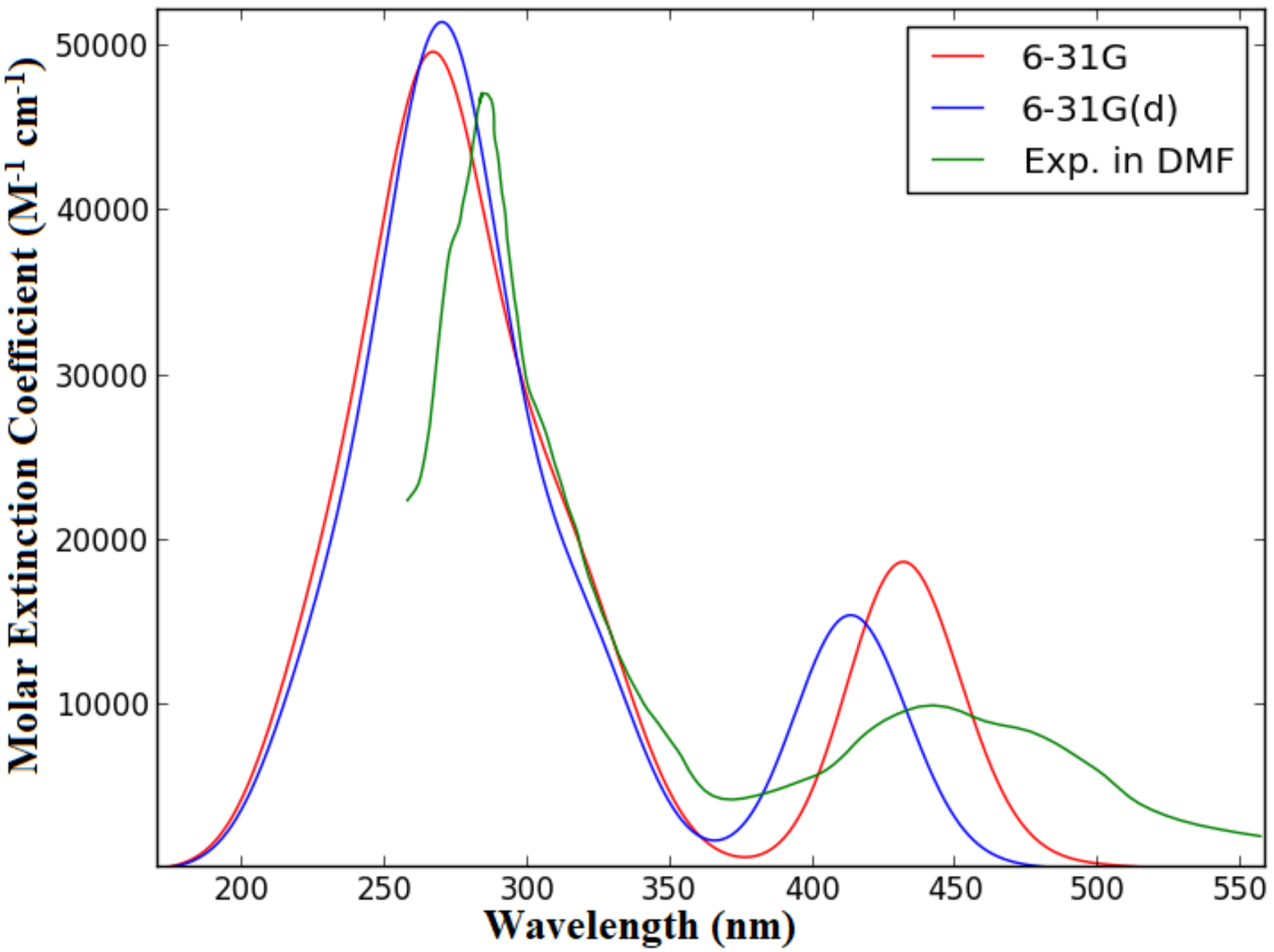}
\end{center}
[Ru(bpy)$_2$(pq)]$^{2+}$
TD-B3LYP/6-31G, TD-B3LYP/6-31G(d), and experimental spectra.
Experimental curve measured in DMF \cite{TVD+87}.

% ================================================
\newpage
\section{Complex {\bf (37)}: [Ru(bpy)$_2$(DMCH)]$^{2+}$}
% ================================================

\begin{center}
   {\bf PDOS}
\end{center}

\begin{center}
\begin{tabular}{cc}
\includegraphics[width=0.4\textwidth]{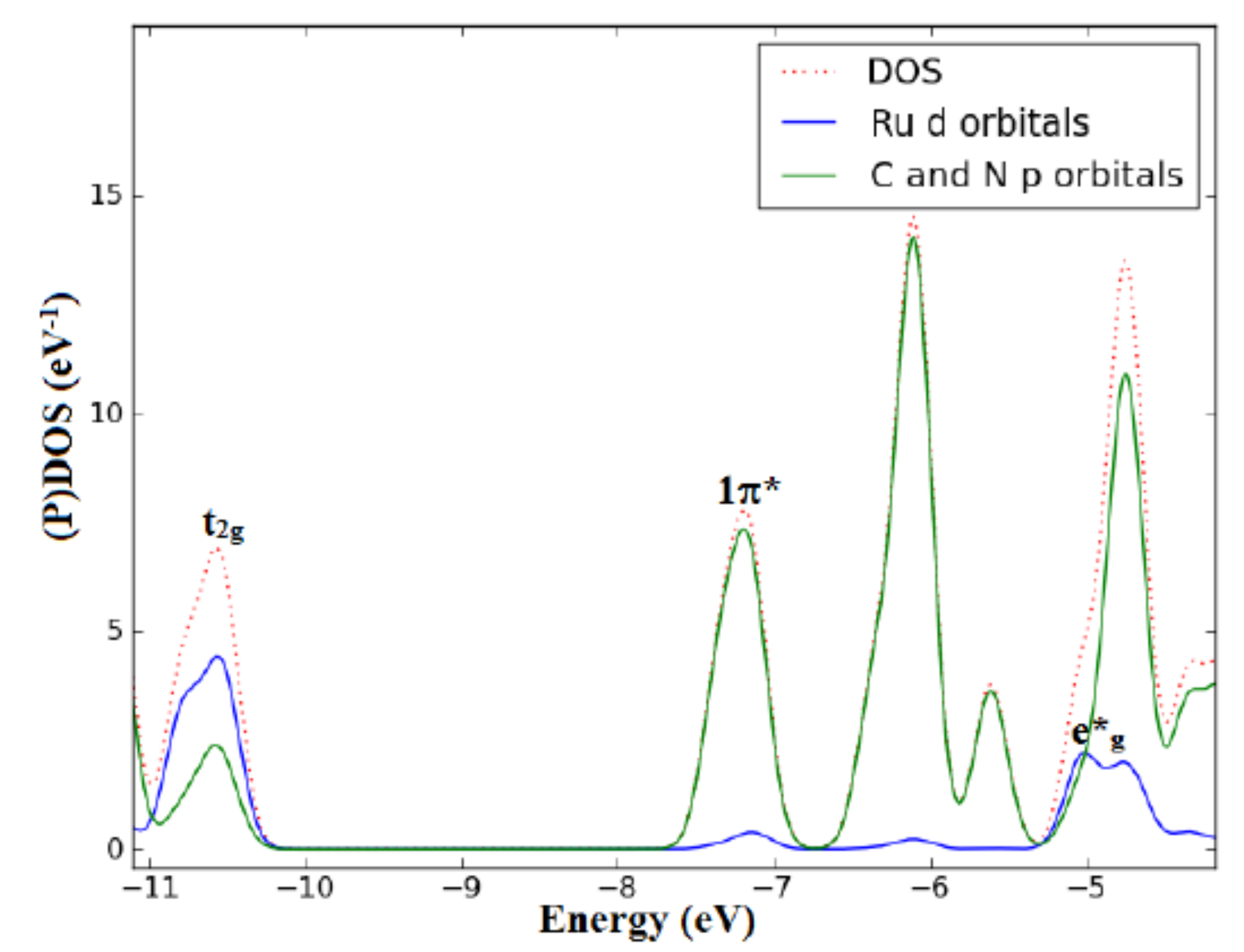} &
\includegraphics[width=0.4\textwidth]{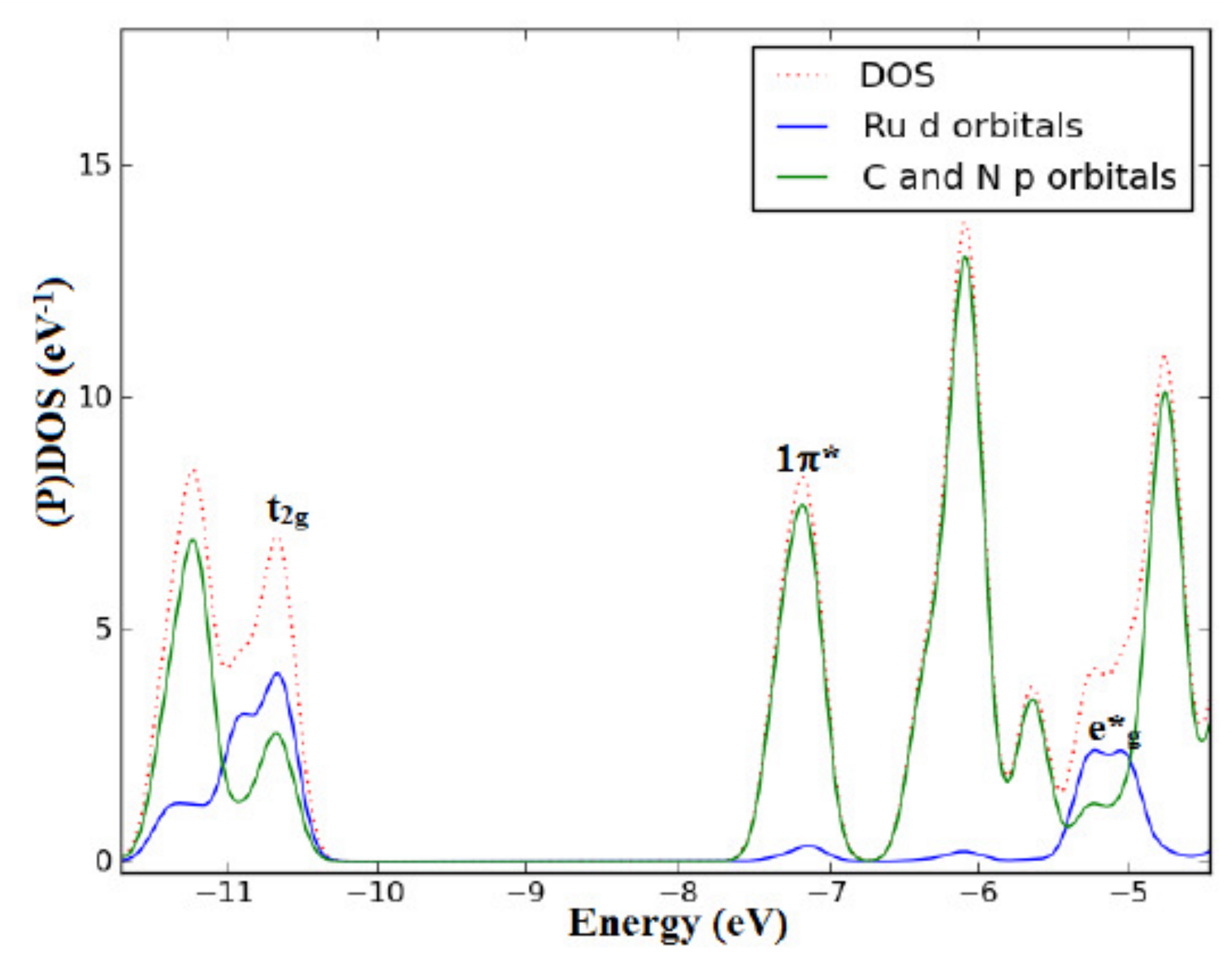} \\
B3LYP/6-31G & B3LYP/6-31G(d) \\
$\epsilon_{\text{HOMO}} = \mbox{-10.50 eV}$ & 
$\epsilon_{\text{HOMO}} = \mbox{-10.61 eV}$ 
\end{tabular}
\end{center}
Total and partial density of states of [Ru(bpy)$_2$(DMCH)]$^{2+}$
partitioned over Ru d orbitals and ligand C and N p orbitals.
% for the 6-31G (left-hand side) and 6-31G* (right-hand side) basis sets.

\begin{center}
   {\bf Absorption Spectrum}
\end{center}

\begin{center}
\includegraphics[width=0.8\textwidth]{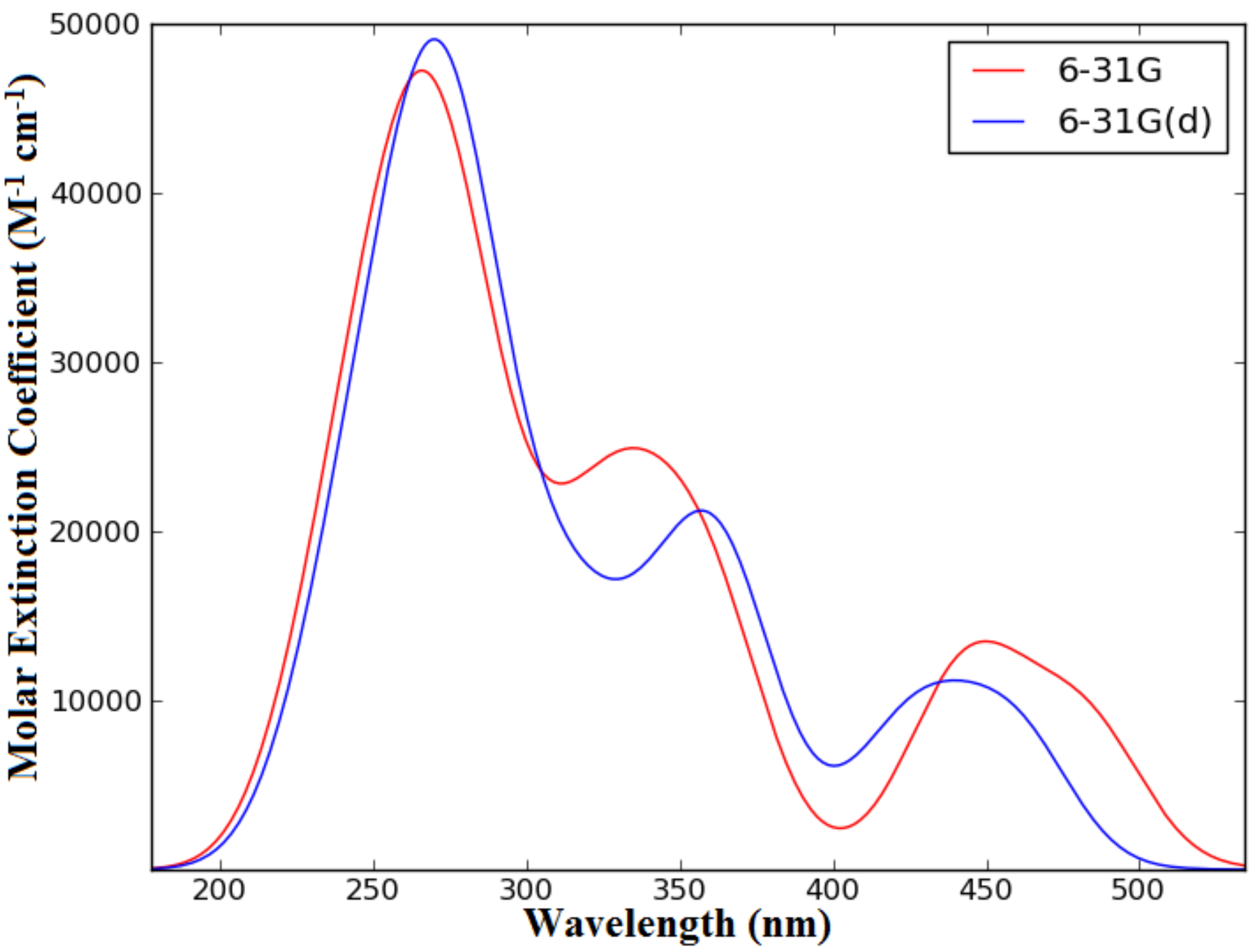}
\end{center}
[Ru(bpy)$_2$(DMCH)]$^{2+}$ 
TD-B3LYP/6-31G and TD-B3LYP/6-31G(d) spectra.

% ================================================
\newpage
\section{Complex {\bf (38)}$^\dagger$: [Ru(bpy)$_2$(OMCH)]$^{2+}$}
% ================================================

% \begin{center}
%    {\bf PDOS}
% \end{center}
% 
% \begin{center}
% \includegraphics[width=0.4\textwidth]{graphics1/framedquestionmark.pdf}
% \includegraphics[width=0.4\textwidth]{graphics1/framedquestionmark.pdf}
% \end{center}
% {\color{red} Do we have this?}

\begin{center}
\begin{tabular}{cc}
B3LYP/6-31G & B3LYP/6-31G(d) \\
$\epsilon_{\text{HOMO}} = \mbox{-10.35 eV}$ & 
$\epsilon_{\text{HOMO}} = \mbox{-10.39 eV}$ 
\end{tabular}
\end{center}

\begin{center}
   {\bf Absorption Spectrum}
\end{center}

\begin{center}
\includegraphics[width=0.8\textwidth]{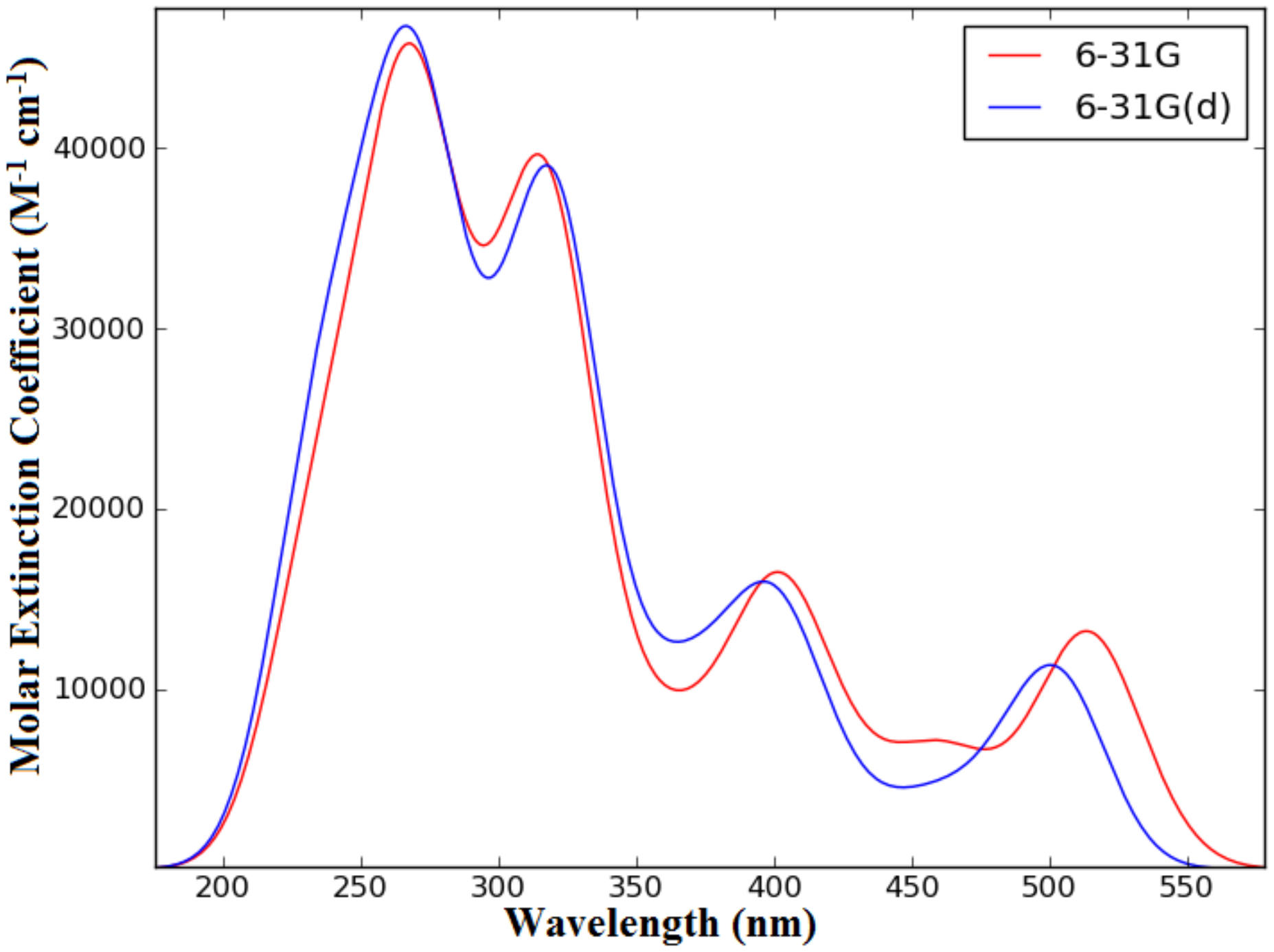}
\end{center}
[Ru(bpy)$_2$(OMCH)]$^{2+}$
TD-B3LYP/6-31G and TD-B3LYP/6-31G(d) spectra.

% ================================================
\newpage
\section{Complex {\bf (39)}$^\dagger$: [Ru(bpy)$_2$(biq)]$^{2+}$}
% ================================================

\begin{center}
   {\bf Absorption Spectrum}
\end{center}

\begin{center}
\begin{tabular}{cc}
B3LYP/6-31G & B3LYP/6-31G(d) \\
$\epsilon_{\text{HOMO}} = \mbox{-10.72 eV}$ & 
$\epsilon_{\text{HOMO}} = \mbox{-10.82 eV}$ 
\end{tabular}
\end{center}

\begin{center}
\includegraphics[width=0.8\textwidth]{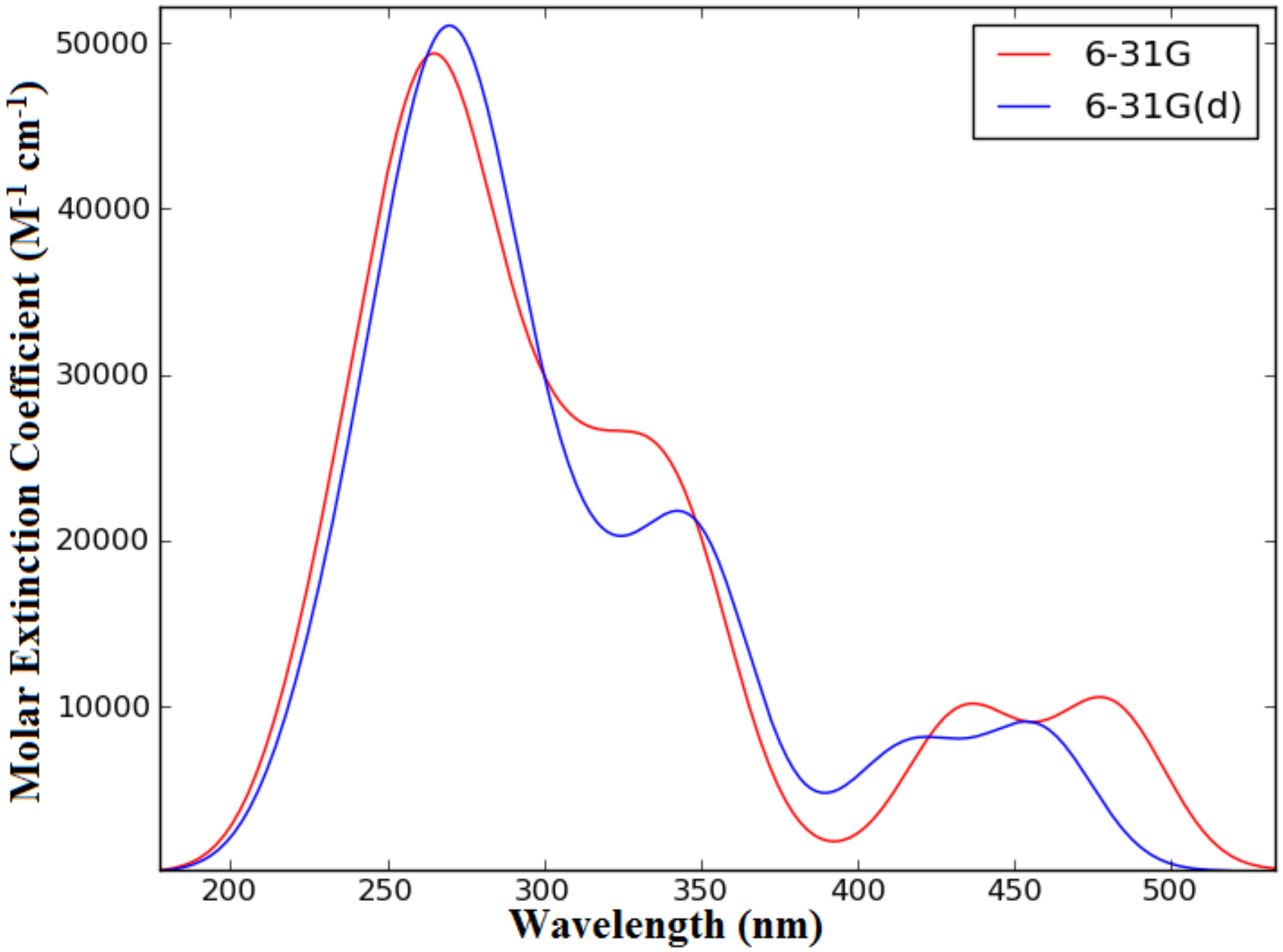}
\end{center}
[Ru(bpy)$_2$(biq)]$^{2+}$
TD-B3LYP/6-31G and TD-B3LYP/6-31G(d) spectra.

% ================================================
\newpage
\section{Complex {\bf (40)}: [Ru(bpy)$_2$(i-biq)]$^{2+}$}
% ================================================

\begin{center}
   {\bf PDOS}
\end{center}

\begin{center}
\begin{tabular}{cc}
\includegraphics[width=0.4\textwidth]{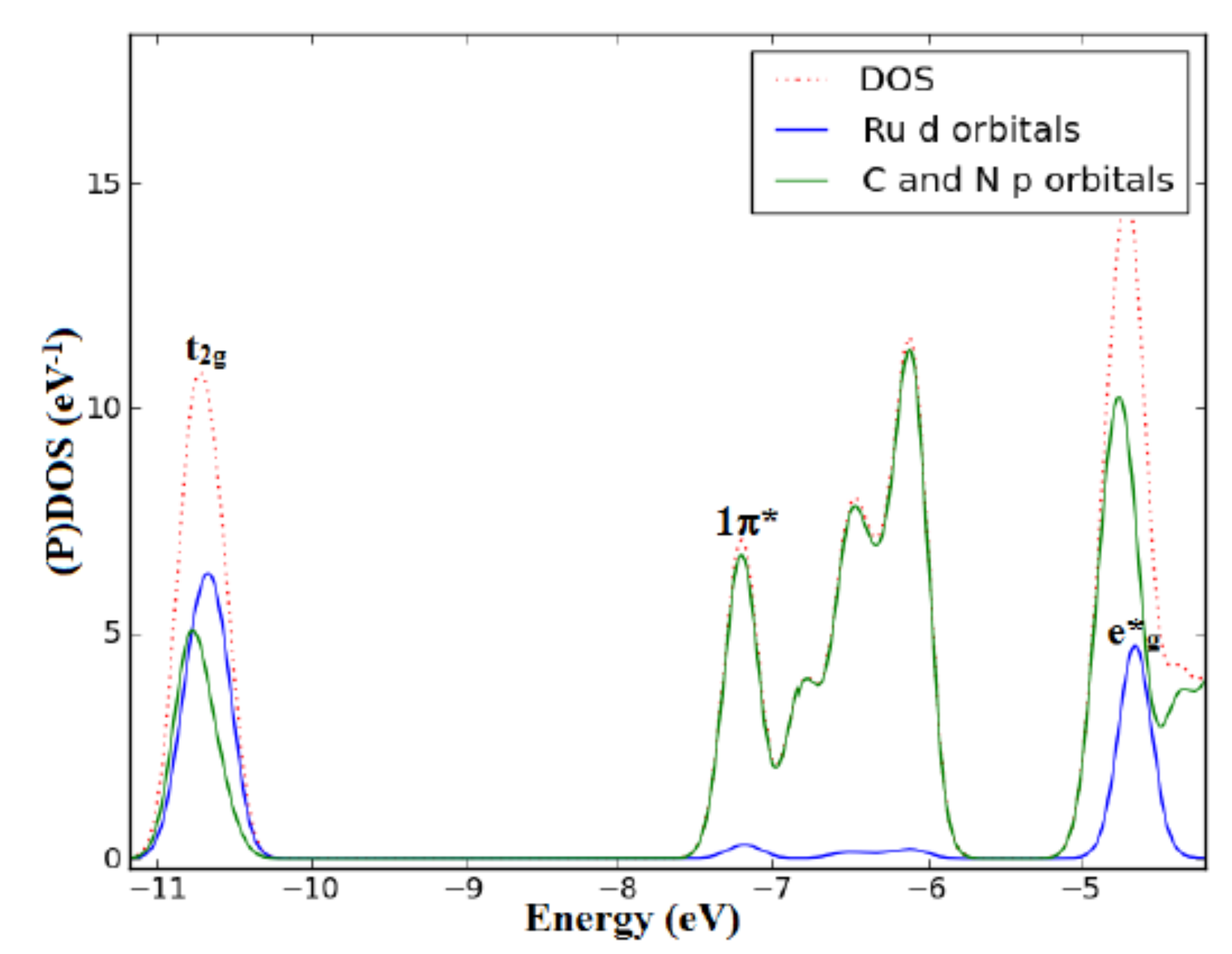} &
\includegraphics[width=0.4\textwidth]{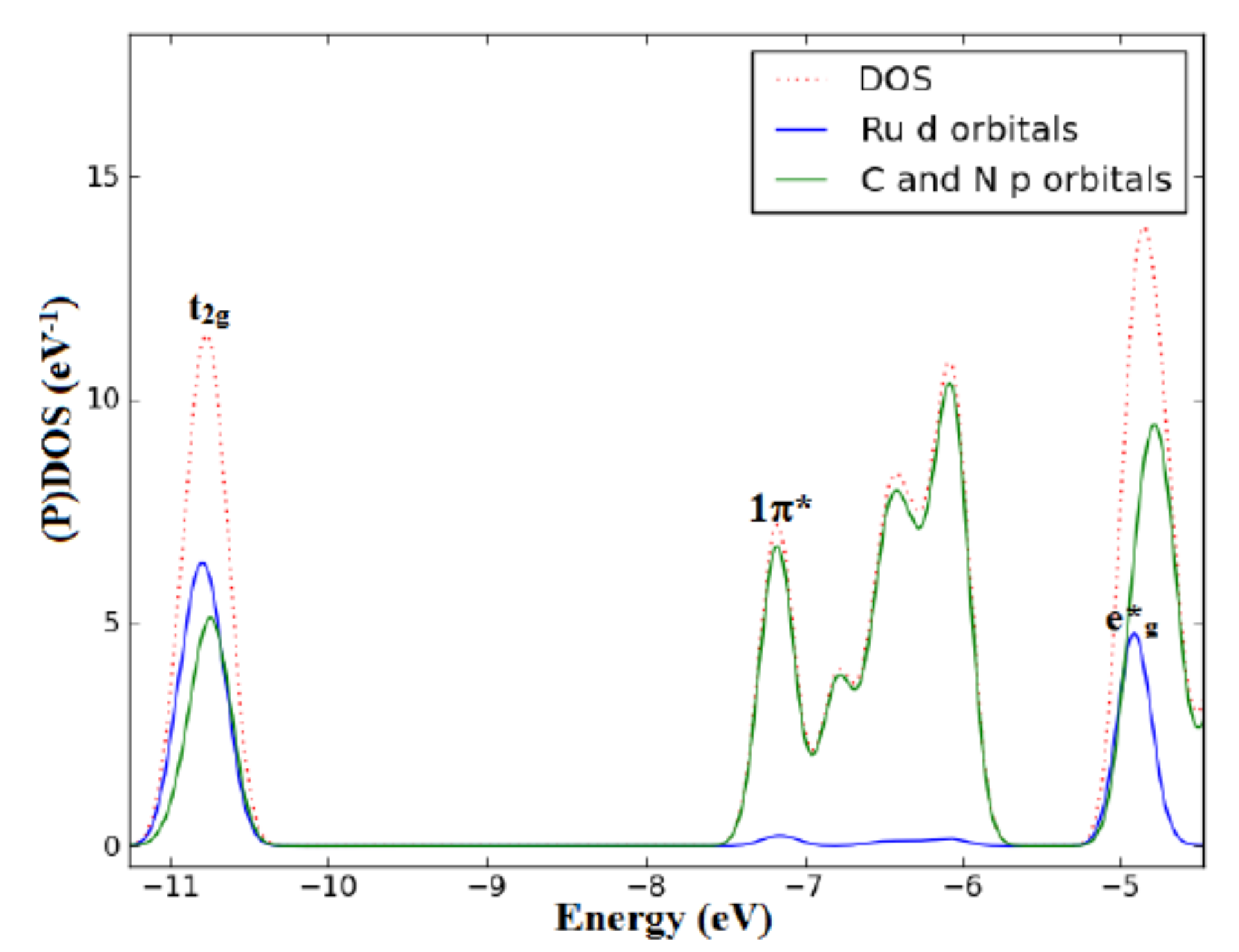} \\
B3LYP/6-31G & B3LYP/6-31G(d) \\
$\epsilon_{\text{HOMO}} = \mbox{-10.58 eV}$ & 
$\epsilon_{\text{HOMO}} = \mbox{-10.67 eV}$ 
\end{tabular}
\end{center}
Total and partial density of states of [Ru(bpy)$_2$(i-biq)]$^{2+}$
partitioned over Ru d orbitals and ligand C and N p orbitals. 
% for the 6-31G (left-hand side) and 6-31G* (right-hand side) basis sets.

\begin{center}
   {\bf Absorption Spectrum}
\end{center}

\begin{center}
\includegraphics[width=0.8\textwidth]{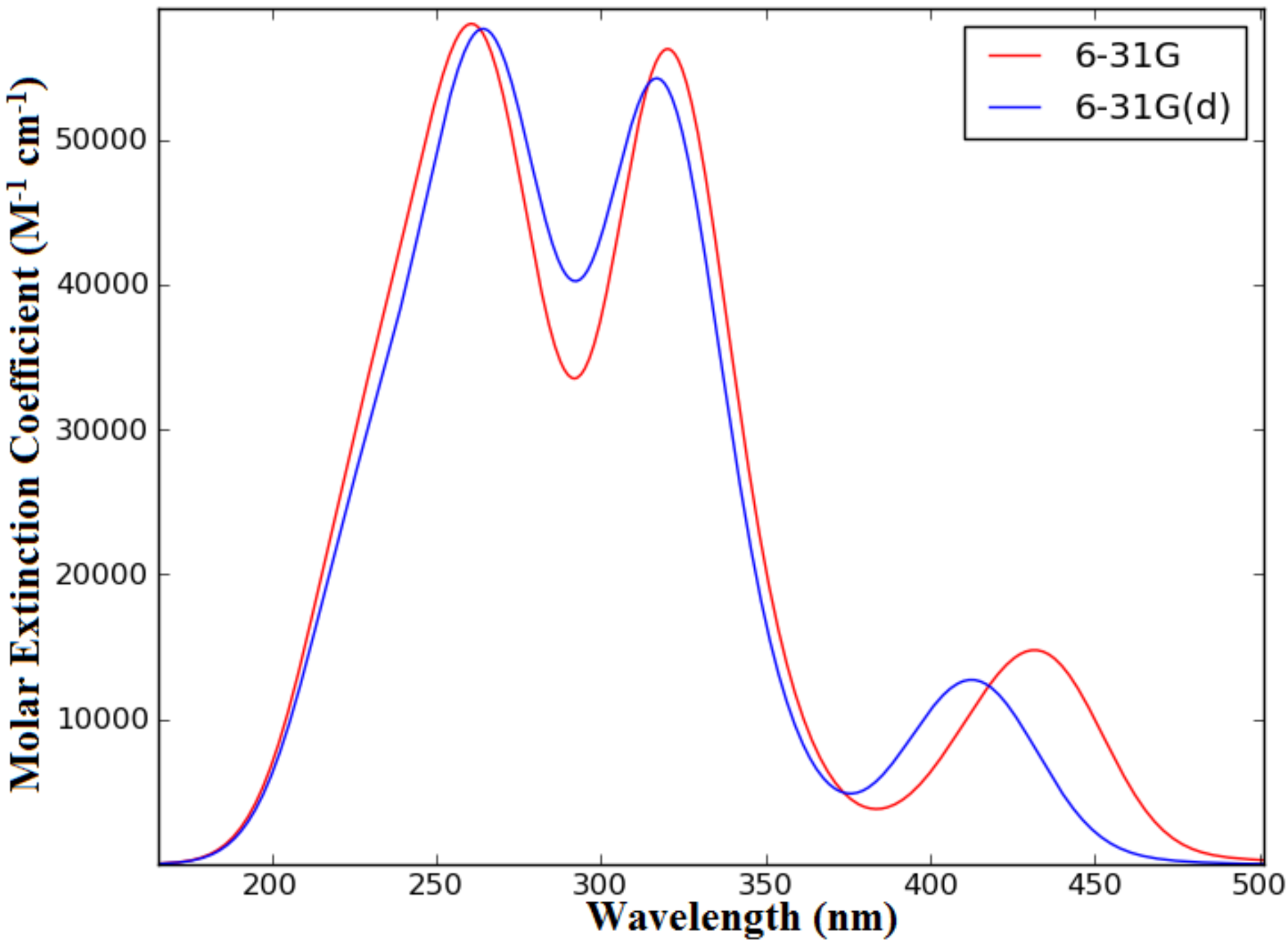}
\end{center}
[Ru(bpy)$_2$(i-biq)]$^{2+}$ 
TD-B3LYP/6-31G and TD-B3LYP/6-31G(d) spectra.

% ================================================
\newpage
\section{Complex {\bf (41)}: [Ru(bpy)$_2$(BL4)]$^{2+}$}
% ================================================

\begin{center}
   {\bf PDOS}
\end{center}

\begin{center}
\begin{tabular}{cc}
\includegraphics[width=0.4\textwidth]{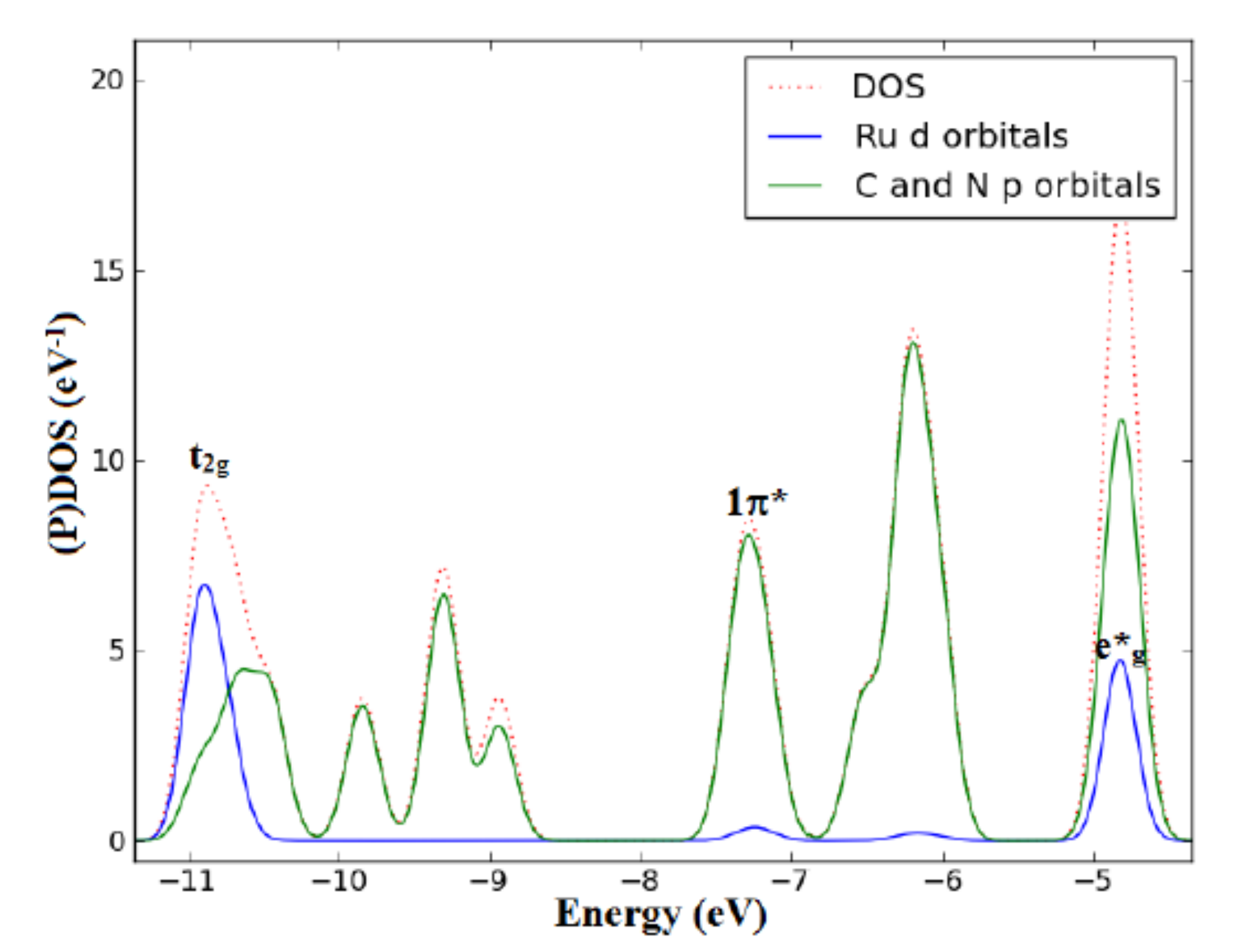} &
\includegraphics[width=0.4\textwidth]{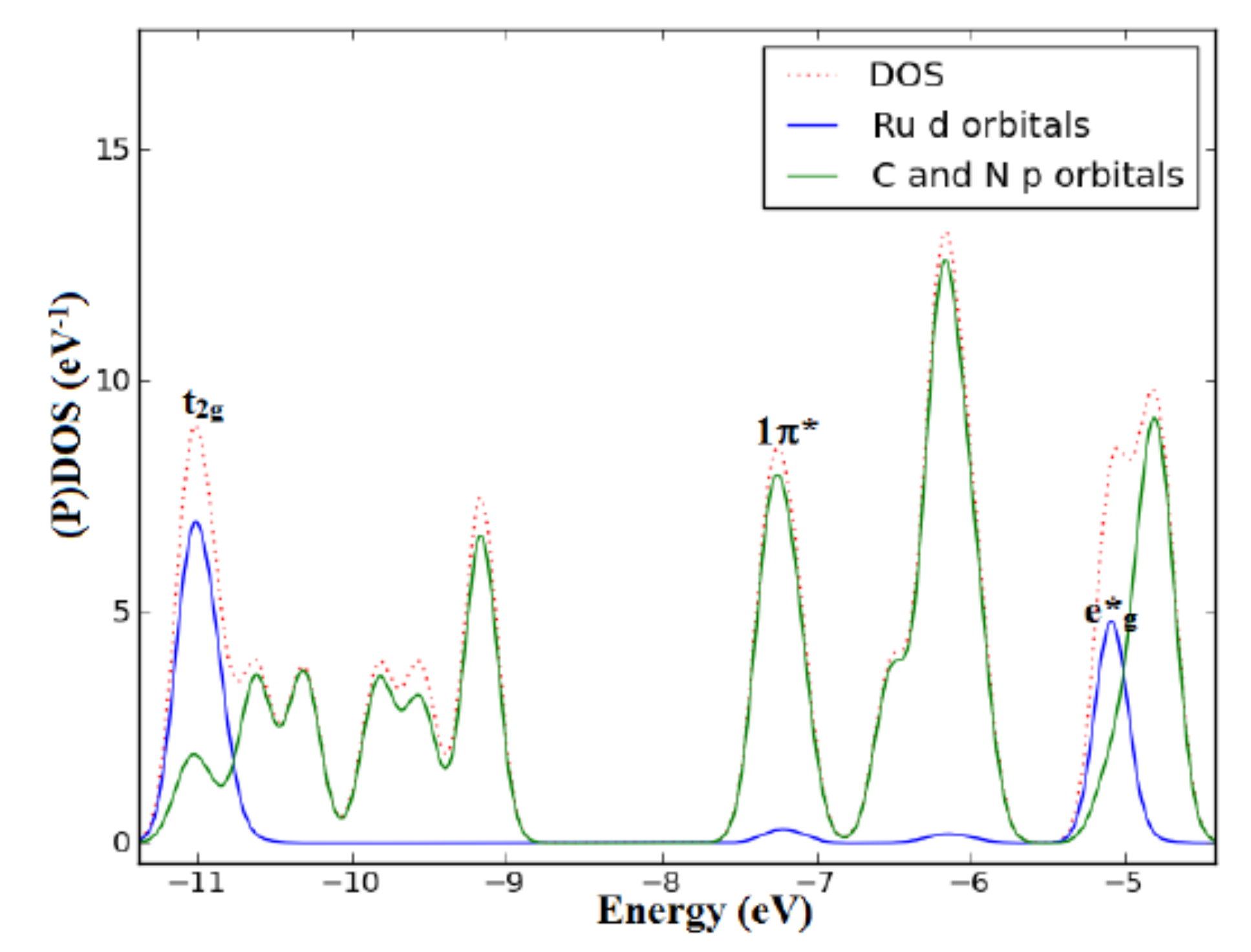} \\
B3LYP/6-31G & B3LYP/6-31G(d) \\
$\epsilon_{\text{HOMO}} = \mbox{-8.94 eV}$ & 
$\epsilon_{\text{HOMO}} = \mbox{-9.15 eV}$ 
\end{tabular}
\end{center}
Total and partial density of states of [Ru(bpy)$_2$(BL4)]$^{2+}$
partitioned over Ru d orbitals and ligand C and N p orbitals. 
% for the 6-31G (left-hand side) and 6-31G* (right-hand side) basis sets.

\begin{center}
   {\bf Absorption Spectrum}
\end{center}

\begin{center}
\includegraphics[width=0.8\textwidth]{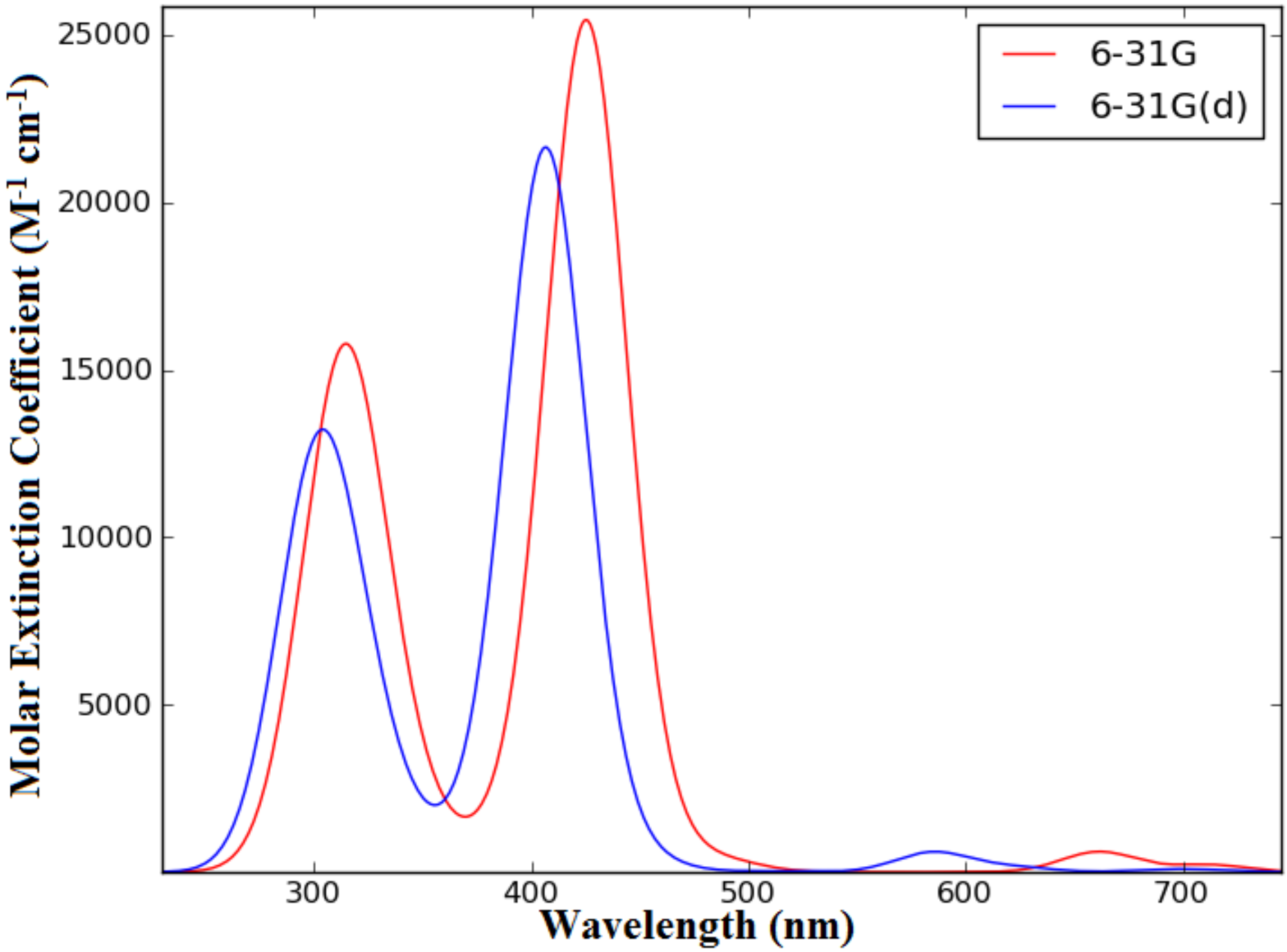}
\end{center}
[Ru(bpy)$_2$(BL4)]$^{2+}$
TD-B3LYP/6-31G and TD-B3LYP/6-31G(d) spectra.

% ================================================
\newpage
\section{Complex {\bf (42)}: [Ru(bpy)$_2$(BL5)]$^{2+}$}
% ================================================

\begin{center}
   {\bf PDOS}
\end{center}

\begin{center}
\begin{tabular}{cc}
\includegraphics[width=0.4\textwidth]{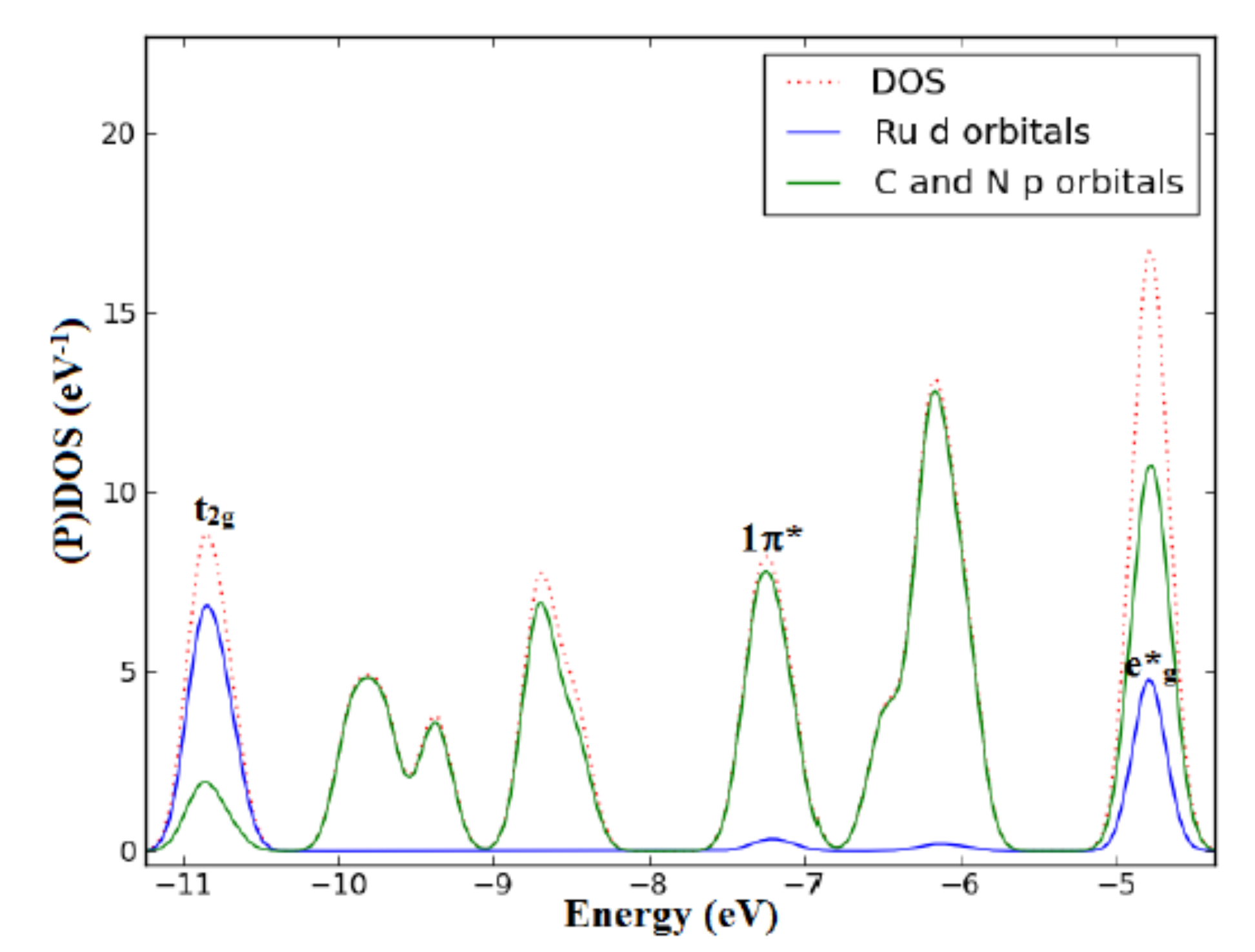} &
\includegraphics[width=0.4\textwidth]{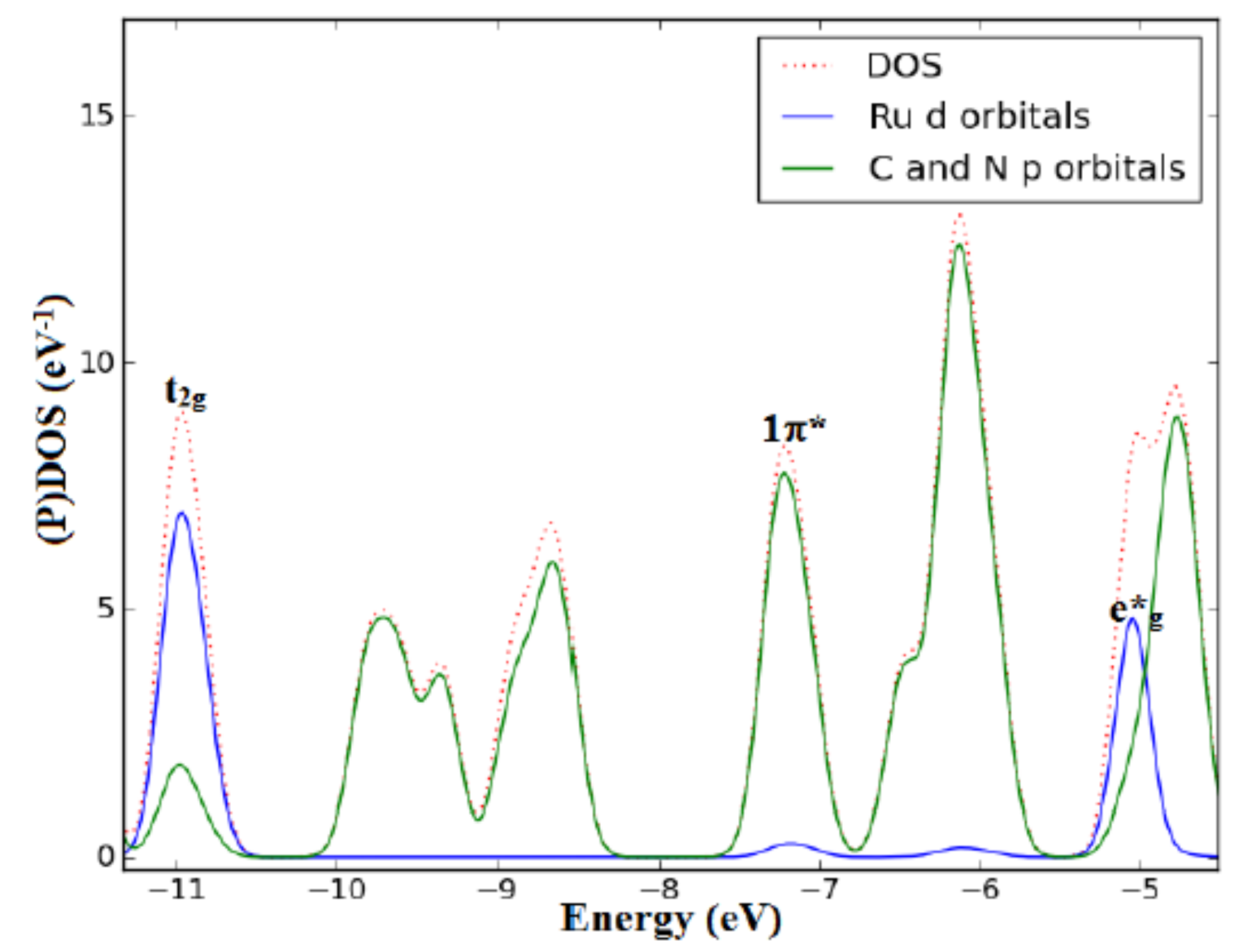} \\
B3LYP/6-31G & B3LYP/6-31G(d) \\
$\epsilon_{\text{HOMO}} = \mbox{-8.48 eV}$ & 
$\epsilon_{\text{HOMO}} = \mbox{-8.59 eV}$ 
\end{tabular}
\end{center}
Total and partial density of states of [Ru(bpy)$_2$(BL5)]$^{2+}$
partitioned over Ru d orbitals and ligand C and N p orbitals. 
% for the 6-31G (left-hand side) and 6-31G* (right-hand side) basis sets.

\begin{center}
   {\bf Absorption Spectrum}
\end{center}

\begin{center}
\includegraphics[width=0.8\textwidth]{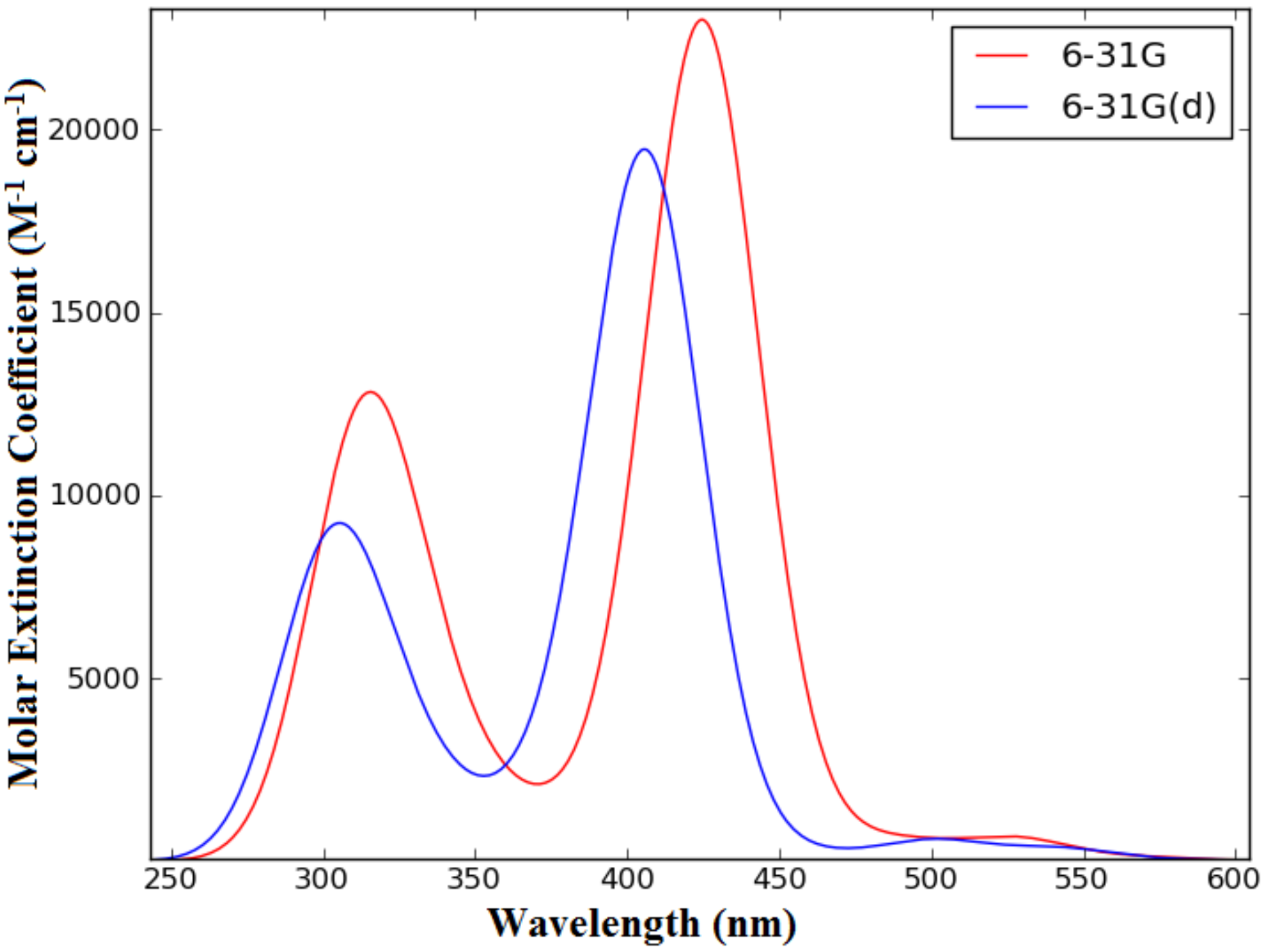}
\end{center}
[Ru(bpy)$_2$(BL5)]$^{2+}$
TD-B3LYP/6-31G and TD-B3LYP/6-31G(d) spectra.

\newpage
\section{Complex {\bf (46)}: [Ru(bpy)(4,4'-DTB-bpy)$_2$]$^{2+}$}
% ================================================

\begin{center}
   {\bf PDOS}
\end{center}

\begin{center}
\begin{tabular}{cc}
\includegraphics[width=0.4\textwidth]{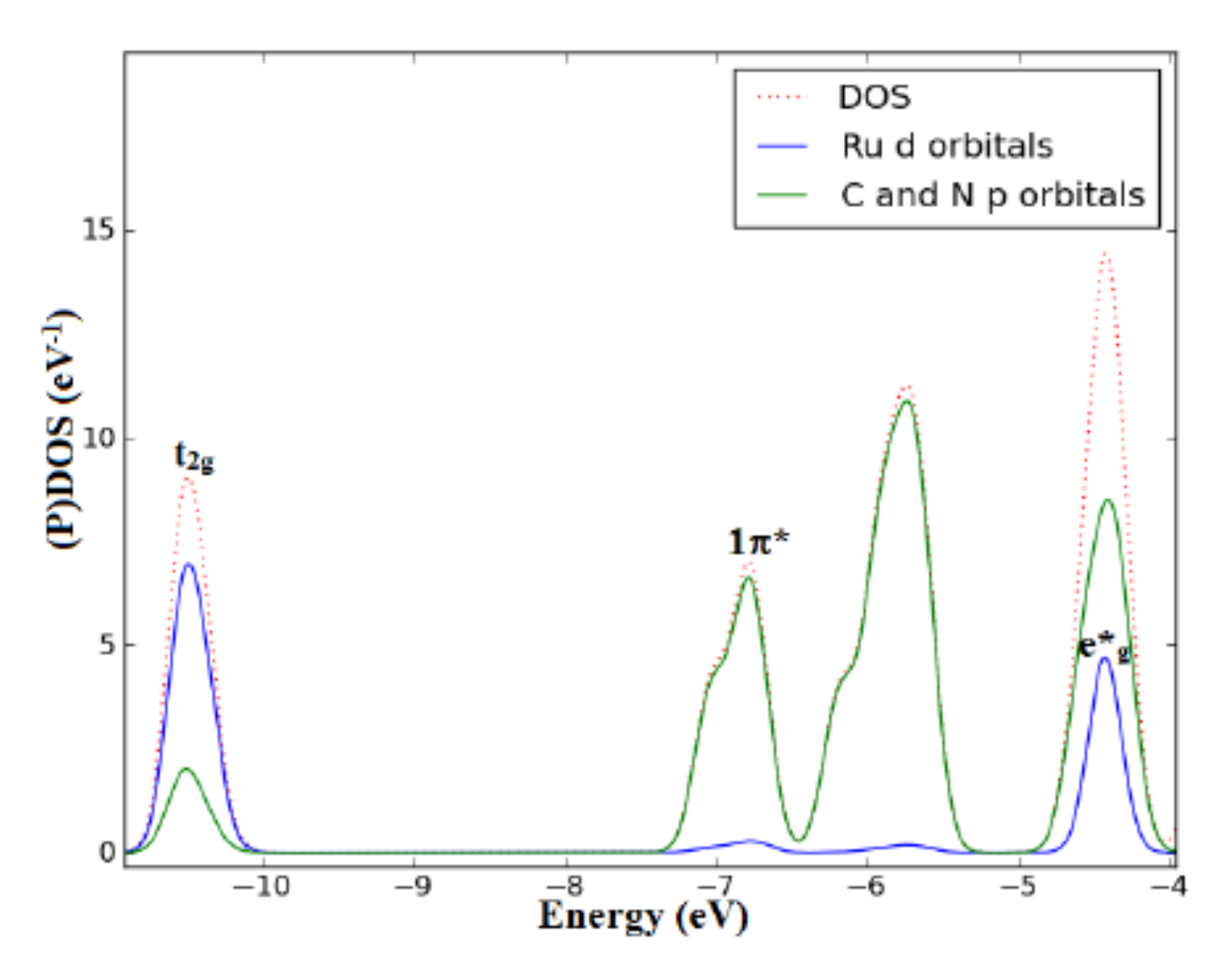} &
\includegraphics[width=0.4\textwidth]{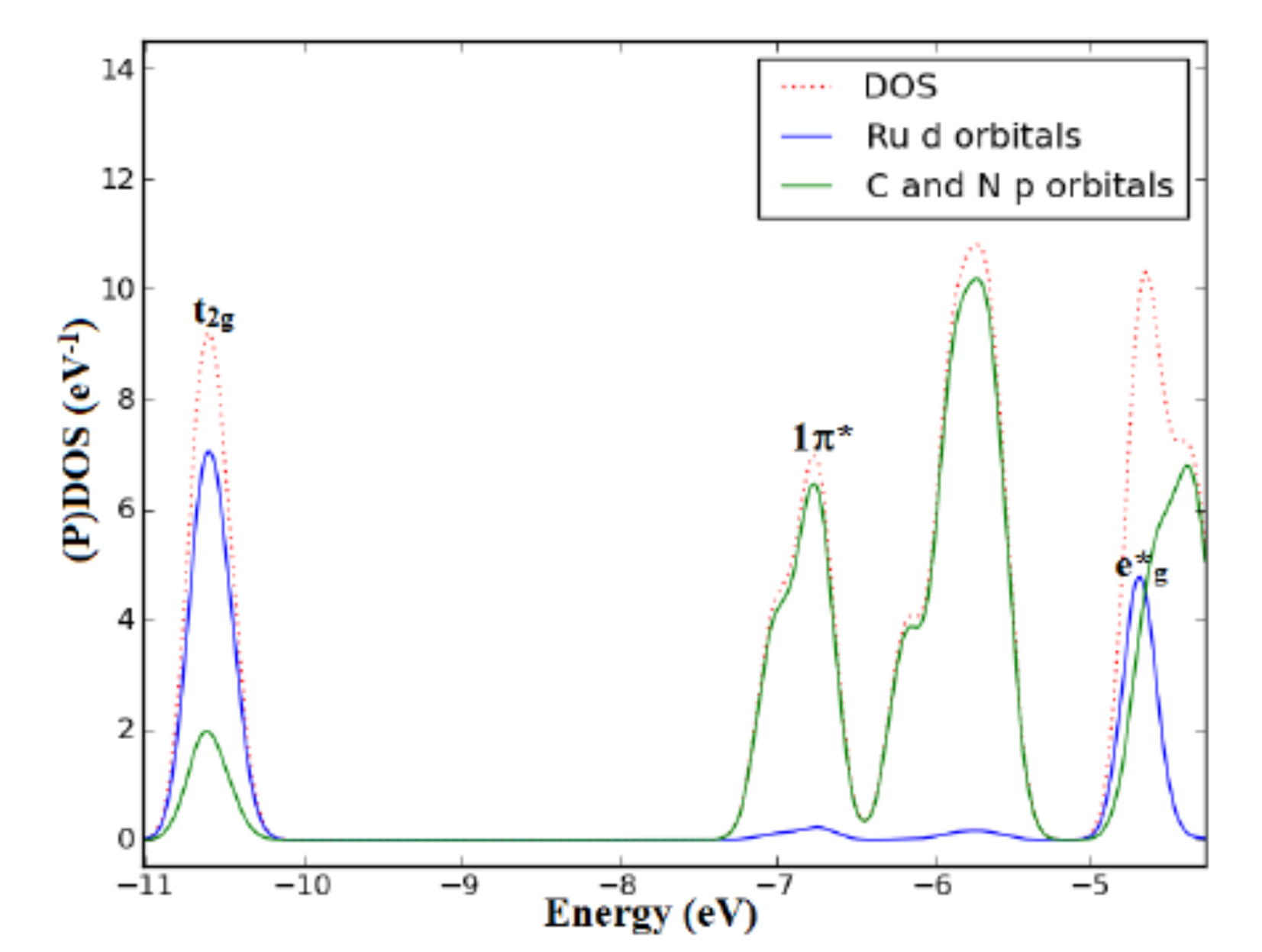} \\
B3LYP/6-31G & B3LYP/6-31G(d) \\
$\epsilon_{\text{HOMO}} = \mbox{-10.36 eV}$ & 
$\epsilon_{\text{HOMO}} = \mbox{-10.49 eV}$ 
\end{tabular}
\end{center}
Total and partial density of states of [Ru(bpy)(4,4'-DTB-bpy)$_2$]$^{2+}$
partitioned over Ru d orbitals and ligand C and N p orbitals.

\begin{center}
   {\bf Absorption Spectrum}
\end{center}

\begin{center}
\includegraphics[width=0.8\textwidth]{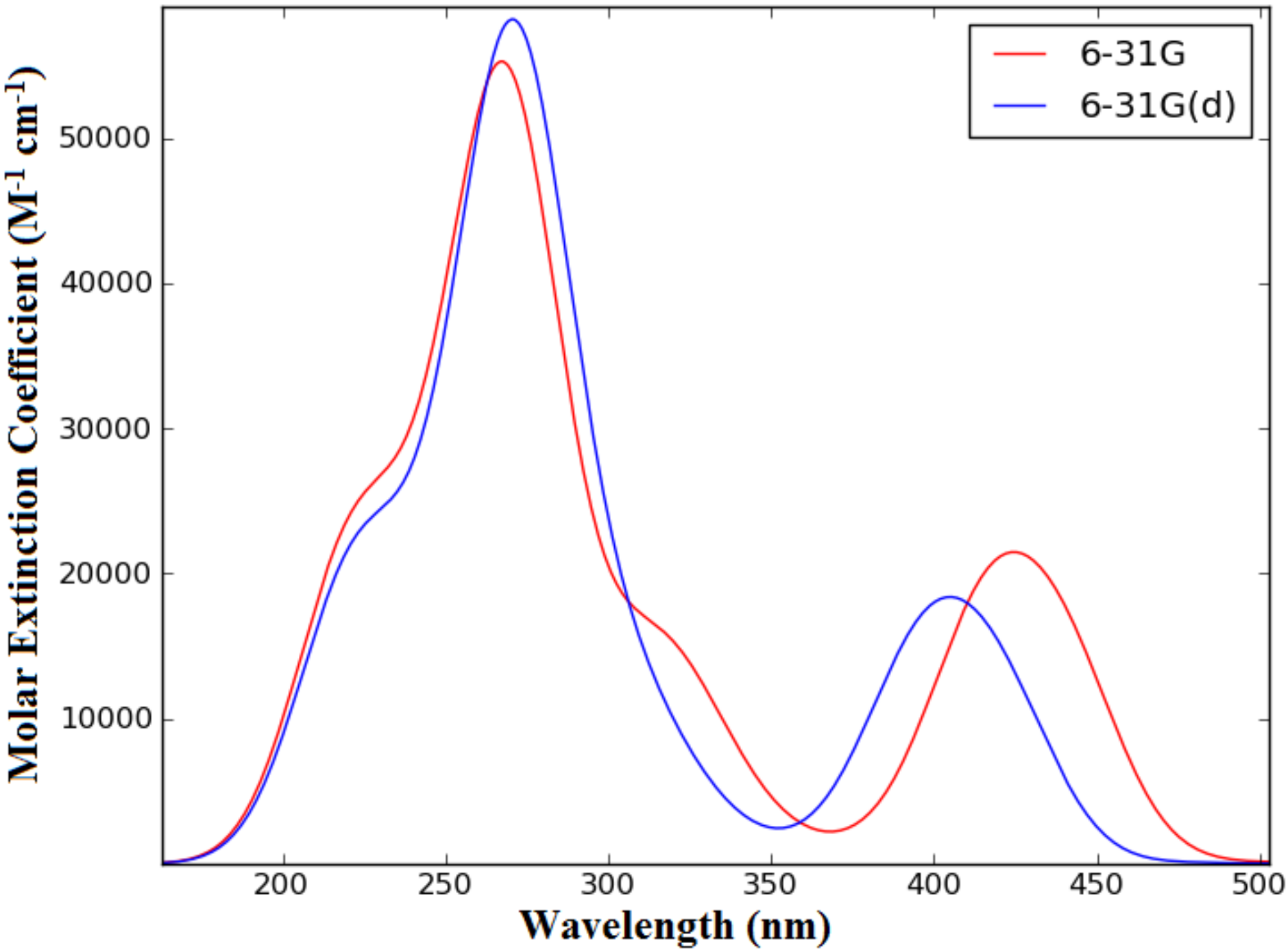}
\end{center}
[Ru(bpy)(4,4'-DTB-bpy)$_2$]$^{2+}$
TD-B3LYP/6-31G and TD-B3LYP/6-31G(d) spectra.

% ================================================
\newpage
\section{Complex {\bf (47)}: [Ru(bpy)(h-phen)]$^{2+}$}
% ================================================

\begin{center}
   {\bf PDOS}
\end{center}

\begin{center}
\begin{tabular}{cc}
\includegraphics[width=0.4\textwidth]{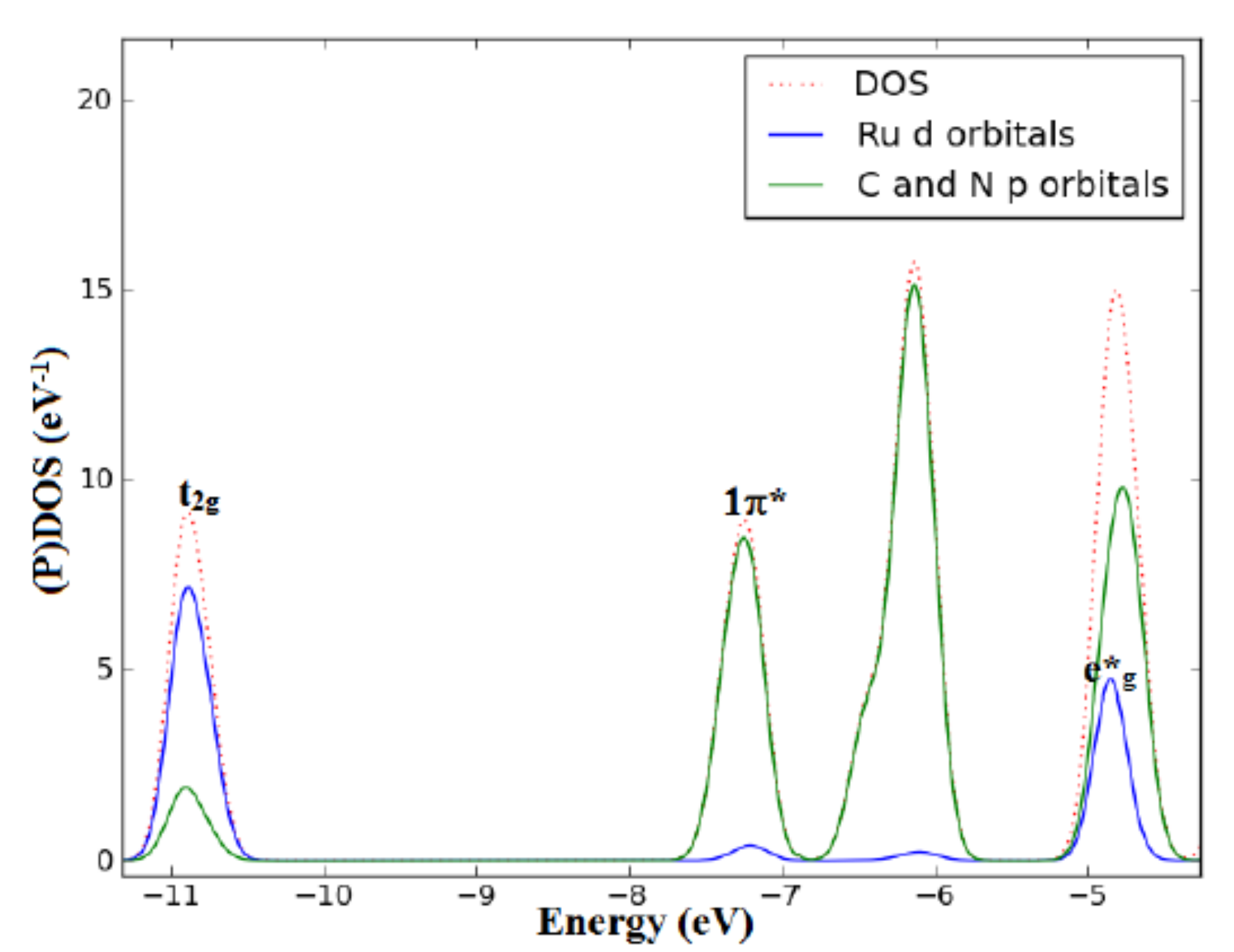} &
\includegraphics[width=0.4\textwidth]{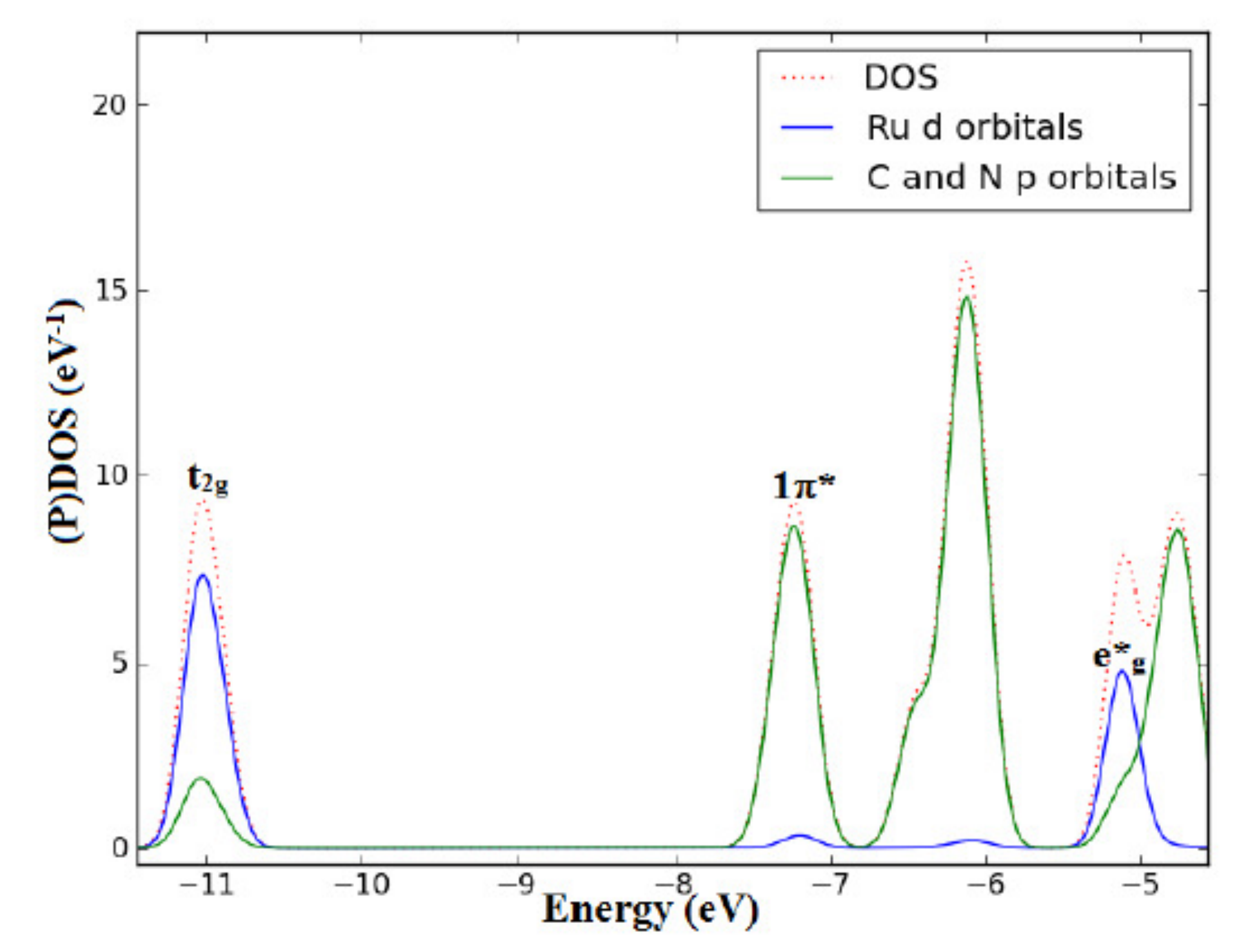} \\
B3LYP/6-31G & B3LYP/6-31G(d) \\
$\epsilon_{\text{HOMO}} = \mbox{-10.77 eV}$ & 
$\epsilon_{\text{HOMO}} = \mbox{-10.91 eV}$ 
\end{tabular}
\end{center}
Total and partial density of states of [Ru(bpy)(h-phen)]$^{2+}$
partitioned over Ru d orbitals and ligand C and N p orbitals. 
% for the 6-31G (left-hand side) and 6-31G* (right-hand side) basis sets.

\begin{center}
   {\bf Absorption Spectrum}
\end{center}

\begin{center}
\includegraphics[width=0.8\textwidth]{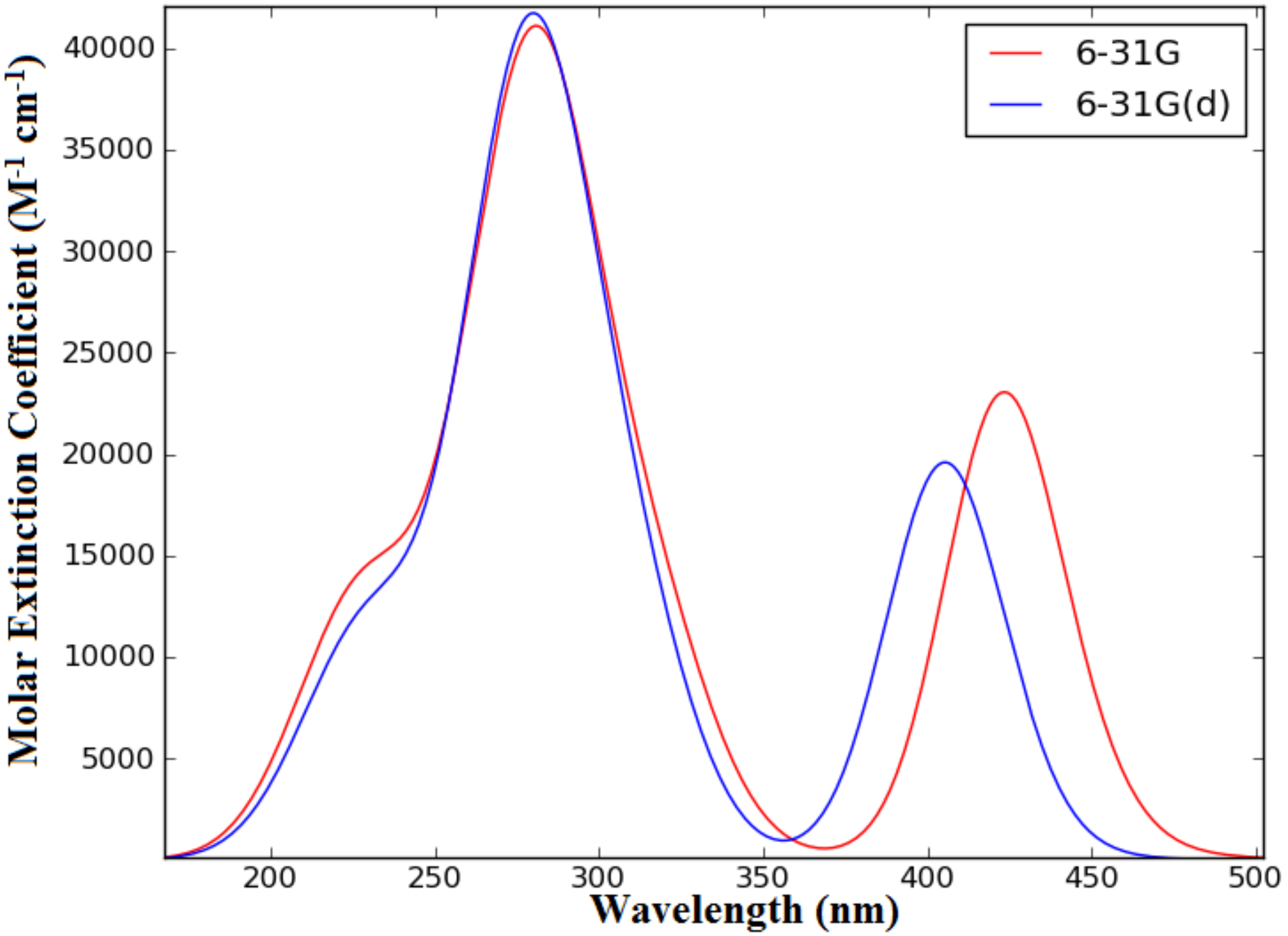}
\end{center}
[Ru(bpy)(h-phen)]$^{2+}$
TD-B3LYP/6-31G and TD-B3LYP/6-31G(d) spectra.

% ================================================
\newpage
\section{Complex {\bf (48)}: [Ru(bpy)(phen)$_2$]$^{2+}$}
% ================================================

\begin{center}
   {\bf PDOS}
\end{center}

\begin{center}
\begin{tabular}{cc}
\includegraphics[width=0.4\textwidth]{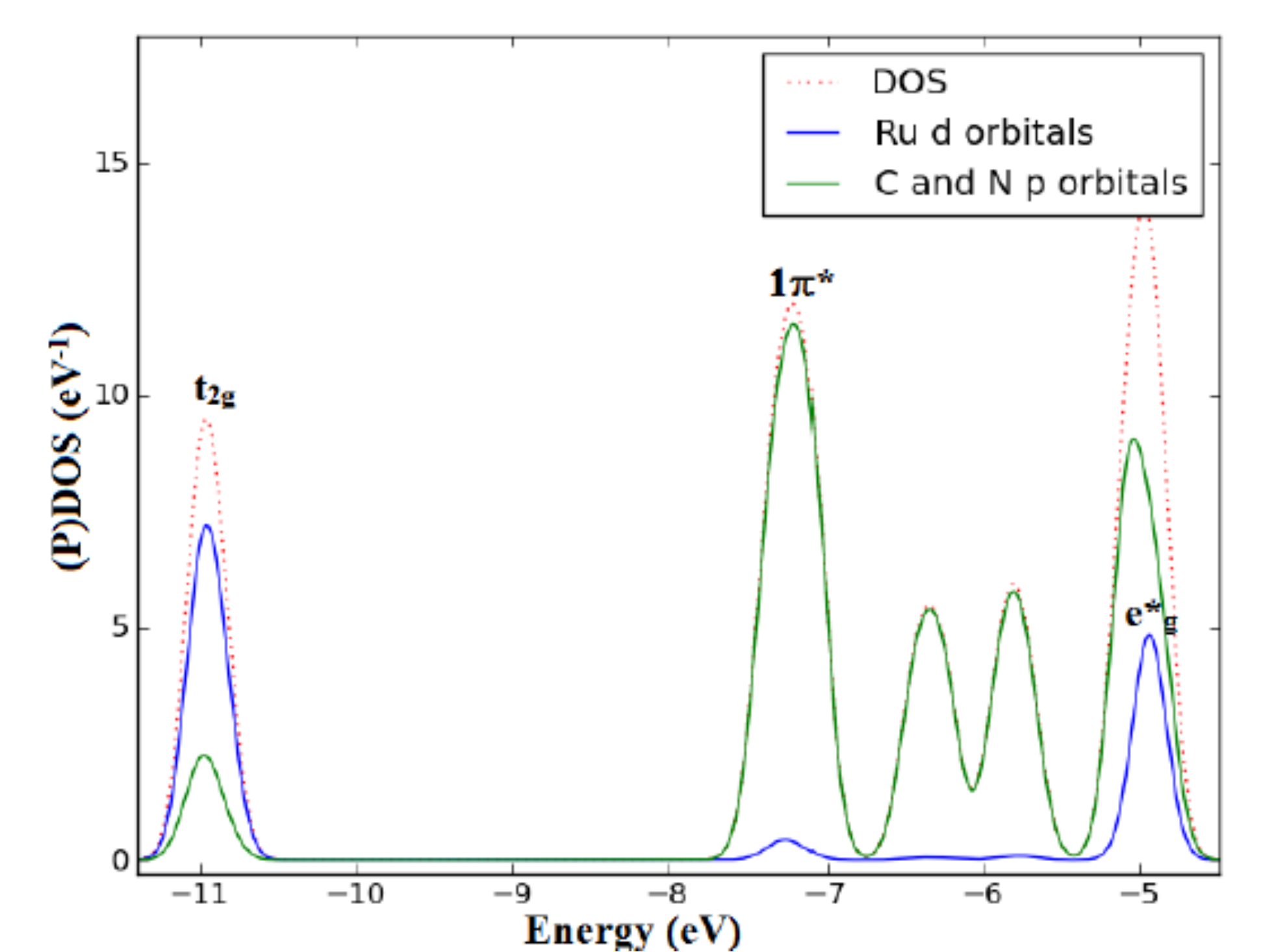} &
\includegraphics[width=0.4\textwidth]{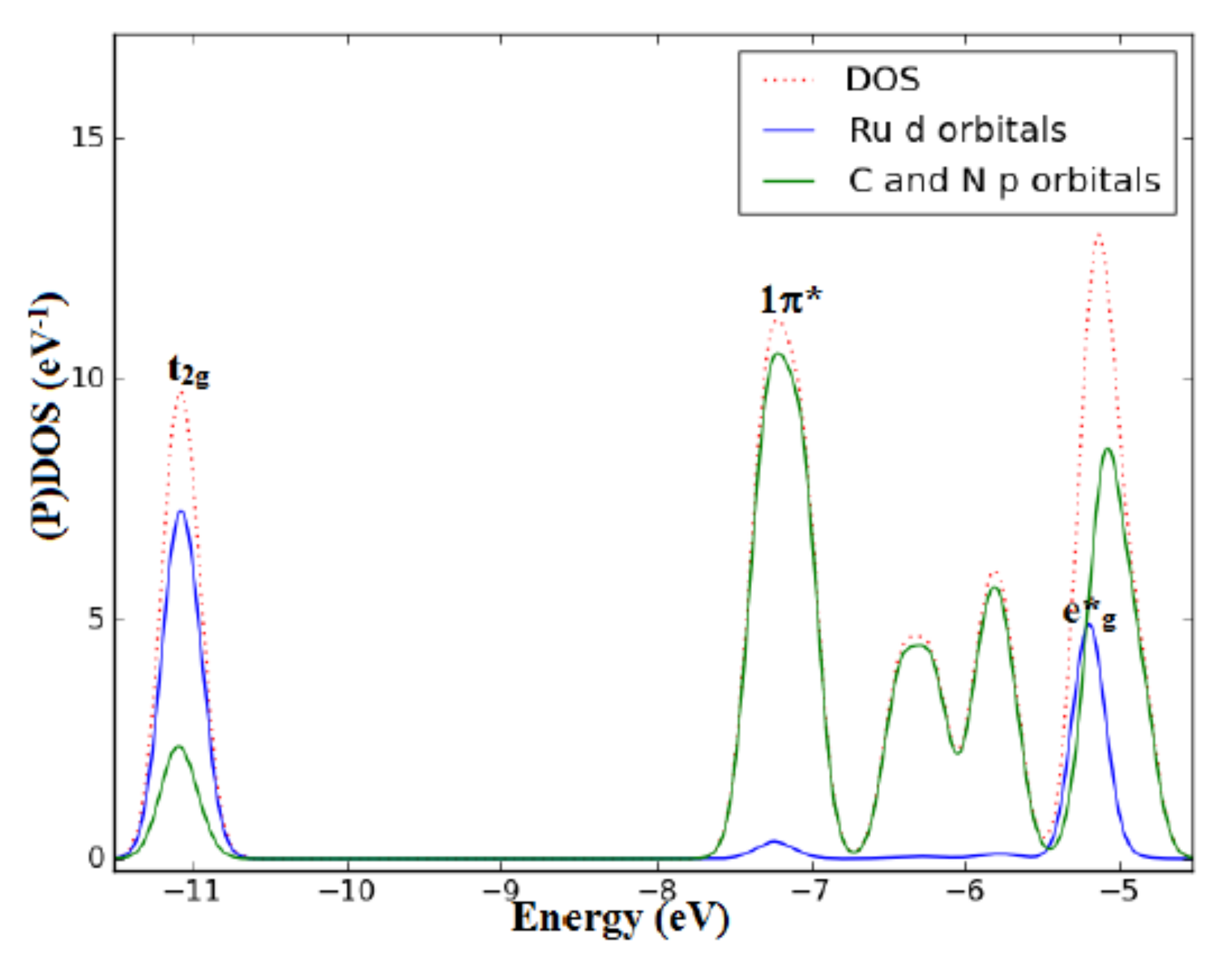} \\
B3LYP/6-31G & B3LYP/6-31G(d) \\
$\epsilon_{\text{HOMO}} = \mbox{-10.86 eV}$ & 
$\epsilon_{\text{HOMO}} = \mbox{-11.00 eV}$ 
\end{tabular}
\end{center}
Total and partial density of states of [Ru(bpy)(phen)$_2$]$^{2+}$
partitioned over Ru d orbitals and ligand C and N p orbitals.

\begin{center}
   {\bf Absorption Spectrum}
\end{center}

\begin{center}
\includegraphics[width=0.8\textwidth]{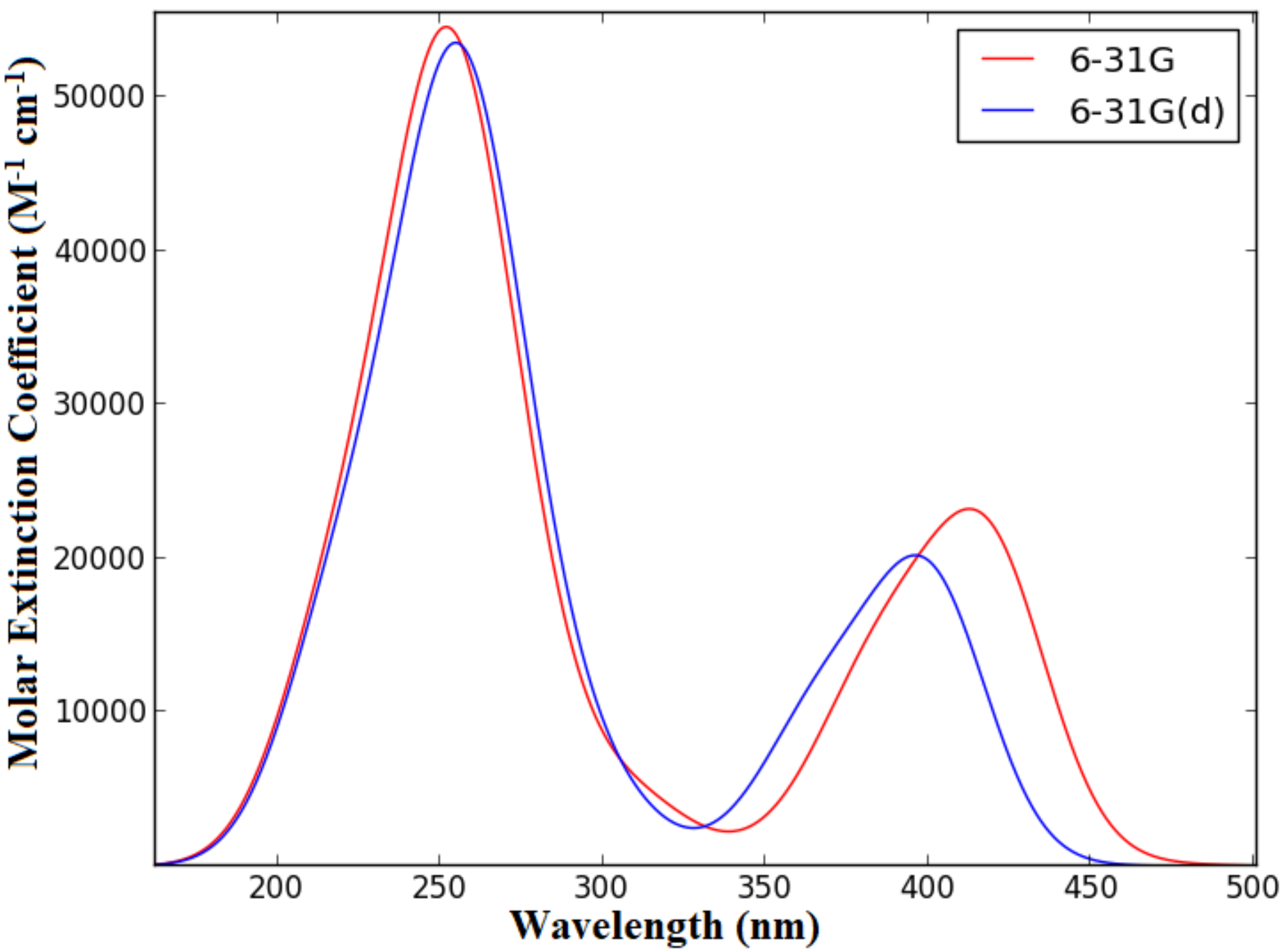}
\end{center}
[Ru(bpy)(phen)$_2$]$^{2+}$
TD-B3LYP/6-31G and TD-B3LYP/6-31G(d) spectra.

% % ================================================
% \newpage
% \section{Complex {\bf (49)}: {\em cis}-[Ru(bpy)(phen)(py)$_2$]$^{2+}$}
% % ================================================
% 
% \begin{center}
%    {\bf PDOS}
% \end{center}
% 
% \begin{center}
% \includegraphics[width=0.4\textwidth]{graphics1/framedquestionmark.pdf}
% \includegraphics[width=0.4\textwidth]{graphics1/framedquestionmark.pdf}
% \end{center}
% {\color{red} Do we have this?}
% 
% \begin{center}
%    {\bf Absorption Spectrum}
% \end{center}
% 
% \begin{center}
% \includegraphics[width=0.4\textwidth]{graphics1/framedquestionmark.pdf}
% \end{center}
% {\color{red} Do we have this?}

% ================================================
\newpage
\section{Complex {\bf (50)}: {\em trans}-[Ru(bpy)(phen)(py)$_2$]$^{2+}$}
% ================================================

\begin{center}
   {\bf PDOS}
\end{center}

\begin{center}
\begin{tabular}{cc}
\includegraphics[width=0.4\textwidth]{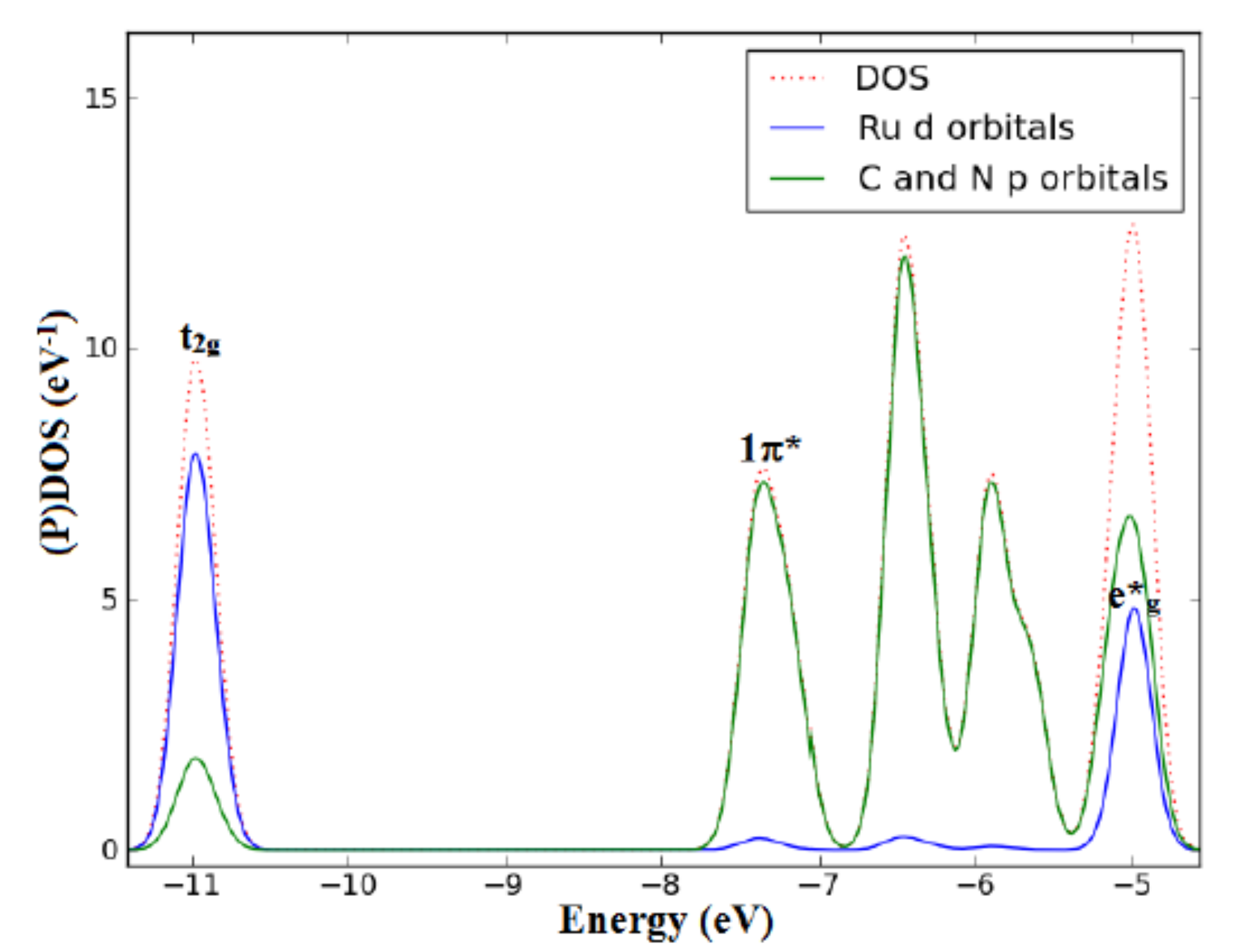} &
\includegraphics[width=0.4\textwidth]{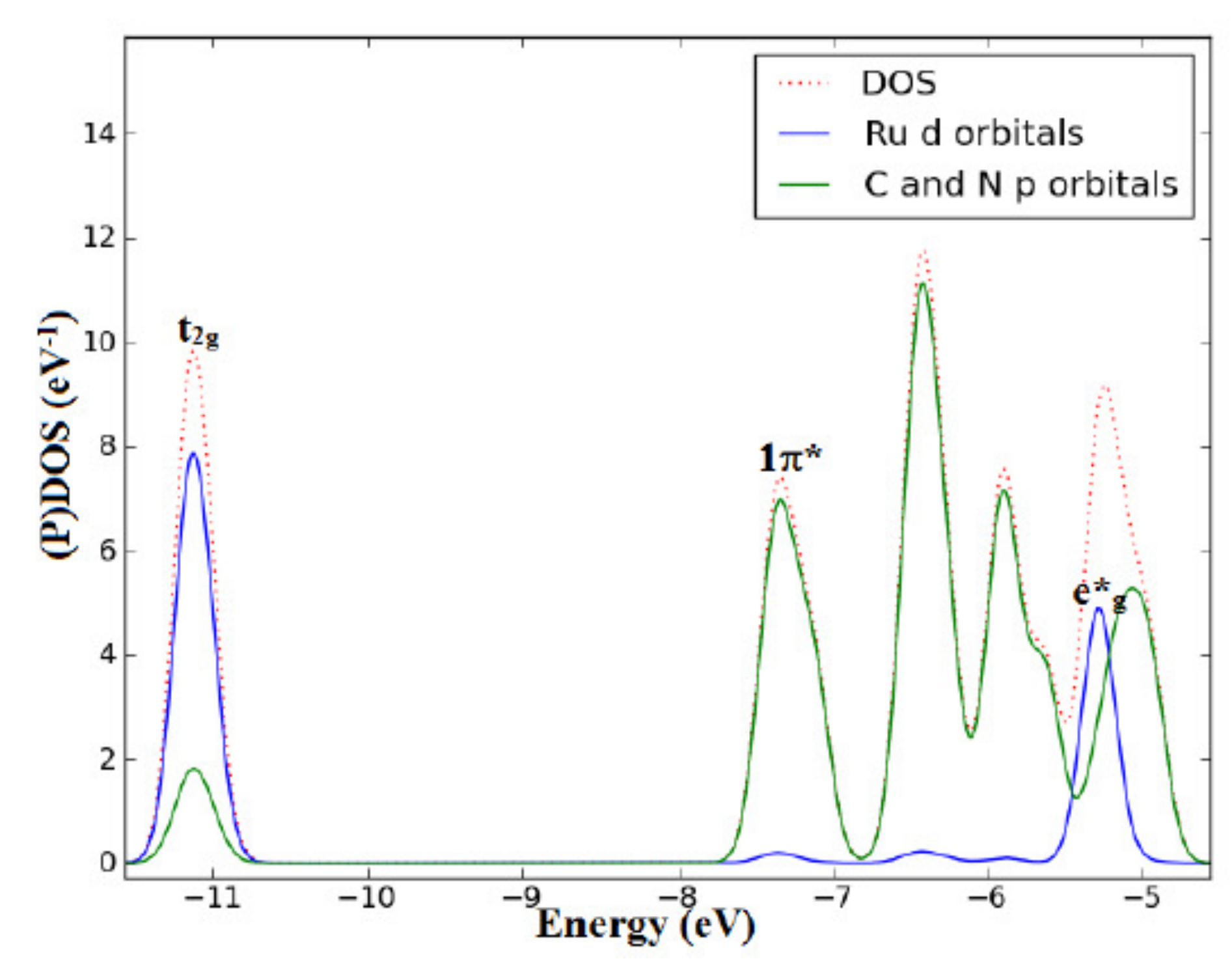} \\
B3LYP/6-31G & B3LYP/6-31G(d) \\
$\epsilon_{\text{HOMO}} = \mbox{-10.89 eV}$ & 
$\epsilon_{\text{HOMO}} = \mbox{-11.04 eV}$ 
\end{tabular}
\end{center}
Total and partial density of states of {\em trans}-[Ru(bpy)(phen)(py)$_2$]$^{2+}$
partitioned over Ru d orbitals and ligand C and N p orbitals. 
% for the 6-31G (left-hand side) and 6-31G* (right-hand side) basis sets.

\begin{center}
   {\bf Absorption Spectrum}
\end{center}

\begin{center}
\includegraphics[width=0.8\textwidth]{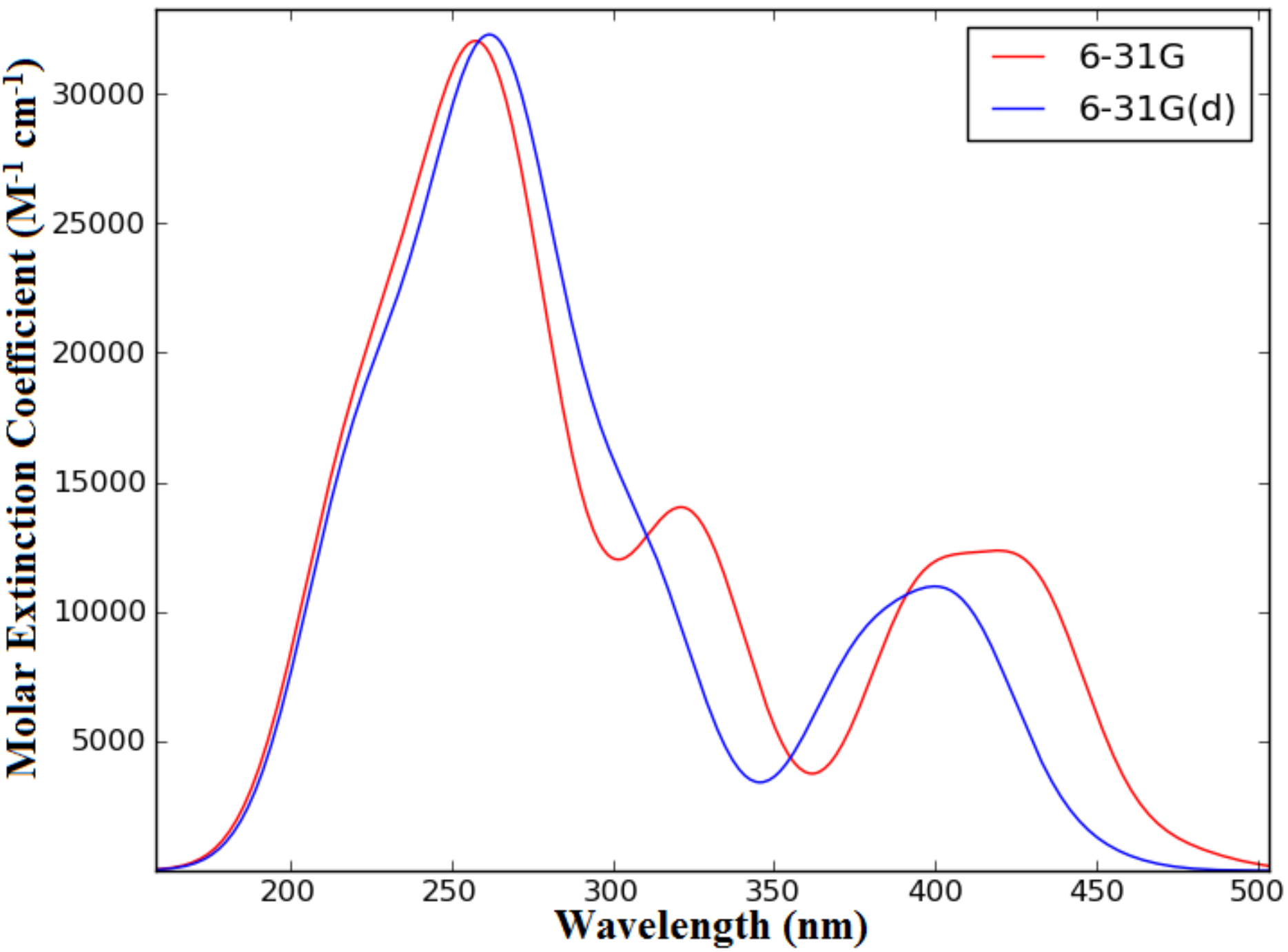}
\end{center}
{\em Trans}-[Ru(bpy)(phen)(py)$_2$]$^{2+}$
TD-B3LYP/6-31G and TD-B3LYP/6-31G(d) spectra.

% % ================================================
% \newpage
% \section{Complex {\bf (51)}: [Ru(bpy)(DIAFO)$_2$]$^{2+}$}
% % ================================================
% 
% {\color{magenta} \sf The geometry optimzation was unsuccessful for this complex.}
% 
% \begin{center}
%    {\bf PDOS}
% \end{center}

% \begin{center}
% \includegraphics[width=0.4\textwidth]{graphics1/framedquestionmark.pdf}
% \includegraphics[width=0.4\textwidth]{graphics1/framedquestionmark.pdf}
% \end{center}
% {\color{red} Do we have this?}

% \begin{center}
%    {\bf Absorption Spectrum}
% \end{center}
 
% \begin{center}
% \includegraphics[width=0.4\textwidth]{graphics1/framedquestionmark.pdf}
% \end{center}
% {\color{red} Do we have this?}

% =========================================-=======
\newpage
\section{Complex {\bf (52)}: [Ru(bpy)(taphen)$_2$]$^{2+}$}
% ================================================

\begin{center}
   {\bf PDOS}
\end{center}

\begin{center}
\begin{tabular}{cc}
\includegraphics[width=0.4\textwidth]{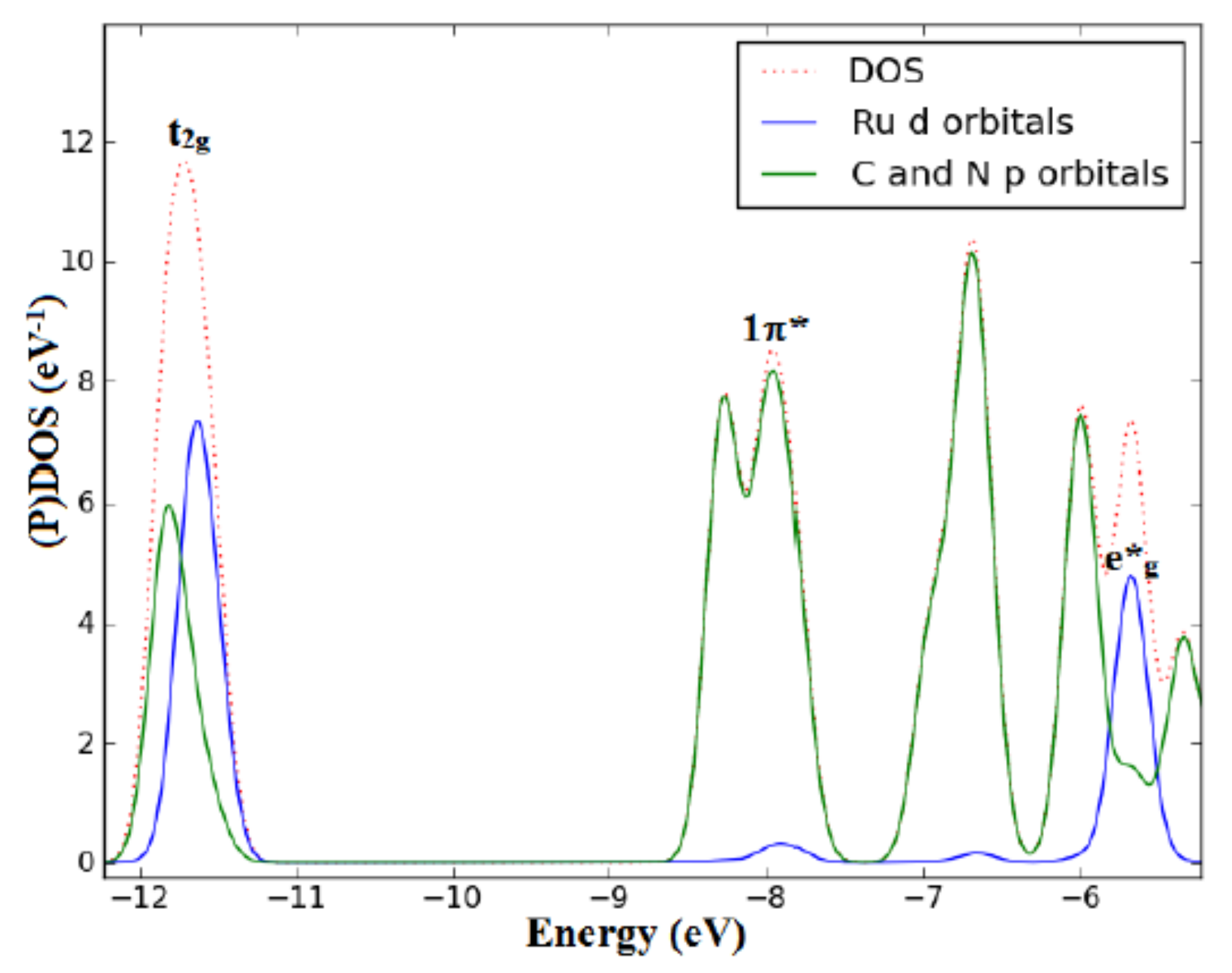} &
\includegraphics[width=0.4\textwidth]{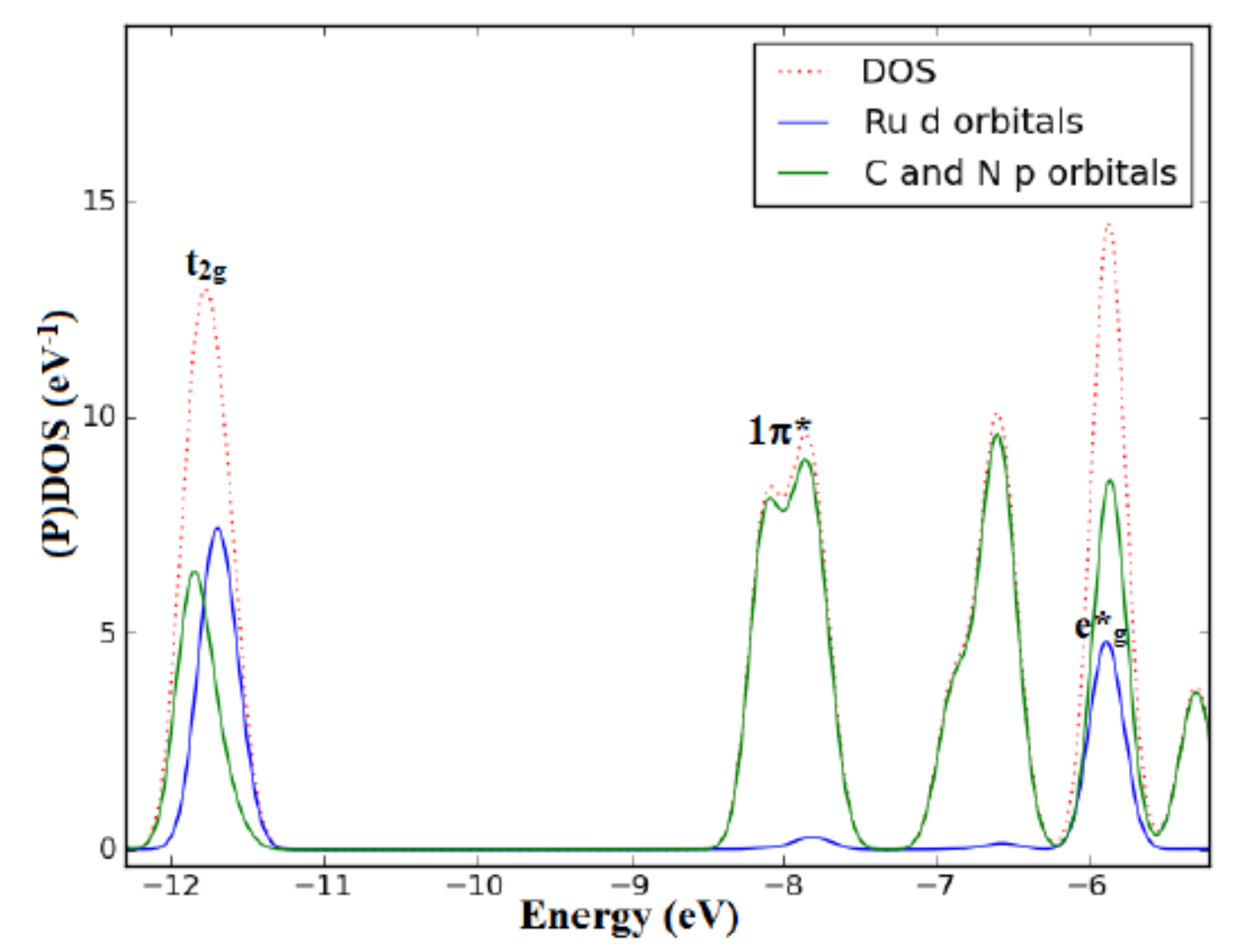} \\
B3LYP/6-31G & B3LYP/6-31G(d) \\
$\epsilon_{\text{HOMO}} = \mbox{-11.53 eV}$ & 
$\epsilon_{\text{HOMO}} = \mbox{-11.60 eV}$ 
\end{tabular}
\end{center}
Total and partial density of states of [Ru(bpy)(taphen)$_2$]$^{2+}$
partitioned over Ru d orbitals and ligand C and N p orbitals.
% for the 6-31G (left-hand side) and 6-31G* (right-hand side) basis sets.

\begin{center}
   {\bf Absorption Spectrum}
\end{center}

\begin{center}
\includegraphics[width=0.8\textwidth]{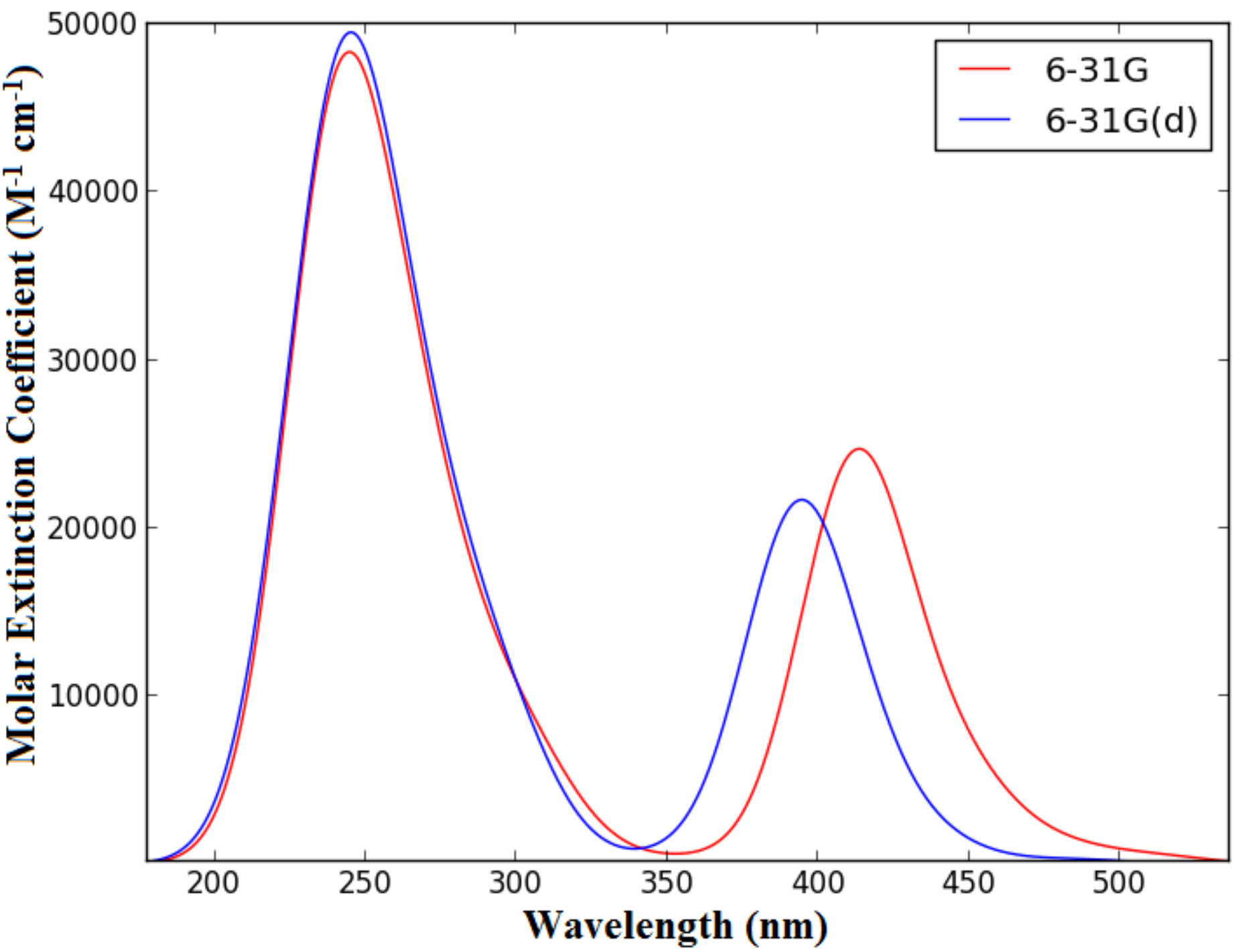}
\end{center}
[Ru(bpy)(taphen)$_2$]$^{2+}$
TD-B3LYP/6-31G and TD-B3LYP/6-31G(d) spectra.

% ================================================
\newpage
\section{Complex {\bf (53)}: [Ru(bpy)(py)$_2$(en)]$^{2+}$}
% ================================================

\begin{center}
   {\bf PDOS}
\end{center}

\begin{center}
\begin{tabular}{cc}
\includegraphics[width=0.4\textwidth]{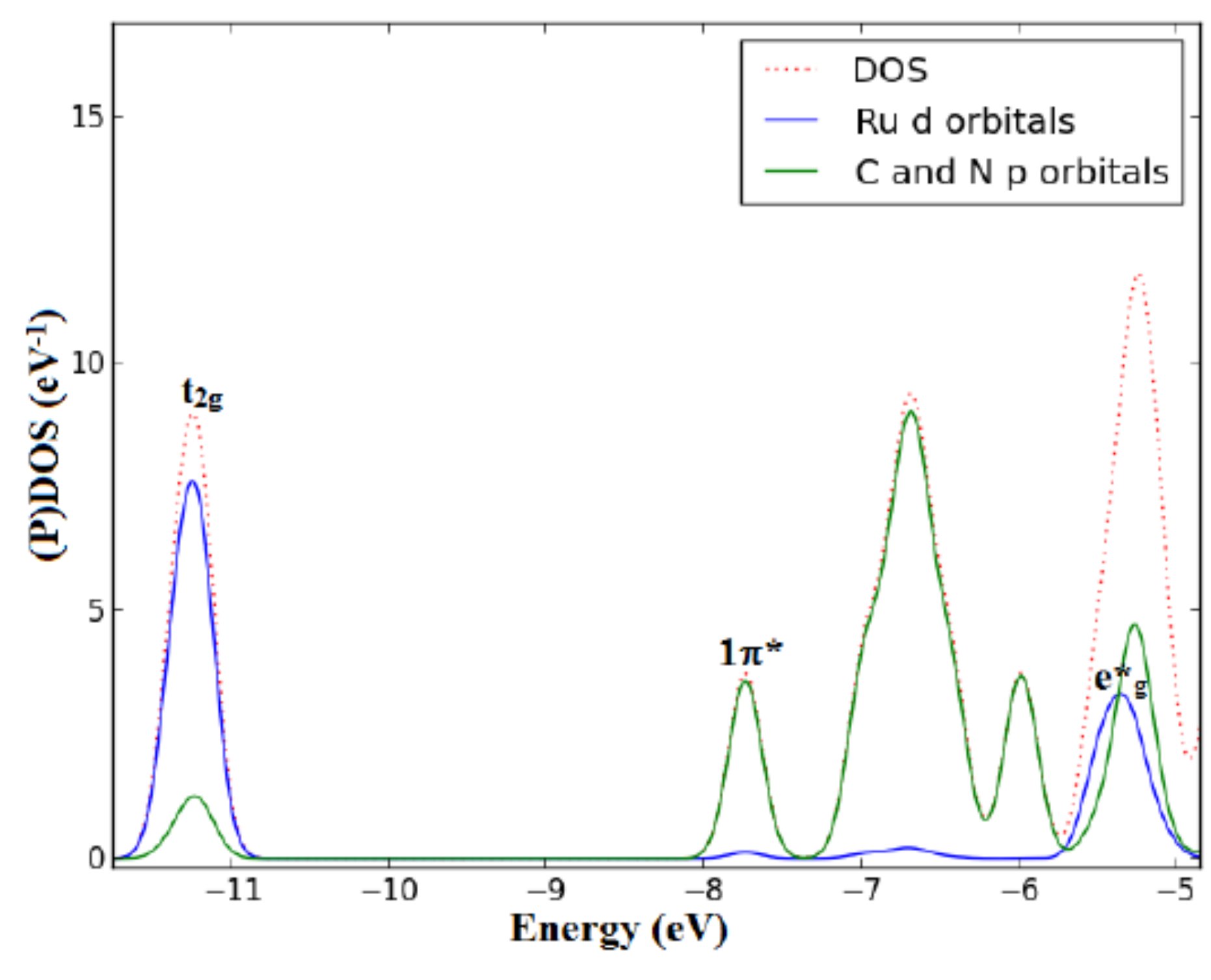} &
\includegraphics[width=0.4\textwidth]{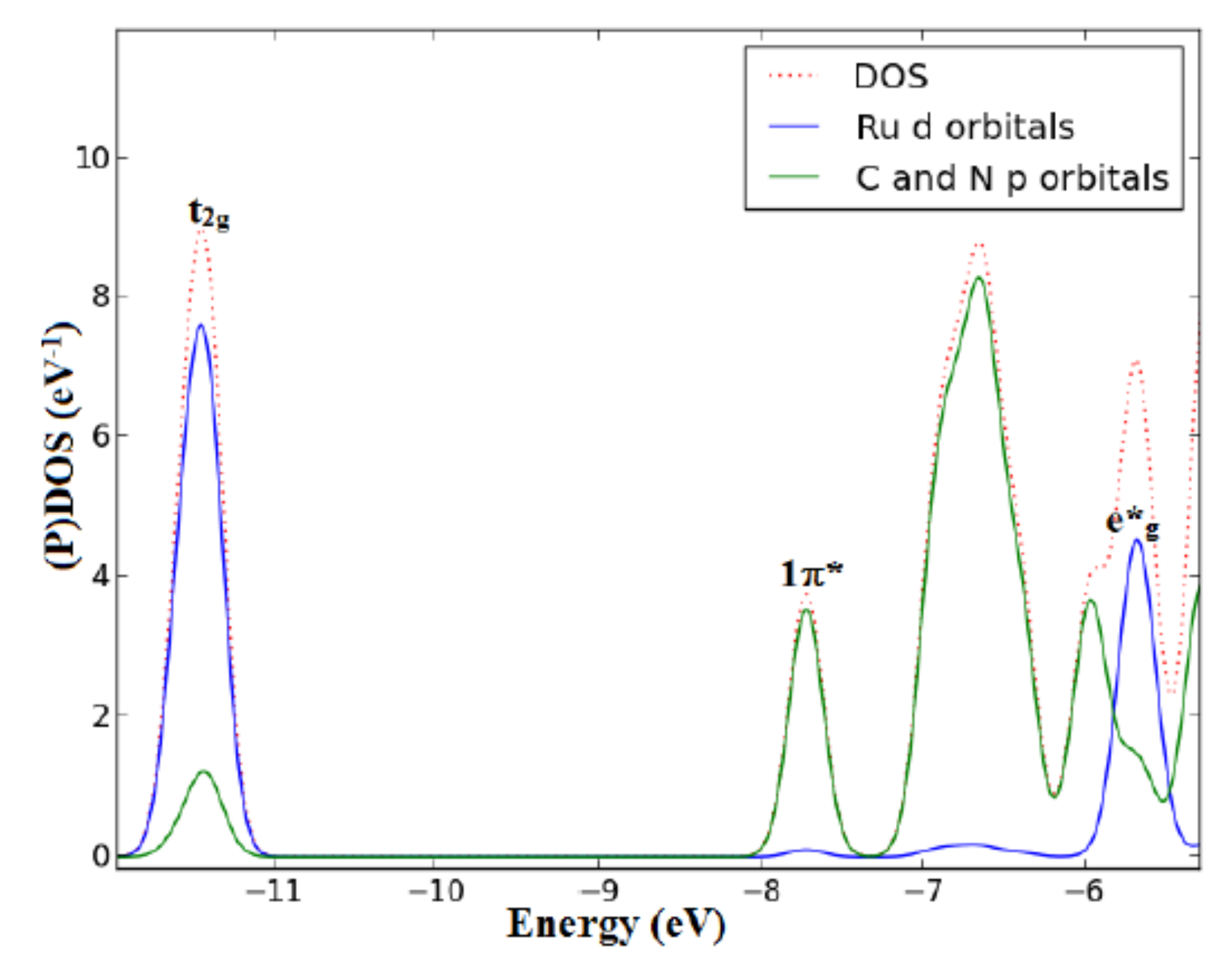} \\
B3LYP/6-31G & B3LYP/6-31G(d) \\
$\epsilon_{\text{HOMO}} = \mbox{-11.19 eV}$ & 
$\epsilon_{\text{HOMO}} = \mbox{-11.39 eV}$ 
\end{tabular}
\end{center}
Total and partial density of states of [Ru(bpy)(py)$_2$(en)]$^{2+}$
partitioned over Ru d orbitals and ligand C and N p orbitals.
% for the 6-31G (left-hand side) and 6-31G* (right-hand side) basis sets.

\begin{center}
   {\bf Absorption Spectrum}
\end{center}

\begin{center}
\includegraphics[width=0.8\textwidth]{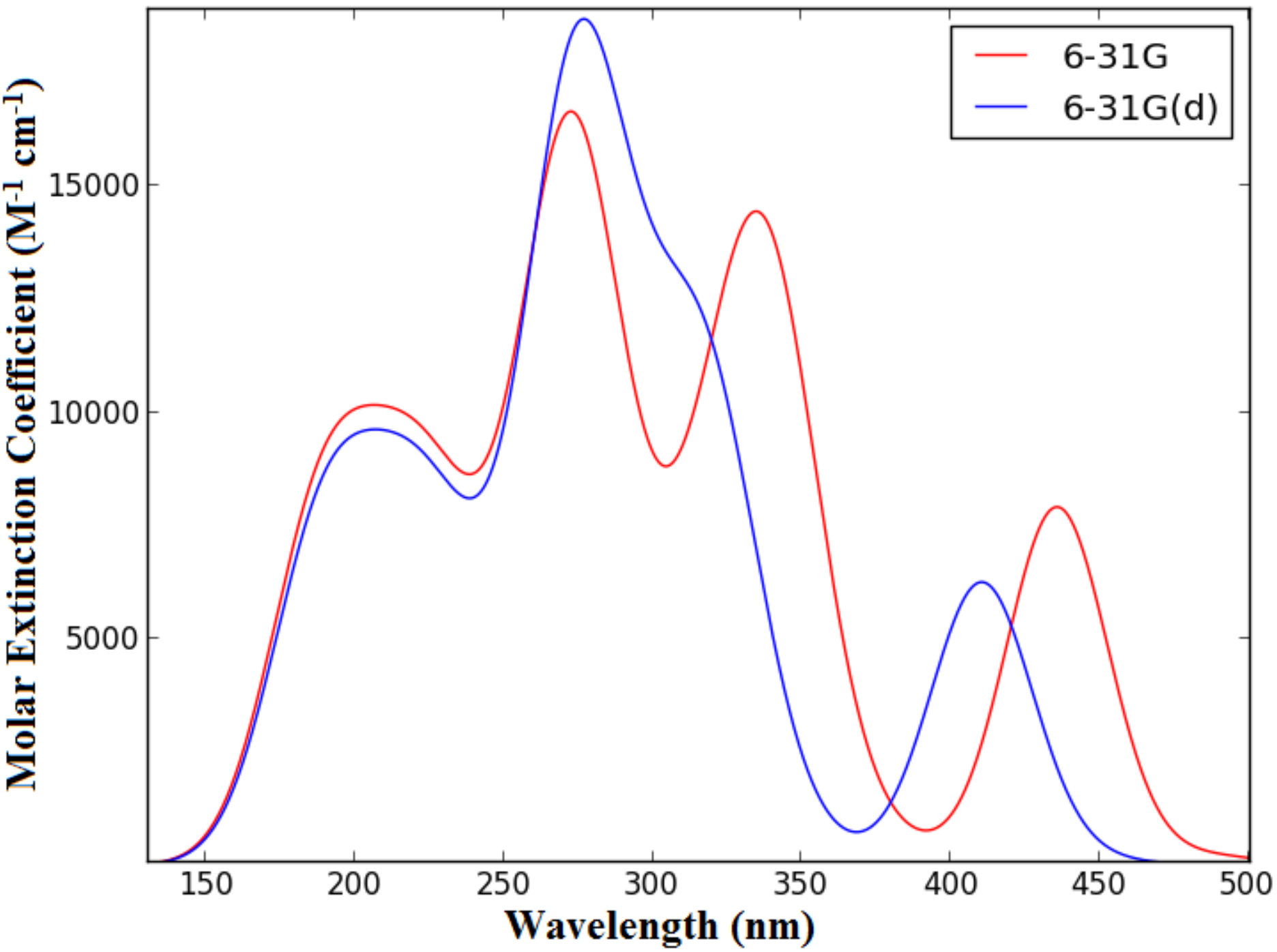}
\end{center}
[Ru(bpy)(py)$_2$(en)]$^{2+}$
TD-B3LYP/6-31G and TD-B3LYP/6-31G(d) spectra.

% % ================================================
% \newpage
% \section{Complex {\bf (54)}: [Ru(bpy)(py)$_3$Cl]$^{+}$}
% % ================================================
% 
% \begin{center}
%    {\bf PDOS}
% \end{center}
% 
% \begin{center}
% \includegraphics[width=0.4\textwidth]{graphics1/framedquestionmark.pdf}
% \includegraphics[width=0.4\textwidth]{graphics1/framedquestionmark.pdf}
% \end{center}
% {\color{magenta} PDOS could not be calculated for complexes containing Cl.}
% 
% \begin{center}
%    {\bf Absorption Spectrum}
% \end{center}
% 
% \begin{center}
% \includegraphics[width=0.4\textwidth]{graphics1/framedquestionmark.pdf}
% \end{center}
% {\color{red} Do we have this?}

% ================================================
\newpage
\section{Complex {\bf (55)}: [Ru(bpy)(py)$_4$]$^{2+}$}
% ================================================

\begin{center}
   {\bf PDOS}
\end{center}

\begin{center}
\begin{tabular}{cc}
\includegraphics[width=0.4\textwidth]{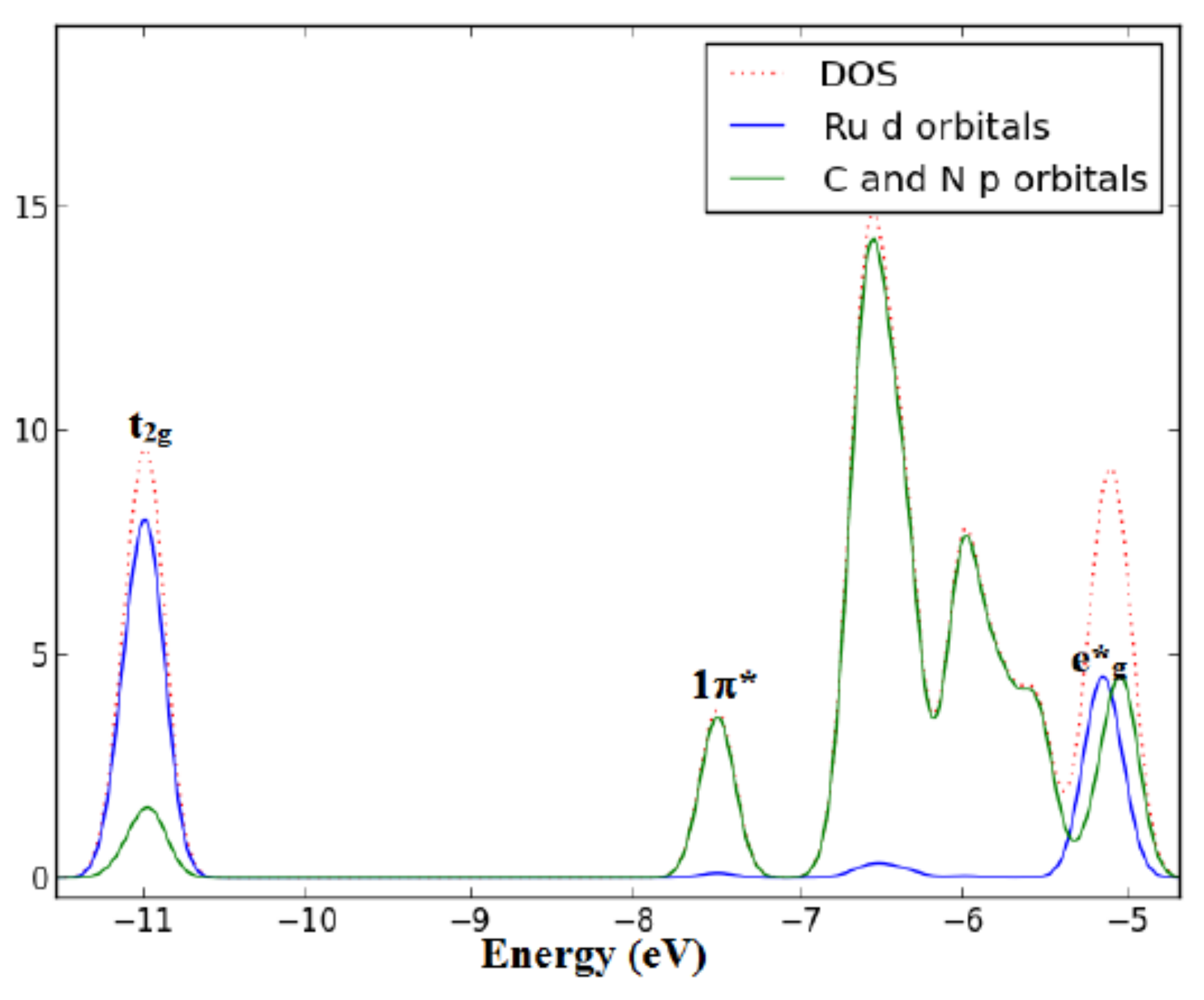} &
\includegraphics[width=0.4\textwidth]{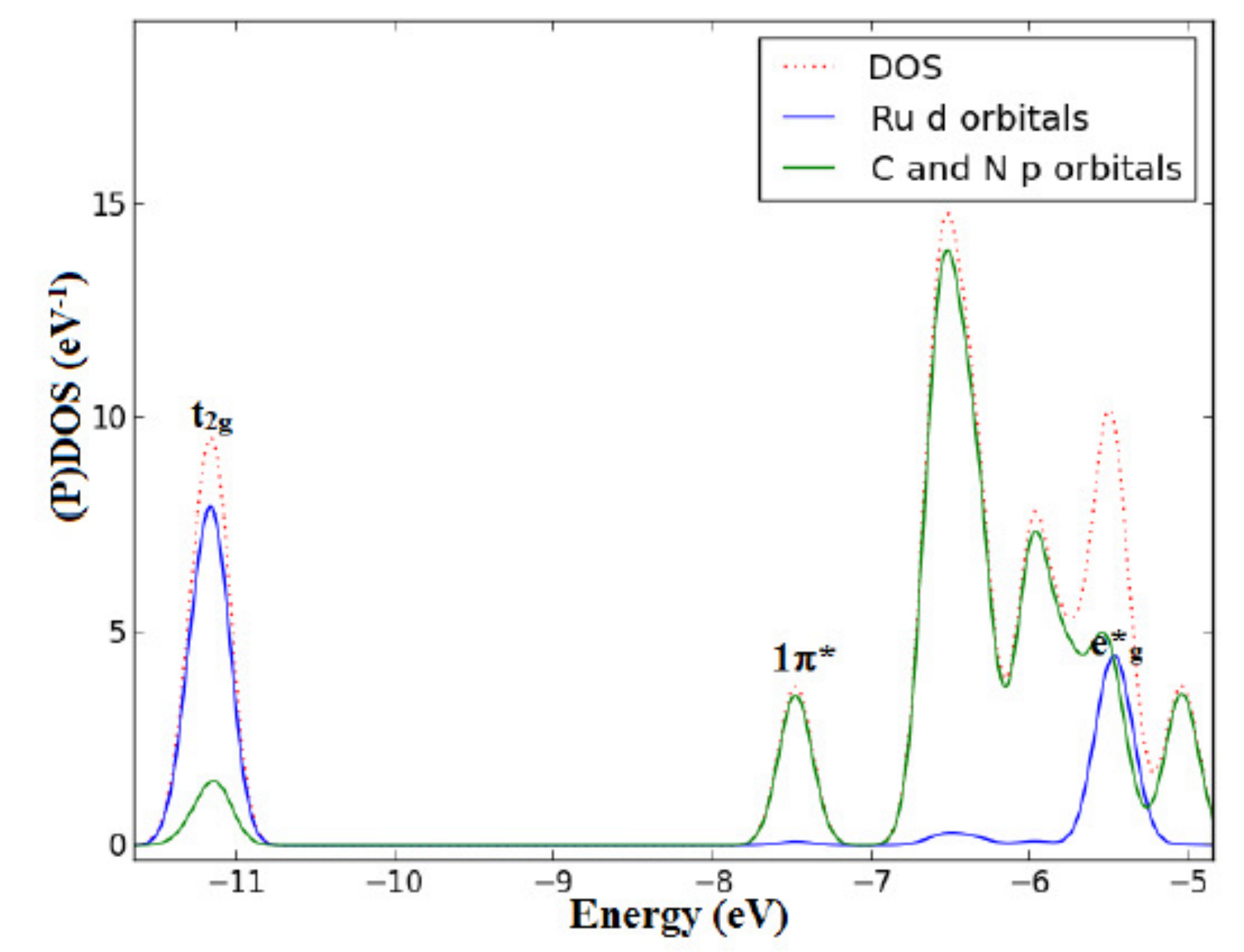} \\
B3LYP/6-31G & B3LYP/6-31G(d) \\
$\epsilon_{\text{HOMO}} = \mbox{-10.96 eV}$ & 
$\epsilon_{\text{HOMO}} = \mbox{-11.13 eV}$ 
\end{tabular}
\end{center}
Total and partial density of states of [Ru(bpy)(py)$_4$]$^{2+}$
partitioned over Ru d orbitals and ligand C and N p orbitals.
% for the 6-31G (left-hand side) and 6-31G* (right-hand side) basis sets.

\begin{center}
   {\bf Absorption Spectrum}
\end{center}

\begin{center}
\includegraphics[width=0.8\textwidth]{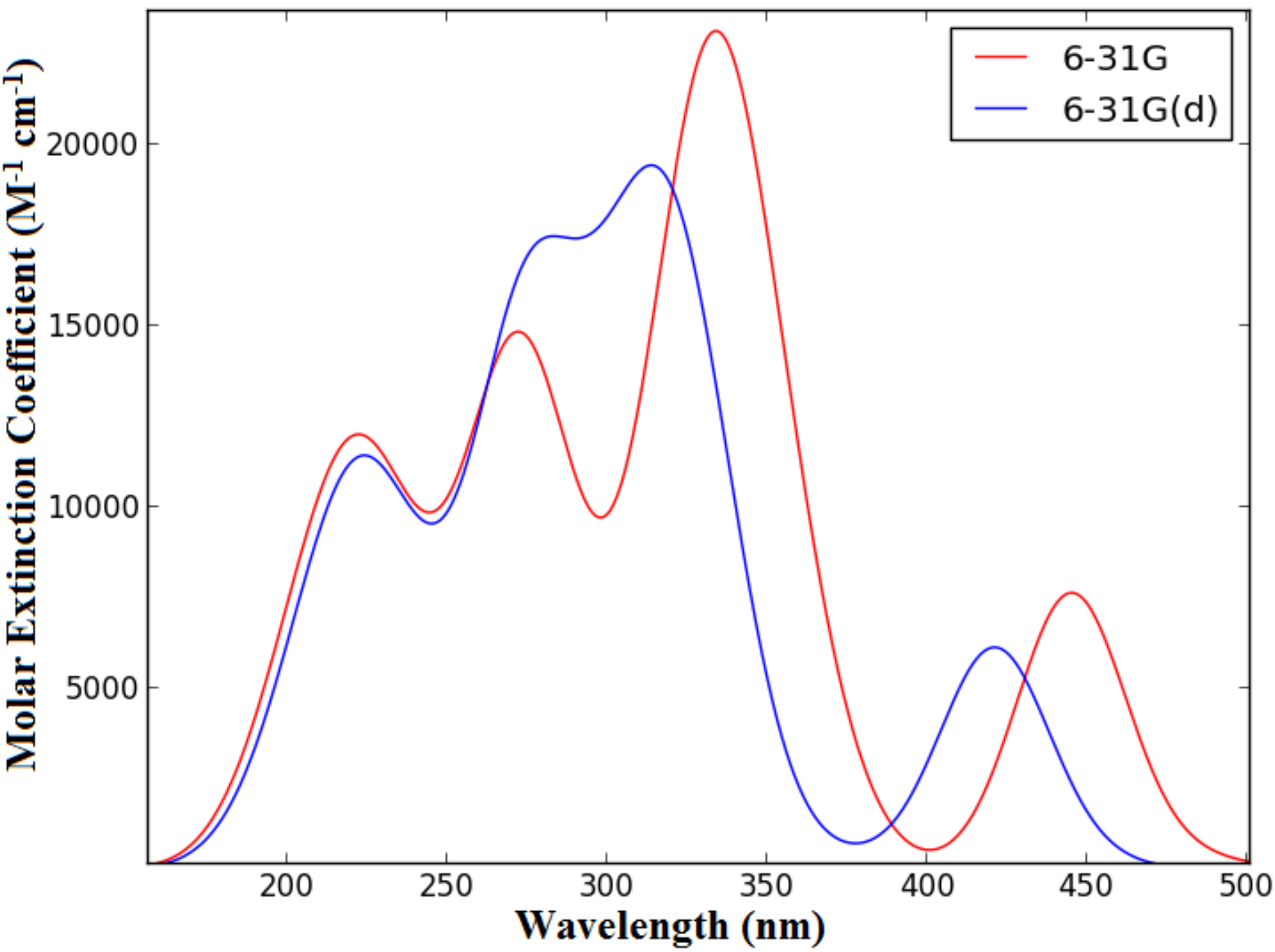}
\end{center}
[Ru(bpy)(py)$_4$]$^{2+}$
TD-B3LYP/6-31G and TD-B3LYP/6-31G(d) spectra.

% ================================================
\newpage
\section{Complex {\bf (56)}: [Ru(bpy)(py)$_2$(PMA)]$^{2+}$}
% ================================================

\begin{center}
   {\bf PDOS}
\end{center}

\begin{center}
\begin{tabular}{cc}
\includegraphics[width=0.4\textwidth]{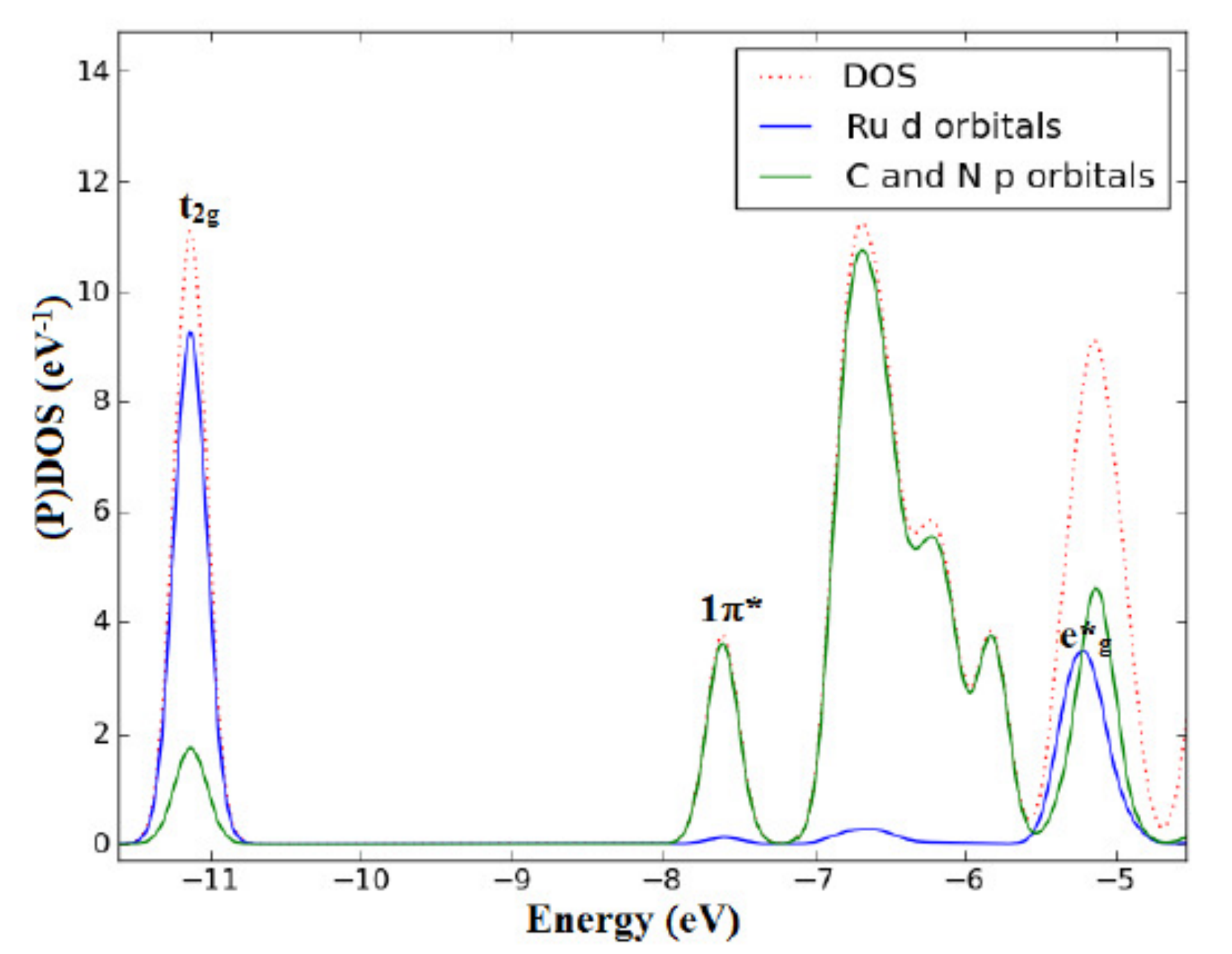} &
\includegraphics[width=0.4\textwidth]{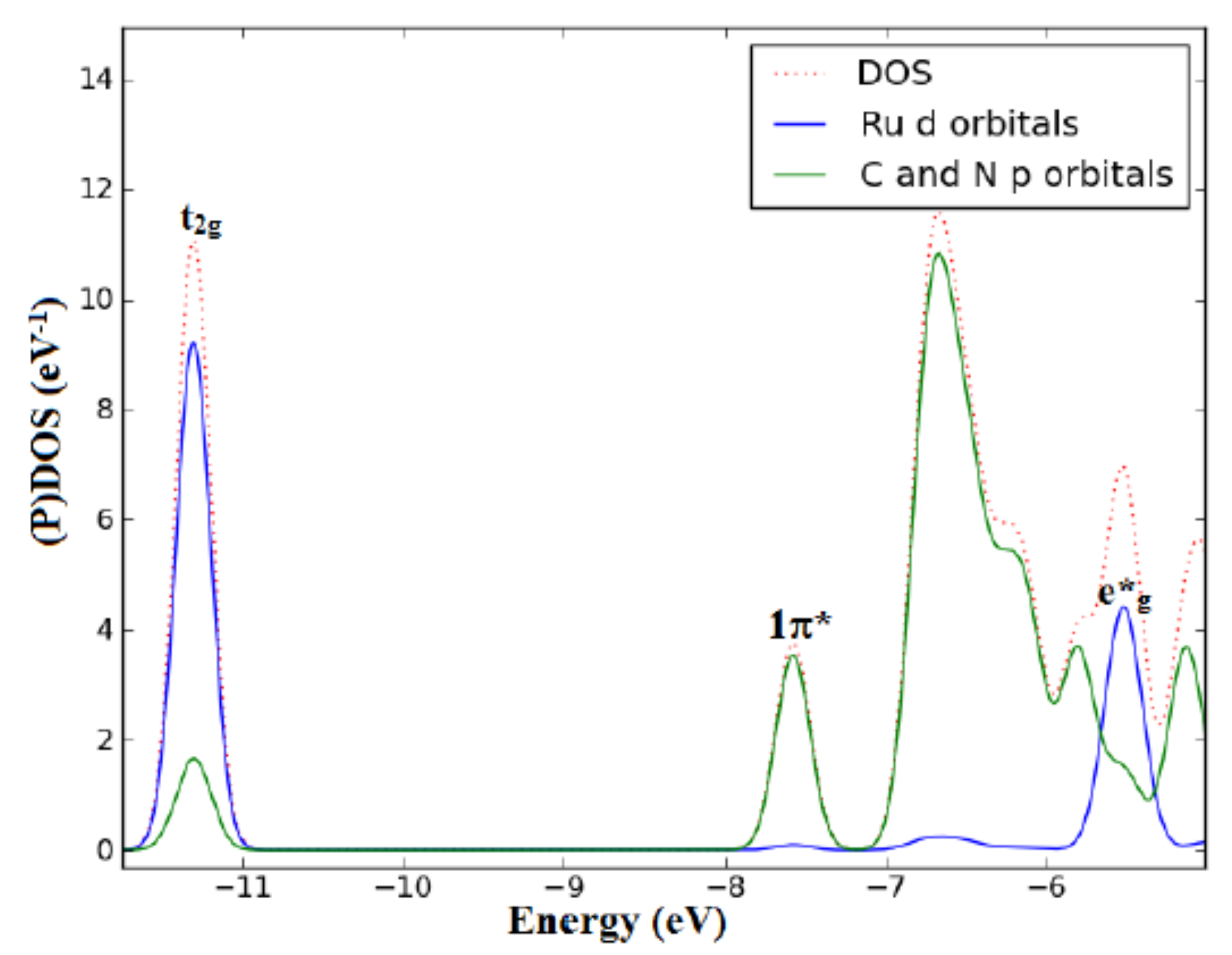} \\
B3LYP/6-31G & B3LYP/6-31G(d) \\
$\epsilon_{\text{HOMO}} = \mbox{-11.12 eV}$ & 
$\epsilon_{\text{HOMO}} = \mbox{-11.29 eV}$ 
\end{tabular}
\end{center}
Total and partial density of states of [Ru(bpy)(py)$_2$(PMA)]$^{2+}$
partitioned over Ru d orbitals and ligand C and N p orbitals. 
% for the 6-31G (left-hand side) and 6-31G* (right-hand side) basis sets.

\begin{center}
   {\bf Absorption Spectrum}
\end{center}

\begin{center}
\includegraphics[width=0.8\textwidth]{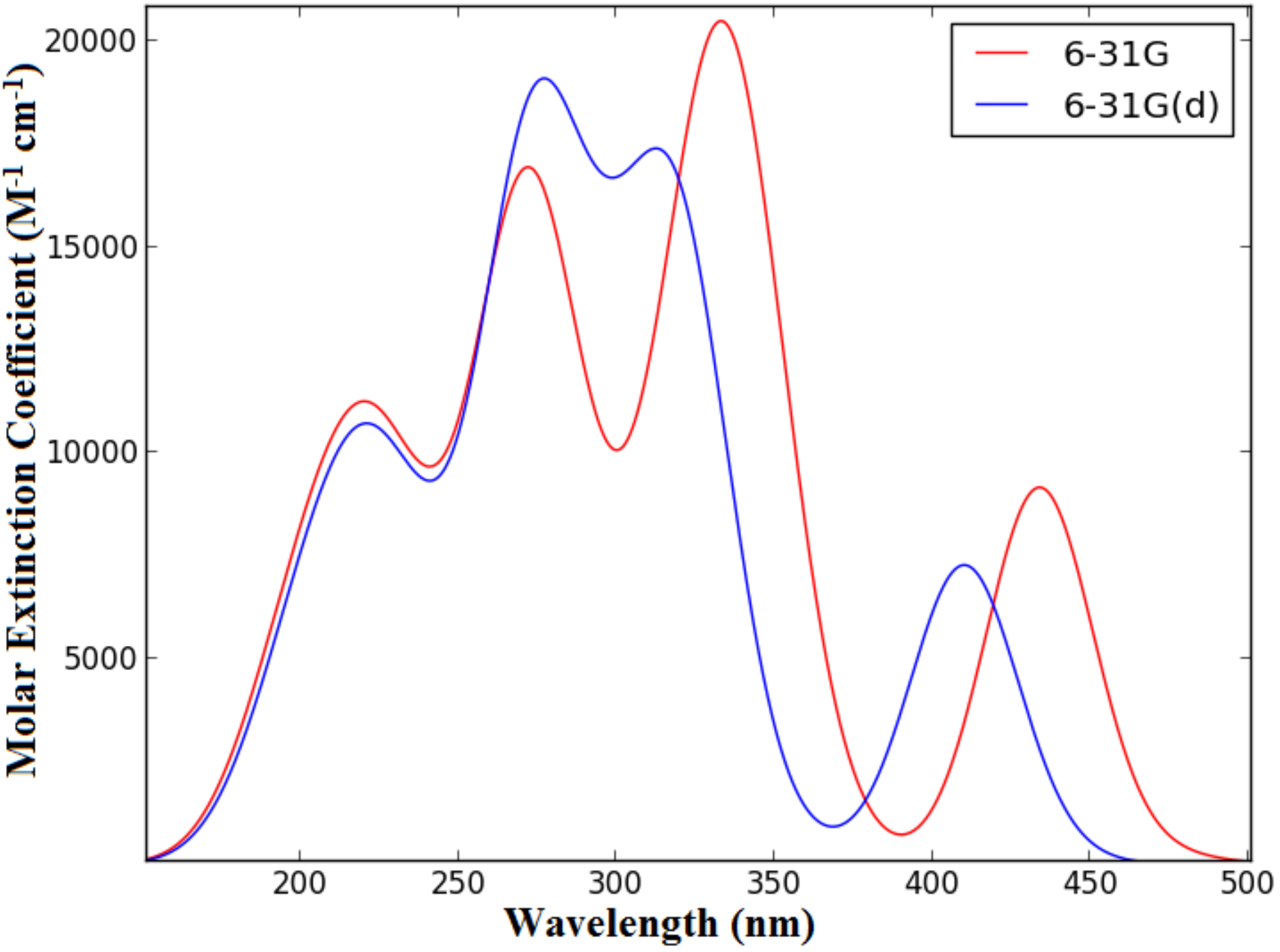}
\end{center}
[Ru(bpy)(py)$_2$(PMA)]$^{2+}$
TD-B3LYP/6-31G and TD-B3LYP/6-31G(d) spectra.

% ================================================
\newpage
\section{Complex {\bf (57)}: [Ru(bpy)(py)$_2$(2-AEP)]$^{2+}$}
% ================================================

\begin{center}
   {\bf PDOS}
\end{center}

\begin{center}
\begin{tabular}{cc}
\includegraphics[width=0.4\textwidth]{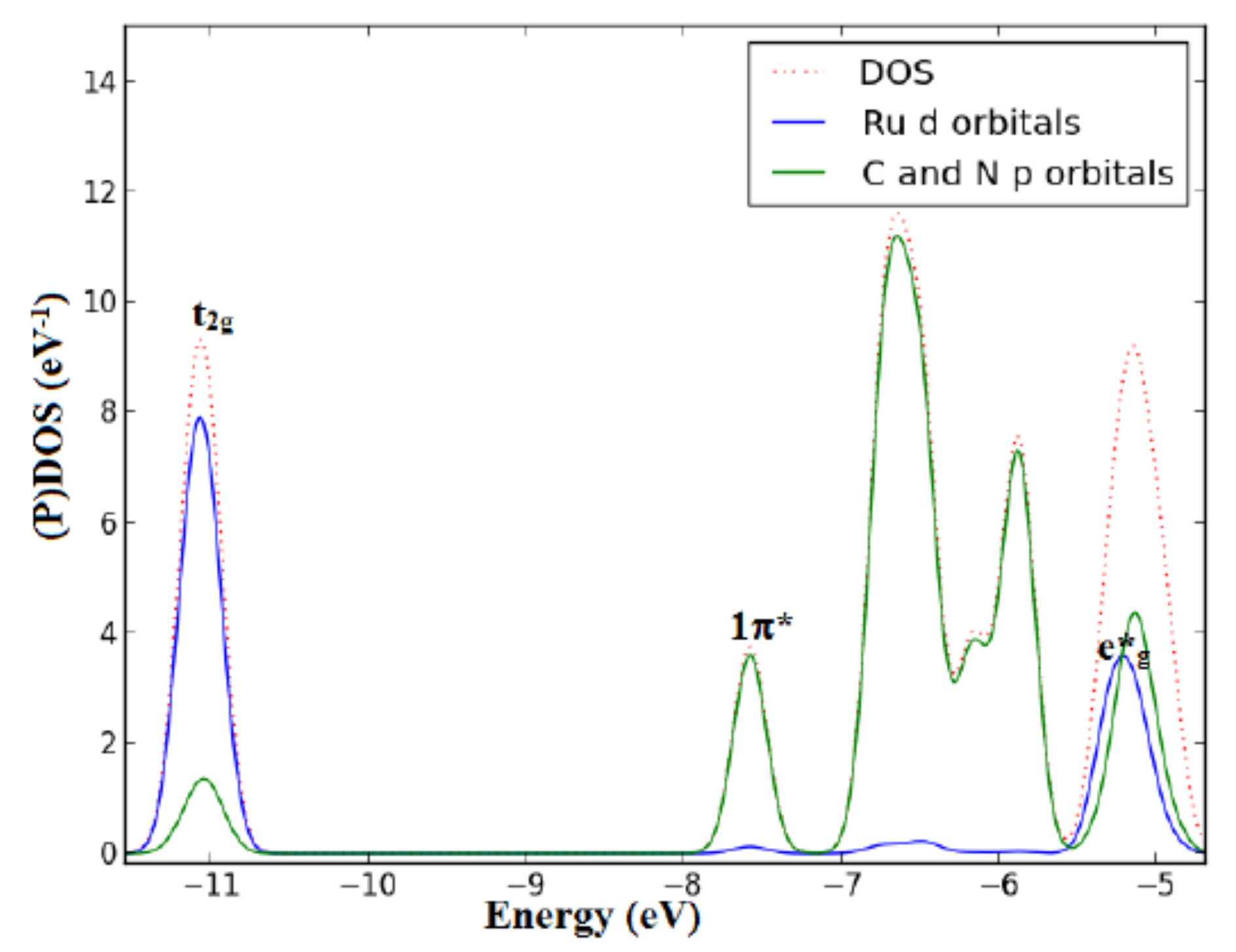} &
\includegraphics[width=0.4\textwidth]{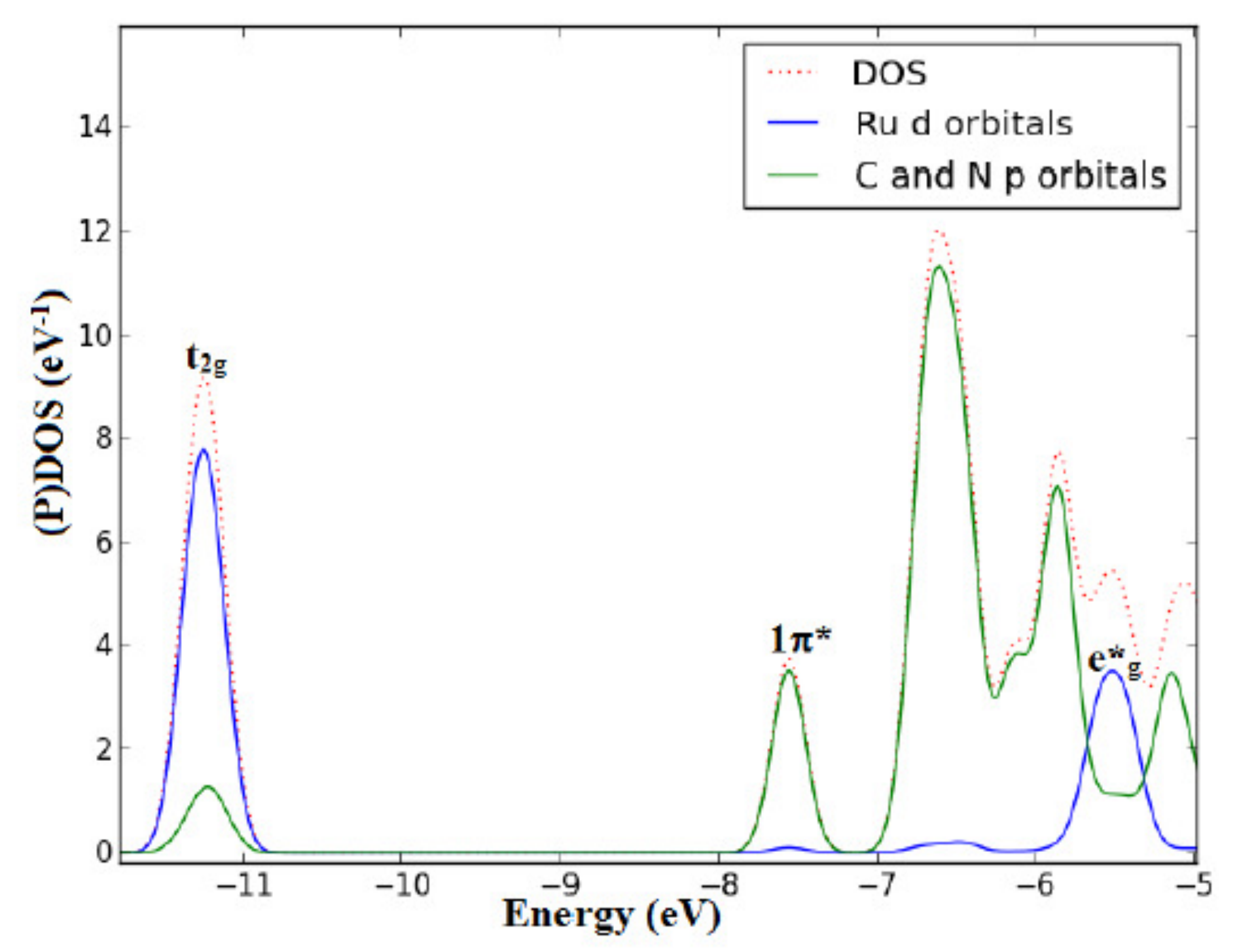} \\
B3LYP/6-31G & B3LYP/6-31G(d) \\
$\epsilon_{\text{HOMO}} = \mbox{-10.99 eV}$ & 
$\epsilon_{\text{HOMO}} = \mbox{-11.17 eV}$ 
\end{tabular}
\end{center}
Total and partial density of states of [Ru(bpy)(py)$_2$(2-AEP)]$^{2+}$
partitioned over Ru d orbitals and ligand C and N p orbitals. 
% for the 6-31G (left-hand side) and 6-31G* (right-hand side) basis sets.

\begin{center}
   {\bf Absorption Spectrum}
\end{center}

\begin{center}
\includegraphics[width=0.8\textwidth]{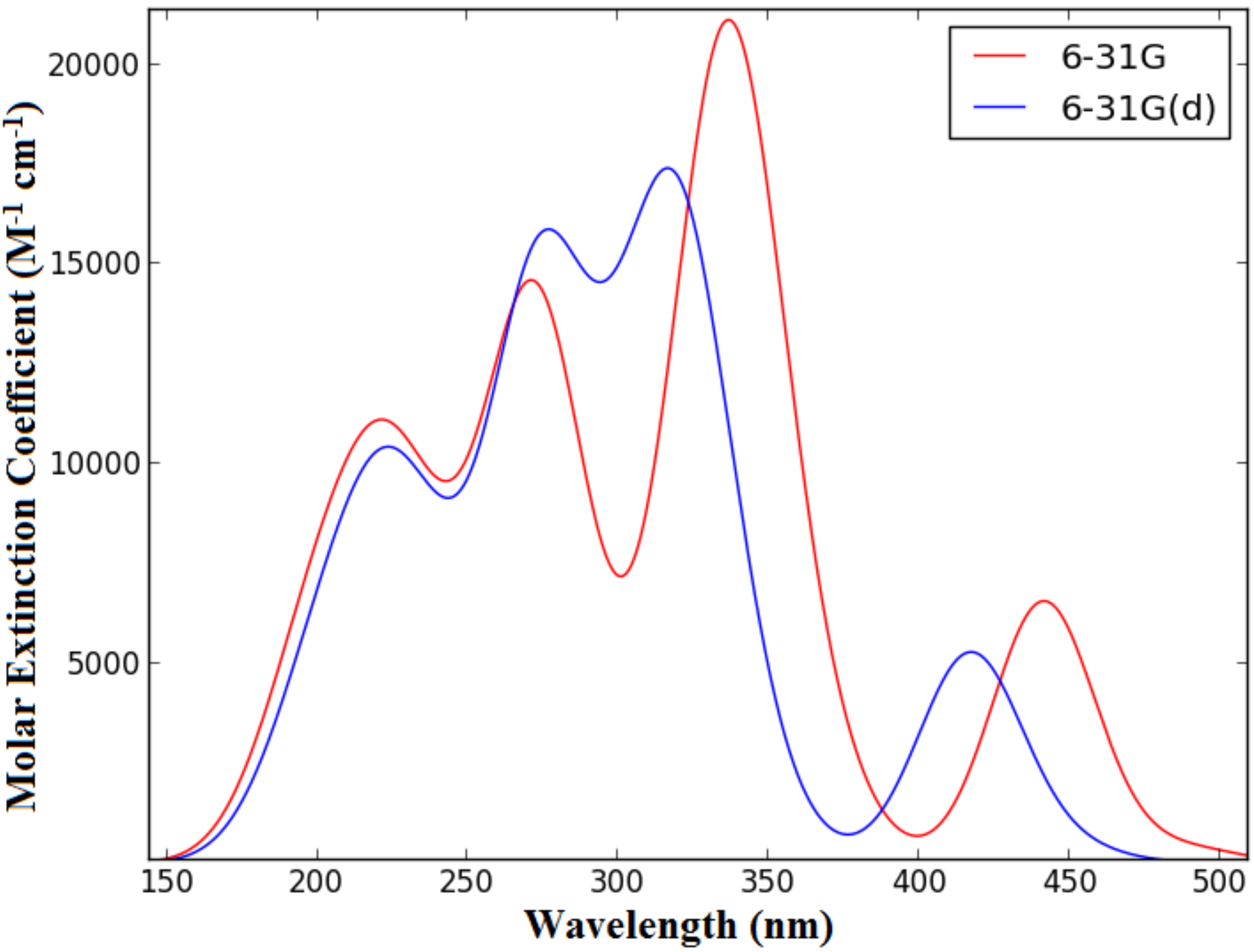}
\end{center}
[Ru(bpy)(py)$_2$(2-AEP)]$^{2+}$
TD-B3LYP/6-31G and TD-B3LYP/6-31G(d) spectra.

% ================================================
\newpage
\section{Complex {\bf (58)}: [Ru(bpy)(PMA)$_2$]$^{2+}$}
% ================================================

\begin{center}
   {\bf PDOS}
\end{center}

\begin{center}
\begin{tabular}{cc}
\includegraphics[width=0.4\textwidth]{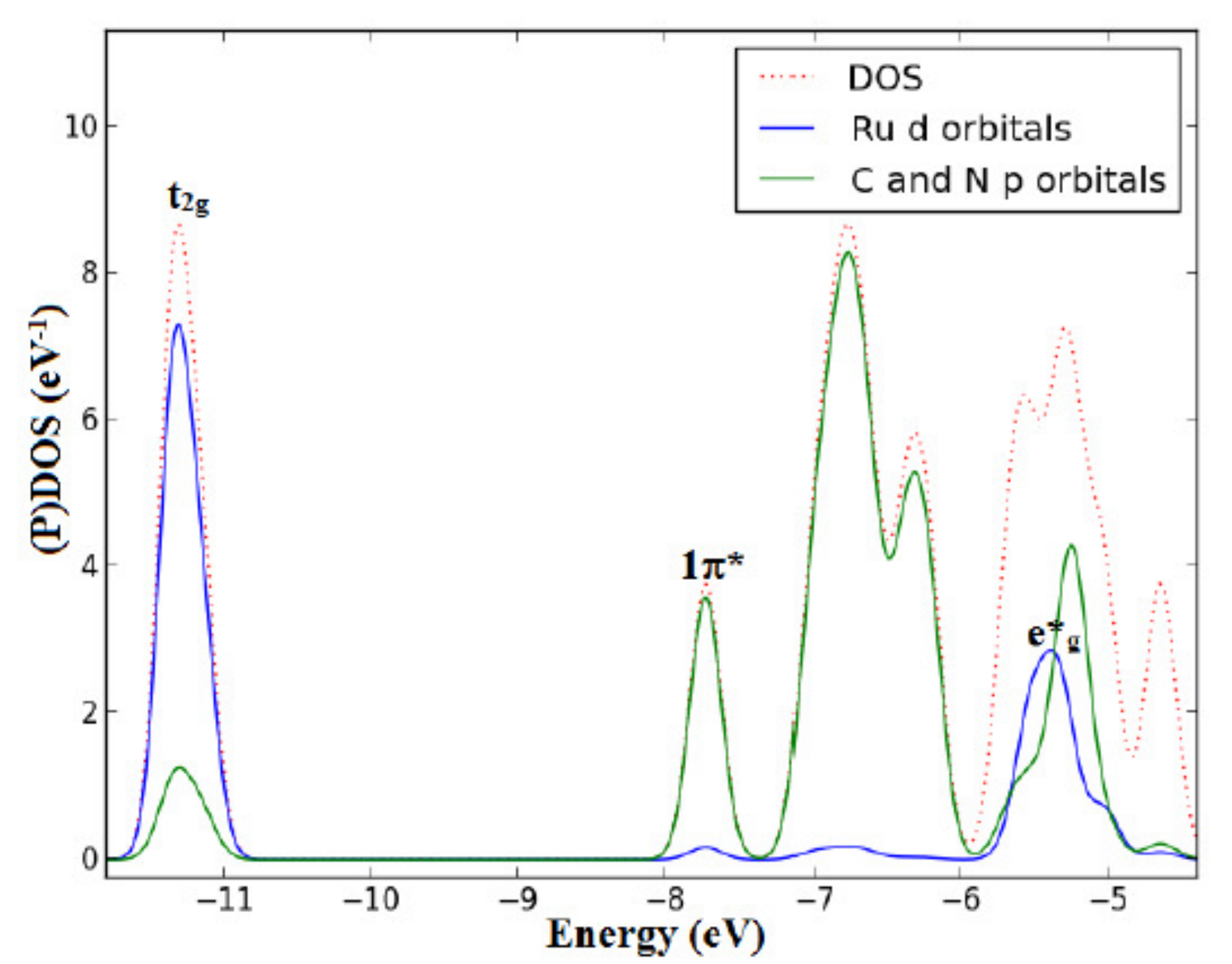} &
\includegraphics[width=0.4\textwidth]{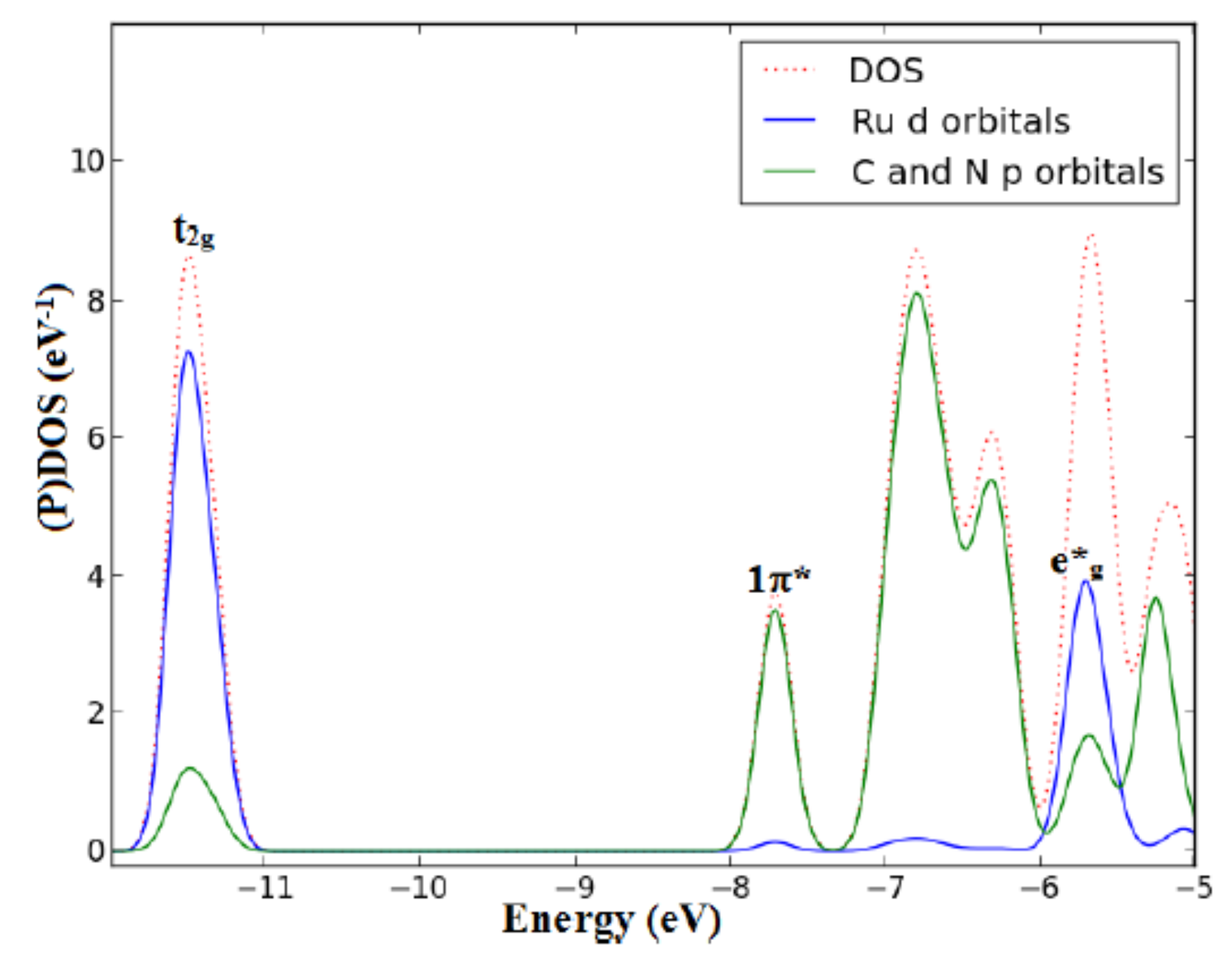} \\
B3LYP/6-31G & B3LYP/6-31G(d) \\
$\epsilon_{\text{HOMO}} = \mbox{-11.17 eV}$ & 
$\epsilon_{\text{HOMO}} = \mbox{-11.35 eV}$ 
\end{tabular}
\end{center}
Total and partial density of states of [Ru(bpy)(PMA)$_2$]$^{2+}$
partitioned over Ru d orbitals and ligand C and N p orbitals.
% for the 6-31G (left-hand side) and 6-31G* (right-hand side) basis sets.

\begin{center}
   {\bf Absorption Spectrum}
\end{center}

\begin{center}
\includegraphics[width=0.8\textwidth]{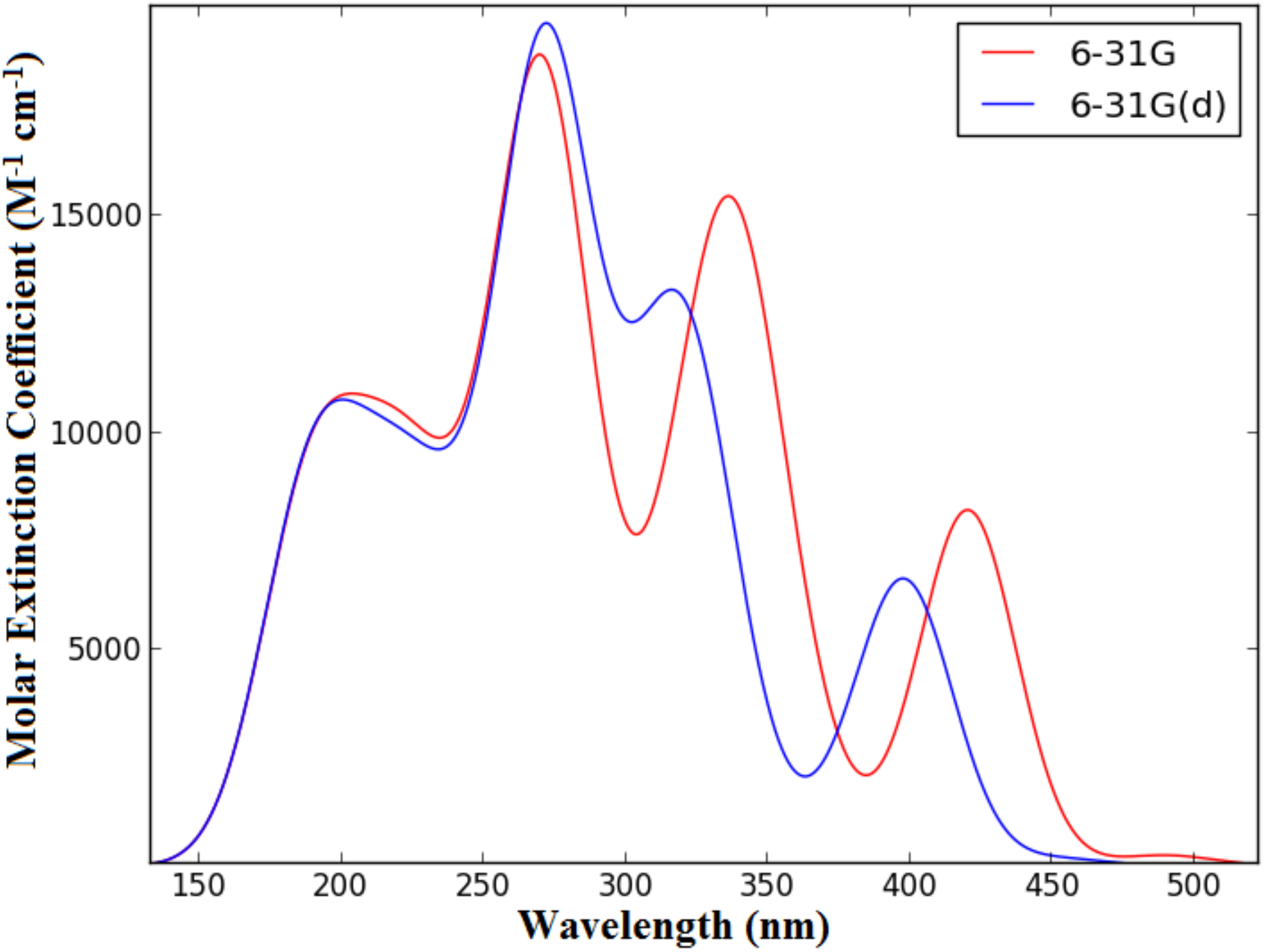}
\end{center}
[Ru(bpy)(PMA)$_2$]$^{2+}$
TD-B3LYP/6-31G and TD-B3LYP/6-31G(d) spectra.

% % ================================================
% \newpage
% \section{Complex {\bf (59)}: [Ru(bpy)(pq)$_2$]$^{2+}$}
% % ================================================
% 
% \begin{center}
%    {\bf PDOS}
% \end{center}
% 
% \begin{center}
% \includegraphics[width=0.4\textwidth]{graphics1/framedquestionmark.pdf}
% \includegraphics[width=0.4\textwidth]{graphics1/framedquestionmark.pdf}
% \end{center}
% {\color{red} Do we have this?}
% 
% \begin{center}
%    {\bf Absorption Spectrum}
% \end{center}
% 
% \begin{center}
% \includegraphics[width=0.4\textwidth]{graphics1/framedquestionmark.pdf}
% \end{center}
% {\color{red} Do we have this?}

% ================================================
\newpage
\section{Complex {\bf (60)}: [Ru(bpy)(DMCH)$_2$]$^{2+}$}
% ================================================

\begin{center}
   {\bf PDOS}
\end{center}

\begin{center}
\begin{tabular}{cc}
\includegraphics[width=0.4\textwidth]{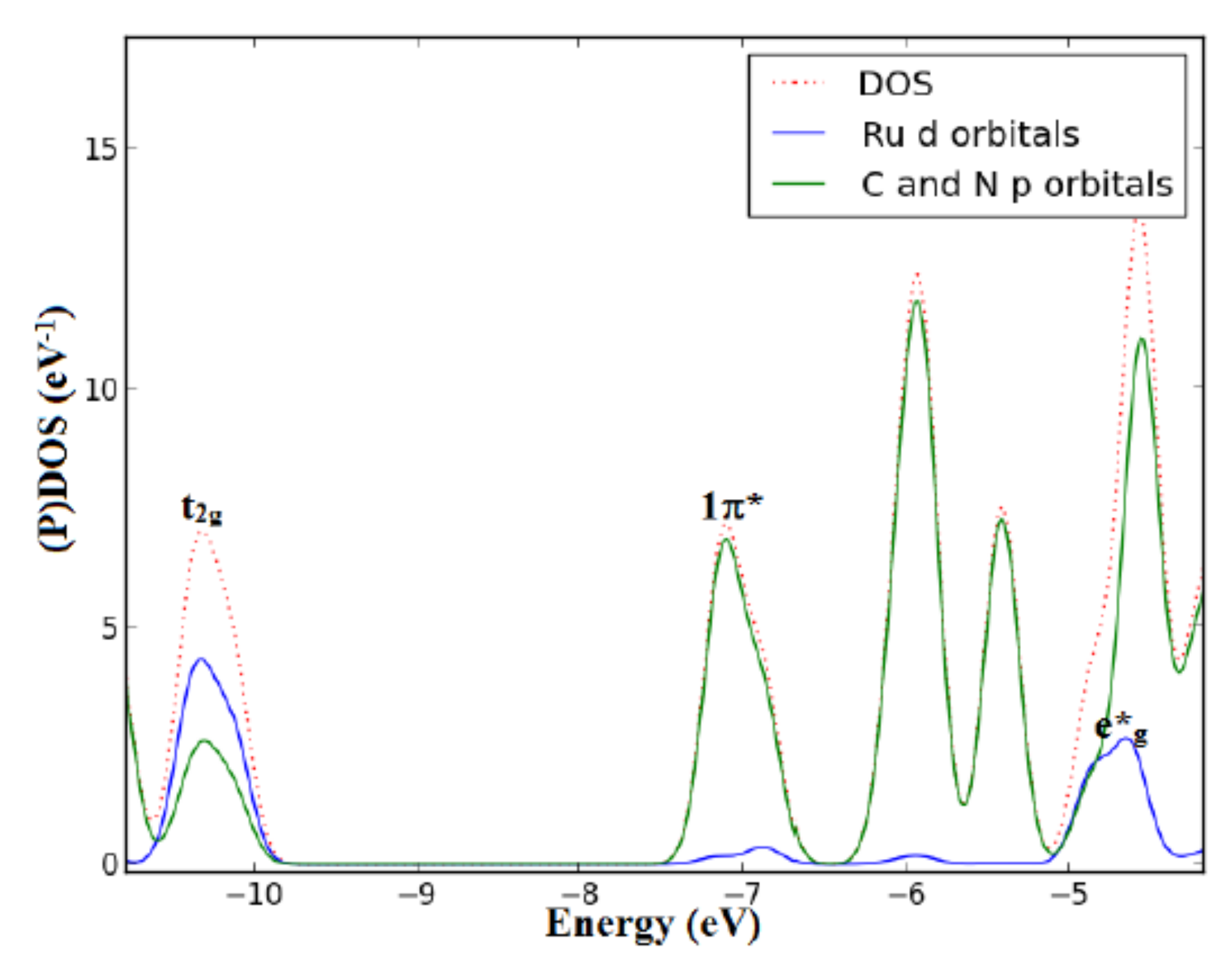} &
\includegraphics[width=0.4\textwidth]{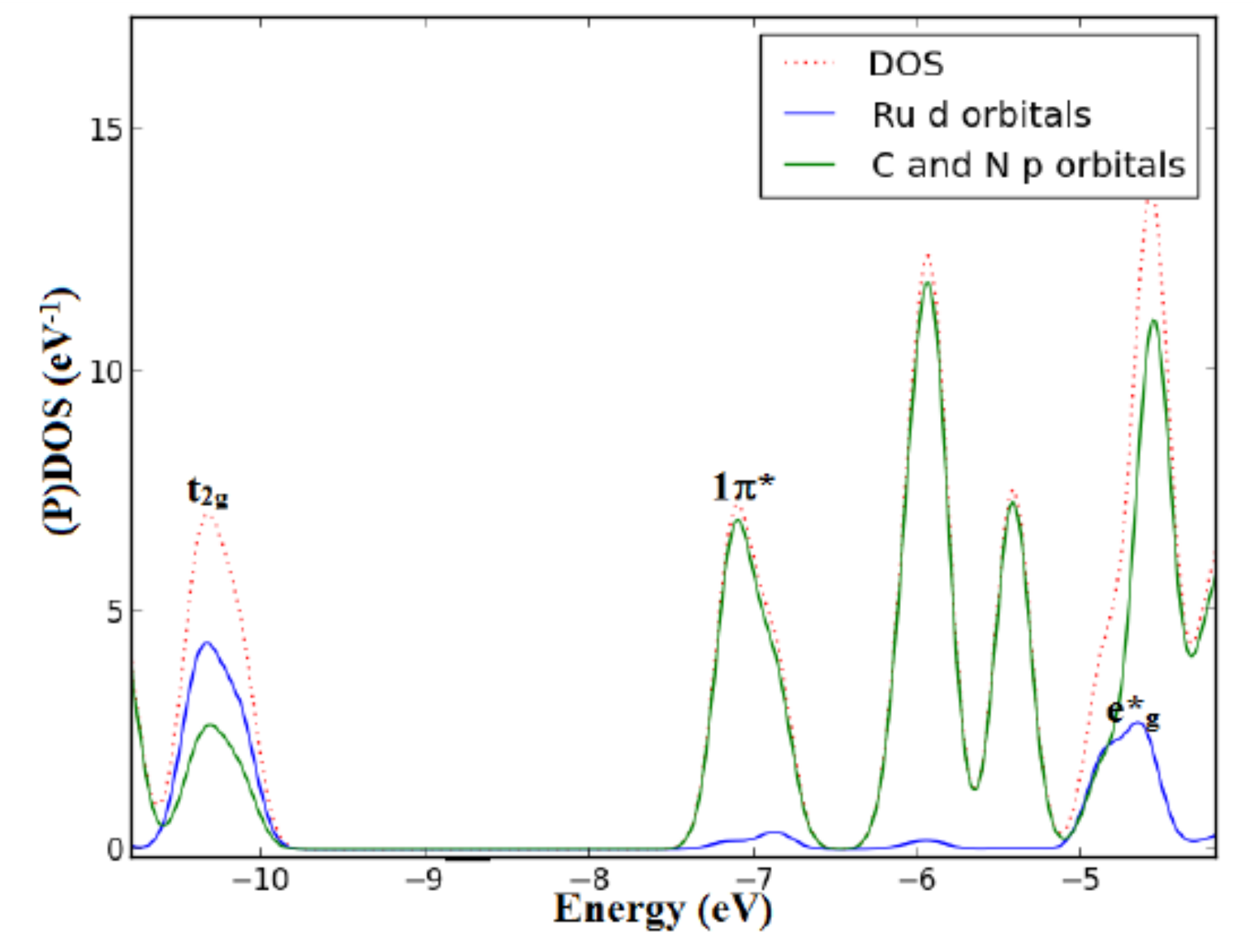} \\
B3LYP/6-31G & B3LYP/6-31G(d) \\
$\epsilon_{\text{HOMO}} = \mbox{-10.13 eV}$ & 
$\epsilon_{\text{HOMO}} = \mbox{-10.23 eV}$ 
\end{tabular}
\end{center}
Total and partial density of states of [Ru(bpy)(DMCH)$_2$]$^{2+}$ 
partitioned over Ru d orbitals and ligand C and N p orbitals.
% for the 6-31G (left-hand side) and 6-31G* (right-hand side) basis sets.

\begin{center}
   {\bf Absorption Spectrum}
\end{center}

\begin{center}
\includegraphics[width=0.8\textwidth]{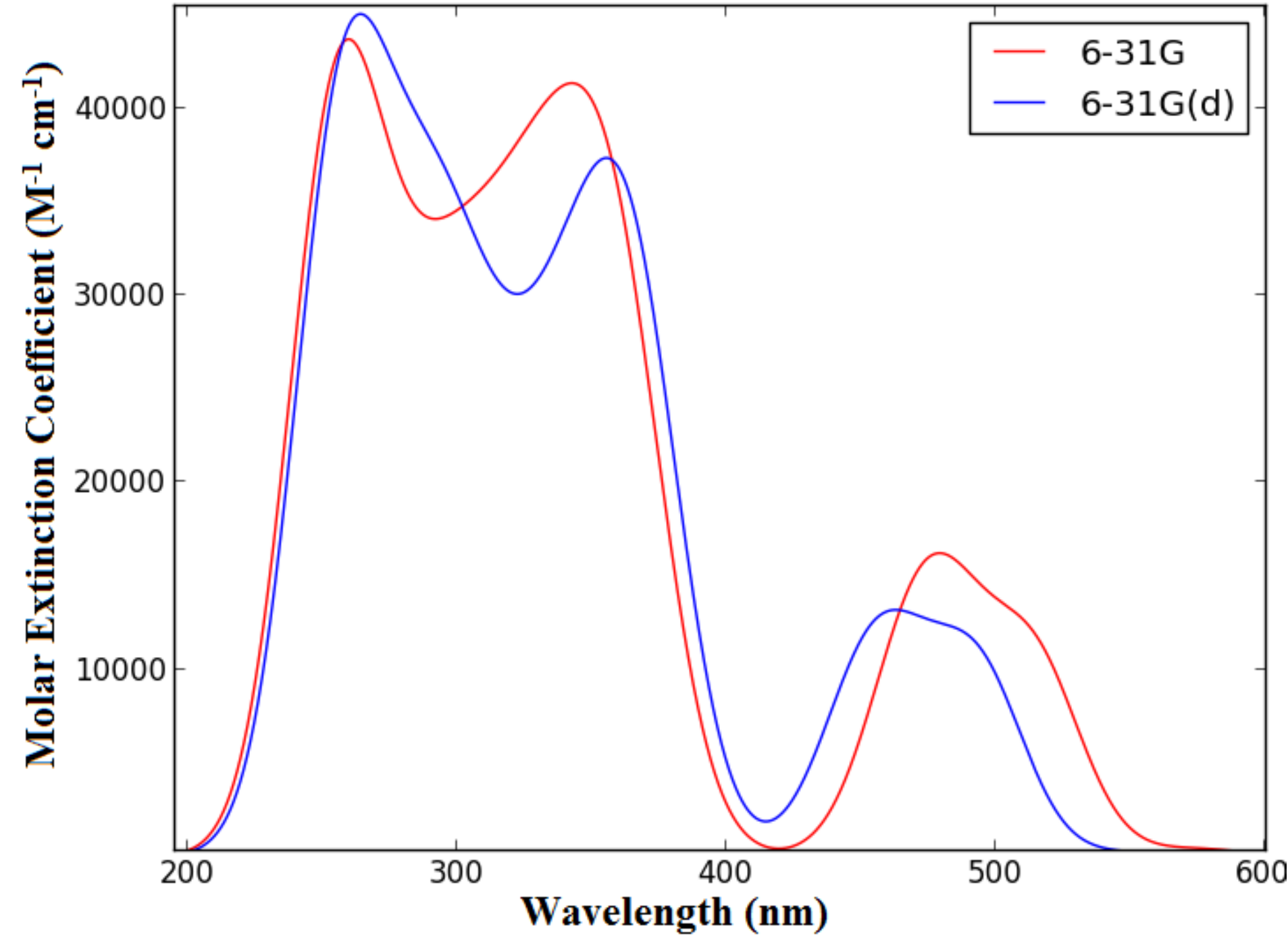}
\end{center}
[Ru(bpy)(DMCH)$_2$]$^{2+}$
TD-B3LYP/6-31G and TD-B3LYP/6-31G(d) spectra.

% ================================================
\newpage
\section{Complex {\bf (61)}: [Ru(bpy)(biq)$_2$]$^{2+}$}
% ================================================

\begin{center}
   {\bf PDOS}
\end{center}

\begin{center}
\begin{tabular}{cc}
\includegraphics[width=0.4\textwidth]{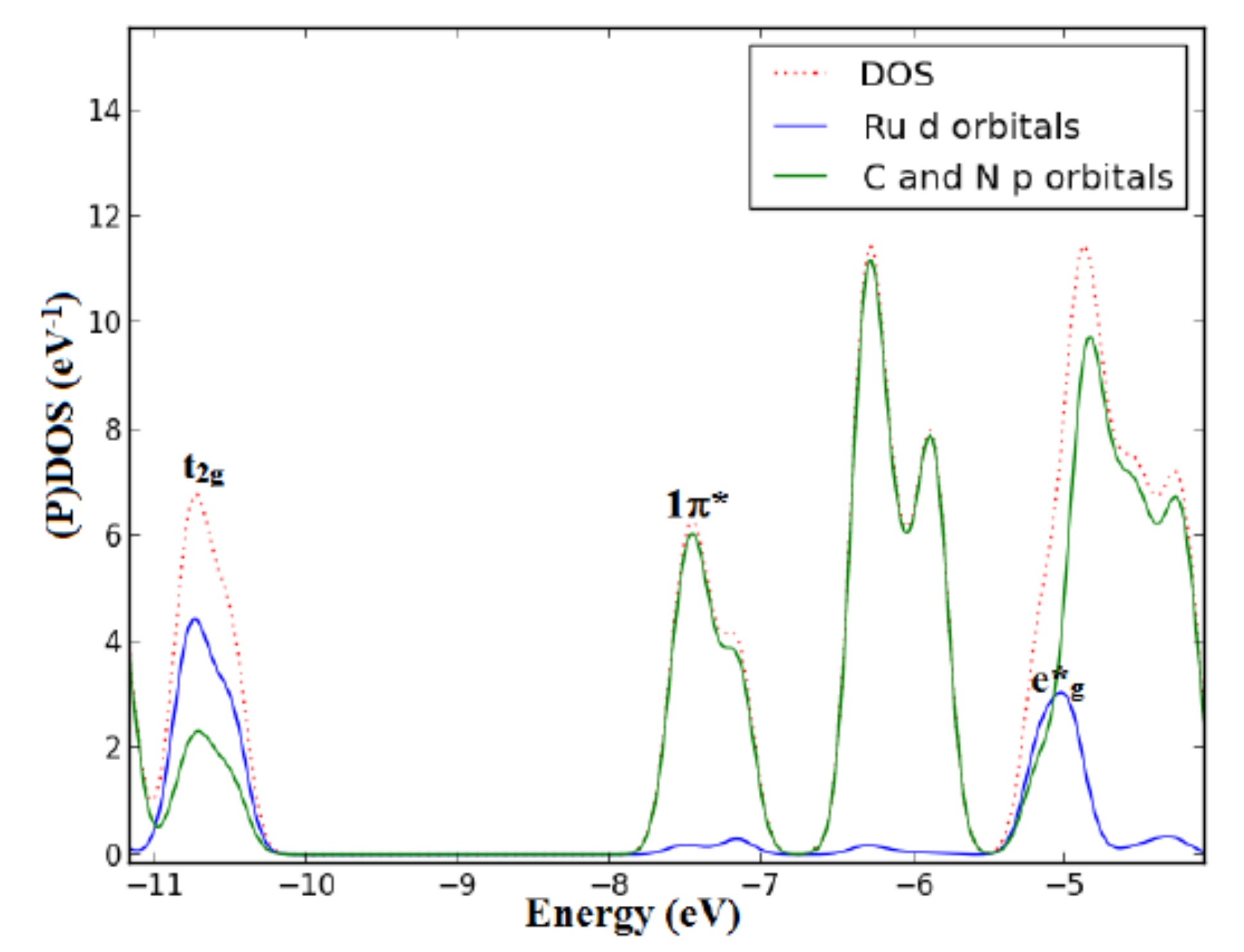} &
\includegraphics[width=0.4\textwidth]{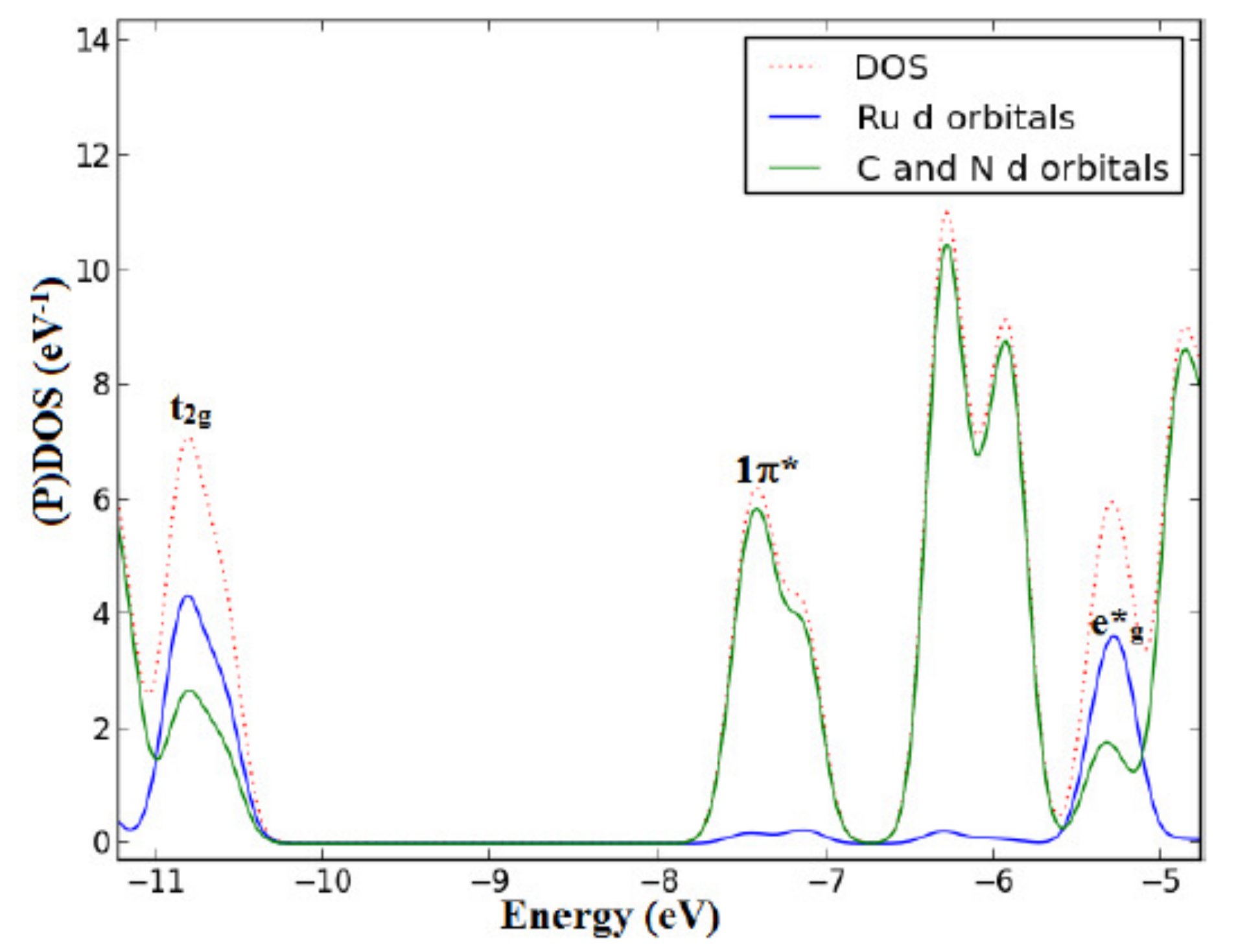} \\
B3LYP/6-31G & B3LYP/6-31G(d) \\
$\epsilon_{\text{HOMO}} = \mbox{-10.50 eV}$ & 
$\epsilon_{\text{HOMO}} = \mbox{-10.59 eV}$ 
\end{tabular}
\end{center}
Total and partial density of states of [Ru(bpy)(biq)$_2$]$^{2+}$
partitioned over Ru d orbitals and ligand C and N p orbitals. 
% for the 6-31G (left-hand side) and 6-31G* (right-hand side) basis sets.

\begin{center}
   {\bf Absorption Spectrum}
\end{center}

\begin{center}
\includegraphics[width=0.8\textwidth]{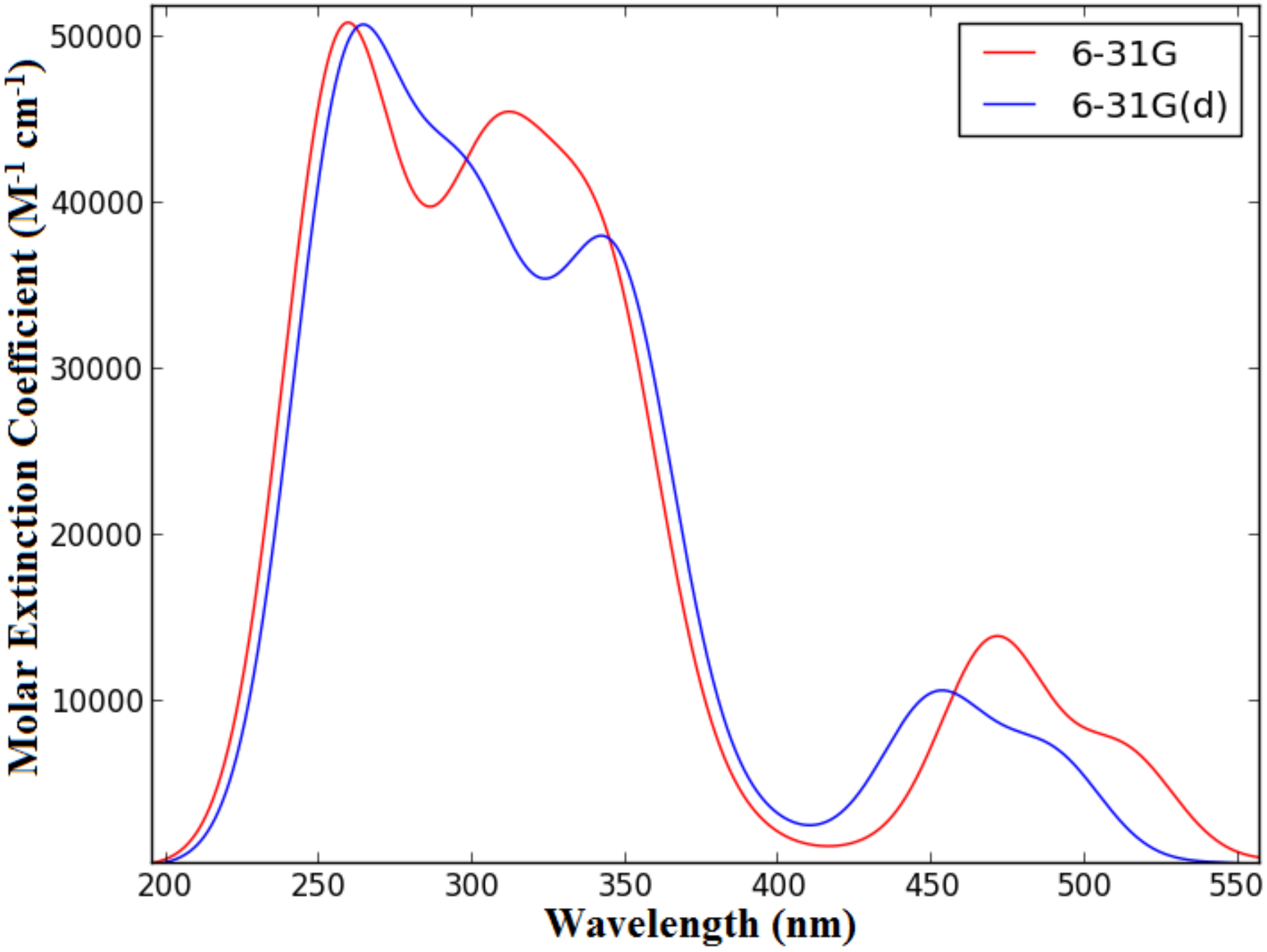}
\end{center}
[Ru(bpy)(biq)$_2$]$^{2+}$
TD-B3LYP/6-31G and TD-B3LYP/6-31G(d) spectra.

% % ================================================
% \newpage
% \section{Complex {\bf (62)}: [Ru(bpy)(i-biq)$_2$]$^{2+}$}
% % ================================================
% 
% \begin{center}
%    {\bf PDOS}
% \end{center}
% 
% \begin{center}
% \includegraphics[width=0.4\textwidth]{graphics1/framedquestionmark.pdf}
% \includegraphics[width=0.4\textwidth]{graphics1/framedquestionmark.pdf}
% \end{center}
% {\color{red} Do we have this?}
% 
% \begin{center}
%    {\bf Absorption Spectrum}
% \end{center}
% 
% \begin{center}
% \includegraphics[width=0.4\textwidth]{graphics1/framedquestionmark.pdf}
% \end{center}
% {\color{red} Do we have this?}

% ================================================
\newpage
\section{Complex {\bf (63)}$^\dagger$: [Ru(bpy)(trpy)Cl]$^{+}$}
% ================================================

% \begin{center}
%    {\bf PDOS}
% \end{center}
% 
% \begin{center}
% \includegraphics[width=0.4\textwidth]{graphics1/framedquestionmark.pdf}
% \includegraphics[width=0.4\textwidth]{graphics1/framedquestionmark.pdf}
% \end{center}
% {\color{magenta} PDOS could not be calculated for complexes containing Cl.}
\begin{center}
\begin{tabular}{cc}
B3LYP/6-31G & B3LYP/6-31G(d) \\
$\epsilon_{\text{HOMO}} = \mbox{-7.76 eV}$ & 
$\epsilon_{\text{HOMO}} = \mbox{-7.78 eV}$ 
\end{tabular}
\end{center}

\begin{center}
   {\bf Absorption Spectrum}
\end{center}

\begin{center}
\includegraphics[width=0.8\textwidth]{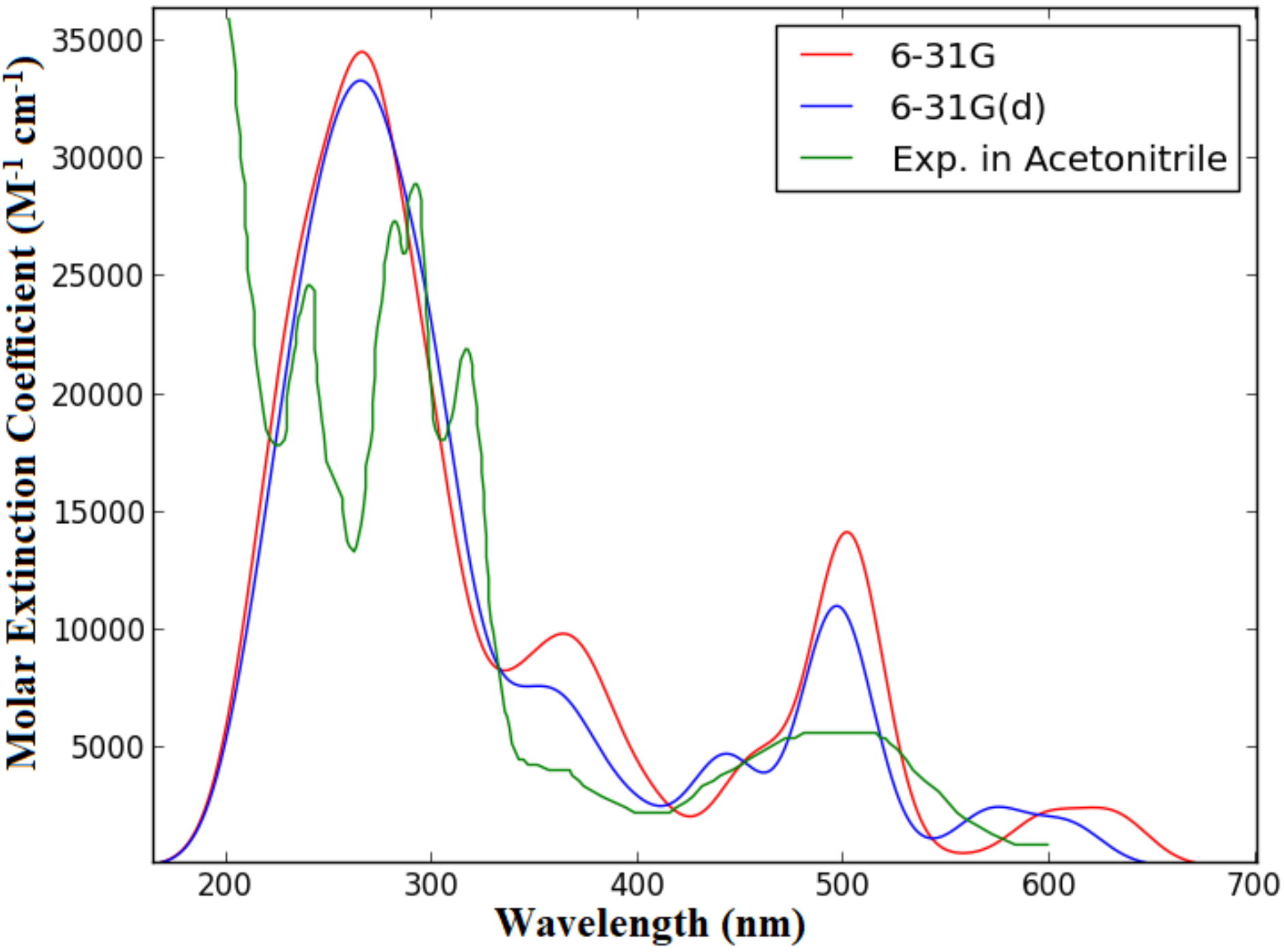}
\end{center}
[Ru(bpy)(trpy)Cl]$^{+}$
TD-B3LYP/6-31G, TD-B3LYP/6-31G(d), and experimental spectra.
Experimental curve measured in acetonitrile at room temperature 
\cite{LNZ+13}.

% ================================================
\newpage
\section{Complex {\bf (64)}*: [Ru(bpy)(trpy)(CN)]$^{+}$}
% ================================================

\begin{center}
   {\bf PDOS}
\end{center}

\begin{center}
\begin{tabular}{cc}
\includegraphics[width=0.4\textwidth]{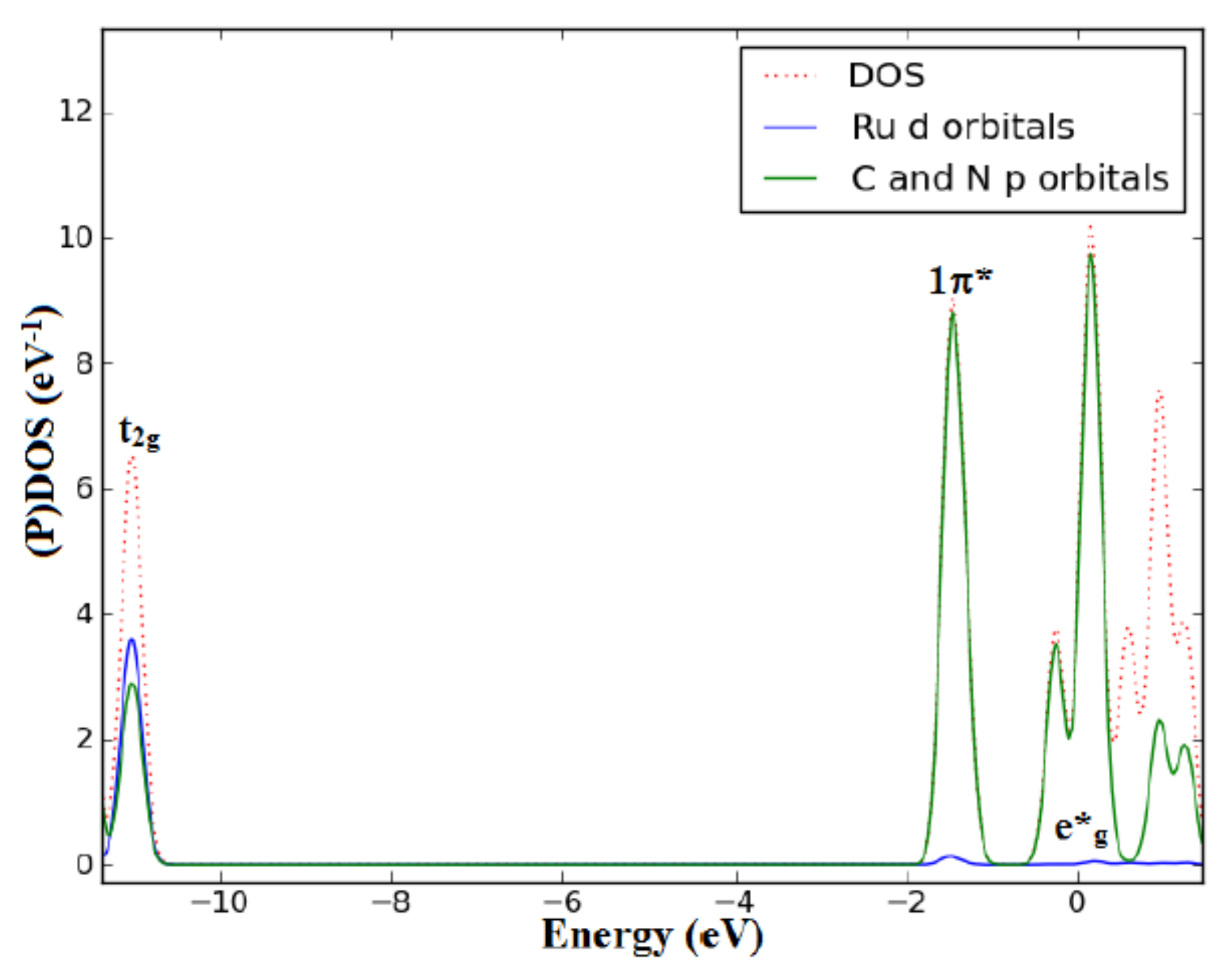} &
\includegraphics[width=0.4\textwidth]{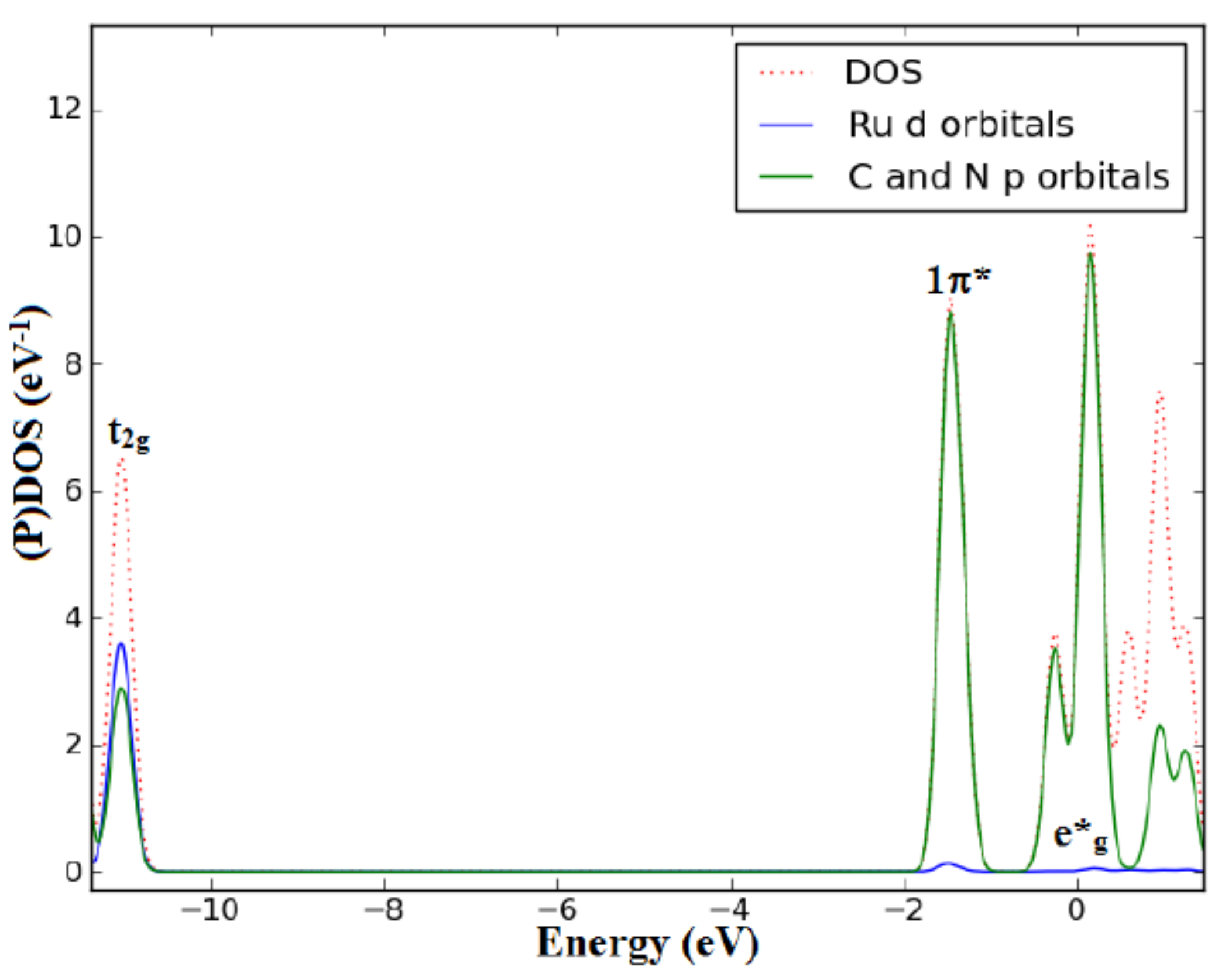} \\
B3LYP/6-31G & B3LYP/6-31G(d) \\
$\epsilon_{\text{HOMO}} = \mbox{-10.97 eV}$ & 
$\epsilon_{\text{HOMO}} = \mbox{-11.15 eV}$ 
\end{tabular}
\end{center}
Total and partial density of states of [Ru(bpy)(trpy)(CN)]$^{+}$
partitioned over Ru d orbitals and ligand C and N p orbitals.
% for the 6-31G (left-hand side) and 6-31G* (right-hand side) basis sets.

\begin{center}
   {\bf Absorption Spectrum}
\end{center}

\begin{center}
\includegraphics[width=0.8\textwidth]{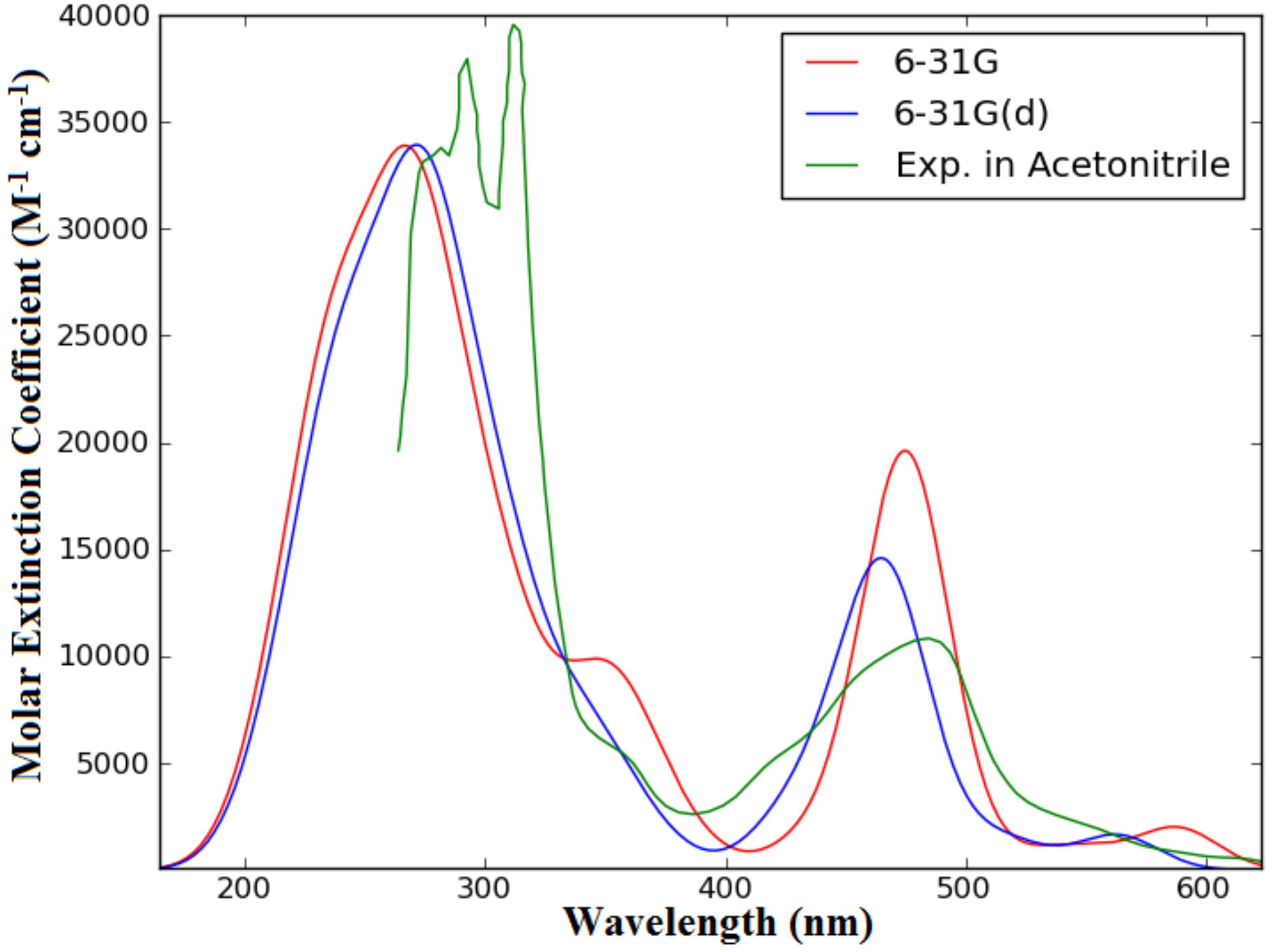}
\end{center}
[Ru(bpy)(trpy)(CN)]$^{+}$
TD-B3LYP/6-31G, TD-B3LYP/6-31G(d), and experimental spectra.
Experiment in acetonitrile \cite{CAY+12}.

% % ================================================
% \newpage
% \section{Complex {\bf (65)}: [Ru(4-n-bpy)$_3$]$^{2+}$}
% % ================================================
% 
% \begin{center}
%    {\bf PDOS}
% \end{center}
% 
% \begin{center}
% \includegraphics[width=0.4\textwidth]{graphics1/framedquestionmark.pdf}
% \includegraphics[width=0.4\textwidth]{graphics1/framedquestionmark.pdf}
% \end{center}
% {\color{red} Do we have this?}
% 
% \begin{center}
%    {\bf Absorption Spectrum}
% \end{center}
% 
% \begin{center}
% \includegraphics[width=0.4\textwidth]{graphics1/framedquestionmark.pdf}
% \end{center}
% {\color{red} Do we have this?}

% ================================================
\newpage
\section{Complex {\bf (66)}*: [Ru(6-m-bpy)$_3$]$^{2+}$}
% ================================================

\begin{center}
   {\bf PDOS}
\end{center}

\begin{center}
\begin{tabular}{cc}
\includegraphics[width=0.4\textwidth]{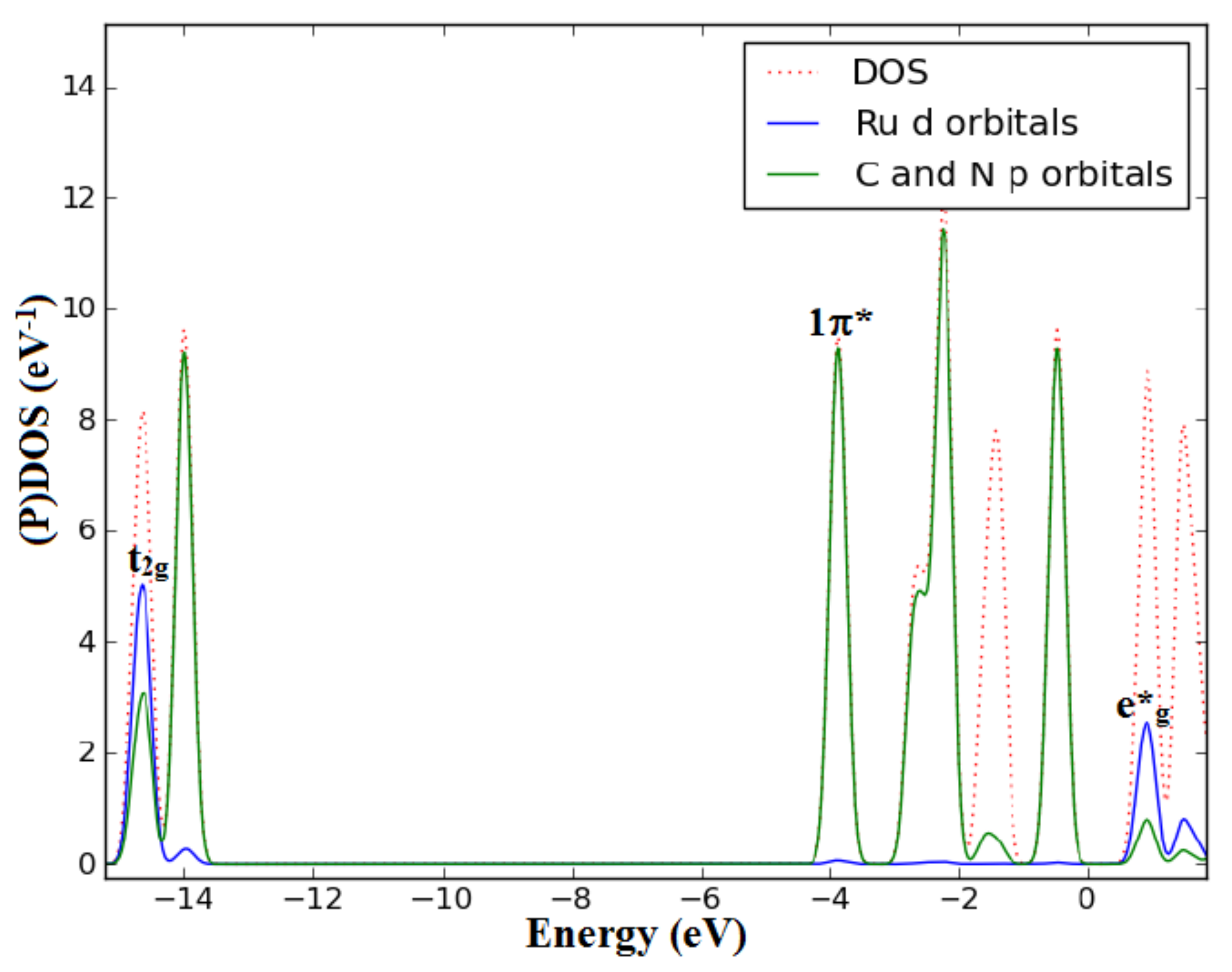} &
\includegraphics[width=0.4\textwidth]{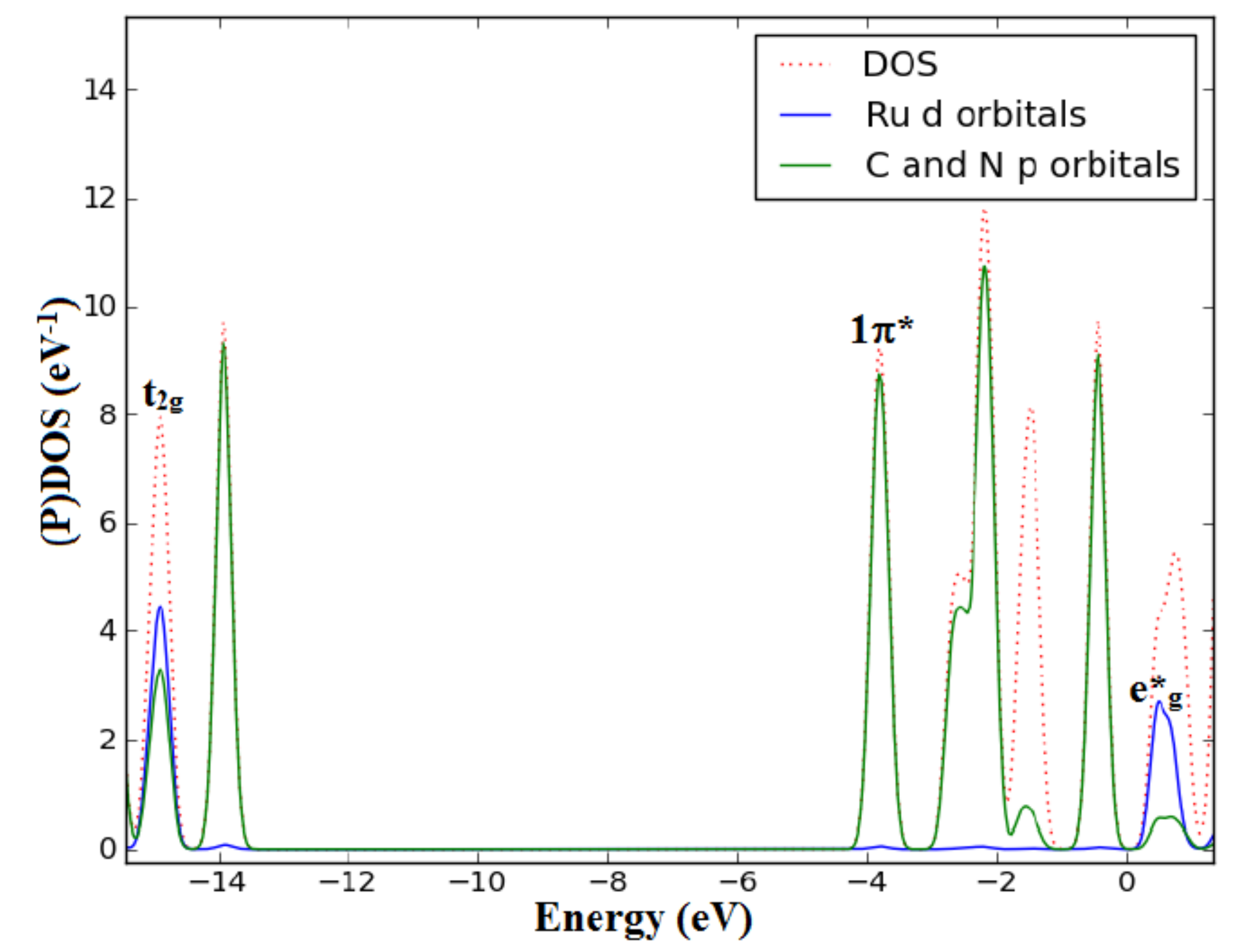} \\
B3LYP/6-31G & B3LYP/6-31G(d) \\
$\epsilon_{\text{HOMO}} = \mbox{-13.92 eV}$ & 
$\epsilon_{\text{HOMO}} = \mbox{-13.84 eV}$ 
\end{tabular}
\end{center}
Total and partial density of states of [Ru(6-m-bpy)$_3$]$^{2+}$
partitioned over Ru d orbitals and ligand C and N p orbitals.
% for the 6-31G (left-hand side) and 6-31G* (right-hand side) basis sets.

\begin{center}
   {\bf Absorption Spectrum}
\end{center}

\begin{center}
\includegraphics[width=0.8\textwidth]{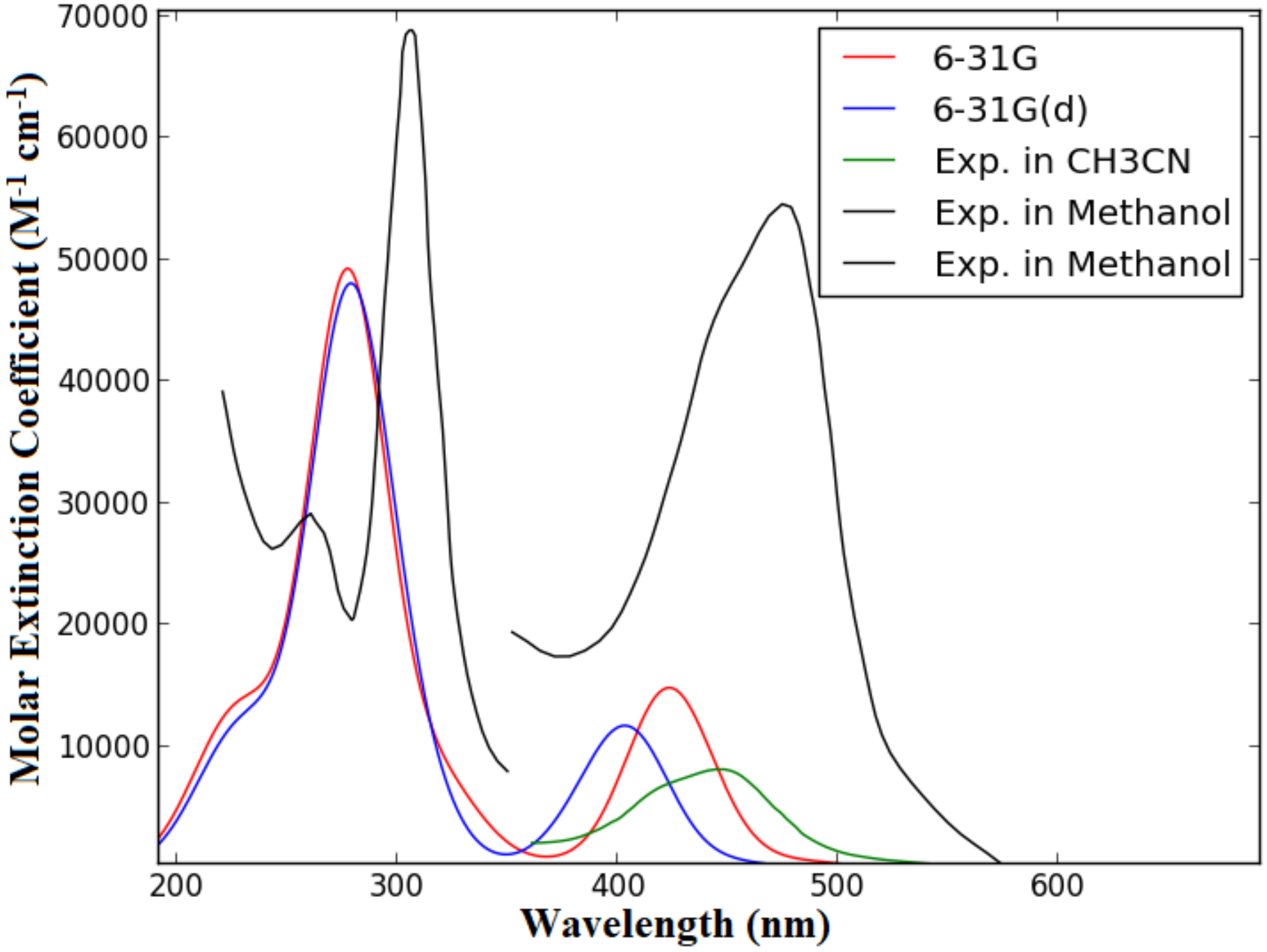}
\end{center}
[Ru(6-m-bpy)$_3$]$^{2+}$
TD-B3LYP/6-31G, TD-B3LYP/6-31G(d), and experimental spectra.
Experimental curves measured in acetonitrile \cite{SML+13} and methanol \cite{FKS80},
both at room temperature.

% ================================================
\newpage
\section{Complex {\bf (67)}*: [Ru(3,3'-dm-bpy)$_3$]$^{2+}$}
% ================================================

\begin{center}
   {\bf PDOS}
\end{center}

\begin{center}
\begin{tabular}{cc}
\includegraphics[width=0.4\textwidth]{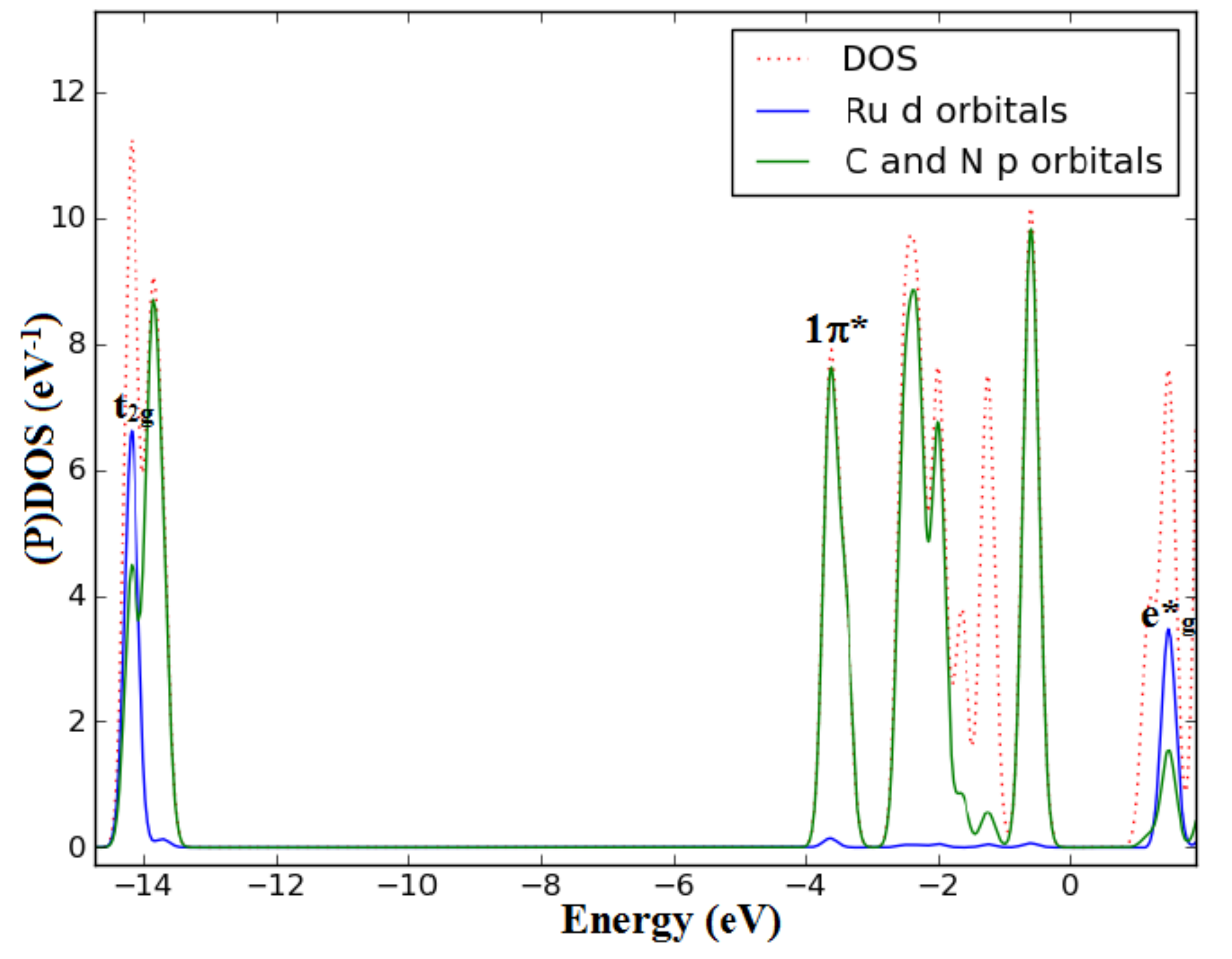} &
\includegraphics[width=0.4\textwidth]{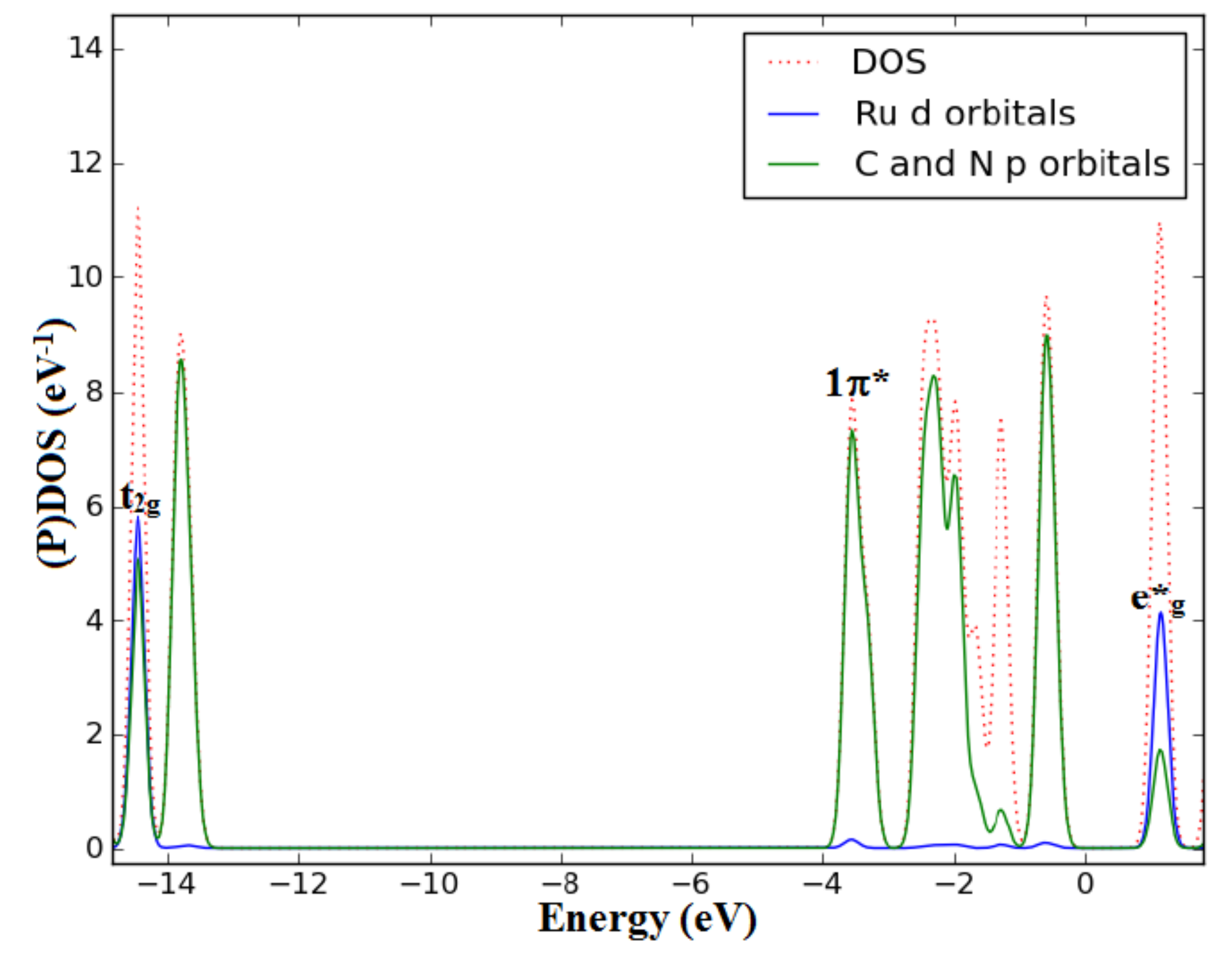} \\
B3LYP/6-31G & B3LYP/6-31G(d) \\
$\epsilon_{\text{HOMO}} = \mbox{-13.72 eV}$ & 
$\epsilon_{\text{HOMO}} = \mbox{-13.67 eV}$ 
\end{tabular}
\end{center}
Total and partial density of states of [Ru(3,3'-dm-bpy)$_3$]$^{2+}$
partitioned over Ru d orbitals and ligand C and N p orbitals.
% for the 6-31G (left-hand side) and 6-31G* (right-hand side) basis sets.

\begin{center}
   {\bf Absorption Spectrum}
\end{center}

\begin{center}
\includegraphics[width=0.8\textwidth]{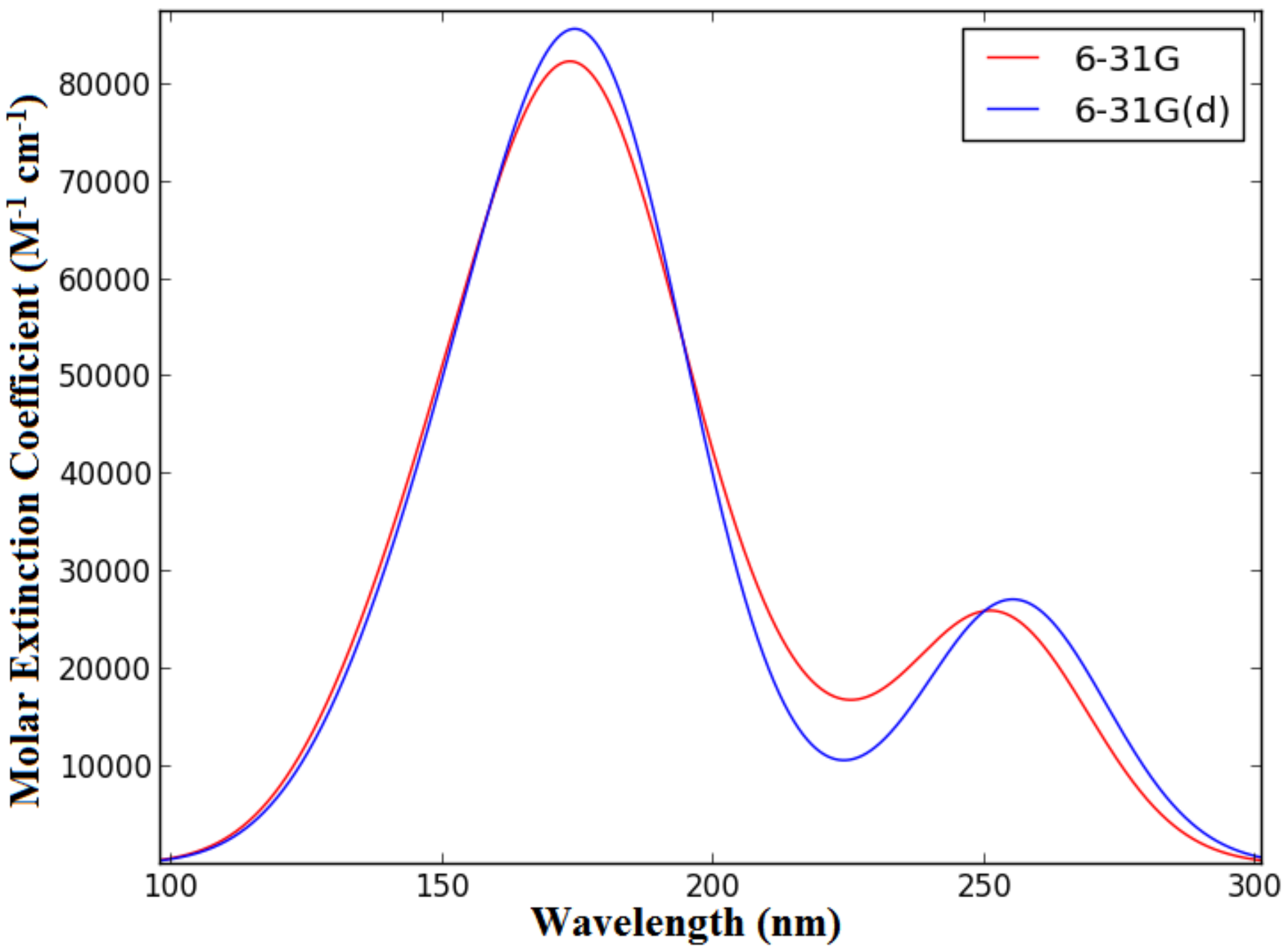}
\end{center}
[Ru(3,3'-dm-bpy)$_3$]$^{2+}$
TD-B3LYP/6-31G and TD-B3LYP/6-31G(d) spectra.

% % ================================================
% \newpage
% \section{Complex {\bf (68)}*: [Ru(3,3'-dm-bpy)$_2$(phen)]$^{2+}$}
% % ================================================
% 
% \begin{center}
%    {\bf PDOS}
% \end{center}
% 
% \begin{center}
% \includegraphics[width=0.4\textwidth]{graphics1/cmplx068_pdos_631g.pdf}
% \includegraphics[width=0.4\textwidth]{graphics1/cmplx068_pdos_631gd.pdf}
% \end{center}
% Total and partial density of states of [Ru(3,3'-dm-bpy)$_2$(phen)]$^{2+}$
% partitioned over Ru d orbitals and ligand C and N p orbitals for the 6-31G 
% (left-hand side) and 6-31G* (right-hand side) basis sets.
% 
% \begin{center}
%    {\bf Absorption Spectrum}
% \end{center}
% 
% \begin{center}
% \includegraphics[width=0.4\textwidth]{graphics1/framedquestionmark.pdf}
% \end{center}
% {\color{red} Do we have this?}

% ================================================
\newpage
\section{Complex {\bf (69)}: [Ru(3,3'-dm-bpy)(phen)$_2$]$^{2+}$}
% ================================================

\begin{center}
   {\bf PDOS}
\end{center}

\begin{center}
\begin{tabular}{cc}
\includegraphics[width=0.4\textwidth]{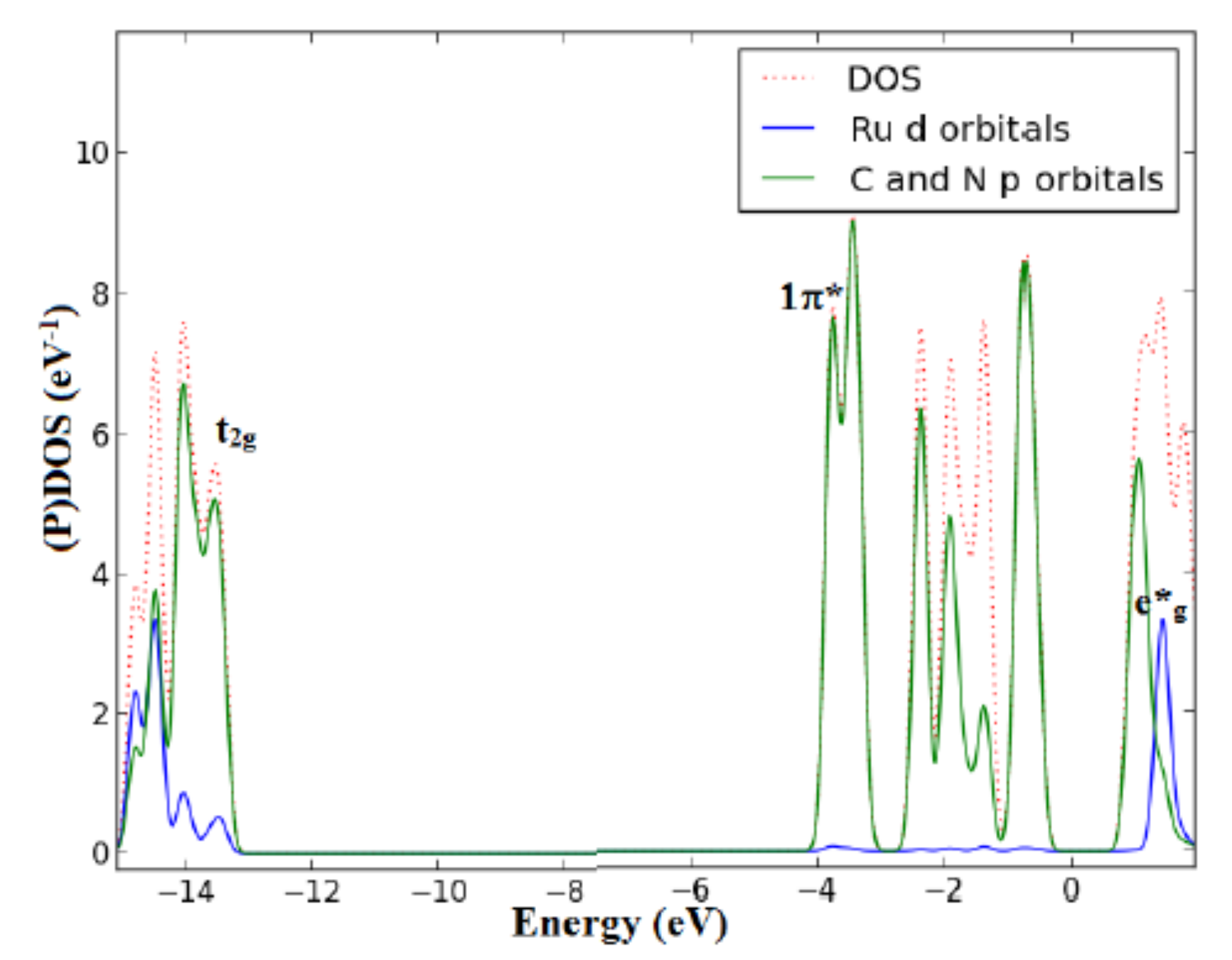} &
\includegraphics[width=0.4\textwidth]{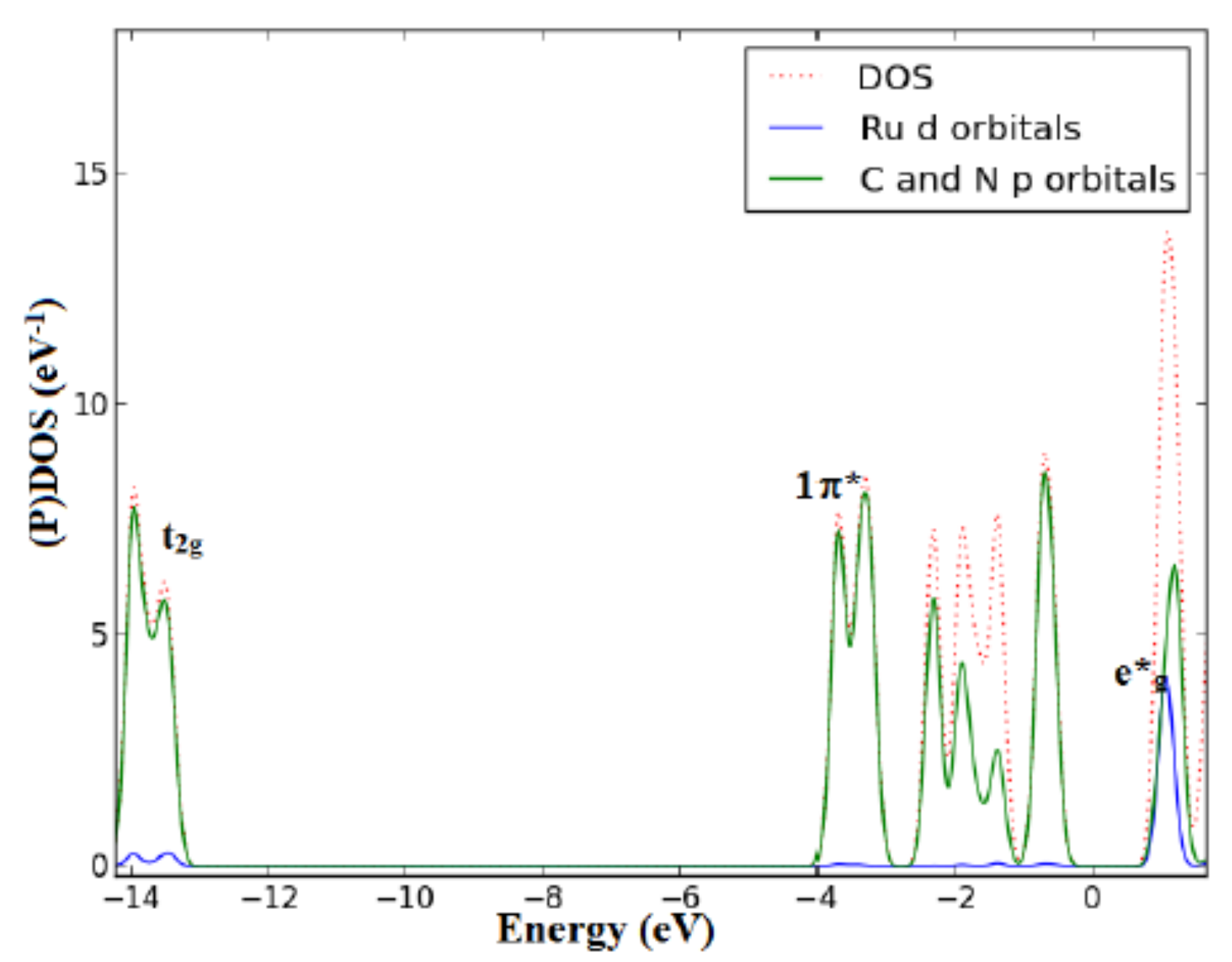} \\
B3LYP/6-31G & B3LYP/6-31G(d) \\
$\epsilon_{\text{HOMO}} = \mbox{-13.45 eV}$ & 
$\epsilon_{\text{HOMO}} = \mbox{-13.45 eV}$ 
\end{tabular}
\end{center}
Total and partial density of states of [Ru(3,3'-dm-bpy)(phen)$_2$]$^{2+}$
partitioned over Ru d orbitals and ligand C and N p orbitals. 
% for the 6-31G (left-hand side) and 6-31G* (right-hand side) basis sets.

\begin{center}
   {\bf Absorption Spectrum}
\end{center}

\begin{center}
\includegraphics[width=0.8\textwidth]{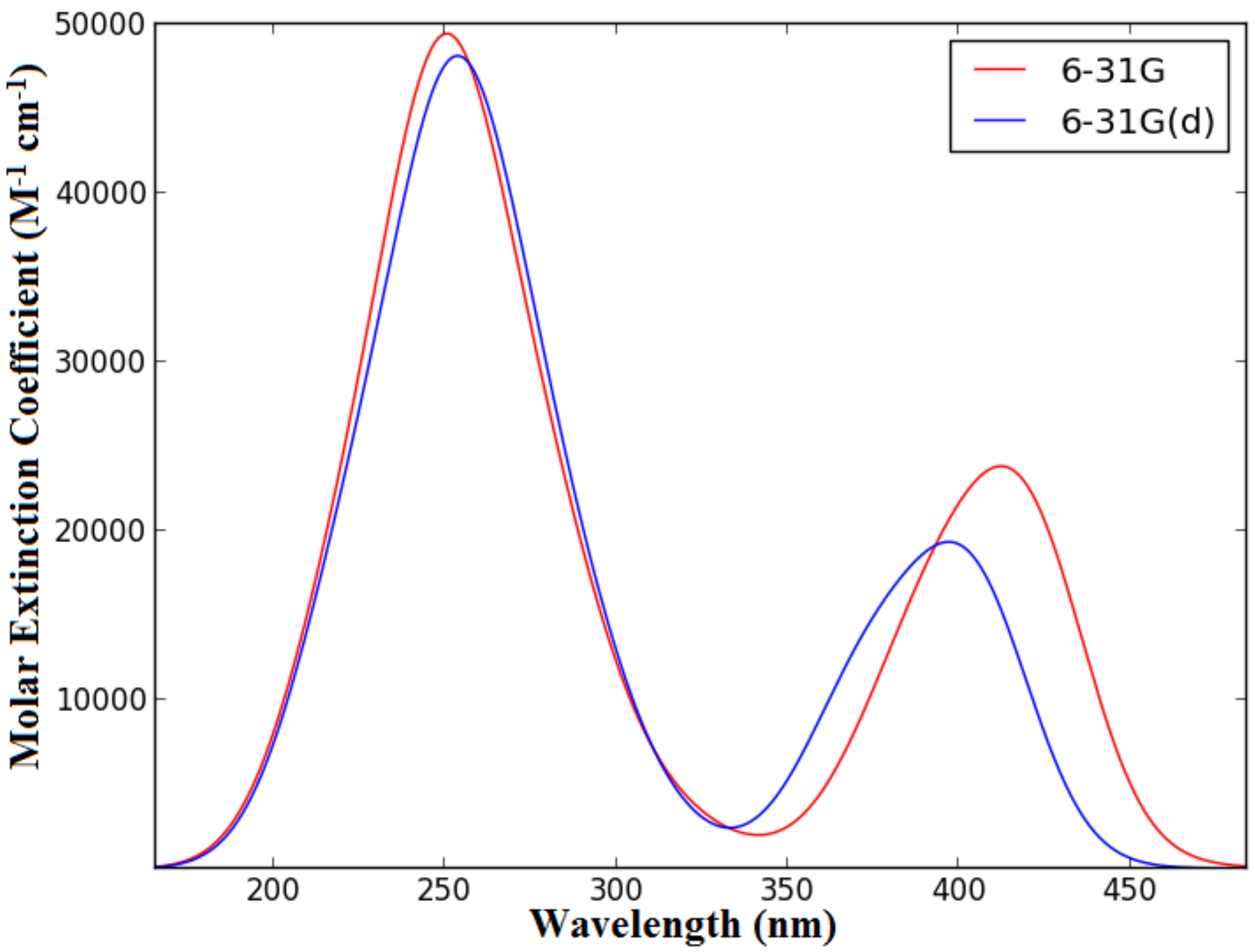}
\end{center}
[Ru(3,3'-dm-bpy)(phen)$_2$]$^{2+}$
TD-B3LYP/6-31G and TD-B3LYP/6-31G(d) spectra.

% ================================================
\newpage
\section{Complex {\bf (70)}: [Ru(4,4'-dm-bpy)$_3$]$^{2+}$}
% ================================================

\begin{center}
   {\bf PDOS}
\end{center}

\begin{center}
\begin{tabular}{cc}
\includegraphics[width=0.4\textwidth]{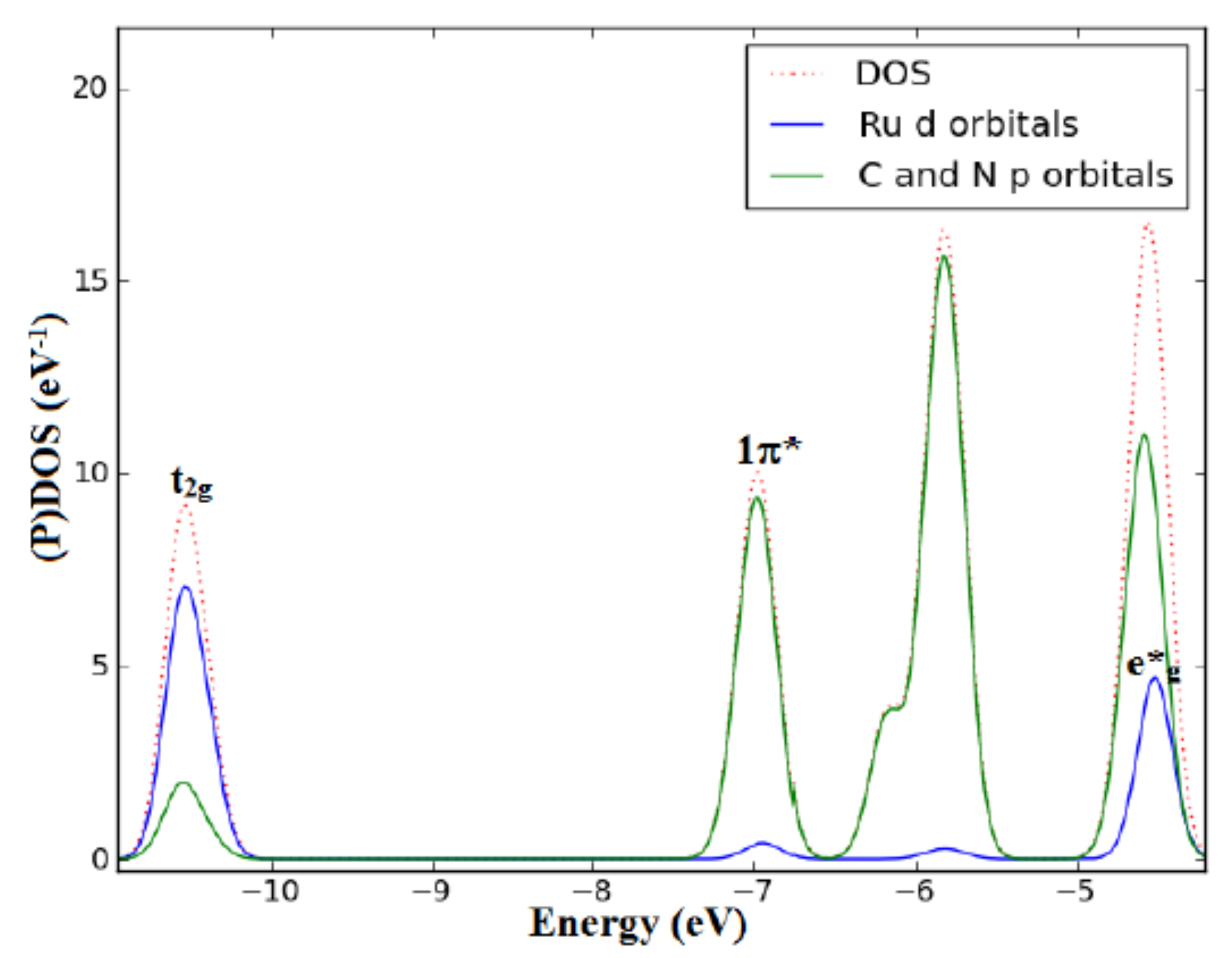} &
\includegraphics[width=0.4\textwidth]{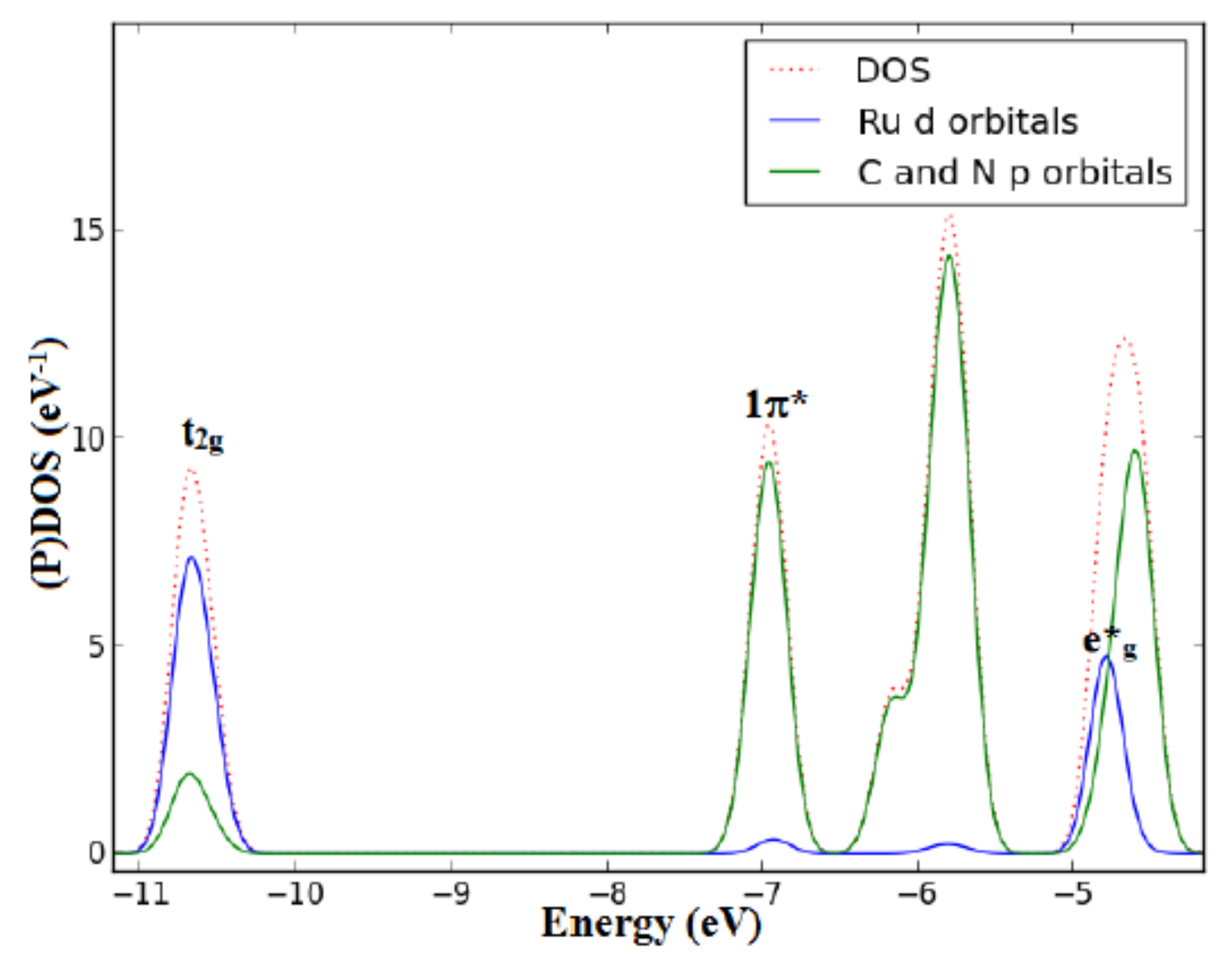} \\
B3LYP/6-31G & B3LYP/6-31G(d) \\
$\epsilon_{\text{HOMO}} = \mbox{-10.42 eV}$ & 
$\epsilon_{\text{HOMO}} = \mbox{-10.56 eV}$ 
\end{tabular}
\end{center}
Total and partial density of states of [Ru(3,3'-dm-bpy)$_3$]$^{2+}$
partitioned over Ru d orbitals and ligand C and N p orbitals.
% for the 6-31G (left-hand side) and 6-31G* (right-hand side) basis sets.

\begin{center}
   {\bf Absorption Spectrum}
\end{center}

\begin{center}
\includegraphics[width=0.8\textwidth]{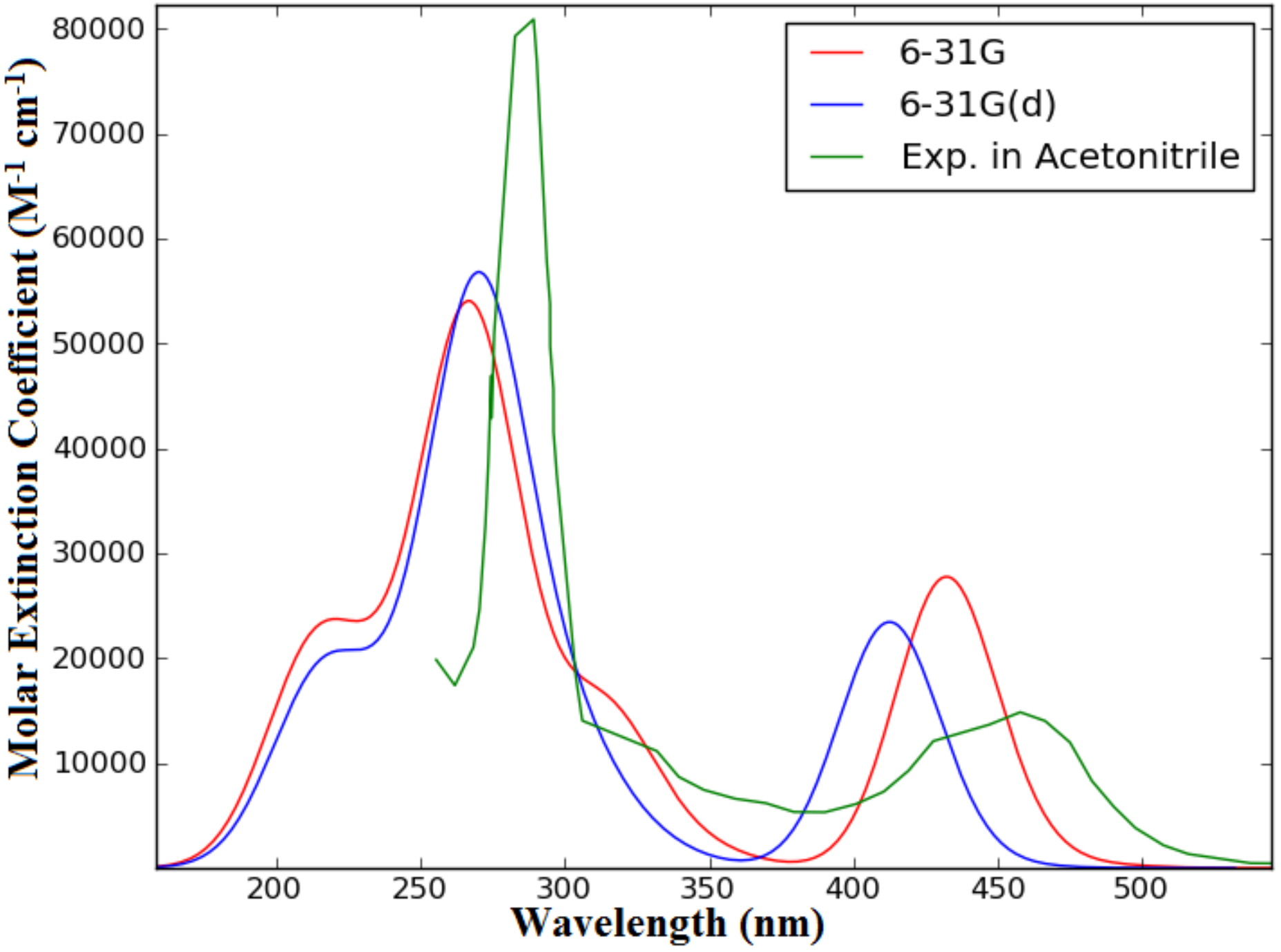}
\end{center}
[Ru(3,3'-dm-bpy)$_3$]$^{2+}$
TD-B3LYP/6-31G, TD-B3LYP/6-31G(d), and experimental spectra.
Experimental spectrum measured in acetonitrile \cite{MMIC03}.

% ================================================
\newpage
\section{Complex {\bf (71)}: [Ru(4,4'-dm-bpy)$_2$(4,7-dhy-phen)]$^{2+}$}
% ================================================

\begin{center}
   {\bf PDOS}
\end{center}

\begin{center}
\begin{tabular}{cc}
\includegraphics[width=0.4\textwidth]{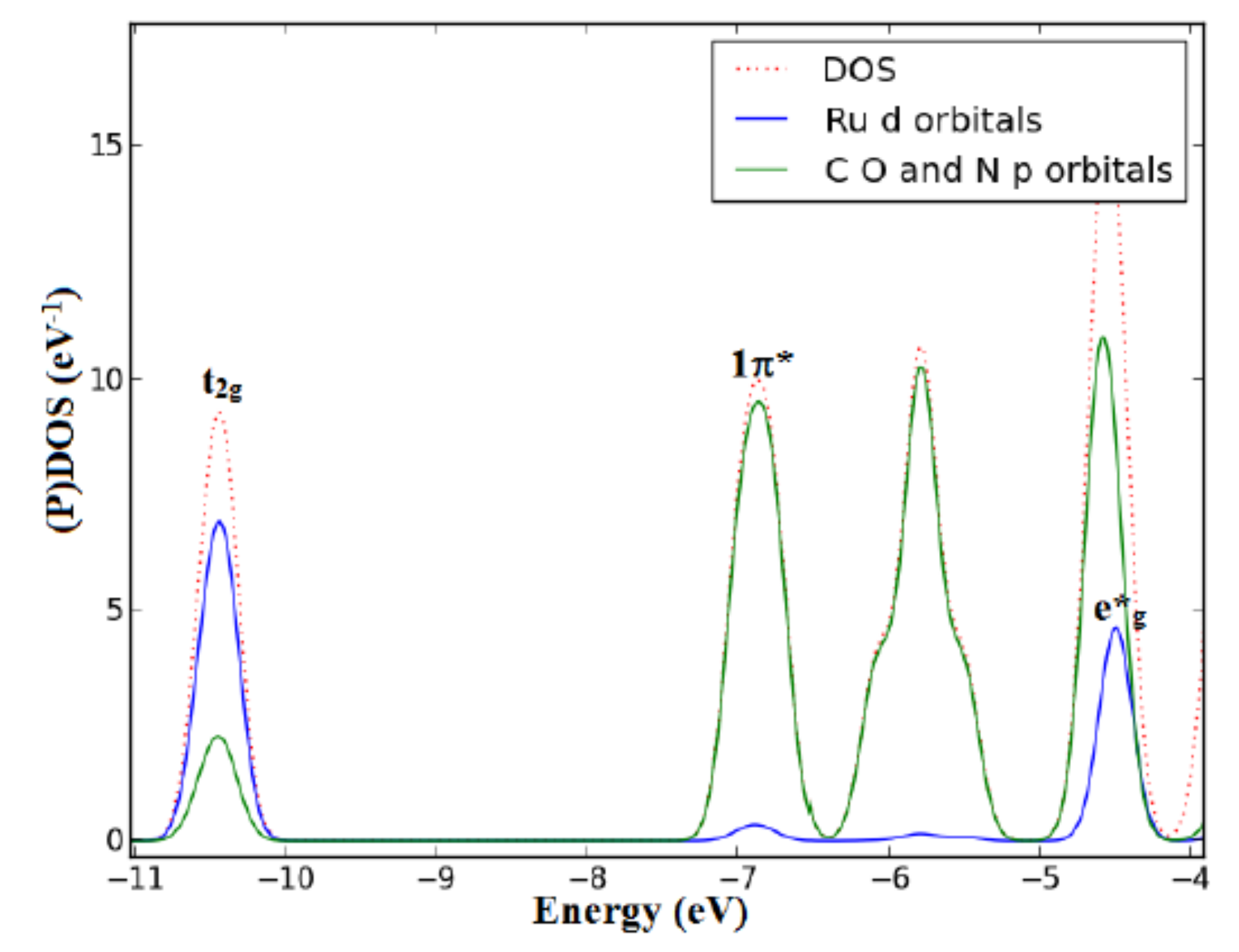} &
\includegraphics[width=0.4\textwidth]{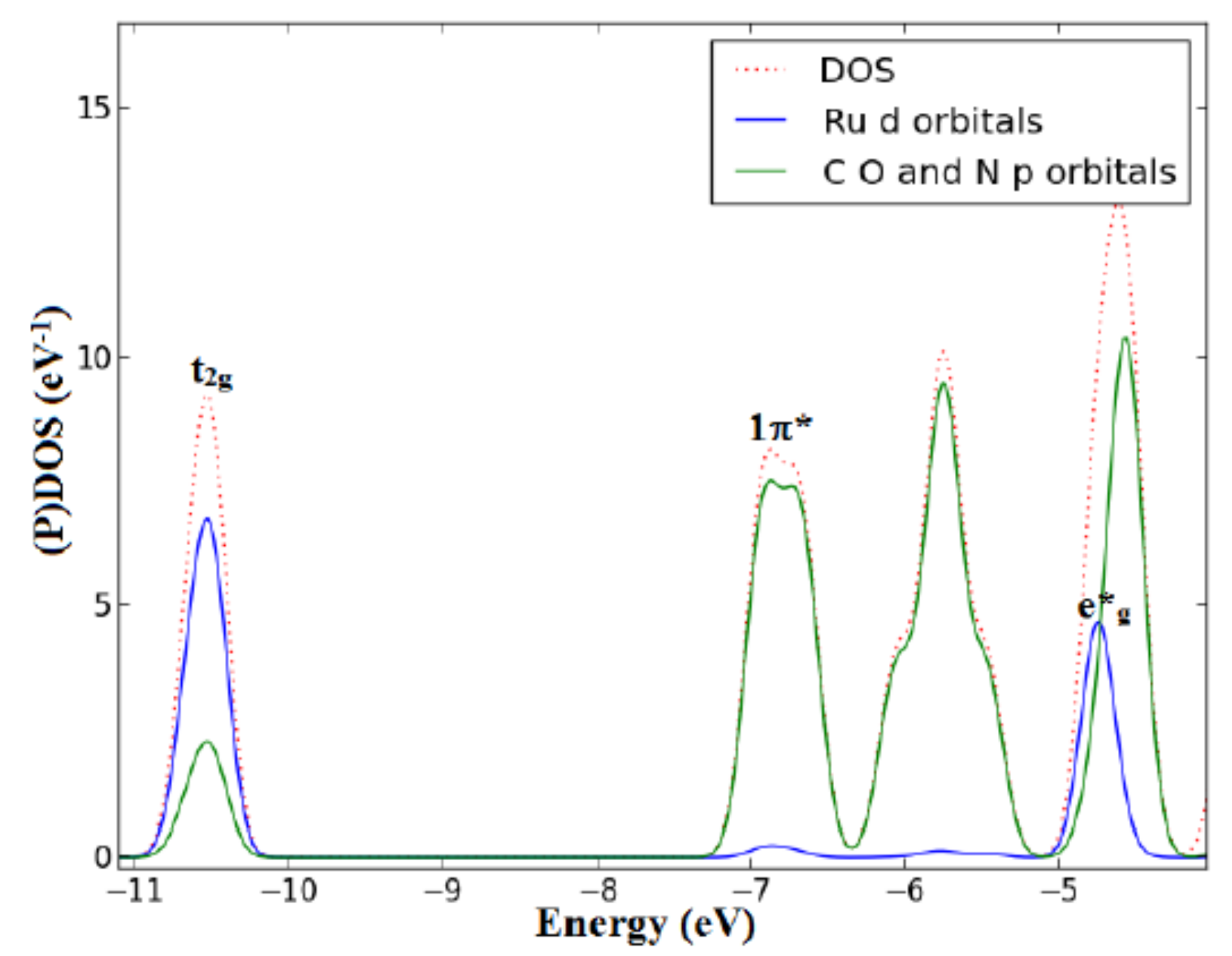} \\
B3LYP/6-31G & B3LYP/6-31G(d) \\
$\epsilon_{\text{HOMO}} = \mbox{-10.38 eV}$ & 
$\epsilon_{\text{HOMO}} = \mbox{-10.48 eV}$ 
\end{tabular}
\end{center}
Total and partial density of states of [Ru(4,4'-dm-bpy)$_2$(4,7-dhy-phen)]$^{2+}$
partitioned over Ru d orbitals and ligand C, O, and N p orbitals. 
% for the 6-31G (left-hand side) and 6-31G* (right-hand side) basis sets.

\begin{center}
   {\bf Absorption Spectrum}
\end{center}

\begin{center}
\includegraphics[width=0.8\textwidth]{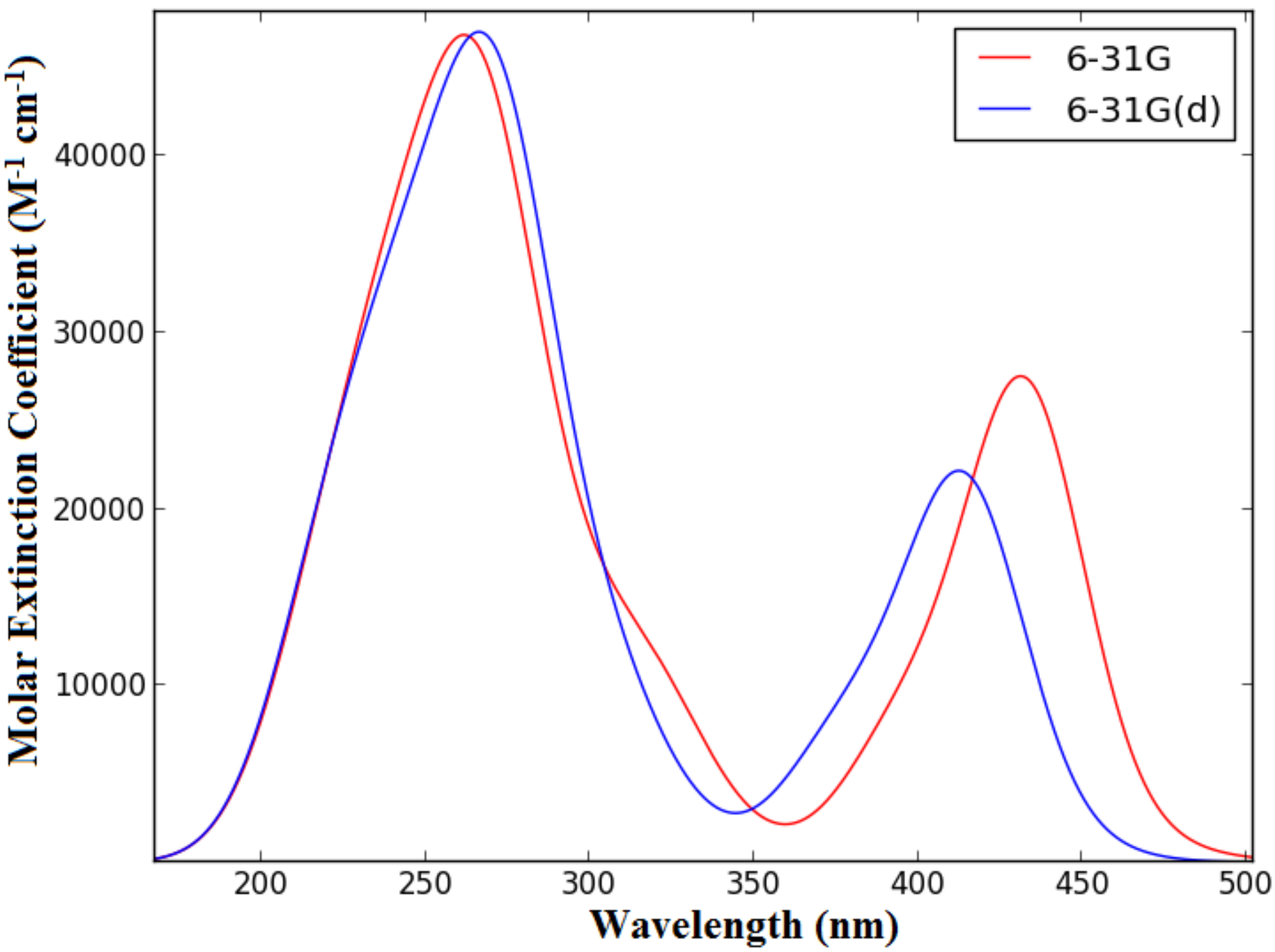}
\end{center}
[Ru(4,4'-dm-bpy)$_2$(4,7-dhy-phen)]$^{2+}$
TD-B3LYP/6-31G and TD-B3LYP/6-31G(d) spectra.

% % ================================================
% \newpage
% \section{Complex {\bf (72)}: [Ru(4,4'-dCl-bpy)$_3$]$^{2+}$}
% % ================================================
% 
% \begin{center}
%    {\bf PDOS}
% \end{center}
% 
% \begin{center}
% \includegraphics[width=0.4\textwidth]{graphics1/framedquestionmark.pdf}
% \includegraphics[width=0.4\textwidth]{graphics1/framedquestionmark.pdf}
% \end{center}
% {\color{magenta} PDOS could not be calculated for complexes containing Cl.}
% 
% \begin{center}
%    {\bf Absorption Spectrum}
% \end{center}
% 
% \begin{center}
% \includegraphics[width=0.4\textwidth]{graphics1/framedquestionmark.pdf}
% \end{center}
% {\color{red} Do we have this?}

% ================================================
\newpage
\section{Complex {\bf (73)}: [Ru(4,4'-dph-bpy)$_3$]$^{2+}$}
% ================================================

\begin{center}
   {\bf PDOS}
\end{center}

\begin{center}
\includegraphics[width=0.4\textwidth]{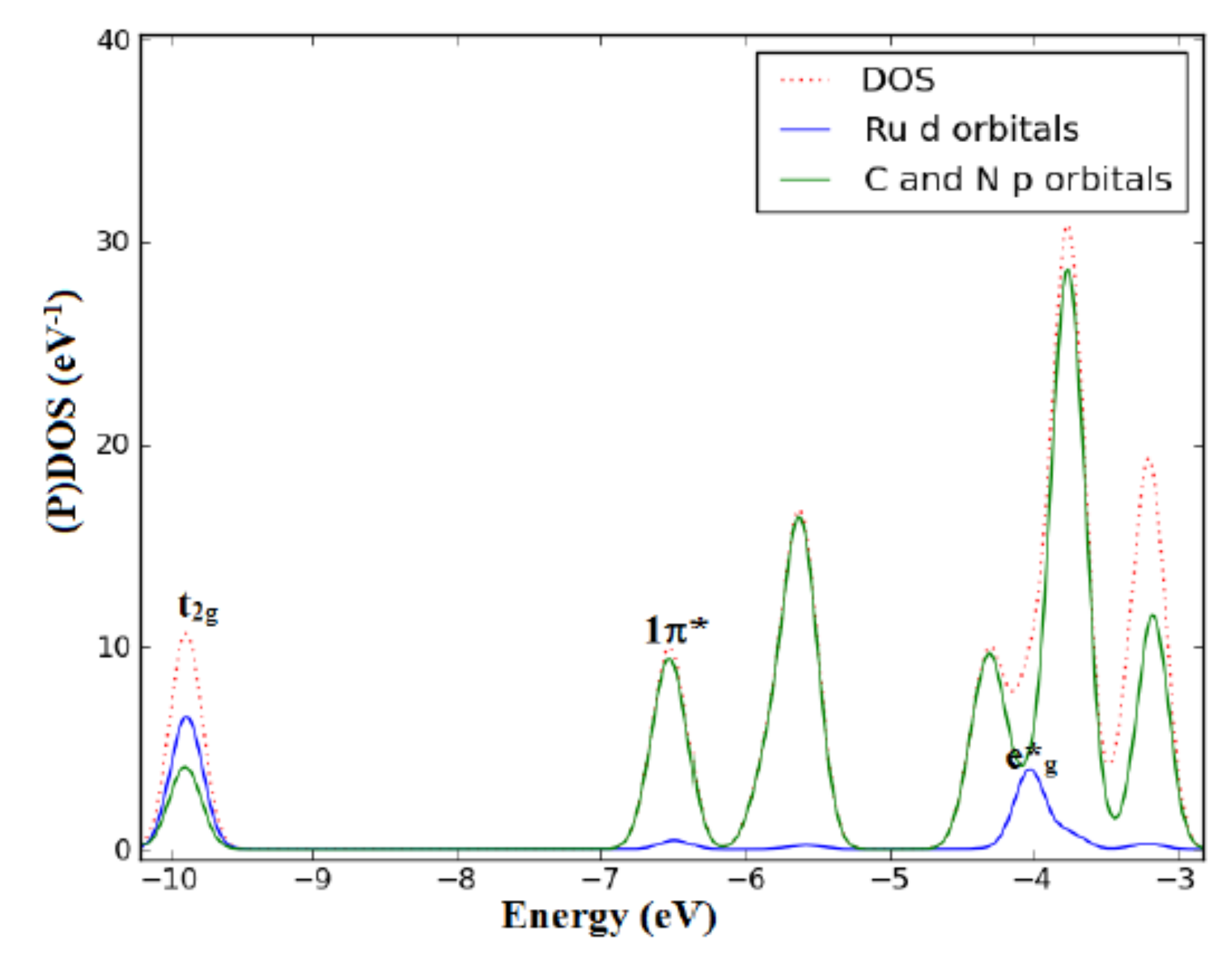}
\\ B3LYP/6-31G \\ $\epsilon_{\text{HOMO}} = \mbox{-9.84 eV}$
\end{center}
Total and partial density of states of [Ru(4,4'-dph-bpy)$_3$]$^{2+}$
partitioned over Ru d orbitals and ligand C and N p orbitals.
% for the 6-31G (left-hand side) and 6-31G* (right-hand side) % {\color{red} Do we have this?}) basis sets.

\begin{center}
   {\bf Absorption Spectrum}
\end{center}

\begin{center}
\includegraphics[width=0.8\textwidth]{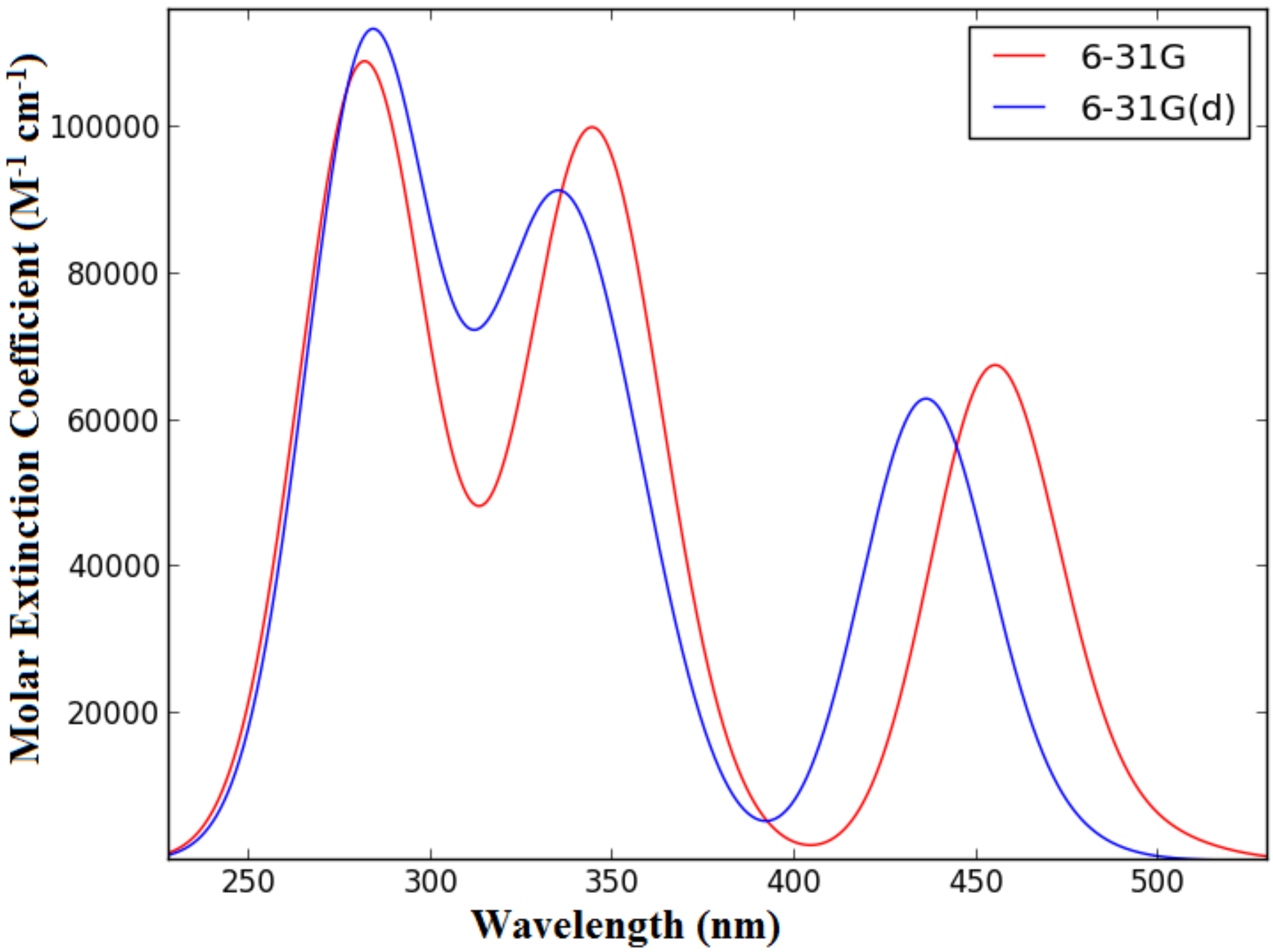}
\end{center}
[Ru(4,4'-dph-bpy)$_3$]$^{2+}$
TD-B3LYP/6-31G and TD-B3LYP/6-31G(d) spectra.

% ================================================
\newpage
\section{Complex {\bf (74)}: [Ru(4,4'-DTB-bpy)$_3$]$^{2+}$}
% ================================================

\begin{center}
   {\bf PDOS}
\end{center}

\begin{center}
\includegraphics[width=0.4\textwidth]{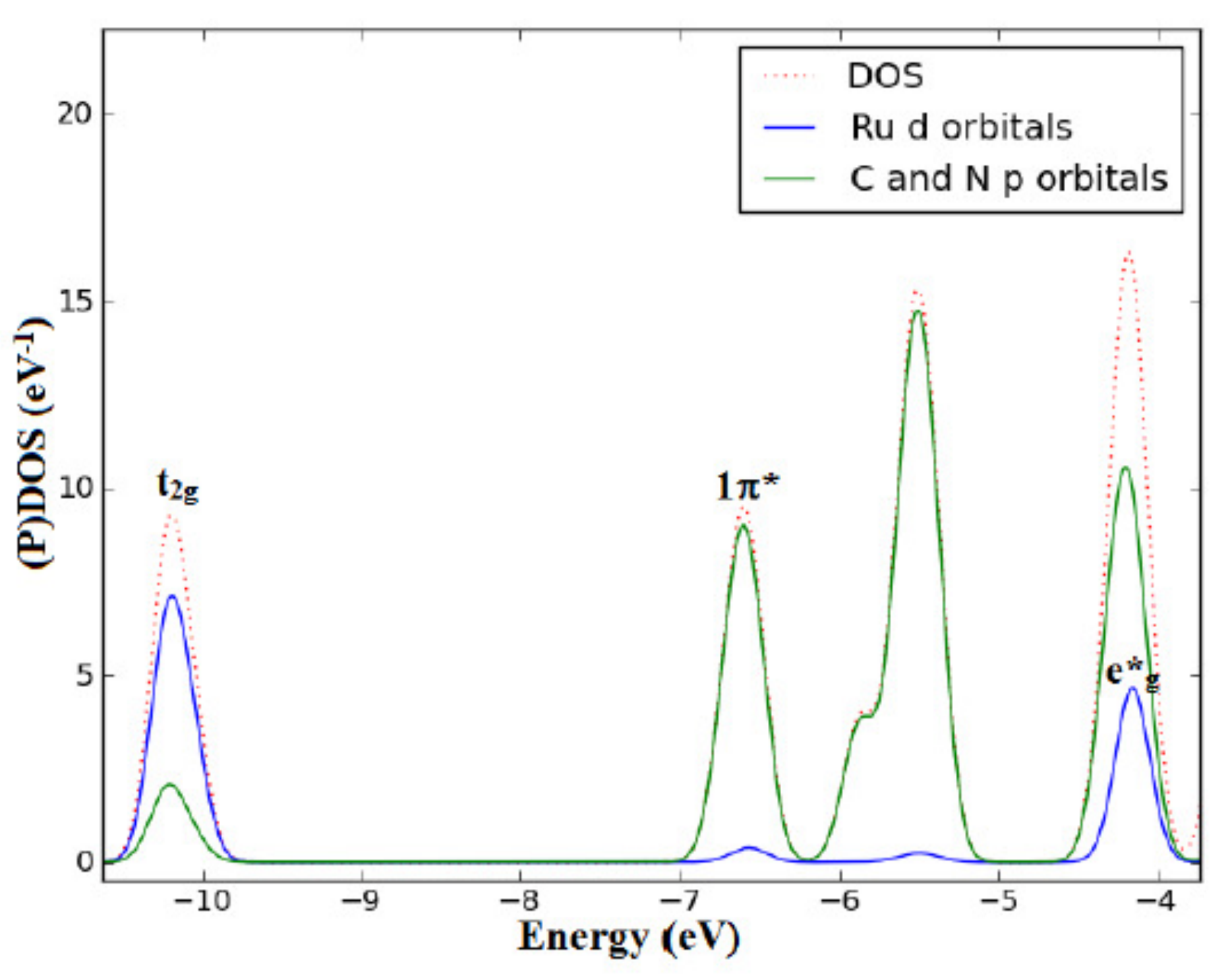}
\\ B3LYP/6-31G \\ $\epsilon_{\text{HOMO}} = \mbox{-10.08 eV}$
\end{center}
Total and partial density of states of [Ru(4,4'-DTB-bpy)$_3$]$^{2+}$
partitioned over Ru d orbitals and ligand C and N p orbitals.
% for the 6-31G (left-hand side) and 6-31G* (right-hand side {\color{red} Do we have this?}) basis sets.

\begin{center}
   {\bf Absorption Spectrum}
\end{center}

\begin{center}
\includegraphics[width=0.8\textwidth]{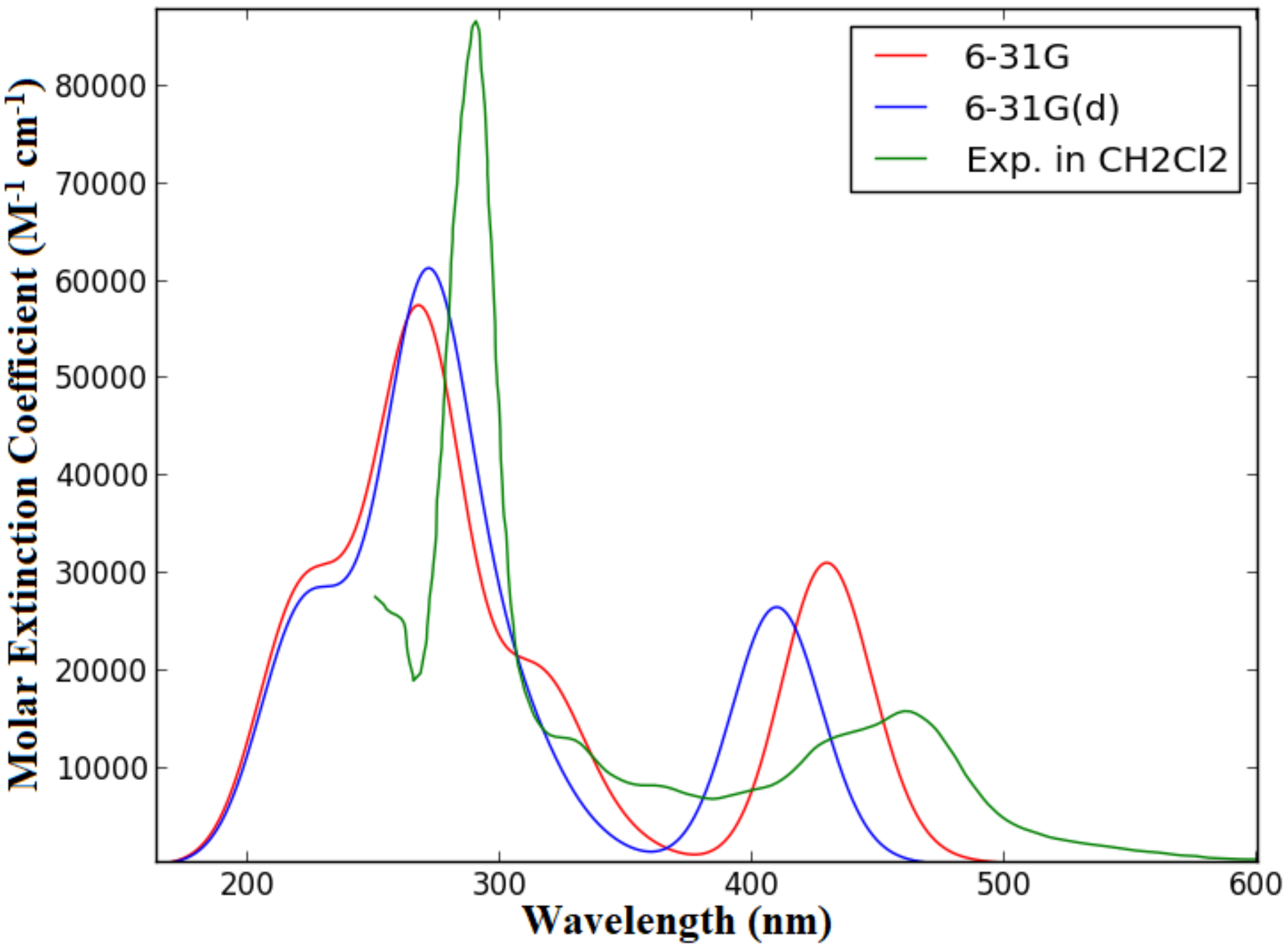}
\end{center}
[Ru(4,4'-DTB-bpy)$_3$]$^{2+}$
TD-B3LYP/6-31G, TD-B3LYP/6-31G(d), and experimental spectra.
Experimental spectrum measured in dichloromethane at room temperature 
\cite{SSG+08}.

% ================================================
\newpage
\section{Complex {\bf (75)}: [Ru(6,6'-dm-bpy)$_3$]$^{2+}$}
% ================================================

\begin{center}
   {\bf PDOS}
\end{center}

\begin{center}
\begin{tabular}{cc}
\includegraphics[width=0.4\textwidth]{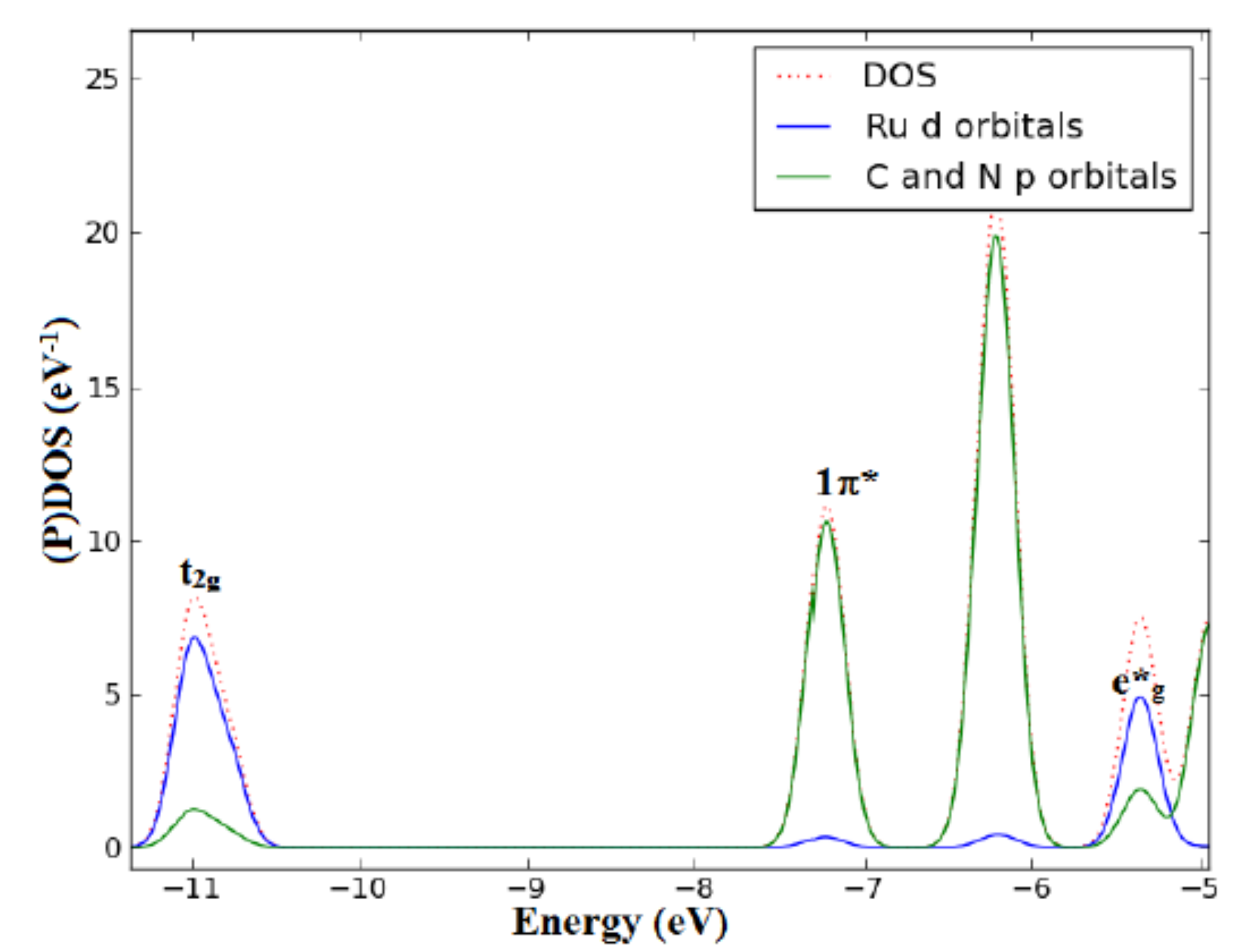} &
\includegraphics[width=0.4\textwidth]{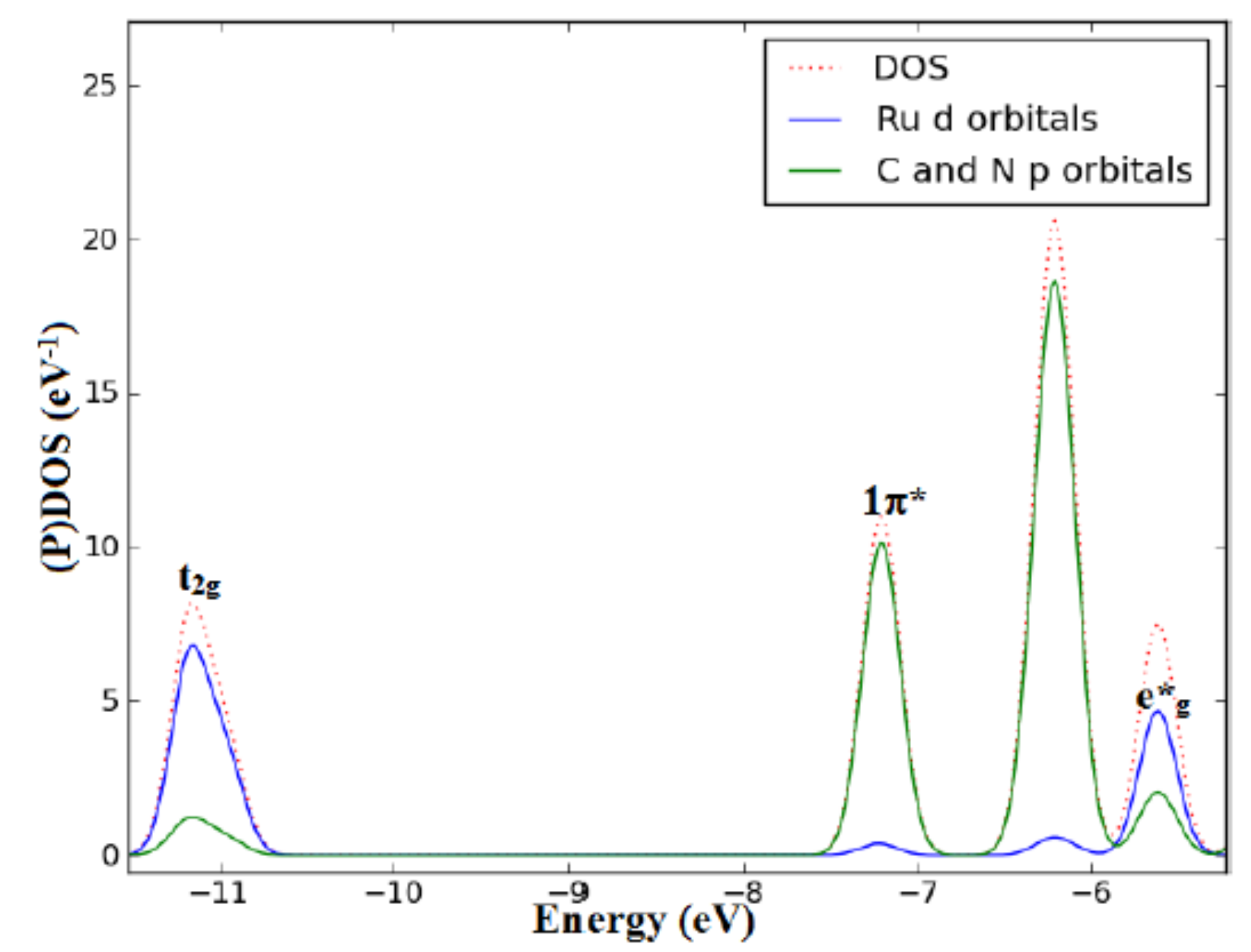} \\
B3LYP/6-31G & B3LYP/6-31G(d) \\
$\epsilon_{\text{HOMO}} = \mbox{-10.80 eV}$ & 
$\epsilon_{\text{HOMO}} = \mbox{-10.97 eV}$ 
\end{tabular}
\end{center}
Total and partial density of states of [Ru(6,6'-dm-bpy)$_3$]$^{2+}$
partitioned over Ru d orbitals and ligand C and N p orbitals.
% for the 6-31G (left-hand side) and 6-31G* (right-hand side) basis sets.

\begin{center}
   {\bf Absorption Spectrum}
\end{center}

\begin{center}
\includegraphics[width=0.8\textwidth]{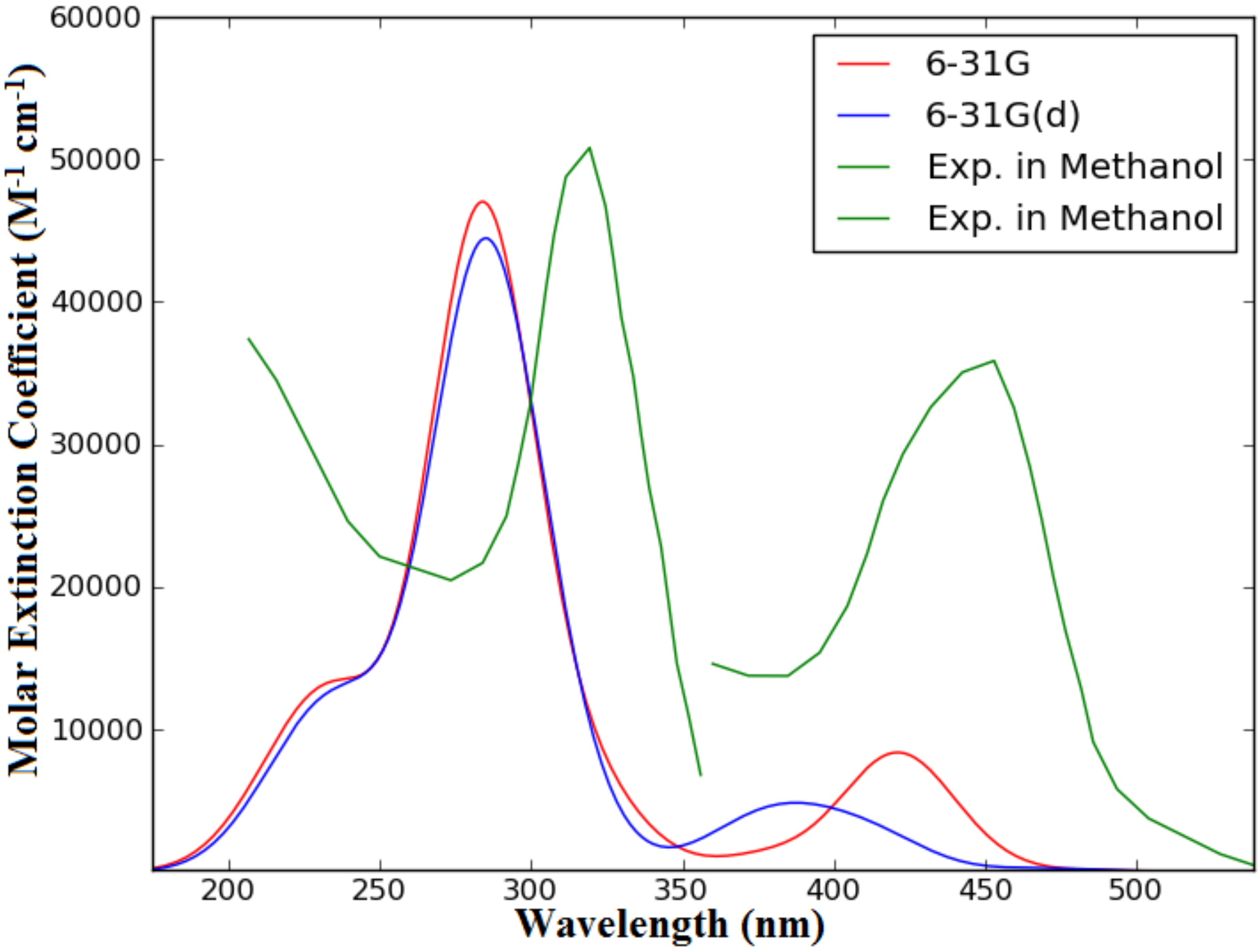}
\end{center}
[Ru(6,6'-dm-bpy)$_3$]$^{2+}$
TD-B3LYP/6-31G, TD-B3LYP/6-31G(d), and experimental spectra.
Experimental spectrum measured in methanol \cite{FKS80}.

% ================================================
\newpage
\section{Complex {\bf (76)}: [Ru(h-phen)$_3$]$^{2+}$}
% ================================================

\begin{center}
   {\bf PDOS}
\end{center}

\begin{center}
\begin{tabular}{cc}
\includegraphics[width=0.4\textwidth]{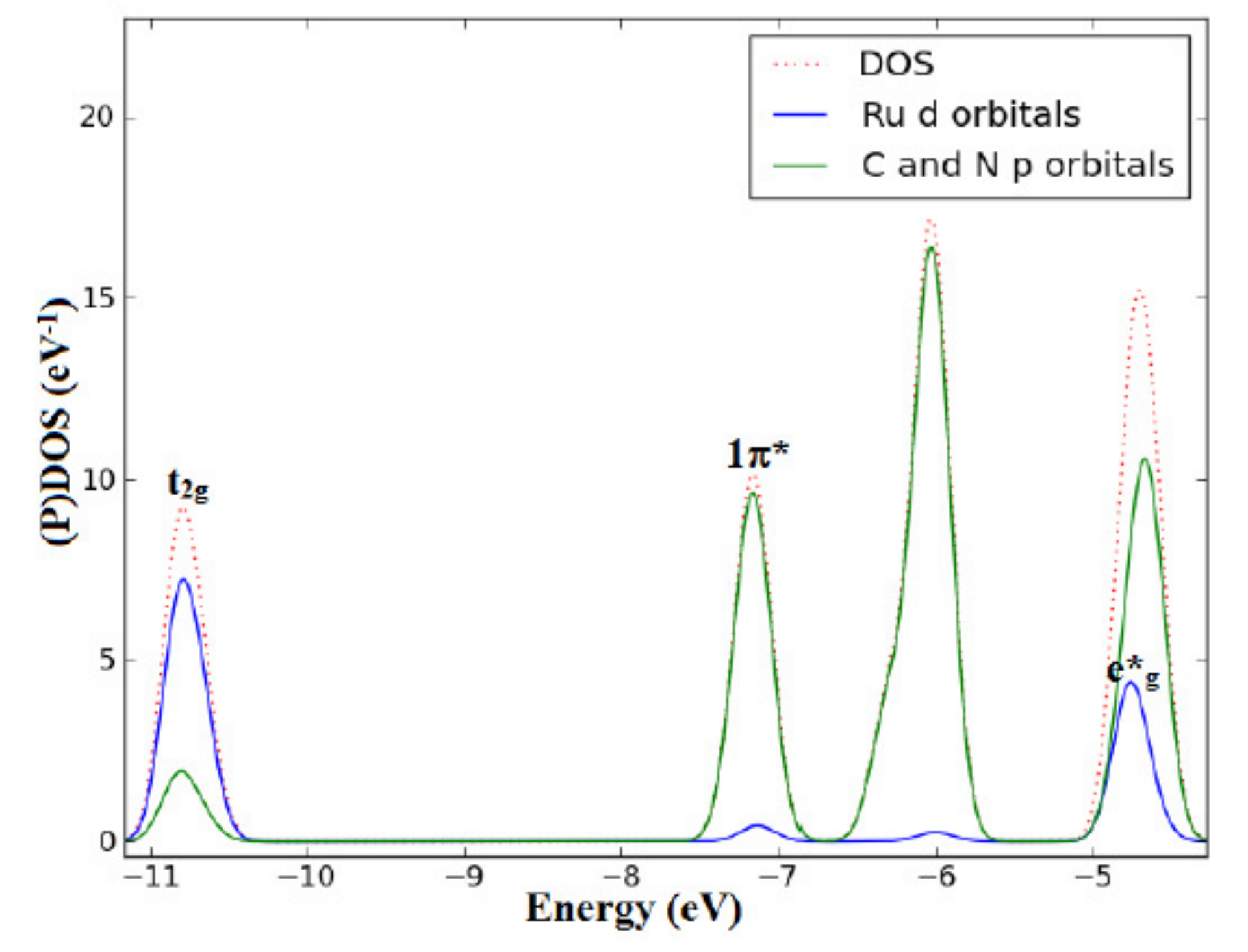} &
\includegraphics[width=0.4\textwidth]{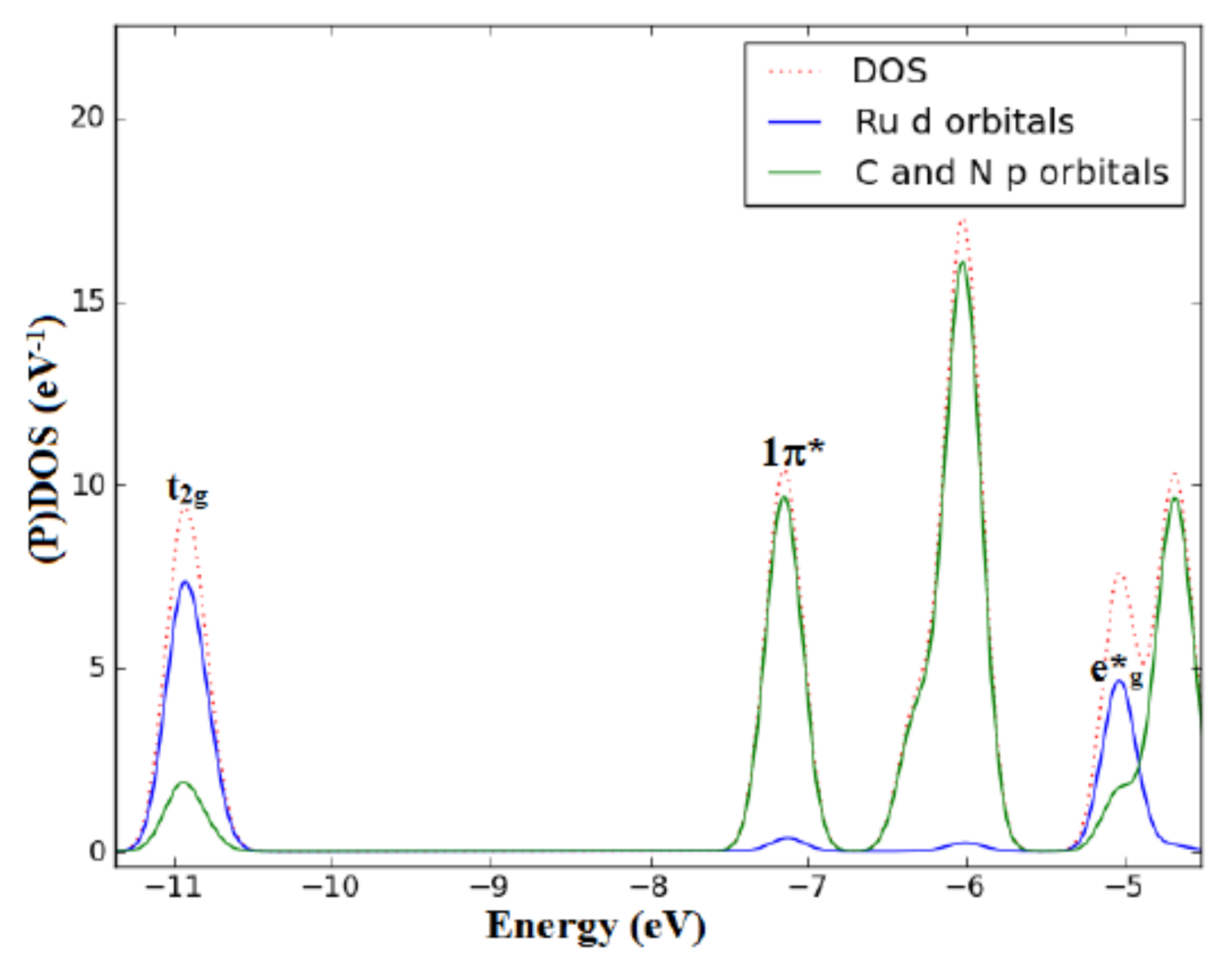} \\
B3LYP/6-31G & B3LYP/6-31G(d) \\
$\epsilon_{\text{HOMO}} = \mbox{-10.68 eV}$ & 
$\epsilon_{\text{HOMO}} = \mbox{-10.82 eV}$ 
\end{tabular}
\end{center}
Total and partial density of states of [Ru(h-phen)$_3$]$^{2+}$
partitioned over Ru d orbitals and ligand C and N p orbitals. 
% for the 6-31G (left-hand side) and 6-31G* (right-hand side) basis sets.

\begin{center}
   {\bf Absorption Spectrum}
\end{center}

\begin{center}
\includegraphics[width=0.8\textwidth]{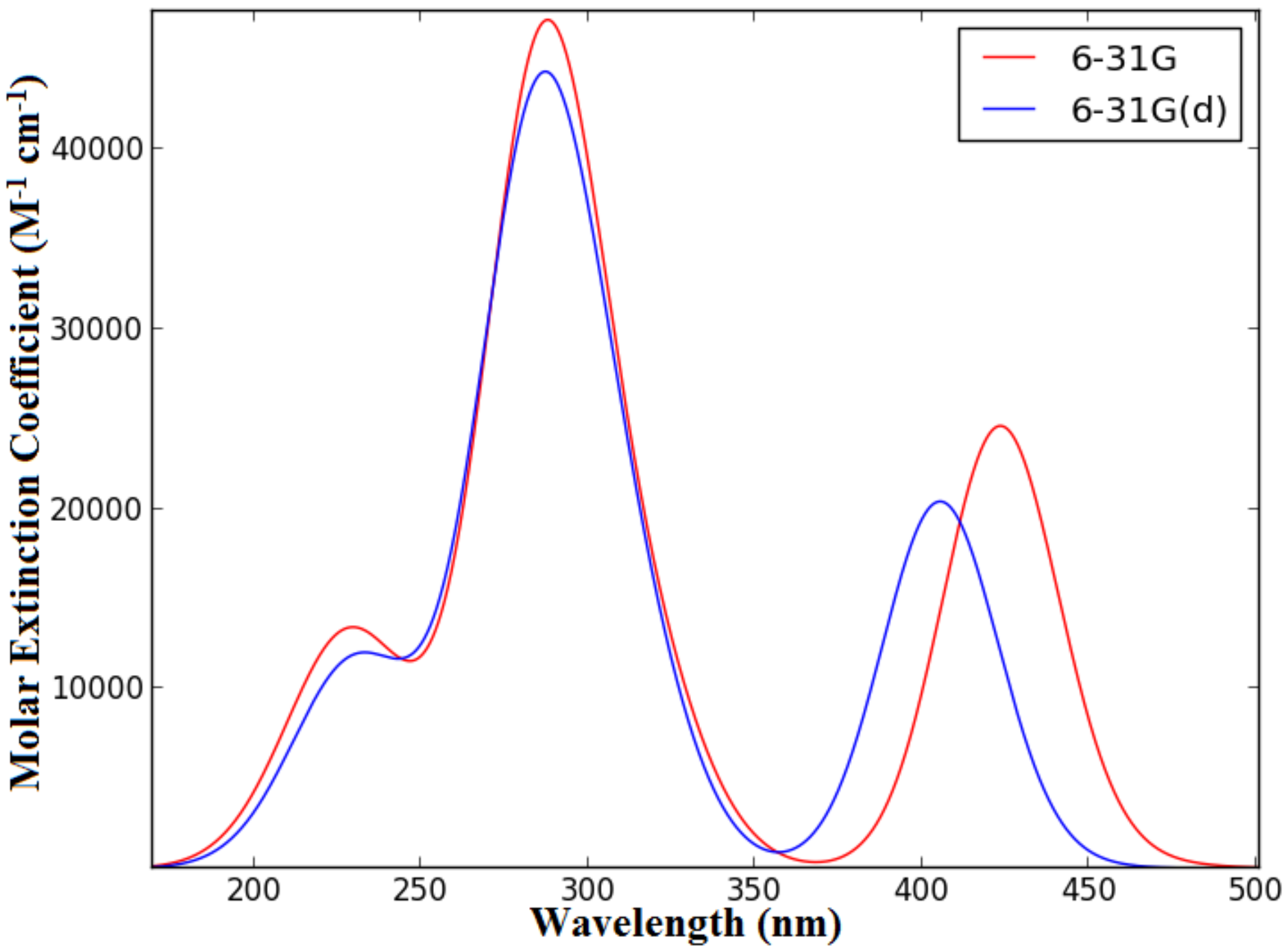}
\end{center}
[Ru(h-phen)$_3$]$^{2+}$
TD-B3LYP/6-31G and TD-B3LYP/6-31G(d) spectra.

% ================================================
\newpage
\section{Complex {\bf (77)}: [Ru(phen)$_3$]$^{2+}$}
% ================================================

\begin{center}
   {\bf PDOS}
\end{center}

\begin{center}
\begin{tabular}{cc}
\includegraphics[width=0.4\textwidth]{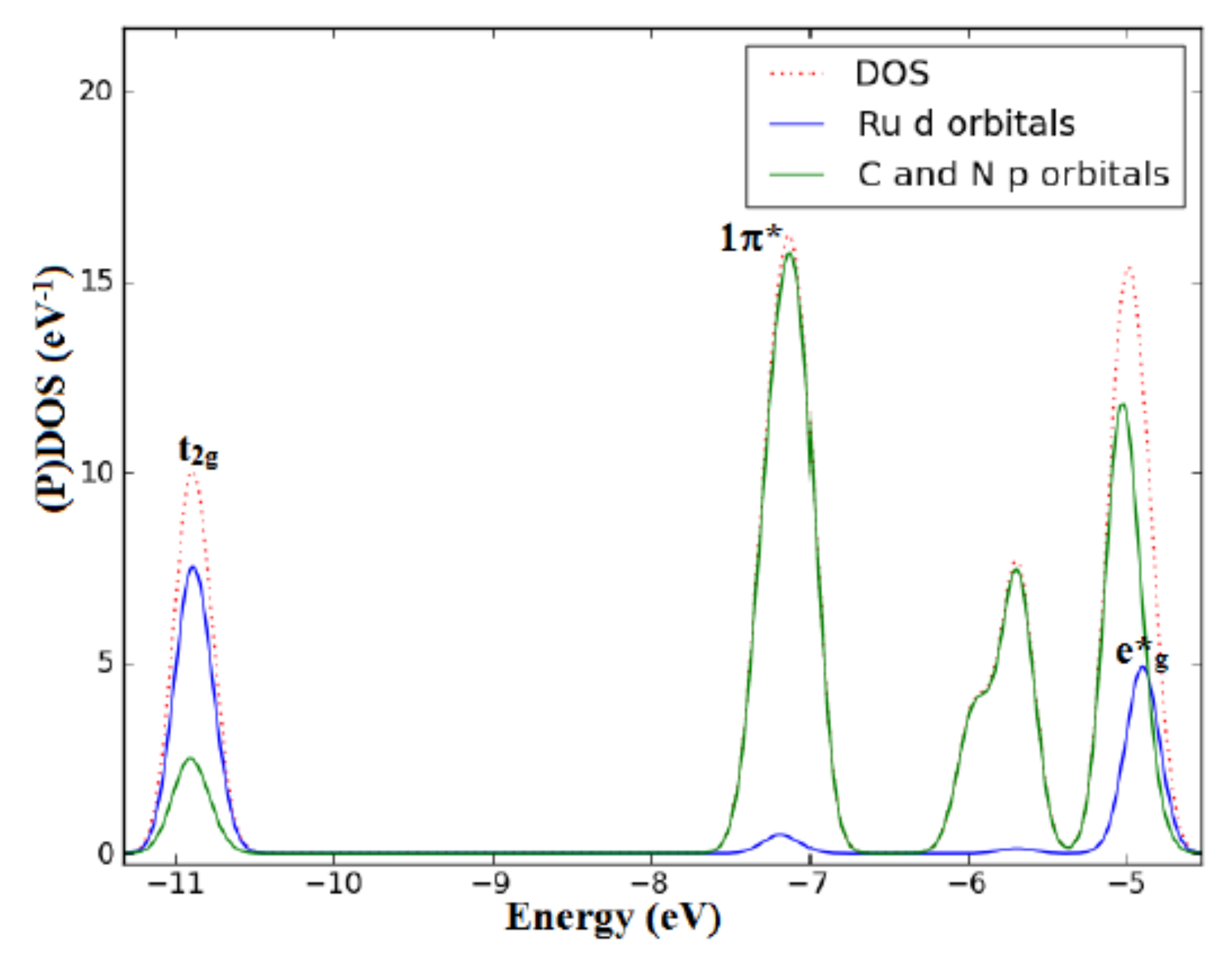} &
\includegraphics[width=0.4\textwidth]{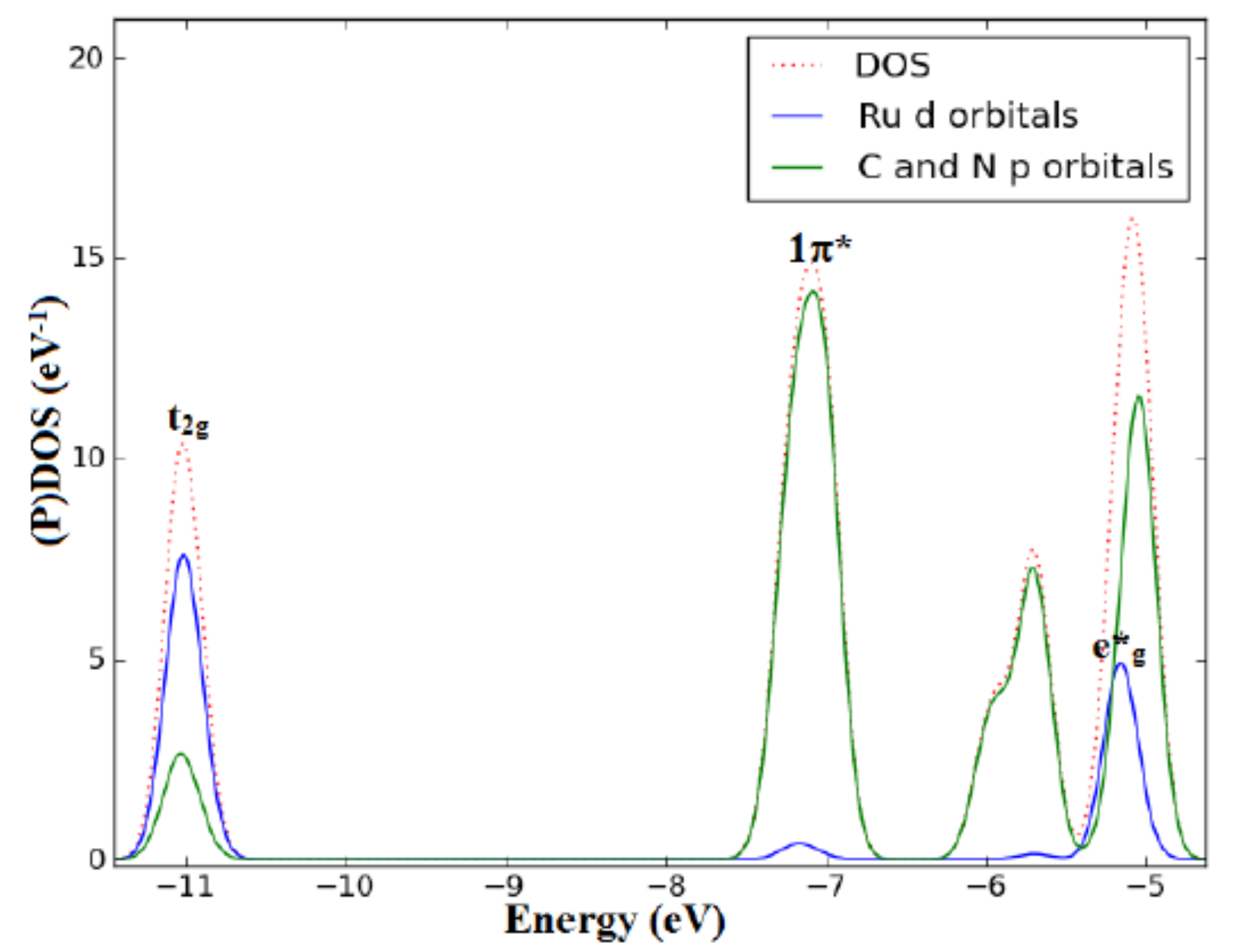} \\
B3LYP/6-31G & B3LYP/6-31G(d) \\
$\epsilon_{\text{HOMO}} = \mbox{-10.82 eV}$ & 
$\epsilon_{\text{HOMO}} = \mbox{-10.95 eV}$ 
\end{tabular}
\end{center}
Total and partial density of states of [Ru(phen)$_3$]$^{2+}$
partitioned over Ru d orbitals and ligand C and N p orbitals. 
% for the 6-31G (left-hand side) and 6-31G* (right-hand side) basis sets.

\begin{center}
   {\bf Absorption Spectrum}
\end{center}

\begin{center}
\includegraphics[width=0.8\textwidth]{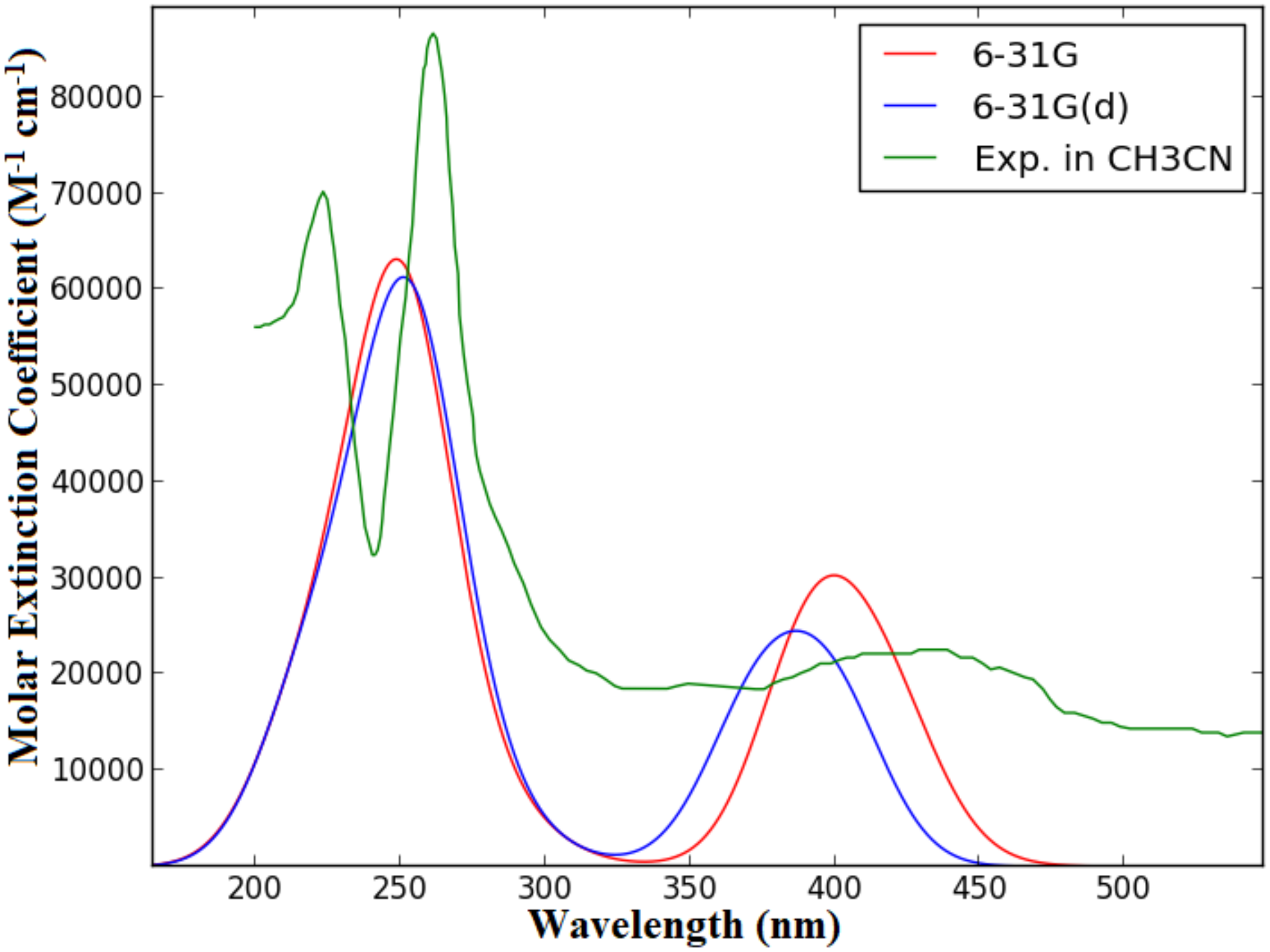}
\end{center}
[Ru(phen)$_3$]$^{2+}$
TD-B3LYP/6-31G, TD-B3LYP/6-31G(d), and experimental spectra.
Experimental spectrum measured in acetonitrile \cite{LRE+04}.

% ================================================
\newpage
\section{Complex {\bf (78)}: [Ru(phen)$_2$(4,7-dhy-phen)]$^{2+}$}
% ================================================

\begin{center}
   {\bf PDOS}
\end{center}

\begin{center}
\begin{tabular}{cc}
\includegraphics[width=0.4\textwidth]{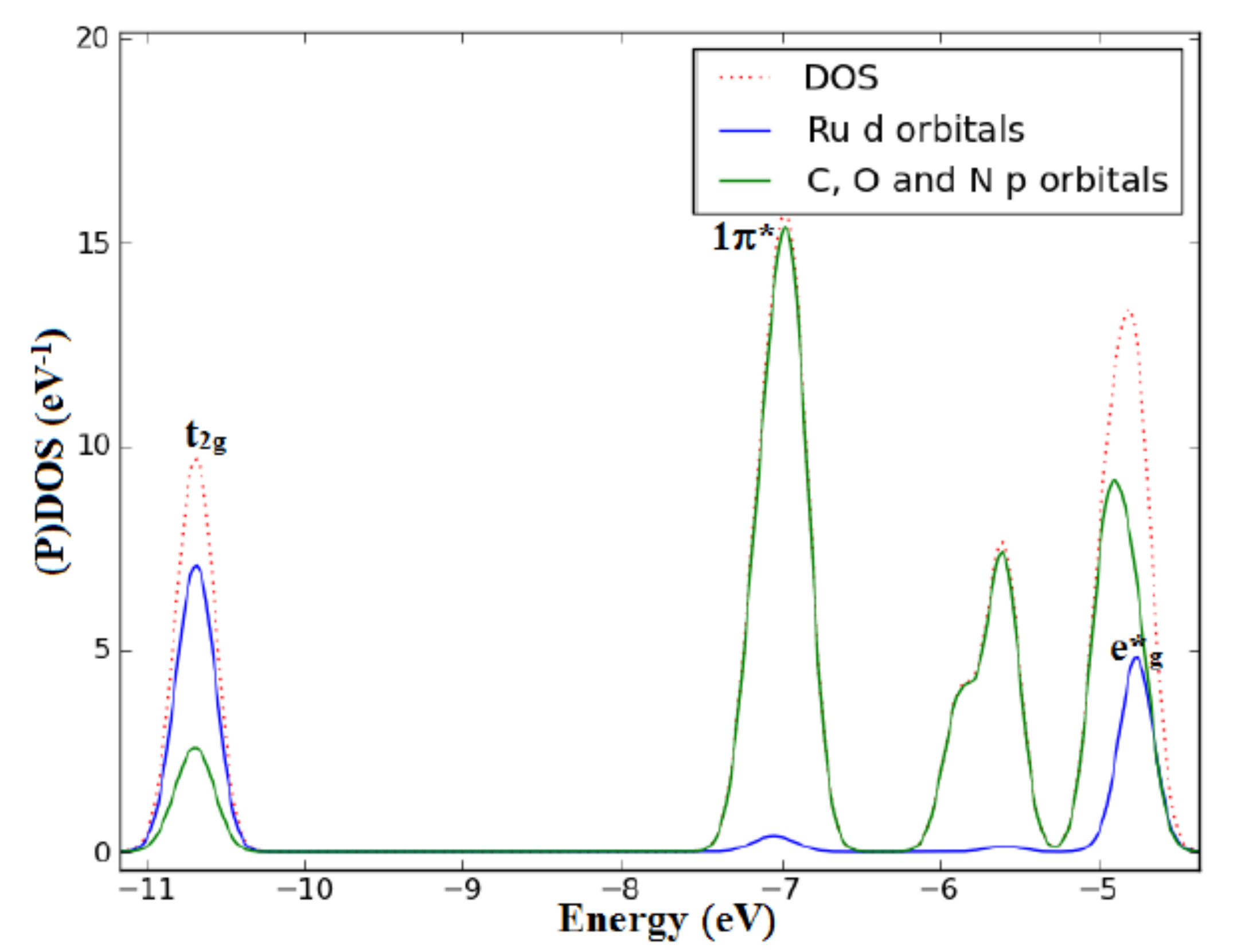} &
\includegraphics[width=0.4\textwidth]{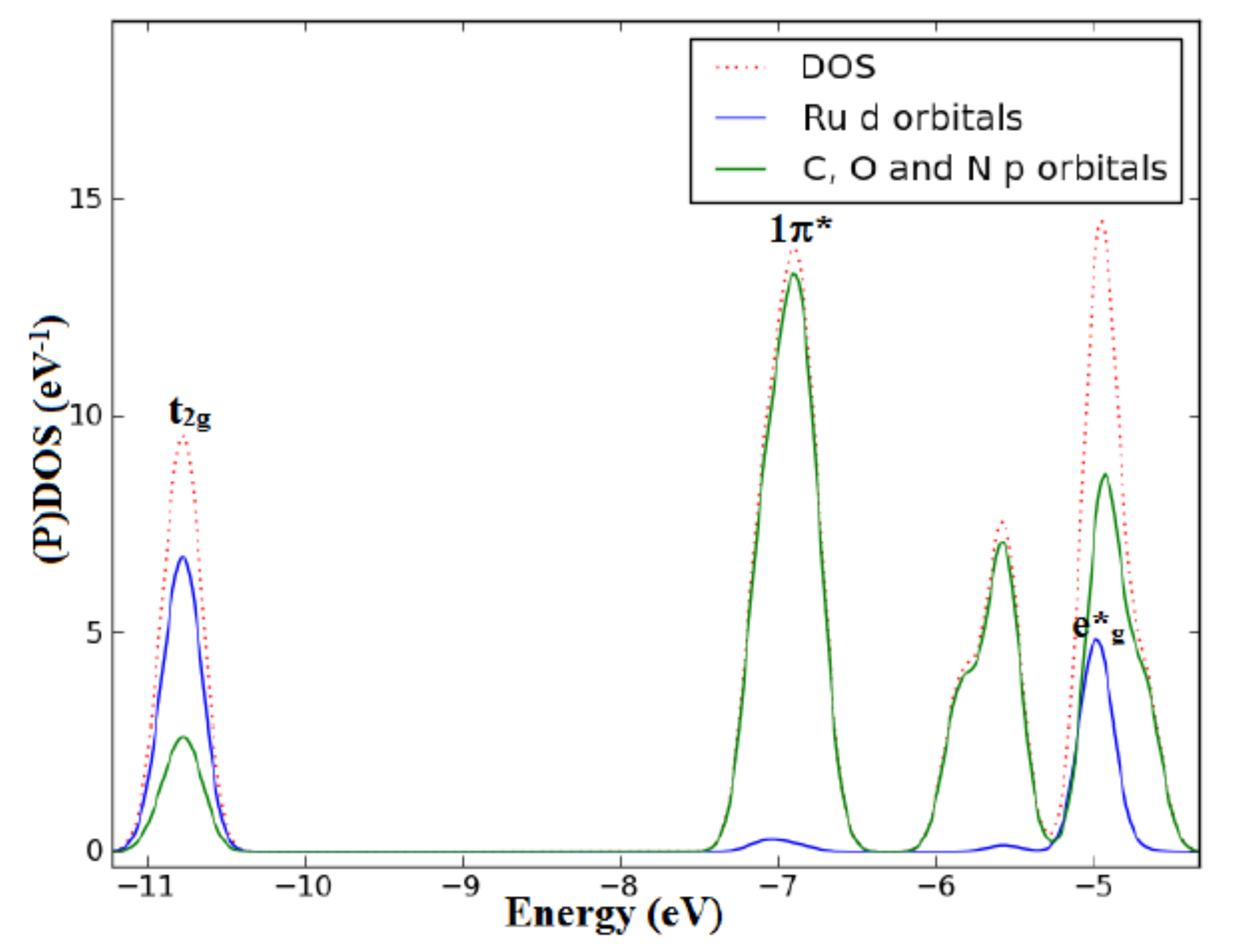} \\
B3LYP/6-31G & B3LYP/6-31G(d) \\
$\epsilon_{\text{HOMO}} = \mbox{-10.63 eV}$ & 
$\epsilon_{\text{HOMO}} = \mbox{-10.72 eV}$ 
\end{tabular}
\end{center}
Total and partial density of states of [Ru(phen)$_2$(4,7-dhy-phen)]$^{2+}$
partitioned over Ru d orbitals and ligand C, O, and N p orbitals. 
% for the 6-31G (left-hand side) and 6-31G* (right-hand side) basis sets.

\begin{center}
   {\bf Absorption Spectrum}
\end{center}

\begin{center}
\includegraphics[width=0.8\textwidth]{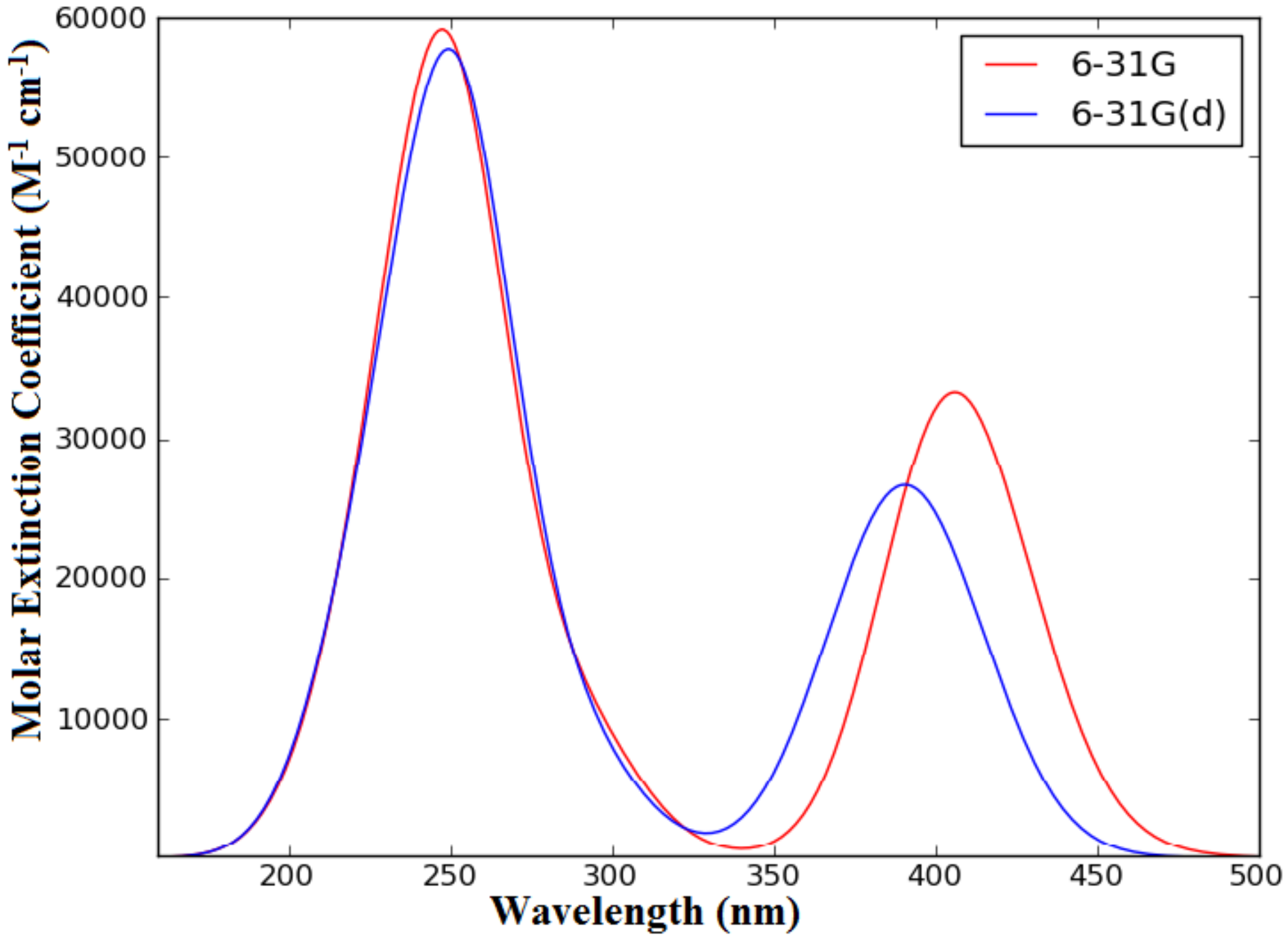}
\end{center}
[Ru(phen)$_2$(4,7-dhy-phen)]$^{2+}$
TD-B3LYP/6-31G and TD-B3LYP/6-31G(d) spectra.

% ================================================
\newpage
\section{Complex {\bf (79)}: [Ru(phen)$_2$(pq)]$^{2+}$}
% ================================================

\begin{center}
   {\bf PDOS}
\end{center}

\begin{center}
\begin{tabular}{cc}
\includegraphics[width=0.4\textwidth]{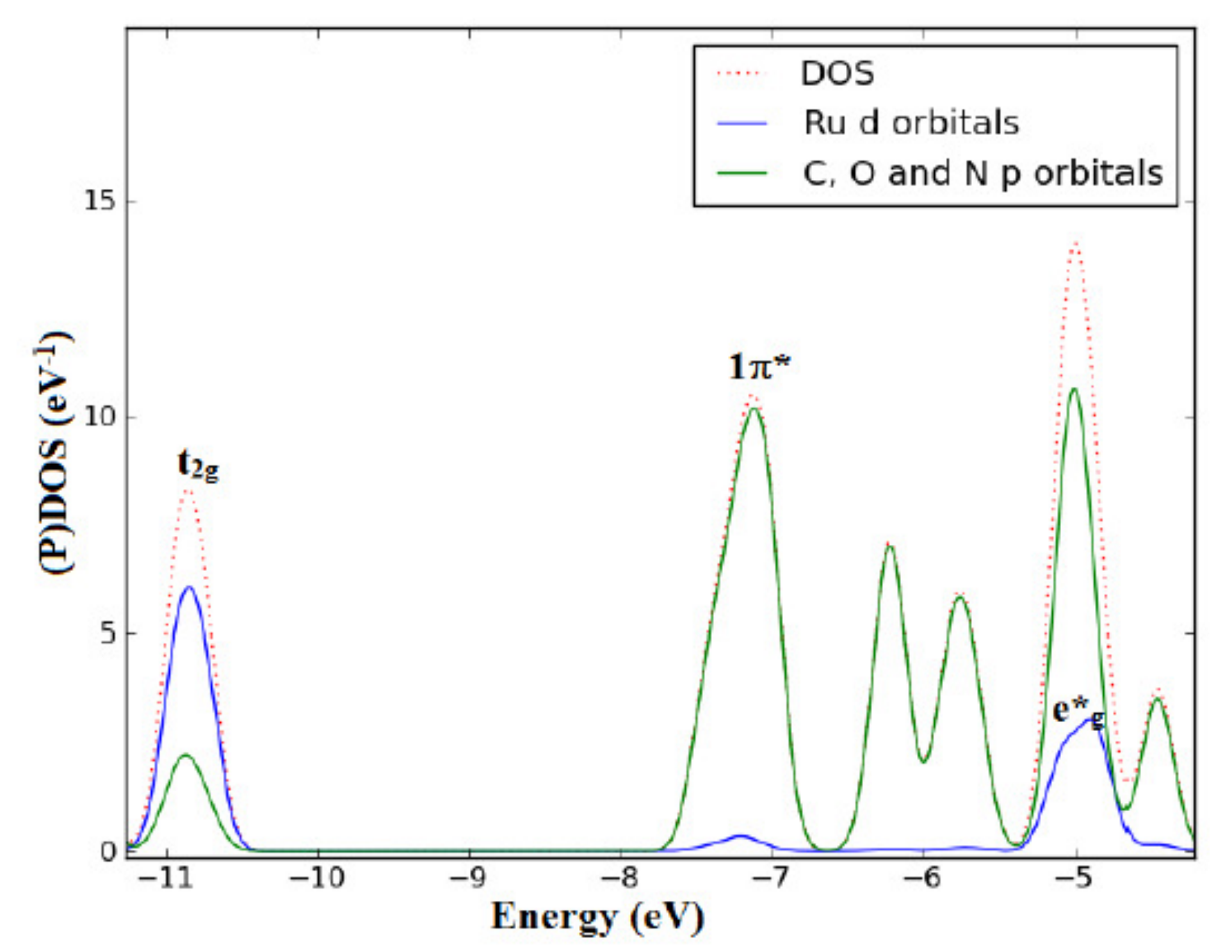} &
\includegraphics[width=0.4\textwidth]{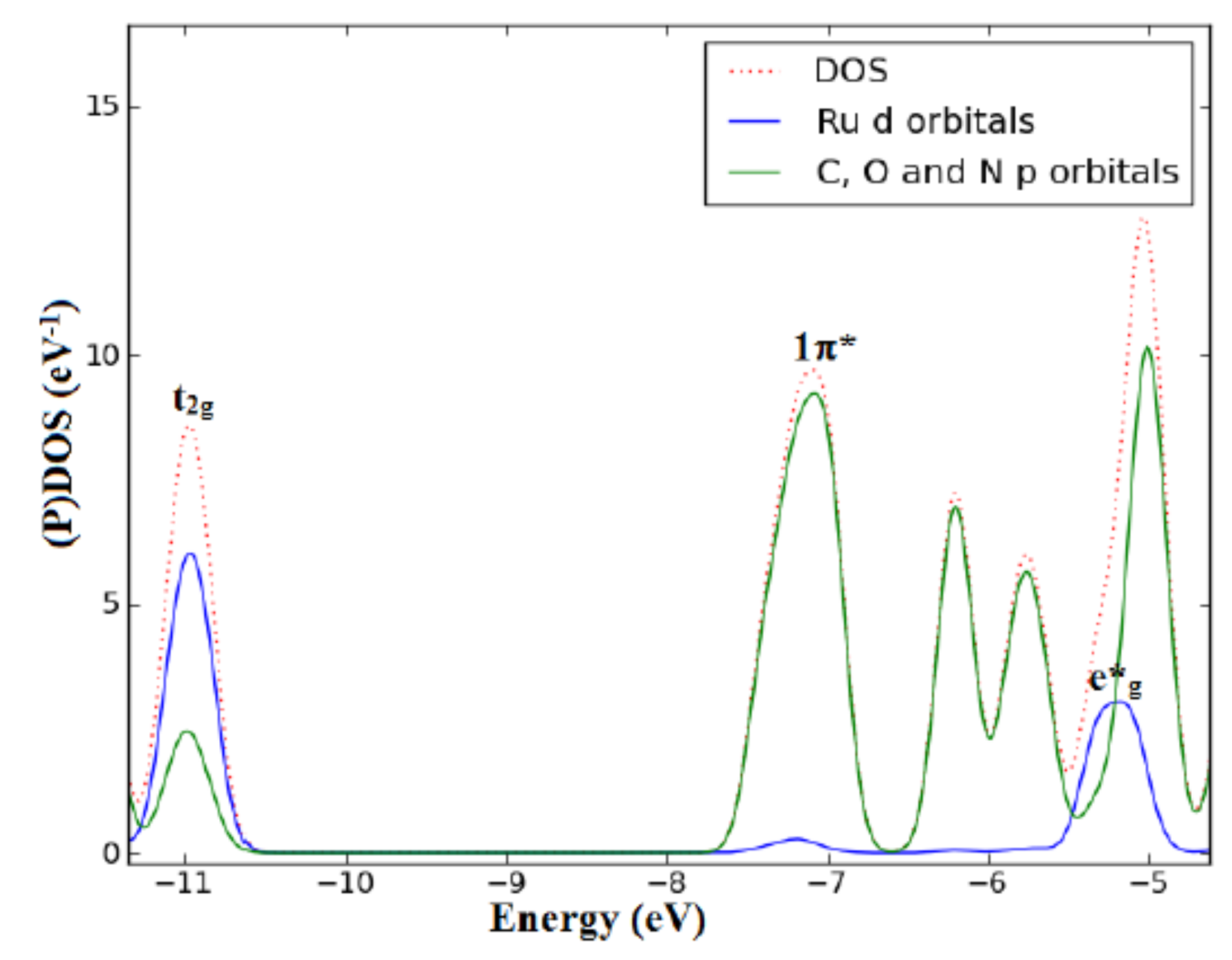} \\
B3LYP/6-31G & B3LYP/6-31G(d) \\
$\epsilon_{\text{HOMO}} = \mbox{-10.74 eV}$ & 
$\epsilon_{\text{HOMO}} = \mbox{-10.87 eV}$ 
\end{tabular}
\end{center}
Total and partial density of states of [Ru(phen)$_2$(pq)]$^{2+}$
partitioned over Ru d orbitals and ligand C and N p orbitals. 
% for the 6-31G (left-hand side) and 6-31G* (right-hand side) basis sets.

\begin{center}
   {\bf Absorption Spectrum}
\end{center}

\begin{center}
\includegraphics[width=0.8\textwidth]{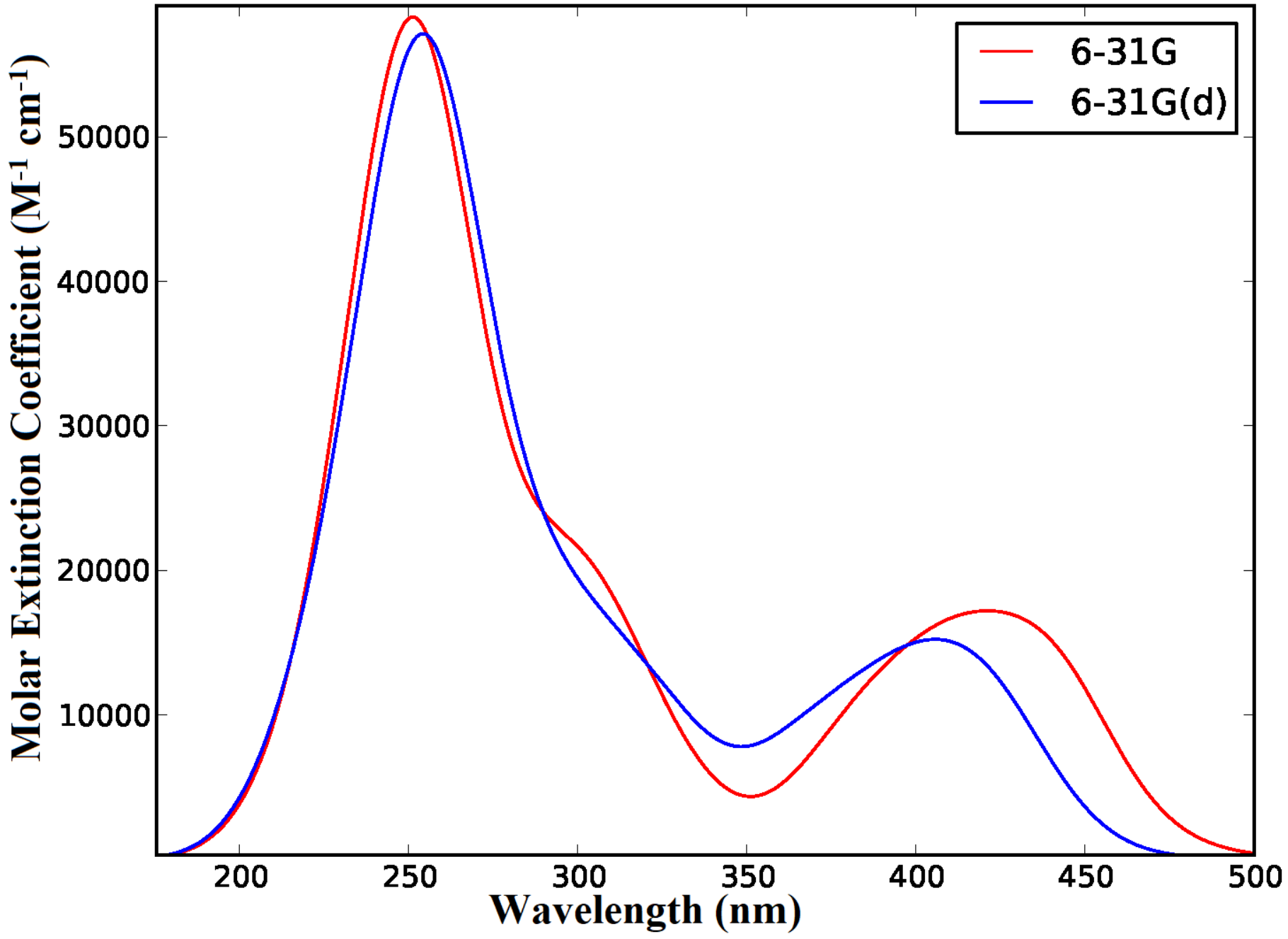}
\end{center}
[Ru(phen)$_2$(pq)]$^{2+}$
TD-B3LYP/6-31G and TD-B3LYP/6-31G(d) spectra.

% ================================================
\newpage
\section{Complex {\bf (80)}: [Ru(phen)$_2$(DMCH)]$^{2+}$}
% ================================================

\begin{center}
   {\bf PDOS}
\end{center}

\begin{center}
\begin{tabular}{cc}
\includegraphics[width=0.4\textwidth]{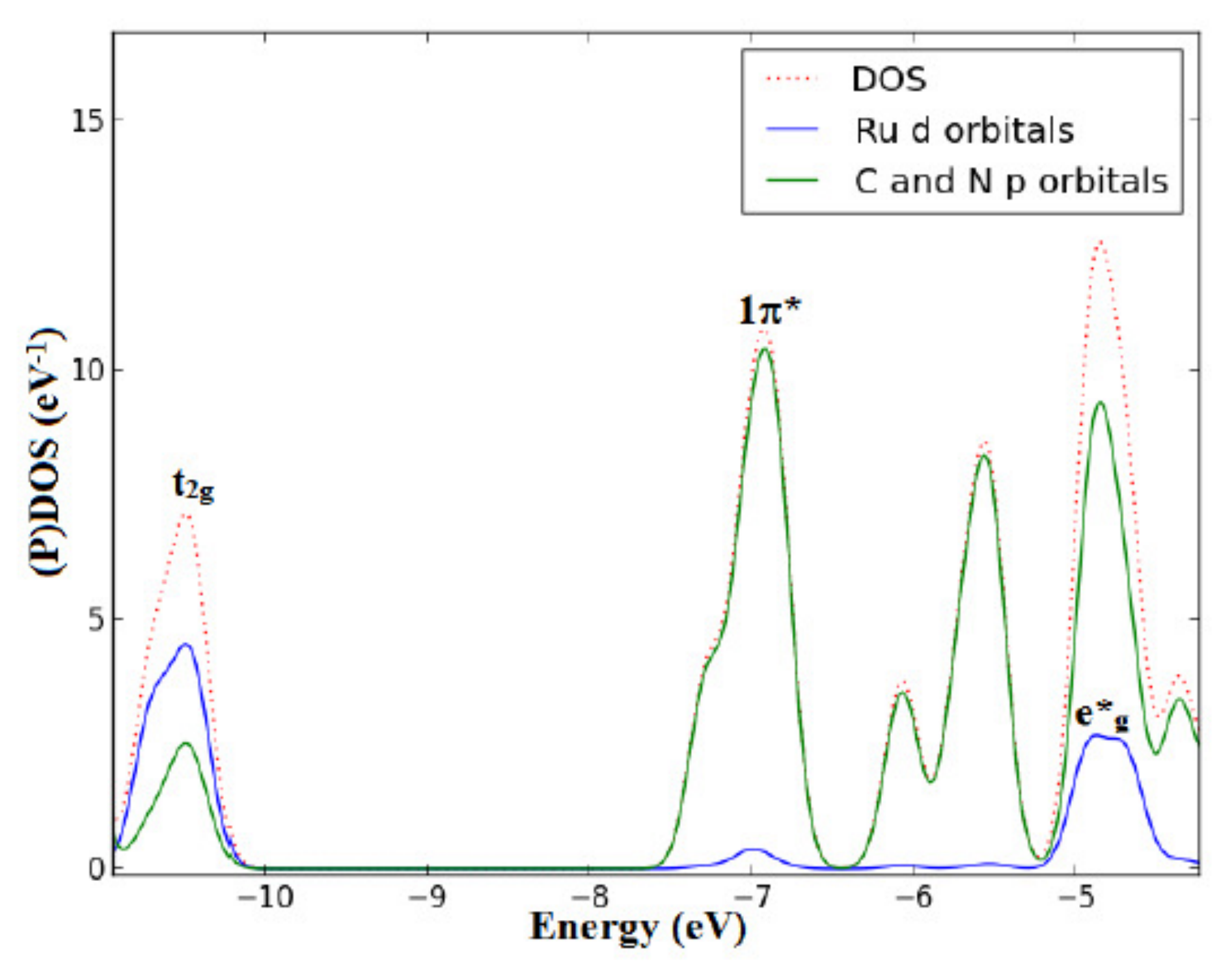} &
\includegraphics[width=0.4\textwidth]{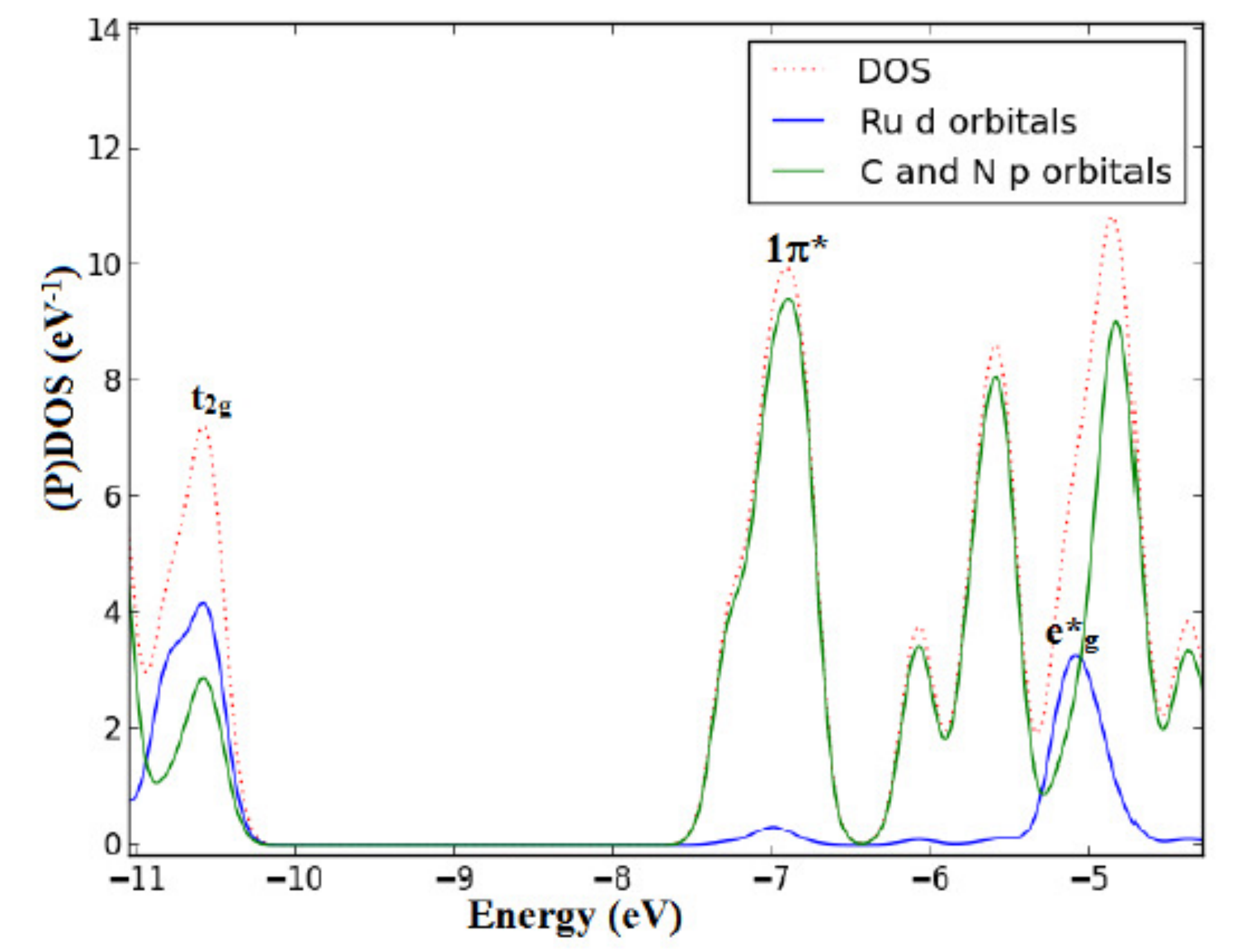} \\
B3LYP/6-31G & B3LYP/6-31G(d) \\
$\epsilon_{\text{HOMO}} = \mbox{-10.42 eV}$ & 
$\epsilon_{\text{HOMO}} = \mbox{-10.52 eV}$ 
\end{tabular}
\end{center}
Total and partial density of states of [Ru(phen)$_2$(DMCH)]$^{2+}$
partitioned over Ru d orbitals and ligand C and N p orbitals. 
% for the 6-31G (left-hand side) and 6-31G* (right-hand side) basis sets.

\begin{center}
   {\bf Absorption Spectrum}
\end{center}

\begin{center}
\includegraphics[width=0.8\textwidth]{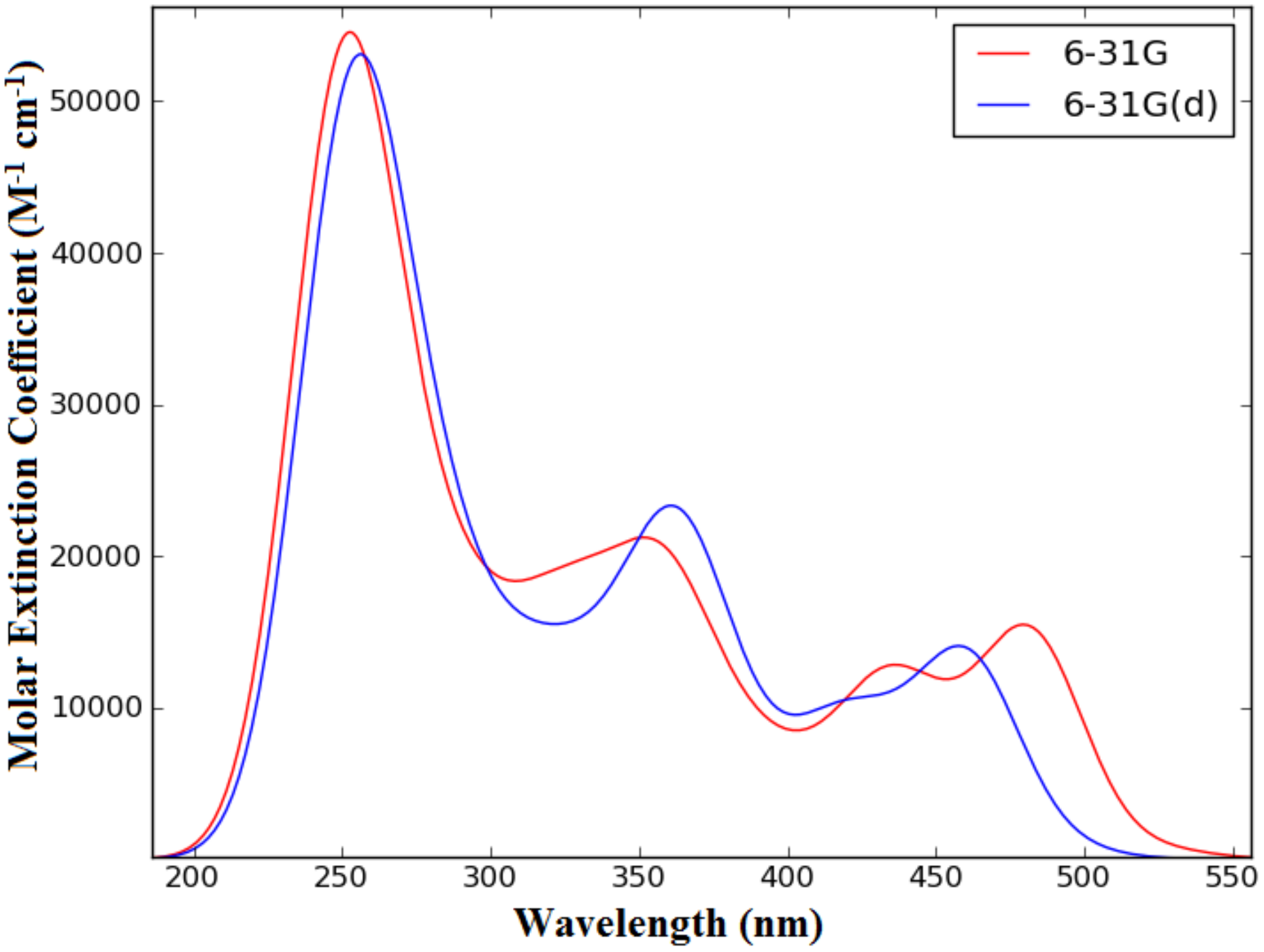}
\end{center}
[Ru(phen)$_2$(DMCH)]$^{2+}$
TD-B3LYP/6-31G and TD-B3LYP/6-31G(d) spectra.

% ================================================
\newpage
\section{Complex {\bf (81)}: [Ru(phen)$_2$(biq)]$^{2+}$}
% ================================================

\begin{center}
   {\bf PDOS}
\end{center}

\begin{center}
\begin{tabular}{cc}
\includegraphics[width=0.4\textwidth]{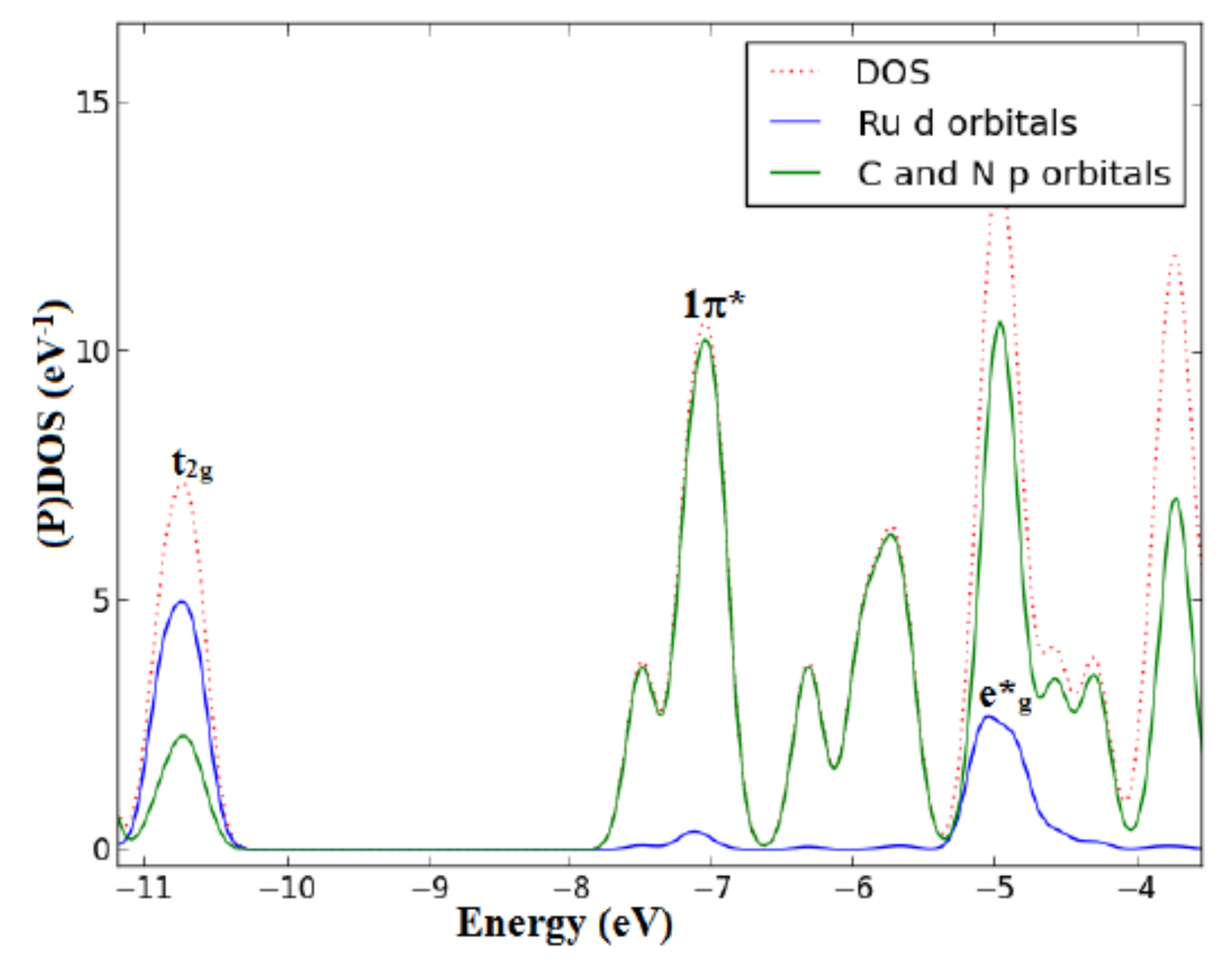} &
\includegraphics[width=0.4\textwidth]{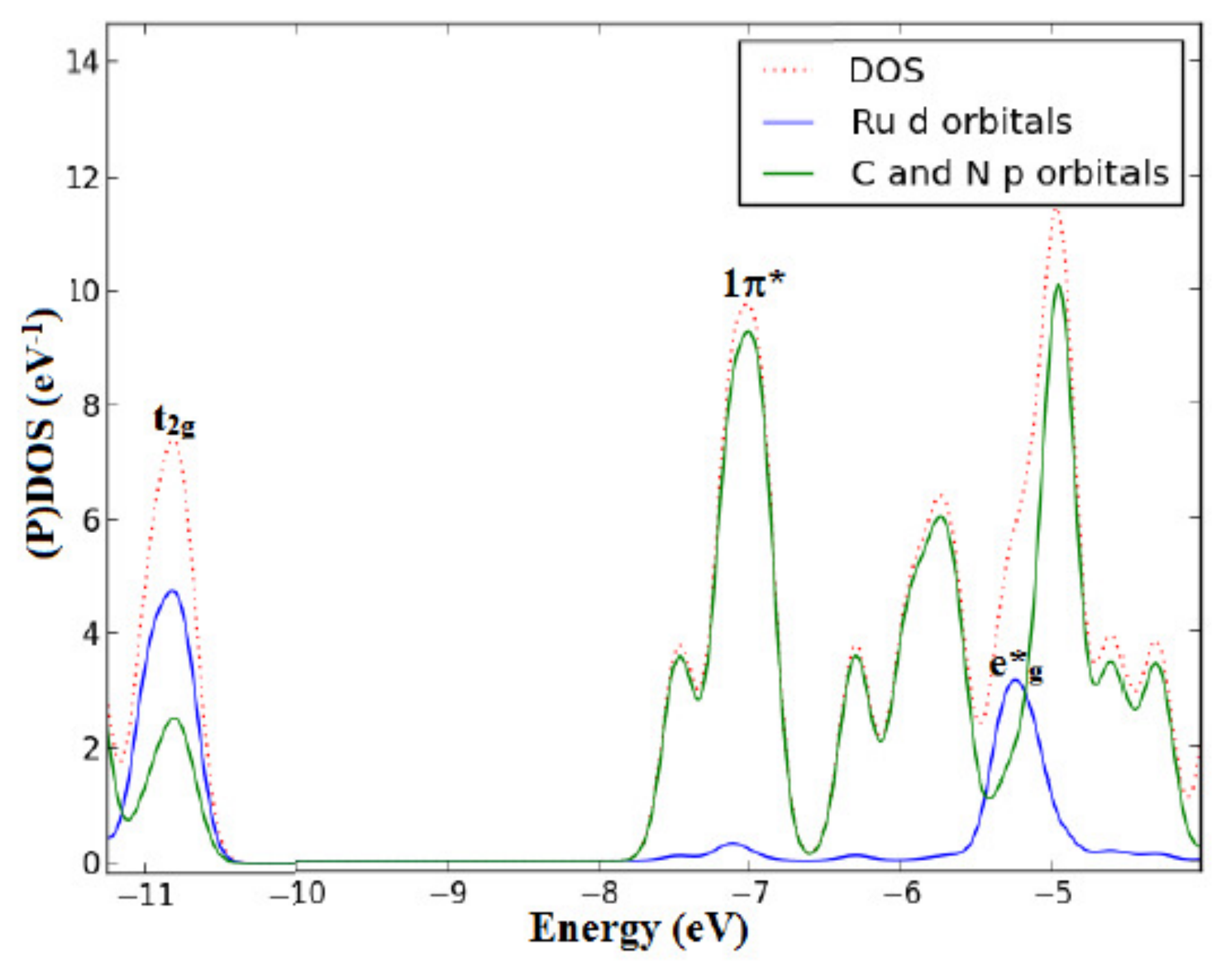} \\
B3LYP/6-31G & B3LYP/6-31G(d) \\
$\epsilon_{\text{HOMO}} = \mbox{-10.62 eV}$ & 
$\epsilon_{\text{HOMO}} = \mbox{-10.73 eV}$ 
\end{tabular}
\end{center}
Total and partial density of states of [Ru(phen)$_2$(biq)]$^{2+}$
partitioned over Ru d orbitals and ligand C and N p orbitals. 
% for the 6-31G (left-hand side) and 6-31G* (right-hand side) basis sets.

\begin{center}
   {\bf Absorption Spectrum}
\end{center}

\begin{center}
\includegraphics[width=0.8\textwidth]{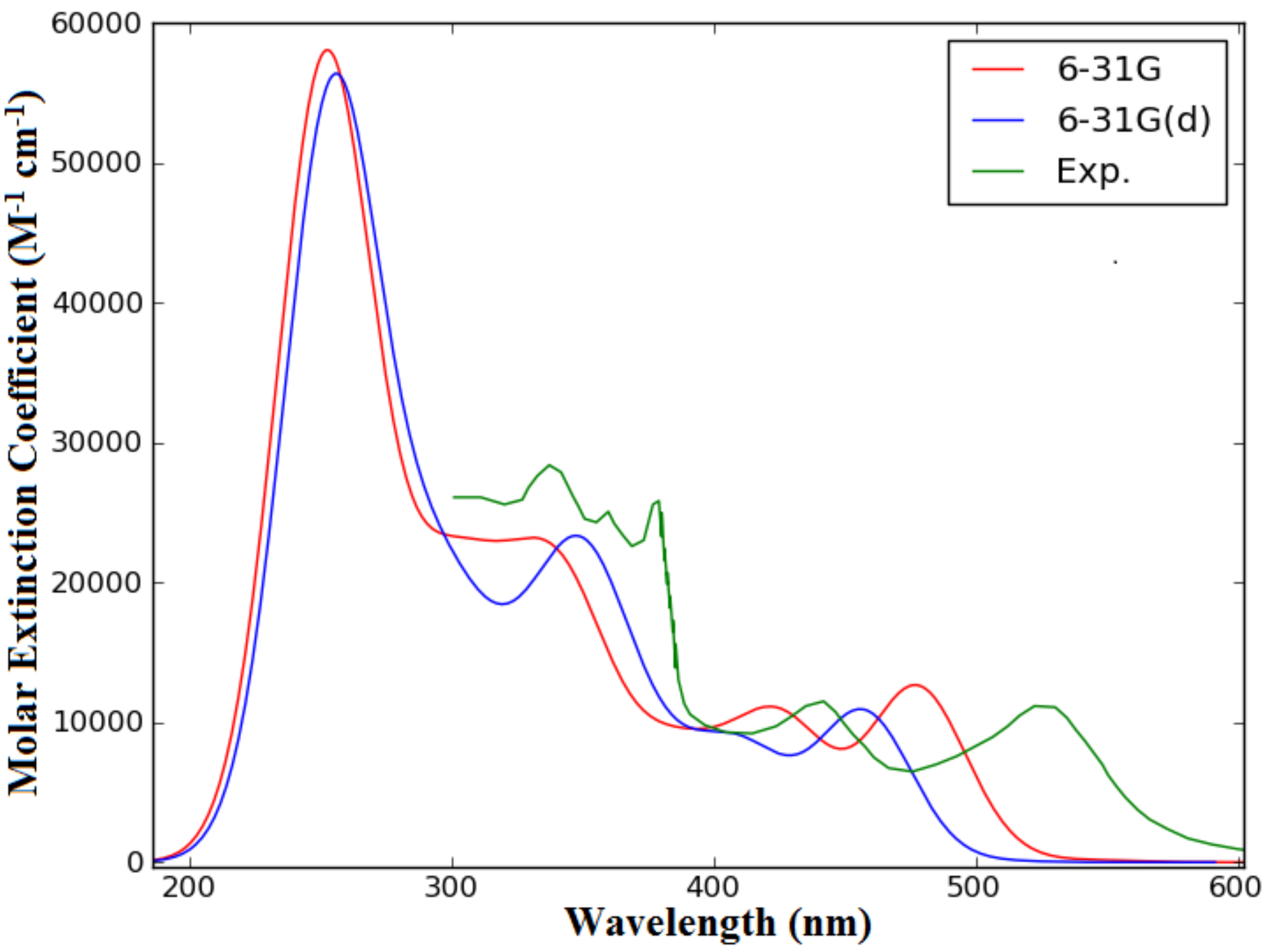}
\end{center}
[Ru(phen)$_2$(biq)]$^{2+}$
TD-B3LYP/6-31G,  TD-B3LYP/6-31G(d), and experimental spectra.
The experimental spectrum is measured in water \cite{WHH+12b}.

% ================================================
\newpage
\section{Complex {\bf (82)}: [Ru(phen)(pq)$_2$]$^{2+}$}
% ================================================

\begin{center}
   {\bf PDOS}
\end{center}

\begin{center}
\begin{tabular}{cc}
\includegraphics[width=0.4\textwidth]{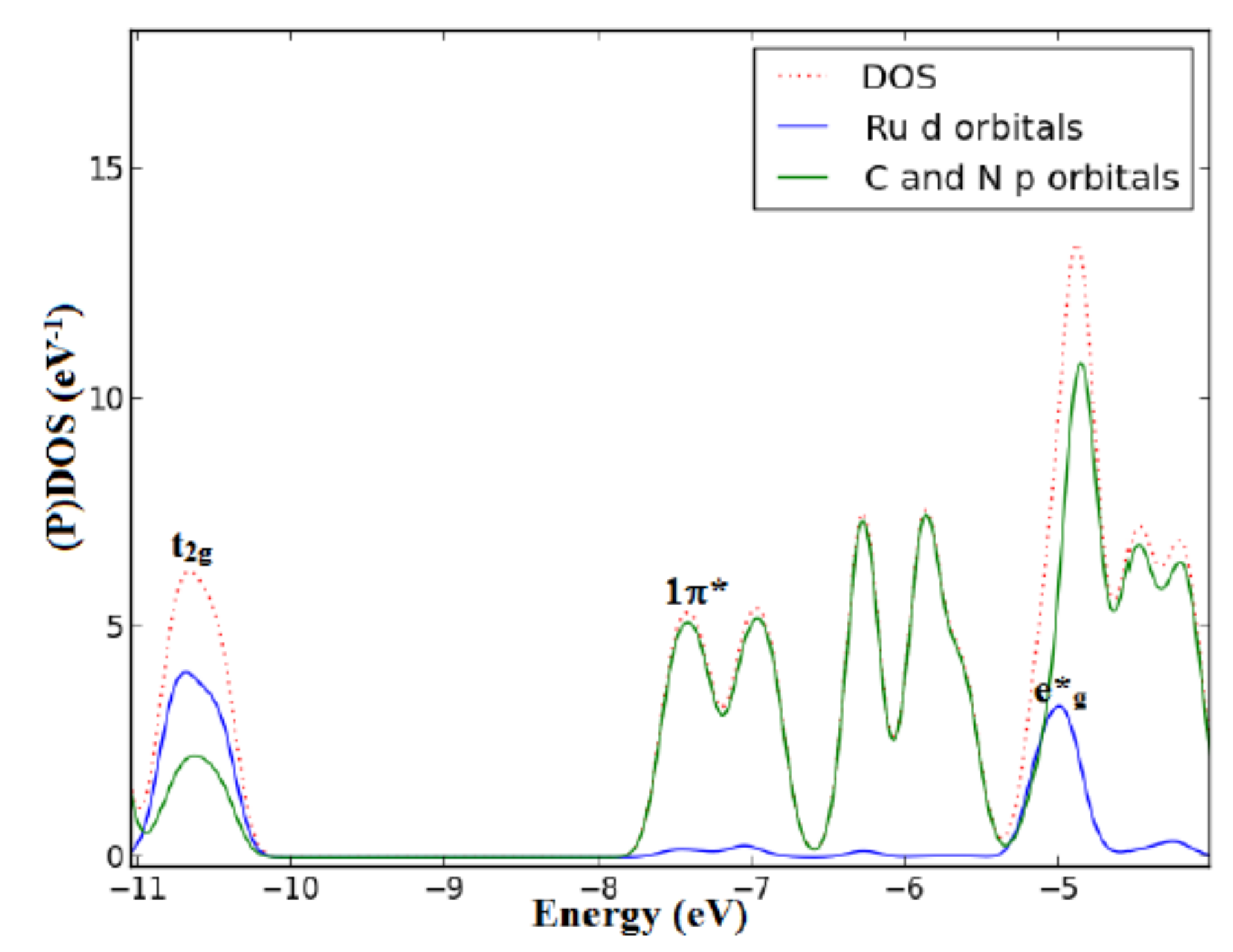} &
\includegraphics[width=0.4\textwidth]{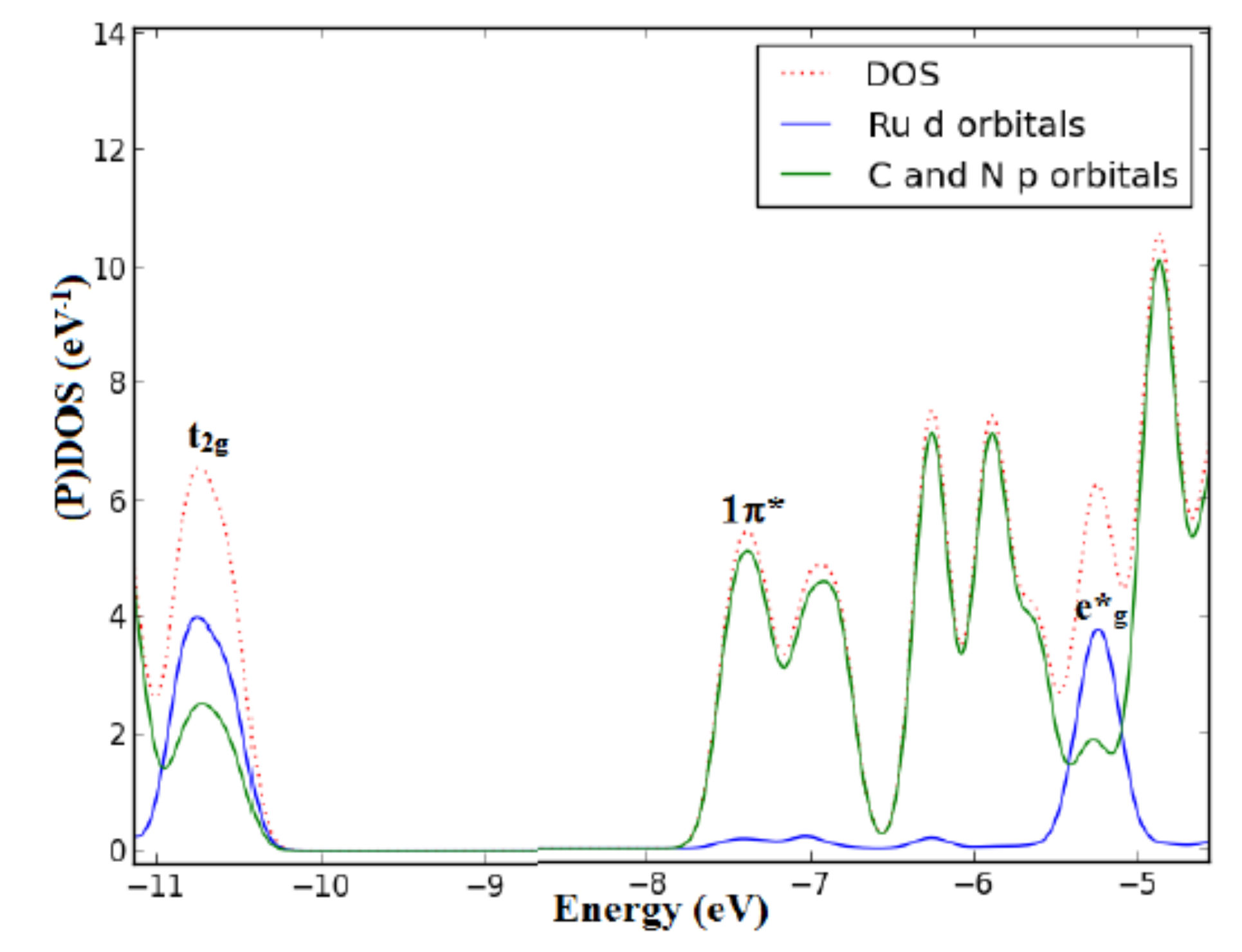} \\
B3LYP/6-31G & B3LYP/6-31G(d) \\
$\epsilon_{\text{HOMO}} = \mbox{-10.69 eV}$ & 
$\epsilon_{\text{HOMO}} = \mbox{-10.80 eV}$ 
\end{tabular}
\end{center}
Total and partial density of states of [Ru(phen)(pq)$_2$]$^{2+}$
partitioned over Ru d orbitals and ligand C and N p orbitals. 
% for the 6-31G (left-hand side) and 6-31G* (right-hand side) basis sets.

\begin{center}
   {\bf Absorption Spectrum}
\end{center}

\begin{center}
\includegraphics[width=0.8\textwidth]{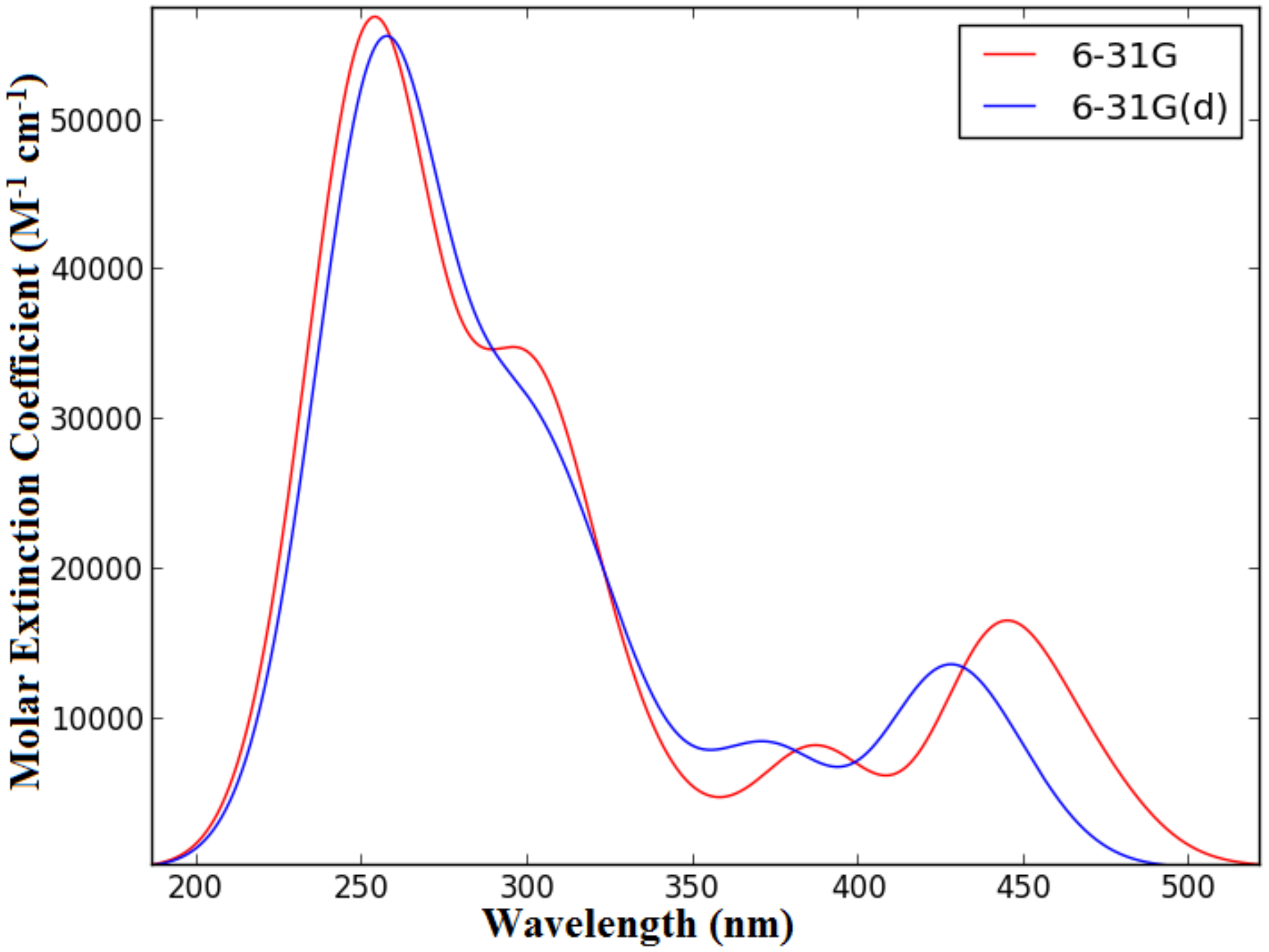}
\end{center}
[Ru(phen)(pq)$_2$]$^{2+}$
TD-B3LYP/6-31G and  TD-B3LYP/6-31G(d) spectra.

% ================================================
\newpage
\section{Complex {\bf (83)}: [Ru(phen)(biq)$_2$]$^{2+}$}
% ================================================

\begin{center}
   {\bf PDOS}
\end{center}

\begin{center}
\begin{tabular}{cc}
\includegraphics[width=0.4\textwidth]{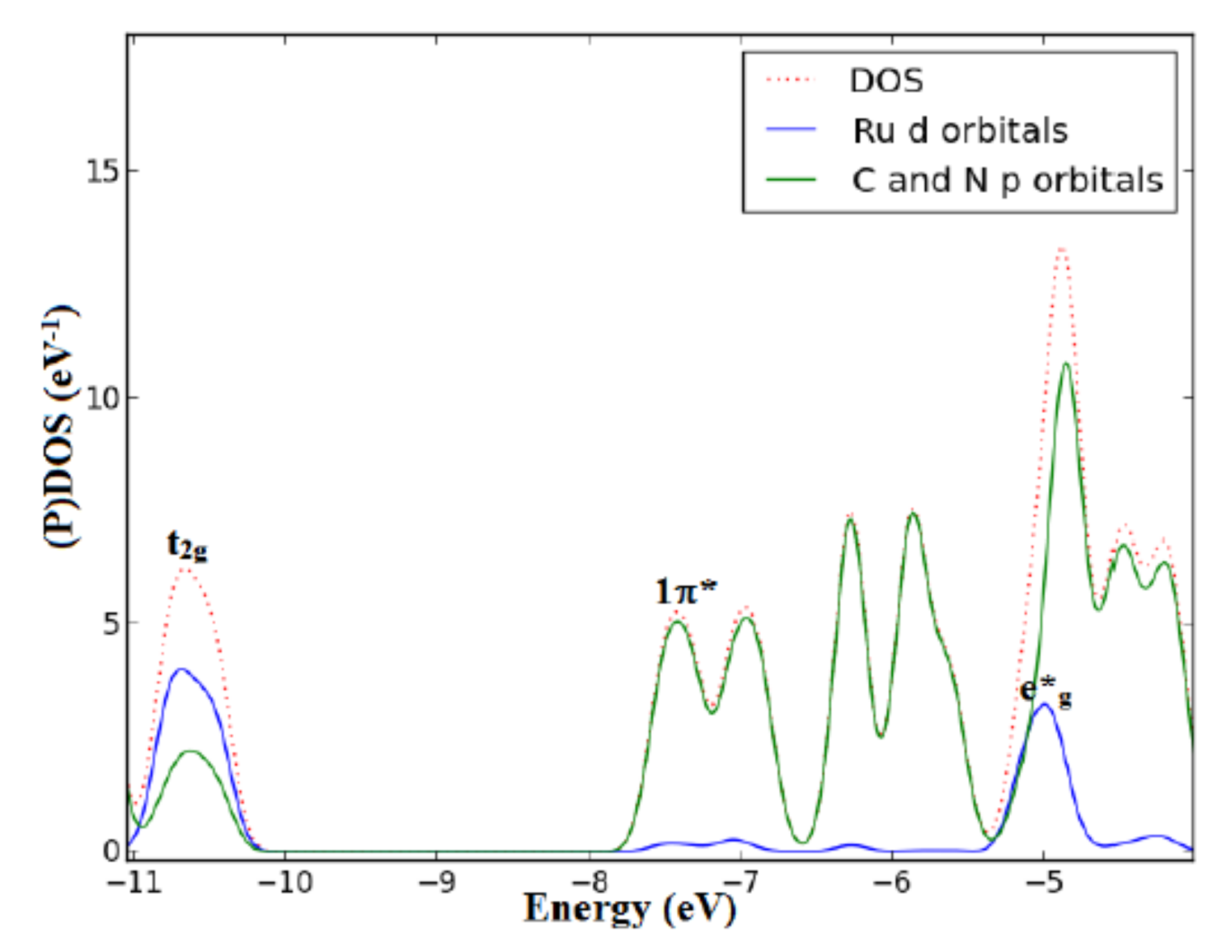} &
\includegraphics[width=0.4\textwidth]{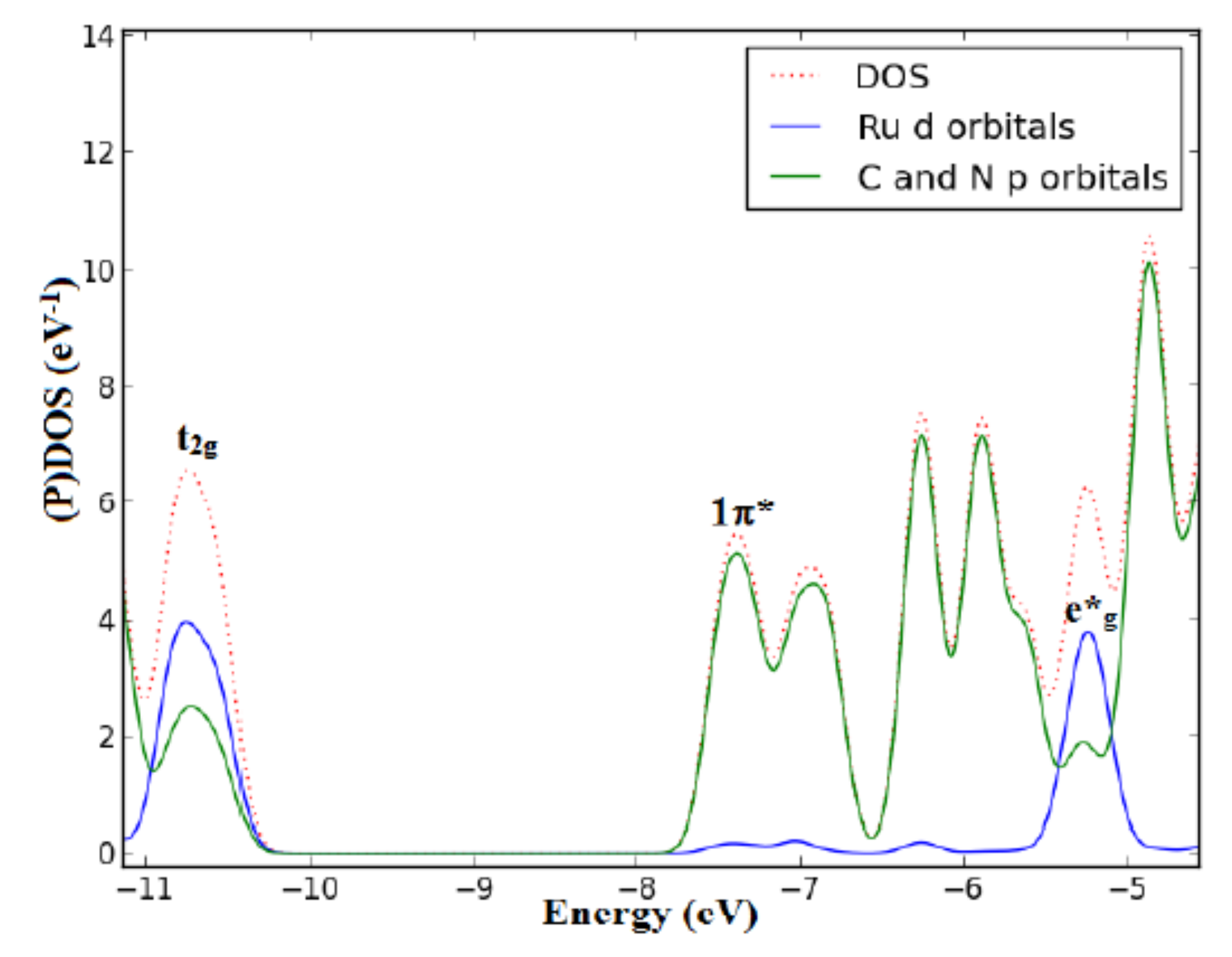} \\
B3LYP/6-31G & B3LYP/6-31G(d) \\
$\epsilon_{\text{HOMO}} = \mbox{-10.46 eV}$ & 
$\epsilon_{\text{HOMO}} = \mbox{-10.55 eV}$ 
\end{tabular}
\end{center}
Total and partial density of states of [Ru(phen)(biq)$_2$]$^{2+}$
partitioned over Ru d orbitals and ligand C and N p orbitals.
% for the 6-31G (left-hand side) and 6-31G* (right-hand side) basis sets.

\begin{center}
   {\bf Absorption Spectrum}
\end{center}

\begin{center}
\includegraphics[width=0.8\textwidth]{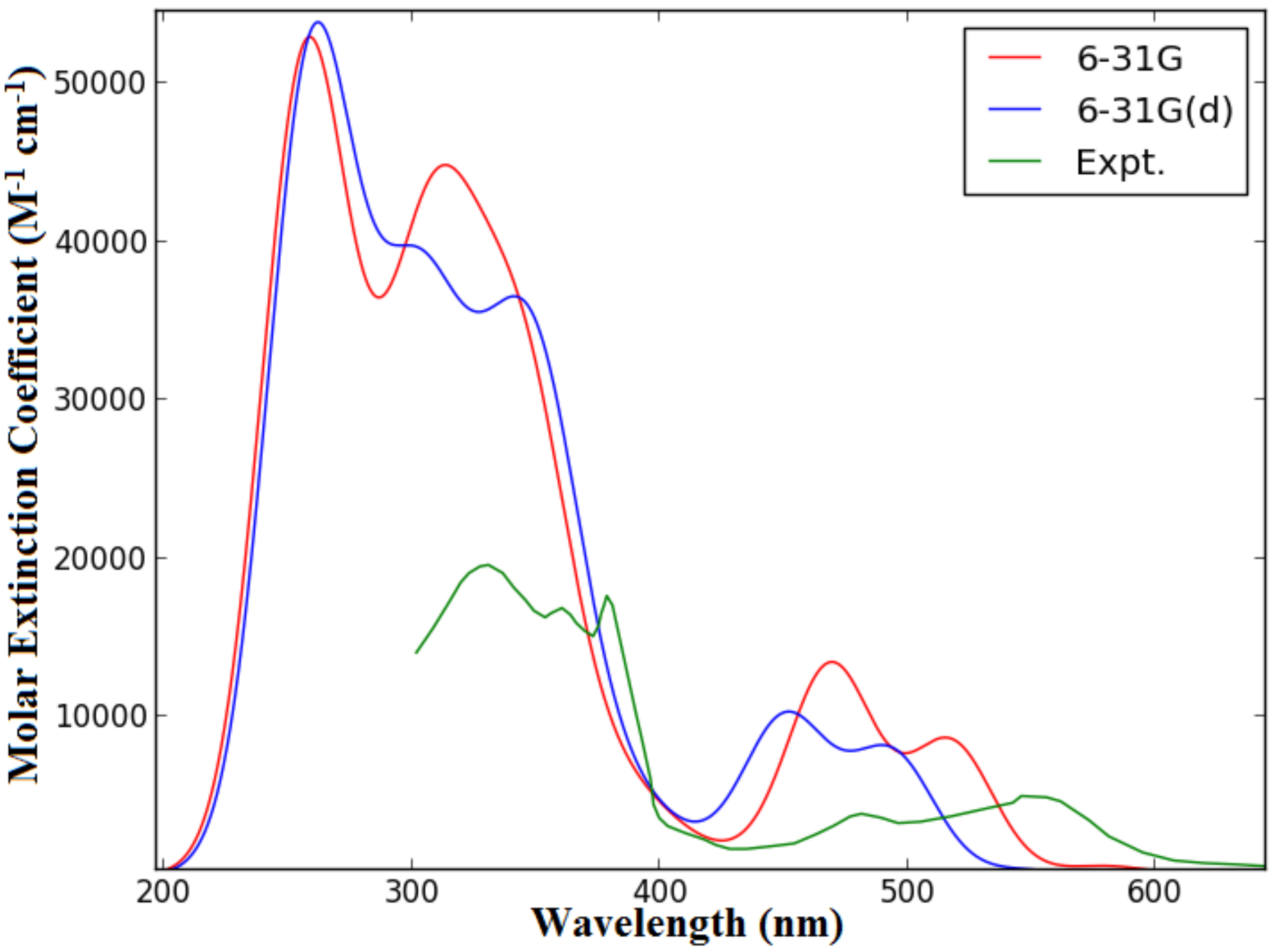}
\end{center}
[Ru(phen)(biq)$_2$]$^{2+}$
TD-B3LYP/6-31G, TD-B3LYP/6-31G(d), and experimental spectra.
Experimental spectrum measured in water \cite{WHH+12b}.

% ================================================
\newpage
\section{Complex {\bf (84)}: [Ru(2-m-phen)$_3$]$^{2+}$}
% ================================================

\begin{center}
   {\bf PDOS}
\end{center}

\begin{center}
\begin{tabular}{cc}
\includegraphics[width=0.4\textwidth]{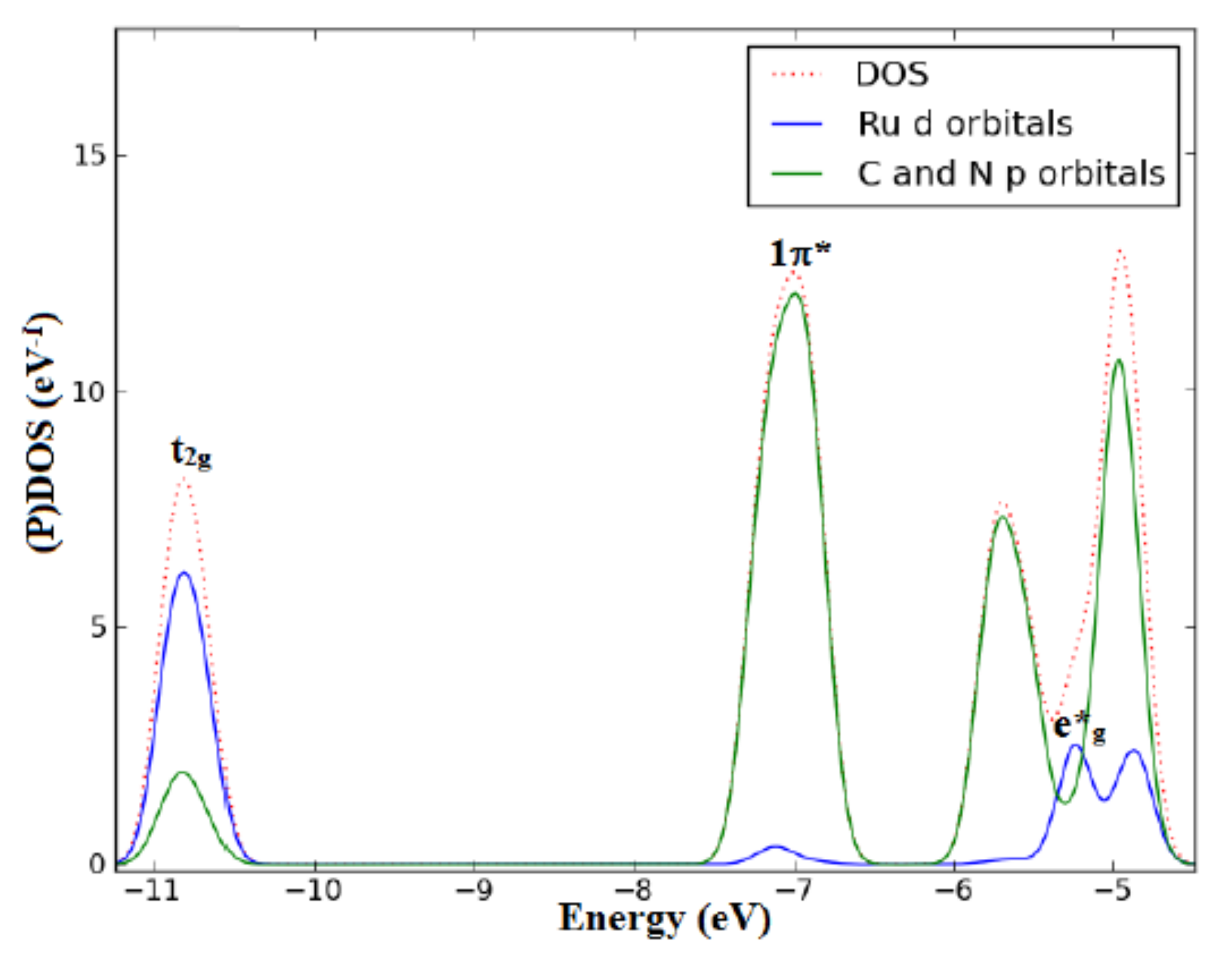} &
\includegraphics[width=0.4\textwidth]{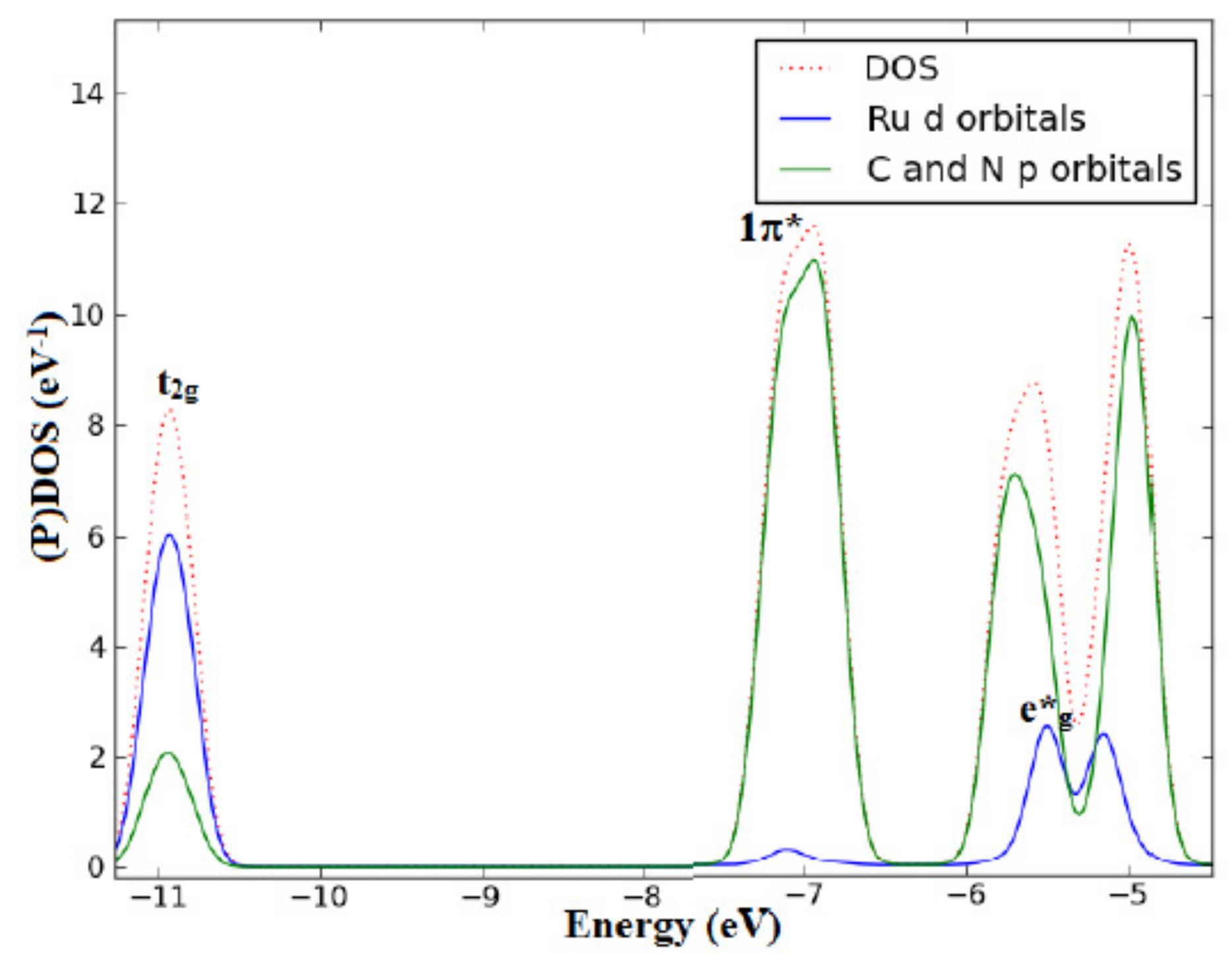} \\
B3LYP/6-31G & B3LYP/6-31G(d) \\
$\epsilon_{\text{HOMO}} = \mbox{-10.68 eV}$ & 
$\epsilon_{\text{HOMO}} = \mbox{-10.83 eV}$ 
\end{tabular}
\end{center}
Total and partial density of states of [Ru(2-m-phen)$_3$]$^{2+}$
partitioned over Ru d orbitals and ligand C and N p orbitals.
% for the 6-31G (left-hand side) and 6-31G* (right-hand side) basis sets.

\begin{center}
   {\bf Absorption Spectrum}
\end{center}

\begin{center}
\includegraphics[width=0.8\textwidth]{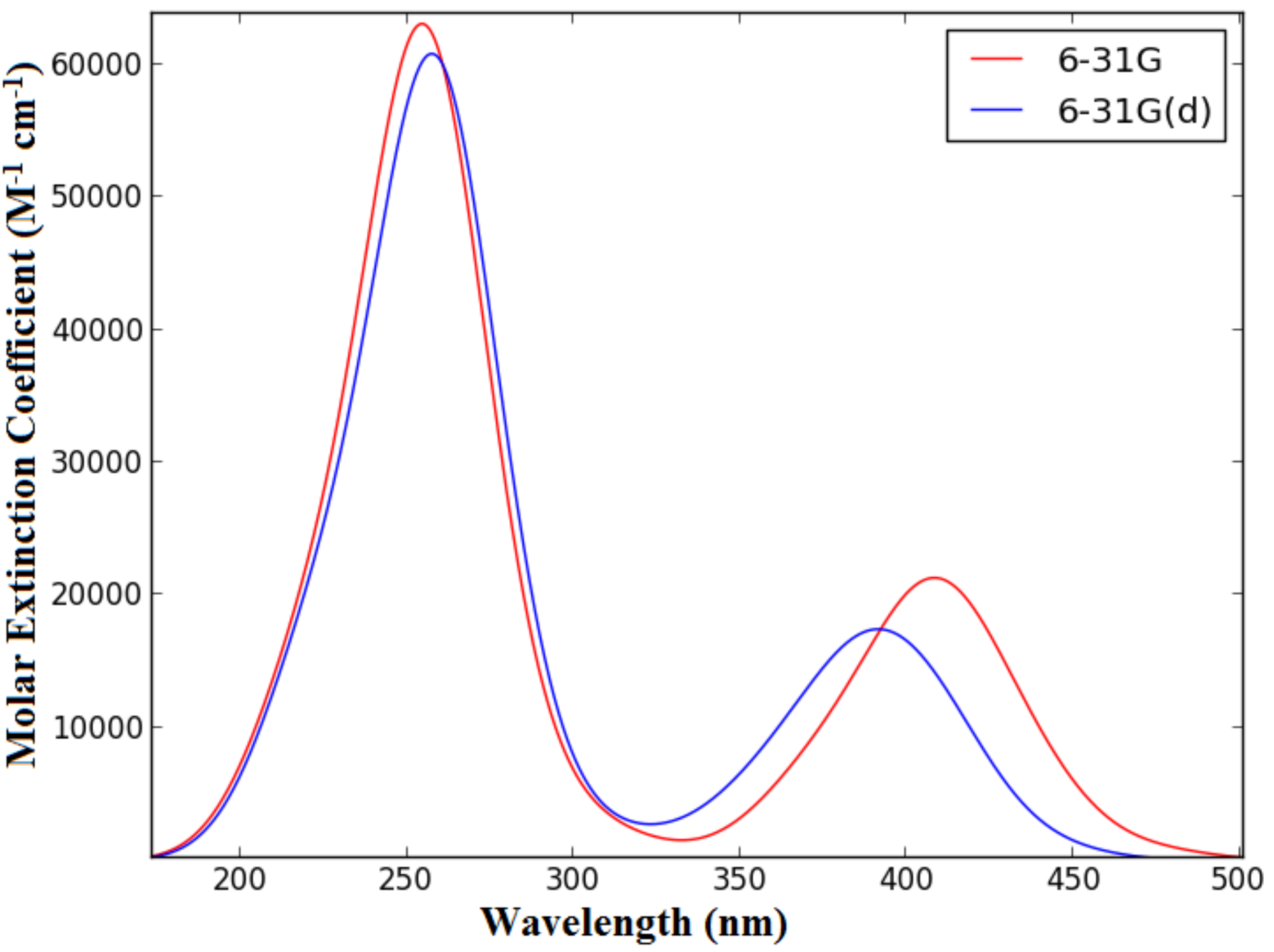}
\end{center}
[Ru(2-m-phen)$_3$]$^{2+}$
TD-B3LYP/6-31G and TD-B3LYP/6-31G(d) spectra.

% ================================================
\newpage
\section{Complex {\bf (85)}: [Ru(2,9-dm-phen)$_3$]$^{2+}$}
% ================================================

\begin{center}
   {\bf PDOS}
\end{center}

\begin{center}
\begin{tabular}{cc}
\includegraphics[width=0.4\textwidth]{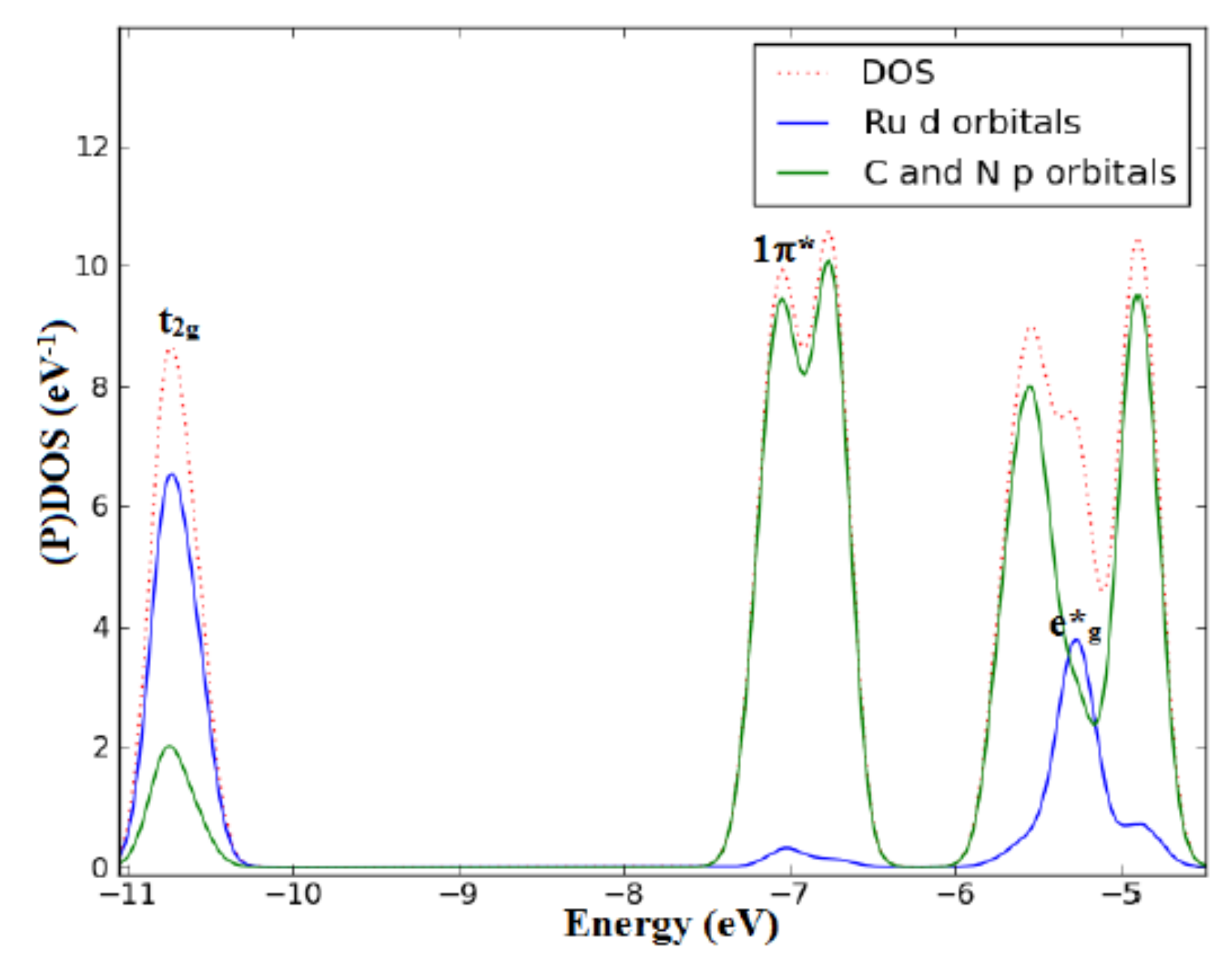} &
\includegraphics[width=0.4\textwidth]{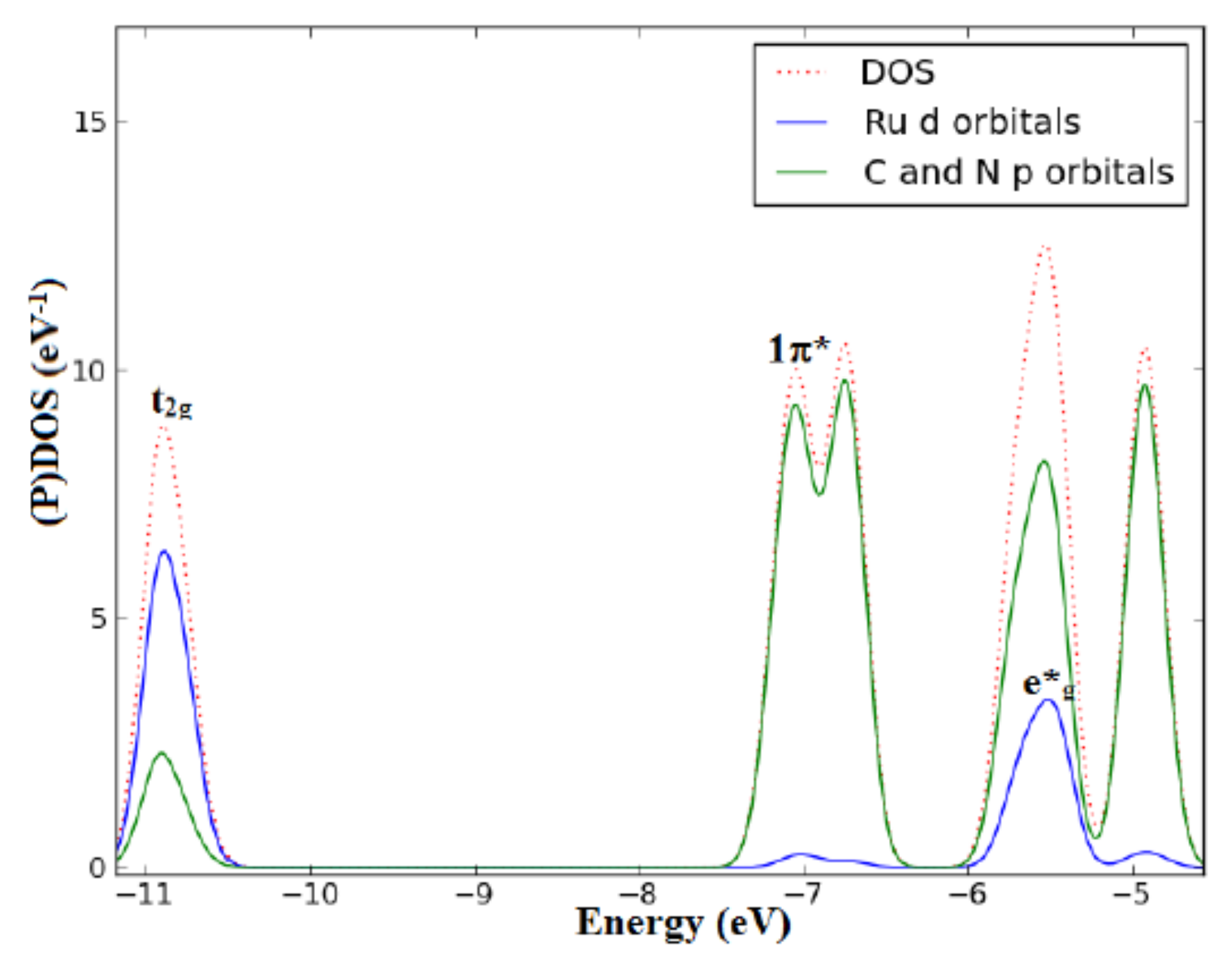} \\
B3LYP/6-31G & B3LYP/6-31G(d) \\
$\epsilon_{\text{HOMO}} = \mbox{-10.59 eV}$ & 
$\epsilon_{\text{HOMO}} = \mbox{-10.75 eV}$ 
\end{tabular}
\end{center}
Total and partial density of states of [Ru(2,9-dm-phen)$_3$]$^{2+}$
partitioned over Ru d orbitals and ligand C and N p orbitals. 
% for the 6-31G (left-hand side) and 6-31G* (right-hand side) basis sets.

\begin{center}
   {\bf Absorption Spectrum}
\end{center}

\begin{center}
\includegraphics[width=0.8\textwidth]{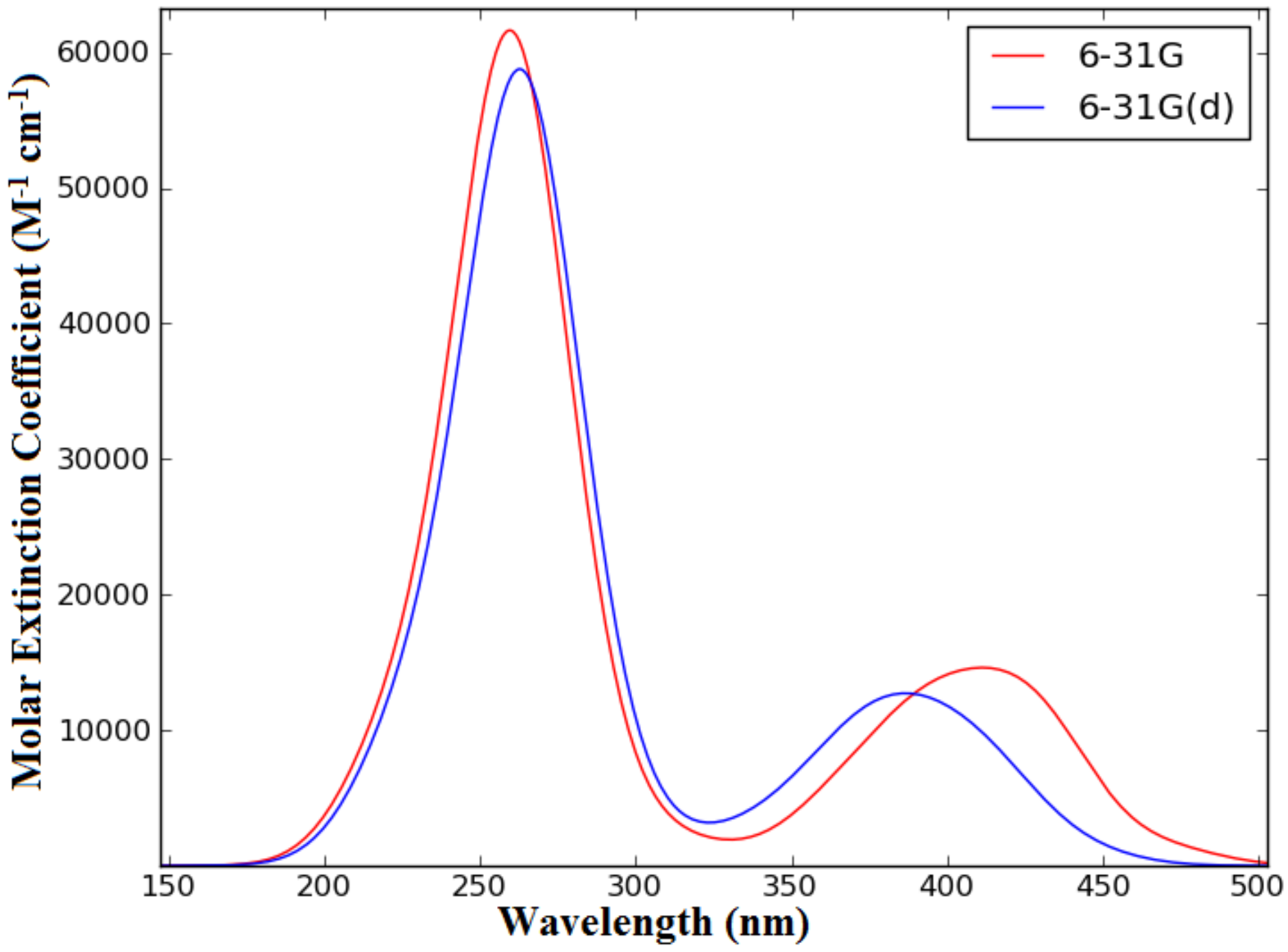}
\end{center}
[Ru(2,9-dm-phen)$_3$]$^{2+}$
TD-B3LYP/6-31G and TD-B3LYP/6-31G(d) spectra.

% ================================================
\newpage
\section{Complex {\bf (86)}: [Ru(4,7-Ph$_2$-phen)$_3$]$^{2+}$}
% ================================================

\begin{center}
   {\bf PDOS}
\end{center}

\begin{center}
\includegraphics[width=0.4\textwidth]{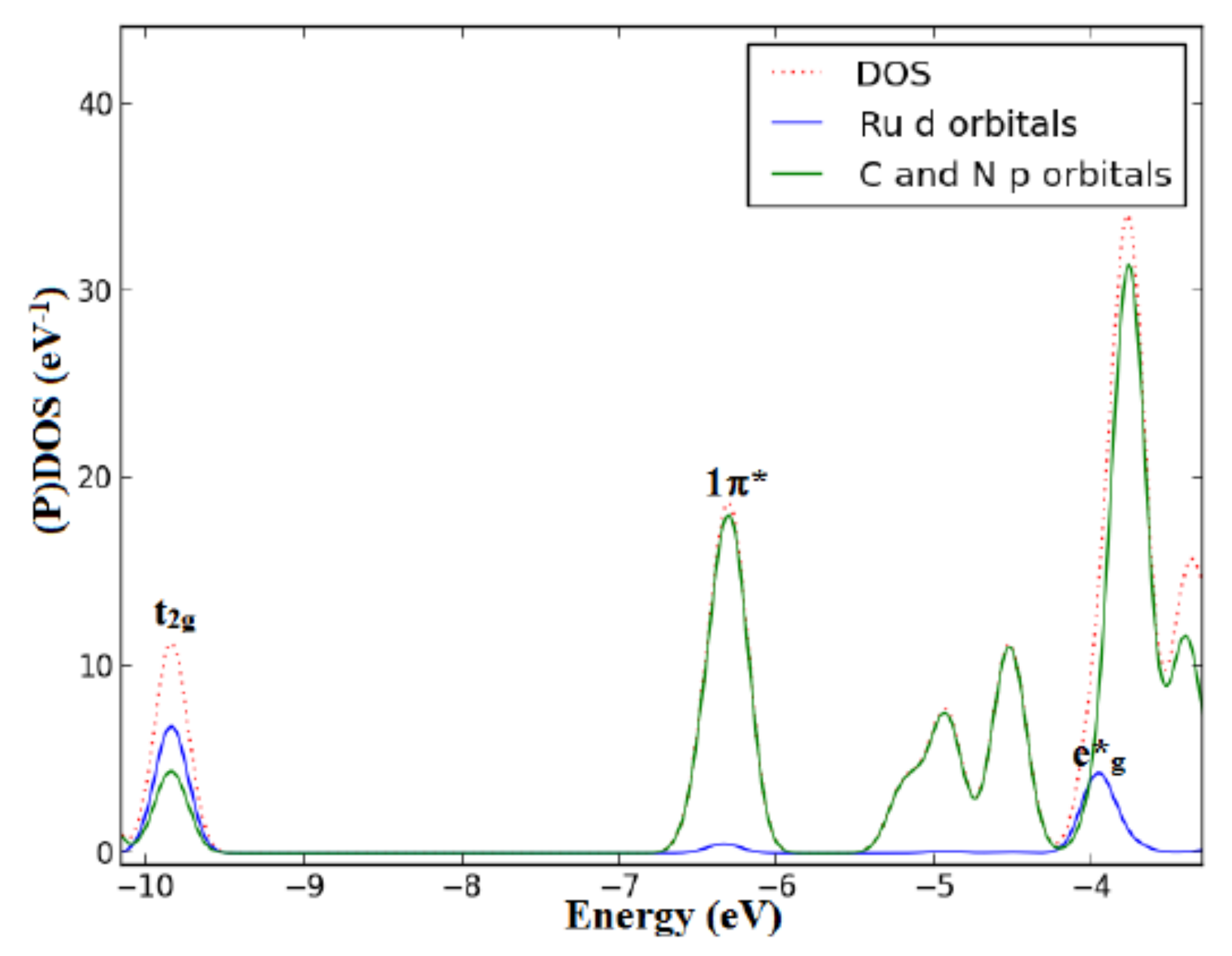}
\\ B3LYP/6-31G \\ $\epsilon_{\text{HOMO}} = \mbox{-9.84 eV}$ 
\end{center}
Total and partial density of states of [Ru(4,7-Ph$_2$-phen)$_3$]$^{2+}$
partitioned over Ru d orbitals and ligand C and N p orbitals.
% for the 6-31G (left-hand side) basis set. 
% and 6-31G* (right-hand side {\color{red} \sf Do we have this?}) basis sets.

\begin{center}
   {\bf Absorption Spectrum}
\end{center}

\begin{center}
\includegraphics[width=0.8\textwidth]{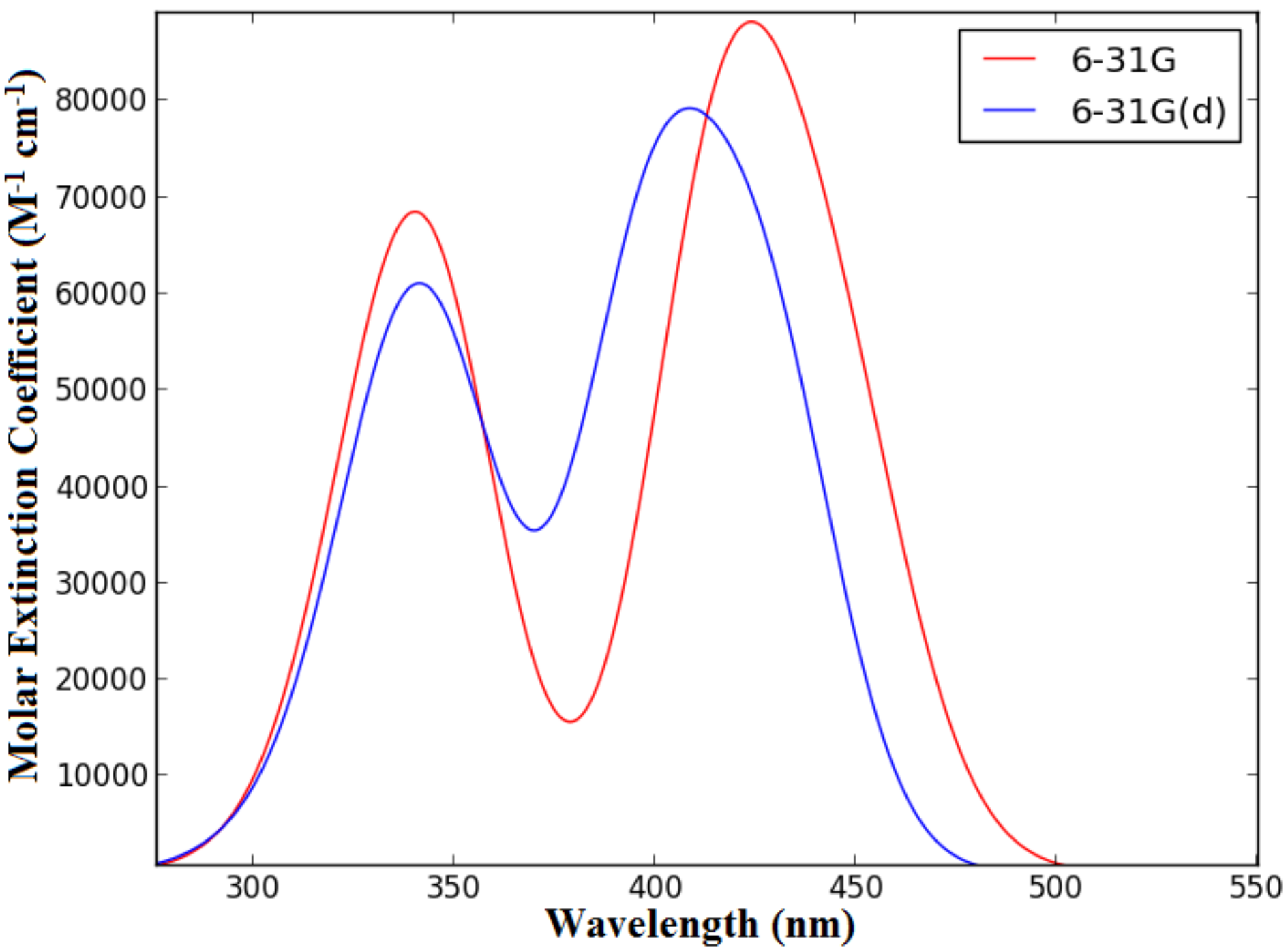}
\end{center}
[Ru(4,7-Ph$_2$-phen)$_3$]$^{2+}$
TD-B3LYP/6-31G and TD-B3LYP/6-31G(d) spectra.

% ================================================
\newpage
\section{Complex {\bf (87)}: [Ru(4,7-dhy-phen)(tm1-phen)$_2$]$^{2+}$}
% ================================================

\begin{center}
   {\bf PDOS}
\end{center}

\begin{center}
\begin{tabular}{cc}
\includegraphics[width=0.4\textwidth]{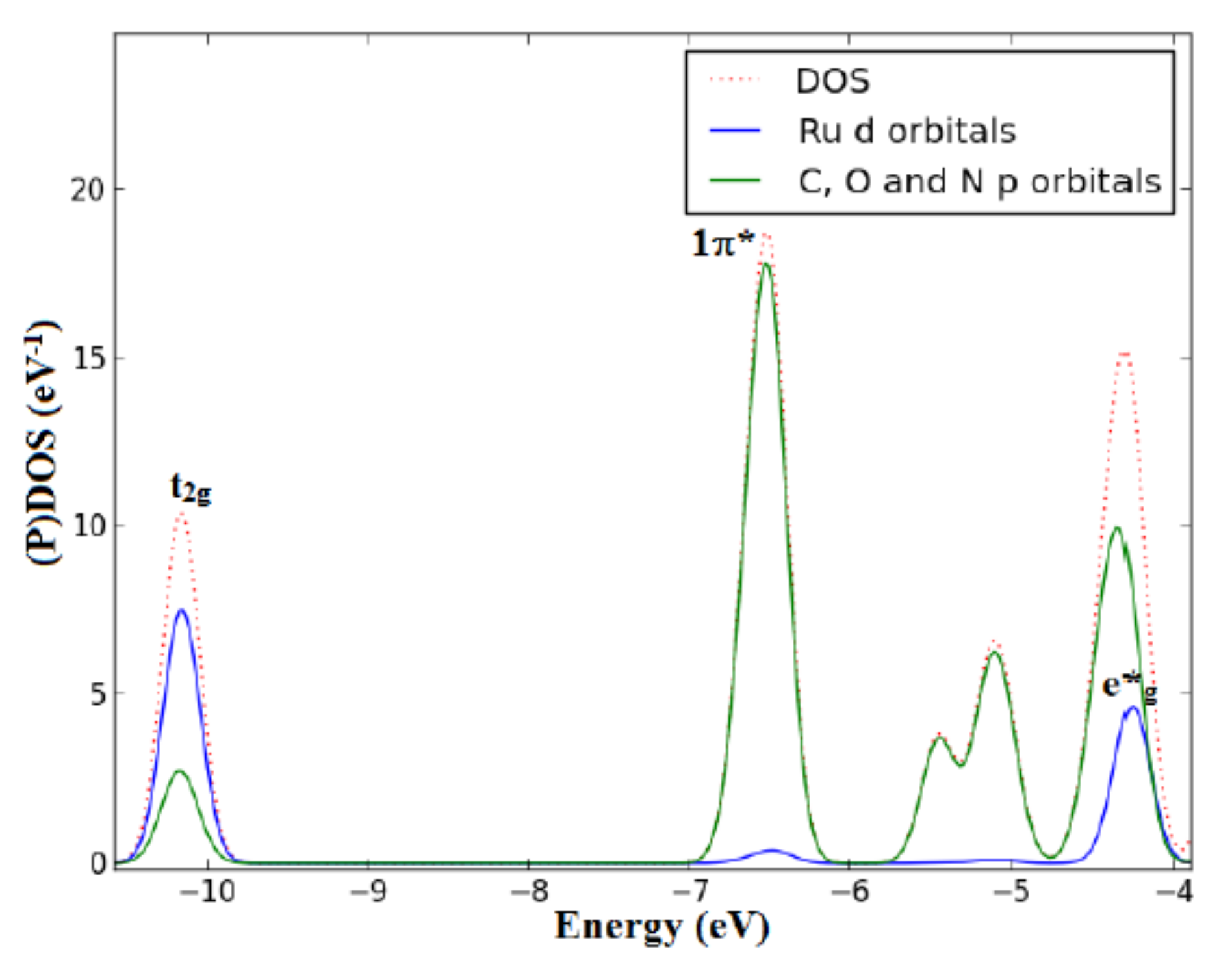} &
\includegraphics[width=0.4\textwidth]{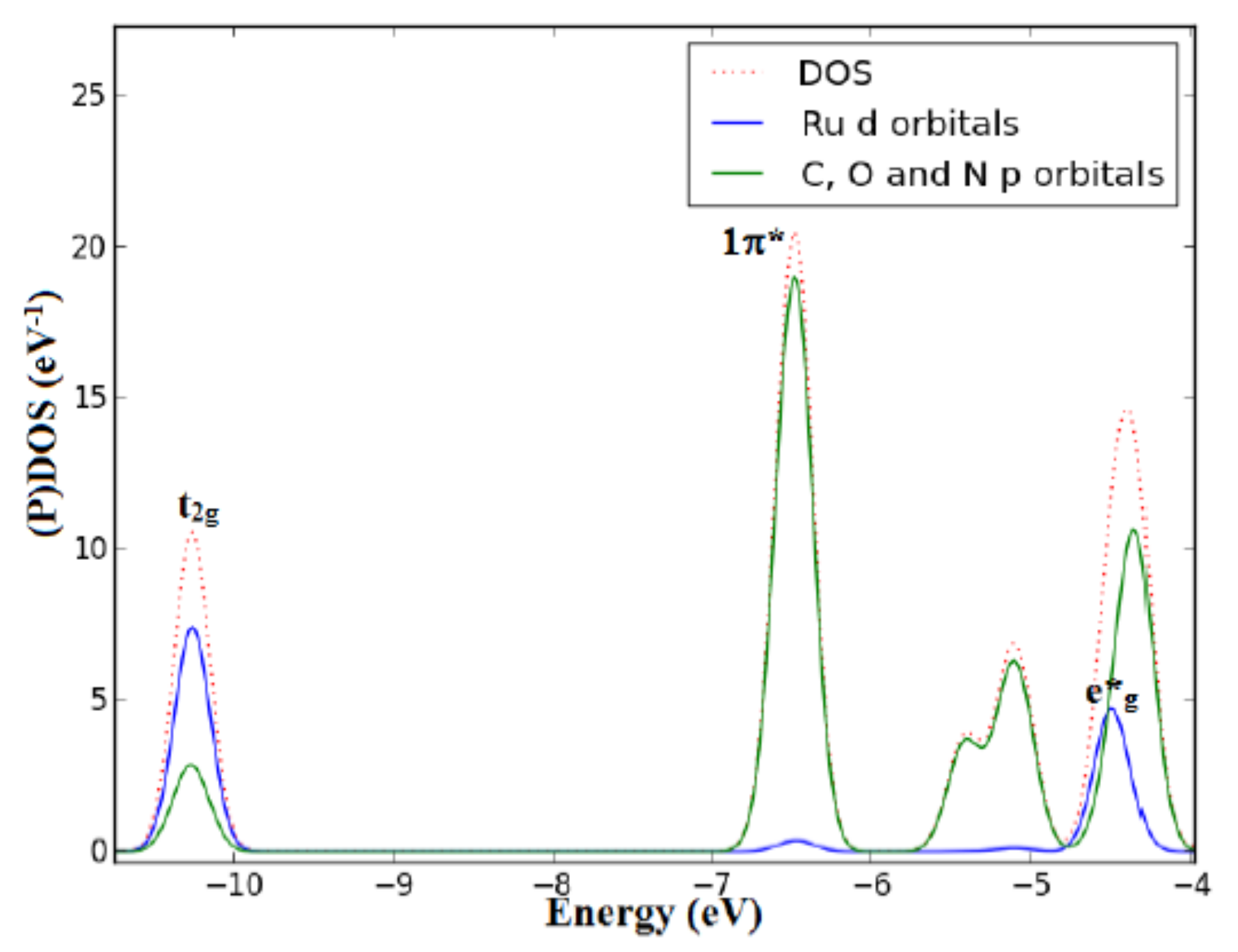} \\
B3LYP/6-31G & B3LYP/6-31G(d) \\
$\epsilon_{\text{HOMO}} = \mbox{-10.11 eV}$ & 
$\epsilon_{\text{HOMO}} = \mbox{-10.23 eV}$ 
\end{tabular}
\end{center}
Total and partial density of states of [Ru(4,7-dhy-phen)(tm1-phen)$_2$]$^{2+}$
partitioned over Ru d orbitals and ligand C, O, and N p orbitals.
% for the 6-31G (left-hand side) and 6-31G* (right-hand side) basis sets.

\begin{center}
   {\bf Absorption Spectrum}
\end{center}

\begin{center}
\includegraphics[width=0.8\textwidth]{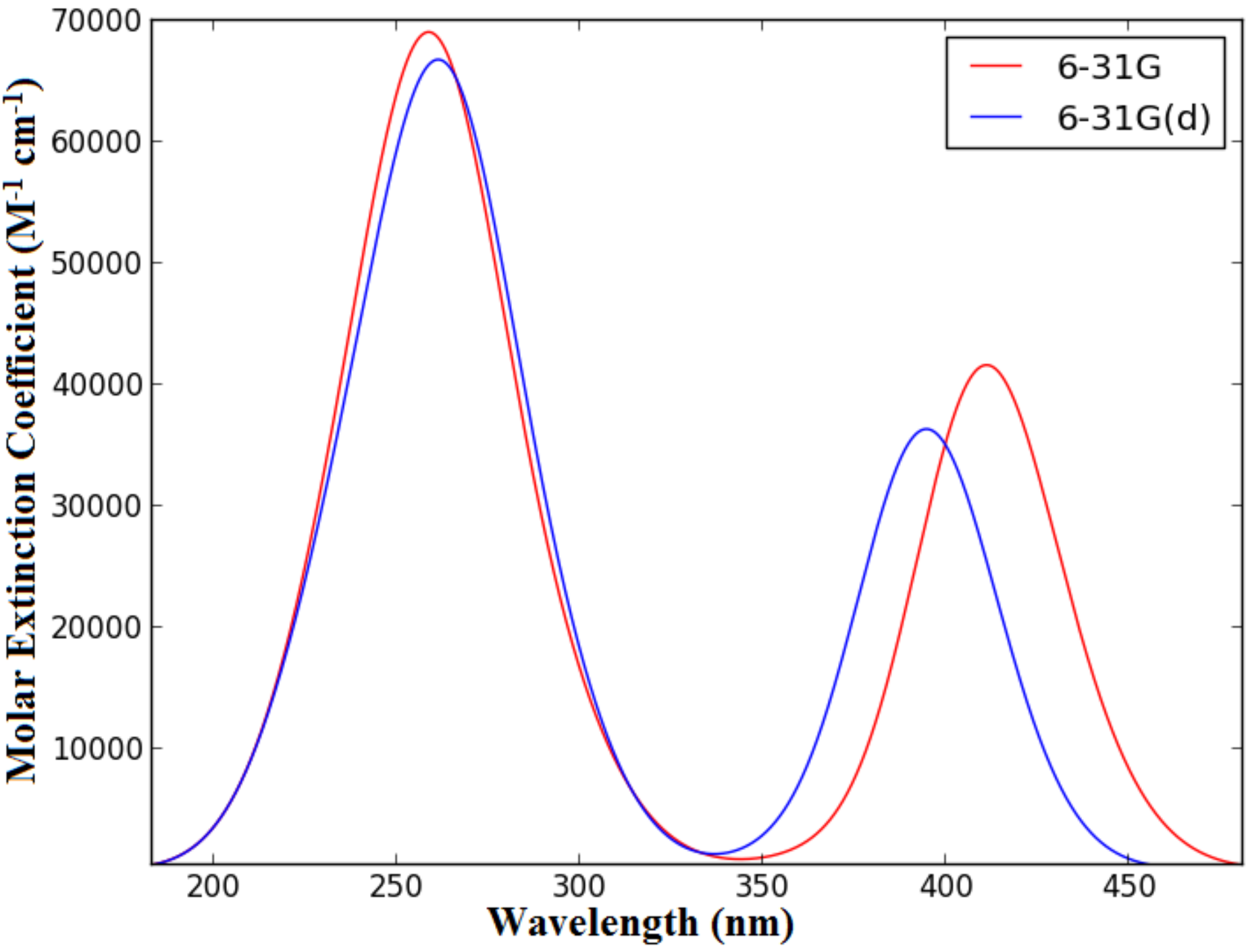}
\end{center}
[Ru(4,7-dhy-phen)(tm1-phen)$_2$]$^{2+}$
TD-B3LYP/6-31G and TD-B3LYP/6-31G(d) spectra.

% ================================================
\newpage
\section{Complex {\bf (88)}*: [Ru(DPA)$_3$]$^{-}$}
% ================================================

\begin{center}
   {\bf PDOS}
\end{center}

\begin{center}
\begin{tabular}{cc}
\includegraphics[width=0.4\textwidth]{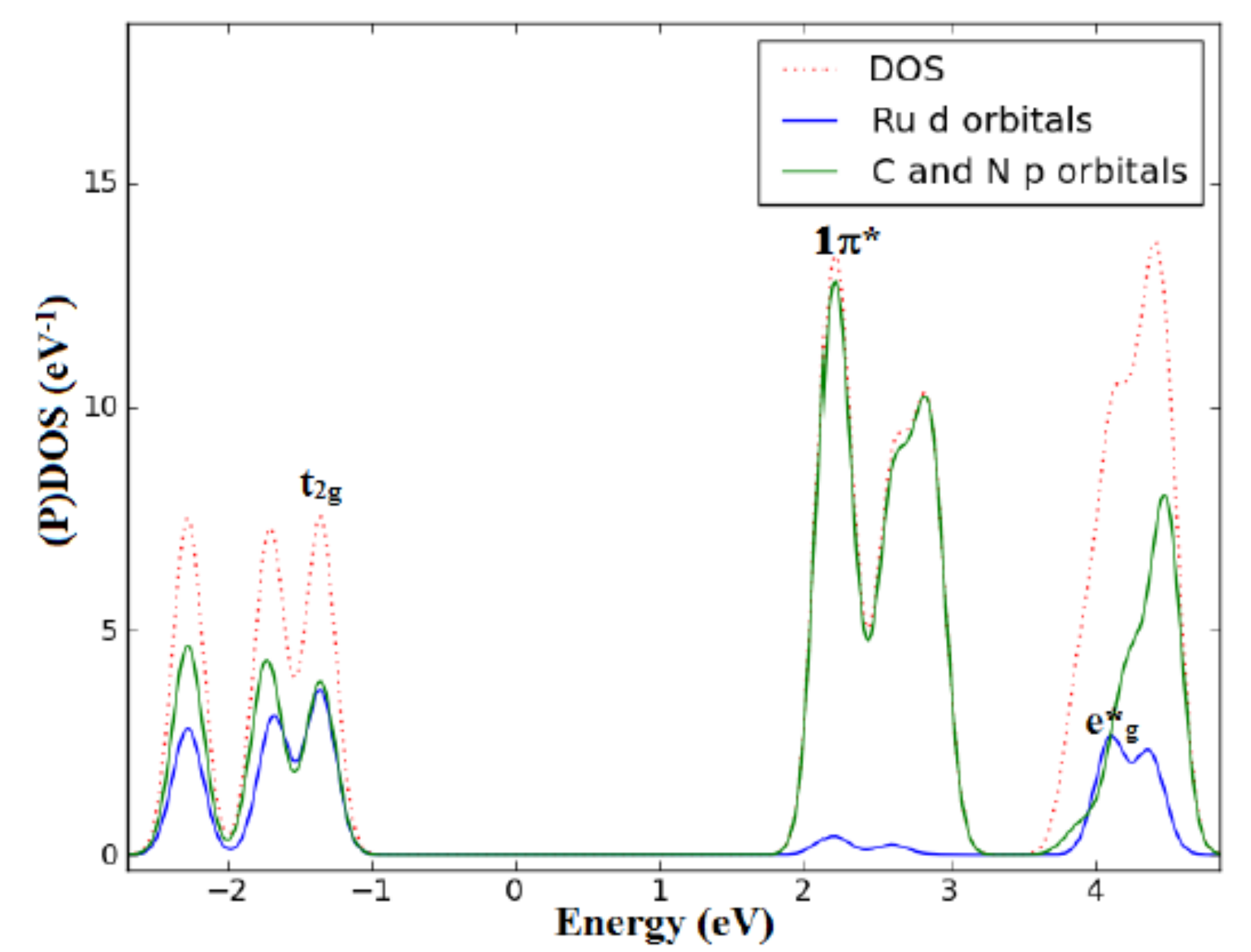} &
\includegraphics[width=0.4\textwidth]{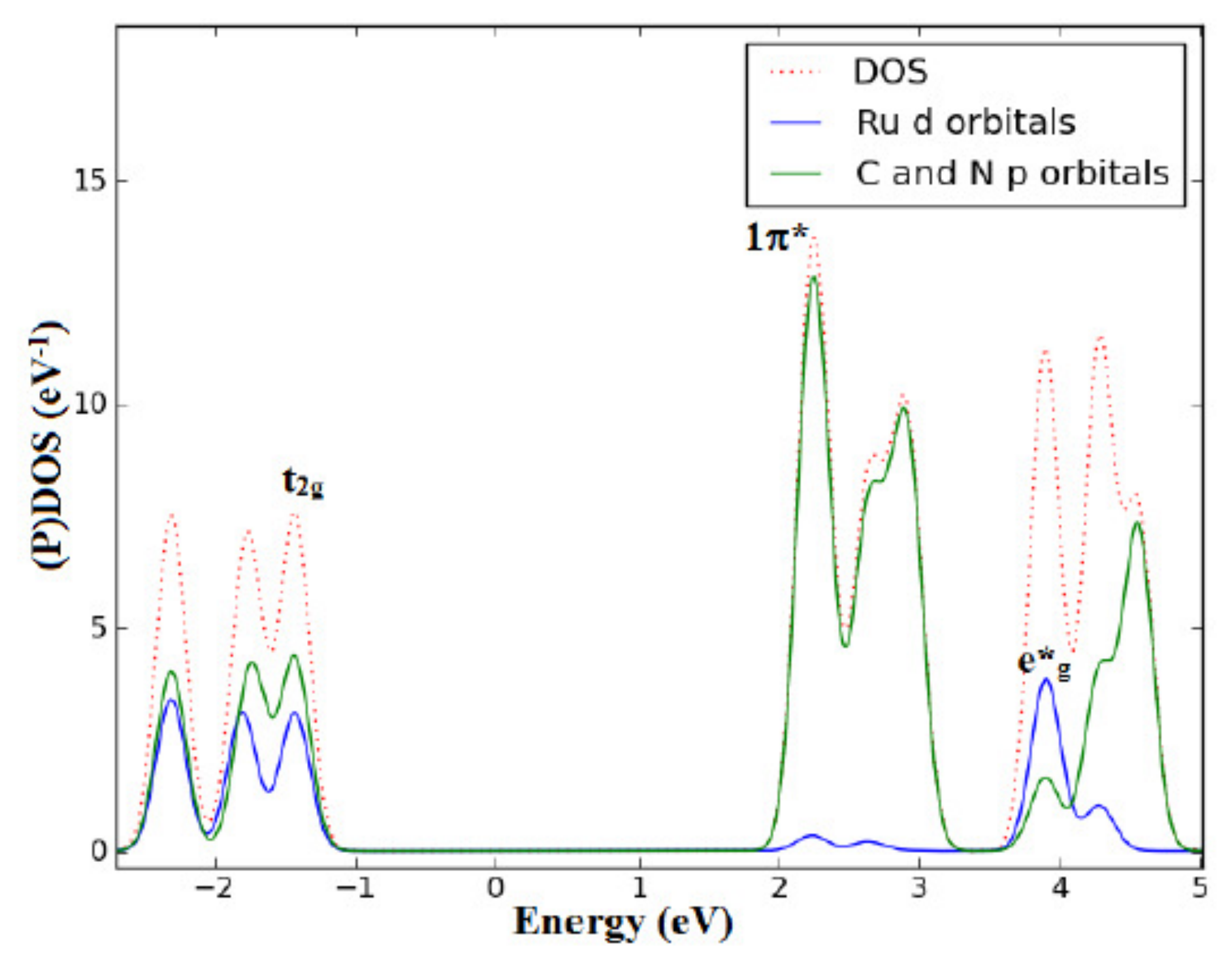} \\
B3LYP/6-31G & B3LYP/6-31G(d) \\
$\epsilon_{\text{HOMO}} = \mbox{-1.35 eV}$ & 
$\epsilon_{\text{HOMO}} = \mbox{-1.43 eV}$ 
\end{tabular}
\end{center}
Total and partial density of states of [Ru(DPA)$_3$]$^{-}$
partitioned over Ru d orbitals and ligand C and N p orbitals.
% for the 6-31G (left-hand side) and 6-31G* (right-hand side) basis sets.

\begin{center}
   {\bf Absorption Spectrum}
\end{center}

\begin{center}
\includegraphics[width=0.8\textwidth]{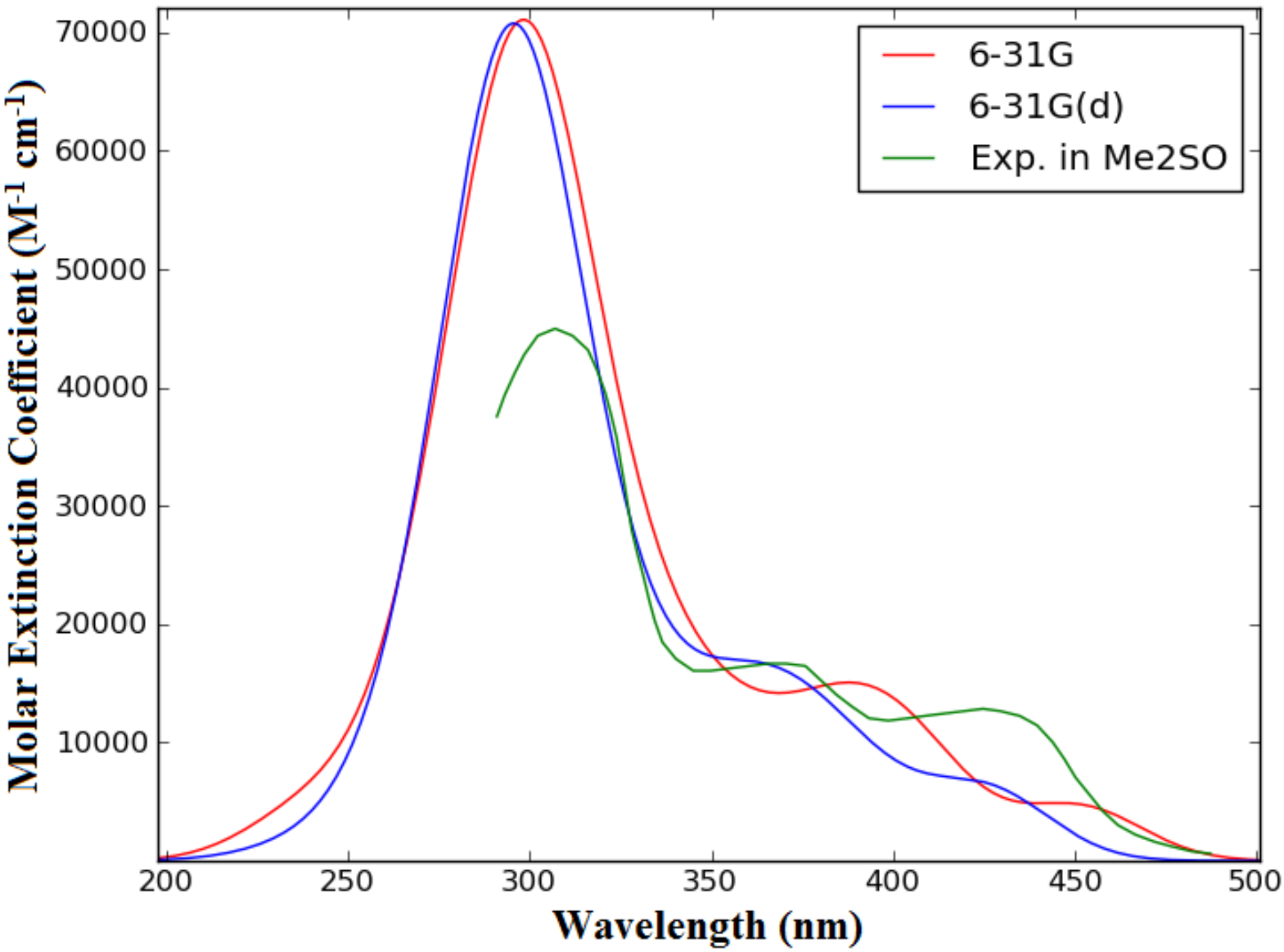}
\end{center}
[Ru(DPA)$_3$]$^{-}$
TD-B3LYP/6-31G, TD-B3LYP/6-31G(d), and experimental spectra.
Experimental spectrum measured in dimethyl sulfoxide \cite{SD82}.

% ================================================
\newpage
\section{Complex {\bf (89)}: [Ru(DPA)(DPAH)$_2$]$^{+}$}
% ================================================

\begin{center}
   {\bf PDOS}
\end{center}

\begin{center}
\begin{tabular}{cc}
\includegraphics[width=0.4\textwidth]{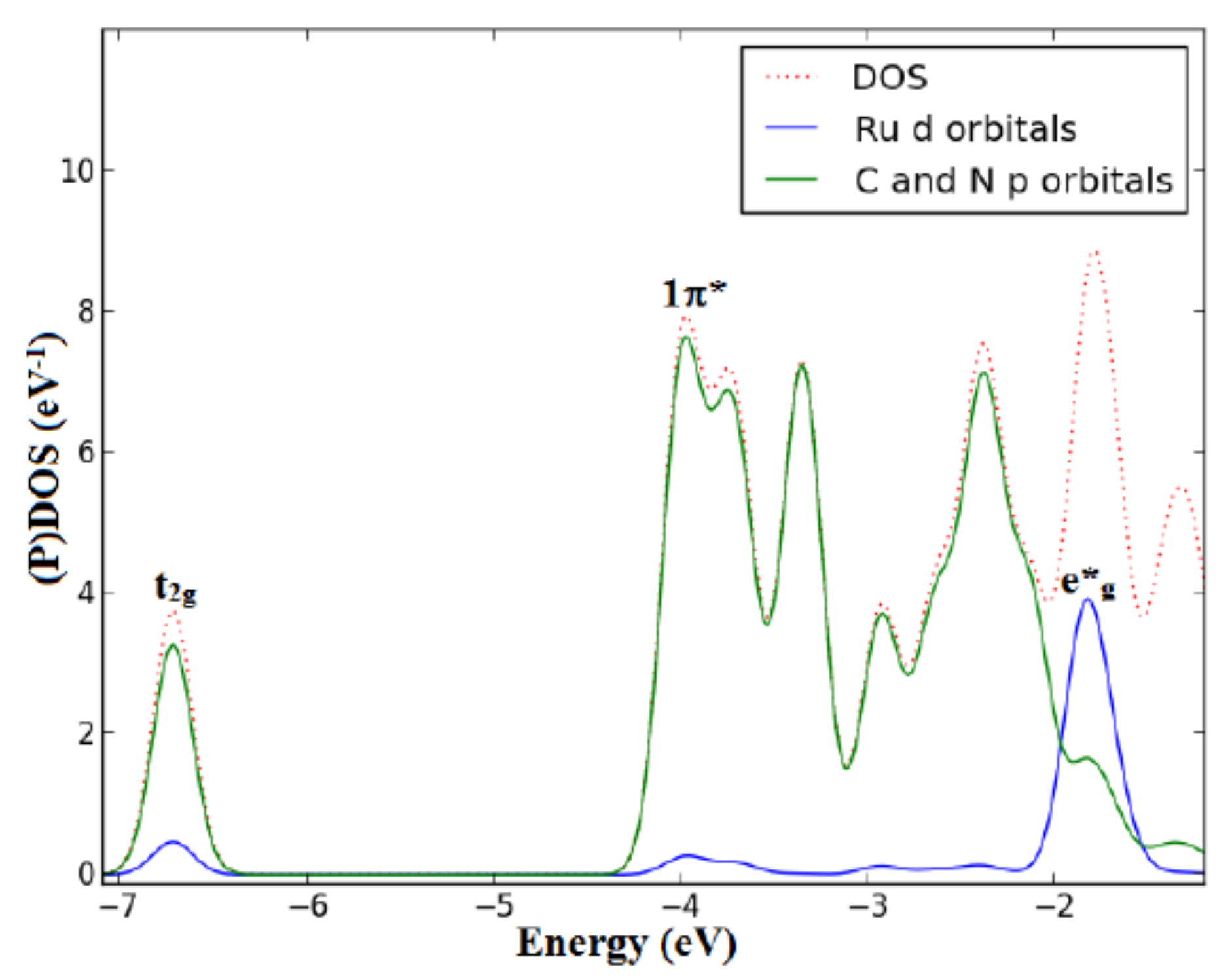} &
\includegraphics[width=0.4\textwidth]{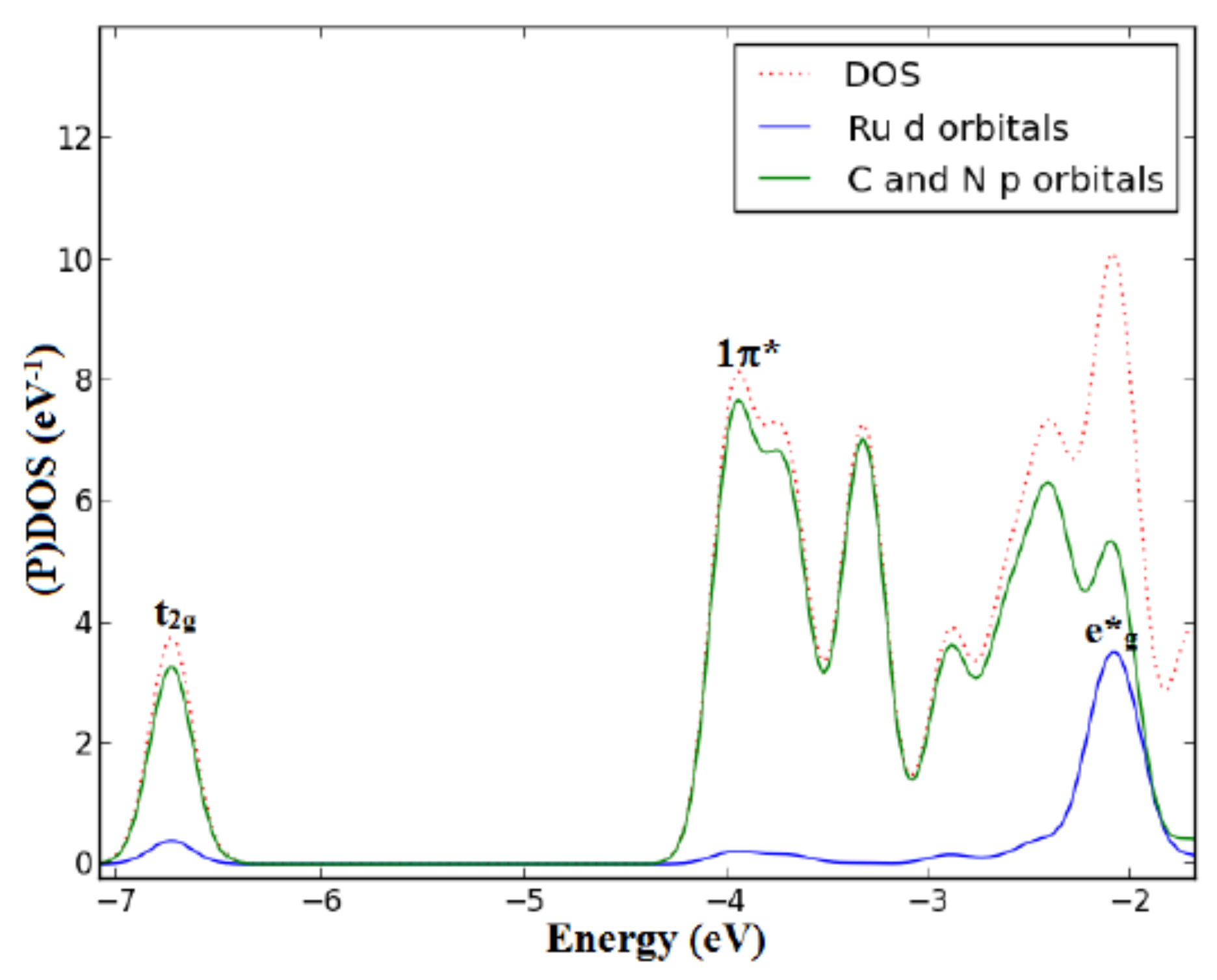} \\
B3LYP/6-31G & B3LYP/6-31G(d) \\
$\epsilon_{\text{HOMO}} = \mbox{-6.71 eV}$ & 
$\epsilon_{\text{HOMO}} = \mbox{-6.73 eV}$ 
\end{tabular}
\end{center}
Total and partial density of states of [Ru(DPA)(DPAH)$_2$]$^{+}$
partitioned over Ru d orbitals and ligand C and N p orbitals.
% for the 6-31G (left-hand side) and 6-31G* (right-hand side) basis sets.

\begin{center}
   {\bf Absorption Spectrum}
\end{center}

\begin{center}
\includegraphics[width=0.8\textwidth]{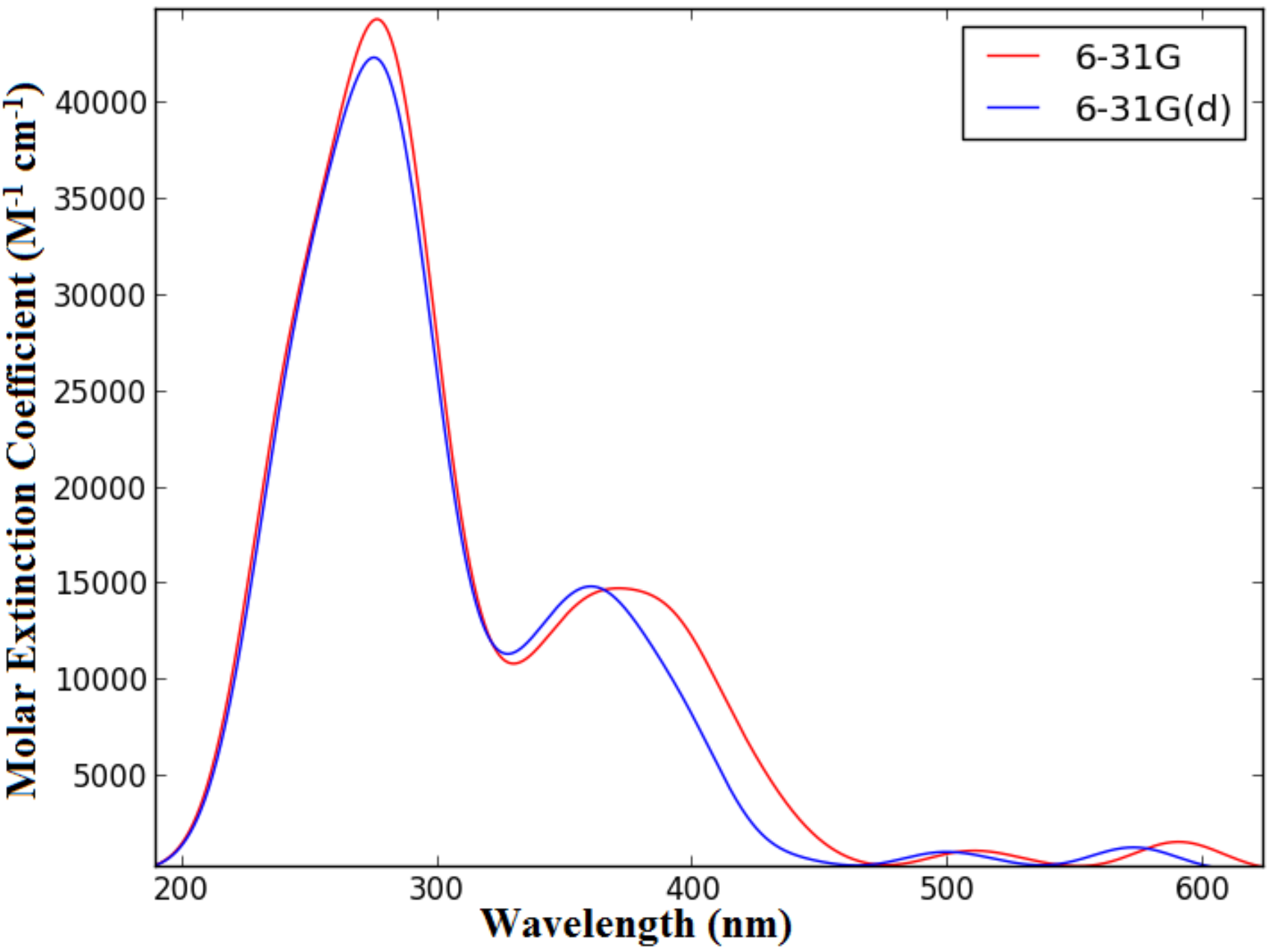}
\end{center}
[Ru(DPA)(DPAH)$_2$]$^{+}$
TD-B3LYP/6-31G and TD-B3LYP/6-31G(d) spectra.

% ================================================
\newpage
\section{Complex {\bf (90)}: [Ru(DPAH)$_3$]$^{2+}$}
% ================================================

\begin{center}
   {\bf PDOS}
\end{center}

\begin{center}
\begin{tabular}{cc}
\includegraphics[width=0.4\textwidth]{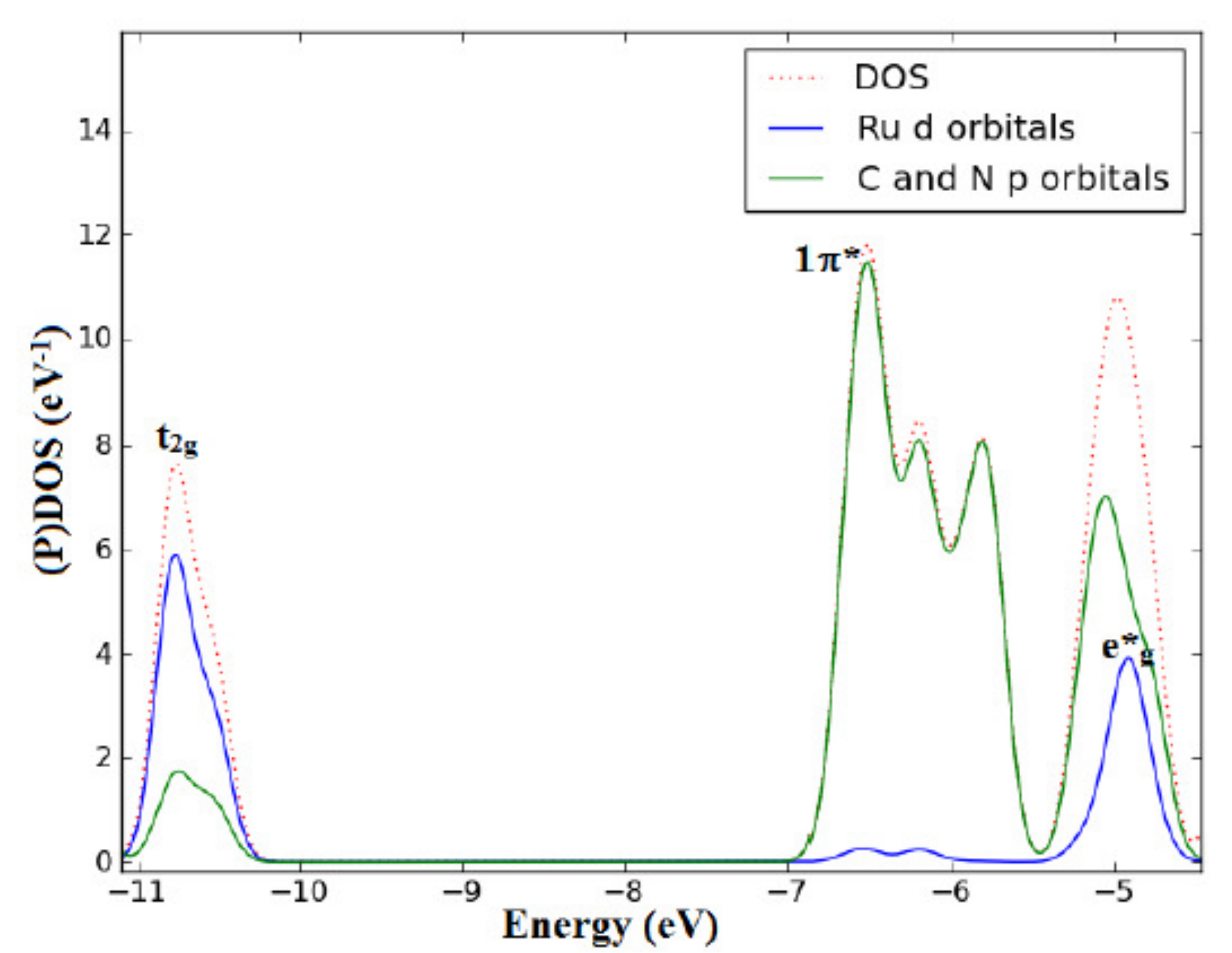} &
\includegraphics[width=0.4\textwidth]{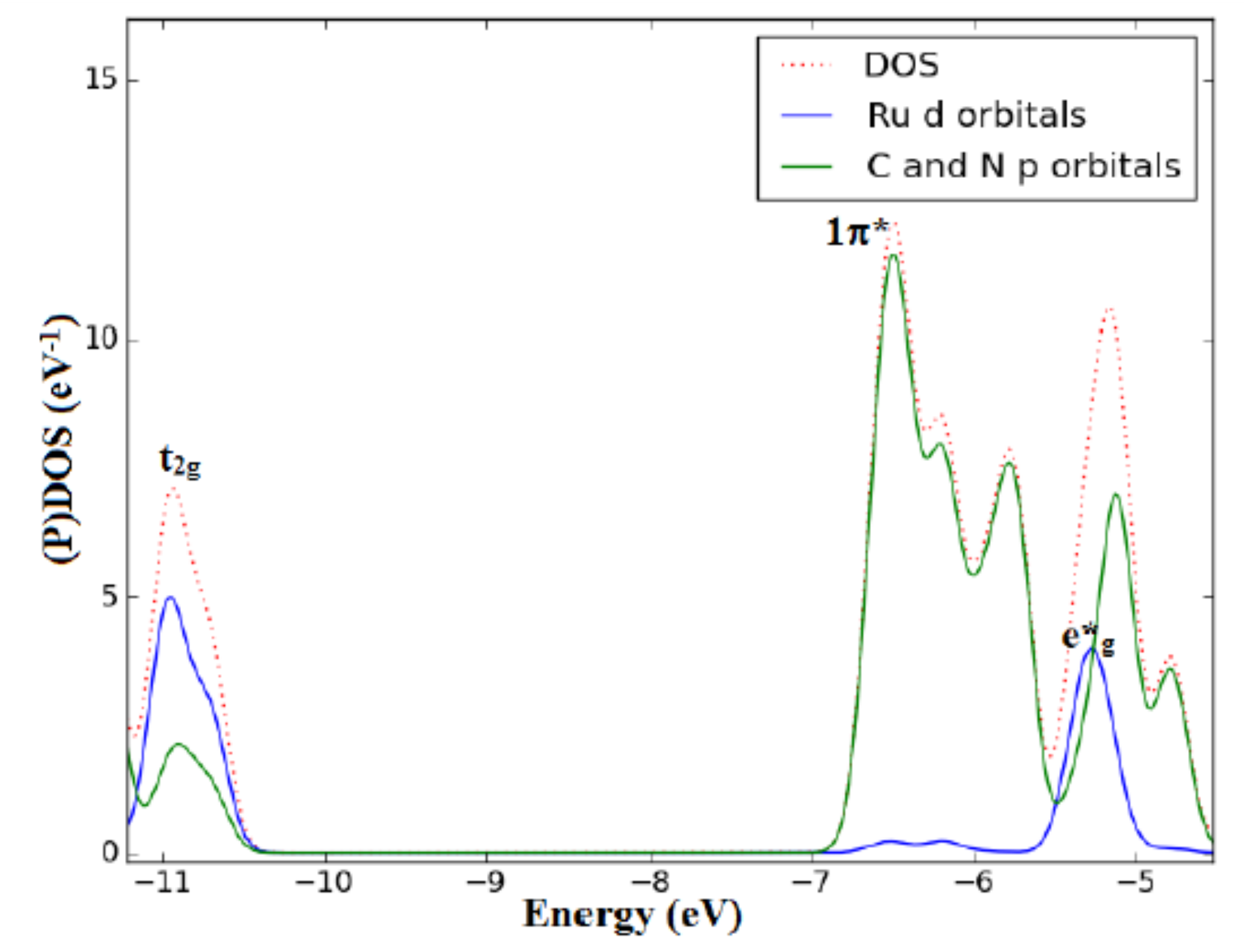} \\
B3LYP/6-31G & B3LYP/6-31G(d) \\
$\epsilon_{\text{HOMO}} = \mbox{-10.53 eV}$ & 
$\epsilon_{\text{HOMO}} = \mbox{-10.70 eV}$ 
\end{tabular}
\end{center}
Total and partial density of states of [Ru(DPAH)$_3$]$^{2+}$
partitioned over Ru d orbitals and ligand C and N p orbitals.
% for the 6-31G (left-hand side) and 6-31G* (right-hand side) basis sets.

\begin{center}
   {\bf Absorption Spectrum}
\end{center}

\begin{center}
\includegraphics[width=0.8\textwidth]{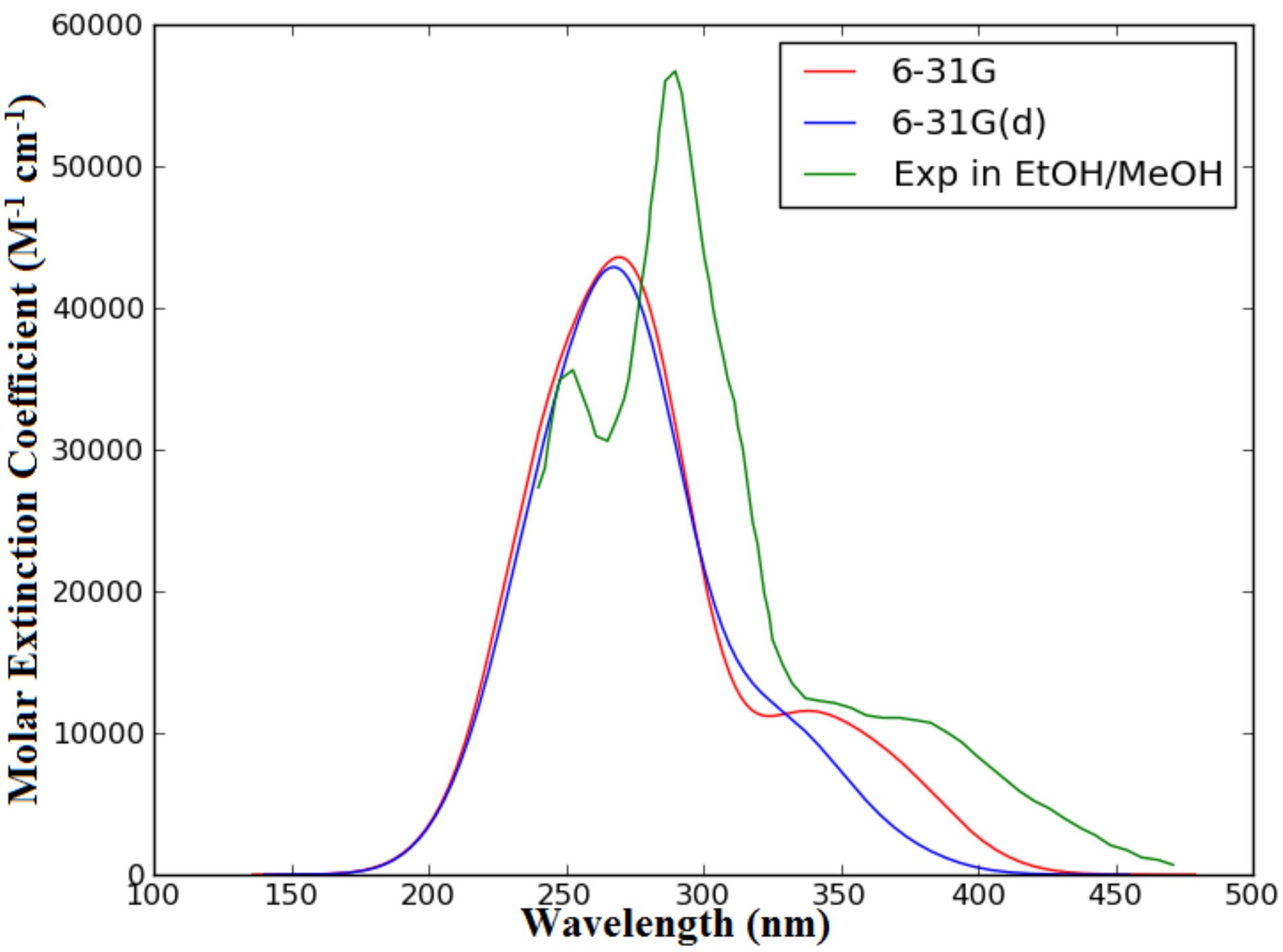}
\end{center}
[Ru(DPAH)$_3$]$^{2+}$
TD-B3LYP/6-31G, TD-B3LYP/6-31G(d), and experimental spectra.
Experimental spectrum measured in a mixture of methanol and ethanol
\cite{SD82}.

% ================================================
\newpage
\section{Complex {\bf (91)}: [Ru(Azpy)$_3$]$^{2+}$}
% ================================================

\begin{center}
   {\bf PDOS}
\end{center}

\begin{center}
\begin{tabular}{cc}
\includegraphics[width=0.4\textwidth]{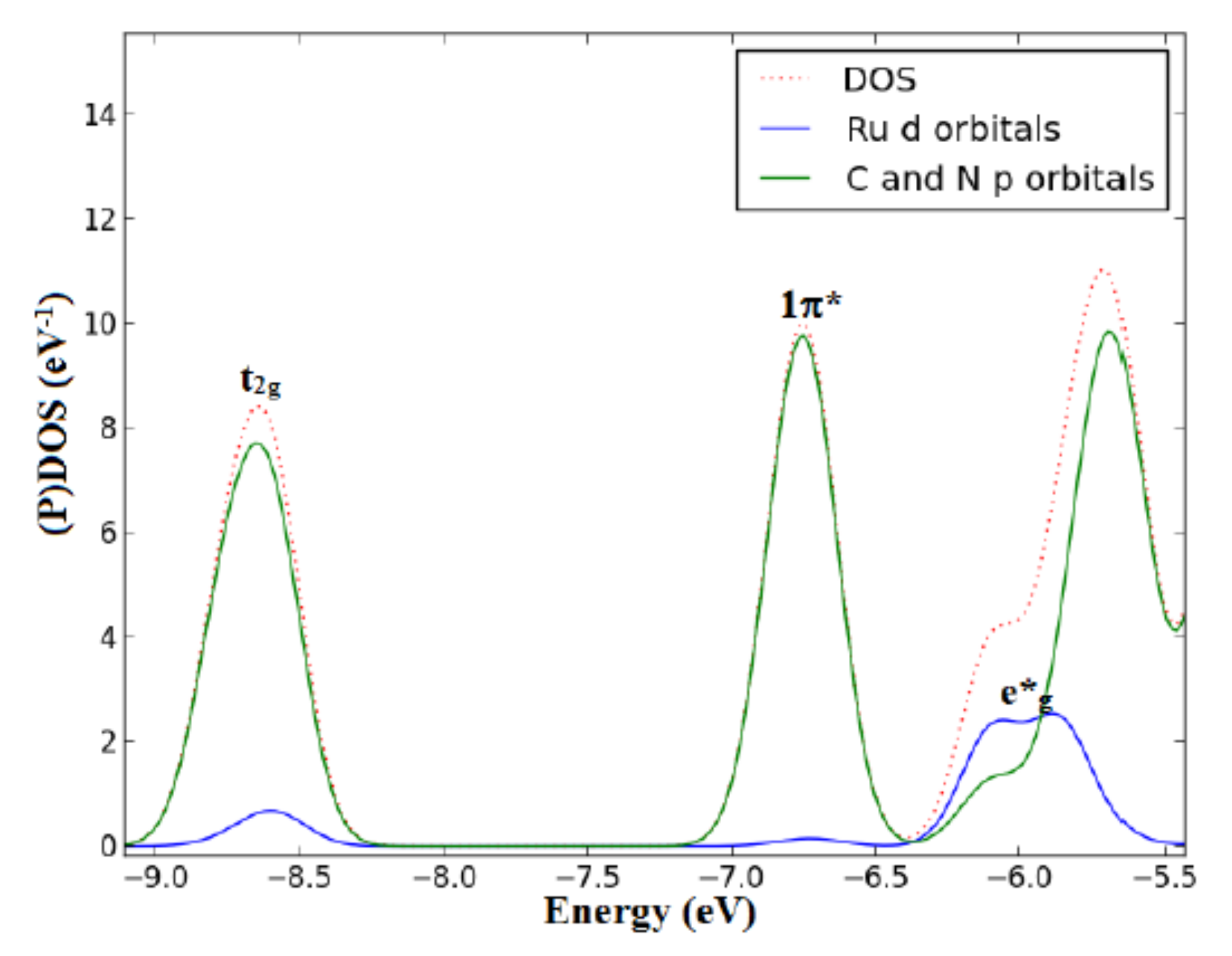} &
\includegraphics[width=0.4\textwidth]{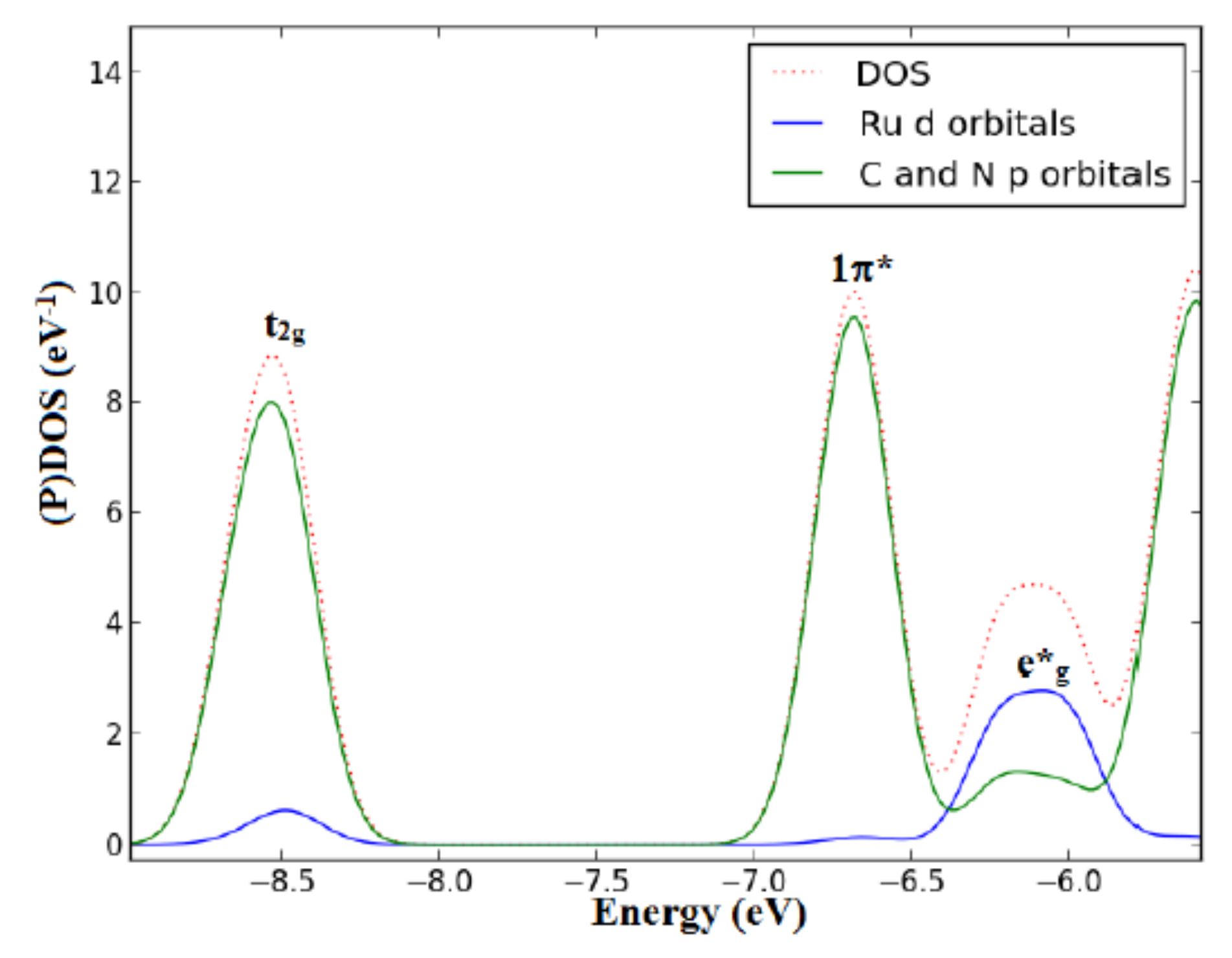} \\
B3LYP/6-31G & B3LYP/6-31G(d) \\
$\epsilon_{\text{HOMO}} = \mbox{-11.66 eV}$ & 
$\epsilon_{\text{HOMO}} = \mbox{-11.67 eV}$ 
\end{tabular}
\end{center}
Total and partial density of states of [Ru(Azpy)$_3$]$^{2+}$
partitioned over Ru d orbitals and ligand C and N p orbitals. 
% for the 6-31G (left-hand side) and 6-31G* (right-hand side) basis sets.

\begin{center}
   {\bf Absorption Spectrum}
\end{center}

\begin{center}
\includegraphics[width=0.8\textwidth]{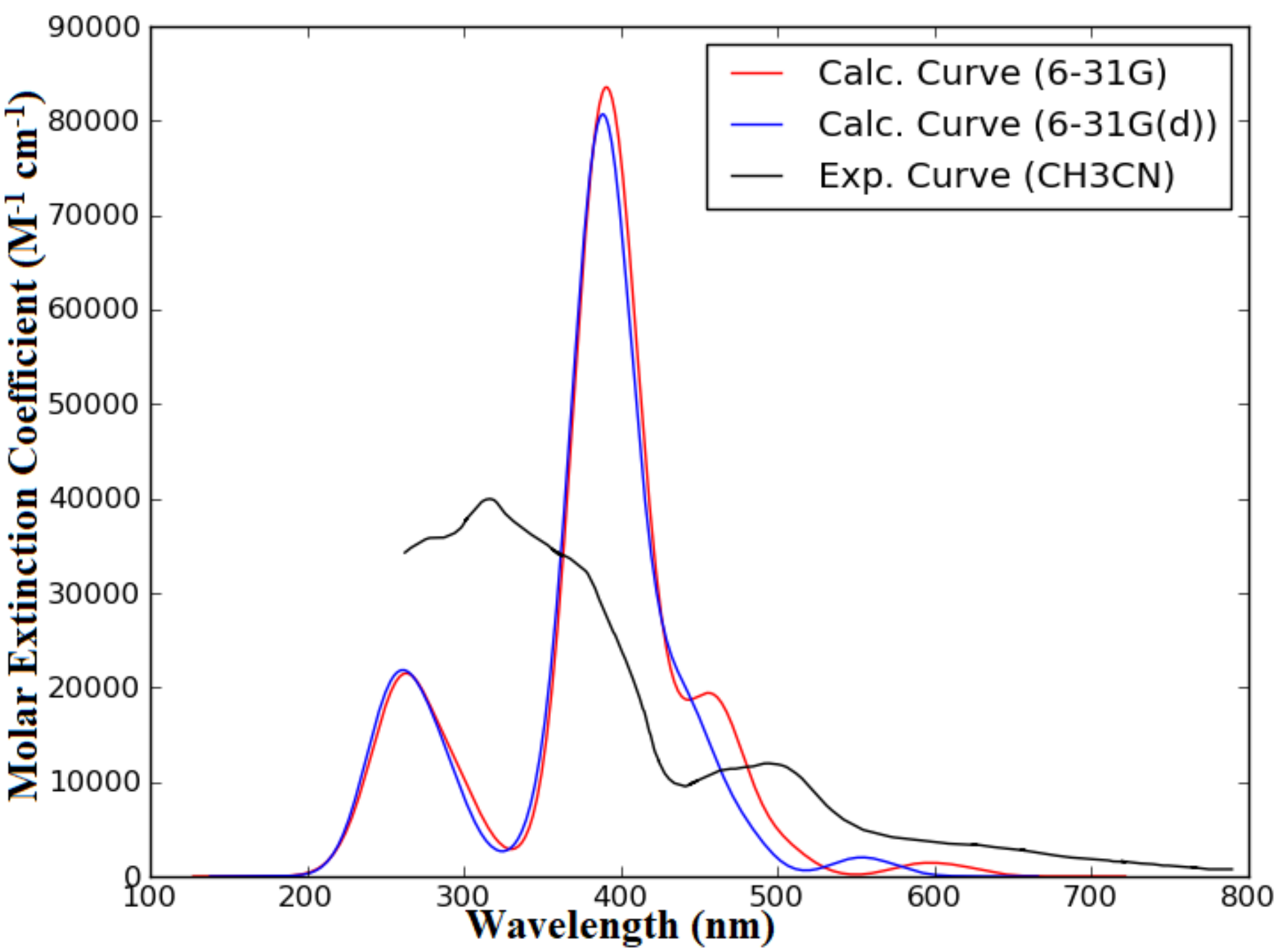}
\end{center}
[Ru(Azpy)$_3$]$^{2+}$
TD-B3LYP/6-31G, TD-B3LYP/6-31G(d), and experimental spectra.
Experimental spectrum measured in acrylonitrile \cite{LRE+04}.

% ================================================
\newpage
\section{Complex {\bf (92)}: [Ru(NA)$_3$]$^{2+}$}
% ================================================

\begin{center}
   {\bf PDOS}
\end{center}

\begin{center}
\begin{tabular}{cc}
\includegraphics[width=0.4\textwidth]{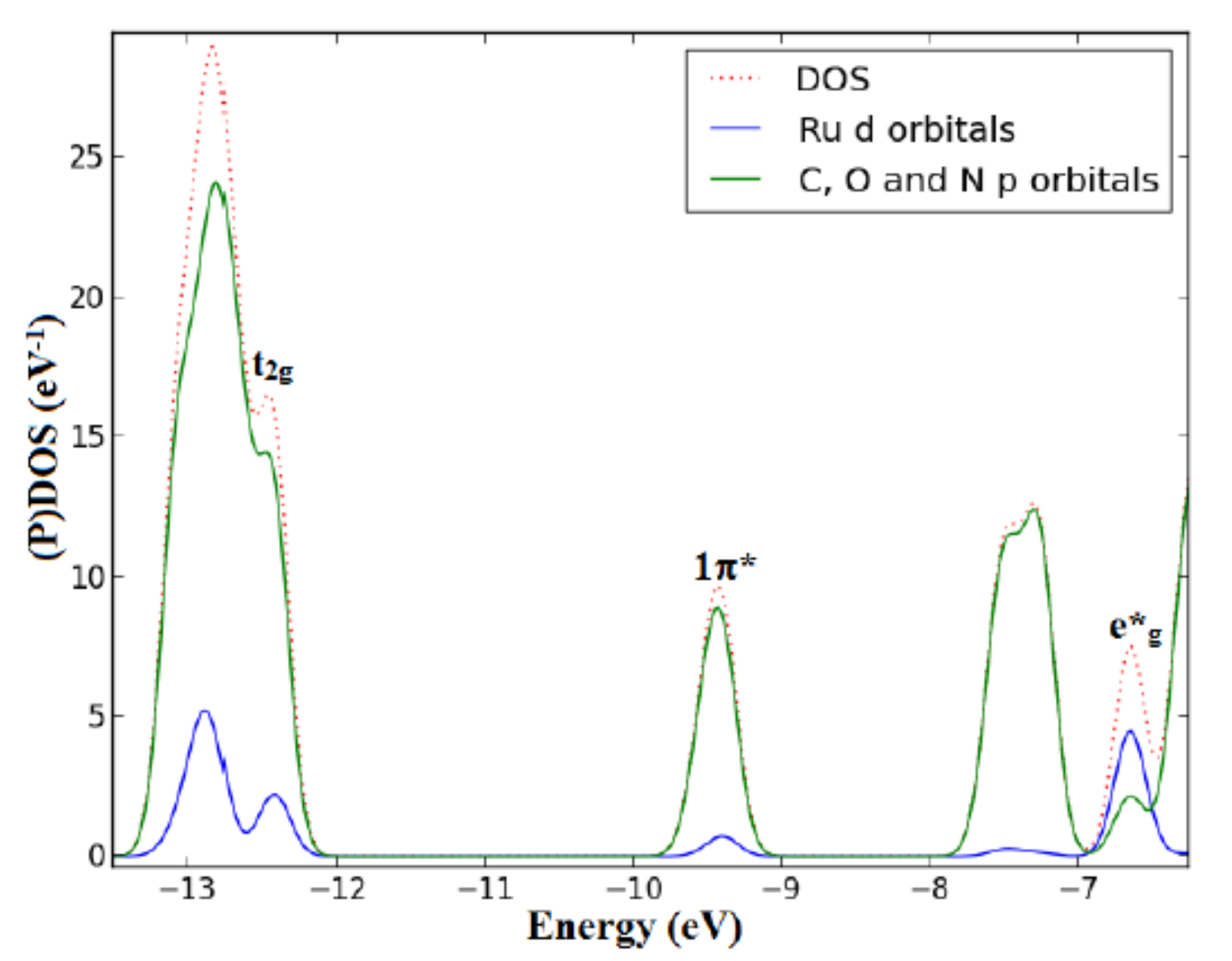} &
\includegraphics[width=0.4\textwidth]{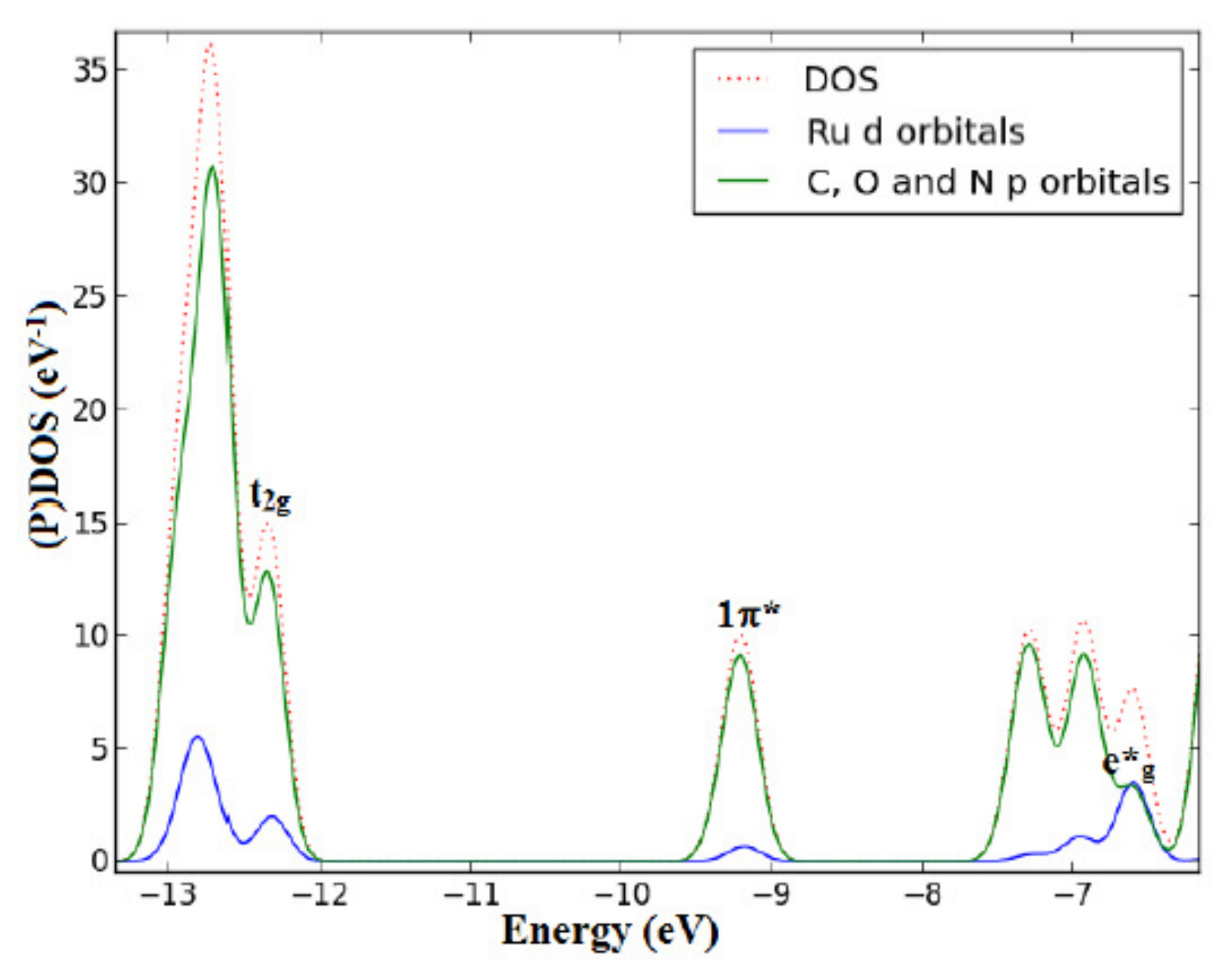} \\
B3LYP/6-31G & B3LYP/6-31G(d) \\
$\epsilon_{\text{HOMO}} = \mbox{-12.42 eV}$ & 
$\epsilon_{\text{HOMO}} = \mbox{-12.33 eV}$ 
\end{tabular}
\end{center}
Total and partial density of states of [Ru(NA)$_3$]$^{2+}$ 
partitioned over Ru d orbitals and ligand C, O, and N p orbitals.
% for the 6-31G (left-hand side) and 6-31G* (right-hand side) basis sets.

\begin{center}
   {\bf Absorption Spectrum}
\end{center}

\begin{center}
\includegraphics[width=0.8\textwidth]{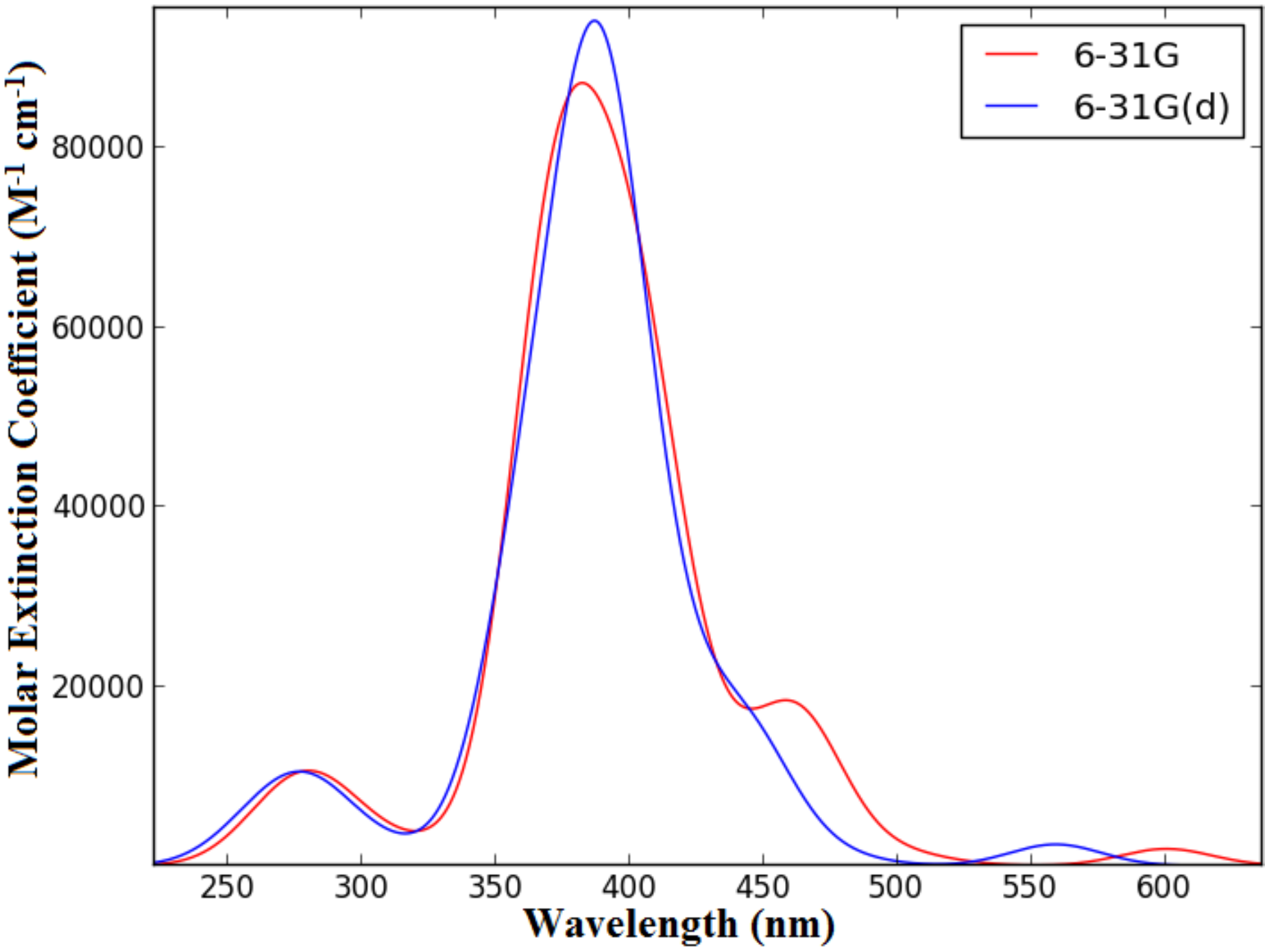}
\end{center}
[Ru(NA)$_3$]$^{2+}$
TD-B3LYP/6-31G and TD-B3LYP/6-31G(d) spectra.

% ================================================
\newpage
\section{Complex {\bf (93)}: [Ru(hpiq)$_3$]$^{2+}$}
% ================================================

\begin{center}
   {\bf PDOS}
\end{center}

\begin{center}
\begin{tabular}{cc}
\includegraphics[width=0.4\textwidth]{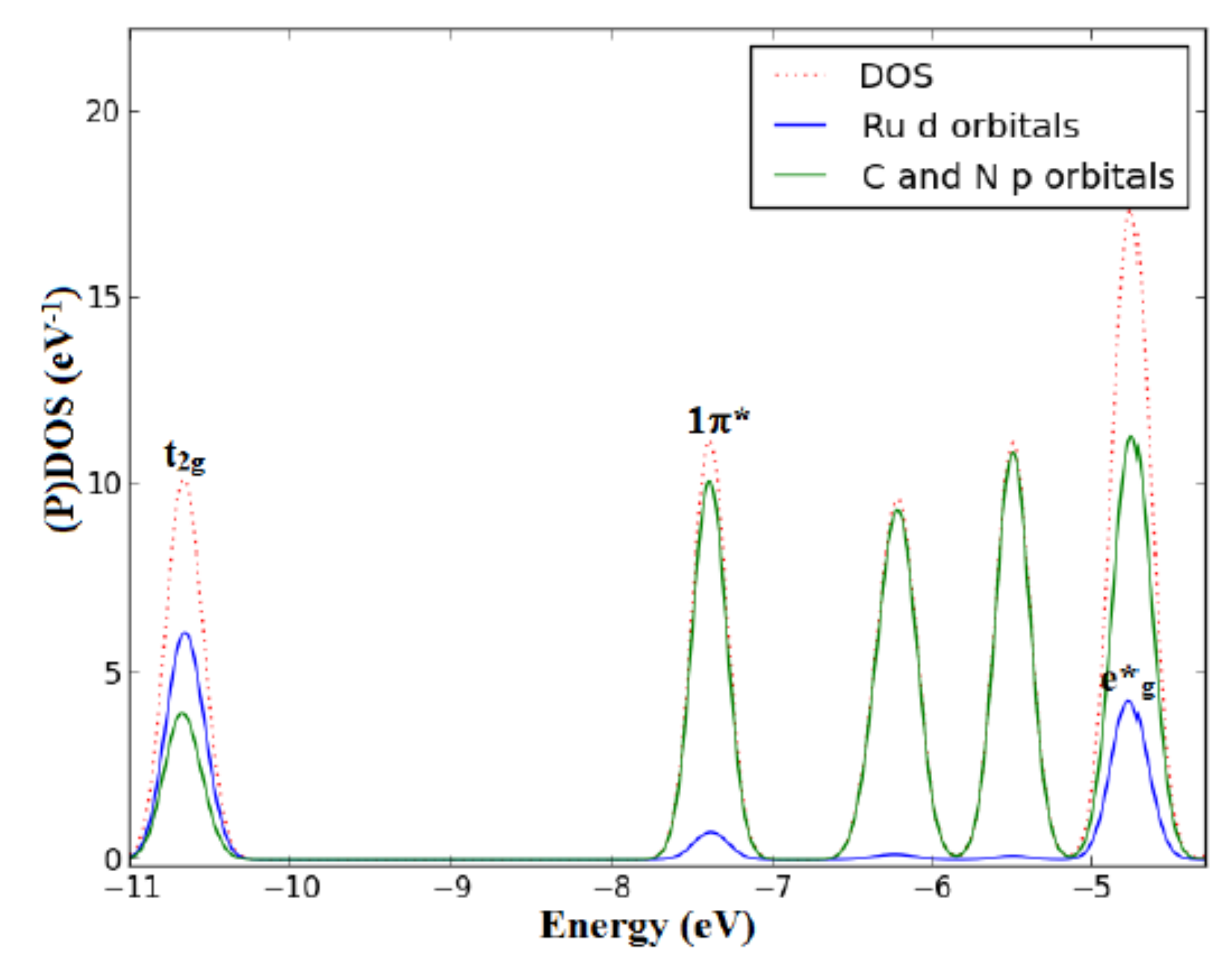} &
\includegraphics[width=0.4\textwidth]{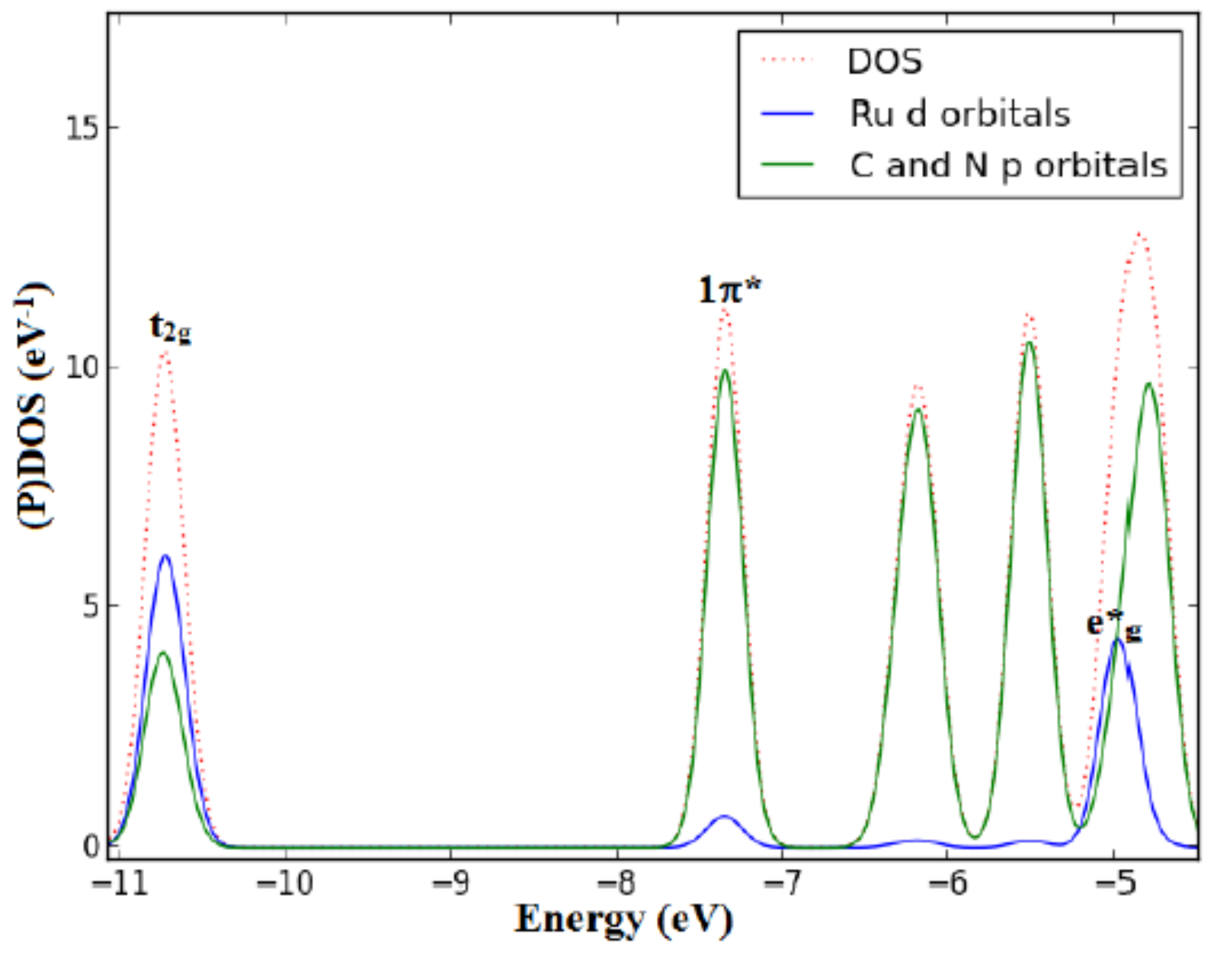} \\
B3LYP/6-31G & B3LYP/6-31G(d) \\
$\epsilon_{\text{HOMO}} = \mbox{-10.60 eV}$ & 
$\epsilon_{\text{HOMO}} = \mbox{-10.67 eV}$ 
\end{tabular}
\end{center}
Total and partial density of states of [Ru(hpiq)$_3$]$^{2+}$
partitioned over Ru d orbitals and ligand C and N p orbitals. 
% for the 6-31G (left-hand side) and 6-31G* (right-hand side) basis sets.

\begin{center}
   {\bf Absorption Spectrum}
\end{center}

\begin{center}
\includegraphics[width=0.8\textwidth]{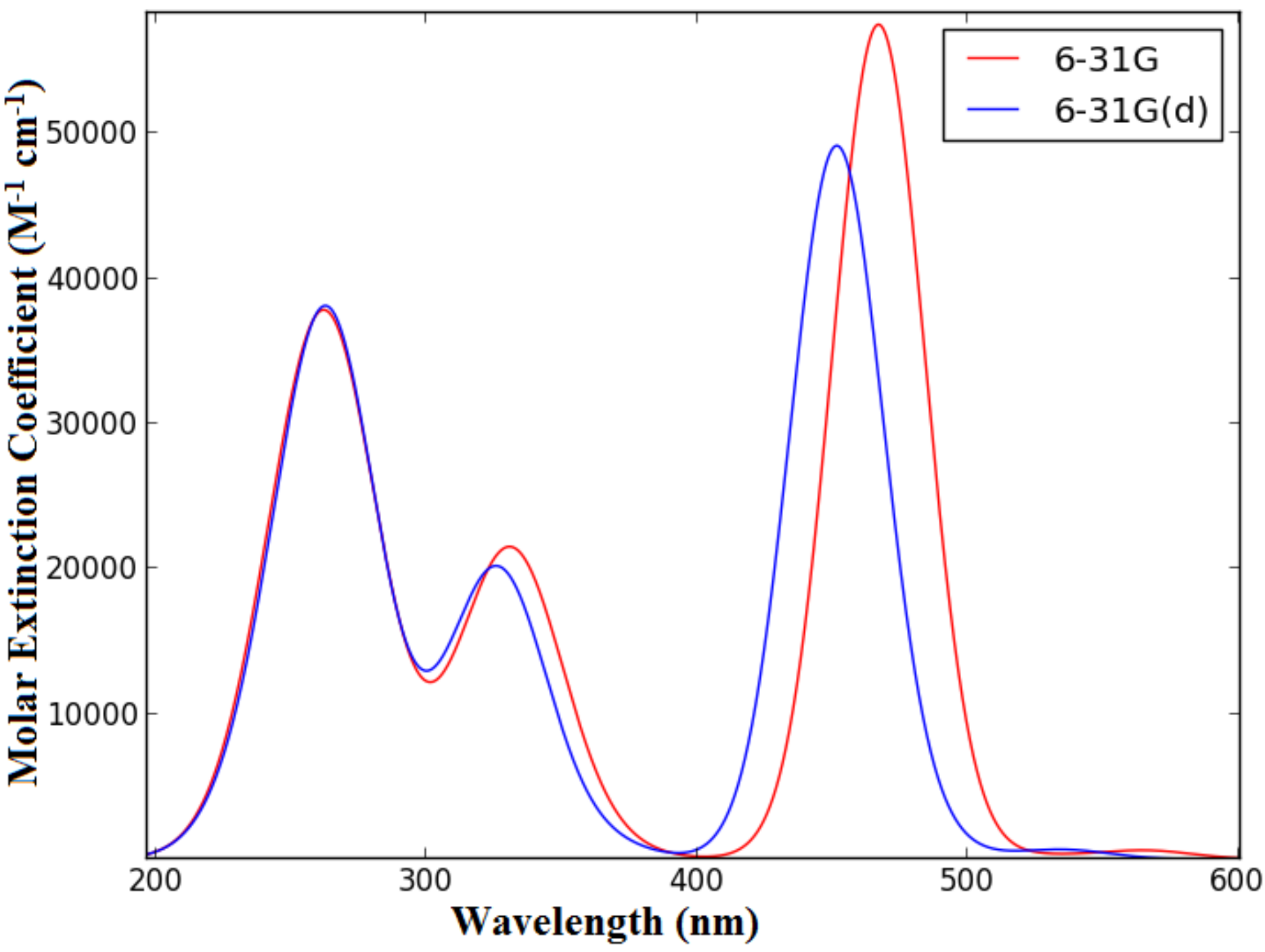}
\end{center}
[Ru(hpiq)$_3$]$^{2+}$
TD-B3LYP/6-31G and TD-B3LYP/6-31G(d) spectra.

% ================================================
\newpage
\section{Complex {\bf (94)}: [Ru(pq)$_3$]$^{2+}$}
% ================================================

\begin{center}
   {\bf PDOS}
\end{center}

\begin{center}
\begin{tabular}{cc}
\includegraphics[width=0.4\textwidth]{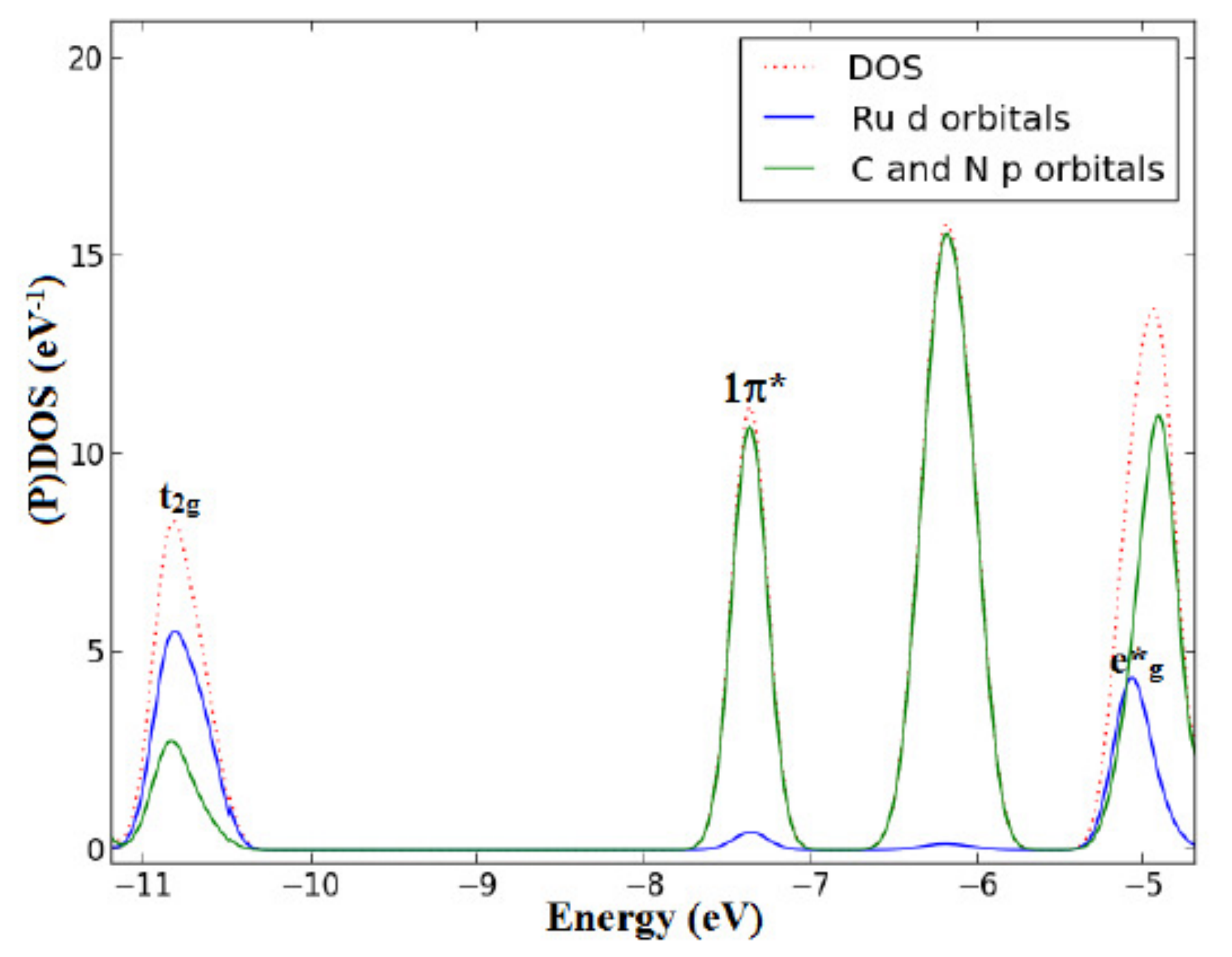} &
\includegraphics[width=0.4\textwidth]{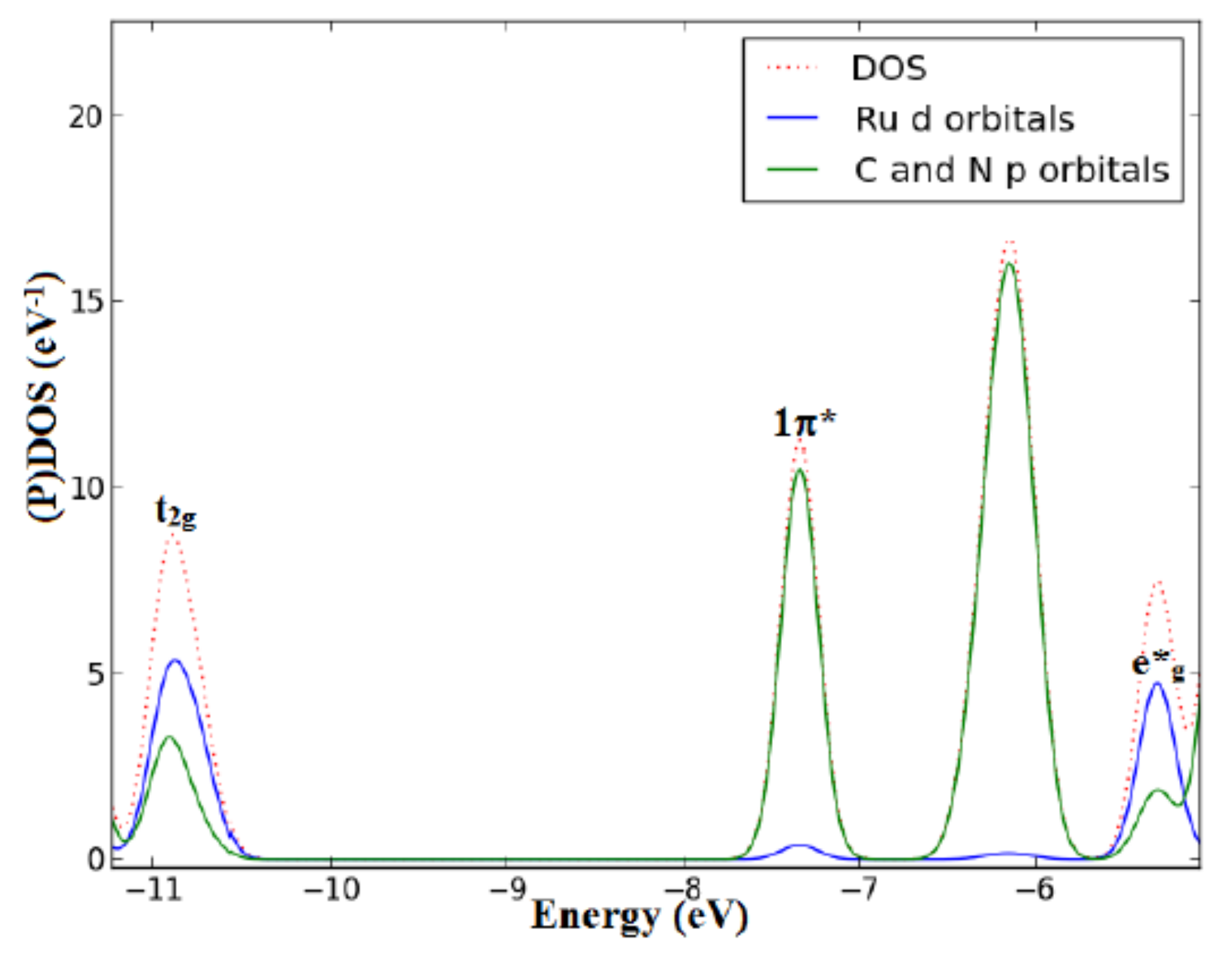} \\
B3LYP/6-31G & B3LYP/6-31G(d) \\
$\epsilon_{\text{HOMO}} = \mbox{-10.65 eV}$ & 
$\epsilon_{\text{HOMO}} = \mbox{-10.77 eV}$ 
\end{tabular}
\end{center}
Total and partial density of states of [Ru(pq)$_3$]$^{2+}$
partitioned over Ru d orbitals and ligand C and N p orbitals. 
% for the 6-31G (left-hand side) and 6-31G* (right-hand side) basis sets.

\begin{center}
   {\bf Absorption Spectrum}
\end{center}

\begin{center}
\includegraphics[width=0.8\textwidth]{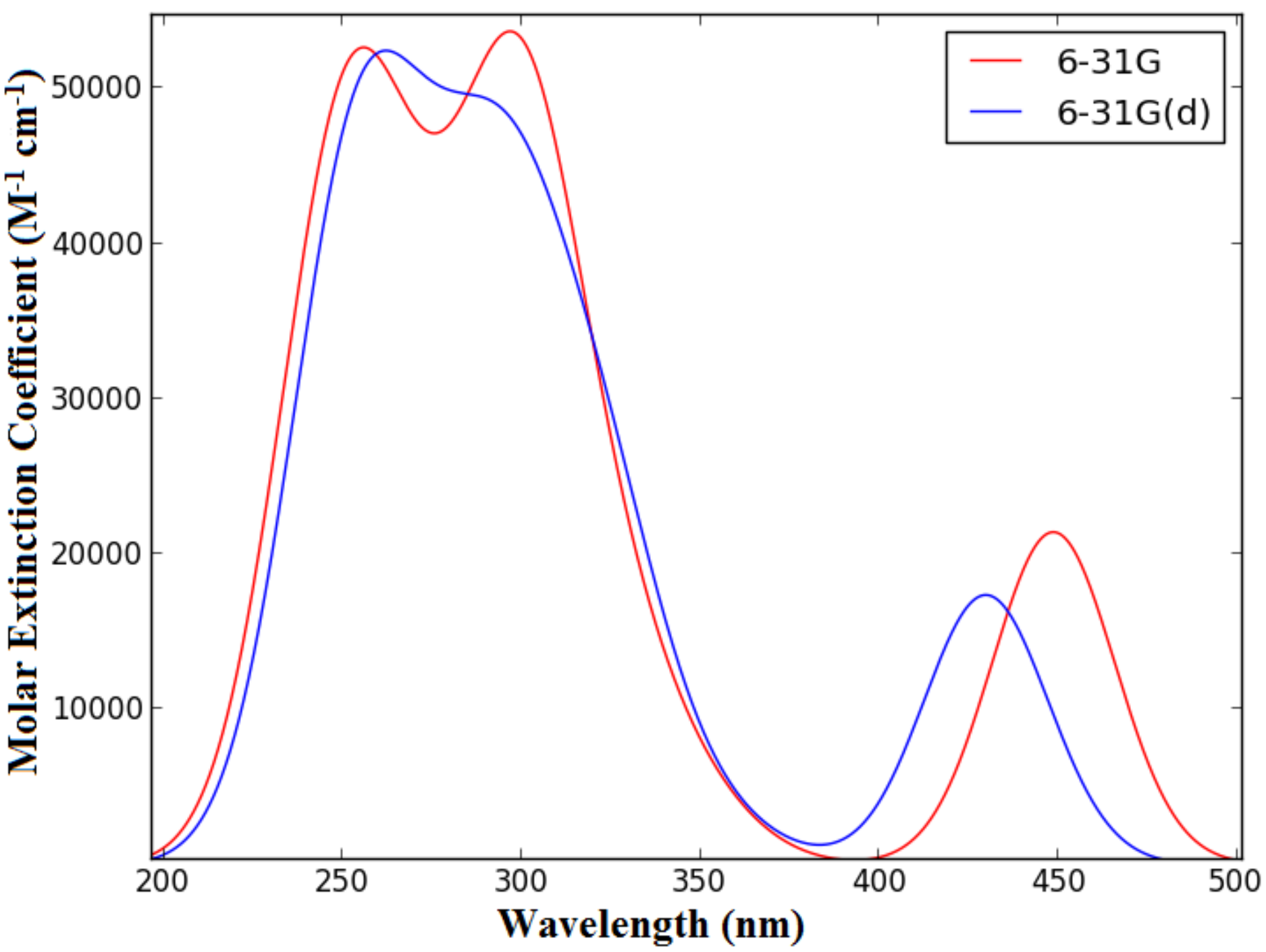}
\end{center}
[Ru(pq)$_3$]$^{2+}$
TD-B3LYP/6-31G and TD-B3LYP/6-31G(d) spectra.

% ================================================
\newpage
\section{Complex {\bf (95)}: [Ru(pq)$_2$(biq)]$^{2+}$}
% ================================================

\begin{center}
   {\bf PDOS}
\end{center}

\begin{center}
\begin{tabular}{cc}
\includegraphics[width=0.4\textwidth]{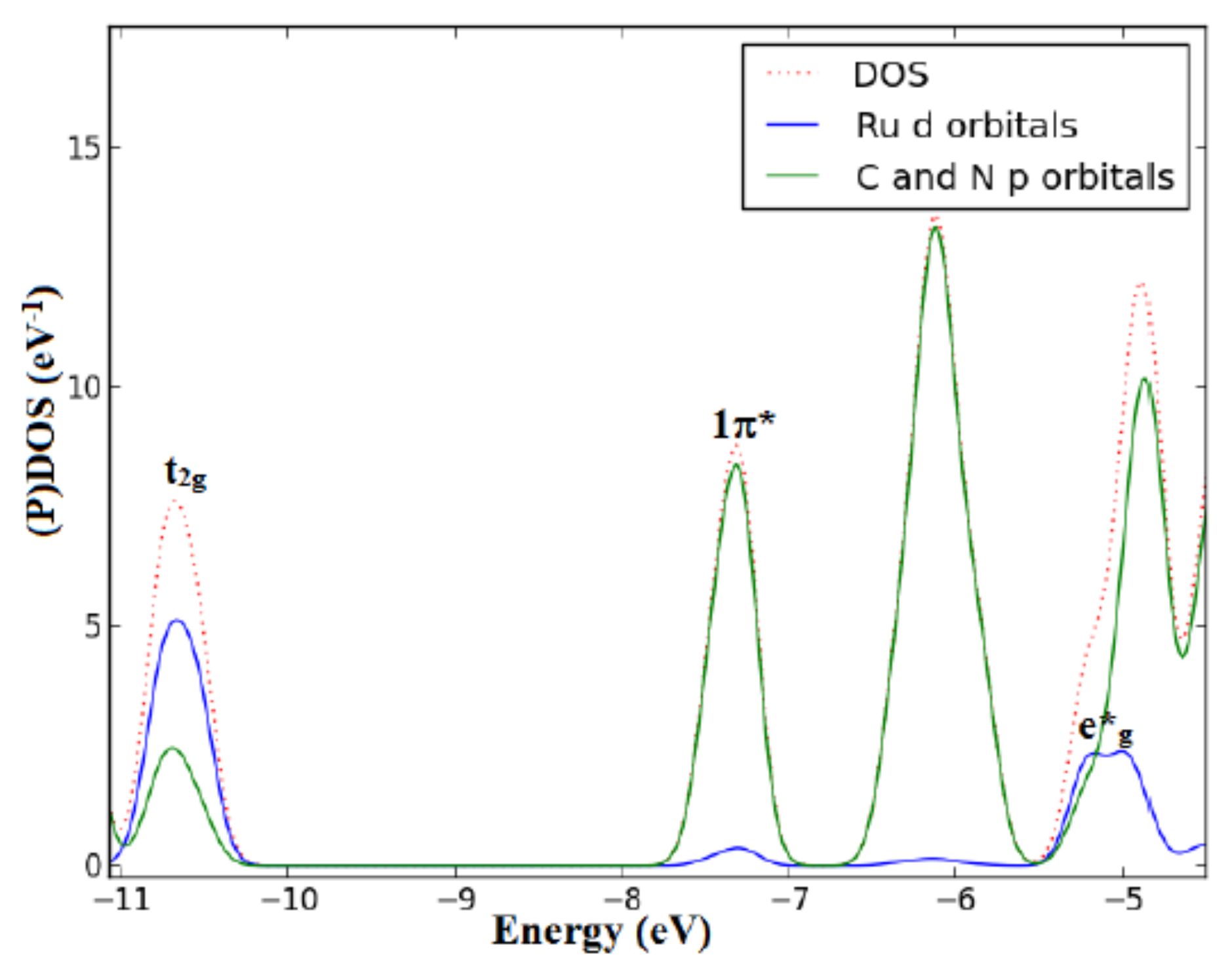} &
\includegraphics[width=0.4\textwidth]{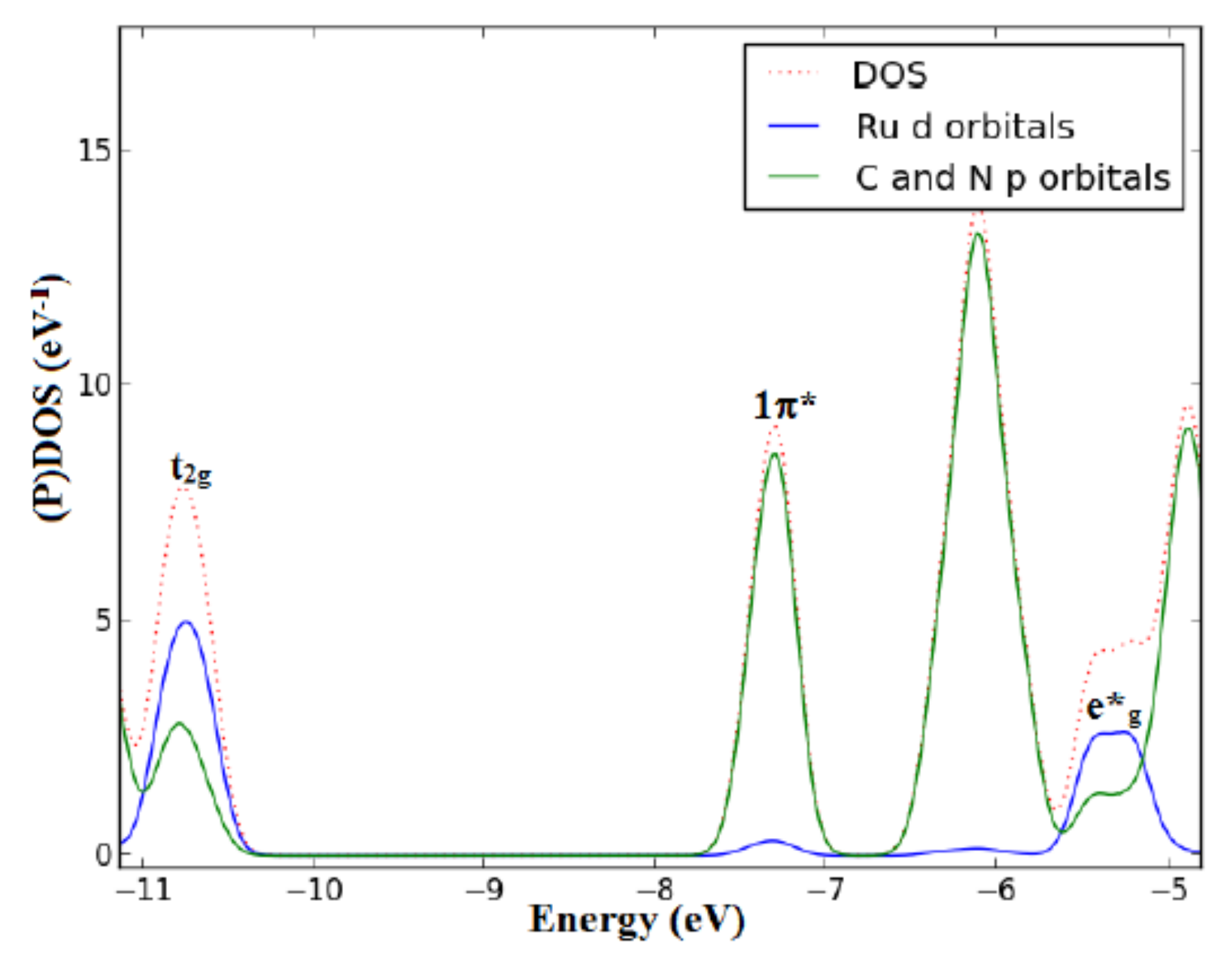} \\
B3LYP/6-31G & B3LYP/6-31G(d) \\
$\epsilon_{\text{HOMO}} = \mbox{-10.54 eV}$ & 
$\epsilon_{\text{HOMO}} = \mbox{-10.64 eV}$ 
\end{tabular}
\end{center}
Total and partial density of states of [Ru(pq)$_2$(biq)]$^{2+}$
partitioned over Ru d orbitals and ligand C and N p orbitals.
% for the 6-31G (left-hand side) and 6-31G* (right-hand side) basis sets.

\begin{center}
   {\bf Absorption Spectrum}
\end{center}

\begin{center}
\includegraphics[width=0.8\textwidth]{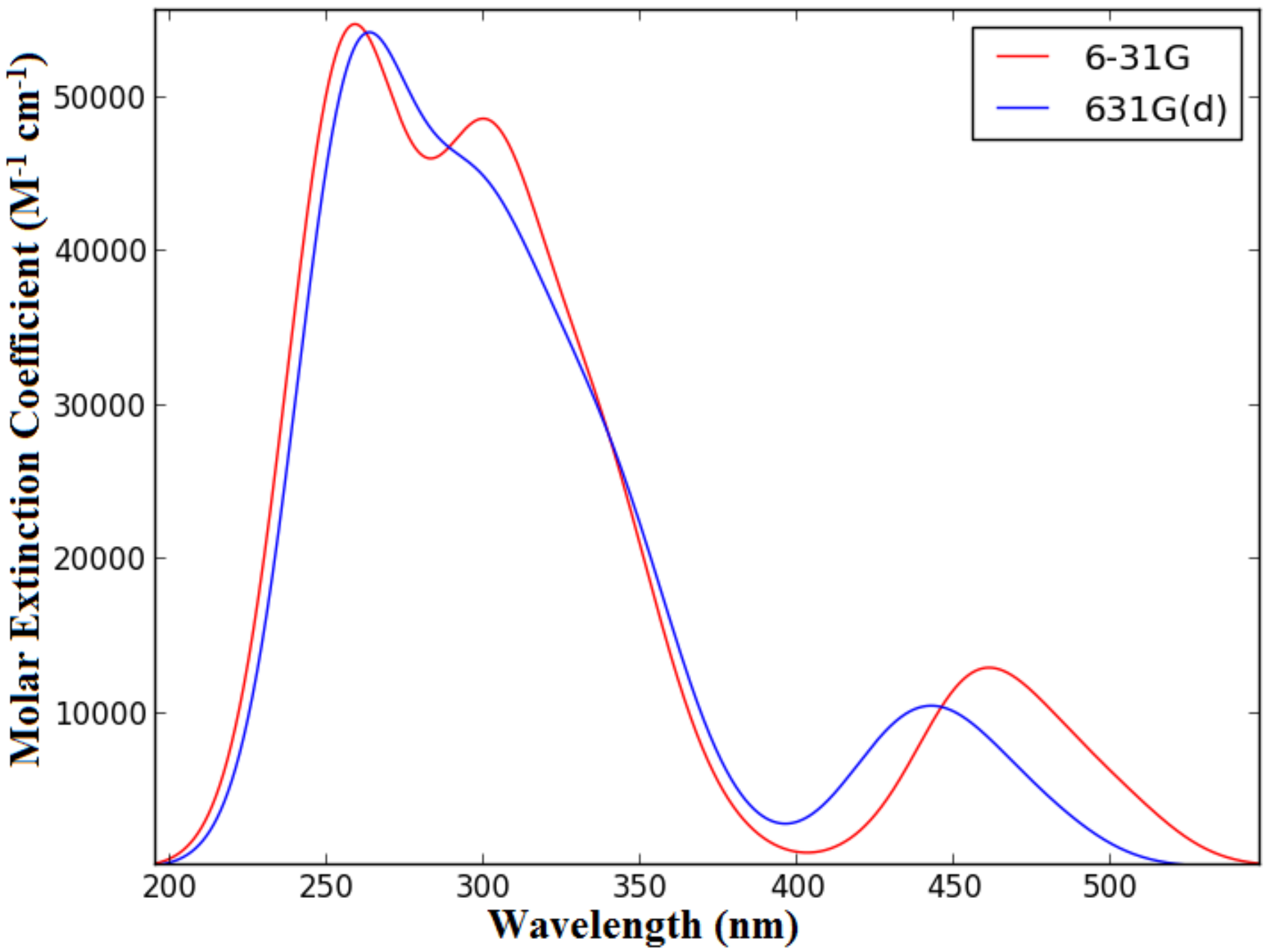}
\end{center}
[Ru(pq)$_2$(biq)]$^{2+}$
TD-B3LYP/6-31G and TD-B3LYP/6-31G(d) spectra.

% ================================================
\newpage
\section{Complex {\bf (96)}: [Ru(pq)(biq)$_2$]$^{2+}$}
% ================================================

\begin{center}
   {\bf PDOS}
\end{center}

\begin{center}
\begin{tabular}{cc}
\includegraphics[width=0.4\textwidth]{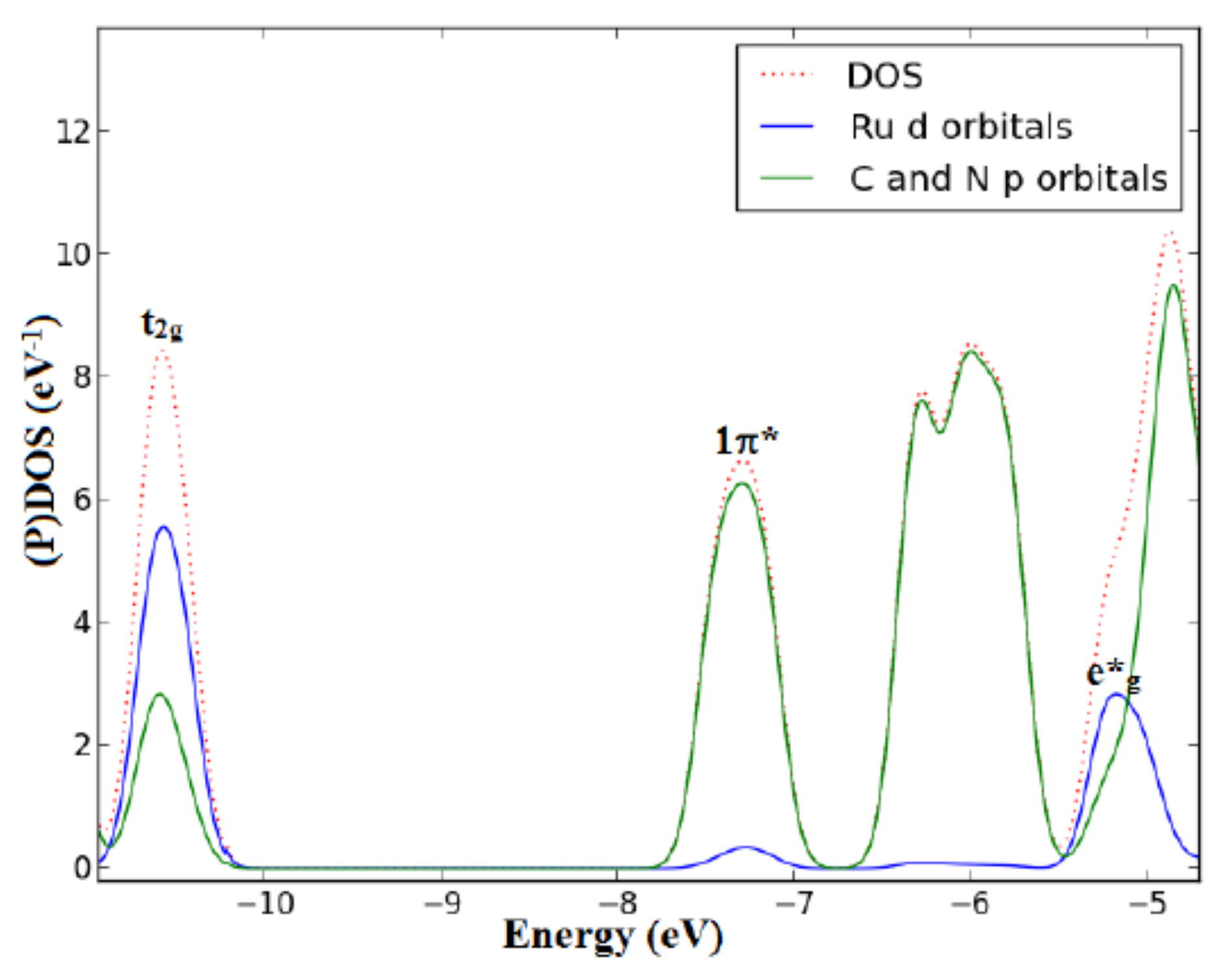} &
\includegraphics[width=0.4\textwidth]{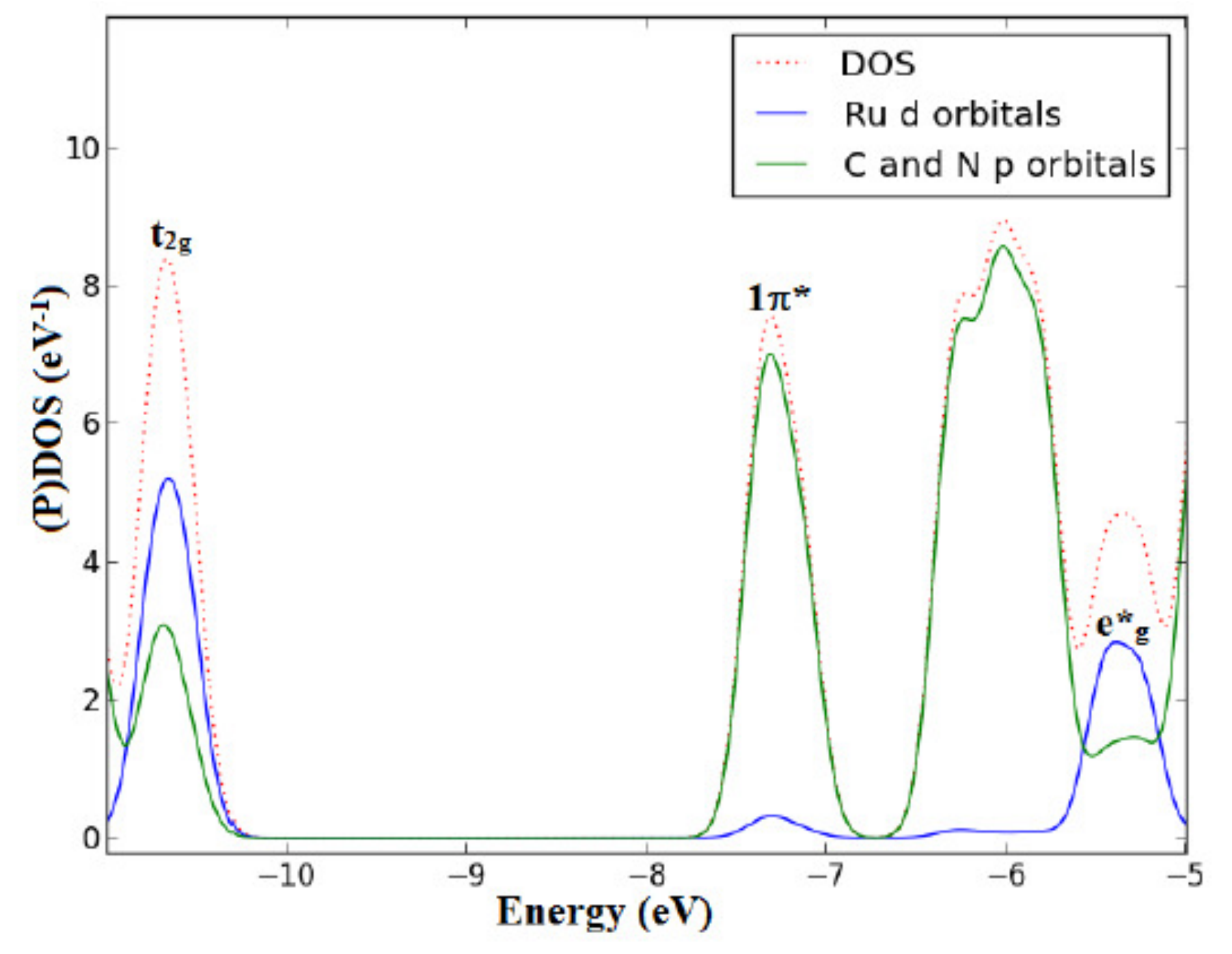} \\
B3LYP/6-31G & B3LYP/6-31G(d) \\
$\epsilon_{\text{HOMO}} = \mbox{-10.45 eV}$ & 
$\epsilon_{\text{HOMO}} = \mbox{-10.55 eV}$ 
\end{tabular}
\end{center}
Total and partial density of states of [Ru(pq)(biq)$_2$]$^{2+}$
partitioned over Ru d orbitals and ligand C and N p orbitals.
% for the 6-31G (left-hand side) and 6-31G* (right-hand side) basis sets.

\begin{center}
   {\bf Absorption Spectrum}
\end{center}

\begin{center}
\includegraphics[width=0.8\textwidth]{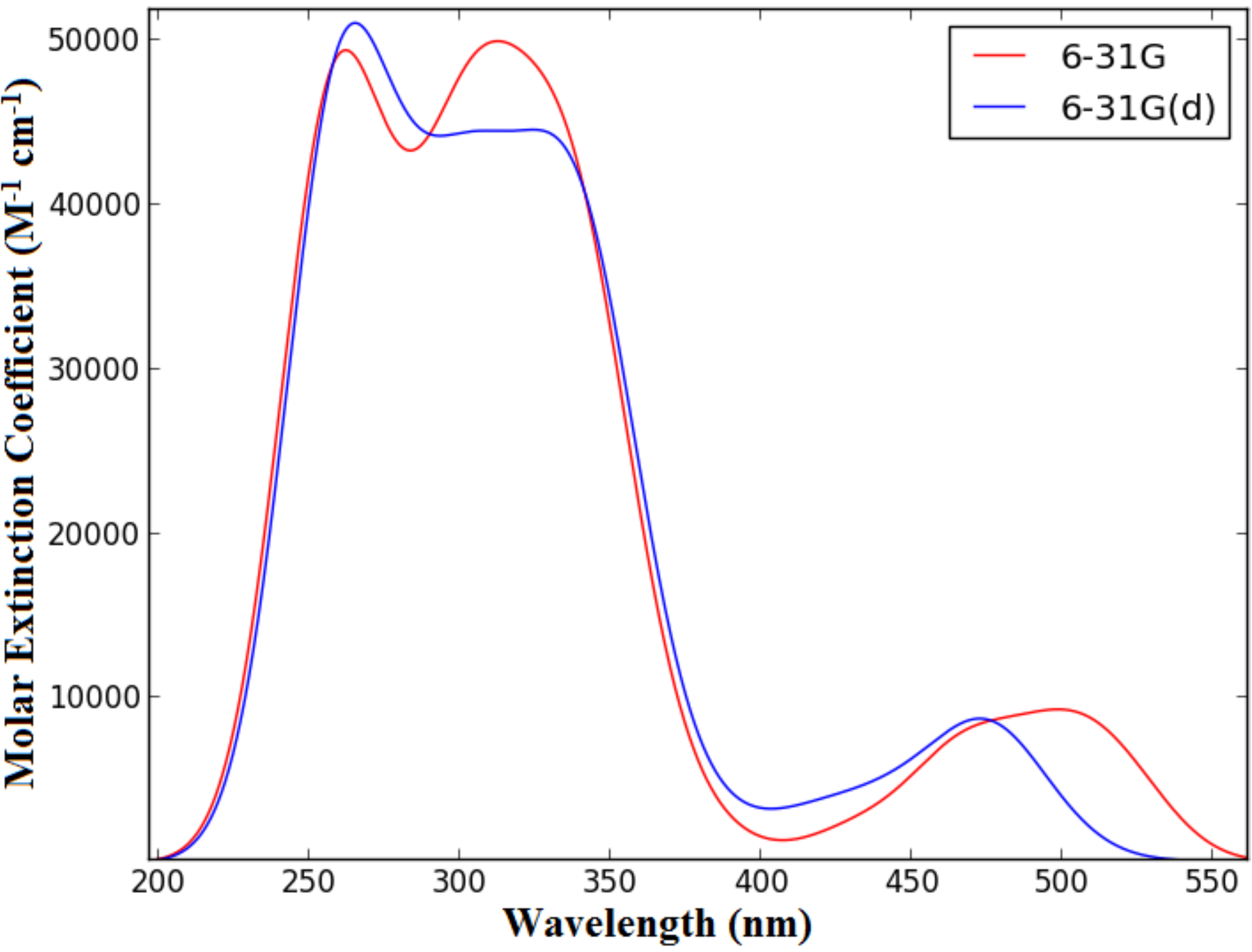}
\end{center}
[Ru(pq)(biq)$_2$]$^{2+}$
TD-B3LYP/6-31G and TD-B3LYP/6-31G(d) spectra.

% ================================================
\newpage
\section{Complex {\bf (97)}: [Ru(pynapy)$_3$]$^{2+}$}
% ================================================

\begin{center}
   {\bf PDOS}
\end{center}

\begin{center}
\begin{tabular}{cc}
\includegraphics[width=0.4\textwidth]{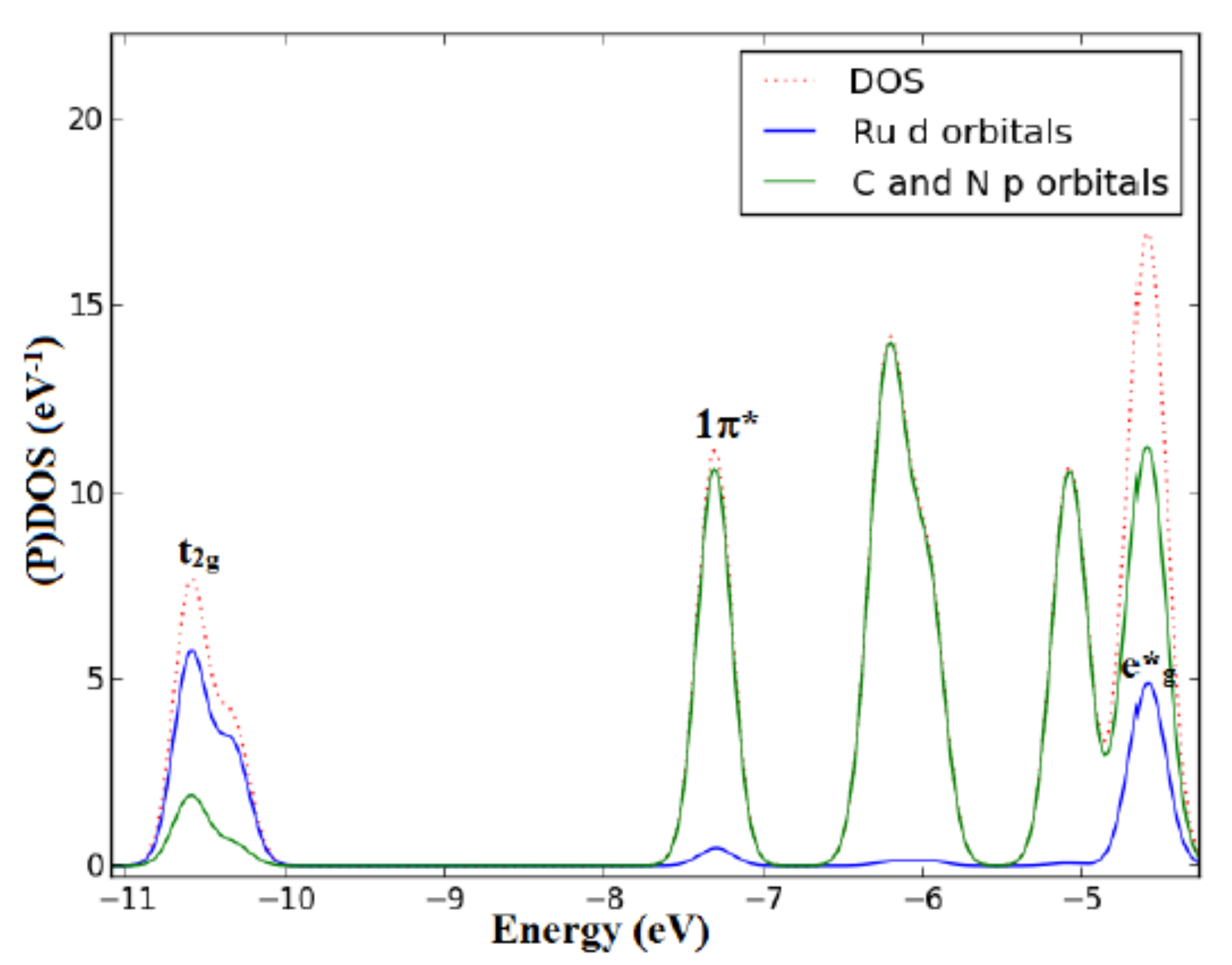} &
\includegraphics[width=0.4\textwidth]{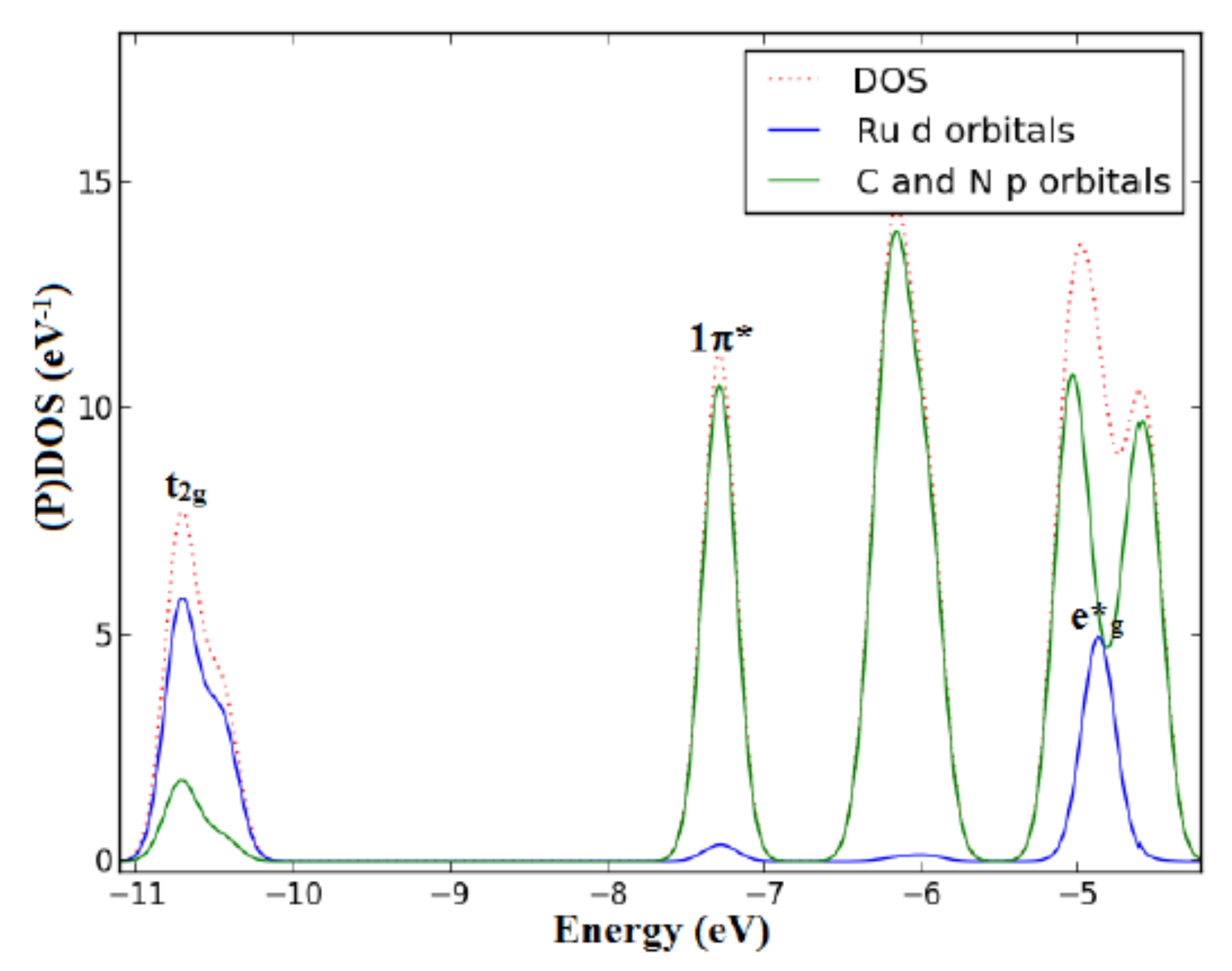} \\
B3LYP/6-31G & B3LYP/6-31G(d) \\
$\epsilon_{\text{HOMO}} = \mbox{-10.34 eV}$ & 
$\epsilon_{\text{HOMO}} = \mbox{-10.46 eV}$ 
\end{tabular}
\end{center}
Total and partial density of states of [Ru(pynapy)$_3$]$^{2+}$
partitioned over Ru d orbitals and ligand C and N p orbitals. 
% for the 6-31G (left-hand side) and 6-31G* (right-hand side) basis sets.

\begin{center}
   {\bf Absorption Spectrum}
\end{center}

\begin{center}
\includegraphics[width=0.8\textwidth]{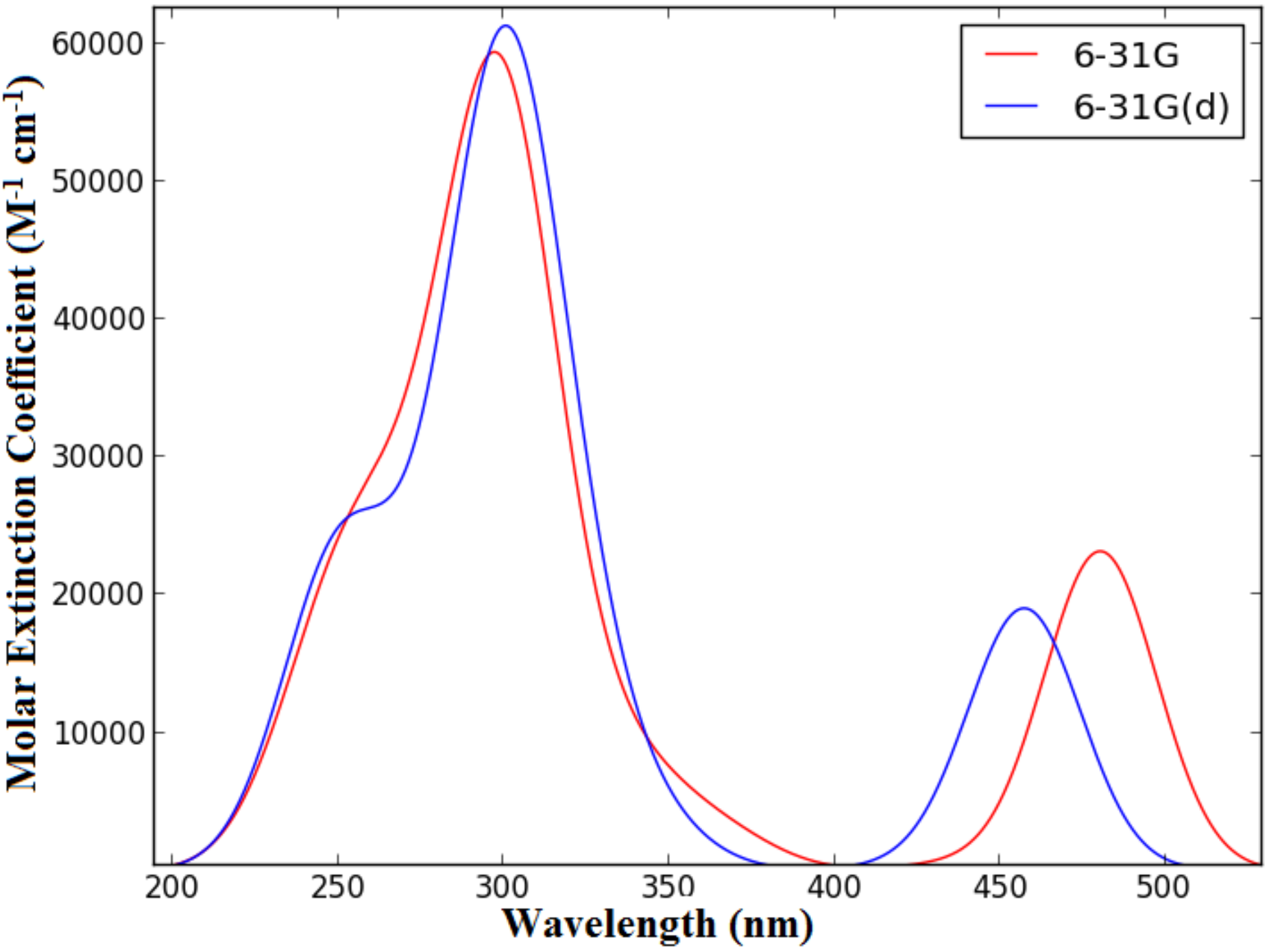}
\end{center}
[Ru(pynapy)$_3$]$^{2+}$
TD-B3LYP/6-31G and TD-B3LYP/6-31G(d) spectra.

% ================================================
\newpage
\section{Complex {\bf (98)}$^\dagger$: [Ru(DMCH)$_2$Cl$_2$]}
% ================================================

\begin{center}
\begin{tabular}{cc}
B3LYP/6-31G & B3LYP/6-31G(d) \\
$\epsilon_{\text{HOMO}} = \mbox{-4.36 eV}$ & 
$\epsilon_{\text{HOMO}} = \mbox{-4.32 eV}$ 
\end{tabular}
\end{center}
% \begin{center}
%    {\bf PDOS}
% \end{center}
% 
% \begin{center}
% \includegraphics[width=0.4\textwidth]{graphics1/framedquestionmark.pdf}
% \includegraphics[width=0.4\textwidth]{graphics1/framedquestionmark.pdf}
% \end{center}
% {\color{magenta} PDOS could not be calculated for complexes containing Cl.}

\begin{center}
   {\bf Absorption Spectrum}
\end{center}

\begin{center}
\includegraphics[width=0.8\textwidth]{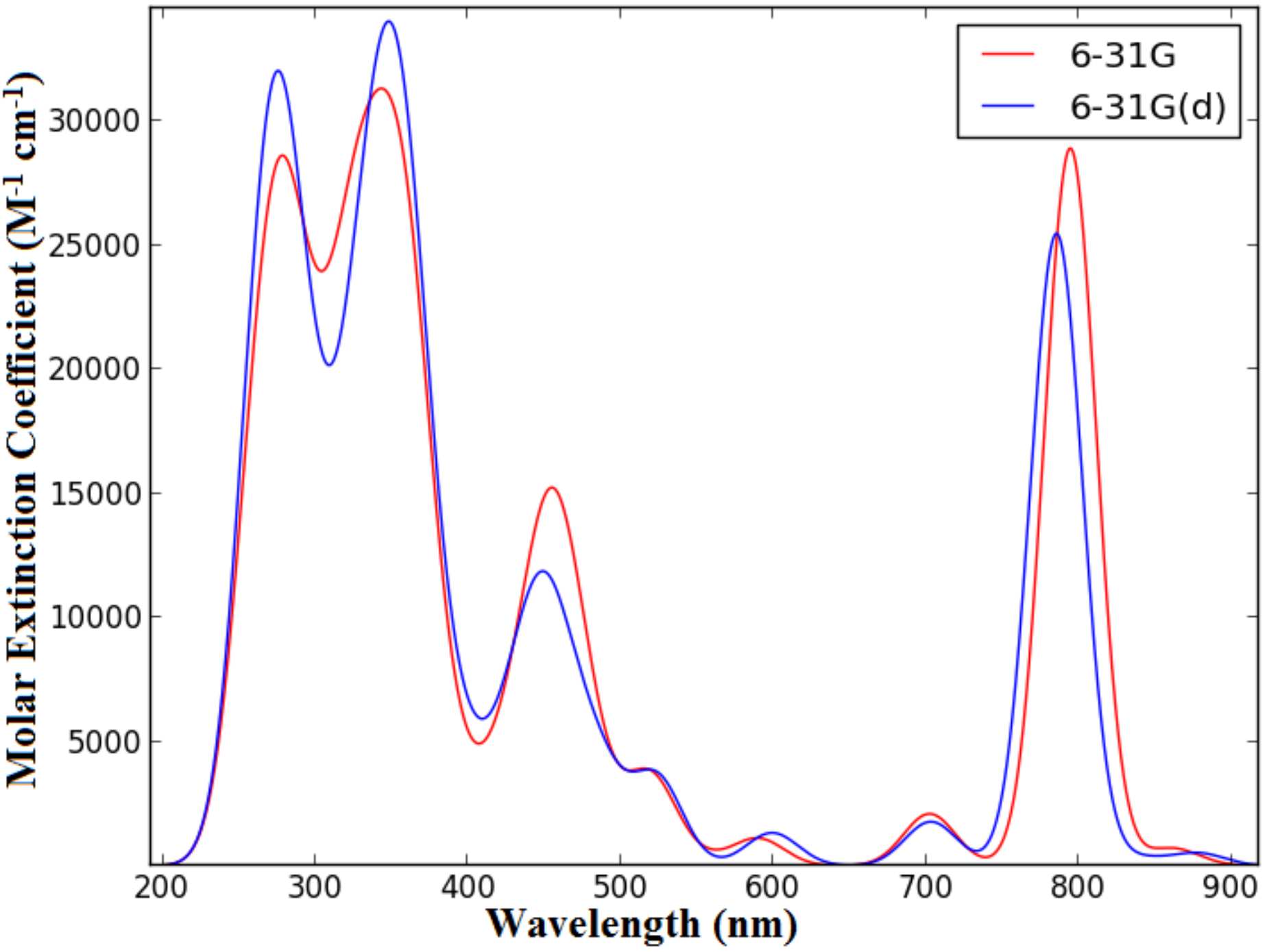}
\end{center}
[Ru(DMCH)$_2$Cl$_2$]
TD-B3LYP/6-31G and TD-B3LYP/6-31G(d) spectra.

% ================================================
\newpage
\section{Complex {\bf (99)}*: [Ru(DMCH)$_2$(CN)$_2$]}
% ================================================

\begin{center}
   {\bf PDOS}
\end{center}

\begin{center}
\begin{tabular}{cc}
\includegraphics[width=0.4\textwidth]{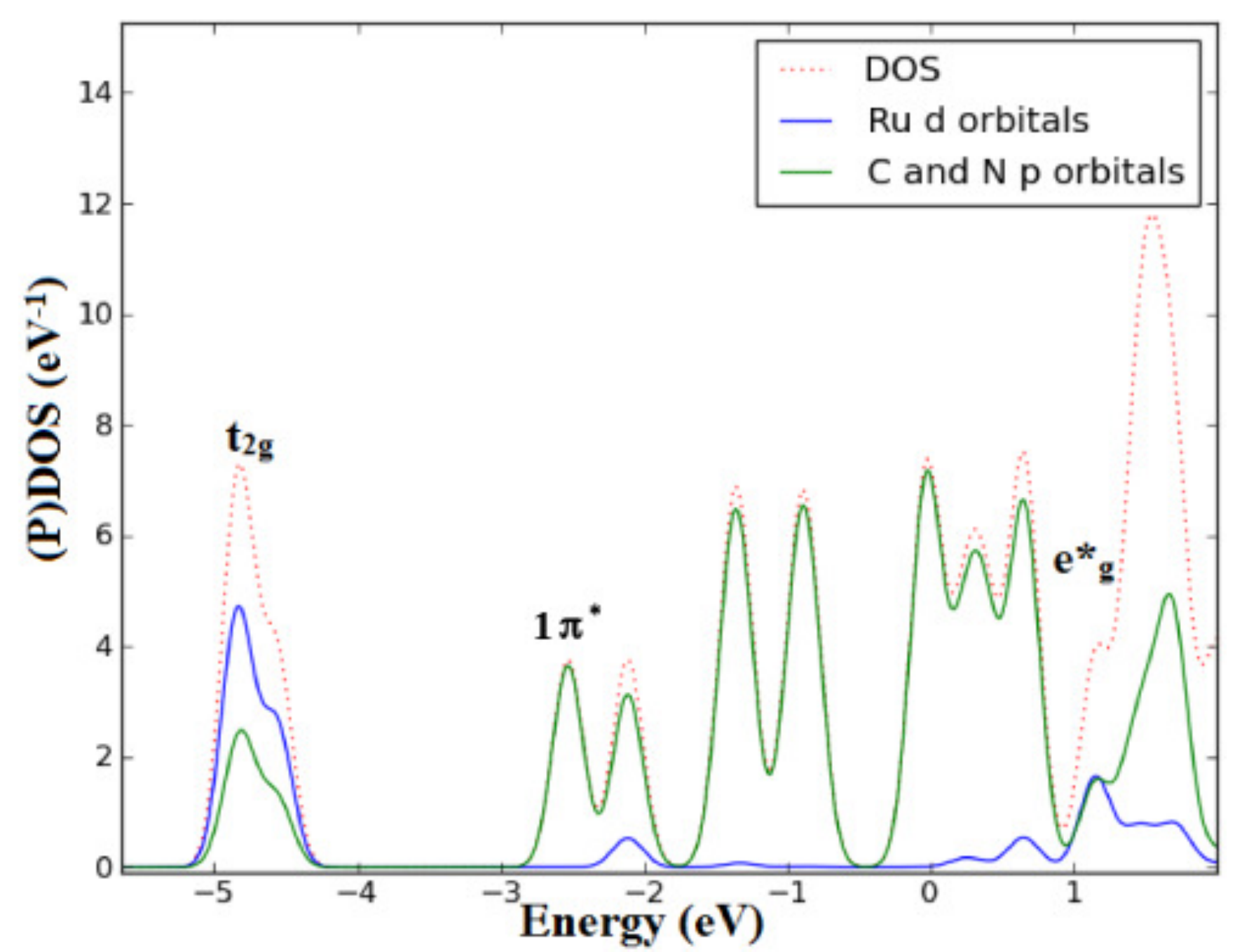} &
\includegraphics[width=0.4\textwidth]{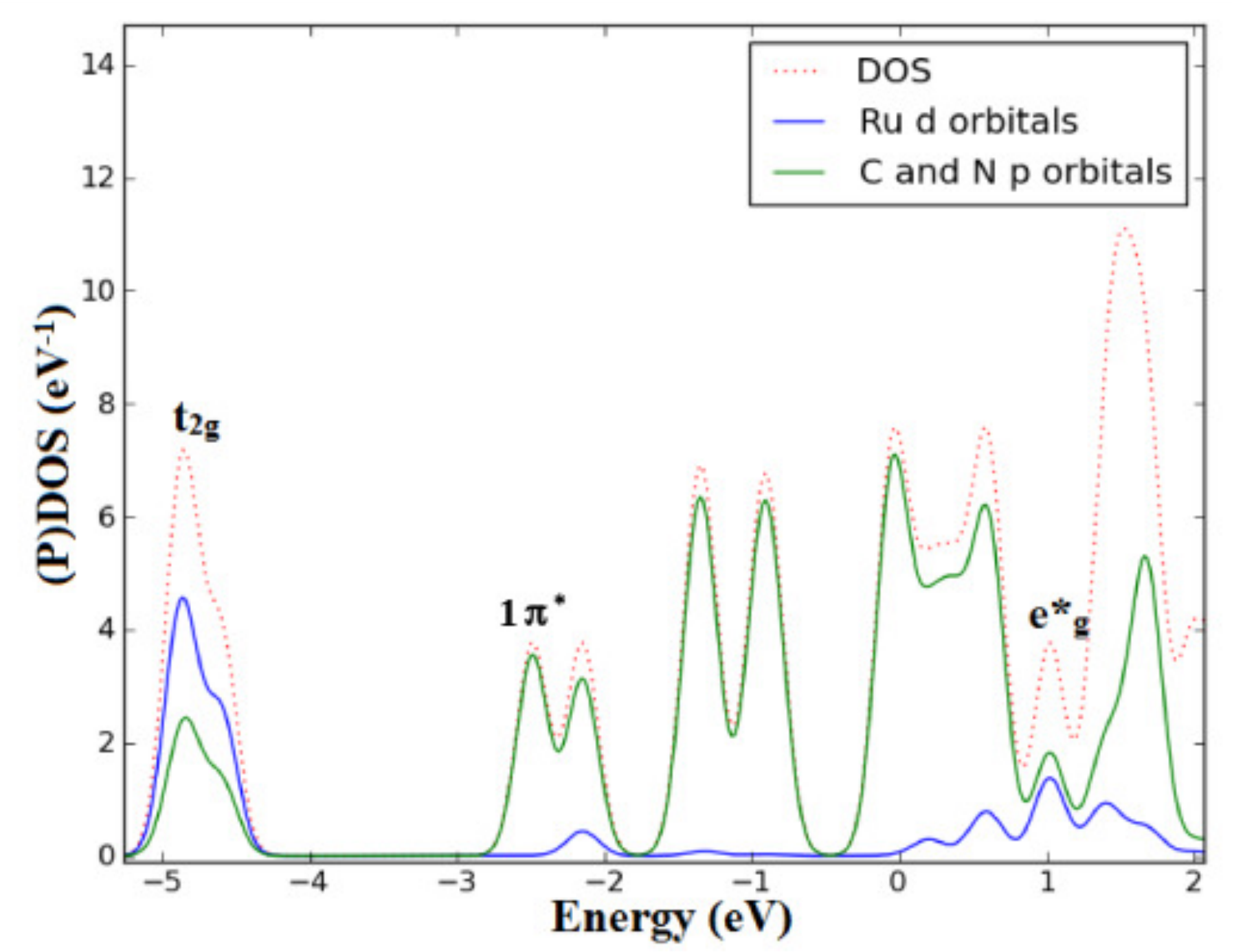} \\
B3LYP/6-31G & B3LYP/6-31G(d) \\
$\epsilon_{\text{HOMO}} = \mbox{-4.56 eV}$ & 
$\epsilon_{\text{HOMO}} = \mbox{-4.60 eV}$ 
\end{tabular}
\end{center}
Total and partial density of states of [Ru(DMCH)$_2$(CN)$_2]$ partitioned 
over Ru d orbitals and ligand C and N p orbitals.
% for the 6-31G (left-hand side) and 6-31G(d) (right-hand side) basis sets.

\begin{center}
   {\bf Absorption Spectrum}
\end{center}

\begin{center}
\includegraphics[width=0.8\textwidth]{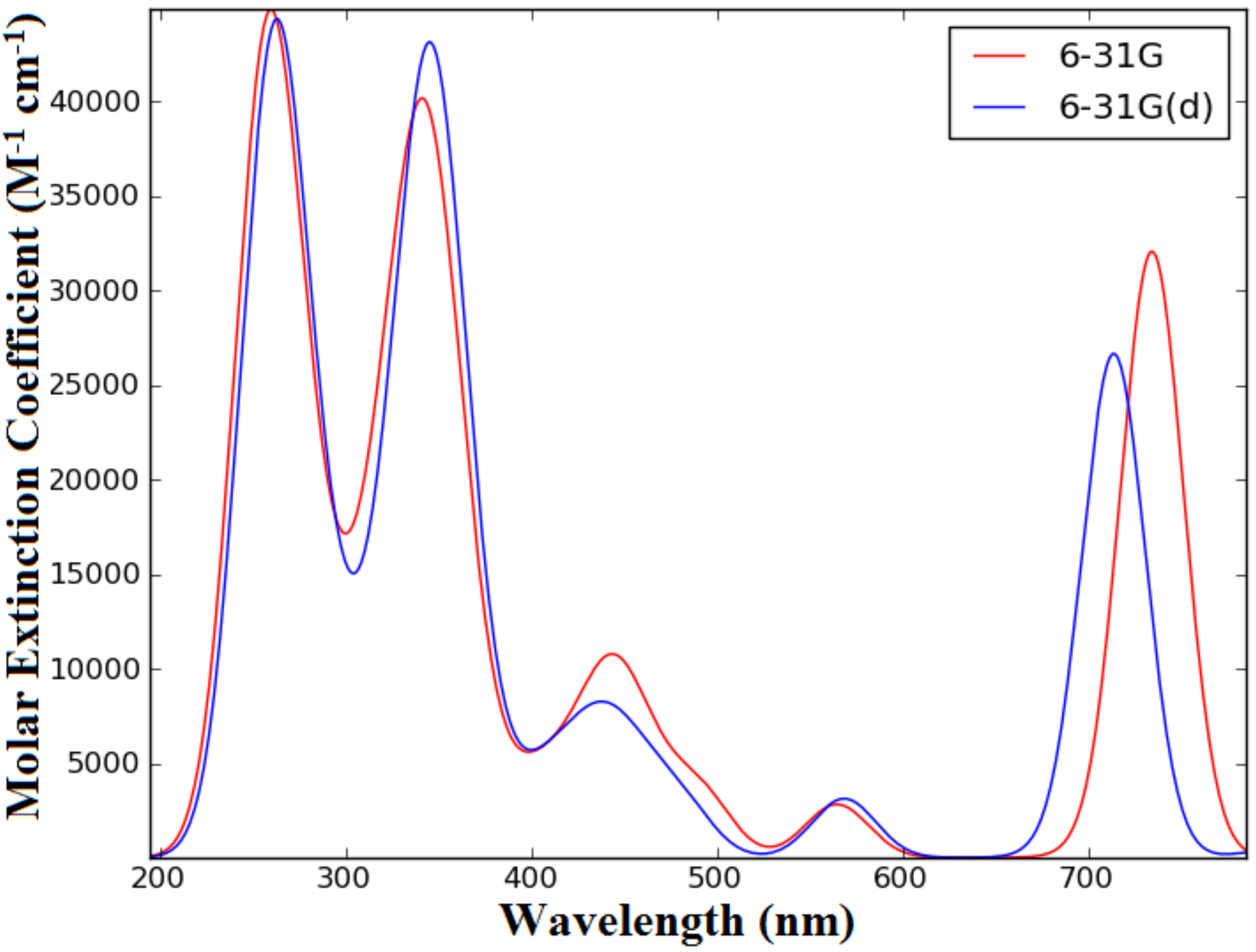}
\end{center}
[Ru(DMCH)$_{2}$(CN)$_{2}$]
TD-B3LYP/6-31G and TD-B3LYP/6-31G(d) spectra.

% ================================================
\newpage
\section{\, Complex {\bf (100)}: [Ru(DMCH)$_3$]$^{2+}$}
% ================================================

\begin{center}
   {\bf PDOS}
\end{center}

\begin{center}
\includegraphics[width=0.4\textwidth]{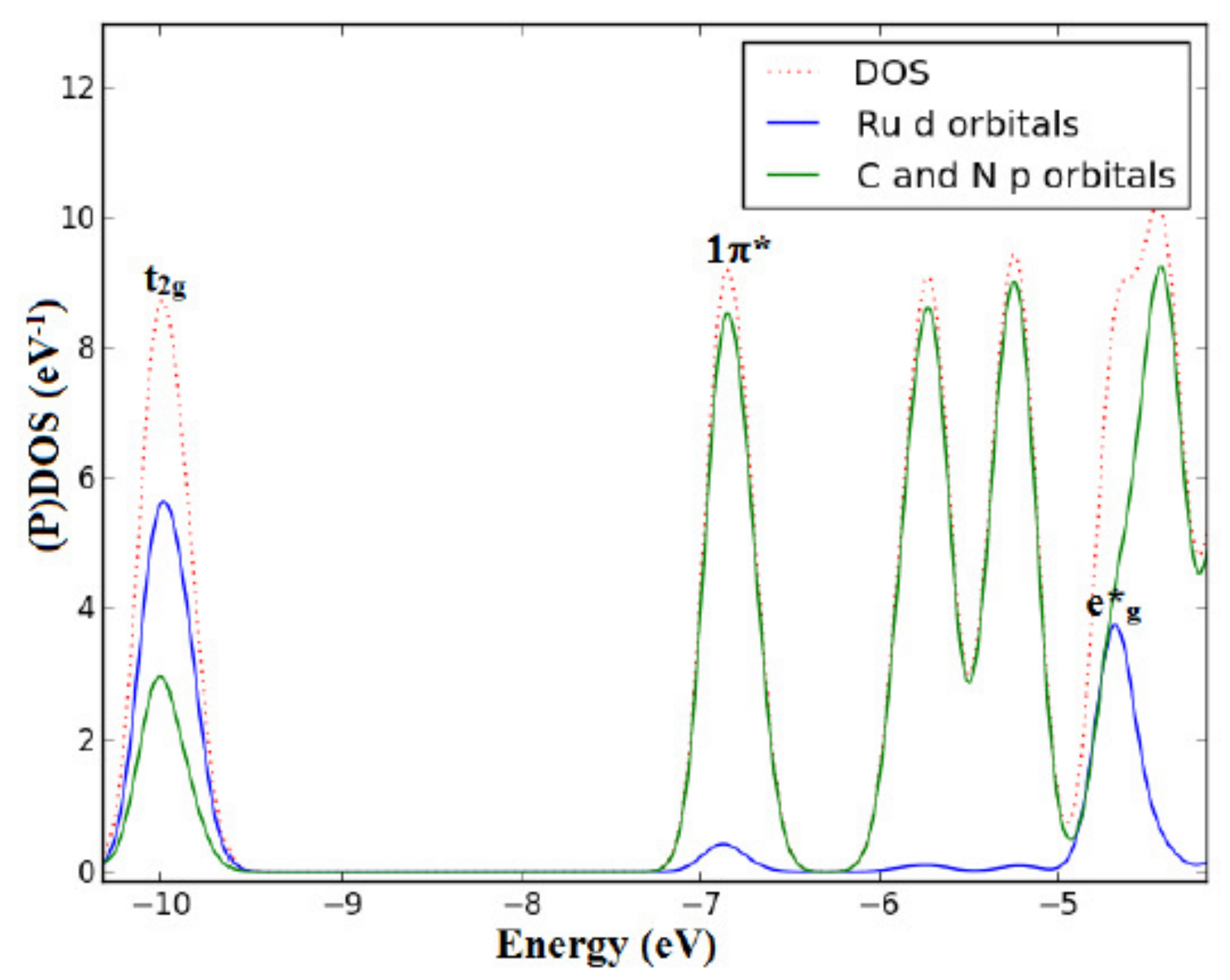}
\\ 6-31G \\ $\epsilon_{\text{HOMO}} = \mbox{-9.87 eV}$
\end{center}
Total and partial density of states of [Ru(DMCH)$_3$]$^{2+}$
partitioned over Ru d orbitals and ligand C and N p orbitals. 
% for the 6-31G (left-hand side) basis set. 

% and 6-31G(d) (right-hand side {\color{red} \sf Do we have this?}) basis sets.

\begin{center}
   {\bf Absorption Spectrum}
\end{center}

\begin{center}
\includegraphics[width=0.8\textwidth]{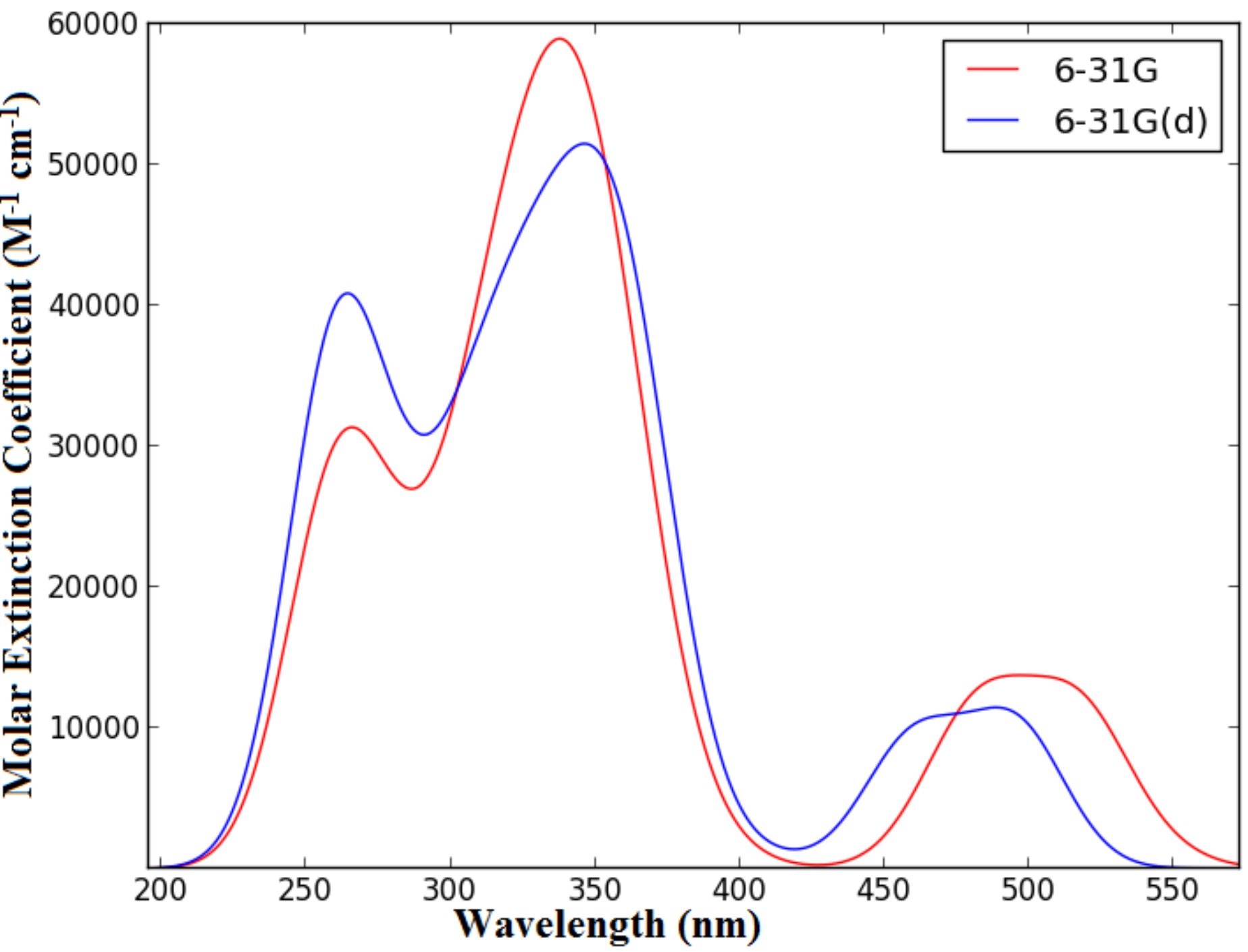}
\end{center}
[Ru(DMCH)$_3$]$^{2+}$
TD-B3LYP/6-31G and TD-B3LYP/6-31G(d) spectra.

% ================================================
\newpage
\section{\, Complex {\bf (101)}: [Ru(dinapy)$_3$]$^{2+}$}
% ================================================

\begin{center}
   {\bf PDOS}
\end{center}

\begin{center}
\includegraphics[width=0.4\textwidth]{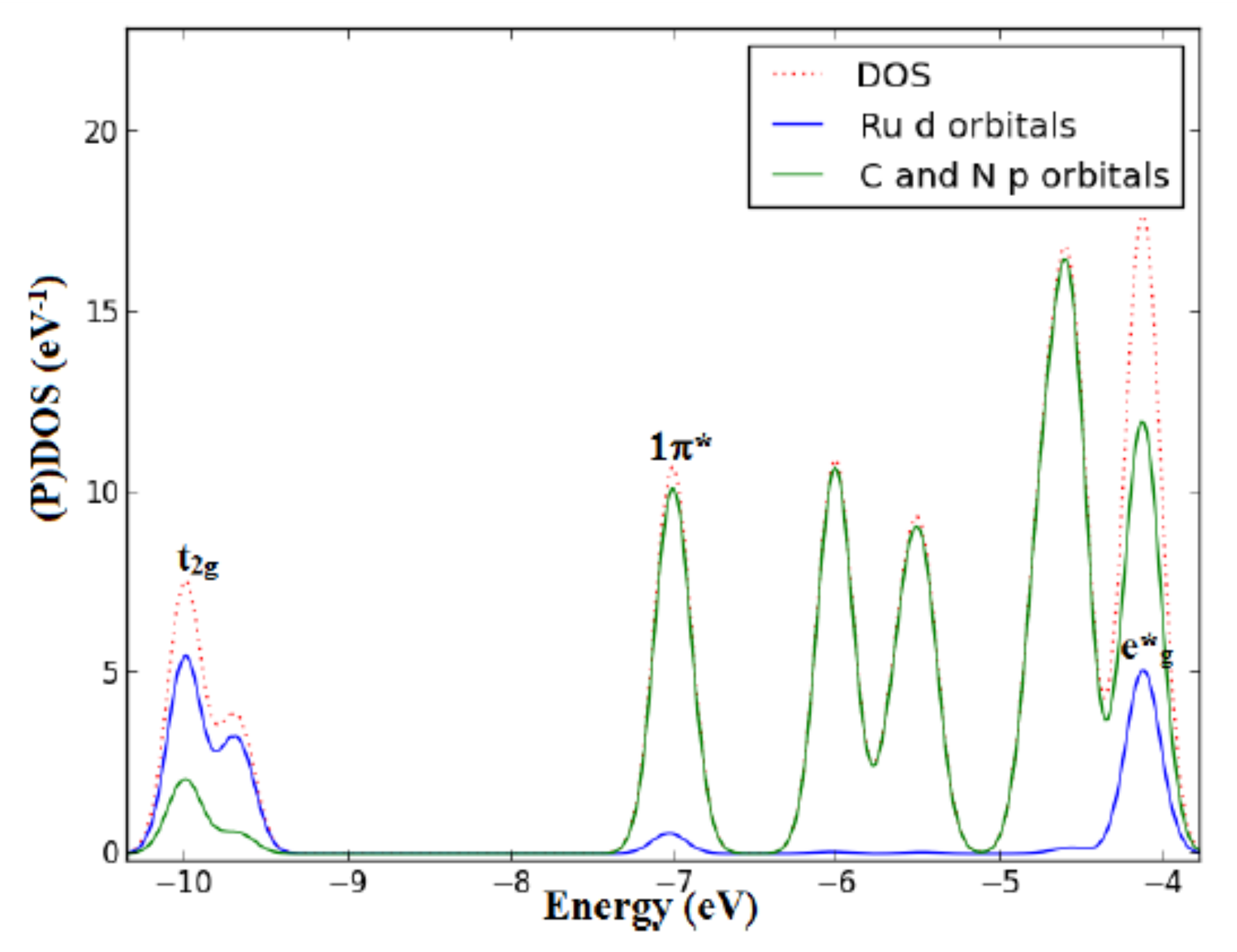}
\\ 6-31G \\ $\epsilon_{\text{HOMO}} = \mbox{-9.69 eV}$
\end{center}
Total and partial density of states of [Ru(dinapy)$_3$]$^{2+}$
partitioned over Ru d orbitals and ligand C and N p orbitals.
% for the 6-31G (left-hand side) basis set. 
% and 6-31G(d) (right-hand side {\color{red} \sf Do we have this?}) basis sets.

\begin{center}
   {\bf Absorption Spectrum}
\end{center}

\begin{center}
\includegraphics[width=0.8\textwidth]{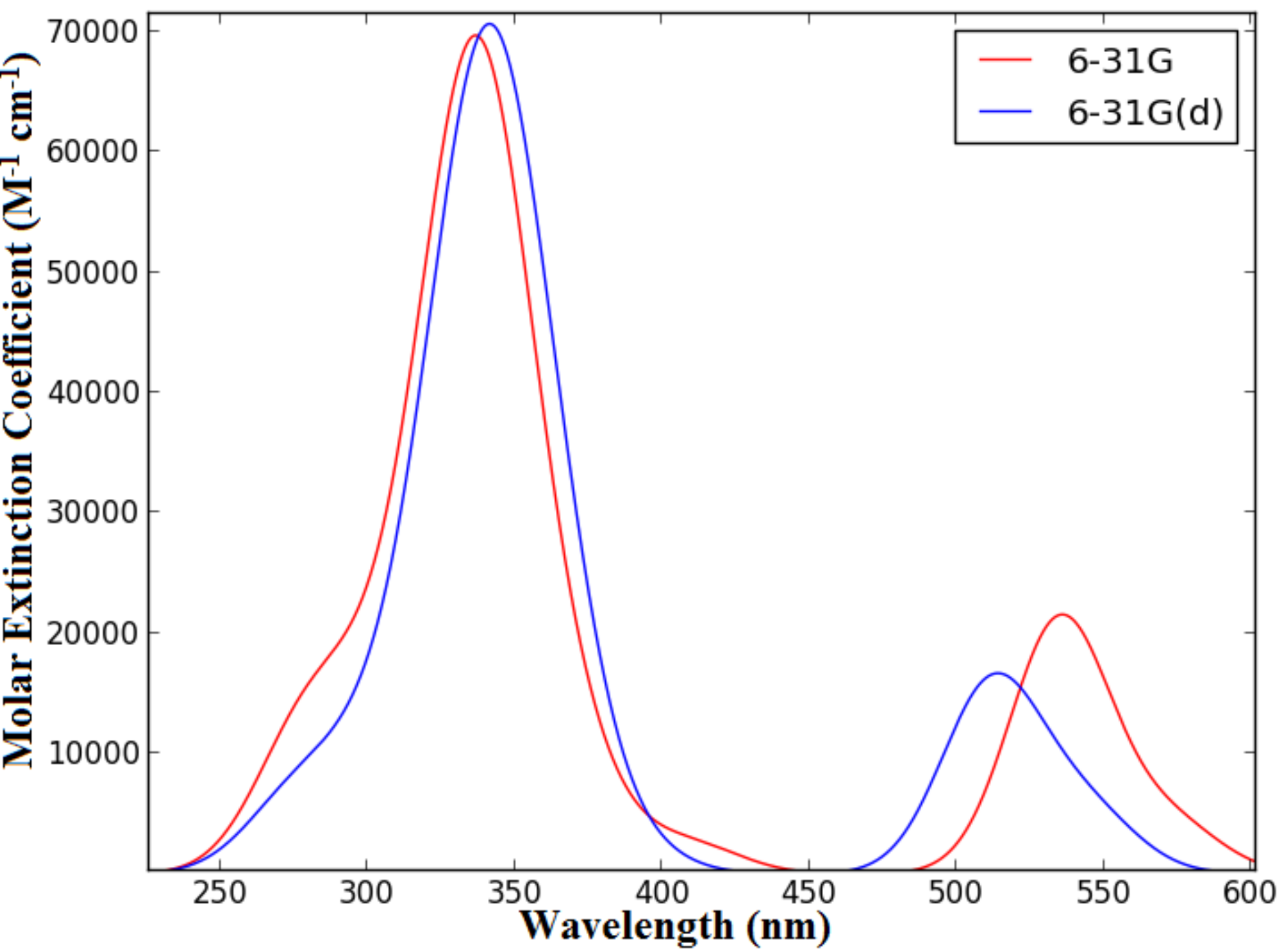}
\end{center}
[Ru(dinapy)$_3$]$^{2+}$
TD-B3LYP/6-31G and TD-B3LYP/6-31G(d) spectra.

% ================================================
\newpage
\section{\, Complex {\bf (102)}$^\dagger$: [Ru(biq)$_2$Cl$_2$]}
% ================================================

% \begin{center}
%    {\bf PDOS}
% \end{center}
% 
% \begin{center}
% \includegraphics[width=0.4\textwidth]{graphics1/framedquestionmark.pdf}
% \includegraphics[width=0.4\textwidth]{graphics1/framedquestionmark.pdf}
% \end{center}
% {\color{magenta} PDOS could not be calculated for complexes containing Cl.}

\begin{center}
\begin{tabular}{cc}
B3LYP/6-31G & B3LYP/6-31G(d) \\
$\epsilon_{\text{HOMO}} = \mbox{-4.80 eV}$ & 
$\epsilon_{\text{HOMO}} = \mbox{-4.74 eV}$ 
\end{tabular}
\end{center}

\begin{center}
   {\bf Absorption Spectrum}
\end{center}

\begin{center}
\includegraphics[width=0.8\textwidth]{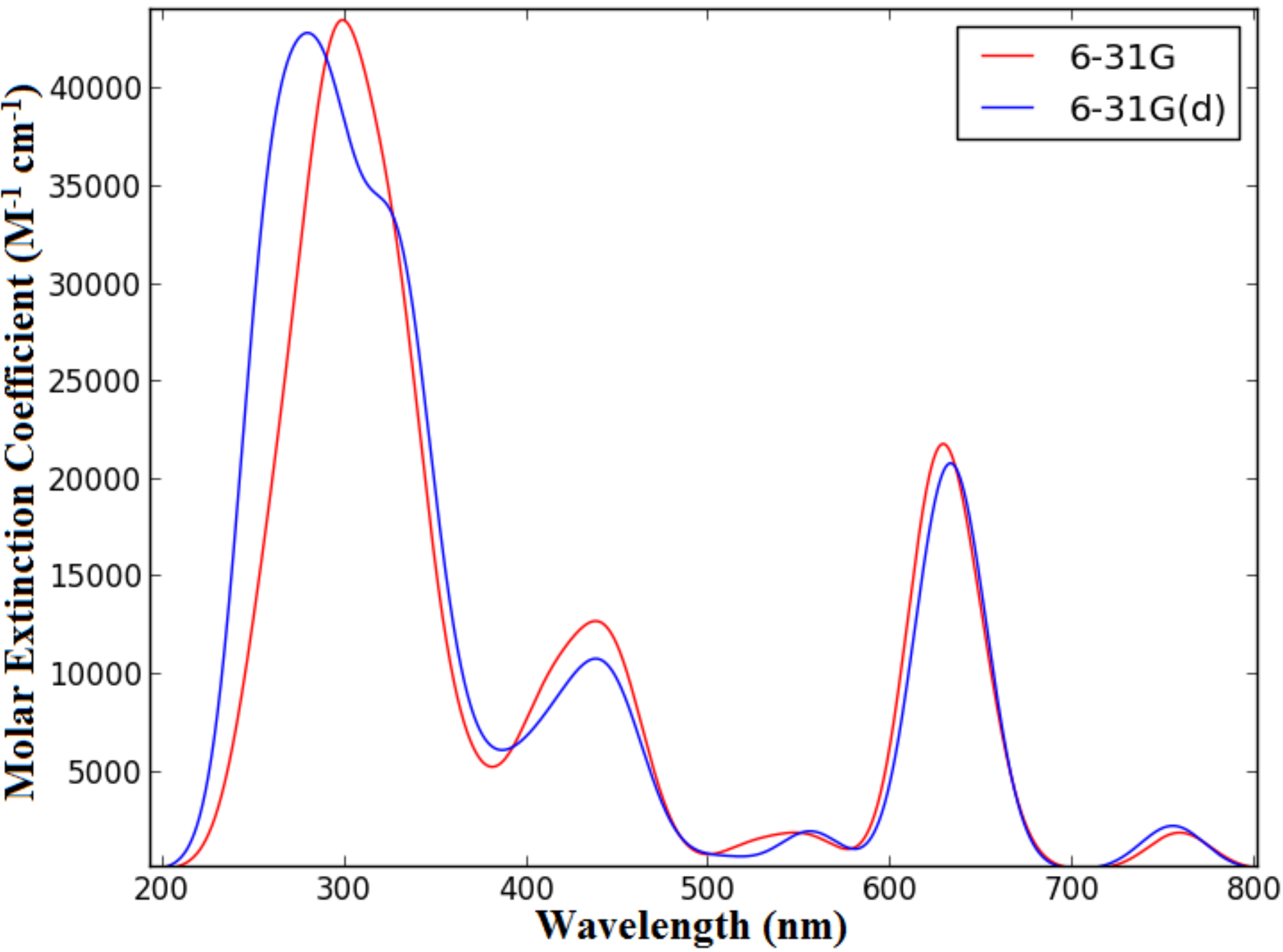}
\end{center}
[Ru(biq)$_2$Cl$_2$]
TD-B3LYP/6-31G and TD-B3LYP/6-31G(d) spectra.

% ================================================
\newpage
\section{\, Complex {\bf (103)}*: [Ru(biq)$_2$(CN)$_2$]}
% ================================================

\begin{center}
   {\bf PDOS}
\end{center}

\begin{center}
\begin{tabular}{cc}
\includegraphics[width=0.4\textwidth]{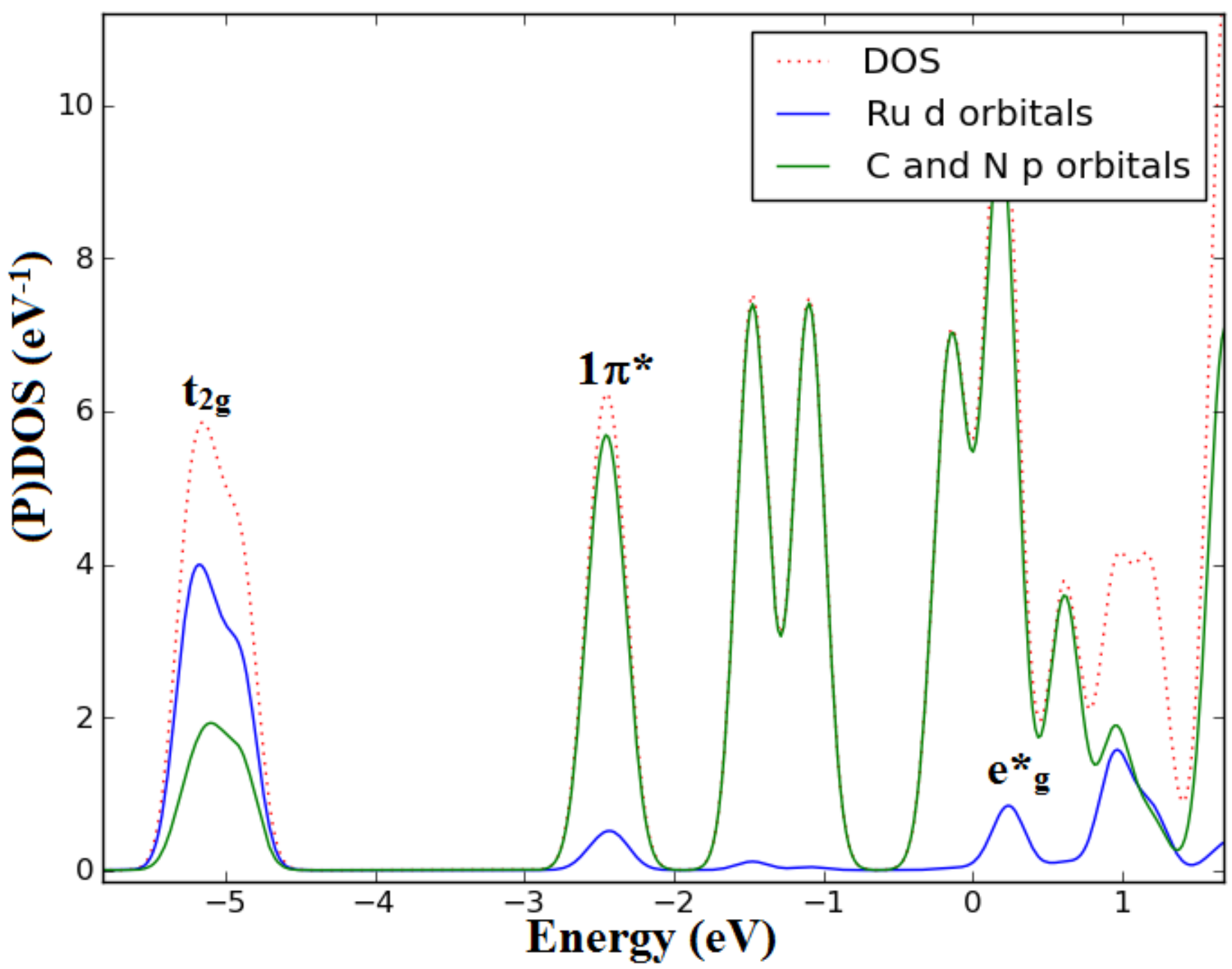} &
\includegraphics[width=0.4\textwidth]{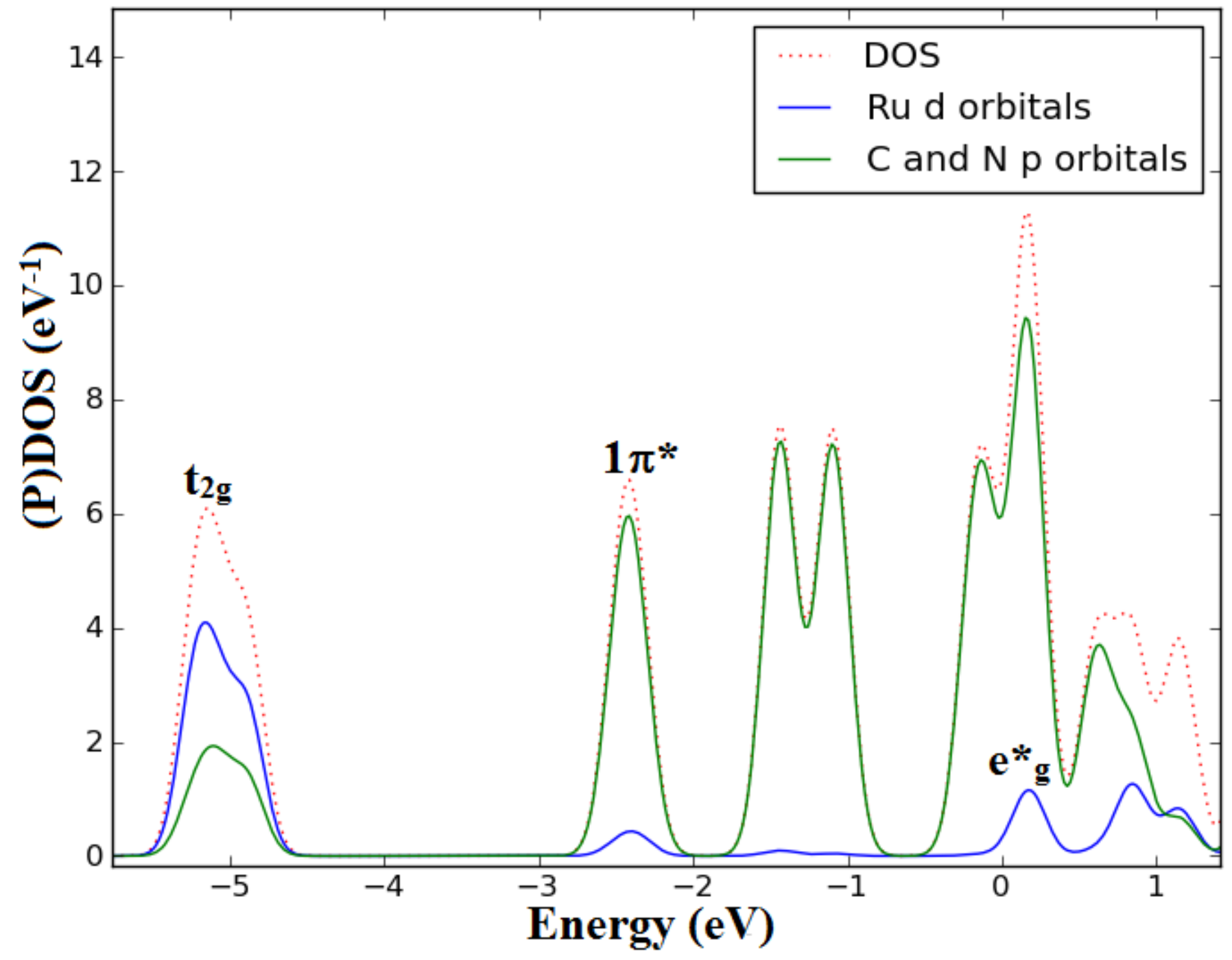} \\
B3LYP/6-31G & B3LYP/6-31G(d) \\
$\epsilon_{\text{HOMO}} = \mbox{-4.89 eV}$ & 
$\epsilon_{\text{HOMO}} = \mbox{-4.88 eV}$ 
\end{tabular}
\end{center}
Total and partial density of states of [Ru(biq)$_{2}$(CN)$_{2}$] partitioned 
over Ru d orbitals and ligand C and N p orbitals.
% for the 6-31G (left-hand side) and 6-31G(d) (right-hand side) basis sets.

\begin{center}
   {\bf Absorption Spectrum}
\end{center}

\begin{center}
\includegraphics[width=0.8\textwidth]{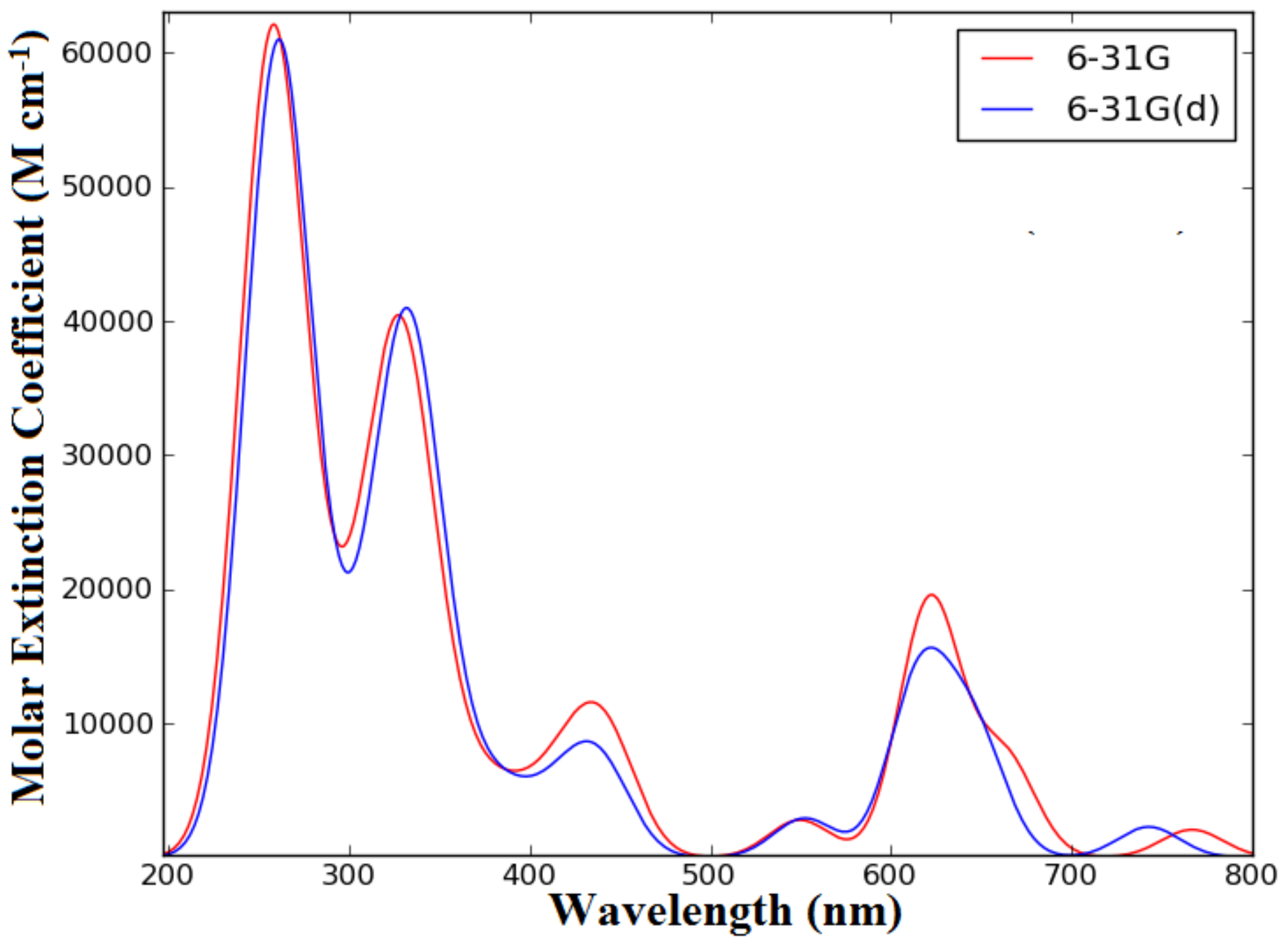}
\end{center}
[Ru(biq)$_{2}$(CN)$_{2}$]
TD-B3LYP/6-31G and TD-B3LYP/6-31G(d) spectra.

% ================================================
\newpage
\section{\, Complex {\bf (104)}: [Ru(biq)$_3$]$^{2+}$}
% ================================================

\begin{center}
   {\bf PDOS}
\end{center}

\begin{center}
\begin{tabular}{cc}
\includegraphics[width=0.4\textwidth]{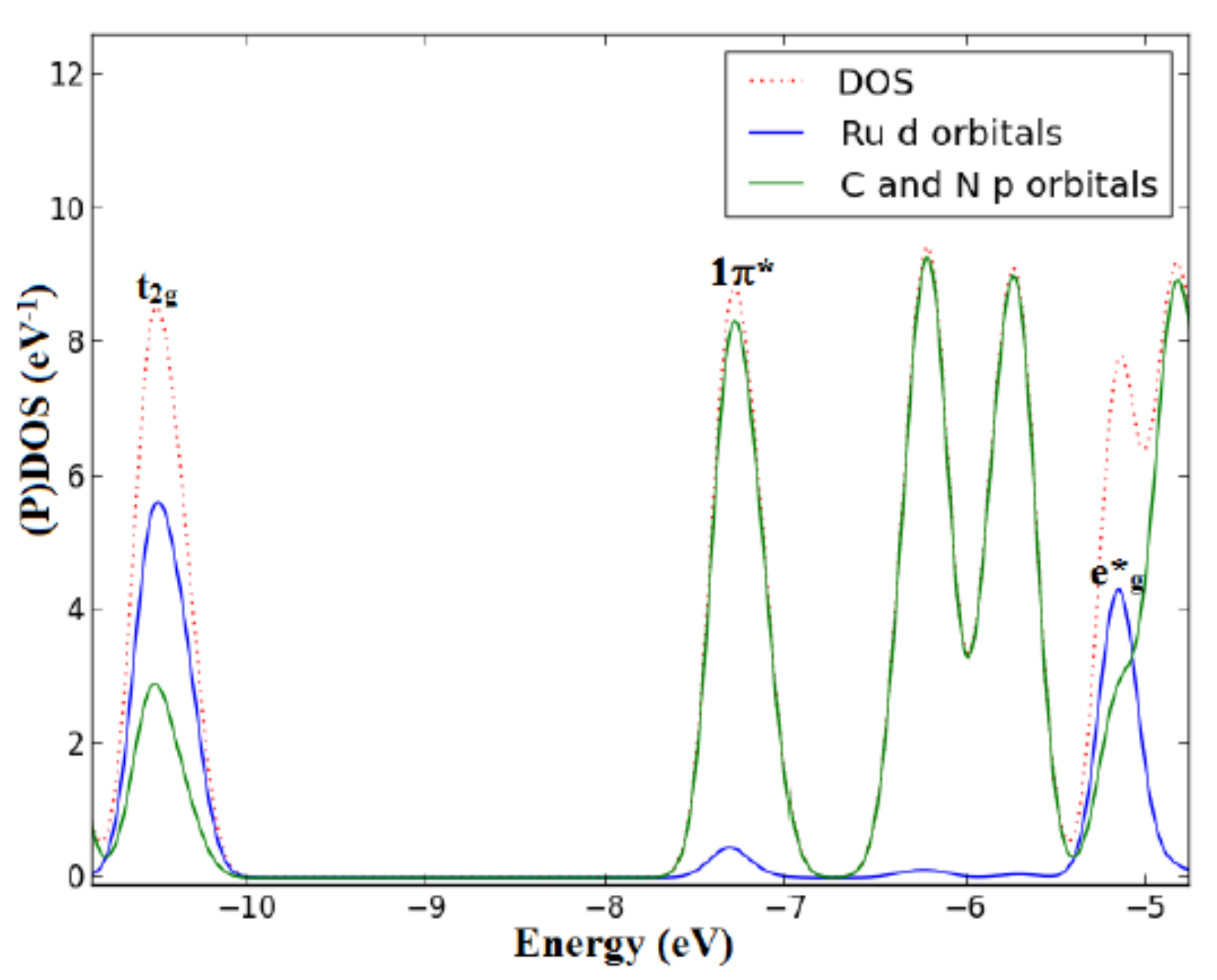} &
\includegraphics[width=0.4\textwidth]{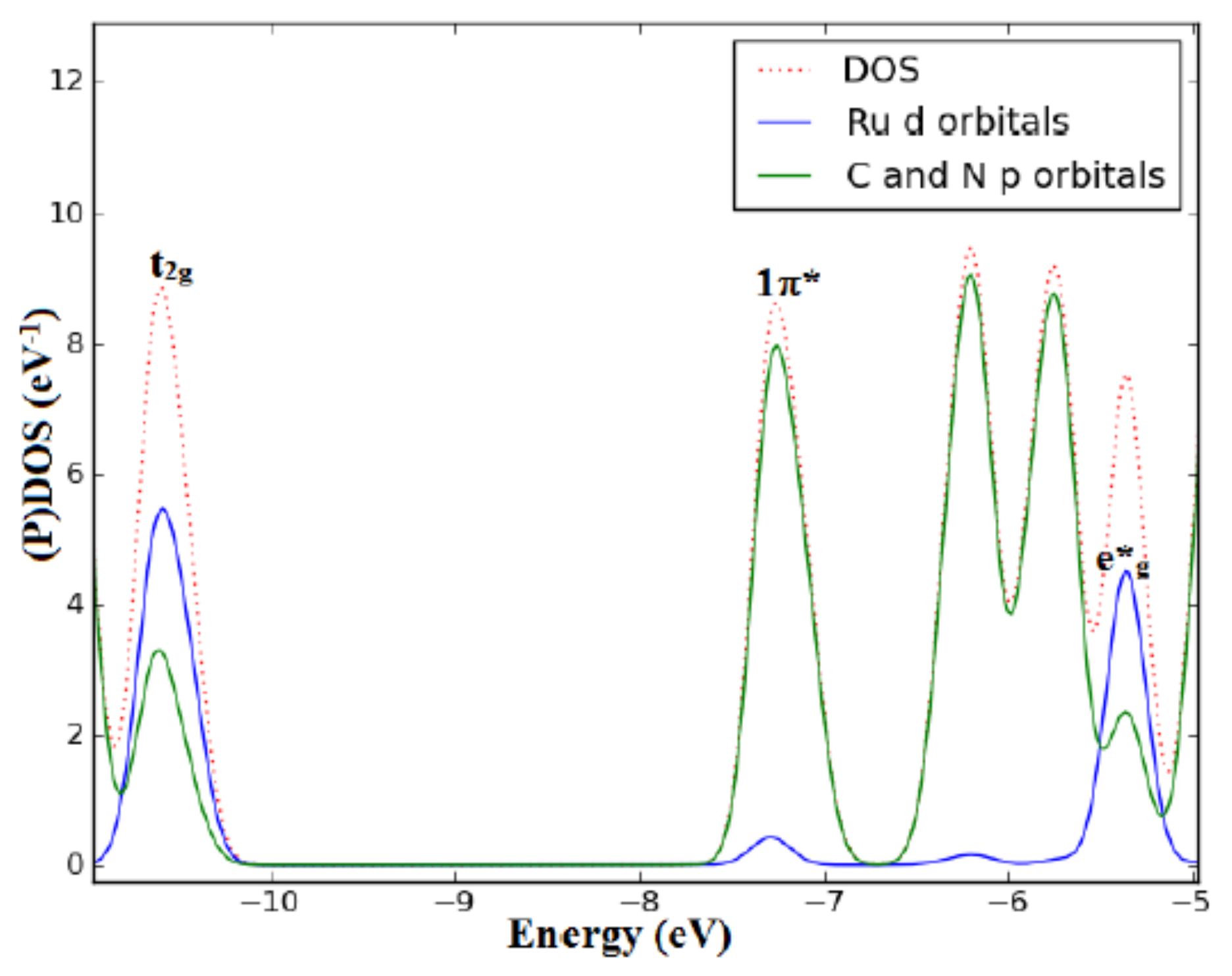} \\
B3LYP/6-31G & B3LYP/6-31G(d) \\
$\epsilon_{\text{HOMO}} = \mbox{-10.36 eV}$ & 
$\epsilon_{\text{HOMO}} = \mbox{-10.46 eV}$ 
\end{tabular}
\end{center}
Total and partial density of states of [Ru(biq)$_3$]$^{2+}$ partitioned 
over Ru d orbitals and ligand C and N p orbitals.
% for the 6-31G (left-hand side) and 6-31G(d) (right-hand side) basis sets.

\begin{center}
   {\bf Absorption Spectrum}
\end{center}

\begin{center}
\includegraphics[width=0.8\textwidth]{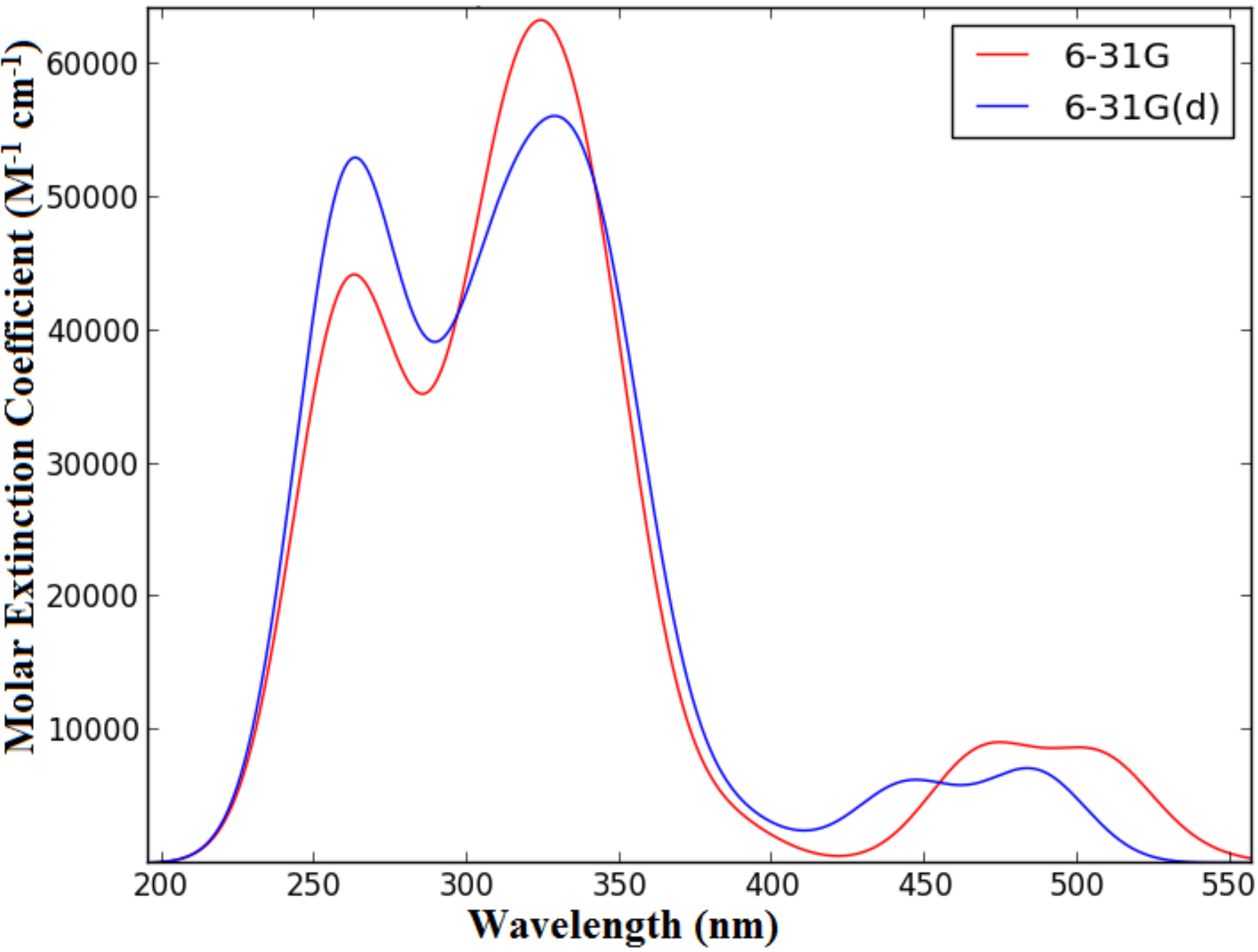}
\end{center}
[Ru(biq)$_3$]$^{2+}$
TD-B3LYP/6-31G and TD-B3LYP/6-31G(d) spectra.

% ================================================
\newpage
\section{\, Complex {\bf (105)}$^\dagger$: [Ru(i-biq)$_2$Cl$_2$]}
% ================================================

\begin{center}
\begin{tabular}{cc}
B3LYP/6-31G & B3LYP/6-31G(d) \\
$\epsilon_{\text{HOMO}} = \mbox{-4.30 eV}$ & 
$\epsilon_{\text{HOMO}} = \mbox{-4.27 eV}$ 
\end{tabular}
\end{center}

\begin{center}
   {\bf Absorption Spectrum}
\end{center}

\begin{center}
\includegraphics[width=0.8\textwidth]{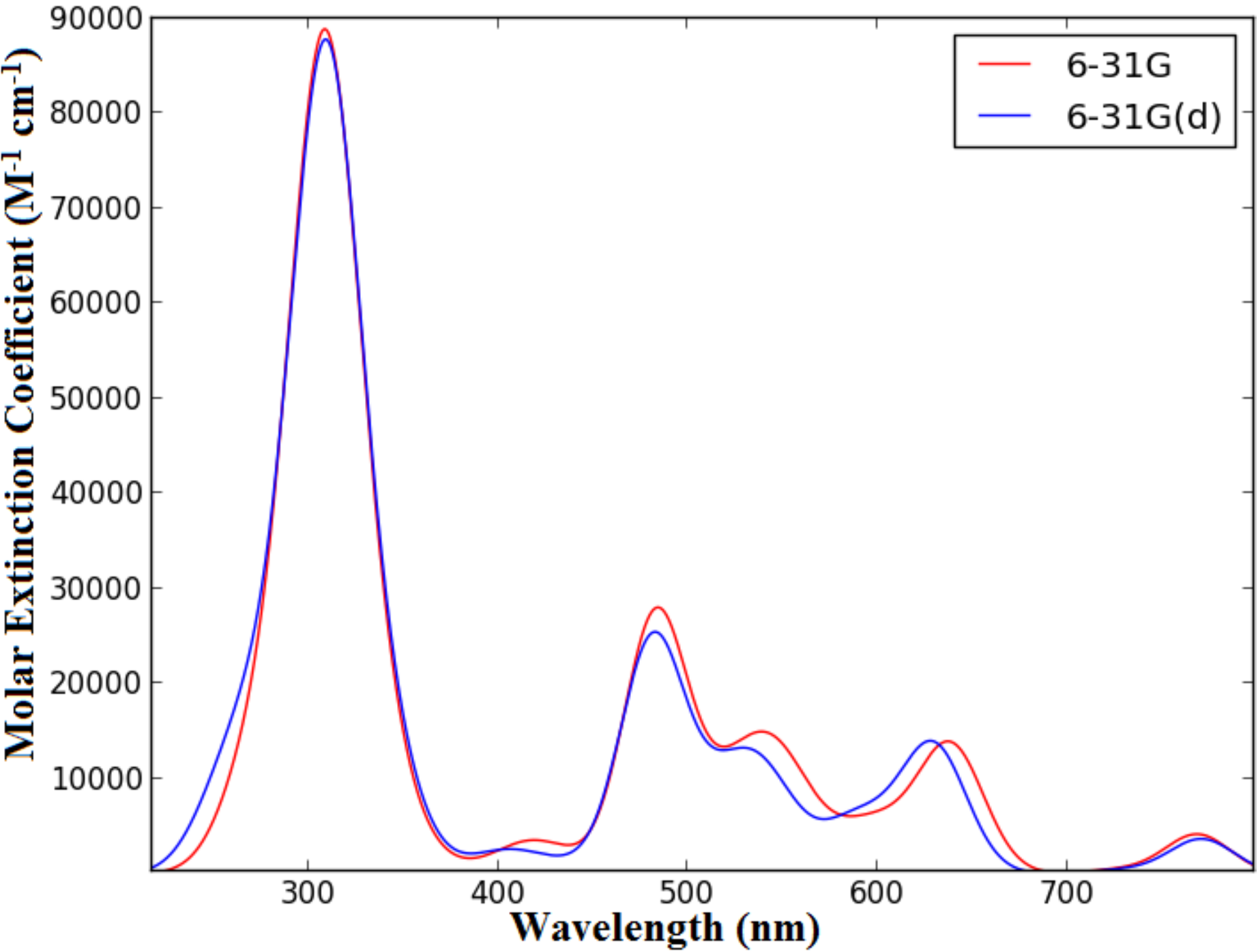}
\end{center}
[Ru(i-biq)$_2$Cl$_2$]
TD-B3LYP/6-31G and TD-B3LYP/6-31G(d) spectra.

% ================================================
\newpage
\section{\, Complex {\bf (106)}*: [Ru(i-biq)$_2$(CN)$_2$]}
% ================================================

\begin{center}
   {\bf PDOS}
\end{center}

\begin{center}
\begin{tabular}{cc}
\includegraphics[width=0.4\textwidth]{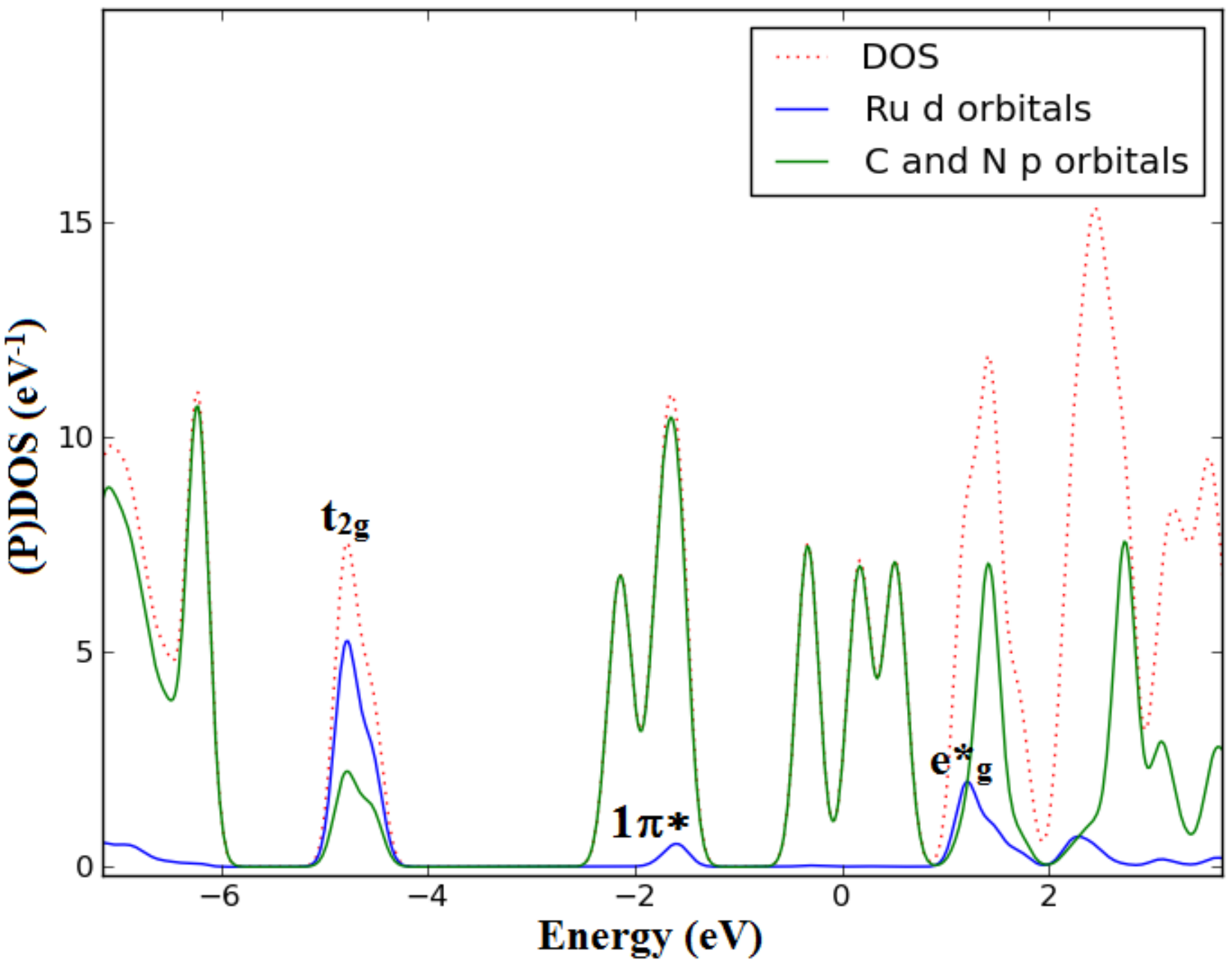} &
\includegraphics[width=0.4\textwidth]{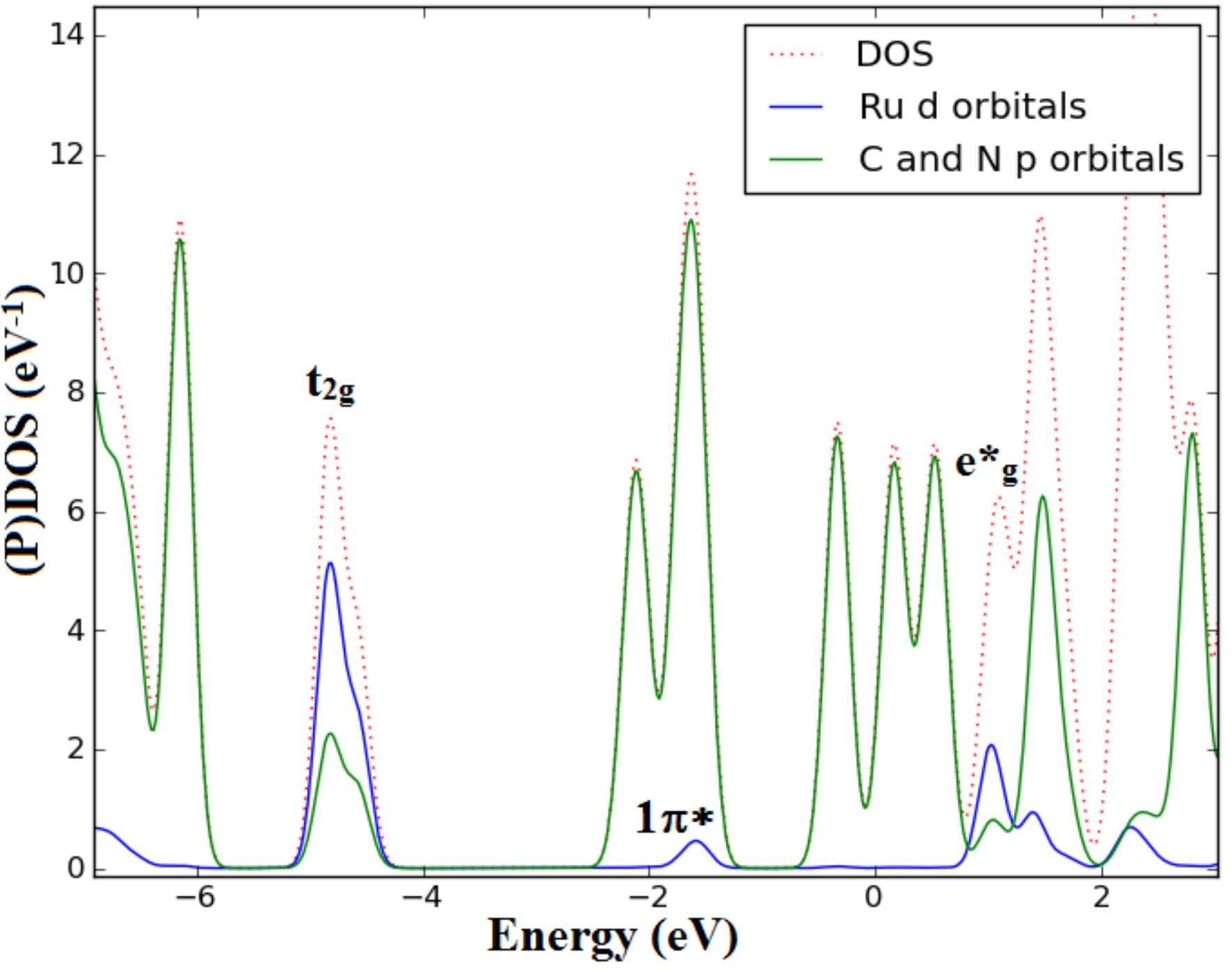} \\
B3LYP/6-31G & B3LYP/6-31G(d) \\
$\epsilon_{\text{HOMO}} = \mbox{-4.54 eV}$ & 
$\epsilon_{\text{HOMO}} = \mbox{-4.57 eV}$ 
\end{tabular}
\end{center}
Total and partial density of states of [Ru(i-biq)$_{2}$(CN)$_{2}$] partitioned 
over Ru d orbitals and ligand C and N p orbitals.
% for the 6-31G (left-hand side) and 6-31G(d) (right-hand side) basis sets.

\begin{center}
   {\bf Absorption Spectrum}
\end{center}

\begin{center}
\includegraphics[width=0.8\textwidth]{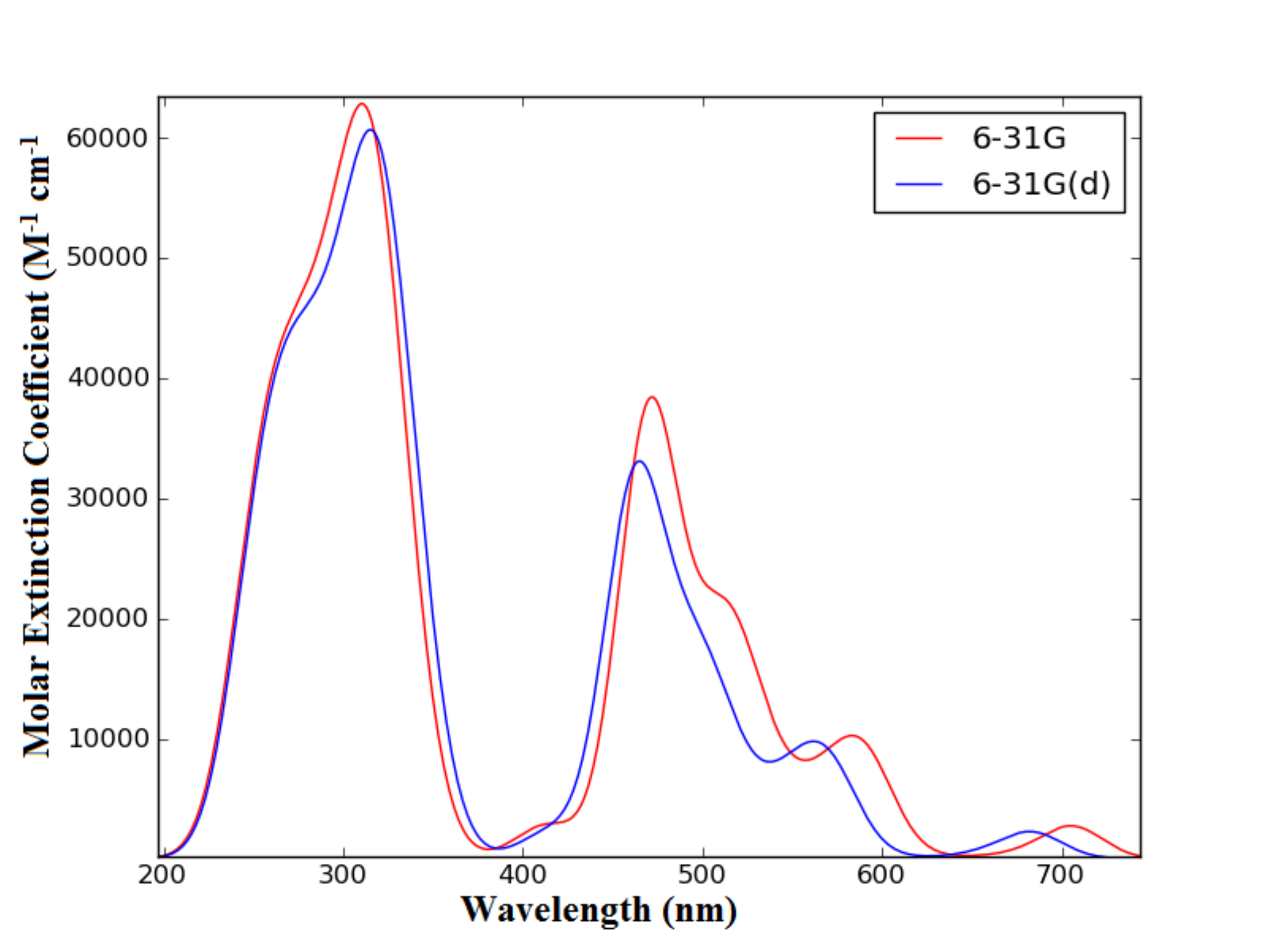}
\end{center}
[Ru(i-biq)$_{2}$(CN)$_{2}$]
TD-B3LYP/6-31G and TD-B3LYP/6-31G(d) spectra.

% ================================================
\newpage
\section{\, Complex {\bf (107)}: [Ru(i-biq)$_3$]$^{2+}$}
% ================================================

\begin{center}
   {\bf PDOS}
\end{center}

\begin{center}
\begin{tabular}{cc}
\includegraphics[width=0.4\textwidth]{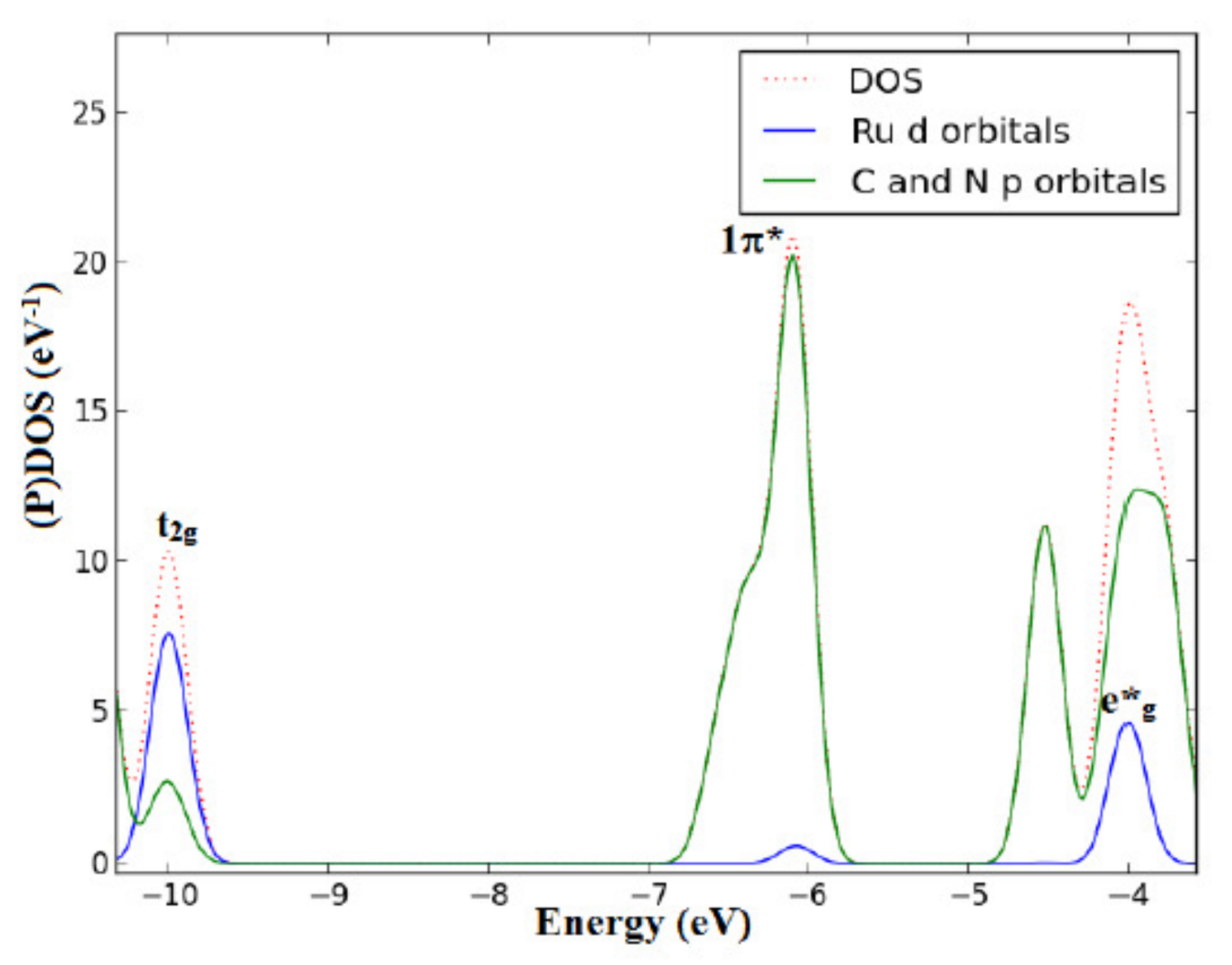} &
\includegraphics[width=0.4\textwidth]{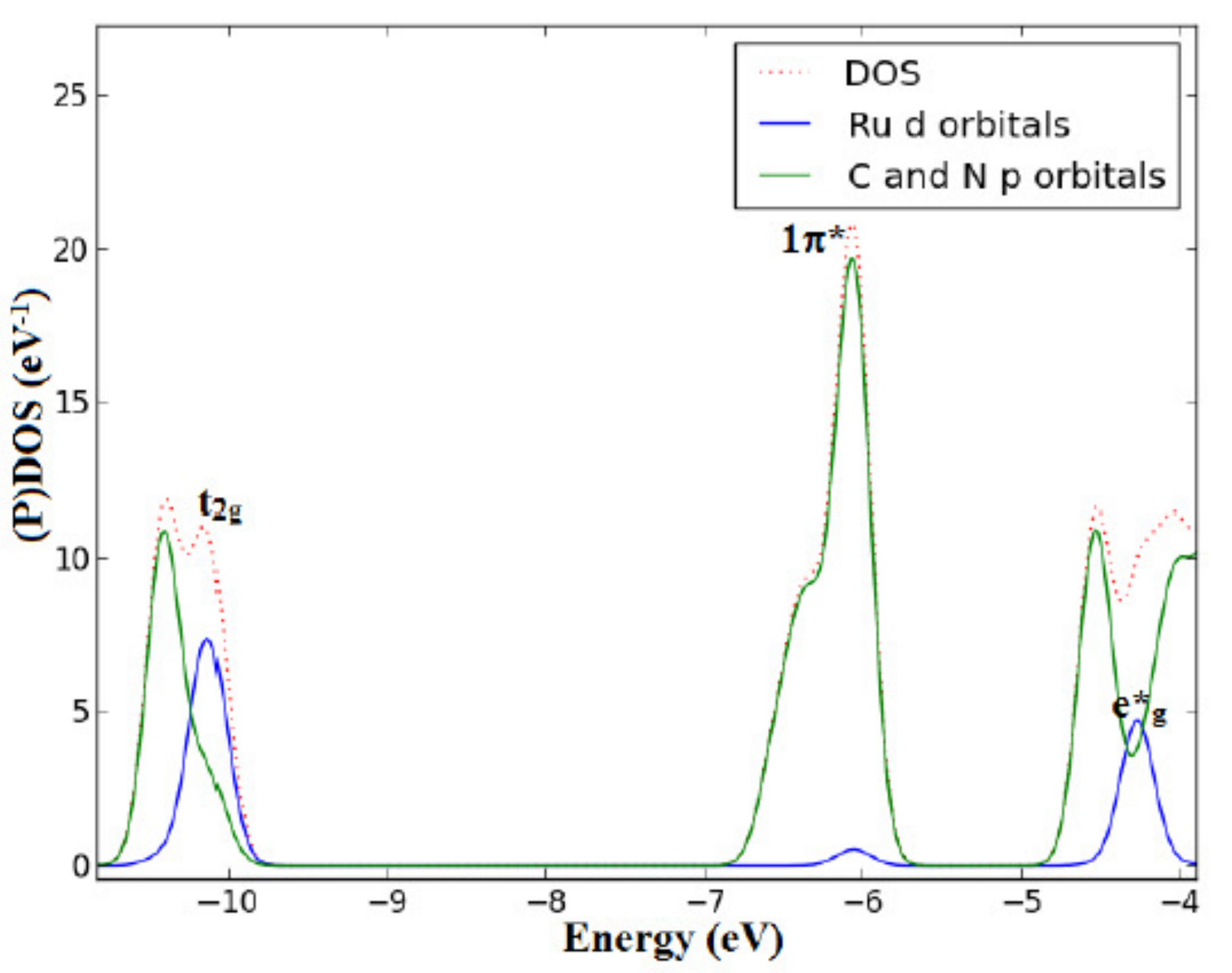} \\
B3LYP/6-31G & B3LYP/6-31G(d) \\
$\epsilon_{\text{HOMO}} = \mbox{-9.95 eV}$ & 
$\epsilon_{\text{HOMO}} = \mbox{-10.07 eV}$ 
\end{tabular}
\end{center}
Total and partial density of states of [Ru(i-biq)$_3$]$^{2+}$
partitioned 
over Ru d orbitals and ligand C and N p orbitals.
% for the 6-31G (left-hand side) and 6-31G(d) (right-hand side) basis sets.

\begin{center}
   {\bf Absorption Spectrum}
\end{center}

\begin{center}
\includegraphics[width=0.8\textwidth]{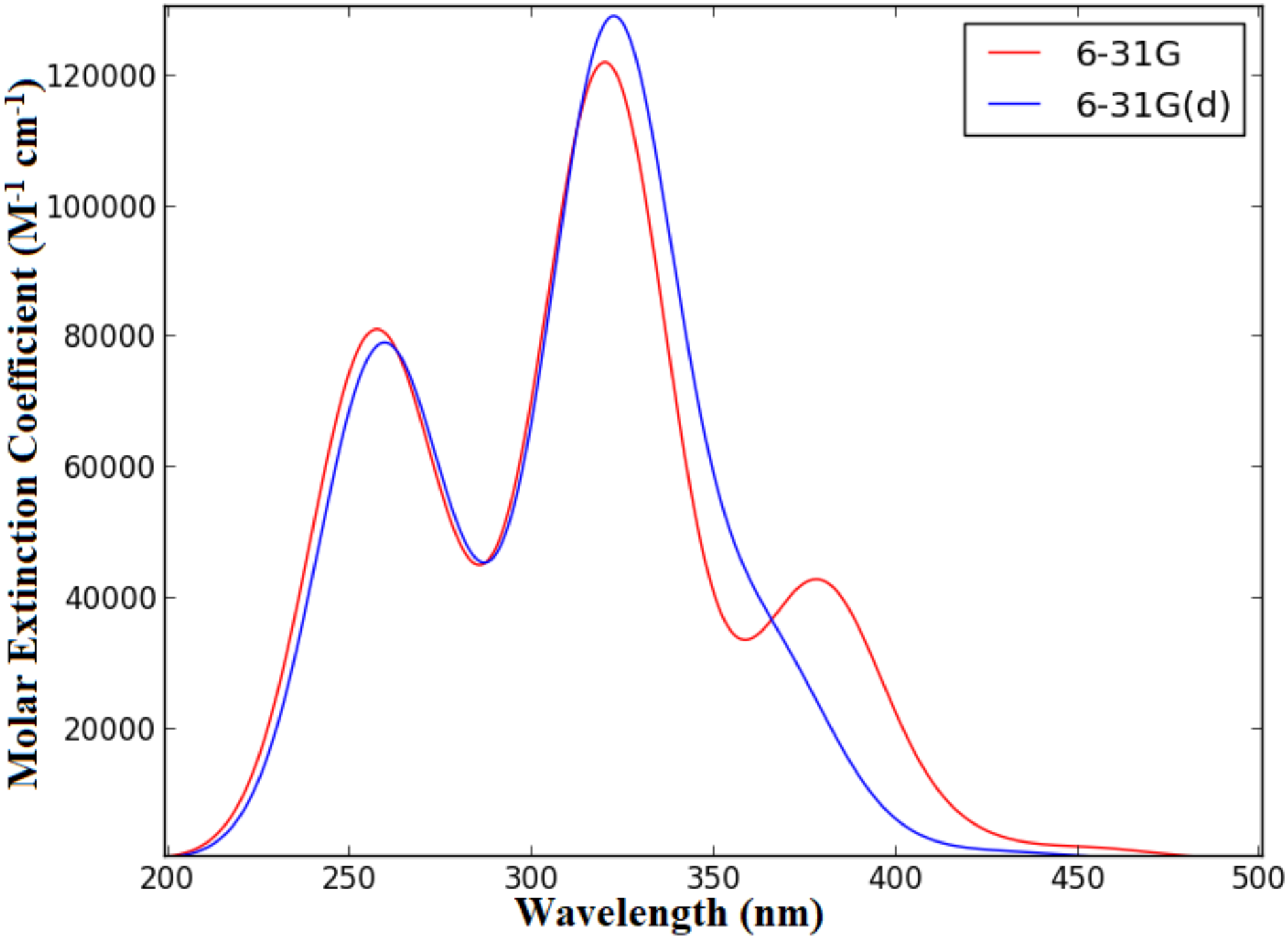}
\end{center}
[Ru(i-biq)$_3$]$^{2+}$
TD-B3LYP/6-31G and TD-B3LYP/6-31G(d) spectra.

% ================================================
\newpage
\section{\, Complex {\bf (108)}: [Ru(trpy)$_2$]$^{2+}$}
% ================================================

% {\color{magenta} \sf This compound was not in Denis' latest version 
% of his supplementary information.}

\begin{center}
   {\bf PDOS}
\end{center}

\begin{center}
\begin{tabular}{cc}
\includegraphics[width=0.4\textwidth]{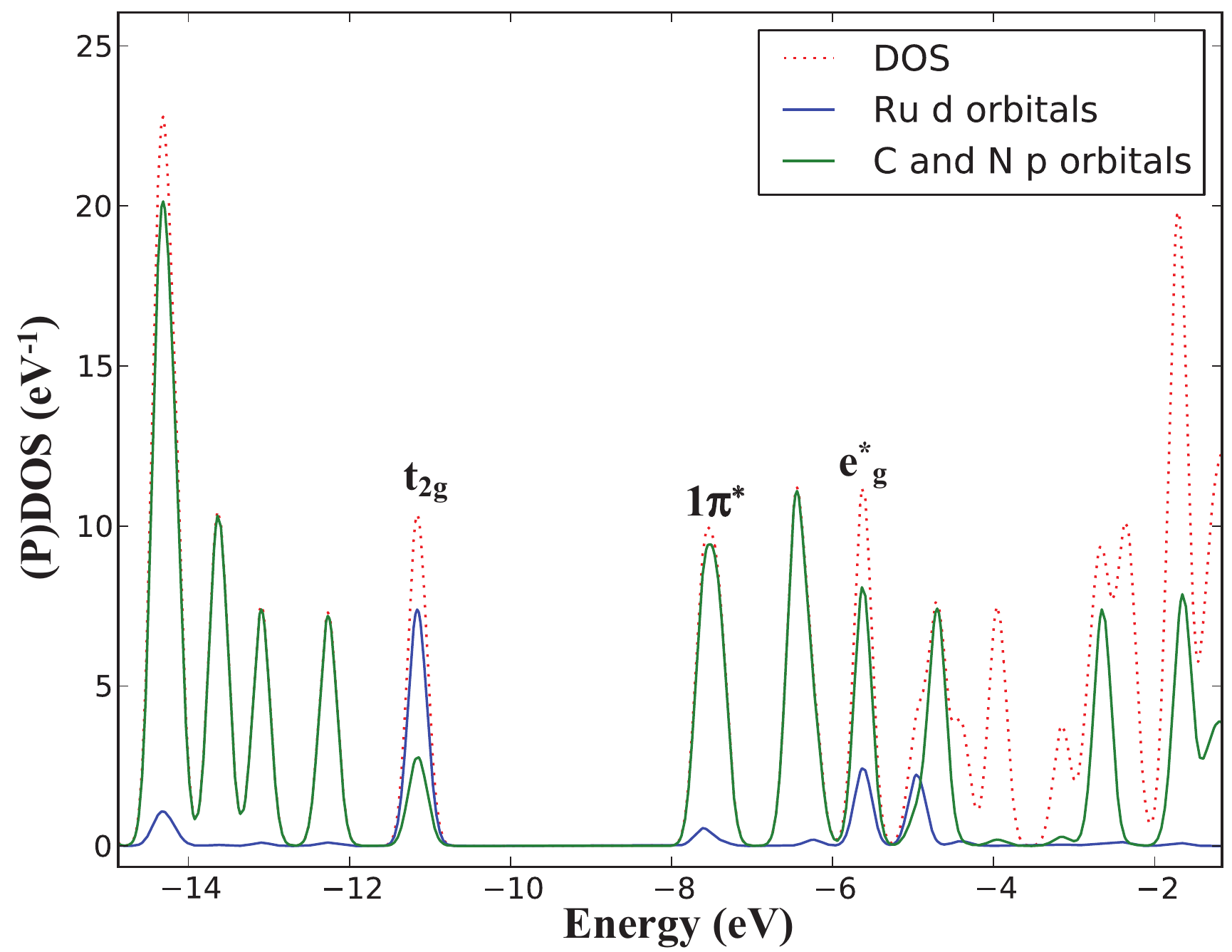} &
\includegraphics[width=0.4\textwidth]{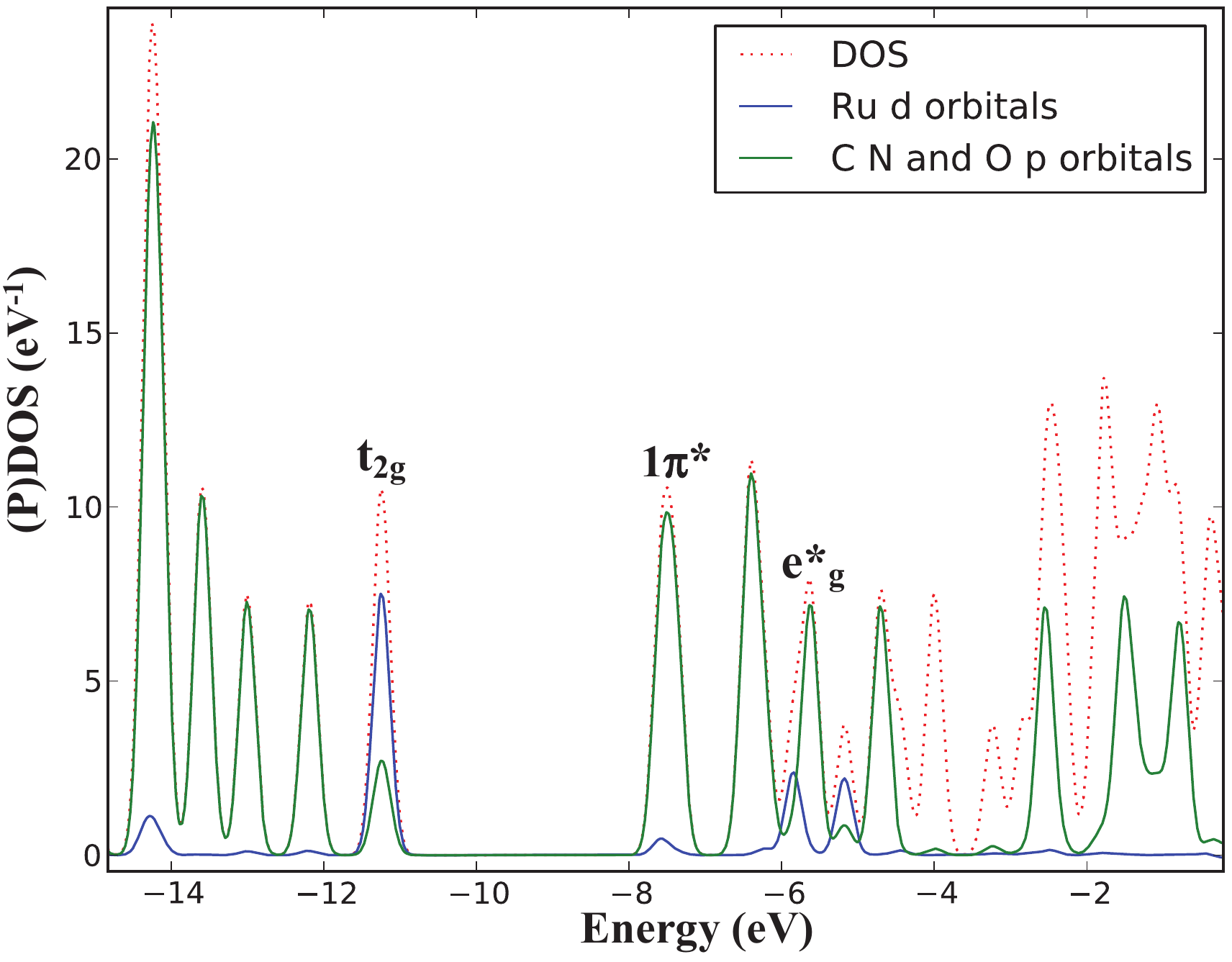} \\
B3LYP/6-31G & B3LYP/6-31G(d) \\
$\epsilon_{\text{HOMO}} = \mbox{-11.09 eV}$ & 
$\epsilon_{\text{HOMO}} = \mbox{-11.19 eV}$ 
\end{tabular}
\end{center}
Total and partial density of states of [Ru(trpy)$_2]^{2+}$ partitioned over 
Ru d orbitals and ligand C and N p orbitals.
% for the 6-31G (left-hand side) and 6-31G(d) (right-hand side) basis sets.

\begin{center}
   {\bf Absorption Spectrum}
\end{center}

\begin{center}
\includegraphics[width=0.8\textwidth]{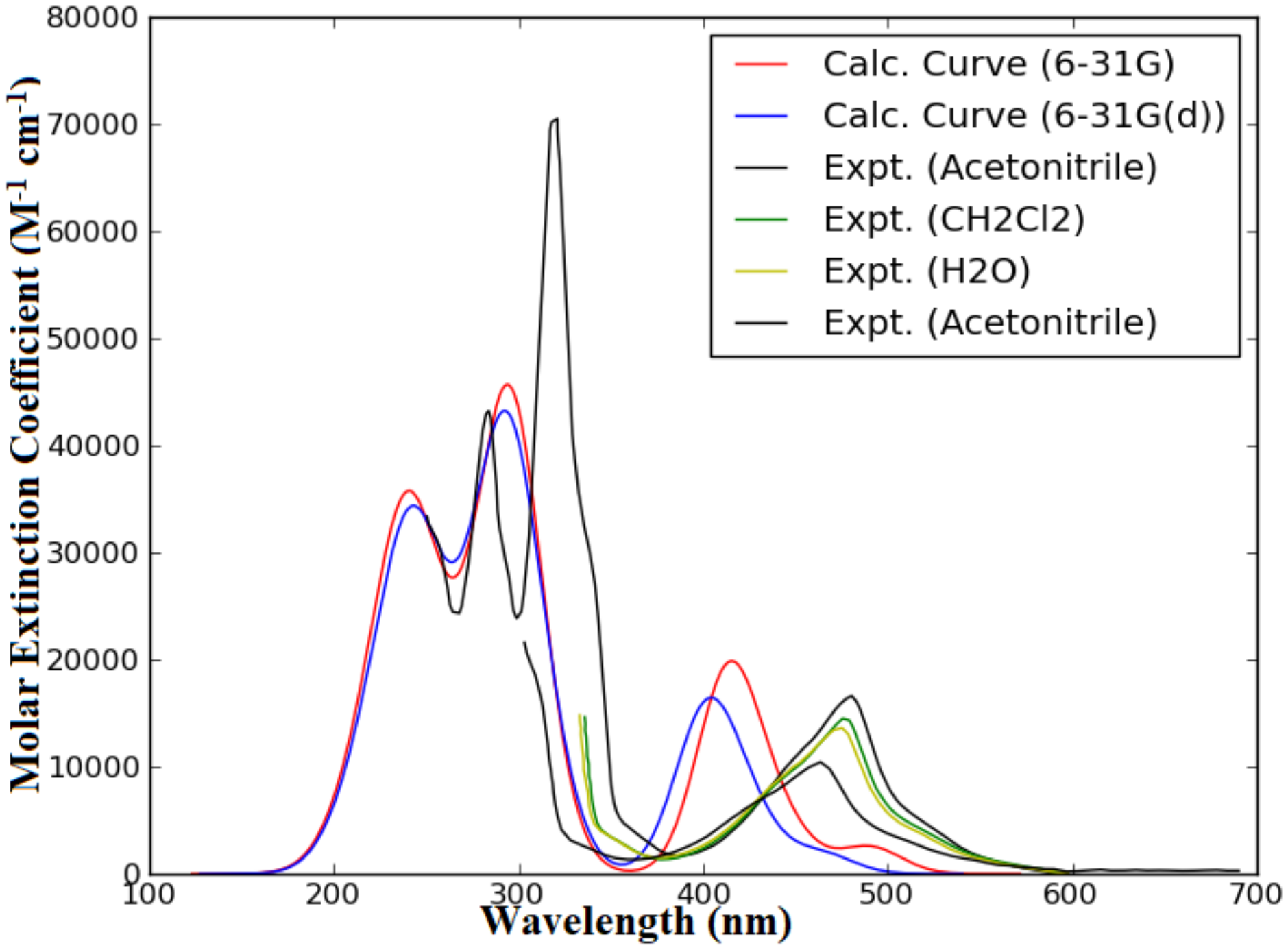}
\end{center}
[Ru(trpy)$_2$]$^{2+}$
TD-B3LYP/6-31G, TD-B3LYP/6-31G(d), and experimental spectra.
Experimental spectra measured at 294K in acetronitrile\cite{HVD12,SCC+94} and
at 298K in water (H$_2$O) and dichloromethane (CH$_2$Cl$_2$)\cite{JCD+09}.

% ================================================
\newpage
\section{\, Complex {\bf (109)}: [Ru(tro)$_2$]$^{2+}$}
% ================================================

\begin{center}
   {\bf PDOS}
\end{center}

\begin{center}
\begin{tabular}{cc}
\includegraphics[width=0.4\textwidth]{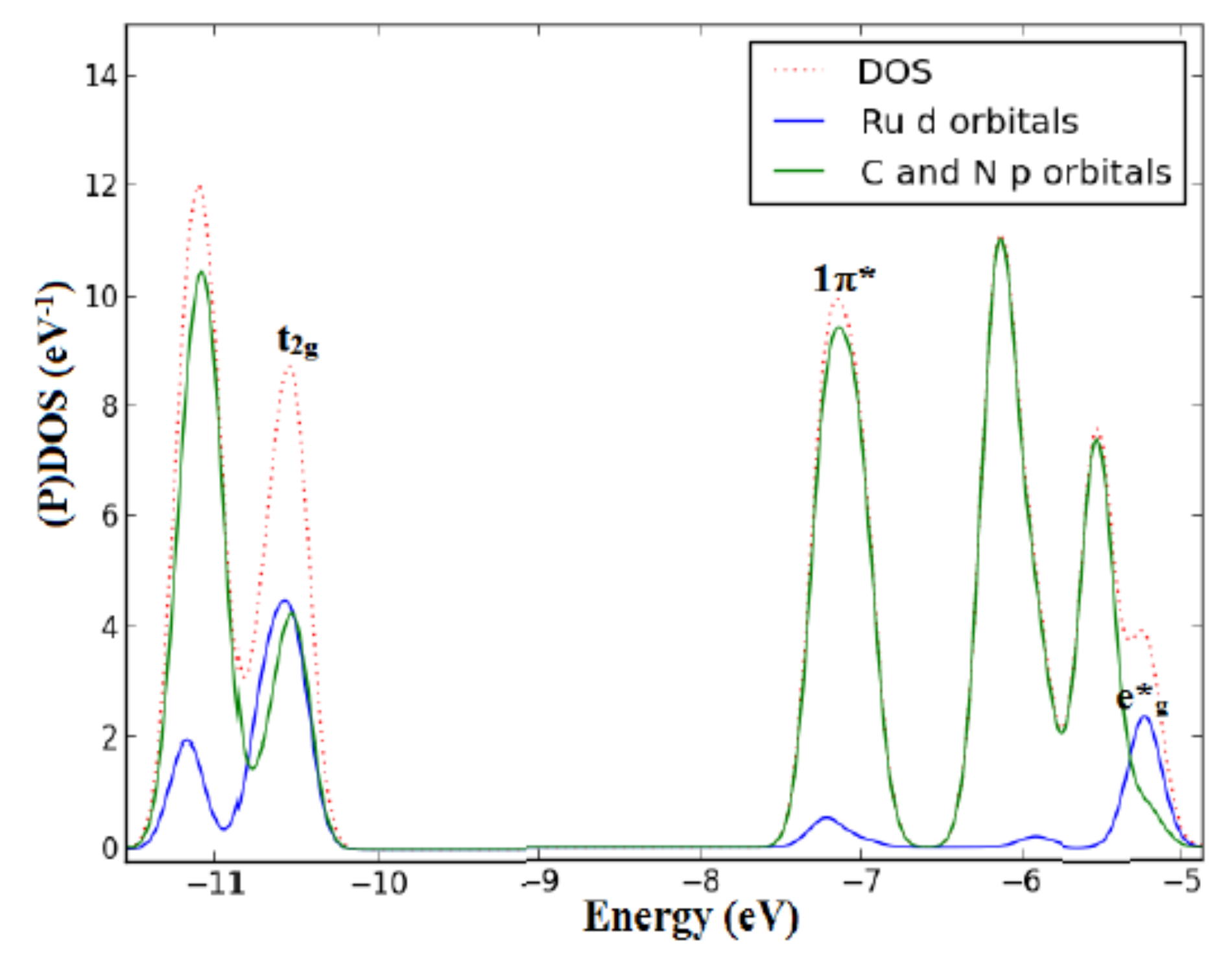} &
\includegraphics[width=0.4\textwidth]{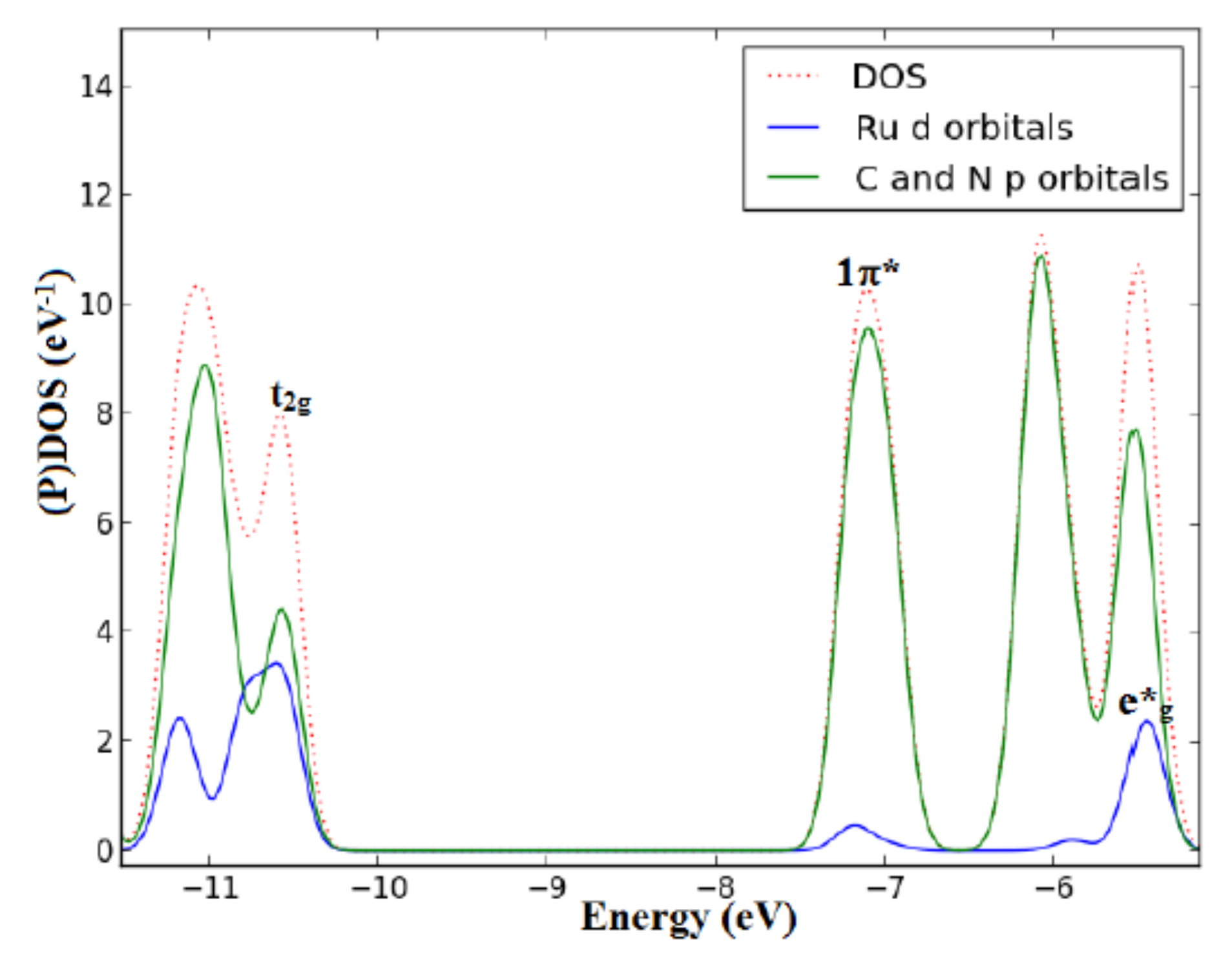} \\
B3LYP/6-31G & B3LYP/6-31G(d) \\
$\epsilon_{\text{HOMO}} = \mbox{-10.53 eV}$ & 
$\epsilon_{\text{HOMO}} = \mbox{-10.57 eV}$ 
\end{tabular}
\end{center}
Total and partial density of states of [Ru(tro)$_2$]$^{2+}$
partitioned 
over Ru d orbitals and ligand C and N p orbitals.
% for the 6-31G (left-hand side) and 6-31G(d) (right-hand side) basis sets.

\begin{center}
   {\bf Absorption Spectrum}
\end{center}

\begin{center}
\includegraphics[width=0.8\textwidth]{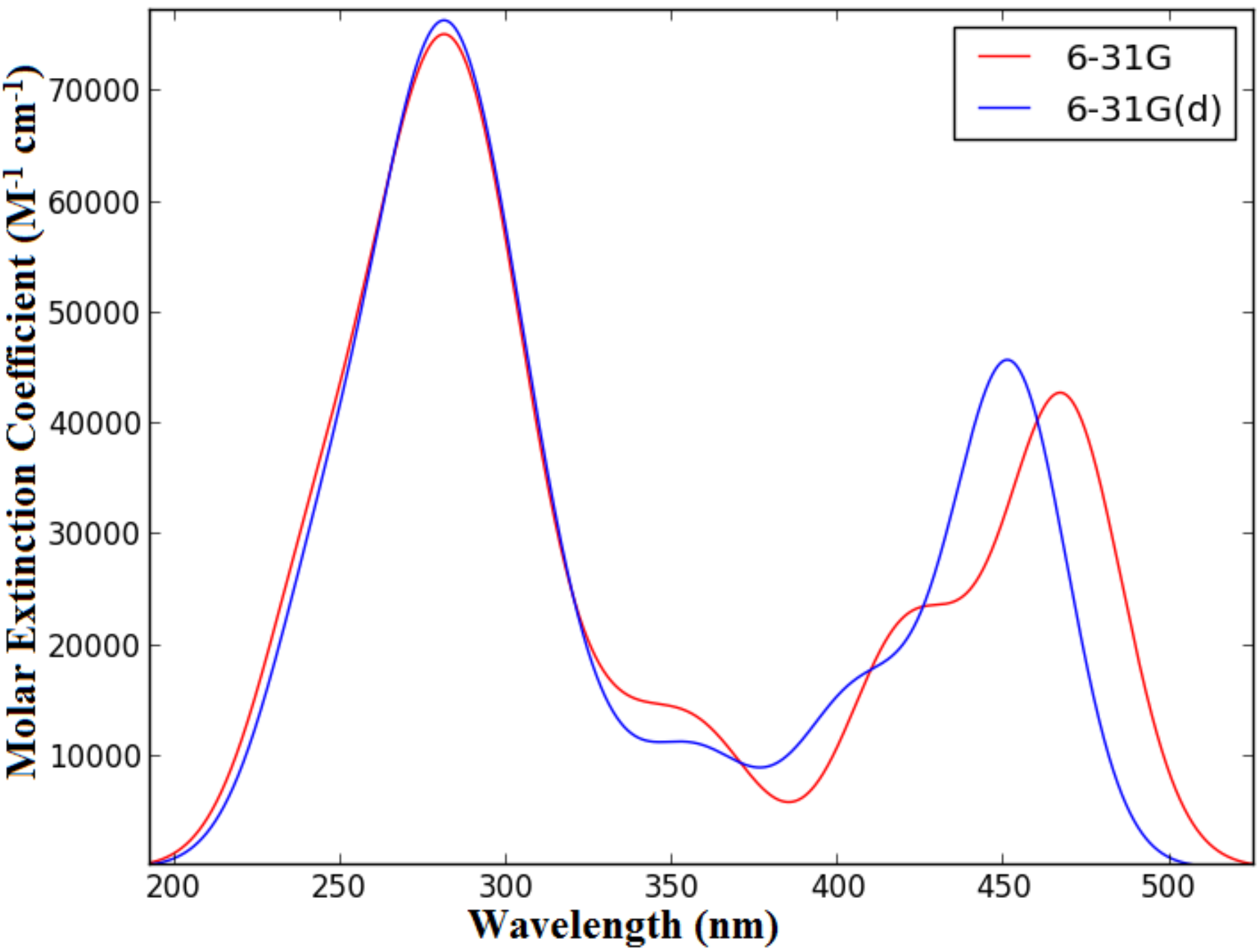}
\end{center}
[Ru(tro)$_2$]$^{2+}$
TD-B3LYP/6-31G and TD-B3LYP/6-31G(d) spectra.

% ================================================
\newpage
\section{\, Complex {\bf (110)}: [Ru(tsite)$_2$]$^{2+}$}
% ================================================

\begin{center}
   {\bf PDOS}
\end{center}

\begin{center}
\includegraphics[width=0.4\textwidth]{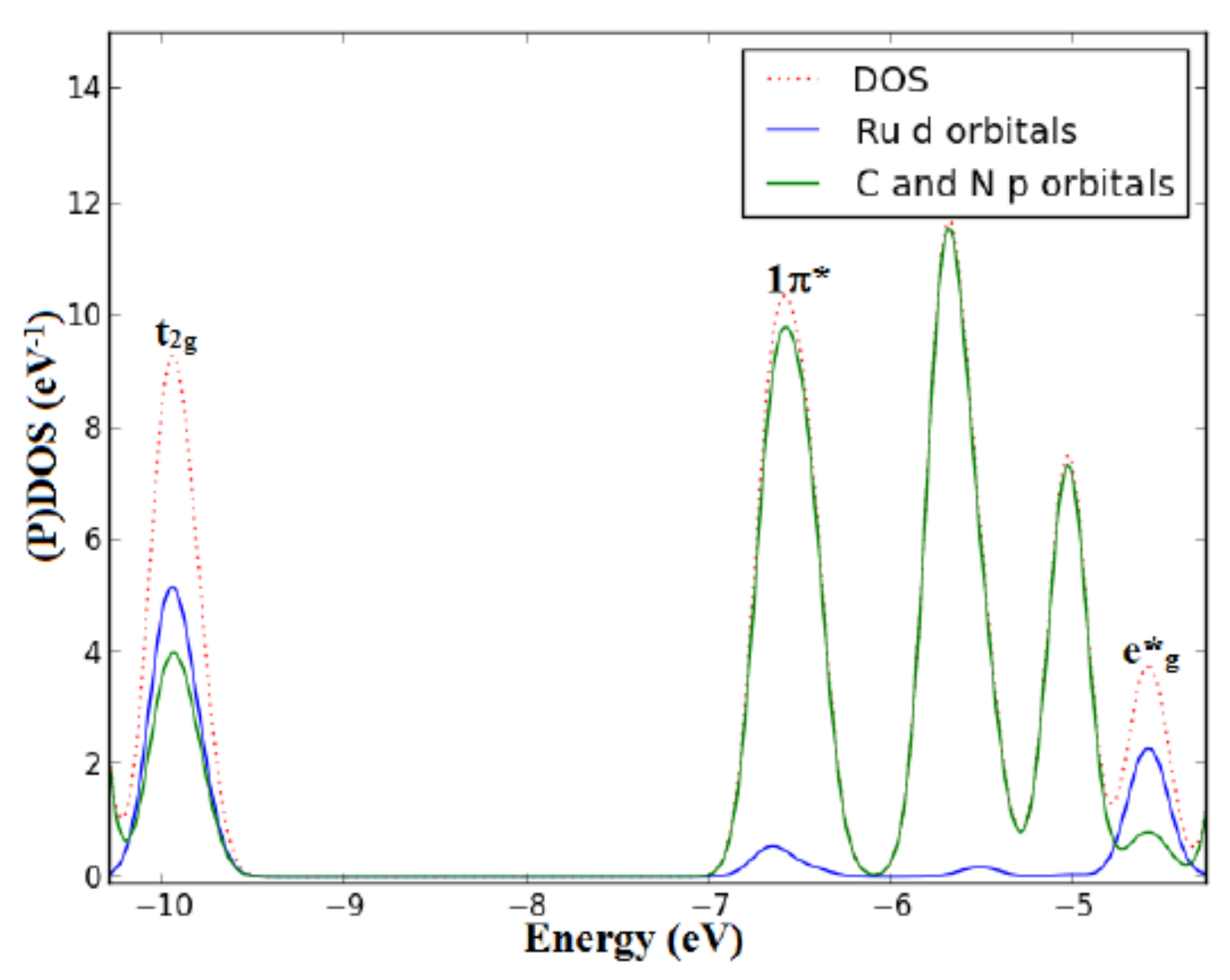}
\\ 6-31G \\ $\epsilon_{\text{HOMO}} = \mbox{-9.84 eV}$
\end{center}
Total and partial density of states of [Ru(tsite)$_2$]$^{2+}$
partitioned 
over Ru d orbitals and ligand C and N p orbitals.
%  for the 6-31G (left-hand side) and 6-31G(d) (right-hand side) 
% {\color{red} \sf Do we have this?}) basis sets.

\begin{center}
   {\bf Absorption Spectrum}
\end{center}

\begin{center}
\includegraphics[width=0.8\textwidth]{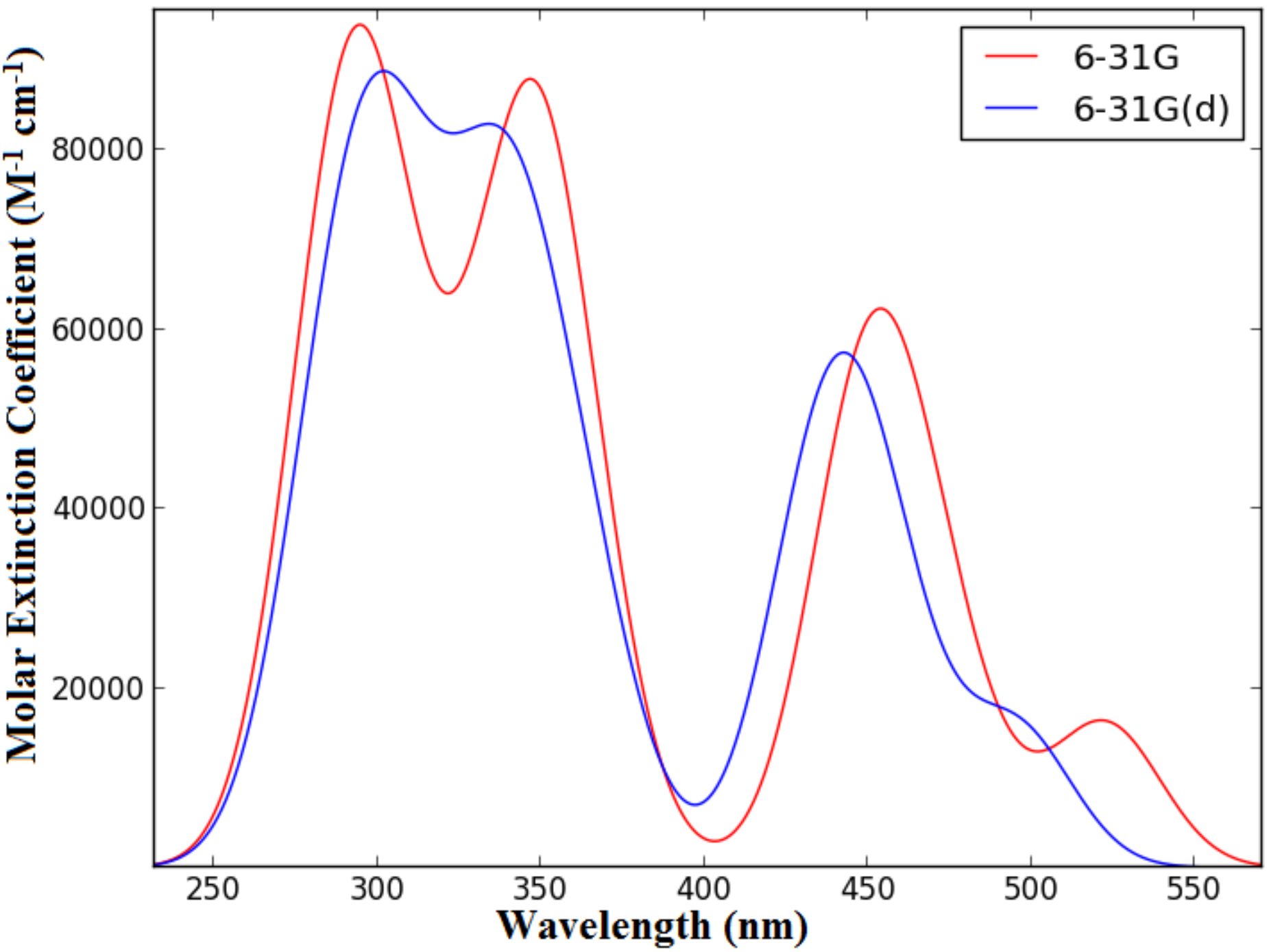}
\end{center}
[Ru(tsite)$_2$]$^{2+}$
TD-B3LYP/6-31G and TD-B3LYP/6-31G(d) spectra.

% ================================================
\newpage
\section{\, Complex {\bf (111)}*: [Ru(dqp)$_2$]$^{2+}$}
% ================================================

\begin{center}
   {\bf PDOS}
\end{center}

\begin{center}
\begin{tabular}{cc}
\includegraphics[width=0.4\textwidth]{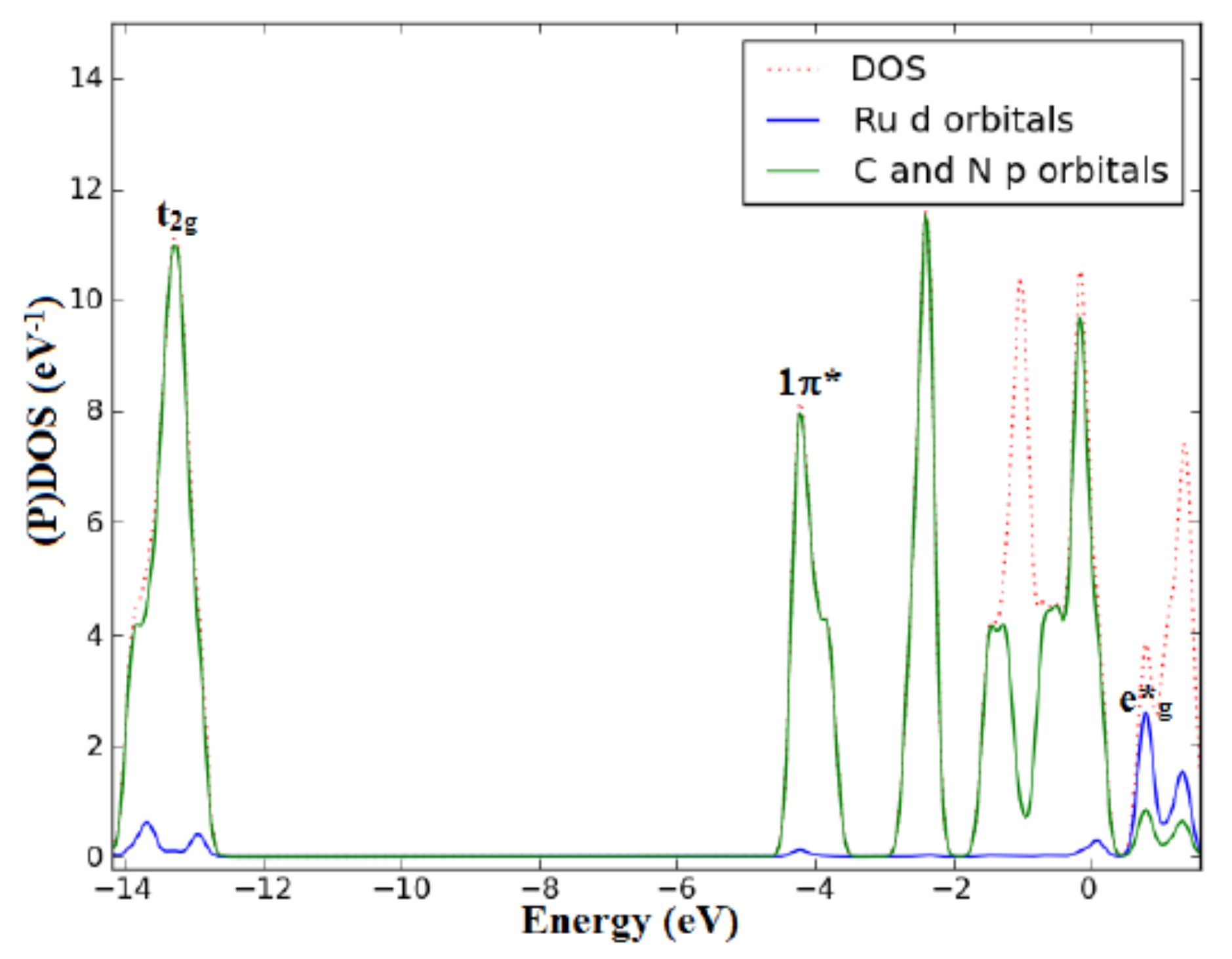} &
\includegraphics[width=0.4\textwidth]{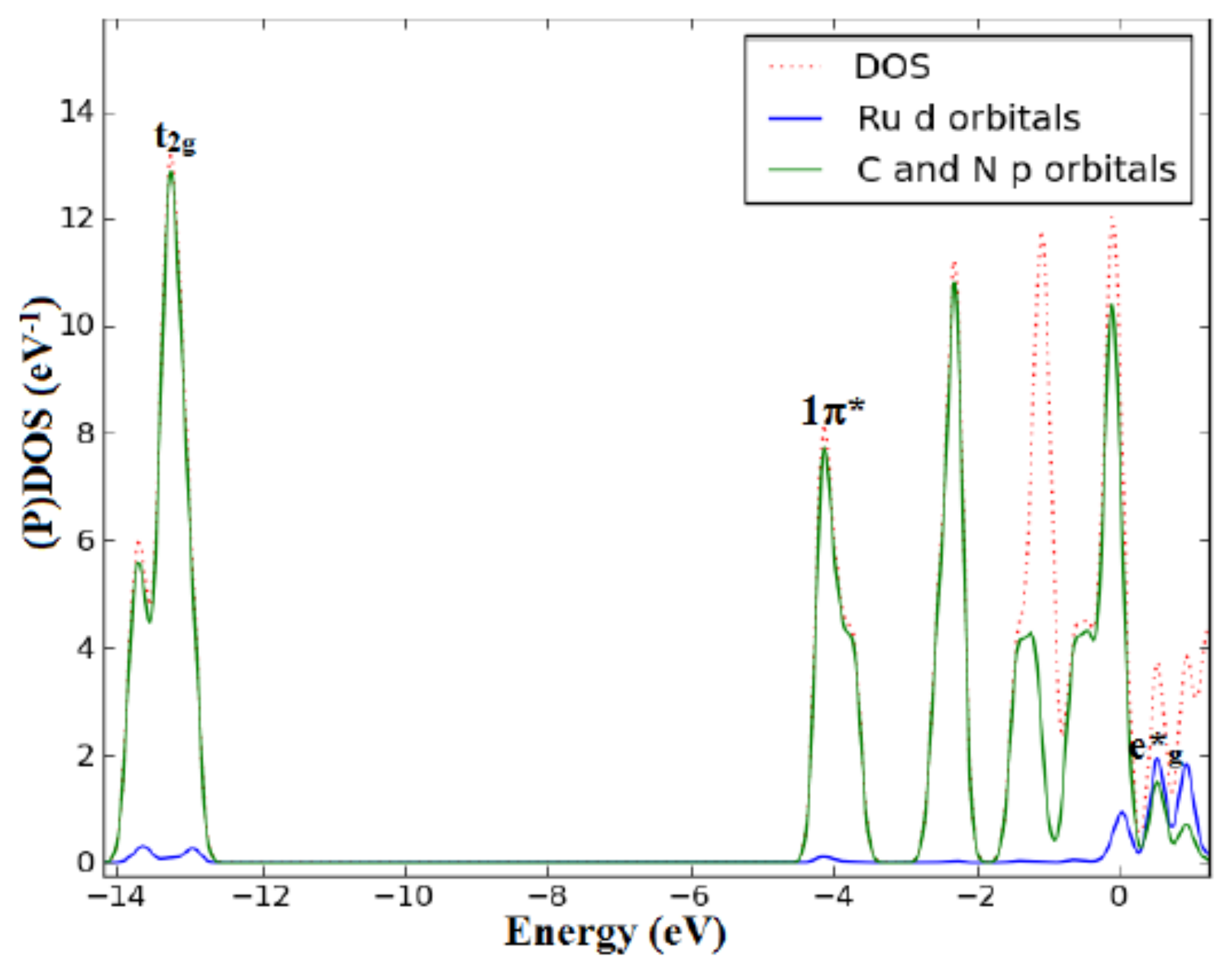} \\
B3LYP/6-31G & B3LYP/6-31G(d) \\
$\epsilon_{\text{HOMO}} = \mbox{-4.25 eV}$ & 
$\epsilon_{\text{HOMO}} = \mbox{-12.94 eV}$ 
\end{tabular}
\end{center}
\begin{center}
\end{center}
Total and partial density of states of [Ru(dqp)$_2$]$^{2+}$
partitioned 
over Ru d orbitals and ligand C and N p orbitals.
% for the 6-31G (left-hand side) and 6-31G(d) (right-hand side) basis sets.

\begin{center}
   {\bf Absorption Spectrum}
\end{center}

\begin{center}
\includegraphics[width=0.8\textwidth]{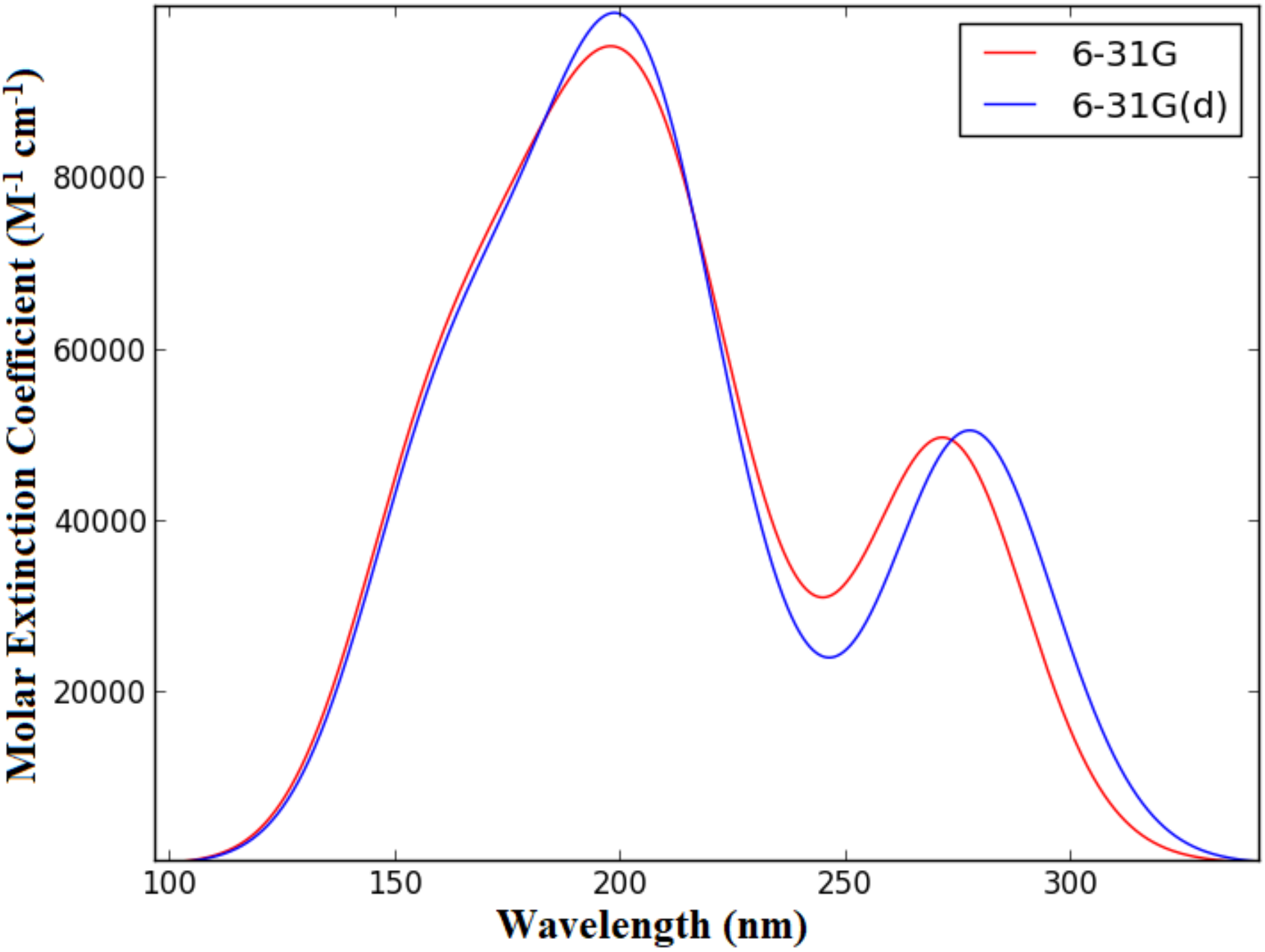}
\end{center}
[Ru(dqp)$_2$]$^{2+}$
TD-B3LYP/6-31G and TD-B3LYP/6-31G(d) spectra.

% ==================================================
\section*{REFERENCES}
% \bibliographystyleb{myaip}
% \bibliographyb{refs}
\bibliographystyle{myaip}
\bibliography{refs}

% -----------------------------------------------